\documentclass[useAMS,usenatbib]{mn2e}

\usepackage{graphicx}
\usepackage[small]{caption}
\usepackage{epstopdf}
\usepackage{float}
\usepackage{multirow}
\usepackage{amssymb}
\usepackage{amsmath}
\usepackage{bm}
\usepackage{subfig}
\usepackage{booktabs}
\usepackage{xspace}
\usepackage{pdflscape}
\usepackage{enumitem}
\usepackage{longtable}
\usepackage{threeparttable}
\usepackage{threeparttablex}
\newcommand{\vunit}{\mbox{\,km\,s$^{-1}$}\xspace}
\newcommand{\fekab}{Fe\,K$\alpha$[K$\beta$]\xspace}
\newcommand{\brga}{broad(GA)\xspace}
\newcommand{\brdl}{broad(DL)\xspace}
\newcommand{\feka}{Fe\,K$\alpha$\xspace}
\newcommand{\fekb}{Fe\,K$\beta$\xspace}
\newcommand{\hea}{He$\alpha$\xspace}
\newcommand{\ka}{K$\alpha$\xspace}
\newcommand{\lya}{Ly$\alpha$\xspace}
\newcommand{\fexxv}{Fe\,\textsc{xxv}\xspace}
\newcommand{\fexxvabs}{He$\alpha$\xspace}
\newcommand{\fexxvi}{Fe\,\textsc{xxvi}\xspace}
\newcommand{\fexxviabs}{Ly$\alpha$\xspace}
\newcommand{\fexxvxxvi}{Fe\,\textsc{xxv--xxvi}\xspace}
\newcommand{\blendabs}{He$\alpha$--Ly$\alpha$\xspace}
\newcommand{\fix}{$^{f}$}
\newcommand{\na}{--}
\newcommand{\vout}{v_{\rm out}\xspace}
\newcommand{\suzaku}{\emph{Suzaku}\xspace}
\newcommand{\xmm}{\emph{XMM-Newton}\xspace}

\newcommand{\dchidof}{\Delta\chi^{2}/\Delta\nu}
\newcommand{\xstar}{{\tt xstar}\xspace}

\newcommand{\chandra}{\emph{Chandra}\xspace}
\newcommand{\chandrahetg}{\emph{Chandra}/HETG\xspace}
\newcommand{\chandraletg}{\emph{Chandra}/LETG\xspace}

\newcommand{\reflionx}{{\tt reflionx}\xspace}
\newcommand{\pexrav}{{\tt pexrav}\xspace}
\newcommand{\pexriv}{{\tt pexriv}\xspace}
\newcommand{\mytorus}{{\tt MYTorus}\xspace}
\newcommand{\xillver}{{\tt xillver}\xspace}
\newcommand{\xspec}{{\sc xspec}\xspace}
\newcommand{\bbody}{{\tt bbody}\xspace}

\newcommand{\zxipcf}{{\tt zxipcf}\xspace}

\newcommand{\nh}{N_{\rm H}}
\newcommand{\logxi}{\log (\xi/\rm{erg\,cm\,s}^{-1})}
\newcommand{\lognh}{\log (N_{\rm H}/\rm{cm}^{-2})}
\newcommand{\logvout}{\log(\vout/\vunit)}

\bibpunct{[}{]}{,}{a}{,}{;}
\defcitealias{tombesi:2010a}{T10A}
\defcitealias{tombesi:2010b}{T10B}
\defcitealias{tombesi:2011b}{T11}
\defcitealias{tombesi:2012}{T12}
\newcommand{\tombesiA}{\citetalias{tombesi:2010a}\xspace}
\newcommand{\tombesiBLRG}{\citetalias{tombesi:2010b}\xspace}
\newcommand{\tombesiB}{\citetalias{tombesi:2011b}\xspace}

\title[The Suzaku view of highly-ionised outflows in AGN.]{The \suzaku view of highly-ionised outflows in AGN: I -- Statistical detection and global absorber properties}

\author[Jason Gofford et al.]{Jason~Gofford$^{1}$\thanks{E-mail: j.gofford@keele.ac.uk}, James~N.~Reeves$^{1,5}$, Francesco~Tombesi$^{2,3}$, Valentina~Braito$^{4,8}$, \newauthor T. Jane Turner$^{5}$, Lance~Miller$^{6}$, Massimo~Cappi$^{7}$\\
\\
$^{1}$ Astrophysics Group, Keele University, Keele, ST5 5BG, UK \\
$^{2}$ X-ray Astrophysics Laboratory and CRESST, NASA/GSFC, Greenbelt, MD 20771, USA \\
$^{3}$ Department of Astronomy, University of Maryland, College Park, MD 20742, USA \\
$^{4}$ INAF-Osservatorio Astronomico di Brera, Via Bianchi 46, I-23807 Merate, Italy \\
$^{5}$ Department of Physics, University of Maryland Baltimore County, Baltimore, MD 21250, USA \\
$^{6}$ Department of Physics, Oxford University, Denys Wilkinson Building, Keble Road, Oxford, OX1 3RH, UK \\
$^{7}$ INAF-IASF Bologna, Via Gobetti 101, I-40129, Bologna, Italy \\
$^{8}$ Department of Physics and Astronomy, University of Leicester, University Road, Leicester, LE1 7RH, UK
}

\begin{document}

\date{Accepted 22 November 2012. Received 22 November 2012; in original form 14 October 2012}

\pagerange{\pageref{firstpage}--\pageref{}} \pubyear{?}

\maketitle

\label{firstpage}

\begin{abstract}
We present the results of a new spectroscopic study of Fe\,K-band absorption in Active Galactic Nuclei (AGN). Using data obtained from the \suzaku public archive we have performed a statistically driven blind search for \fexxv~\hea and/or \fexxvi~\lya absorption lines in a large sample of 51 type $1.0-1.9$ AGN. Through extensive Monte Carlo simulations we find that statistically significant absorption is detected at $E\gtrsim6.7$\,keV in 20/51 sources at the $P_{\rm MC}\geq95\%$ level, which corresponds to $\sim40\%$ of the total sample. In all cases, individual absorption lines are detected independently and simultaneously amongst the two (or three) available XIS detectors which confirms the robustness of the line detections. The most frequently observed outflow phenomenology consists of two discrete absorption troughs corresponding to \fexxv~\hea and \fexxvi~\lya at a common velocity shift. From \xstar fitting the mean column density and ionisation parameter for the Fe\,K absorption components are $\lognh\approx23$ and $\logxi\approx4.5$, respectively. Measured outflow velocities span a continuous range from $<1,500$\,km\,s$^{-1}$ up to $\sim100,000$\,km\,s$^{-1}$, with mean and median values of $\sim0.1$\,c and $\sim0.056$\,c, respectively. The results of this work are consistent with those recently obtained using \xmm and independently provides strong evidence for the existence of very highly-ionised circumnuclear material in a significant fraction of both radio-quiet and radio-loud AGN in the local universe. 
\end{abstract}



\section{Introduction}
\label{sec:intro}
Observational evidence for outflows and winds in Active Galactic Nuclei (AGN) is seen in multiple energy regimes; ranging from the prominent radio-jets seen in radio-loud sources, to the broad absorption lines observed in Broad Absorption Line (BAL) quasars, through to the photoionised `warm absorber' which is frequently observed in the soft X-rays (e.g., \citealt{blustin:2005, mckernan:2007, crenshaw&kraemer&george:2003}). Gravitational micro-lensing studies have shown that the primary X-ray emission region in AGN is on the order of a few tens of gravitational radii ($R_{\rm g}=GM/c^{2}$) in size (\citealt{morgan:2008, chartas:2009, dai:2010}) and so the spectral analysis of absorption features imprinted on the X-ray continuum by circumnuclear material in an AGN is a powerful diagnostic of the physical conditions of the environment in the vicinity of the central Super Massive Black Hole (SMBH), of the dynamics and kinematics of the outflowing material, its chemical composition and its ionisation state. Understanding how such winds are formed, their physical characteristics and their energetic output is of vital importance when it comes to studying how the interplay between the accretion- and ejection-flows at small radii can affect the host galaxy on larger scales. (e.g., \citealt{ferrarese:2000, king:2003, sazonov:2005, ciotti:2009})

In the X-ray regime the presence of photoionised material in AGN is well established, with at least 50\% of objects showing direct spectroscopic evidence for discrete absorption lines and photoelectric edges in their soft-band ($E<3$\,keV) spectra (e.g., \citealt{reynolds:1997, crenshaw&kraemer&george:2003, blustin:2005}), with typical line of sight velocities ranging from a few hundred to around a thousand km\,s$^{-1}$ (\citealt{blustin:2005, mckernan:2007}). Detailed modelling at high spectral resolution with photoionisation codes such as \xstar (\citealt{kallman:2004}) often finds the soft-band absorber to be described by column densities and ionisation parameters\footnote{The ionisation parameter is defined as $\xi=L_{\rm ion}/nR^{2}$ (\citealt{tarter:1969}), where $L_{\rm ion}$ is the $1-1000$ Rydberg ionising luminosity, $n$ is the electron density and $R$ is the distance of the ionising source from the absorbing clouds.} in the range of $\lognh\sim20-23$ and $\logxi\sim0-3$, respectively. These parameters imply that the warm absorbers are typically located on fairly large distances from the central black hole, and perhaps associated with a wind originating from the putative parsec scale torus (\citealt{blustin:2005}) or with the latter stages of an accretion disc wind which has propagated out to larger radii (\citealt{proga&kallman:2004}). By virtue of their low outflow velocities the soft X-ray warm absorber is thought to only have a weak feedback effect in their host galaxy. Indeed, the mechanical power imparted by individual warm absorption components is very low, typically $\lesssim0.01\%$ of an AGN's bolometric luminosity ($L_{\rm bol}$) (e.g., \citealt{blustin:2005}), which is significantly lower than the $\sim0.5\%$ of $L_{\rm bol}$ thought necessary for feedback to affect the host galaxy (\citealt{hopkins&elvis:2010}). However, \cite{crenshaw&kraemer:2012} have recently shown that this $\sim0.5\%$ threshold can be exceeded provided the mechanical power is integrated over all UV and X-ray absorption components, at least in the case of a few moderate-luminosity local AGN.

More recently the higher throughput and larger effective area offered by \xmm and \suzaku at higher X-ray energies ($E\sim5-10$\,keV) has shown that absorption, specifically in the form of very highly-ionised resonant absorption lines associated with the K-shell (1s$\to$2p) transitions of \fexxv and \fexxvi, is also manifested in the hard X-ray spectrum of a significant fraction of local AGN. While evidence for such absorption lines was initially confined to detailed studies of individual sources (e.g., \citealt{reeves:2003, pounds:2003b, risaliti:2005, turner:2008, cappi:2009, giustini:2011}) the recent systematic archival \xmm study conducted by \cite{tombesi:2010a, tombesi:2011a, tombesi:2012} has shown that Fe\,{\sc xxv-xxvi} absorption lines are present in the X-ray spectra of $\gtrsim40\%$ of radio-quiet AGN in the local universe ($z<0.1$). Moreover, such outflows have also been detected in a small sample of local Broad Line Radio Galaxies (BLRGs; \citealt[hereafter `T10B']{tombesi:2010b}) which thus suggests that they may represent an important addition to the commonly held AGN unification model (e.g., \citealt{antonucci:1993, urry&padovani:1995}. 

In comparison to the soft-band absorbers these hard X-ray absorbers generally have much more extreme parameters, with $\lognh\approx23-24$ and $\logxi\approx3-6$, and their outflow velocities relative to the host galaxy can reach mildly relativistic values. While alternative explanations have been posited in the literature, e.g., resonant absorption by highly-ionised material in an co-rotating and optically-thin plasma above the accretion disc (\citealt{gallo:2011}), the large inferred velocities -- combined with the short time-scale variability sometimes exhibited by the absorption features -- point to an origin more likely associated with a wind which is launched from the surface of the accretion disc itself (e.g., \citealt{pounds:2003b, reeves:2009, gofford:2011, tombesi:2012}). In this scenario the inferred mass outflow rates for disc-winds are often comparable to those of the matter which accretes onto the central black hole and the consequent mechanical power can also be a sizeable fraction (i.e., $\geq$ few\%) of an AGN's bolometric luminosity (e.g., \citealt{chartas:2002, pounds:2003b, gibson:2005, reeves:2009, gofford:2011, tombesi:2012}) making them a promising means of linking the small- and large-scale processes at play in galactic evolution over cosmic time.

\subsection{Why \suzaku?}
\label{subsec:why_suzaku}
In this work we use archival \suzaku observations of a large sample of AGN to characterise the properties of highly-ionised outflows in the Fe\,K band. To date, the most comprehensive study of these outflows has been conducted by \citet[hereafter~\tombesiA]{tombesi:2010a} who performed a systematic narrow-band (i.e., $3.5-10.5$\,keV) analysis of 42 sources (with 101 observations) obtained from the \xmm archive; using a simple baseline model to describe the AGN continuum in the Fe\,K band. The baseline model consisted of a power-law, narrow Gaussians and, where required, neutral absorption to approximate for any spectral curvature. This phenomenological approach yielded a fit to the $4-10$\,keV energy band of most sources which was sufficient to enable the systematic search of \fexxv~\hea and \fexxvi~\lya absorption lines without needing to take additional spectral complexity into account. Moreover, \tombesiA found that this approach  resulted in continuum parameters which were largely consistent with those found by authors who conducted a more thorough fit using the entire \xmm bandpass. However, while this approach is suitable for most sources it is important to note that in those which have more complex X-ray spectra, e.g., those which are very heavily absorbed or those with strong hard excesses, using a narrow-band fit can yield a model which is a poor representation of the data when extrapolated to consider the higher and lower energies. The only way to overcome this limitation is by performing a detailed broadband analysis. \suzaku is currently the only X-ray observatory which offers a sufficiently broad bandpass (i.e., $0.6-50$\,keV) to allow for the effects of soft band absorption, the soft-excess {\it and} the Compton reflection component to be constrained simultaneously. This makes it the ideal instrument to confidently constrain the underlying continuum and to robustly assess for the presence of highly-ionised outflows in AGN.

\section{Sample Selection}
\label{sec:sample_selection}
The sample was selected from the Data Archive and Transmission System\footnote{http://darts.jaxa.jp/astro/suzaku/} (DARTS; \citealt{tamura:2004}) which contains all publicly available \suzaku observations categorised by the classification of the target source. An initial sample of AGN was drawn by positionally cross-correlating the targeted pointing coordinates of all publicly available (as of the end of December 2011) \suzaku observations of extragalactic compact sources against the known positions of sources contained in the VERONCAT catalogue of Quasars \& AGN \citep{veron-cetty:2010}. The VERONCAT has an extensive list of local AGN, all of which have been quantitatively classified based upon their optical properties using the criteria introduced by \cite{winkler:1992}. Observations of non-AGN which were matched by virtue of their having a similar position to a known AGN on the sky, such as those of extragalactic Ultra-luminous X-ray Sources (ULXs; \citealt{feng&soria:2011}) or X-ray bright supernovae, were systematically excluded. Only those observations with exposures long enough to ensure that the net source counts between $2-10$\,keV in the source rest-frame exceeded $\sim$15,000 were retained (typical exposures were $\gtrsim50$\,ks) such that a narrow (i.e., $EW=30$\,eV) absorption features was detectable at 95\% from Monte Carlo simulations at source rest-frame energies of up to $8-9$\,keV (see Section~\ref{subsec:Monte_carlo_simulations}). In the case of the high red-shift quasar APM\,08279+5255, which shows evidence for absorption lines at rest-frame energies of $E>10$\,keV (e.g., \citealt{saez:2009, chartas:2009, saez:2011}) the measure of net source counts was instead taken for the entire XIS bandpass (i.e., $0.6-10$\,keV in the observer frame) because the Fe\,K features are shifted to lower energies due to the high red-shift of the AGN.

As highly-ionised outflows are thought to originate at relatively small distances from the central nucleus (e.g., \citealt{gofford:2011, tombesi:2011a, tombesi:2012}, hereafter `T11' and `T12', respectively) it is important that the primary continuum emission from the central nucleus, rather than that which is reprocessed/scattered by circumnuclear material out of line of sight, is directly observed so that such outflows can be detected. To this end we exclude all Type-2 sources to make sure that \textit{all} sources were optically-thin to X-rays below 10\,keV.
\begin{figure}
\begin{center}
\includegraphics[width=8cm]{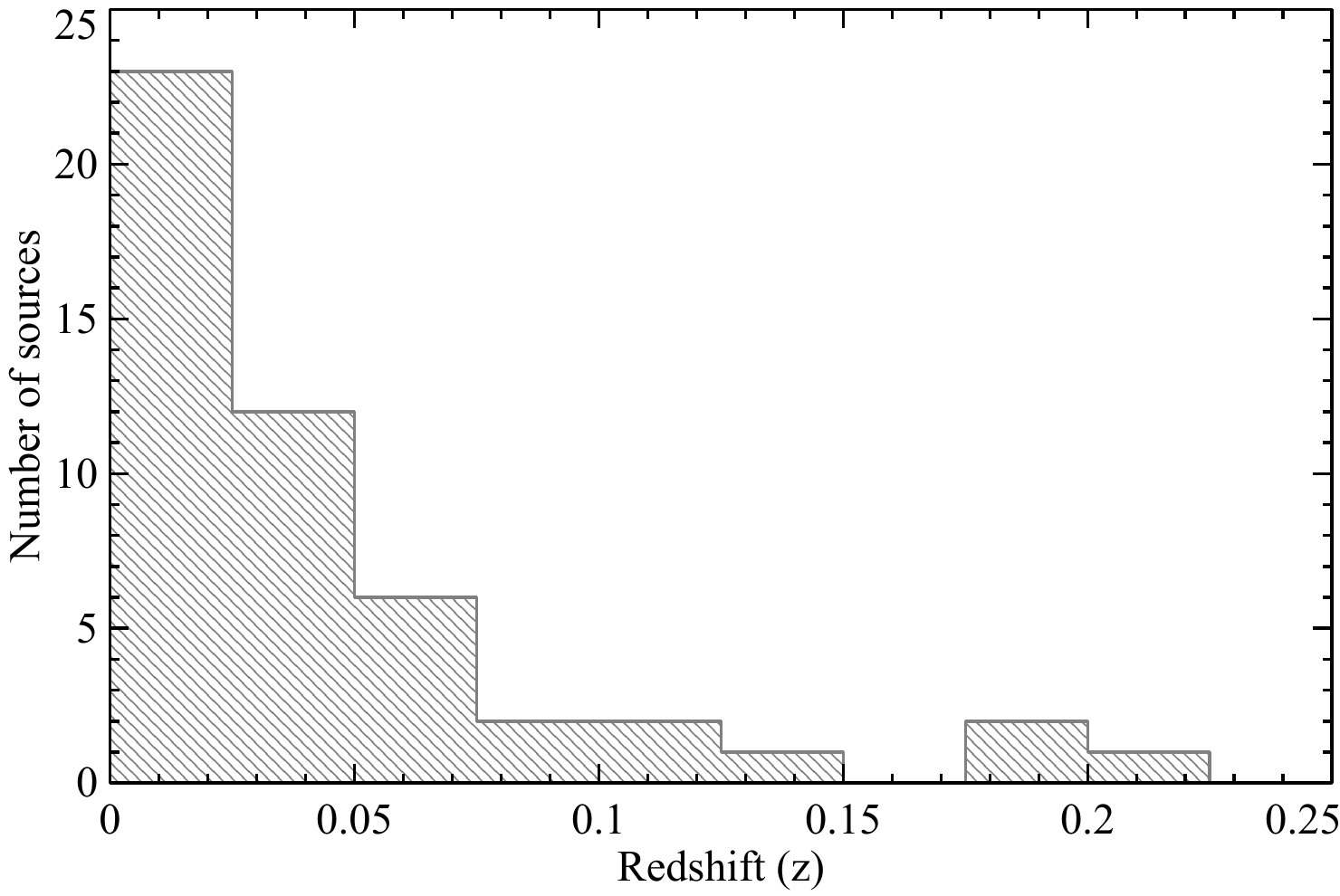}
\caption{\small Histogram showing the distribution of sources included in this work. The high red-shift QSO APM\,08279+5255 ($z=3.91$) has been omitted for scaling purposes}
\label{fig:red-shifts}
\end{center}
\end{figure}
Therefore only those sources with a Type $1.0-1.9$ orientation to the line of sight, as per the classifications listed on the {\it NASA/IPAC Extragalactic Database} (hereafter NED), in the VERONCAT catalogue itself and through literary sources were included. Note that we conservatively include the radio-quiet AGN ESO\,103-G035 ($z=0.01329$) which, despite being classified as Type-$2.0$ in the VERONCAT and on NED, is often regarded as a Type-1.9 Seyfert in the literature (e.g., \citealt{warwick:1988}) by virtue of the presence of a moderately broad H$\alpha$ line in its optical spectrum \citep{phillips:1979}. 
\begin{figure}
\begin{center}
\includegraphics[width=8cm]{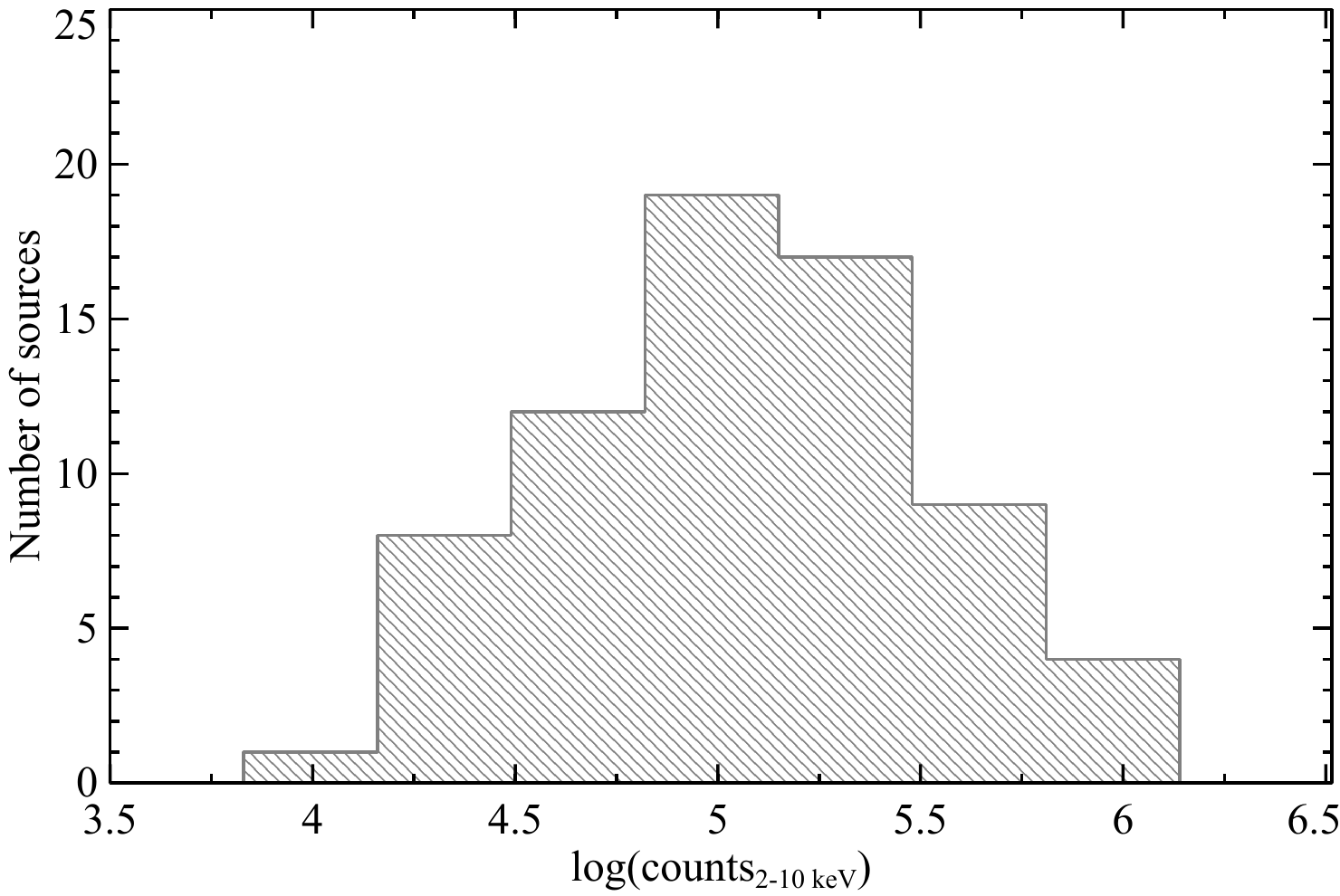}
\caption{\small Histogram showing the logarithm of the total front-illuminated XIS $2-10$\,keV counts in all fitted spectra. In stacked spectra (see table \ref{tab:observation_details}) the total co-added counts are considered here, rather than the counts in each individual sequence.}
\label{fig:logcounts}
\end{center}
\end{figure}
Observation details for all sources included in the heterogeneously selected sample are listed in Appendix A. There are 99 observations of 51 AGN spanning a wide range of spectral types and radio-properties. As shown in Table~\ref{table:source_classification_number} the sample is dominated by low-moderate inclination Seyfert $1.0-1.5$ galaxies (28/51; 34/51 if Narrow Line Seyfert 1s are included), and contains comparatively few high-inclination (Type $1.8-1.9$) Seyferts (6/51). There are 6 radio-loud sources in the sample, including all 5 of the Broad Line Radio Galaxies (BLRGs) included in the \tombesiBLRG outflow case study, and 5 QSOs. The distributions of source red-shift and total $2-10$\,keV counts (for the XIS-FI detector) are shown in Figure~\ref{fig:red-shifts} and Figure~\ref{fig:logcounts}, respectively. The AGN are predominantly local, with $\sim90\%$ of the sample being located at a red-shift of $z\lesssim0.1$, but also includes more distant objects such as PDS\,456 ($z=0.1840$), 1H\,0419-577 ($z=0.1040$), PKS\,0558-504 ($z=0.1372$) and PBC\,J0839.7-1214 ($z=0.198$). The gravitationally lensed quasar APM\,08279+5545 is by far the most distant object in the sample with a red-shift of $z=3.91$. As shown in Figure~\ref{fig:logcounts} the total $2-10$\,keV counts for the fitted spectra span two orders of magnitude but are approximately normally distributed about the peak and mean number of $10^{5}$ counts.
\begin{table}
\begin{center}
\begin{minipage}{4cm}
\caption{Source classifications}
\begin{tabular}{l c}
	\toprule
	Source classification & Number \\
	\midrule
	Sy $1.0-1.2$ & 17 \\
	Sy $1.5$ & 11 \\
	Sy $1.8-1.9$ & 6 \\
	NLSy1 & 6 \\
	BLRG & 6 \\
	QSO & 5 \\
	\midrule
	{\bf Total} & {\bf 51}\\
	\bottomrule
\end{tabular}
\label{table:source_classification_number}
\end{minipage}
\end{center}
\end{table}

\section{Data reduction}
\subsection{XIS Reduction}
There are four X-ray Imaging Spectrometers (XIS) CCD cameras aboard \suzaku which cover the energy range of 0.3--12.0\,keV. The XIS\,0, XIS\,2 and XIS\,3 are front-illuminated, while the XIS\,1 is back-illuminated and provides superior sensitivity below $\sim1$\,keV but a lower effective area and a higher background at harder X-ray energies. The XIS\,2 suffered a charge leak in November 2006\footnote{http://heasarc.gsfc.nasa.gov/docs/suzaku/news/xis2.html} and is therefore only available in observations taken before this date. All spectra were reduced using v6.11 of the {\it HEASoft}\footnote{http://heasarc.nasa.gov/lheasoft/} suite following the process outlined in the \suzaku ABC Data Reduction Guide\footnote{ftp://legacy.gsfc.nasa.gov/suzaku/doc/suzaku\_abc\_guide.pdf}. Events were selected using ASCA X-ray grades 0, 2, 3, 4, and 6, adopting the standard screening criteria such that data were excluded if taken: (1) within 436\,s of passage through the South Atlantic Anomaly (SAA), (2) within an Earth elevation angle (ELV) $<5^{\circ}$, and/or (3) with Earth day-time elevation angles $<20^{\circ}$. Hot and flickering pixels were removed from the XIS images using the {\sc cleansis} script. Source spectra were extracted from within circular regions of radius $1.5\arcmin \leq r < 3.0\arcmin$ to ensure good coverage of the source events, while background spectra were typically extracted from offset circles of the same radius with care taken to avoid the chip corners containing the Fe$^{55}$ calibration sources. For intrinsically faint sources the background spectra were extracted from circles larger than those of the source and the ratio of source/background area was accounted for with an appropriate background scaling factor. Redistribution matrices and ancillary response files for each observation were generated using the tasks {\sc xisrmfgen} and {\sc xissimarfgen}, respectively. 

Where possible spectra obtained from the front-illuminated XIS detectors where combined into a single source spectrum (hereafter XIS-FI) using \textsc{mathpha} in order to maximize signal-to-noise ($S/N$) in the Fe\,K band. Most of the observations in the sample have data for at least the XIS\,0 and XIS\,3 with a further 22 also having data for the XIS\,2 (see Table~\ref{tab:observation_details}). For SWIFT\,J2127.4+5654 (OBSID~702122010) only data for the front-illuminated XIS\,3 camera is available because the XIS\,0 was not operational during the observation. Spectra for the back-illuminated XIS\,1 (hereafter XIS-BI) were reduced using the same process as outlined above and were analysed simultaneously as a separate input spectrum. All XIS spectra were grouped to have a minimum of 50 counts per energy bin to enable the use of the $\chi^{2}$ fit statistic. Net XIS exposures and the total background subtracted $2-10$\,keV count rates (in the source rest-frame), for both the co-added XIS-FI and the XIS-BI detectors, are listed in Table~\ref{tab:observation_details}. 

\subsection{HXD/PIN Reduction}
HXD/PIN spectra were also reduced using the method outlined in the \suzaku Data Reduction guide and, again, processed according to the screening criteria described previously. As the HXD/PIN is a collimating rather than an imaging instrument the contribution of both the instrumental Non X-ray Background (NXB) and the Cosmic X-ray Background (CXB) need to be independently accounted for when estimating the total background. The NXB was generated using the appropriate response and tuned event files (background model D) for each observation. Common good time intervals (GTIs) in the event and background files were selected using {\sc mgtime} and spectral files were extracted using {\sc xselect}. PIN detector dead-time was accounted for using the {\sc hxddtcor} task and the NXB background exposure was increased by $\times 10$ to reduce the effects of photon noise. The CXB contribution was simulated using the form of \cite{boldt:1987}, combined with the NXB to form a total background spectrum using {\sc mathpha}, which was subsequently subtracted from the source spectrum within {\sc xspec}. 

All background subtracted PIN spectra were binned to {\it at least} the $3\sigma$ level above background (typically $>5\sigma$) to enable the use of $\chi^{2}$ fit statistic. Hard X-ray faint and/or high background observations where source count rates were $<4\%$ of the PIN total were not considered in our analysis. Only APM\,08279+5255 and PDS\,456, amounting to a total of five observations, meet this criteria (see Table~\ref{tab:observation_details}). As with the XIS data, the final PIN exposures and the total background subtracted rest-frame $15-50$\,keV counts are listed in Table~\ref{tab:observation_details}.


\section{Analysis}
\label{sec:analysis}

\subsection{Spectral Fitting}  
\label{subsec:spectralmodelling}
A detailed broadband spectral analysis of all sources was conducted using version 12.6.0q of the \xspec spectral fitting package. All spectral models are modified by the appropriate column density of Galactic absorption using values taken from the Leiden/Argentine/Bonn Survey of Galactic HI (\citealt{kalberla:2005}) which were obtained from the on-line version of the $N_{H}$ FTOOL\footnote{http://heasarc.gsfc.nasa.gov/cgi-bin/Tools/w3nh/w3nh.pl}. The values of Galactic absorption for each source are listed in Table \ref{tab:observation_details}. XIS-FI data were typically considered between $0.6-10.0$\,keV, while XIS-BI were only included between $0.6-5.0$\,keV due to decreasing S/N in the Fe\,K band which could hamper absorption line detection. All XIS spectra were excluded between $1.6-2.1$\,keV due to uncertainties in calibration associated with the Si\,K detector edge. Where available, PIN data were included to cover at least the $15-40$\,keV energy range. 

A constant multiplicative factor was included in all models to account for the XIS/PIN cross-normalization whose value depends not only on the nominal pointing of the observation, but also the version of the \suzaku pipeline with which the data has been processed. The current XIS/PIN ratios suitable for version~2 processed data are $1.16$($1.18$) for the XIS(HXD) nominal pointing positions\footnote{http://www.astro.isas.jaxa.jp/suzaku/doc/suzakumemo/suzakumemo-2007-11.pdf\label{calibration_memo}}. However, the cross-normalization was up to $\sim5-6$\% lower for data processed with version~1 of the pipeline, corresponding to a XIS/PIN ratio of $1.09(1.13)$\footnotemark[\value{footnote}] in the XIS(PIN) nominal pointing positions. While the difference is only small the additional uncertainty can have a considerable effect on the continuum parameters at $E>10$\,keV, particularly in hard X-ray bright sources with a high PIN count rate where the model can become driven by the hard X-ray band. To account for any systematic effects associated with the instrumental cross-calibration we therefore allow the constant parameter in each model to vary $\pm5\%$ about the values suggested by the \suzaku team to take into account any systematic errors. 

There are 20 AGN in the sample which have two or more \suzaku observations. In APM\,08279+5255, IC\,4329A, MCG\,-6-30-15, Mrk\,841, NGC\,5506, NGC\,5548 and PKS\,0558-504 the different observations are similar in spectral shape which allows them to be co-added using the appropriate relative weighting factors to take into account differences in individual exposures; with the resultant time-averaged spectra for these sources being used in all subsequent analyses. 1H\,0419-577 and NGC\,2992 both had observations which were taken in different \suzaku nominal pointing positions which could influence co-adding. However, effective area was always consistent to within $\pm10\%$ and so the spectra were still co-added using the mean of the response files. Any additional systematic uncertainty introduced was adequately accounted for by the variable XIS/PIN cross-normalisation. In 3C\,120 and Mrk\,509, which both showed notable variability between observations, spectra were jointly fit depending on the extent of their spectral variability between epochs. In 3C\,120 we followed the analysis method of \citet{kataoka:2007} and \citet{tombesi:2010b} by co-adding OBSIDs~700001020, 70001030 and 70001040, which are all of a similar spectral shape and flux level, and jointly fitted them with OBSID~700001010 which has a more prominent underlying soft-excess. In Mrk\,509, which is well known for having a strong soft-excess (\citealt{mehdipour:2011}), OBSID~701093010, the stacked OBSIDs~701093020, 701093030, 701093040, and OBSID~705025010 were fitted simultaneously to account for the observed variability in the soft X-ray band. 

The remaining nine sources with more than one observation (3C\,111, Fairall\,9, Mrk\,766, NGC\,1365, NGC\,3227, NGC\,3516, NGC\,3783, NGC\,4051 and PDS\,456) showed strong spectral variability in both the shape of the spectrum and/or drastic changes in flux state which made co-adding impractical. In these sources the available spectra were fitted simultaneously and a model was constructed to describe the observed spectral variability with as few additional free parameters as possible. In NGC\,1365, OBSIDs~702047010 and 705031010 are dominated by very deep Fe\,K absorption lines. These lines are not present in OBSID~705031020, possibly due to the source dropping into a quasi-scattering-dominated state (see \citealt{maiolino:2010} for details of the variability patterns in this source). For simplicity we therefore only simultaneously fitted OBSID~702047010 and OBSID~705031010 in NGC\,1365, and fit OBSID~705031020 separately.

The $\chi^{2}$ minimization technique is used throughout this work; with all statistical errors quoted to the $90\%$ confidence level ($\Delta\chi^{2}=2.71$ for one parameter of interest). Where the statistical significance of components is quoted in terms of a $\Delta\chi^{2}$ value the component in question has been removed from the model and the data refitted to ensure that the order in which components are added to the model has no influence on the derived statistics. When referring to statistical changes to a fit a positive $\Delta\chi^{2}$ denotes a worsening of the fit, while a negative $\Delta\chi^{2}$ indicates a statistical improvement. Positive outflow velocities correspond to a net blue-shift relative to the systematic of the host galaxy, while a negative velocity indicates a net red-shift.

\subsection{Model Construction}
\label{subsec:model_components_construction}
All spectra were first fitted with a power-law modified solely by Galactic absorption. Additional components were added to the model and retained provided their significance exceeded the $>99\%$ confidence level by the F-test. Any emission lines in the soft X-ray band were fitted with narrow Gaussians ($\sigma\sim5$\,eV). In sources with multiple spectra we initially found a broadband fit to the observation with the highest flux; later observations were then added sequentially and the continuum/absorption parameters were allowed to to vary independently until a simultaneous fit to all spectra had been achieved using as few free parameters as possible. Once a statistically acceptable fit to the broad-band spectrum of each source had been found, we fitted any necessary emission components and then performed a systematic and methodical search for Fe\,K absorption lines between $5-10$\,keV using energy-intensity plane contour-plots (Section~\ref{subsec:searching_for_absorption_lines}) and detailed Monte Carlo simulations (Section~\ref{subsec:Monte_carlo_simulations}). 

The modelling approach for individual spectral components are outlined below, with the continuum parameters for each source being noted in Table~\ref{table:continuum_parameters} and those for the warm absorber being listed in Tables~D2 and D3 for single- and multi-epoch spectra, respectively. A description of the underlying modelling assumptions and associated caveats is presented in Appendix~C.

\subsubsection{Warm absorption I: Fully covering}
\label{subsubsec:warm_absorption}
Depending on the properties of the intervening material (such as its ionisation state and column density) absorption by circumnuclear material can add significant spectral curvature to the observed X-ray spectrum and can therefore have a direct effect on the continuum and line parameters measured in broadband models. In this work we model warm absorption components using a suite of \xstar (v. 2.2.1bc) tables which are all generated assuming input values which are `typical' for local Seyfert galaxies\footnote{Absorption grids are described by: an illuminating photon index of $\Gamma=2.1$, a gas density of $n=10^{10}$\,cm$^{-3}$, a micro-turbulent velocity of $v_{\rm turb}=100$\,km\,s$^{-1}$, and an integrated model luminosity of $L=10^{44}$\,erg\,s$^{-1}$ between $1-1000$\,Rydbergs.} (e.g., \tombesiB). The resultant \xstar tables cover a wide range of parameter space in terms of column densities [$10^{18} < \lognh \leq 10^{24}$], and ionisation parameter [$-3 < \logxi \leq 6$] which makes them well suited for accounting for all manner of warm absorption. 

Fully-covering warm absorption zones are included in models where necessary to achieve a good fit to the soft X-ray band; in some cases more than one absorption zone is needed. In these cases the column density and ionisation parameter of each zone is allowed to vary independently, and represents an absorption geometry which consists of multiple layers of gas. At the energy resolution of the XIS CCDs the bound-bound absorption lines required to constrain the outflow velocities of individual soft X-ray absorption components are unresolved and all absorption zones are therefore fitted as stationary in the source rest-frame (i.e., fixed outflow velocities of $\vout=0$\,km\,s$^{-1}$). Allowing the outflow velocity of the soft X-ray absorber to vary always has a negligible effect on the reported Fe\,K absorption line parameters. 

\subsubsection{Warm absorption II: Partial covering}
\label{subsubsec:partial_covering_absorption}
We also consider the possibility that the sight-line to a source is partially covered. In this absorption geometry a fraction $f_{\rm cov}<1$ of the source flux is absorbed with the remaining $1-f_{\rm cov}$ leaking through the absorption layer. For simplicity we account for partially-covering absorption layers using a customized version of the \zxipcf model\footnote{The standard \zxipcf uses a specific grid with the following parameters: $\Gamma=2.2$, $n=10^{10}$\,cm$^{-3}$, $v_{\rm turb}=200$\,km\,s$^{-1}$, $L=10^{44}$\,erg\,s$^{-1}$} which models the partially-covering absorption by partially-ionised gas (see \citealt{reeves&done:2008}) without needing to use complicated nested power-laws and \xstar tables. As in \zxipcf the free parameters in our customized table model are column density ($N_{\rm H}$), ionisation parameter ($\log\xi$), covering fraction ($f_{\rm cov}$) and red-shift relative to the observer ($z$). The model is based on the same tables as discussed in Section~\ref{subsubsec:warm_absorption} and therefore has a slightly lower turbulent velocity than \zxipcf, at $v_{\rm turb}=100$\,km\,s$^{-1}$, but covers the same parameter space in terms of  column density and ionisation parameter. 

Partially-covering absorption can have a strong effect on the observed continuum with moderate column densities of material ($N_{\rm H}\sim10^{23}$\,cm$^{-2}$) adding considerable spectral curvature at $E<10$\,keV (\citealt{reeves:2004, risaliti:2005, braito:2011,turner:2011}). Bound-free transitions in similar column density material can also fit broad residual emission profiles in the Fe\,K band (\citealt{inoue:2003, miller:2008, miller:2009, tatum:2012a}), and partial-covering by Compton-thick material ($N_{\rm H}\gtrsim10^{24}$\,cm$^{-2}$) can reduce the observed emission below $10$\,keV with the true intrinsic continuum only becoming apparent at higher energies as a `hard excess' of emission relative to that expected from standard reflection models (\citealt{reeves:2009, turner:2009, risaliti:2009}; \citealt{tatum:2012b}). 

We include partial-covering absorbers in our models if and when they are required by the data at the $P_{\rm F}\geq99\%$ confidence level and a satisfactory fit to the data could not be achieved using solely fully-covering absorption; several sources appear to need multiple partially-covering absorption zones which suggests the presence of a clumpy stratified absorber along the line of sight. Again, all absorber parameters are listed in Tables~\ref{tab:warmabs_single} and \ref{tab:warmabs_multi} in Appendix~\ref{appendix:model_parameters}. We note that in some circumstances the need of a high column density partially covering component can be contingent on the means with which the underlying reflection component is modelled leading to some model degeneracies. However, as we discuss in Appendix~\ref{modelling_complexities}, regardless of whether the hard X-ray data is fitted with reflection or not, partial-covering has little measurable effect on the parameters measured for any highly-ionised absorption line systems.

\subsubsection{The Soft-excess}
\label{subsubsec:the_soft_excess}
Relative to the low energy extrapolation of the power-law continuum in the $2-10$\,keV band the X-ray spectrum of AGN often show a smooth increase in emitted flux below $\sim1$\,keV (\citealt{turner:1988, porquet:2004a}).  The thermal temperature of this `soft-excess' suggests it is unlikely to be the direct emission from a standard accretion disc without additional reprocessing (\citealt{sobolewska:2007, done:2012}). Alternative explanations posit that the soft-excess may be due to an increase in optical depth associated with circumnuclear O\,{\sc vii-viii} and Fe\,L-shell transitions at $E\lesssim0.7$\,keV which can enhance either the transmitted or reflected flux along the sight-line through smeared absorption (e.g., \citealt{chevallier:2006, sobolewska:2007, done:2007}) or blurred reflection (e.g., \citealt{crummy:2006, nardini:2011a, brenneman:2011, nardini:2012}) effects. Furthermore, in some sources (e.g., Mrk\,766, \citealt{miller:2007}; MCG\,-6-30-15, \citealt{miller:2008}) the `excess' could simply be a product of complex absorption and just be the manifestation of opacity around $\sim1-2$\,keV.

Regardless of the true physical origin we take a purely phenomenological approach when fitting the soft-excess in this work, and predominantly use the {\tt bbody} model which represents the emission from a constant temperature blackbody. While not necessarily physically motivated, modelling the soft-excess in this manner offers a simple parametrisation of any `excess' soft X-ray emission which is sufficient to get a good handle on the underlying continuum parameters. For completeness we investigate the effects that other ways of modelling the soft-excess can  have on any Fe\,K-band absorption lines in Appendix~\ref{modelling_complexities}. Roughly half of the sources in the sample (24/51; $\sim 47$\%) show evidence for a soft-excess, of which 22/24 are fit with a {\tt bbody} component. In 3C\,120 and Mrk\,509 the soft-excess is very broad and extends beyond that which can be fitted with a simple blackbody. In these sources we instead fit the excess with a second power-law with a softer photon-index.

\subsubsection{Lowly ionised reflection}
\label{subsubsec:compton_reflection}
Cold reflection from large column densities of neutral or lowly ionised material outside of the sight-line can have a strong influence on the observed X-ray spectrum. The strongest observational characteristics of such reflection include the Compton-reflection hump at $\sim30-40$\,keV and the almost ubiquitous Fe\,K$\alpha$ and Fe\,K$\beta$ fluorescence lines at $\sim 6.4$\,keV and $\sim7.06$\,keV (\citealt{nandra:2007, shu:2010}), respectively. Compton down-scattering of both K$\alpha$ line photons and high energy continuum emission also gives rise to a `Compton shoulder' at $\sim6.2$\,keV (e.g., \citealt{matt:2002, yaqoob:2011}) and resonant line emission in the soft X-ray band (e.g., \citealt{ross&fabian:2005, garcia&kallman:2010}) which can further complicate the emergent spectrum.

Naturally, owing to the important effect it can have on the observed X-ray spectrum there are numerous models available for modelling the reflection component (e.g., {\tt pexrav/pexriv}, \citealt{magdziarz&zdziarski:1995}; {\tt pexmon}, \citealt{nandra:2007}; \reflionx, \citealt{ross&fabian:2005}; \xillver\footnote{The \xillver reflection model is not currently available in the public domain and is therefore not considered for use in this work.}, \citealt{garcia&kallman:2010}; \mytorus\footnote{The \mytorus model and documentation are publicly available at: www.mytorus.com}, \citealt{murphy&yaqoob:2009}). In this work we use a combination of \reflionx and \pexrav, both of which are publicly available and extensively used in the literature. The primary reason for this it that because \reflionx interpolates the observed reflection spectrum from a pre-generated grid of table models it is significantly faster at fitting spectra than, for example, \pexrav, \pexriv or {\tt pexmon}, which analytically calculate the Compton reflection spectrum on the fly at each step of the fitting process, or \mytorus which requires the model to be tailored for each individual source. Using \reflionx as our primary means of fitting the reflection spectrum ensures the least time-consuming Monte Carlo simulations which is important when dealing with a large sample of objects such as that considered in this work. Secondly, when fitting reflection continuum without the simultaneous constraint of the Fe\,K$\alpha$ line the reflection fraction, $R$, reported by \pexrav (and \pexriv) can become degenerate with the photon-index of the primary power-law, with a hardening reflection component compensated for by a softer primary continuum. By simultaneously fitting the reflection continuum, the \feka line and the soft X-ray resonance lines, \reflionx is able to overcome these modelling degeneracies which leads to a confident constraint on the contribution of the reflection continuum to the observed spectrum. \reflionx also has the additional advantage of allowing the ionisation state of the reflector, $\xi$, to be a free parameter which enables it to model changes in the \feka emitted flux and Fe\,K-shell edge profile associated with the reflectors ionisation state. We stress, however, that equivalent results are found for the detected Fe\,K absorbers if {\tt pexmon} is used instead, but with the resultant Monte Carlo simulations taking significantly longer to complete which effectively prohibits its uniform use throughout the sample.

We initially fitted all sources with \pexrav to determine the parameters of the \feka and \fekb lines. \pexrav was then replaced with \reflionx, and a systematic search for additional atomic features in the Fe\,K band was conducted. A total of 11 sources have best-fitting reflector Fe abundances which are non-solar, with 4 requiring a slight over-abundance (MCG\,+8-11-11, NCG\,4593, NCG\,7213, NCG\,7469) and 7 with an under-abundance (4C\,+74.26, Fairall\,9, IC\,4329A, IGR\,J16185-5928, Mrk\,335, Mrk\,359, Mrk\,841). These abundances are most likely a by-product of \reflionx assuming a face-on reflection geometry, and is a caveat which is discussed in greater detail in see Appendix~\ref{modelling_complexities}. PDS\,456 and APM\,08279+5255 are not fitted with a reflection component because they lack sufficient counts in the HXD/PIN..

\subsection{Searching for Fe\,K absorption}
\label{subsec:searching_for_absorption_lines}
Once a statistically acceptable fit to the broadband continuum had been found, i.e., including all necessary absorption regions, soft X-ray emission lines and continuum components, we performed a thorough search for additional spectral features between $5-10$\,keV. The method consists of inspecting the $|\Delta\chi^{2}|$ deviations from the best-fit continuum model using inverted contour plots of the energy-intensity plane in the Fe\,K band. The method of calculating the contour plots was adapted from the method outlined in \tombesiA, and was carried out as follows: 
\begin{enumerate}[leftmargin=*, label=(\roman*)]
\item an unresolved ($\sigma=10$\,eV) Gaussian was stepped across the entire $5-10$\,keV energy band of the baseline continuum model in 25\,eV intervals, with normalization allowed to adopt both positive and negative values to probe for spectral lines in both emission and absorption. All of the other spectral components were allowed to be free; 
\item after each step the $\Delta\chi^{2}$ deviation was recorded generating a $\chi^{2}$ distribution of the entire Fe\,K band relative to the baseline continuum model; 
\item confidence contours for the grid of $\chi^{2}$ values were plotted according to $\Delta\chi^{2}$ deviations of $-2.3$, $-4.61$, $-9.21$, $-13.82$, $-18.42$ from the baseline model, which correspond to confidence intervals of $68\%$, $90\%$, $99\%$, $99.9\%$ and $99.99\%$, respectively. A confidence contour corresponding to a $\Delta\chi^{2}=+0.5$ worse fit is also plotted which is intended to denote an approximate level for the continuum. 
\end{enumerate}
The energy-intensity contour plots produced with this method provide a powerful means of searching for additional emission or absorption components present in the Fe\,K band while also visually assessing their energy, intensity and rough statistical requirement relative to the underlying continuum model.

\begin{figure}
\begin{center}
\includegraphics[angle=-90,width=7.7cm]{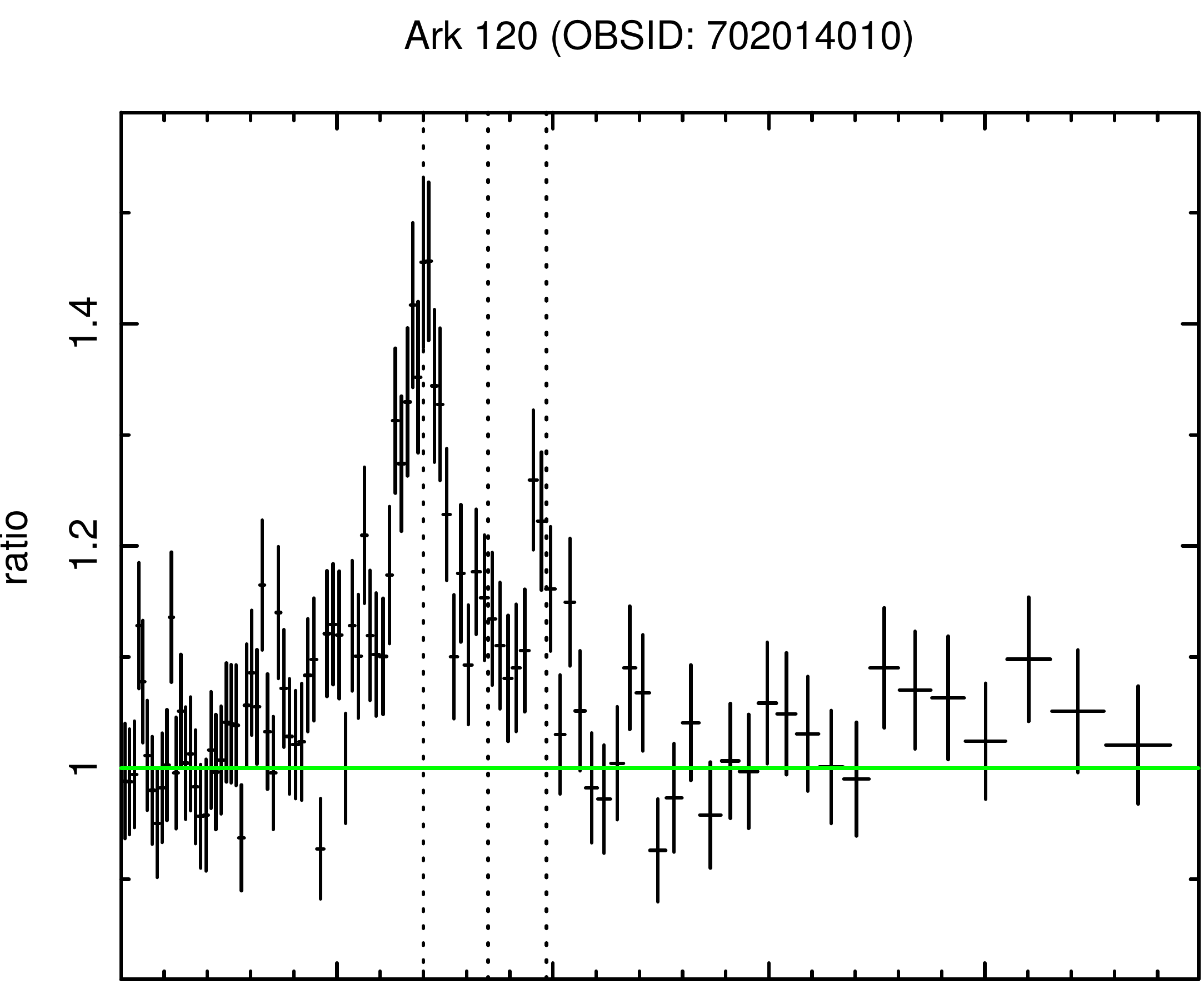}\vspace{-3pt}
\includegraphics[angle=-90,width=7.7cm]{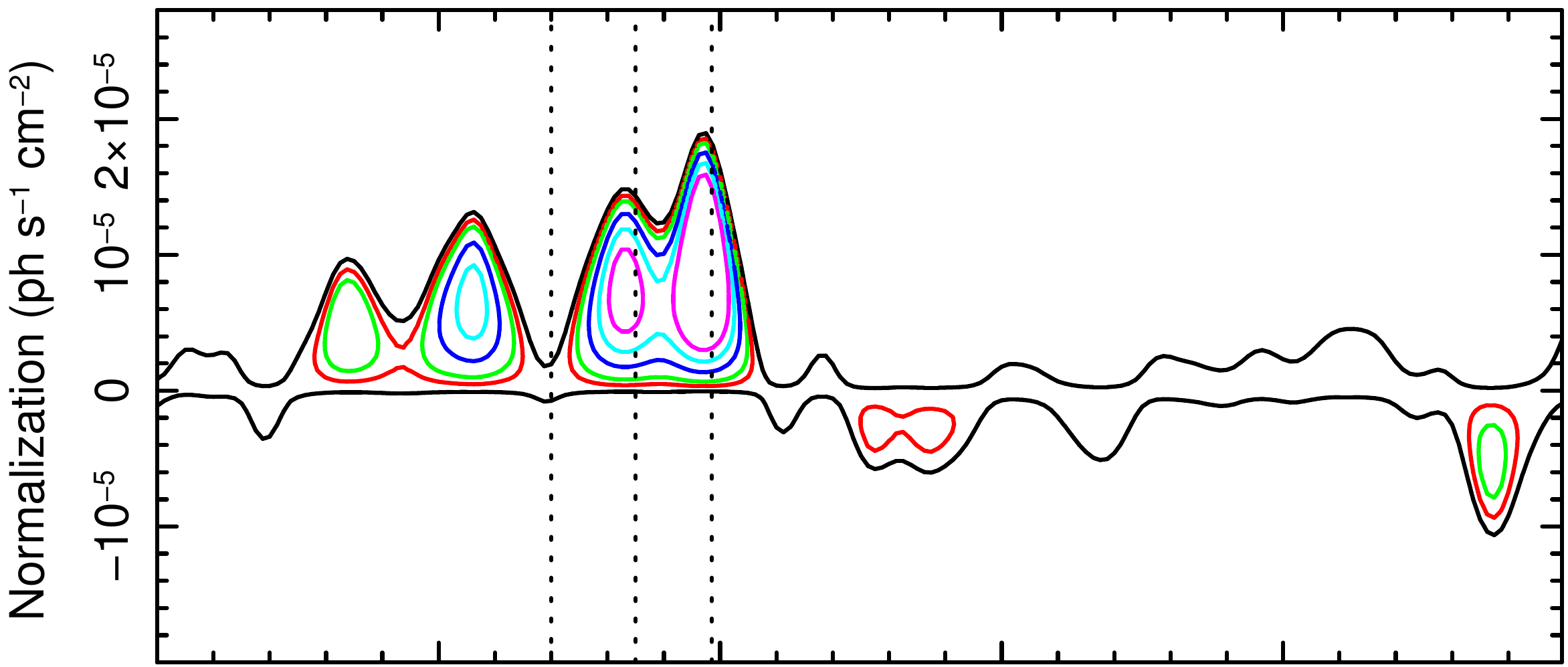}\vspace{-3pt}
\includegraphics[angle=-90,width=7.7cm]{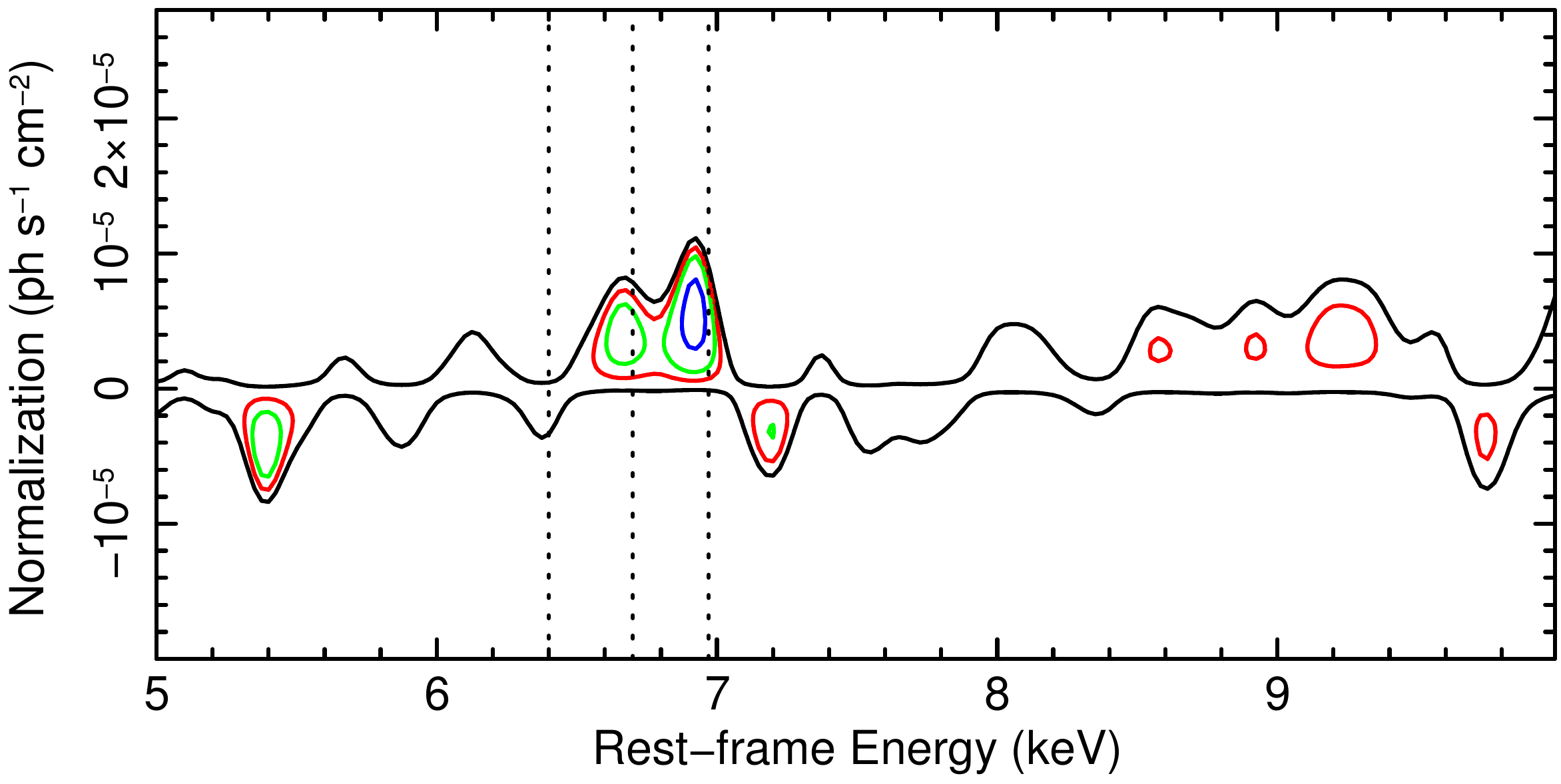}
\caption{\small {\it Top panel:} Ratio plot showing the residuals remaining in the Fe\,K band of Ark\,120 (OBSID~702014010) once all atomic lines have been removed and reflection fitted with {\tt \pexrav}; {\it Middle panel:} confidence contours showing deviations from the best-fit model when the \feka and \fekb lines have been fitted with \reflionx and a narrow Gaussian respectively. The closed significance contours correspond (from outer to inner) to $\Delta\chi^{2}$ improvements of $-2.3$ ($P_{\rm F}=68\%$), $-4.61$ ($P_{\rm F}=90\%$), $-9.21$ ($P_{\rm F}=99\%$), $-13.82$ ($P_{\rm F}=99.9\%$) and $-18.42$ ($P_{\rm F}=99.99\%$) relative to the best-fit continuum model are shown in red, green, blue, cyan and magenta, respectively. The magenta contours in the middle panel indicate that the broad emission residual is significant at the $>99.99\%$ level and that an additional component is required in the model; {\it Bottom panel:} the remaining confidence contours once the broad profile has been fitted. The broad residua are no longer detected but an additional \fexxvi emission line is present at $E\sim6.97$\,keV which is significant at $>99\%$. In all panels the dashed vertical lines indicate the expected rest-frame energies of (from left to right) the \feka fluorescence line, \fexxv~\hea and \fexxvi~\lya, respectively. Colour version available online.}
\label{figure:ark120_rat_cont}
\end{center}
\end{figure}

\begin{figure}
\begin{center}
\includegraphics[angle=-90,width=7.5cm]{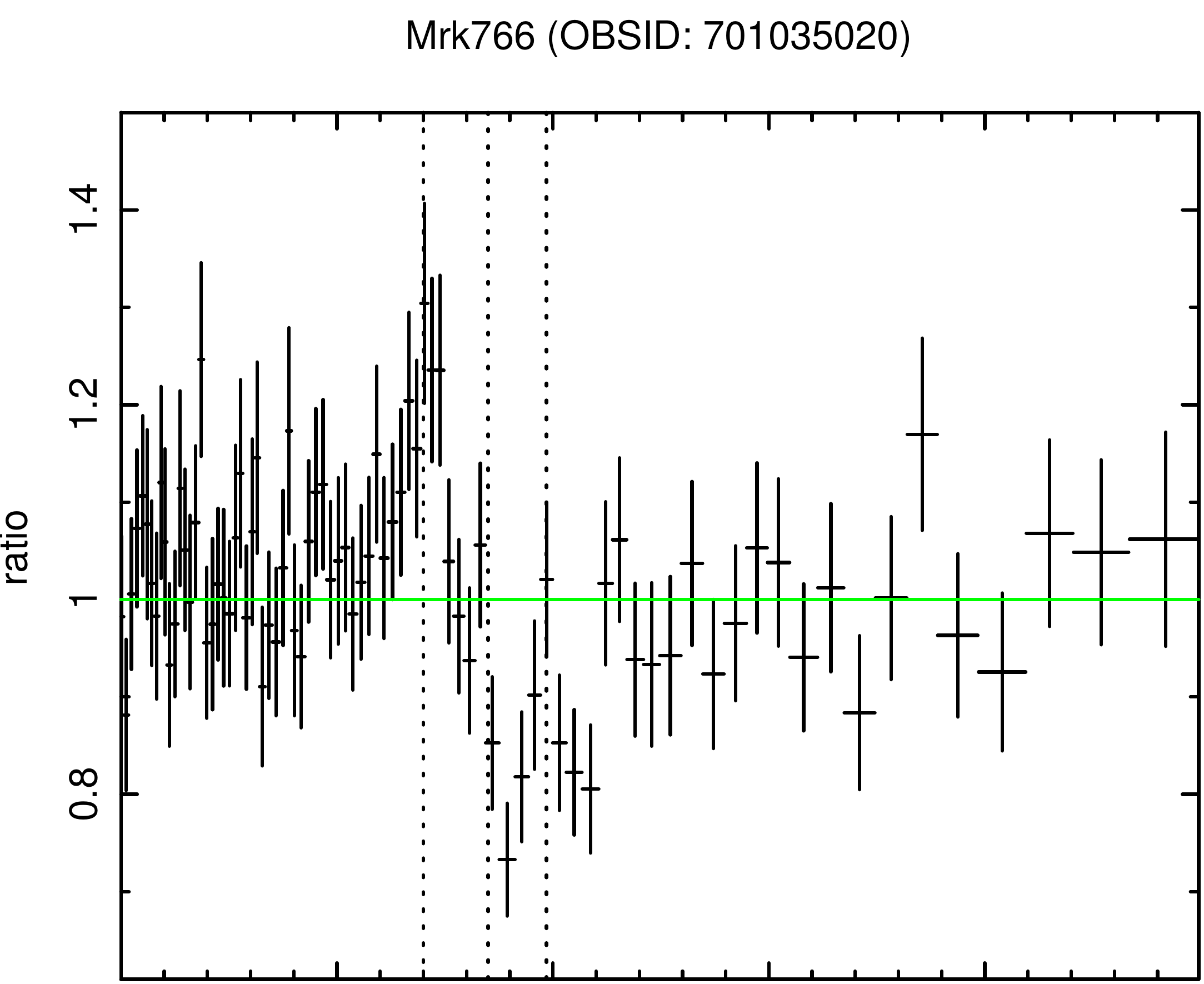}\vspace{-3pt}
\includegraphics[angle=-90,width=7.5cm]{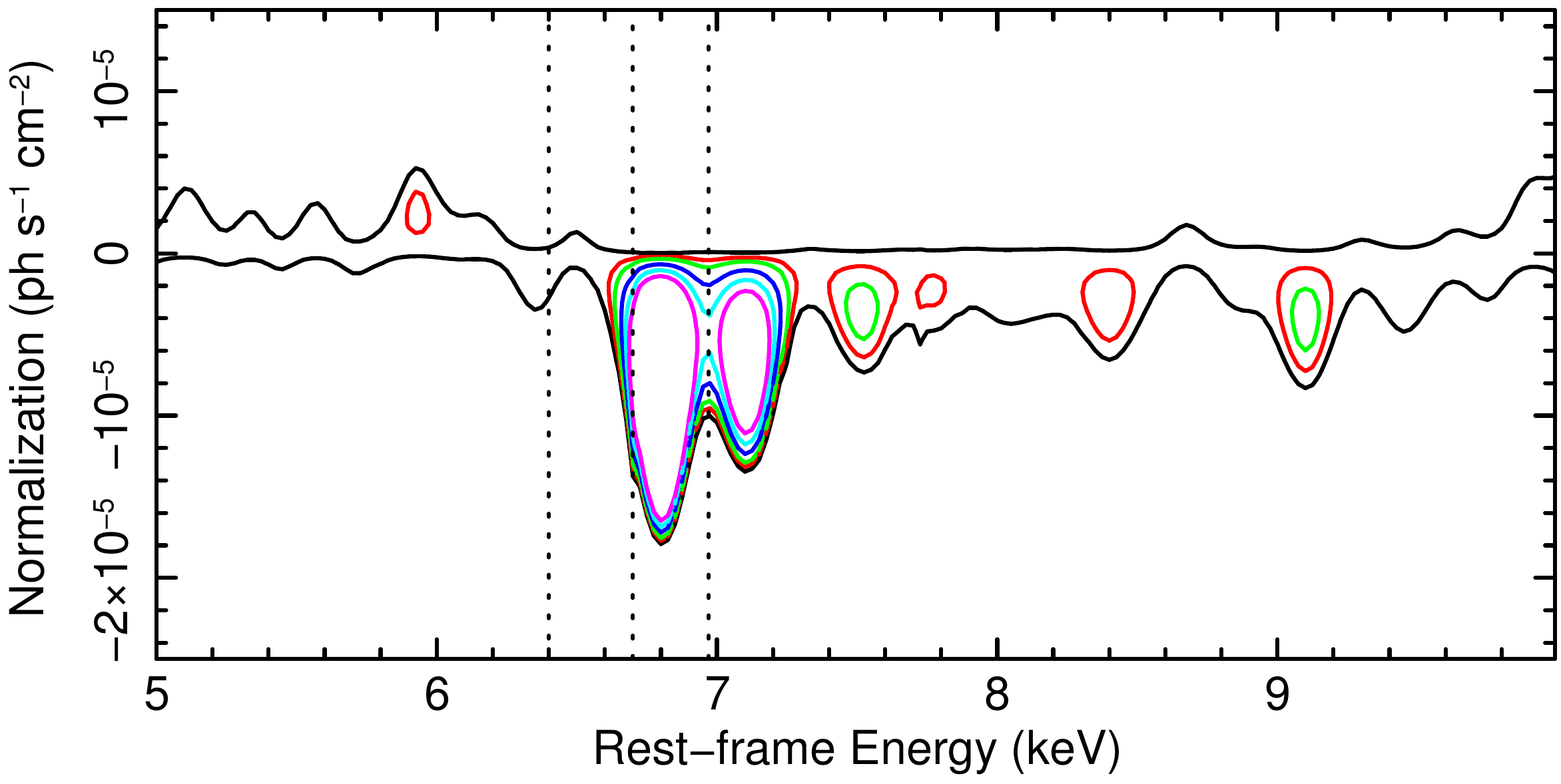}
\caption{\small {\it Top Panel}: As in the case of Figure~\ref{figure:ark120_rat_cont}, but this time for Mrk\,766 (OBSID~701035020); {\it Bottom Panel}: As in the corresponding panel of Figure~\ref{figure:ark120_rat_cont}. There are no ionised emission lines required in Mrk\,766 however two highly significant (each required at $P_{\rm F}>99.99\%$) absorption profiles are clearly detected. The energy of these lines is consistent with \fexxv~\hea and \fexxvi~\lya, respectively, outflowing at $\vout\sim6000$\,km\,s$^{-1}$). Colour version available online.}
\label{figure:mrk766_rat_cont}
\end{center}
\end{figure}

There are a number of atomic features between $6.0-8.0$\,keV which can complicate the detection of absorption line systems. Such features include ionised emission lines from \fexxv and \fexxvi expected at $\sim6.63-6.7$\,keV and $\sim6.97$\,keV, respectively, the Fe\,K-shell edge complex at $\sim7.1$\,keV, and any broad Fe line profile due to GR or scattering effects. It is important to account for these emission residua prior to searching for absorption lines, particularly in sources which show evidence for a broad residual as the broadness of the profile can effectively mask the presence of low velocity absorption systems in the raw data. For a given continuum model we searched for highly-ionised absorption lines in the Fe\,K-band using the following steps:
\begin{enumerate}[leftmargin=*, label=(\roman*)]
\item we first generated an energy-intensity contour plot using the method outlined above;
\item we then inspected the contour plot to determine whether there were any intense ionised emission and absorption lines present in the data with confidence contours of $>99\%$;
\item where there was evidence for a broad emission residual at an F-test significance of at least $P_{\rm F}>99\%$ they were fitted with either a broadened Gaussian (with $\sigma$-width a free parameter) or a {\tt diskline}\footnote{For simplicity, the {\tt diskline} was fitted with an assumed rest-frame energy of $E=6.4$\,keV and a typical emissivity profile of $q=-2.5$ (e.g., \citealt{patrick:2011a}). The inner emission radius ($R_{\rm in}$), outer radius ($R_{\rm out}$) and disk inclination ($i$), were left free to vary.} depending on the asymmetry of the observed profile, and a second intermediate contour plot was generated to determine whether any further components were needed by the model. As before, all other model parameters were allowed to vary freely during this process. If there was no evidence for a broad profile we did not generate an additional contour plot and instead moved directly onto the next step;
\item where narrow emission profiles were detected with a resolved confidence contour of $>99\%$ they were fitted with unresolved ($\sigma=10$\,eV fixed) Gaussian profiles provided they were required by the data at $P_{\rm F}>99\%$ (corresponding to $\Delta\chi^{2}>9.21$ for two parameters of interest); 
\item once all emission profiles had been accounted for we again checked for the presence of any blue-shifted Fe\,K absorption lines at $E>6.7$\,keV. If no absorption lines were present we ended the search at this step, reported the best-fit continuum and Fe\,K emission line parameters in the relevant tables of Appendix C, and moved onto the next observation in the sample; 
\item otherwise, where absorption troughs were clearly detected with a confidence contour of $>99\%$ we parametrised the line(s) using inverted Gaussians with $\sigma$-width initially fixed at either $10$\,eV, $30$\,eV or $100$\,eV depending on which provided the greatest improvement to $\Delta\chi^{2}$. Note that while initially fixed, the line widths were allowed to vary where appropriate leading to four sources with resolved absorption profiles. The key parameters of all detected absorption lines are reported in Table~\ref{table:absorption_line_parameters}. 
\end{enumerate}
This process was carried out on each individually fitted spectrum in the sample; including those which were included in simultaneous fits. In Figures~\ref{figure:ark120_rat_cont} and \ref{figure:mrk766_rat_cont} we show examples of this process applied to Ark\,120 (OBSID~702014010), which shows evidence for a broad asymmetric emission profile and no absorption lines, and Mrk\,766 (OBSID~701035020), which is dominated by \fexxv~\hea and \fexxvi~\lya absorption. The top panel of both figures show the residuals which remain in the Fe\,K band when all atomic lines have been removed from the best-fit continuum model and the reflection component is fitted with \pexrav to highlight the presence of the neutral \feka/\fekb fluorescence lines. The contour plots show the significances of the remaining residuals when the \feka and \fekb lines have been fitted with \reflionx and a narrow Gaussian line. The continuous outer contour corresponds to the $\Delta\chi^{2}=+0.5$ residual as mentioned previously. From outer to inner the closed significance contours corresponding to $\Delta\chi^{2}$ improvements of $-2.3$ ($P_{\rm F}=68\%$), $-4.61$ ($P_{\rm F}=90\%$), $-9.21$ ($P_{\rm F}=99\%$), $-13.82$ ($P_{\rm F}=99.9\%$) and $-18.42$ ($P_{\rm F}=99.99\%$) relative to the best-fit continuum model are shown in red, green, blue, cyan and magenta, respectively. 

There are clear residual profiles present in both sources; particularly Ark\,120 where the positive asymmetric profile and additional \fexxvi emission is apparent in the middle and bottom panels of Figure~\ref{figure:ark120_rat_cont}, respectively. The two absorption profiles in Mrk\,766 -- which have previously been identified as \fexxv~\hea and \fexxvi~\lya by \cite{miller:2007} -- are statistically distinguishable at $>99.99\%$ confidence and are self-consistently fitted with a single highly-ionised region of photoionised absorption (see Section~\ref{subsec:photoionisation_modelling}). Energy-intensity contour plots for all fitted spectra in this sample are included in Figures~\ref{figure:rat_cont1}1 and \ref{figure:rat_cont2} in Appendix~\ref{appendix:contour_plots}. We note that the $\Delta\chi^{2}$ of individual lines and their corresponding significances according to the F-Test as listed in Table~\ref{table:absorption_line_parameters} are taken from a direct spectral fitting of the line profiles themselves, and are not determined directly from the contour plots.

It is also important to note that while we use the F-test as a rough initial gauger of the significance of any line-like profiles in the Fe\,K band we {\it do not claim to robustly detect any absorption lines based solely on their measured $\mathit{\Delta\chi^{2}}$ and corresponding F-Test significances}. Indeed, as has been pointed out by numerous authors in the literature (e.g., \citealt{protassov:2002}; \citealt{porquet:2004b}; \citealt{markowitz:2006}; \tombesiA) the F-test alone might not be an adequate statistical test to determine the detection significance of atomic lines in complex spectral models as it neither takes into account the number of energy resolution bins over a given energy range, or the expected energy of a given atomic line. In cases where there is no {\it a priori} justification for expecting a spectral line at a particular energy and the line search is done over what is essentially an arbitrary energy range, as is often the case when dealing with strongly blue-shifted absorption lines, the F-test can somewhat over-predict the detection probability and when compared to extensive simulations. For these reasons all suspected absorption lines which have an F-test significance of $P_{\rm F}>99\%$ were followed up with extensive Monte Carlo simulations which allows their detection significance to be assessed against random fluctuations and photon noise in the spectrum.

\begin{table*}
	\scriptsize
	\begin{minipage}{170 mm}
	\caption{Gaussian parameters for detected Fe\,K band absorption lines. {\sl Notes}: (1) Source name; (2) \suzaku observation ID; (3) Absorption line identification; (4) Measured line energy in the source rest-frame, in units of keV; (5) Measured line energy-width, in units of eV. Unresolved lines were fit with widths fixed at 10\,eV, 30\,eV or 100\,eV depending upon which yielded the best statistical improvement to the fit; (6) Line intensity, in units of $\times10^{-6}$\,photons\,cm$^{-2}$\,s$^{-1}$; (7) Line equivalent width, in units of eV; (8) Change in fit statistic (and degrees of freedom) when line is removed from the best-fit model; (9) Line significance according to the F-test, in per cent; (10) Monte Carlo significance of individual lines, in per cent; (11) Monte Carlo significance of observing a pair of lines with energy separation corresponding to a \fexxvxxvi pair (see text for details), in per cent. Absorption complexes detected at $P_{\rm MC}\geq99\%$ confidence are listed in bold.}

	\begin{tabular}{@{}lrcrcrcrrrr}

      \toprule
      \multicolumn{1}{c}{Source} & \multicolumn{1}{c}{OBSID} & \multicolumn{1}{c}{Line ID} & \multicolumn{1}{c}{Energy} & 
      \multicolumn{1}{c}{$\sigma$-width} & \multicolumn{1}{c}{Intensity} & \multicolumn{1}{c}{EW} & 
      \multicolumn{1}{c}{$\Delta\chi^{2}/\Delta\nu$} & \multicolumn{1}{c}{$P_{\rm F}$} & \multicolumn{1}{c}{$P_{\rm MC}^{1}$} &
      \multicolumn{1}{c}{$P_{\rm MC}^{2}$}\\[0.5ex]
      \multicolumn{1}{c}{(1)} & \multicolumn{1}{c}{(2)} & \multicolumn{1}{c}{(3)} & \multicolumn{1}{c}{(4)} & \multicolumn{1}{c}{(5)} & 
      \multicolumn{1}{c}{(6)} & \multicolumn{1}{c}{(7)} & \multicolumn{1}{c}{(8)} & \multicolumn{1}{c}{(9)} & \multicolumn{1}{c}{(10)} &
      \multicolumn{1}{c}{(11)}\\
      \midrule

3C\,111 
	& 703034010 & \fexxviabs & $7.24^{+0.04}_{-0.04}$ & 10\fix & $-5.6^{+2.1}_{-2.1}$ & $-26^{+9}_{-9}$ & $16.4/2$ 
				& $>99.99$ & $\mathbf{99.8}$\\
	& 705040020 & \fexxviabs & $7.76^{+0.07}_{-0.04}$ & 10\fix & $-15.2^{+5.1}_{-5.1}$ & $-20^{+10}_{-10}$ & $11.1/2$
				& $99.60$ & $97.8$\\
3C\,390.3 
	& 701060010 & \fexxviabs & $8.08^{+0.06}_{-0.05}$ & 10\fix & $-5.6^{+2.6}_{-2.6}$ & $-21^{+10}_{-10}$ & $12.1/2$ & $>99.99$ & $98.7$\\

4C\,+74.26 
	& 702057010 & \fexxviabs & $8.38^{+0.05}_{-0.07}$ & 10\fix & $-7.2^{+3.4}_{-3.4}$ & $-29^{+14}_{-14}$ & $12.4/2$ & $99.80$ 
				& $96.6$\\

APM\,08279+5255 
	& stacked	& \blendabs(1) & $7.76^{+0.11}_{-0.11}$ & $100^{\it f}$ & $-6.9^{+2.5}_{-2.3}$ & $-24^{+8}_{-8}$ 
				& $23.8/2$ & $>99.99$& $\mathbf{99.7}$\\
 	& 			& \blendabs(2) & $10.97^{+0.16}_{-0.16}$ & $100^{\it f}$ & $-4.4^{+1.9}_{-1.8}$ & $-30^{+15}_{-15}$ 
 				& $18.4/2$ & $>99.99$& $\mathbf{97.3}$\\
CBS\,126 
	& 705042010 & \fexxviabs & $7.04^{+0.04}_{-0.05}$ & 30\fix & $-3.3^{+1.0}_{-1.0}$ & $-77^{+22}_{-22}$ & $27.5/2$ 
				& $>99.99$ & $\mathbf{>99.9}$ \\
ESO\,103-G035
	& 703031010	& \fexxviabs & $7.26^{+0.05}_{-0.05}$ & $104^{+61}_{-55}$ & $-14.7^{+4.8}_{-4.1}$ & $-40^{+14}_{-14}$ 
				& $35.0/3$ & $>99.99$ & $\mathbf{>99.9}$\\
MCG\,-6-30-15 
	& stacked[all] & \fexxvabs & $6.72^{+0.02}_{-0.01}$ & 10\fix & $-7.8^{+1.2}_{-1.2}$ & $-18^{+3}_{-3}$ 
				& $81.6/2$ & $>99.99$ & $\mathbf{>99.9}$ & $\mathbf{>99.9}$\\
	&			   & \fexxviabs & $7.05^{+0.02}_{-0.02}$ & 10\fix & $-5.1^{+1.2}_{-1.2}$ & $-14^{+3}_{-3}$ 
				& $25.7/2$ & $>99.99$ & $\mathbf{>99.9}$ \\
MR\,2251-178
	& 704055010	& Fe\,L--\hea & $7.55^{+0.12}_{-0.12}$ & $193^{+84}_{-62}$ & $-12.9^{+4.1}_{-4.1}$ & $-32^{+10}_{-10}$ 
   				& $26.9/3$ & $>99.99$ & $\mathbf{99.6}$ \\
Mrk\,279 
	& 704031010 & \fexxviabs & $8.88^{+0.06}_{-0.06}$ & $10^{f}$ & $-2.5^{+1.0}_{-1.0}$ & $-73^{+30}_{-30}$ & $15.1/2$ 
				& $99.95$ & $\mathbf{99.1}$ \\
Mrk\,766 
	& 701035010 & \fexxvabs & $7.11^{+0.07}_{-0.07}$ & $10^{f}$ & $-2.8^{+1.6}_{-1.6}$ & $-27^{+15}_{-15}$ & $7.0/2$ 
				& $96.27$ & $71.0$ & $\mathbf{99.8}$\\
 	& 			& \fexxviabs & $7.42^{+0.06}_{-0.06}$ & $10^{f}$ & $-3.9^{+1.6}_{-1.6}$ & $-40^{+17}_{-17}$ & $15.3/2$ 
 				& $99.92$ & $98.4$\\
	& 701035020	& \fexxvabs & $6.80^{+0.02}_{-0.02}$ & $10^{f}$ & $-7.7^{+1.8}_{-1.8}$ & $-60^{+14}_{-14}$ & $50.0/2$ 
				& $>99.99$ & $\mathbf{>99.9}$ & $\mathbf{>99.9}$\\
 	& 			& \fexxviabs & $7.12^{+0.04}_{-0.04}$ & $10^{f}$ & $-5.0^{+1.9}_{-1.9}$ & $-44^{+17}_{-17}$ & $18.9/2$ 
 				& $>99.99$ & $\mathbf{99.8}$ \\
NGC\,1365 
	& 702047010 & \fexxvabs & $6.77^{+0.01}_{-0.01}$ & $65^{+8}_{-8}$ & $-29.9^{+1.6}_{-1.6}$ & $-156^{+8}_{-8}$ 
				& $1123.8/3$ & $>99.99$ & $\mathbf{>99.9}$ & $\mathbf{>99.9}$\\
 	& 			& \fexxviabs & $6.97^{+0.03}_{-0.03}$ & $65^{*}$ & $-8.7^{+1.7}_{-1.7}$ & $-69^{+13}_{-13}$ 
 				& $77.0/3$ & $>99.99$ & $\mathbf{>99.9}$ \\
 	& 			& He$\beta$ & $7.94^{+0.01}_{-0.01}$ & $65^{*}$ & $-26.1^{+1.5}_{-1.5}$ & $-164^{+10}_{-10}$ 
 				& $964.5/3$ & $>99.99$ & $\mathbf{>99.9}$ \\
 	& 			& Ly$\beta$ & $8.38^{+0.03}_{-0.03}$ & $65^{*}$ & $-10.6^{+1.8}_{-1.8}$ & $-87^{+15}_{-15}$ 
 				& $94.3/3$ & $>99.99$ & $\mathbf{>99.9}$ \\
 	& 704031010 & \fexxvabs & $6.71^{+0.02}_{-0.02}$ & 30\fix & $-16.1^{+2.4}_{-2.4}$ & $-67^{+10}_{-10}$ 
 				& $134.9/2$ & $>99.99$ & $\mathbf{>99.9}$ & $\mathbf{>99.9}$\\
 	& 			& \fexxviabs & $7.00^{+0.02}_{-0.02}$ & $30^{*}$ & $-13.4^{+2.0}_{-2.0}$ & $-85^{+14}_{-14}$ 
 				& $90.7/2$ & $>99.99$ & $\mathbf{>99.9}$\\
 	& 			& He$\beta$ & $7.87^{+0.08}_{-0.08}$ & $30^{*}$  & $-4.7^{+2.5}_{-2.5}$ & $-34^{+20}_{-20}$ 
 				& $7.2/2$ & $96.77$ & \na\\
	& 			& Ly$\beta$ & $8.37^{+0.11}_{-0.11}$ & $30^{*}$  & $-5.3^{+2.7}_{-2.7}$ & $-45^{+24}_{-24}$ 
				& $8.5/2$ & $98.23$ & \na\\
NGC\,3227 
	& 703022010 & \fexxvabs & $6.69^{+0.04}_{-0.04}$ & 10\fix & $-9.9^{+3.7}_{-3.7}$ &$-21^{+7}_{-7}$ 
				& $17.5/2$ & $99.99$ & $\mathbf{>99.9}$ & $\mathbf{>99.9}$\\
	& 			& \fexxviabs & $6.95^{+0.03}_{-0.03}$ & 10\fix & $-15.5^{+3.7}_{-3.7}$ & $-38^{+9}_{-9}$ 
				& $42.4/2$ & $>99.99$ & $\mathbf{>99.9}$\\
 	& 703022030 & \fexxvabs & $6.76^{+0.03}_{-0.04}$ & 10\fix & $-8.5^{+3.5}_{-3.5}$ & $-24^{+9}_{-9}$ 
 				& $16.2/2$ & $99.94$ & $\mathbf{99.8}$ & $\mathbf{>99.9}$\\
 	& 			& \fexxviabs & $7.04^{+0.05}_{-0.05}$ & 10\fix & $-6.6^{+3.6}_{-3.6}$ & $-22^{+11}_{-11}$ 
 				& $9.3/3$ & $98.64$ & $91.8$\\
 	& 703022050 & \fexxvabs & $6.76^{+0.07}_{-0.07}$ & 10\fix & $-5.8^{+2.8}_{-2.8}$ & $-19^{+9}_{-8}$ 
 				& $12.0/2$ & $99.74$ & $96.2$ & $\mathbf{>99.9}$\\
 	& 			& \fexxviabs & $7.05^{+0.03}_{-0.03}$ & 10\fix & $-11.0^{+2.8}_{-2.8}$ & $-41^{+9}_{-9}$ 
 				& $41.2/2$ & $>99.99$ & $\mathbf{>99.9}$\\
NGC\,3516 
	& 100031010 & \fexxvabs & $6.74^{+0.02}_{-0.02}$ & $10^{f}$ & $-9.1^{+1.6}_{-1.6}$ & $-25^{+5}_{-5}$ 
				& $65.6/2$ & $>99.99$ & $\mathbf{>99.9}$ & $\mathbf{>99.9}$\\
 	& 			& \fexxviabs & $7.00^{+0.02}_{-0.02}$ & $10^{f}$ & $-6.0^{+1.7}_{-1.7}$ & $-20^{+6}_{-6}$ 
 				& $66.7/2$ & $>99.99$  & $\mathbf{>99.9}$ \\
NGC\,3783 
	& 701033010 & \fexxvabs & $6.69^{+0.04}_{-0.04}$ & 10\fix & $-10.0^{+3.2}_{-3.0}$ & $-16^{+5}_{-6}$ 
				& $30.2/2$ & $>99.99$ & $\mathbf{>99.9}$\\
	& 704063010 & \fexxvabs & $6.71^{+0.02}_{-0.02}$ & 10\fix & $-14.0^{+3.1}_{-3.1}$ & $-20^{+3}_{-5}$ 
				& $56.7/2$ & $>99.99$ & $\mathbf{>99.9}$\\
NGC\,4051 
	& 700004010 & \fexxvabs & $6.77^{+0.04}_{-0.04}$ & $10^{f}$ & $-2.2^{+1.0}_{-1.0}$ & $-21^{+10}_{-10}$ 
				& $9.9/2$ & $99.39$ & $89.4$ & $\mathbf{>99.9}$\\
  	& 			& \fexxviabs & $7.08^{+0.03}_{-0.03}$ & $10^{f}$ & $-3.3^{+1.1}_{-1.1}$ & $-38^{+13}_{-13}$ 
  				& $22.7/2$ & $>99.99$ & $\mathbf{>99.9}$ \\
   	& 703023010 & \fexxvabs & $6.82^{+0.03}_{-0.03}$ & $10^{f}$ & $-4.8^{+1.3}_{-1.3}$ & $-22^{+6}_{-6}$ 
   				& $26.4/2$ & $>99.99$ & $\mathbf{>99.9}$ & $\mathbf{>99.9}$\\
 	& 			& \fexxviabs & $7.11^{+0.02}_{-0.02}$ & $10^{f}$ & $-6.4^{+1.3}_{-1.3}$ & $-33^{+7}_{-7}$ 
 				& $57.5/2$ & $>99.99$ & $\mathbf{>99.9}$ \\
NGC\,4151 
	& 701034010 & \fexxviabs & $7.17^{+0.04}_{-0.04}$ & $92^{+51}_{-41}$ & $-15.9^{+4.7}_{-4.3}$ & $-28^{+8}_{-8}$ 
				& $43.7/3$ & $>99.99$ & $\mathbf{>99.9}$ \\
NGC\,4395 
	& 702001010 & \fexxvabs & $6.63^{+0.07}_{-0.05}$ & 10\fix & $-2.2^{+1.1}_{-1.1}$ & $-35^{+19}_{-19}$ 
				& $10.4/2$ & $>99.99$ & $74.5$ & $\mathbf{>99.9}$\\
	& 			& \fexxviabs & $6.91^{+0.06}_{-0.06}$ & 10\fix & $-2.9^{+1.1}_{1.1}$ & $-55^{+20}_{-20}$ 
				& $18.3/2$ & $>99.99$ & $86.8$\\
NGC\,5506
	& stacked[all] & \fexxviabs & $9.23^{+0.06}_{-0.06}$ & 10\fix & $-11.1^{+4.7}_{-4.7}$ & $-16^{+5}_{-5}$ 
				& $16.2/2$ & $>99.99$ & $\mathbf{99.8}$\\
PDS\,456
	& 701056010 & \blendabs(1) & $9.07^{+0.06}_{-0.06}$ & 100\fix & $-2.5^{+0.8}_{-0.8}$ & $-108^{+35}_{-35}$ 
				& $29.5/2$ & $>99.99$ & $\mathbf{>99.9}$\\
	&			& \blendabs(2) & $9.57^{+0.09}_{-0.09}$ & 100\fix & $-2.0^{+0.8}_{-0.8}$ & $-99^{+41}_{-41}$ 
				& $15.7/2$ & $>99.99$ & $\mathbf{99.8}$\\
 	& 705041010 & \blendabs(1) & $8.58^{+0.09}_{-0.09}$ & 100\fix & $-2.8^{+1.0}_{-1.0}$ & $-118^{+46}_{-46}$ 
 				& $19.5/2$ & $>99.99$ & $96.2$\\
 	&			& \blendabs(2) & $9.03^{+0.09}_{-0.09}$ & 100\fix & $-3.5^{+1.1}_{-1.1}$ & $-169^{+51}_{-51}$ 
 				& $78.6/2$ & $>99.99$ & $\mathbf{>99.9}$\\
S\,J2127.4+5654 
  	& 702122010	& \fexxviabs & $9.04^{+0.05}_{-0.05}$ & $10^{f}$ & $-10.0^{+4.7}_{-4.7}$ & $-60^{+28}_{-28}$ 
  				& $11.9/2$ & $99.78$ & $98.9$ \\

    \bottomrule
	\end{tabular}\\[0.5ex]
	\footnotesize
	$^{f}$ Indicates a parameter was frozen during spectral fitting.
	\label{table:absorption_line_parameters}
	\end{minipage}
\end{table*}

\subsection{Monte Carlo simulations}
\label{subsec:Monte_carlo_simulations}
Such Monte Carlo simulations have been used extensively in the literature to overcome the limitations of the F-test (e.g., \citealt{porquet:2004b}; \citealt{markowitz:2006}; \citealt{miniutti&fabian:2006}; \tombesiA; \tombesiBLRG) and enable the statistical significance of a spectral line to be robustly determined independently of spectral noise and detector effects. The method of Monte Carlo simulation we used follows the same process which was first outlined by \cite{porquet:2004b}, and is almost identical to that used by \tombesiA. The process was carried out with the following steps:

\begin{enumerate}[leftmargin=*, label=(\roman*)]
\item from the null hypothesis model (i.e., the broadband continuum model with all narrow absorption lines in the Fe\,K band removed) we simulated both XIS-FI and XIS-BI spectra using the {\it fakeit} command in \xspec and subtracted the appropriate background files. The simulated spectra had the same exposure as the original data and used the same spectral response files;
\item we then fitted the simulated XIS-FI(XIS-BI) spectra between $0.6-10.0$\,keV($0.6-5.0$\,keV) with the null hypothesis model to produce a new and refined null hypothesis which takes into account the uncertainty in the null hypothesis model itself. 
All continuum parameters bar the photon-index of the primary power-law and its normalization, and the normalization of any \bbody component, were frozen to their best-fit parameters taken from the real data to prevent degeneracies between model components during re-fitting. Any broad profile at Fe\,K had its width fixed to the best-fit value but was allowed to vary in both centroid energy and normalization;
\item from the refined null hypothesis model we generated a second set of simulated XIS-FI and XIS-BI spectra and subtracted the appropriate background files. These second simulated spectra were then fitted with the null hypothesis model and the resultant $\chi^{2}_{\rm null}$ value was recorded;
\item an unresolved ($\sigma=10$\,eV) Gaussian was added to the model at $5$\,keV in the source rest-frame with intensity initially set to zero but left free to vary between both positive and negative values to probe for both emission and absorption lines. The Gaussian line was then sequentially stepped between $5-9.5$\,keV (rest-frame) in 25\,eV increments. After each step the $|\Delta\chi^{2}_{\rm noise}|$ was recorded relative to $\chi^{2}_{\rm null}$;
\item this process was carried out for $T=1000$ simulated spectra per observation yielding a distribution of $|\Delta\chi^{2}_{\rm noise}|$ under the null hypothesis which essentially maps the statistical significance of any deviations from the null hypothesis model which are solely due to random photon noise in the spectrum;
\item the measured significance of the line in the real spectrum $|\Delta\chi^{2}_{\rm line}|$ was then compared to the values in the $|\Delta\chi^{2}_{\rm noise}|$ distribution to assess how many simulated spectra had a random fluctuation with a detection significance over this threshold value. If $N$ simulated spectra have $|\Delta\chi^{2}_{\rm noise}|\geq|\Delta\chi^{2}_{\rm line}|$ then the estimated confidence interval for the observed line from Monte Carlo simulations is then $P_{\rm MC}^{\rm Line}=1-\left(\frac{N_{\rm Line}}{T}\right)$. Moreover, if there are two absorption profiles consistent with a \fexxvxxvi pair we can infer the null probability of both lines simultaneously being a false detection by multiplying the probabilities of each individual line.
\end{enumerate}
Monte Carlo significances for all absorption lines detected in this work are listed in column 10 of Table~\ref{table:absorption_line_parameters}. Absorption complexes with a total $P_{\rm MC}\geq99\%$ are conservatively identified as robustly detected, while those with $95\%\leq P_{\rm MC} < 99\%$ are only listed as marginal detections. The overall detection rate and global significance of Fe\,K-band absorption lines is further discussed in Section~\ref{subsec:line_detection_rate}. 

There are two possible caveats associated with the above Monte Carlo process which both warrant further discussion. First and foremost, the Monte Carlo simulations implicitly assume that the null-hypothesis model is the correct representation of the continuum in a given source, and therefore the Monte Carlo probabilities do not account for the possibility of continuum mis-modelling. Given this possibility it is important to note that we have attempted to achieve a statistically acceptable representation of the broad-band spectrum, so that no obvious broad residuals are present. Care has been taken in the Fe\,K band in particular such that any broad residuals are minimised prior to searching for absorption lines, such that the reduced-$\chi^{2}$ is $\approx1.0$ in all cases. Moreover, and as shown in Appendix B, we note the vast majority of suspected absorption residuals are manifested through discrete narrow dips in the spectrum relative to the best-fit model, whereas any systematic residuals would usually be broader than the instrument resolution.

The second caveat is associated with spectral complexity in the Fe\, K band which can complicate both line identification and spectral
interpretation. There can be significant atomic complexity between $\sim5-7$\,keV, e.g., the narrow \feka and \fekb fluorescence lines, broad underlying Fe\,K lines, and ionised \fexxvxxvi emission lines, and the detection of absorption in this regime can depend strongly on how these features are modelled. There are three sources (4C\,+74.26, MGC\,-6-30- 15, SWIFT\,J2127.4+5654; see Figure B2) in the sample in which both a broad underlying profile and at least one absorption line have been detected. In both 4C+74.26 and SWIFT J2127.4+5654 the absorption lines are detected at a high energies (i.e., $E>8$ keV in the source rest-frame) and the effect of the broad Fe line on the line detection is negligible. In MCG\,-6-30-15 the lines have also been confirmed with other X-ray observatories (\citealt{young:2005}; \citealt{miller:2009}) which suggests that the presence of a broad Fe line does not introduce any significant model systematics. Thus modelling of a broad emission line does not appear to effect the detections of Fe\,K absorption lines in these cases.

\subsection{Consistency Checks}
\label{subsec:consistency_checks}
To further test the robustness of the absorption lines detected in the co-added XIS-FI spectra we performed a series of consistency checks with the individual XIS detectors. If a line is detected in two (or more, if the XIS\,2 is also present) detectors the line is very likely to be a real feature intrinsic to the source spectra rather than an artefact of background subtraction, systematic noise or an associated detector effect. The XIS-BI is not suitable for use as a consistency check as it tends to have a much lower signal-to-noise ratio above around 5\,keV owing to its lower effective area and higher background. Where absorption has been detected at the $P_{\rm MC}\geq95\%$ level we fitted each individual background-subtracted XIS\,0, XIS\,2 (where present) and XIS\,3 spectrum with the best-fitting model to the co-added XIS-FI spectrum. We then fitted a Gaussian absorption line, with rest-frame energy and normalization left free to vary, at the energy where the absorption line is detected in the XIS-FI spectrum, refitted the data, and noted the resultant line parameters for each XIS detector. We note that in SWIFT\,J2127.4+5654 this consistency check could not be conducted as only the XIS\,3 spectrum was available.  
\begin{figure}
\begin{center}
\includegraphics[width=8cm]{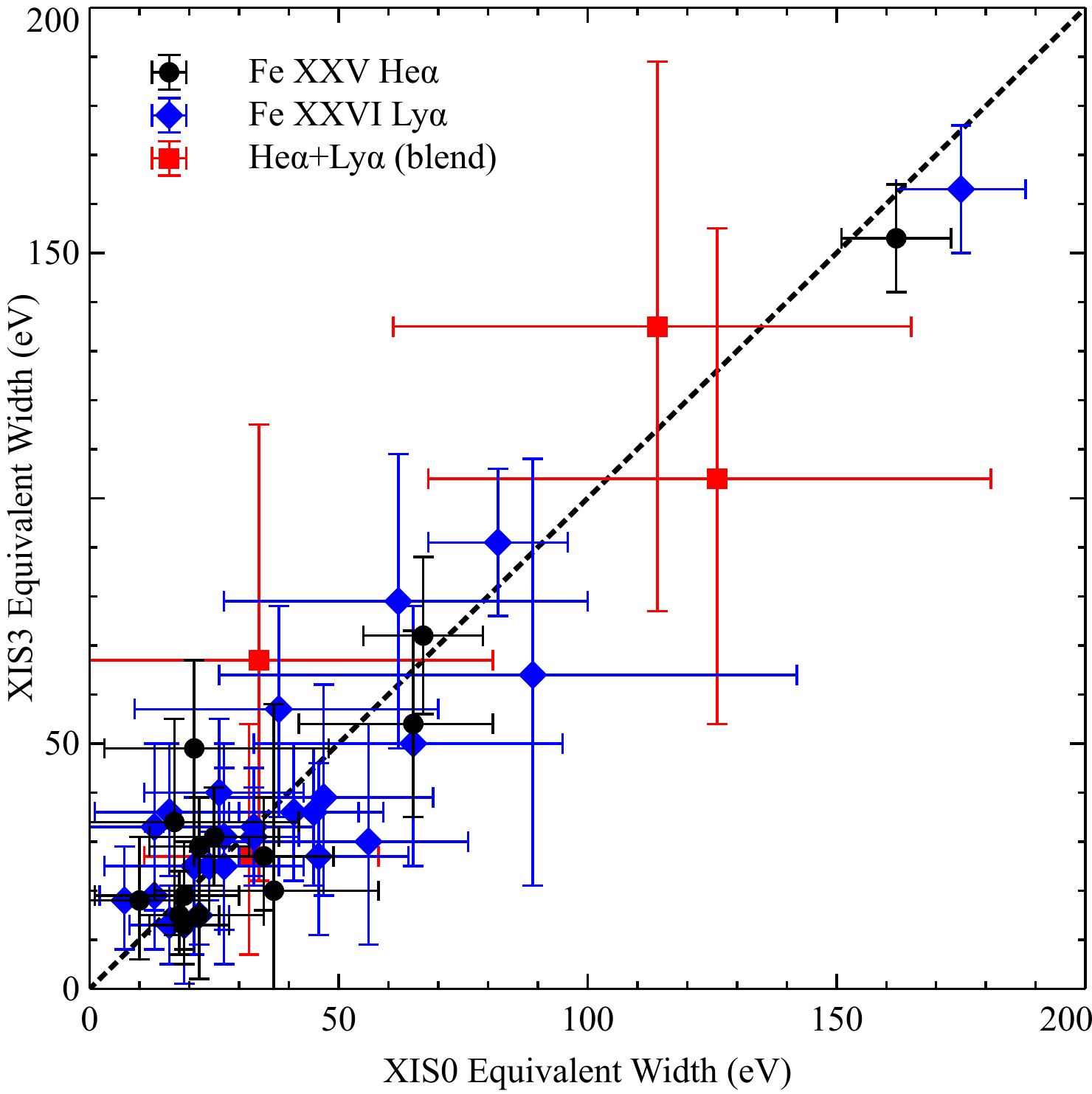}
\end{center}
\caption{\small Consistency check comparing the absorption line equivalent width measured with the XIS\,0 versus the XIS\,3 along with the $90\%$ error bars. The diagonal dashed line represents the location where the $EW$ is equal in both detectors. The circle (black), diamond (blue) and square (red) markers show the $EW$ of the \fexxv~\hea, \fexxvi~\lya and blended \fexxvxxvi lines, respectively. For sources with lines detected in more than one observation the mean $EW$ for each line in each detector and associated errors are plotted. Where available this consistency check was also carried out with the XIS\,2 which yielded  results which are entirely consistent with those displayed here. Colour version available online.}
\label{fig:consistency_check}
\end{figure}

In all sources the residuals detected in the individual XIS spectra have Gaussian parameters consistent with those found for the co-added FI spectrum. In Figure~\ref{fig:consistency_check} we show a comparison of absorption line equivalent width ($EW$), along with the 90\% error bars, as measured independently with the XIS\,0 and XIS\,3 detectors. The dashed diagonal line represents the position where the $EW$ is equal in each of the detectors; where the $EW$ is consistent at the 90\% level the error bars will overlap this line. In this plot \fexxv~\hea, \fexxvi~\lya and blended \fexxvxxvi lines are plotted separately, even in the cases where the lines are found to be part of a \fexxv pair. We find that the mean $EW$s measured with the XIS\,0 and XIS\,3 detectors are consistent at the 90\% level (i.e., the error bars cross the diagonal line in Figure~\ref{fig:consistency_check}) in all sources. Where possible this analysis was also carried out using the XIS\,2 spectra versus both the XIS\,0 and XIS\,3. Again, in all cases the mean parameters of the lines were always consistent at the 90\% level. The consistency check strongly suggests that the observed lines are real features and are not due to detector effects or due to background subtraction.

\subsection{Photoionisation modelling}
\label{subsec:photoionisation_modelling}
All absorption complexes detected at a $P_{\rm MC}\geq95\%$ confidence level were then fitted with the \xstar photoionisation code to probe the likely physical properties of the absorbing material, such as the column density $N_{\rm H}$, ionisation parameter $\xi$, and the red-shift of the absorber relative to the observer $z_{\rm o}$. Probing these parameters is important as not only do they allow the mean properties of the absorbing material to be determined but, through the use of simple geometric assumptions, they also permit an order-of-magnitude assessment of an absorbers likely radial distance from the ionising source, the mass outflow rate, and their global energetic output. Detailed discussion regarding the the absorber kinematics will be presented in a companion paper (Gofford et al., in prep).

In this work the absorption lines were fitted with an \xstar table that had an assumed illuminating continuum of $\Gamma=2.1$ and a micro-turbulent velocity ($v_{\rm turb}$) of $1000$\,km\,s$^{-1}$, which roughly corresponds to the full-width at half-maximum ($FWHM$) velocity width of a $\sigma=10$\,eV \fexxvi~\lya absorption line at a rest-frame energy of $6.97$\,keV. In cases where the lines were resolved (i.e., $\sigma>10$\,eV) we used \xstar tables with a $v_{\rm turb}$ values which closer matched the measured $FWHM$ velocity width of the observed profile we used; we found that tables with $v_{\rm turb}$ equal to $1000$\,km\,s$^{-1}$, $3000$\,km\,s$^{-1}$, $5000$\,km\,s$^{-1}$ and $10000$\,km\,s$^{-1}$ were sufficient to fit all systems in the sample. In 17/20 sources with significantly detected absorption lines all of the parameters of the \xstar table were allowed to vary freely. In APM\,08279+5255 and the joint spectra fits to NGC\,1365 and PDS\,456 some parameters were tied to prevent degeneracies between the column density and ionisation parameter (see Table~\ref{tab:hixiabs}). The absorber red-shift ($z_{o}$) as measured from the spectrum using \xstar is given in the observer frame. This is related to the intrinsic absorber red-shift in the source rest-frame ($z_{a}$) and the cosmological red-shift of the source ($z_{c}$) through the relation: $(1+z_{o})=(1+z_{a})(1+z_{c})$. From this, the intrinsic velocity of the outflow relative to the source ($\vout$) can then be calculated from the relativistic Doppler formula which ensures that the relativistic effects associated with both high red-shift sources and high-velocity outflows are correctly taken into account when inferring absorber outflow velocities relative to the source rest-frame.

\subsubsection{Line identifications}
\label{line_identifcations}
Before discussing the results of the \xstar fitting it is important that the {\it a priori} assumption that the absorption lines detected at $E\gtrsim6.6$\,keV are due to the velocity shifted resonance lines of \fexxv and \fexxvi is justified. Indeed, while the \ka transitions of \fexxv and \fexxvi, which are expected at mean rest-frame energies of $6.697$\,keV and $6.966$\,keV, respectively, are expected to be the strongest lines in the $\approx6.5-7.0$\,keV energy interval, there are several other atomic features at higher energies which may complicate the identification of blue-shifted absorption systems at $E>7$\,keV. For example, the K-shell edges from the various ionised species of Fe are found above 7\,keV, with energies ranging from $7.1$\,keV for neutral Fe up to $9.3$\,keV for \fexxvi. In the case of low-moderate ionisation Fe the K-edge is accompanied by higher order resonance line structure which can give the edge a subtle curved profile in CCD spectra rather than it simply being an abrupt drop in flux (\citealt{kallman:2004}). Furthermore, given its proximity to the rest-frame energy of the Fe\,{\sc xxv-xxvi} transitions the neutral edge at $7.1$\,keV could have an influence on the detection of lower velocity systems. 

Of the sources included in the sample CBS\,126, Mrk\,766 (OBSIDs 701035010, 701035020), NGC\,3227 (OBSIDs 703022010, 703022020, 703022030), NGC\,4051 (OBSIDs 700004010, 703023010) and NGC\,4151 all have at least one significantly detected absorption line at a rest-frame energy which is consistent at the 90\% level with the neutral Fe\,K-shell edge from the reflection component. The absorption in Mrk\,766, NGC\,3227, and NGC\,4051 is manifested by two lines with a common velocity shift equal to that expected for \fexxv~\hea and \fexxvi~\lya and are therefore unlikely to be affected by the presence of an edge. However, the absorption in CBS\,126 and NGC\,4151, which only comprises a single detected profile, could possibly be affected. Even so, given that the K-shell edge structure is already self-consistently accounted for by the \reflionx model and in \xstar, suggests that the residual profiles detected near the Fe\,K-shell edge in these sources is a real additional component rather than the residuals left by an inadequately fitted edge. Alternatively, the residuals could be due to a partially covering absorber with a low outflow velocity. We investigate this possibility further in Appendix~\ref{modelling_complexities}. Finally, we note that while NGC\,5506 shows evidence for both a highly ionised \fexxv emission line at $\sim6.63$\,keV the absorption trough detected at $\sim9.2$\,keV is not consistent with the \fexxv~K-shell edge and the two features are unlikely to be directly associated.  

In addition, there are a few other complications which could have an effect on the identification of blue-shifted absorption profiles at $E>7$\,keV. In particular, because different combinations of \xstar parameters (i.e., $N_{\rm H}$, $\log\xi$) can yield equivalent solutions to the $\chi^{2}$ distribution at differing red-shifts there can be a level of degeneracy when identifying the ion responsible for a discrete absorption trough. This effect is not particularly significant in sources where there are two absorption lines at a common velocity shift because the absorber is determined by the joint constraint of fitting both profiles, but it can become important in the instances where the absorption is manifested through a single trough. In these cases it can be difficult to determine whether an absorption line is due to, for example, \fexxv~\hea or \fexxvi~\lya, which can therefore influence the inferred velocity of the outflow and hence any inferred absorber energetics.

There are 10 sources (11 observations) in the sample in which a single absorption trough is detected at $E>7$\,keV at the $P_{\rm MC}\geq95\%$ significance (see Table~\ref{table:absorption_line_parameters}). In each of these observations we conducted a search for alternative \xstar solutions by stepping the red-shift of the highly-ionised \xstar table through the Fe\,K band of each spectrum. This enabled the $\chi^{2}$ minima for different \xstar solutions at different velocity-shifts, for various combinations of column density and ionisation parameter, to be mapped. This process is analogous to that used by \tombesiB and is useful when it come to ascertaining whether there were any alternative fits to the absorption lines, and to check for degeneracies between line identifications.

\begin{figure}
\includegraphics[width=8cm]{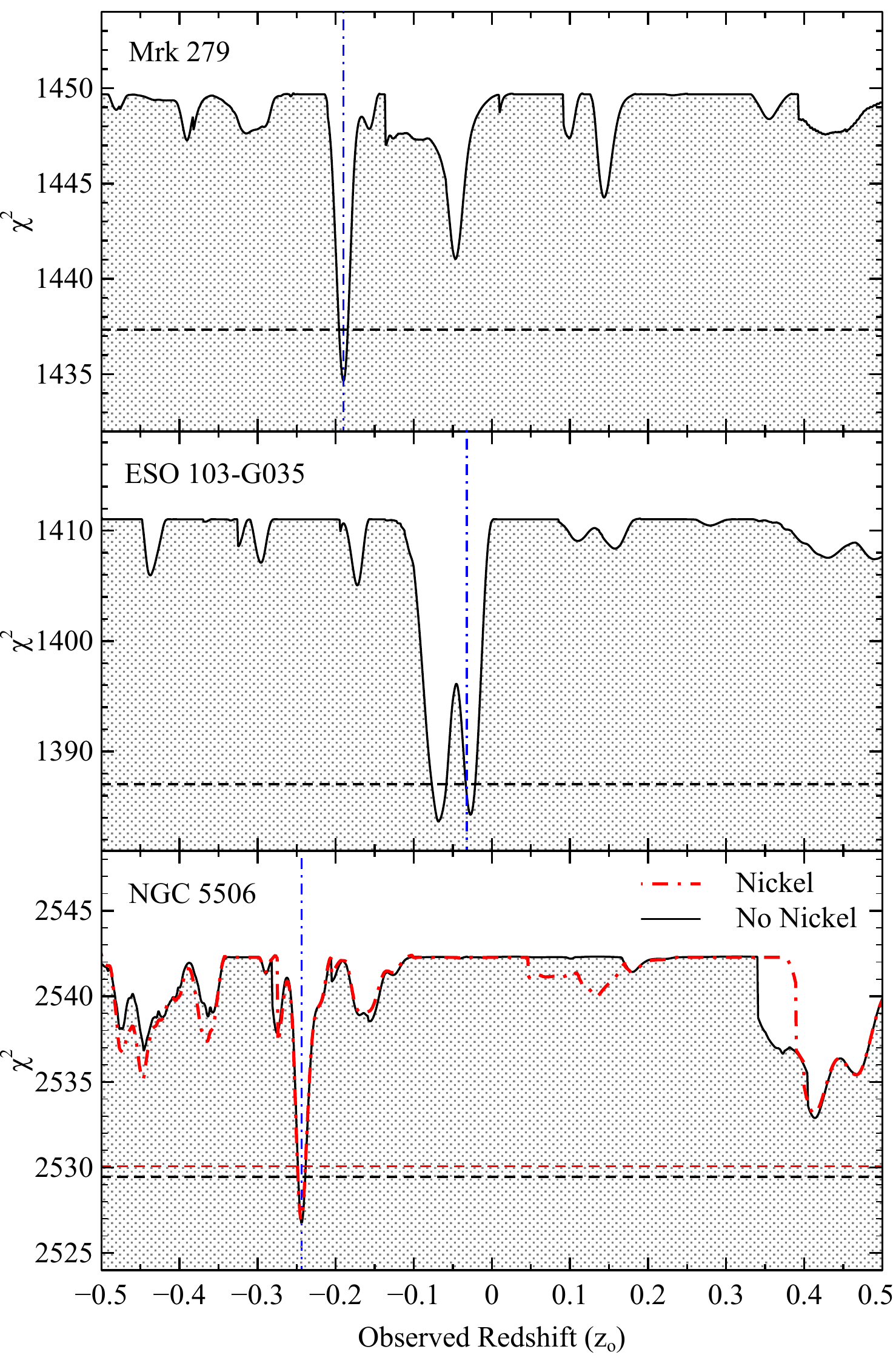}
\caption{Example plots showing the $\chi^{2}$ statistic versus observed red-shift of the \xstar absorber ($z_{o}$). Plots are shown for Mrk\,279 (top panel), ESO\,103-G035 (middle panel) and NGC\,5506 (bottom panel). In all panels the black solid line show the $\chi^{2}$ confidence contour and underlying $\chi^{2}$ distribution. In the bottom panel the black and red solid lines show the confidence contours for \xstar grids both with (dot-dashed red) and without (solid black) Nickel included at Solar abundances. In all panels the horizontal dashed lines show the 90\% confidence level for one interesting parameter, while the vertical dash-dotted line shows the best-fitting absorber red-shift. Colour version available online.}\label{chi2_comp}.
\end{figure}

For each source where only a single line is detected we took the best-fitting continuum model (including any necessary soft-band absorbers) and froze all model parameters bar those of the highly-ionised absorption table and the normalisation of the primary power-law. To map both blue and red-shifted $\chi^{2}$ minima the highly-ionised \xstar table was then stepped in $\Delta z_{o}=10^{-3}$ increments between $-0.5 \leq z_{o} < 0.5$, and contour plots produced after each run. Example contour plots for Mrk\,279 and ESO\,103-G035, which illustrate the cases of non-degenerate and degenerate \xstar solutions, respectively, are shown in Figure~\ref{chi2_comp}. In Mrk\,279 (Figure~\ref{chi2_comp}, top panel) the stepping process yields a single valid \xstar solution at a velocity shift corresponding to \fexxvi~\lya. However, in ESO\,103-G035 (Figure~\ref{chi2_comp}, middle panel) there are two degenerate $\chi^{2}$ minima found which are statistically equivalent at the 90\% level (i.e., $\Delta\chi^{2}\leq2.71$). The measured parameters imply that the lowest velocity solution is associated with \fexxvi~\ka, while the higher velocity solution is likely due \fexxv~\hea with some contamination from lower ionised species of iron. In this scenario it is difficult to unambiguously identify the responsible Fe ion and therefore gauge the appropriate outflow velocity. 

Of the 10 sources in which the Fe\,K absorber is manifested by a single trough only 4 show evidence for degenerate \xstar solutions (3C\,390.3, 4C\,+74.26, ESO\,103-G035 and NGC\,4151). For these cases we report the mean \xstar parameters as measured from all valid solutions in Table~\ref{tab:hixiabs} to account for the uncertainty in line identifications, with the associated errors being taken as half of the range between the absolute and minimum values. The remaining 7 sources have only a single $\chi^{2}$ minimum which corresponds to the line identification reported in Table~\ref{table:absorption_line_parameters}. In sources with degenerate \xstar solutions we report the most conservative line identification (i.e., associated with \fexxvi~\lya) in Table~\ref{table:absorption_line_parameters}.

\subsubsection{The influence of Nickel}
\label{nickel}
In principle, the presence of Nickel could also complicate our line identifications above $7.1$\,keV. In particular, the \ka lines of Ni\,{\sc xxvii} and Ni\,{\sc xxviii}, which are expected at rest-frame energies of $E\sim7.78$\,keV and $E\sim8.09$\,keV, respectively, could offer a more energetically conservative identification for the highest energy absorption lines by virtue of requiring a lower blue-shifted velocity. A total of six sources have absorption lines detected at energies in the regime where Nickel could complicate line identifications (Table~\ref{table:absorption_line_parameters}). Because of uncertainties in the atomic rates of Nickel\footnote{http://heasarc.gsfc.nasa.gov/xstar/docs/html/node74.html} its abundance is set to zero in \xstar by default which means that any additional solutions to the $\chi^{2}$ distribution due to it would not be picked up during the red-shift stepping. In order to robustly search for plausible \xstar solutions which may be associated with Nickel we therefore re-generated the \xstar tables using the same assumed parameters as before (see Section~\ref{subsec:photoionisation_modelling}), but this time with Nickel included at solar abundances, and again searched for alternative \xstar solutions using the method outlined previously.

The bottom panel of Figure~\ref{chi2_comp} shows a comparison of the contour plots obtained for NGC\,5506 when Nickel is set to zero or Solar in the \xstar table. As expected given the $Z_{\rm Ni}/Z_{\rm Fe}\sim0.05$ abundance ratio at solar abundances (i.e., \citealt{grevesse&sauval:1998}), in terms of valid \xstar solutions for NGC\,5506 there are no tangible differences between the two grids with \fexxvi~\lya being the only valid identification in both cases. Similar is also true for the other 5 sources with absorption lines detected at $E\gtrsim7.78$\,keV, with no valid solutions corresponding to Nickel being found in any of the cases. Moreover, it should be noted that in order to achieve the measured equivalent width of the absorption lines the required column density would be unreasonably large, i.e., $\lognh>24$, which would absorb the observed continuum far beyond that which is observed in the spectra.

Therefore, bar the extraordinary case where it is $\gtrsim20$ times over-abundant relative to solar values, Ni has a negligible influence on the Fe\,K band and identifying the observed absorption lines with K-shell resonance lines of Fe is secure. We report the best-fitting \xstar parameters, including the measured outflow velocities relative to the host galaxy, in Table~\ref{tab:hixiabs}. The statistical significances as per the $\Delta\chi^{2}$ and the F-test are also reported; in all cases the addition of an \xstar grid to fit the observed lines improves the fit by at least the 99\% level by the F-test. Note that since some sources have either more than one absorption complex detected in a single spectrum (i.e., APM\,08279+5255 and PDS\,456), or have absorption detected in more than one epoch (i.e., Mrk\,766), we also report the mean absorption parameters on a per source basis. The mean values are used in all subsequent analysis to prevent sources with multiple Fe\,K absorber detection from over-weighting the resultant parameter distributions which are discussed in Section~\ref{subsec:absorber_properties}.

\begin{table*}
	\footnotesize
	\begin{minipage}{134mm}
		\caption{{\textsc XSTAR} parameters for Fe\,K absorbers. {\sl Notes:} (1) Source name;
					 (2) \suzaku observation ID;
					 (3) Logarithm of absorber column density, in units of $\rm{cm^{-2}}$;
					 (4) Logarithm of the ionisation parameter, in units of $\rm{erg\,cm\,s^{-1}}$;
					 (5) Measured absorber outflow velocity, in units of $v/c$. Negative outflow velocities indicate a net red-shift;
					 (6) Change in $\chi^{2}/\nu$ when absorber removed from the best-fit model;
					 (7) Corresponding absorber significance according to the F-Test.}
		
		\begin{tabular}{@{}lrllrrr}

	\toprule
	\multicolumn{1}{c}{Source} & \multicolumn{1}{c}{OBSID} & \multicolumn{1}{c}{$\log N_{\rm H}$} & \multicolumn{1}{c}{$\log\xi$} & 
	\multicolumn{1}{c}{$v_{\rm out}$} & \multicolumn{1}{c}{$\Delta\chi/\Delta\nu$} & \multicolumn{1}{c}{$P_{\rm F}$} \\
	\multicolumn{1}{c}{(1)} & \multicolumn{1}{c}{(2)} & \multicolumn{1}{c}{(3)} & \multicolumn{1}{c}{(4)} & \multicolumn{1}{c}{(5)} & 
	\multicolumn{1}{c}{(6)} & \multicolumn{1}{c}{(7)} \\
	\midrule
	
	3C\,111 			& 703034010 & $23.18^{+0.19}_{-0.24}$ & $4.63^{+0.22}_{-0.18}$ & $0.039\pm0.005$ & $18.7/3$ 
						& $99.97$ \\[0.5ex]
						& 705040010 & $22.43^{+0.20}_{-0.31}$ & $4.13^{+0.27}_{-0.14}$ & $0.105\pm0.008$ & $11.6/3$ & $99.08$ \\[0.5ex]
						& $\langle$mean$\rangle$ & $22.95^{+0.28}_{-0.39}$ & $4.45^{+0.35}_{-0.23}$ & $0.072\pm0.005$ & \na & \na \\[0.5ex]
	3C\,390.3$^{\bigtriangleup}$ & 701060010 & $>23.68$ & $>5.46$ & $0.145\pm0.007$ & $11.7/2$ & $>99.11$\\[0.5ex]
	4C\,+74.26$^{\bigtriangleup}$ & 702057010 & $>21.79$ & $4.06^{+0.45}_{-0.45}$ & $0.185\pm0.026$ & $14.6/3$ 
						& $99.76$\\[0.5ex]
	APM\,08279+5255 	& stacked[all] & $23.00^{+0.07}_{-0.16}$ & $3.41^{+0.08}_{-0.08}$ & $0.139\pm0.012$ & $27.3/3$ 
						& $>99.99$\\[0.5ex]
						&		  & $23.00^{*}$ & $3.61^{+0.23}_{-0.14}$ & $0.431\pm0.019$ & $14.3/2$ & $99.59$\\[0.5ex]
						& $\langle$mean$\rangle$ & $23.00^{+0.12}_{-0.12}$ & $4.13^{+0.23}_{-0.17}$ & $0.285\pm0.011$ & \na & \na \\[0.5ex]
	CBS\,126 			& 705042010 & $>23.73$ & $4.77^{+0.26}_{-0.17}$ & $0.012\pm0.006$ & $28.9/3$ & $>99.99$\\[0.5ex]
	ESO\,103-G035$^{\bigtriangleup}$ & 703031010 & $>21.90$ & $4.36^{+1.19}_{-1.19}$ & $0.056\pm0.025$ & $33.2/3$ & $>99.99$\\[0.5ex]
	MCG\,-6-30-15 		& stacked[all] & $22.16^{+0.08}_{-0.08}$ & $3.64^{+0.05}_{-0.06}$ & $0.007\pm0.002$ & $103.0/3$ 
						& $>99.99$\\[0.5ex]
	MR\,2251-178 		& 705041010 & $21.54^{+0.22}_{-0.20}$ & $3.26^{+0.12}_{-0.12}$ & $0.137\pm0.008$ & $42.4/3$ 
						& $>99.99$\\[0.5ex]
	Mrk\,279 			& 704031010 & $23.38^{+0.25}_{-0.31}$ & $4.42^{+0.15}_{-0.27}$ & $0.220\pm0.006$ 
						& $14.5/3$ & $99.77$\\[0.5ex]
	Mrk\,766 			& 701035010 & $22.64^{+0.20}_{-0.27}$ & $4.02^{+0.21}_{-0.13}$ & $0.061\pm0.008$ 
						& $14.5/3$ & $99.71$\\[0.5ex]
						& 701035020 & $22.76^{+0.10}_{-0.13}$ & $3.67^{+0.06}_{-0.06}$ & $0.017\pm0.004$ 
						& $63.2/3$ & $>99.99$\\[0.5ex]
						& $\langle$mean$\rangle$ & $22.70^{+0.15}_{-0.19}$ & $3.86^{+0.13}_{-0.10}$ & $0.039\pm0.006$ & \na & \na\\[0.5ex]
	NGC\,1365 			& 702047010 & $23.92^{+0.03}_{-0.03}$ & $3.88^{+0.06}_{-0.07}$ & $0.014\pm0.001$ & $4125.1/3$ 
						& $>99.99$\\[0.5ex]
						& 705031010 & $23.40^{+0.10}_{-0.12}$ & $3.88^{*}$ & $0.002\pm0.002$ & $191.8/3$ & $>99.99$\\[0.5ex]
						& $\langle$mean$\rangle$ & $23.73^{+0.06}_{-0.06}$ & $3.88^{+0.06}_{-0.07}$ & $0.008\pm0.001$ & \na & \na \\[0.5ex]
	NGC\,3227       	& 703022010 & $22.74^{+0.11}_{-0.13}$ & $3.89^{+0.08}_{-0.11}$ & $<0.002$ & $63.1/3$ 
						& $>99.99$\\[0.5ex]
						& 703022030 & $22.59^{+0.16}_{-0.20}$ & $3.86^{+0.13}_{-0.14}$ & $0.007\pm0.004$ & $21.0/3$ 
						& $>99.99$\\[0.5ex]
						& 703022050 & $22.62^{+0.12}_{-0.14}$ & $3.89^{+0.11}_{-0.12}$ & $0.011\pm0.004$ & $37.9/3$ 
						& $>99.99$\\[0.5ex]
						& $\langle$mean$\rangle$ & $22.66^{+0.13}_{-0.15}$ & $3.88^{+0.19}_{-0.21}$ & $0.005\pm0.004$ & \na & \na \\[0.5ex]
	NGC\,3516 			& 100031010 & $22.56^{+0.14}_{-0.17}$ & $3.84^{+0.11}_{-0.10}$ & $0.004\pm0.002$ & $141.9/3$ 
						& $>99.99$\\[0.5ex]
	NGC\,3783			& 701033010 & $21.75^{+0.15}_{-0.20}$ & $3.50^{+0.13}_{-0.08}$ & $<0.005$ 
						& $16.2/3$ & $99.90$\\[0.5ex]
						& 704063010 & $21.83^{+0.09}_{-0.10}$ & $3.45^{+0.06}_{-0.04}$ & $<0.008$ 
						& $48.2/3$ & $>99.99$\\[0.5ex]
						& $\langle$mean$\rangle$ & $21.79^{+0.16}_{-0.23}$ & $3.48^{+0.14}_{-0.90}$ & $<0.007 $ & \na & \na\\[0.5ex]
	NGC\,4051 			& 700004010 & $22.78^{+0.08}_{-0.09}$ & $4.05^{+0.05}_{-0.05}$ & $0.020\pm0.002$ & $80.3/3$ 
						& $>99.99$\\[0.5ex]
						& 703023010 & $22.80^{+0.11}_{-0.13}$ & $4.94^{+0.09}_{-0.09}$ & $0.015\pm0.002$ & $20.8/3$ 
						& $99.99$\\[0.5ex]
						& $\langle$mean$\rangle$ & $22.79^{+0.10}_{-0.11}$ & $4.00^{+0.07}_{-0.07}$ & $0.018\pm0.001$ & \na & \na \\[0.5ex]			
	NGC\,4151$^{\bigtriangleup}$ & 701034010 & $>21.74$ & $3.69^{+0.64}_{-0.64}$ & $0.055\pm0.023$ 
						& $54.8/3$ & $>99.99$\\[0.5ex]
	NGC\,4395 			& 702001010 & $22.84^{+0.21}_{-0.25}$ & $3.92^{+0.16}_{-0.15}$ & $<0.001$ & $19.1/3$ & $99.97$\\[0.5ex]
	NGC\,5506 			& stacked[all] & $23.22^{+0.19}_{-0.28}$ & $5.04^{+0.29}_{-0.17}$ & $0.246\pm0.006$ & $16.8/3$ 
						& $99.94$\\[0.5ex]
	PDS\,456			& 701056010 & $23.04^{+0.08}_{-0.08}$ & $4.19^{+0.15}_{-0.14}$ & $0.253\pm0.008$ & $20.4/3$ & $>99.99$\\[0.5ex]
						& 			& $23.04^{*}$ & $4.19^{*}$ & $0.292\pm0.009$ & $12.8/1$ & $99.68$\\[0.5ex]
						& 705041010 & $23.04^{*}$ & $3.93^{+0.02}_{-0.02}$ & $0.253^{*}$ & $20.4/3$ 
						& $>99.99$\\[0.5ex]
						&			& $23.04^{*}$ & $3.93^{*}$ & $0.292^{*}$ & $11.6/1$ & $99.85$\\[0.5ex]
						& 	$\langle$mean$\rangle$ 	& $23.04^{+0.08}_{-0.08}$ & $4.06^{+0.15}_{-0.14}$ & $0.273\pm0.006$ & \na & \na\\[0.5ex]
	SW\,J2127.4+5654 	& 702122010 & $22.78^{+0.23}_{-0.34}$ & $4.16^{+0.29}_{-0.13}$ & $0.231\pm0.006$ 
						& $11.5/3$ & $99.24$\\[0.5ex]

	 	\bottomrule
		\end{tabular}\\[0.5ex]
		$^{*}$ indicates a parameter was tied during spectral fitting;\\
		$^{\bigtriangleup}$ denotes sources with \xstar solutions which are degenerate at the 90\% level (i.e., the $\chi^{2}$ statistic for the \fexxv~\hea and \fexxvi~\lya solutions differ by $\Delta\chi^{2}\leq2.71$). In these cases the reported values are averaged over the solutions and the errors are inferred as half the range between the absolute maximum and minimum values. Significances reported in columns (6) and (7) refer to the least significant of the valid solutions. See text for further details.\\
		\label{tab:hixiabs}
	\end{minipage}
\end{table*}


\section{Results}
\label{sec:results}

\begin{table}
\begin{center}
\begin{minipage}{5cm}
\caption{Fraction of detected outflows}
\begin{tabular}{l c c c}
	\toprule
	Line(s) & Sources(spectra)\\
	\midrule 
	\fexxv~\hea & 2\,(3)  \\
	\fexxvi~\lya & 9\,(10)  \\
	\fexxvxxvi & 7\,(12) \\
	Multi $\vout$ & 2\,(3) \\
	\midrule
	{\bf Total} & 20\,(28) \\
	\bottomrule
\end{tabular}
\label{table:outflow_phenomenology}
\end{minipage}
\end{center}
\end{table}

\subsection{Line detection rate and phenomenology}
\label{subsec:line_detection_rate}
A total of 20/51 sources (in 28/73 fitted spectra; both corresponding to $\sim40\%$ of the total sample) show evidence for highly-ionised absorption lines in their \suzaku spectra at a Monte Carlo significance of $P_{\rm MC}\geq95\%$, with 18/20 of these outflows also robustly detected at $P_{\rm MC}\geq99\%$ significance (see Table~\ref{table:absorption_line_parameters}). Of the 28 observations with individually detected Fe\,K absorbers there are 10 which are consistent with having \fexxvi~\lya as the dominant Fe ion, 3 where \fexxv~\hea (and/or lower ionisation species of iron) is the main contributor, 12 absorbers with both \fexxv~\hea and \fexxvi~\lya lines with a common outflow velocity, and a further 3 that have two absorption components with different outflow velocities. Taking into account that some sources have absorption lines detected in more than one observation, on a per-source basis this corresponds to: 2/20 having \fexxv~\hea absorption, $9$ with \fexxvi~\lya absorption\footnote{In reality the number of \fexxvi~\lya systems could be somewhat lower than this value due to the 4 AGN where the absorption is equally well fitted by \fexxv~\hea at a slightly lower velocity (i.e., see Section~\ref{line_identifcations}).}, $7$ with both \fexxv and \fexxvi~\lya absorption, and $2$ having multi-$\vout$ systems (out of 20). Interestingly, 8/9 sources with just a single \fexxvi~\lya absorption line have outflow velocities which exceed the $\vout\sim10000$\,km\,s$^{-1}$ threshold employed by \tombesiA when identifying `UFO' systems, while only 1 of the 6 sources with \fexxvxxvi absorption exhibit a mean velocity which exceed this threshold value. This is consistent with the view that higher ionised outflows originate closer to the central AGN and therefore have a higher outflow velocity, and is a point which will be further discussed in a companion paper (Gofford et al. 2012; in preparation).

\begin{figure}
\begin{center}
\includegraphics[width=8cm]{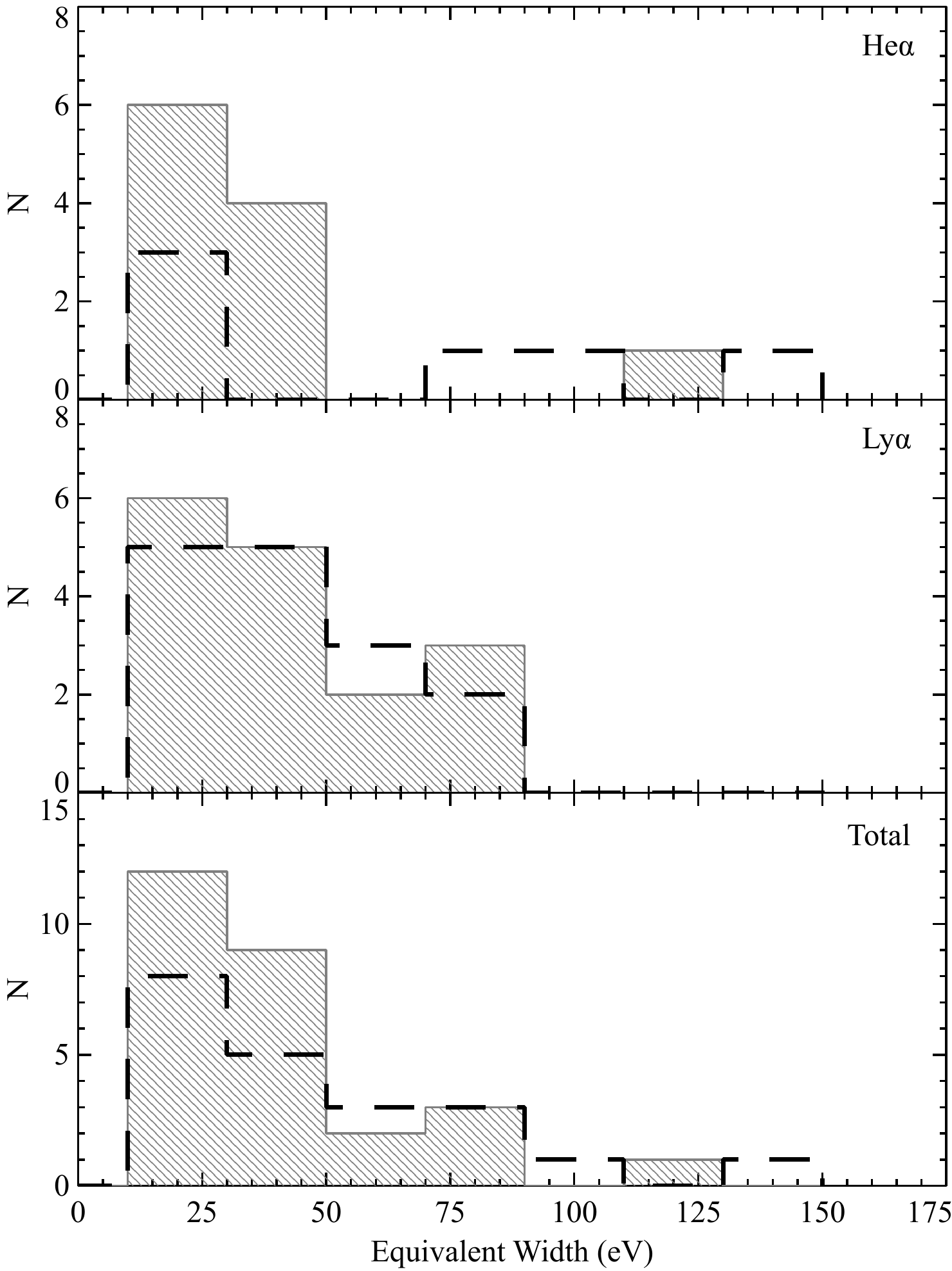}
\caption{\small Histogram showing the mean equivalent widths of detected highly-ionised absorption lines. The top and middle panel show the distributions for \fexxv~\hea and \fexxvi~\lya transitions, respectively, while the bottom panel shows the total distribution of all detected 1s$\rightarrow$2p absorption lines including those which are detected as a blend of the two \ka transitions. For sources with more than one line detected in multiple observations the mean equivalent widths have been used. The grey diagonally shaded area and the dashed black line show the distributions found in this work and with \xmm, respectively. The \xmm distributions independently created using the values listed in table A.3 in T10A.}
\label{fig:ewidth_histogram}
\end{center}
\end{figure}

Histograms of $EW$ for the absorption lines are shown in the top and middle panels of Figure~\ref{fig:ewidth_histogram}, respectively. Note that in these plots the individual line profiles in a \fexxvxxvi pair are considered separately for the sake of clarity, and only those lines which have been individually detected at a Monte Carlo significance of $P_{\rm MC}\geq95\%$ are considered. The histogram in the bottom panel shows the $EW$ distribution for \textit{all} statistically significant absorption lines, including those which are consistent with being a blend of the \fexxvxxvi~\ka transitions. Measured $EW$s span from a few tens of eV up to $\sim130-140$\,eV (Figure~\ref{fig:ewidth_histogram}, bottom panel), in a distribution which is consistent with the curve of growth analysis conducted by \tombesiB. The mean $EW$ of the \fexxv~\hea and \fexxvi~\lya absorption lines are $\sim36$\,eV and $\sim38$\,eV, respectively, with the total mean $EW$ for all detected profiles (i.e., included blended ones) is $\sim40$\,eV. Importantly, the observed distribution of absorption line $EW$s is broadly consistent with that found by \tombesiA using \xmm (marked with a dashed black line in Figure~\ref{fig:ewidth_histogram}).

As noted by \tombesiA a useful quantity to calculate is the global probability that all of the observed lines are due to purely statistical shot noise, which can be done using the binomial distribution. For an event with null-probability $p$ the chance probability of the event $n$ happening in $N$ trials is given by:
\begin{equation}
P(n; N, p) = \frac{N!}{n!(N-n)!} p^{n}(1-p)^{N-n}
\end{equation}
For $n=20$ Fe\,K-band absorption line systems detected in $N=51$ sources at a Monte Carlo significance of $P_{\rm MC}\geq95\%$ the probability of one of these absorption systems being due to shot noise can be taken as $p<0.05$. On a per-source basis the probability of all of the observed absorption systems being associated with noise is then very low, with $P<2\times10^{-13}$, which further reduces to $P<5\times10^{-18}$ when considering the fact that some sources have lines detected in more than one epoch. This suggests that the observed lines are very unlikely to be associated with simple statistical fluctuations in the spectra. Moreover, it is important to remember that because the vast majority of absorption lines are detected at $P_{\rm MC}>95\%$ significance these probabilities only represent conservative lower limits on the global probability of all lines being false detections.

\subsection{Absorber properties}
\label{subsec:absorber_properties}
\begin{figure}
\begin{center}
\includegraphics[width=8cm]{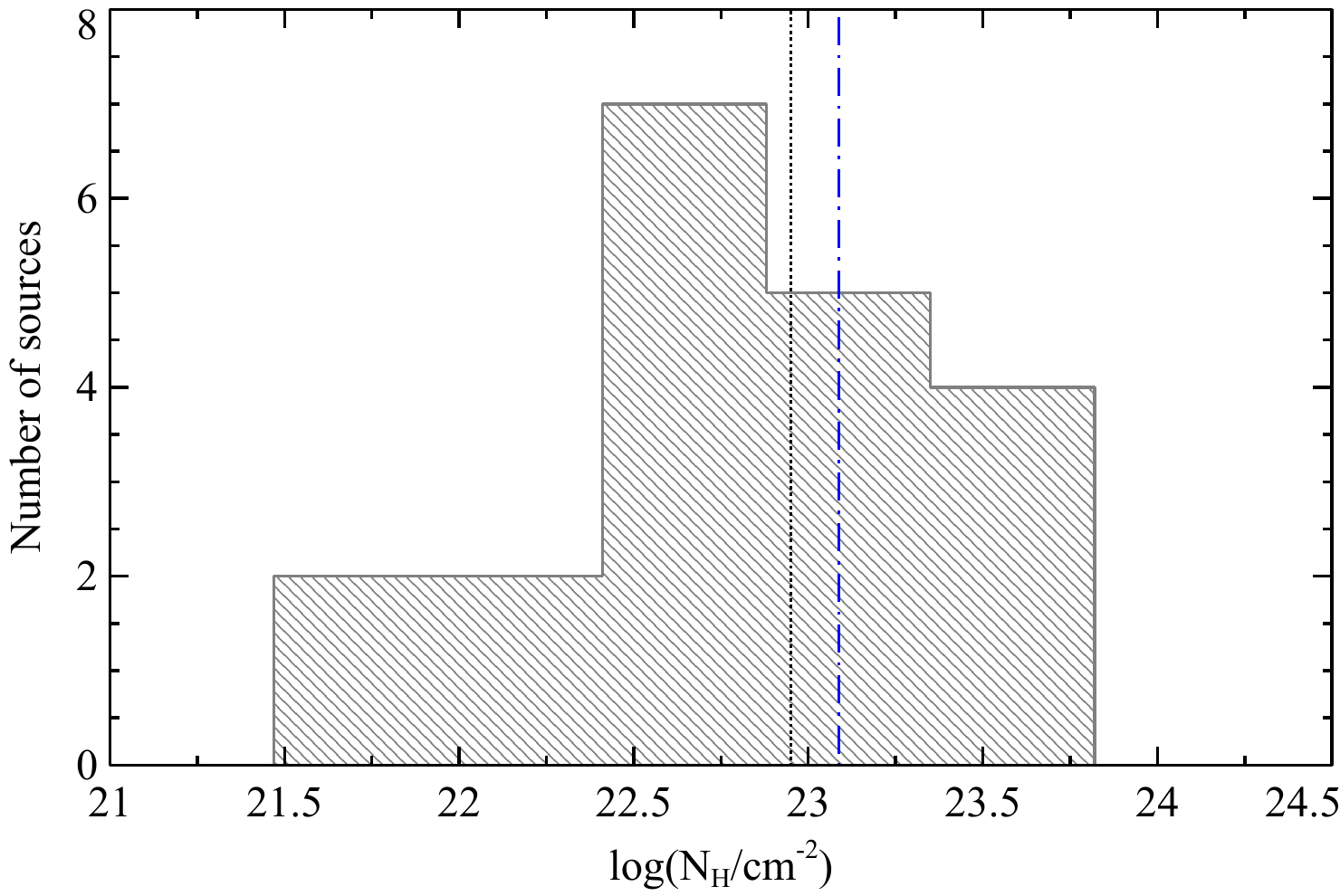}
\includegraphics[width=8cm]{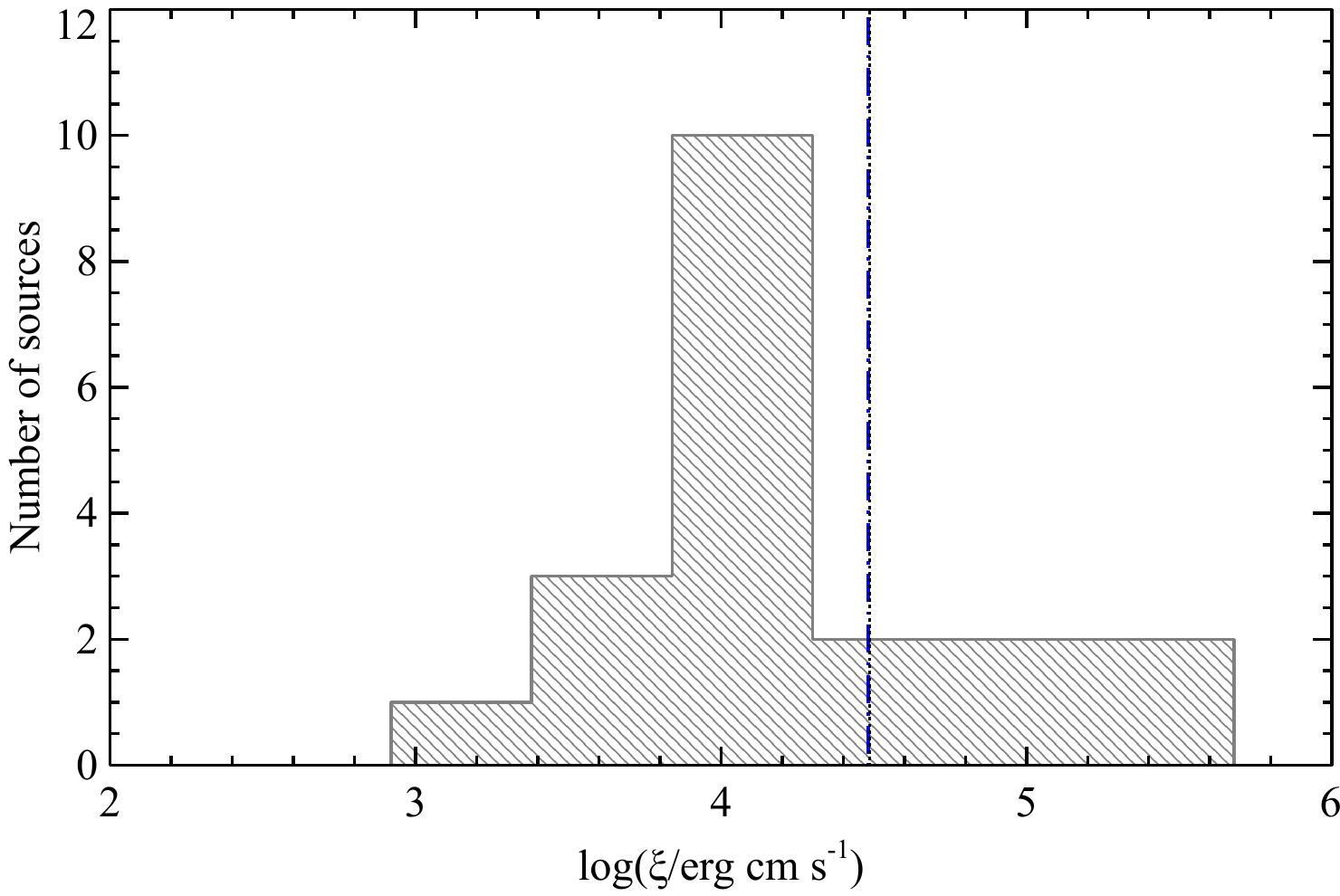}
\includegraphics[width=8cm]{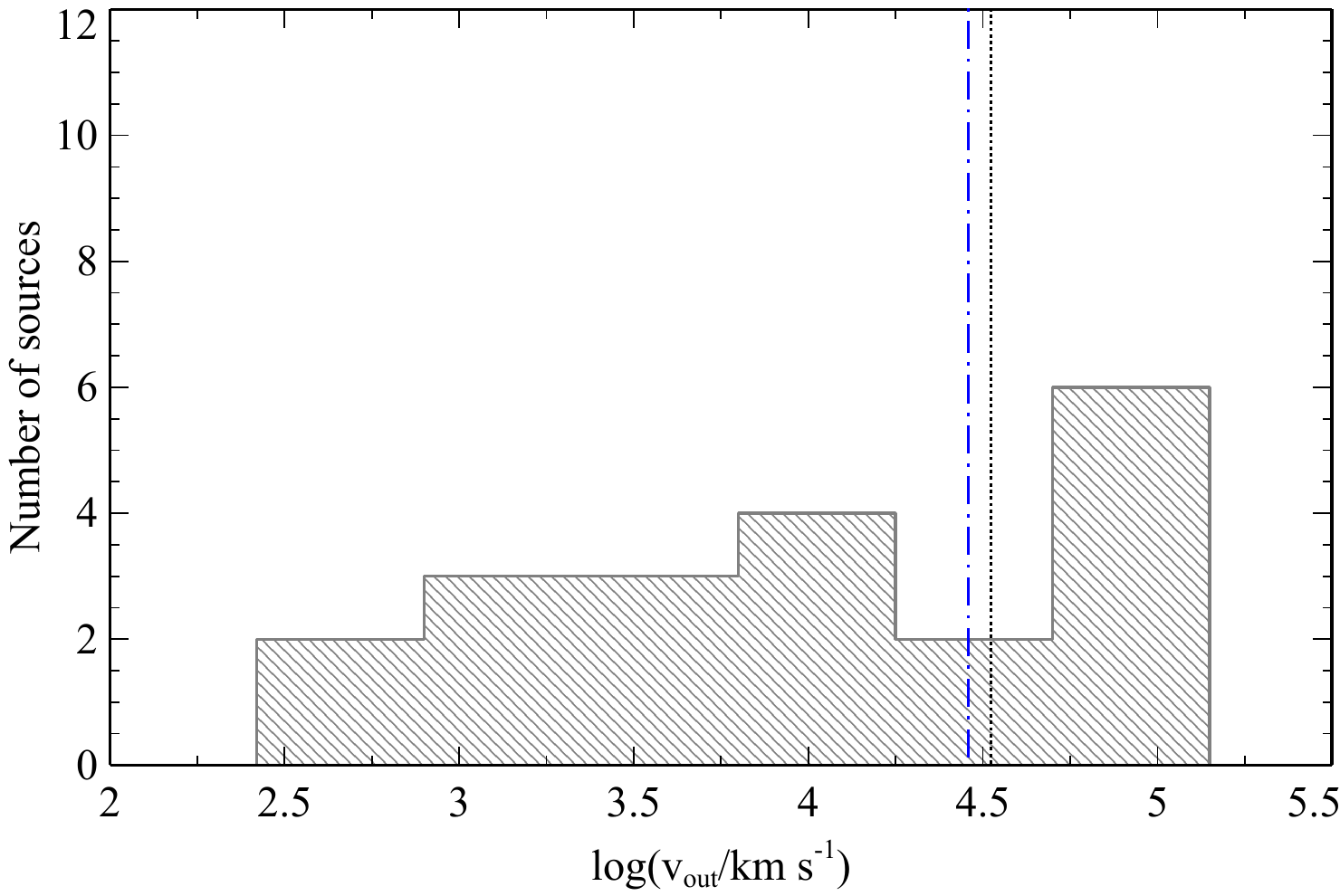}
\caption{\small Histogram showing the overall distributions of mean absorber parameters for each source: (a) logarithm of the mean column density; (b) logarithm of the mean ionisation parameter; (c) logarithm of the mean outflow velocity. 
The dot-dashed (blue) and dotted (black) vertical lines indicate the mean value of the \suzaku and \xmm analyses, respectively.}
\label{fig:absorber_histograms}
\end{center}
\end{figure}

The \xstar parameter distribution as measured with \suzaku, again plotted using the mean absorber parameters averaged over all observations, are shown in Figure~\ref{fig:absorber_histograms}. Overall, the general distributions and mean parameter values are broadly consistent with those found by \tombesiA using \xmm. The Fe\,K absorbers detected with \suzaku cover a wide range of column densities, ranging from $21.5 < \lognh \leq 24.0$, with a peak in the distribution at $\lognh\approx22-23$. As shown by the dot-dashed (blue) and dotted (black) lines, which show the mean $\lognh$ value as found with \suzaku and \xmm, respectively, the mean is $\lognh\approx23$ for both samples. From the middle panel the ionisation parameters are in the interval $2.5 < \logxi \leq 6$ with the significant fraction of \fexxvxxvi pair systems, which persist over only a relatively narrow range in ionisation parameter (see \tombesiB curve of growth analysis), leading to a peak in the distribution at $\logxi \approx 4$. Again, as shown by the vertical lines the mean ionisation parameter in both samples is almost identical at $\logxi \approx 4.5$.
The detection of relatively low-ionisation material in the Fe\,K band, i.e., $\logxi\sim2.5-3.0$, is particularly interesting and suggests that high velocity absorption could feasibly be detected at softer X-ray energies through weak, moderately ionised, iron lines. Moreover, the detection of a small fraction of absorbers with $\logxi\geq5$ in both this work and in the \tombesiA sample raises the possibility that material may be present in some sources which is so highly ionised that even iron is not detectable through spectroscopy. If this is the case then the fraction of sources with Fe\,K absorption ($\sim40\%$) may represent a lower limit on the number of sources with intrinsic nuclear outflows along the line-of-sight.

The $\logvout$ distribution (bottom panel) appears to be relatively continuous over a broad range of velocities; ranging from as low as $\vout<1,500$\,km\,s$^{-1}$ up to $\vout\approx100,000$\,km\,s$^{-1}$. The median outflow velocity is $\vout\sim17,000$\,km\,s$^{-1}$ ($\sim0.056$\,c) Ninety per cent of the detected outflows have $\vout\geq1,500$\,km\,s$^{-1}$ which makes the absorption detected at Fe\,K almost systematically faster than the traditional soft X-ray warm absorber. Only NGC\,3227, NGC\,4395 and NGC\,3783 have Fe\,K outflows which are consistent with having no outflow velocity. From Table~\ref{table:velocity_comparison_table} the distributions of outflow velocity appear to be very similar between both the \suzaku and \xmm samples, with the bulk of outflows in both samples having $\vout>10,000$\,km\,s$^{-1}$. The \suzaku sample does appear to have slightly more low-intermediate velocity systems but, given the low number statistics involved, the differences are probably not significant. A interesting possibility is that both \suzaku and \xmm may be subject to an instrumental bias against the detection of low-velocity absorption systems. At the energy resolution of the EPIC and XIS CCDs the presence of \fexxv and/or \fexxvi emission lines could mask the signature of low-velocity absorption systems and thereby introduce a selection effect against their detection. The only way to reveal the presence of such absorption systems would be through future observations with, for example, the calorimeter aboard \textit{Astro-H}, which would have sufficient resolution to distinguish between individual emission and absorption components at low velocity. 

Even so, the similarity between all of the measured \suzaku and \xmm parameter distributions can be quantitatively assessed using the Kolmogorov-Smirnov two-sample test (K-S Test) which uses the maximum differences between the cumulative fractional distribution of two sets of data to determine the probability that they are both drawn from the same parent sample. In this case such a test can be used to quantify the level at which the column density, ionisation parameter and outflow velocity distributions as measured with both \suzaku and \xmm are in agreement. For a null hypothesis that each of the \suzaku and \xmm distributions are drawn from the same parent sample we are unable to conclusively rule out the null hypothesis in any of the three cases at greater than the 90\% confidence level.

\begin{table}
\begin{center}
\begin{minipage}{7cm}
\caption{Outflow velocity comparison}
\begin{tabular}{l c c}
	\toprule
	Velocity (km\,s$^{-1}$) & \suzaku & \xmm\\
	\midrule 
	No outflow & 2/20 & 2/19 \\
	$0<\vout\leq10,000$ & 6/20 & 2/19 \\
	$\vout>10,000$ & 12/20 & 15/19 \\
	$\vout\geq30,000$ & 8/20 & 9/19 \\
	\bottomrule
\end{tabular}
\label{table:velocity_comparison_table}
\end{minipage}
\end{center}
\end{table}

\section{Discussion}
\subsection{Detailed sample comparison}
\label{sourcebysource}

\subsubsection{Radio-quiet sources}
\label{radio_quiet_sources}
In this work we have detected Fe\,K outflows in 17 radio-quiet AGN and 3 in radio-loud. Of the detections in radio-quiet sources, 9/17 also have had observations included in the \xmm outflow sample; with 6/9 of these cases having Fe\,K outflows confirmed by the \tombesiA analysis (i.e., Mrk\,766, NGC\,4051, NGC\,3516, NGC\,3783, NGC\,4151, Mrk\,279). In the remaining 3 sources, namely NGC\,5506, MCG\,-6-30-15 and NGC\,3227, the \suzaku outflow detections are not confirmed by \tombesiA. Even so, given that in the case of NGC\,3227 the Fe\,K absorption lines are detected at $P_{\rm MC}>99.9\%$ significance in 3 separate \suzaku observations, and in MCG\,-6-30-15 the \fexxv~\hea and \fexxvi~\lya absorption lines present are both statistically resolved at $P_{\rm MC}>99.9\%$ confidence, the \suzaku line detections alone still imply very robust absorber detections in both of these sources. A further 3 sources with absorption detected in their \suzaku spectra have had statistically significant absorption detected in \xmm observations which were not included in the \tombesiA outflow sample (NGC\,1365, \citealt{risaliti:2005}; PDS\,456, \citealt{reeves:2003, reeves:2009}; APM\,08279+5255, \citealt{chartas:2009}). Furthermore, weak hard-band absorption has been noted in the \xmm spectra of MCG\,-6-30-15 by several authors (e.g., \citealt{fabian:2002, vaughan:2004, nandra:2007, miller:2008}). In particular, \cite{nandra:2007} reported the presence of moderately significant (i.e., just below $99\%$ confidence) \fexxv~\hea absorption in XMM\_OBSID: 00297401010, but this detection was not significantly replicated by \tombesiA and highlights the need for detailed broad-band spectral models when assessing for the presence of highly-ionised absorption lines.

Therefore a total of 9/17 ($\sim53\%$) of the sources with outflows detected by \suzaku have also previously been detected in \xmm data. In addition to these, the absorbers in MR\,2251-178 and MCG\,-6-30-15 are also corroborated on the basis of their \chandrahetg data. \cite{gibson:2005} detected a resolved \fexxv~\hea absorption line at $E\sim7.25$\,keV in the HETG spectrum MR\,2251-178 ($\vout\sim13,000$\vunit) while \cite{young:2005} found variable absorption lines due to \fexxv and \fexxvi~\lya in MCG\,6-30-15 with $\vout\sim2,000$\vunit. In both cases the outflow velocities are consistent with those found in our \suzaku analysis, and the detection in the HETG spectrum of MCG\,-6-30-15 overcomes the ambiguities which remained on the basis of the \xmm data and independently corroborates the \suzaku line detections. Fe\,K absorption has also been reported in the \chandra spectrum of APM\,08279+5255 (\citealt{chartas:2002}) making it the only source in the sample to have its Fe\,K outflow independently detected in all 3 observatories at a high significance level.

Thus, a total of 11/17 ($\sim65\%$) of the sources with Fe\,K outflows detected with \suzaku have also had outflows reported elsewhere in the literature with either \xmm or \chandra. Of the remaining 6 sources, SWIFT\,J2127.4+5654 has been observed by \xmm twice: once in 2009 ($\sim24$\,ks) and once in 2010 ($\sim131$\,ks), although neither observation has been published at the time of writing. CBS\,126 have not been observed by either \xmm or \chandra, while ESO\,103-G035 only has a 13\,ks \xmm observation available which is insufficient to test for Fe\,K absorption. This leaves only 3 AGN, NGC\,5506, NGC\,3227 and NGC\,4395, with sufficient observations available which do not have independent confirmations with other observatories; although, as discussed previously, NGC\,3227 does have multiple detections with \suzaku which suggests a robust detection. 

\subsubsection{Radio-loud sources}
\label{radio_loud_sources}
There are 6 Broad-line radio galaxies (BLRGs) in our \suzaku sample. This includes all of the sources which were part of the \tombesiBLRG radio-loud UFO case study (i.e., 3C~111, 3C~120, 3C~445, 3C~390.3 and 3C~382), as well as 4C+74.26 which was not part of their analysis. \tombesiBLRG state that 3/5 of the BLRGs in their sample, namely 3C~111, 3C~120 and 3C~390.3, show evidence for highly-ionised absorption lines in their \suzaku spectra; with that found in 3C~111 also being confirmed in more recent subsequent observations (\citealt{tombesi:2011b})

We are able to confirm the absorption lines detected in both 3C~111 and 3C~390.3 at a similar confidence level. However, while we are unable to directly confirm the line detection reported in 3C~120 we note that by adding two narrow ($\sigma=10$\,eV) Gaussians at the rest-frame energies for the \fexxv~\hea and \fexxvi~\lya lines reported by \tombesiBLRG ($7.25$\,keV and $7.54$\,keV, respectively) we can place lower limits of $EW_{\rm He\alpha}>-8$\,eV and $EW_{\rm Ly\alpha}>-9$\,eV for the \fexxv~\hea and \fexxvi~\lya absorption lines, respectively. These limits are consistent with those reported by \tombesiBLRG at the 90\% level. Furthermore, we have also detected a high-velocity outflow ($\vout\sim0.22$\,c) in the BLRG 4C+72.26 and note that an Fe\,K outflow is also detected in the \chandraletg data for 3C\,445 (\citealt{reeves:2010}; see also \citealt{braito:2011}). Thus the number of BLRGs with high-velocity outflows could tentatively rise to 4/6 if observations from multiple observatories are included, suggesting that such outflows could be an important component in radio-loud sources. It is important to note, however, that the current sample size of 7 \suzaku observed radio-loud AGN is insufficient for a statistical study into the prevalence of Fe\,K absorption in radio-loud sources. That being said, Tombesi et al. (2012; in prep) are currently performing a detailed analysis on a large sample of $\sim30$ radio-loud AGN observed using \xmm and \suzaku which will robustly assess for the prevalence of Fe\,K absorption in these objects.

\subsubsection{Recent Suzaku samples}
\label{suzaku_samples}
\cite{patrick:2012} have recently published the analysis of another large sample of \suzaku selected AGN, and is another sample with which we can compare our results. While the primary aims of the \cite{patrick:2012} study was to assess the properties of Fe\,K emission lines and black hole spin, the authors also report the detection of highly-ionised Fe\,K absorption in 14/46 ($\sim30$\%) of their objects. The two samples have 43 AGN in common. This includes 15 AGN in which we have detected Fe\,K absorption, of which 11 are confirmed by the \cite{patrick:2012} analysis. The four objects where our outflow detections are not corroborated are: 3C\,390.3, 4C+74.26, NGC\,5506 and SWIFT\,J2127.4+5654. The outflows that we detect in three of these objects (3C\,390.3, 4C+74.26 and SWIFT\,J2127.4+5654) are only marginal detections from Monte Carlo simulations, i.e., $95\% \leq P_{MC} < 99\%$, which suggests that the lines may not be visually apparent in the raw data without a statistically driven analysis. The outflow in NGC\,5506 is detected with $P_{MC}=99.8\%$ which suggests that the discrepancy is likely a result of the method in which the datasets were analysed. Indeed, our detection of \fexxvi~\lya in this source is taken from the time-averaged (stacked) \suzaku spectrum while \cite{patrick:2012} analysed the observations separately. The outflow in NGC\,5506 may be intrinsically weak in the individual \suzaku epochs, and only become apparent when they are time-averaged due to higher $S/N$.

An interesting outcome of the \cite{patrick:2012} analysis is that they report outflow detections in three sources (3C\,445, NGC\,5548 and PG\,1211+143) which we do not find in this work. Outflows in 3C\,445 and PG\,1211+143 been detected with other observatories (3C\,445: \citealt{reeves:2010}, PG\,1211+143: \citealt{pounds:2003b}) and they are known to be intrinsically variable, being weakest in their respective \suzaku observations (\citealt{braito:2011}, \citealt{reeves:2008}). In NGC\,5548 we note that while there is a weak absorption trough detected at the energy expected for \fexxv~\hea (see appropriate entry in Appendix~\ref{appendix:contour_plots}) it falls below our detection criteria. Similar is also true for the \suzaku line detection in PG\,1211+143, which is only detected at 90\% significance. However, when compared to the sample of \cite{patrick:2012} the broader outcomes are entirely consistent. Both analyses find that there is a strong peak in the ionisation parameter distribution at $\logxi\sim4-4.5$, with the associated velocities spanning a continuous range from $\sim0 \lesssim \vout \lesssim 100,000$\,km\,s$^{-1}$. Therefore, the two samples are in complete agreement when it comes to the overall properties of the Fe\,K absorbers.

\subsection{Evidence for complex variability and absorber structure?}
\label{discussion:evidence_for_variability}
In addition to broad-band spectral variability there is also compelling evidence for intricate variability in the Fe\,K absorber itself. In a number of cases this variability can be simultaneously and self-consistently fitted using \xstar absorption tables which strongly implies an atomic origin rather than an association with spectral fluctuations and photon statistics. In NGC\,3227 transient Fe\,K absorption lines are detected in 3 of the 6 \suzaku epochs, with the line parameters varying on a $\Delta t\sim2$\,week time-scale. Similar is also true in Mrk\,766, PDS\,456 and NGC\,1365, which all exhibit absorption line variability albeit over a longer (i.e., $\Delta t\sim$year) baseline. In both Mrk\,766 and PDS\,456 the absorber appears to change in ionisation states between the \suzaku epochs, with the outflow in Mrk\,766 also showing a decrease in outflow velocity. Here, we discuss both sources in greater detail.
 
\begin{figure}
\begin{center}
\includegraphics[width=8.5cm]{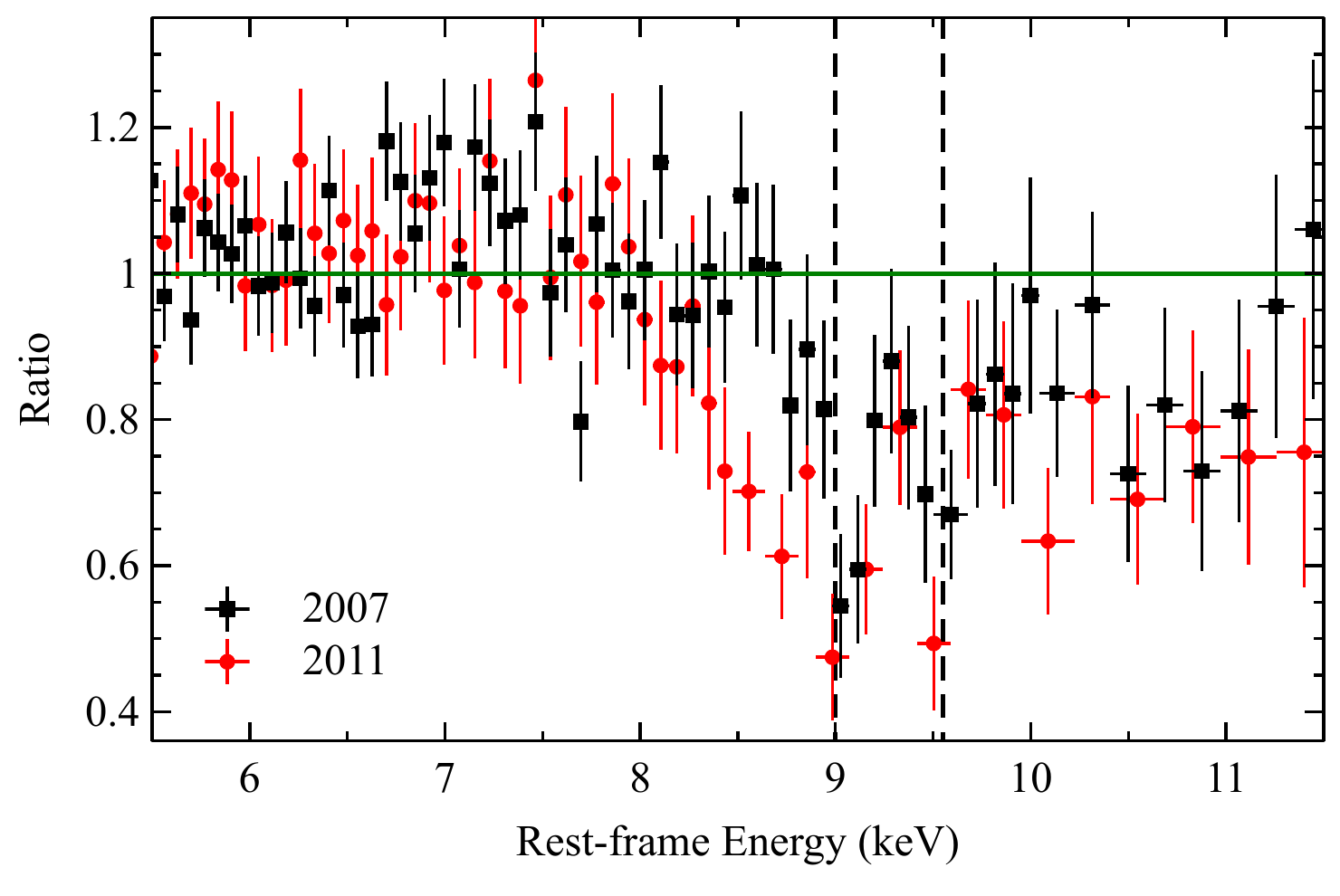}
\caption{\small Ratio plot comparison of the 2007 (black squares) and 2011 (red circles) \suzaku observations of PDS\,456 in the Fe\,K band. The dashed vertical lines indicate the position of the two blended \fexxvxxvi absorption troughs present in both spectra, at rest-frame energies of $\sim9.0$ and $\sim9.6$\,keV, respectively. There is a noticeable broadening in the left hand absorption profile in the 2011 observation which is likely caused by a drop in absorber ionisation state, and an associated increase in the \fexxv~\hea contribution to the absorption profile. See text and Reeves et al. (2012, in prep) for further details. Colour version available online.}
\label{fig:pds456}
\end{center}
\end{figure}

Mrk\,766 and PDS\,456 have both long been known to harbour Fe\,K absorbers (Mrk\,766: \citealt{pounds:2003a, turner:2007, miller:2008, risaliti:2011}. PDS\,456: \citealt{reeves:2000, reeves:2003, reeves:2009}). The range of rest-frame energies for the absorption profiles measured with \suzaku in each source are consistent with those previously published in the literature. In Mrk\,766 we find a common velocity shift of $\vout\sim6,000-18,000$\vunit for the \fexxv~\hea and \fexxvi~\lya lines, which is consistent with the $v_{\rm out,\textsc{xmm}}\sim3,000-16,000$\vunit reported previously in \xmm data (\citealt{miller:2007, turner:2007, risaliti:2011}).  Similarly, from our simultaneous analysis of the 2007 and unpublished 2011 \suzaku observations of PDS\,456 we find two absorption troughs which persist across both sequences (see Figure~\ref{fig:pds456}). These troughs -- which are likely due to blends of the \fexxv~\hea and \fexxvi~\lya lines owing to their breadth -- have a mean inferred $\langle \vout \rangle \sim 0.3$\,c, which is consistent with that reported by \cite{reeves:2009} in their initial analysis of the 2007 observation. 

An interesting property of the outflows in both of these sources is that the parameters of the hard-band absorber appears to be linked to both the incident source flux, and to the broad-band absorption characteristics of the X-ray spectrum. On the basis of detailed time- and flux-resolved spectroscopy of Mrk\,766 using \xmm, \cite{turner:2007} noted that the $EW$ of the \fexxvi~\lya absorption line could be correlated with the intrinsic source flux, with the correlation not easily accounted for by simply appealing to changes in the ionisation state of a constant column density absorber in response to changes in the incident continuum. This provided evidence for complex absorber structure in Mrk\,766. Further evidence for Fe\, K absorber complexity was suggested by \cite{risaliti:2011} who, from a re-analysis of the same \xmm data, found that the presence of the highly-ionised absorption profiles appeared to be contingent on the source being eclipsed by a number of high column density, i.e., $\lognh\gtrsim23$, partially-covering clouds with a low ionisation state. This led the authors to propose a scenario where multiple `cometary' absorption clouds with a large gradient in both column density and ionisation state were moving across the line of sight, thus giving rise to the varying absorption characteristics present in different time-slices (e.g., $\vout$). 

Our \suzaku analyses yields further evidence for a complex and dynamic structure in hard-band absorbers. In Mrk\,766 the $EW$ of the \fexxv~\hea and \fexxvi~\lya lines present in the \suzaku spectrum are statistically consistent at the 90\% level but, as shown in the appropriate entries in Appendix~B, there is a visual shift in the dominant iron ion of the absorber, with \fexxvi~\lya ($EW_{\rm{\fexxvi}}=-40\pm17$\,eV) appearing strongest in OBSID~701035010 and \fexxv~\hea ($EW_{\rm{\fexxv}}=-60\pm14$\,eV) strongest in OBSID~701035020, which is possibly associated with the $\Delta\logxi\sim0.4$ change in ionisation parameter between the two observations. Furthermore, and as mentioned previously, the decrease in ionisation is also met with a decrease in measured outflow velocity and the onset of strong \fexxv emission which could be associated with a cloud/clump of material moving outside of the sight-line. A similar effect is also apparent in PDS\,456 where, despite there being no measurable change in the relative velocity of the two blended \fexxvxxvi absorption complexes, the absorber ionisation state decreases by $\Delta\logxi\sim0.26$ in the 2011 observation relative to 2007 which leads to a slight visual broadening of the absorption profiles due to an increased \fexxv~\hea contribution in the blended absorption line profile. 
Interestingly, the decreased ionisation state of the absorbers in both Mrk\,766 and PDS\,456 occurs when the covering fraction of the soft X-ray absorber is at its highest which implies a dynamic clumpy structure to the outflow. Moreover, as noted by both \cite{turner:2007} and \cite{behar:2010}, respectively, the opposite is also true in both sources, i.e., when the sources are in their least absorbed state, the discrete highly-ionised absorption lines are not significantly present in the X-ray spectra of either source. This is consistent with the conclusions reached by \cite{risaliti:2011} in the context of stratified, partially-covering, absorption clumps needing to be present along the line of sight, and close to the continuum source, for highly-ionised absorption lines to be observed. 
 
Additional evidence for this supposition is provided through consideration of our spectral models to other sources in the sample. It is intriguing to note that statistically acceptable (i.e., $\chi^{2}_{\rm reduced}\approx1.0$) partially covering absorption models are obtained for a curiously large fraction of the sources with detected outflows (12/20; $\sim60$\%) which further suggests that there may be a link between complex absorption geometries and highly-ionised absorption lines. One possibility is where the partially covering components represent denser clumps of material in an inhomogeneous highly-ionised wind from the accretion disc, similar to the case reported in Mrk\,766 (\citealt{miller:2007, turner:2007, risaliti:2011}). Indeed, inhomogeneous winds with stratified or filamentary ionisation and density structure are expected as a natural consequence of the accretion process and are ubiquitously seen in both hydrodynamic (e.g., \citealt{proga&kallman:2004, kurosawa:2009, sim:2010}) and magneto-hydrodynamic (e.g., \citealt{ohsuga:2009, ohsuga:2011}) simulations of accretion discs. It is therefore not particularly surprising to find a possible link between clumpy absorption dominated models and sources with high-velocity outflows. 

\subsection{On the claimed publication bias}
\label{discussion:publication_bias}
The transient and variable nature of Fe\,K absorption lines has in part lead to their true veracity to be questioned in the literature. In particular, \cite{vaughan&uttley:2008} suggested that there may be a publication bias at play in the reporting of both red- and blue-shifted features in the Fe\,K band, with only those observations with the strongest line detections being reported in the literature. Through a plot of `$EW-\rm{error}\langle EW \rangle$' \cite{vaughan&uttley:2008} showed that the $EW/\rm{error\langle EW \rangle}$ ratio remained relatively constant (i.e., lines with larger $EW$ have correspondingly larger 90\% errors) over a wide range of $EW$s in their sample of narrow lines in 38 sources collected from the literature. The authors also noted that since the vast majority of reported detections (prior to 2008, at least) were of relatively low significance (i.e., typically around $2-3\sigma$) suggested that the lines may be spurious, and more consistent with merely being the strongest natural fluctuations in otherwise featureless spectra than real atomic features. Moreover, \cite{vaughan&uttley:2008} suggested that the conspicuous absence of any lines in the upper left quadrant of the $EW-\rm{error}\langle EW \rangle$ plot (which would correspond to stronger lines having smaller uncertainties) implied that the detection of the narrow velocity shifted lines was in some way inversely correlated to the statistical quality of the observation. Observations with longer exposures and better photon statistics only ever showed the weakest lines, which further enforced the possibility of a bias in the published observations.
\begin{figure}
\begin{center}
\includegraphics[width=8cm]{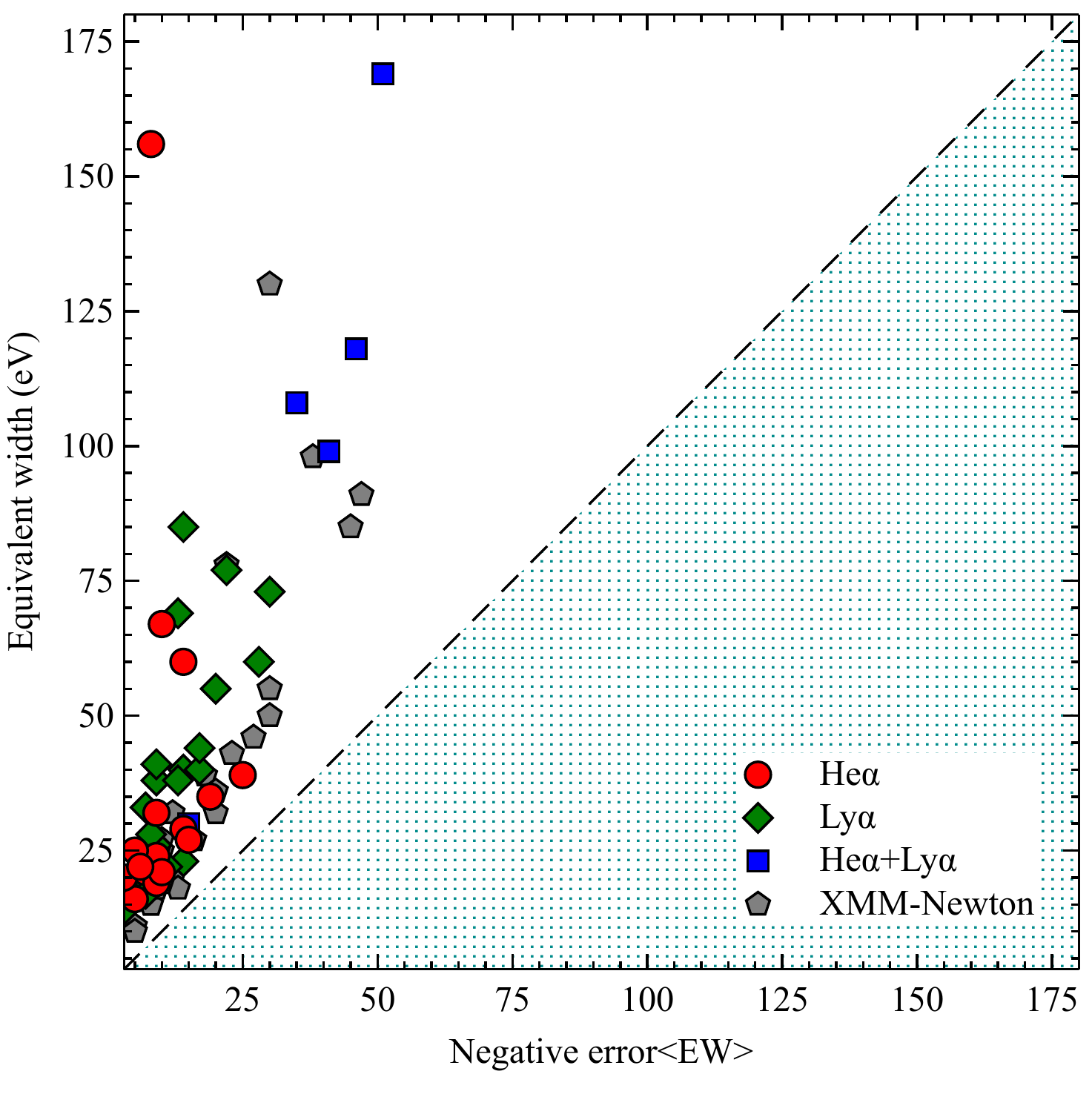}
\end{center}
\caption{\small Plot of the measured equivalent width of the blue-shifted absorption lines with respect to their negative 90\% error. The circles (red), diamonds (green), and squares (blue) correspond to the \fexxv~\hea, \fexxvi~\lya and blended \fexxvxxvi lines, respectively. Analogous values for the blue-shifted absorption lines found in the \tombesiA \xmm sample are shown as pentagons (grey). The diagonal dashed line shows the expected trend if $EW=\rm{-error}\langle EW \rangle$ and the dotted (dark-cyan) area shows the area of non-detection (see text for further details). Colour version available online.}
\label{fig:errew_ew}
\end{figure}

However, these points were addressed through the systematic search conducted by \tombesiA using \xmm, at least in the context of blue-shifted Fe\,K absorption lines. By uniformly and comprehensively searching for such features in a complete sample of AGN observations, carefully reporting the fraction of detections to null detections, and assessing the statistical significance of any detected absorption lines through Monte Carlo simulations, \tombesiA were able to robustly assess the fraction of AGN in the local universe that have Fe\,K outflows in a way which overcame any publication biases and accounted for the possibility of random fluctuations in the source spectra. Our work with the \suzaku outflow sample complements the findings of \tombesiA, and lends further weight to the assertion that such outflows are an important intrinsic feature of the AGN X-ray spectrum in a large fraction of sources. In Figure~\ref{fig:errew_ew} we replicate the \cite{vaughan&uttley:2008} $EW/\rm{error\langle EW \rangle}$ plot\footnote{Note that the axes in this plot use a linear scale rather than the logarithmic one as used by \cite{vaughan&uttley:2008}. As noted by \tombesiA the use of a logarithmic scale visually compresses the data points towards the EW=error$\langle EW \rangle$ threshold of non-detection.} used by \tombesiA (see their Figure~8) to include data points corresponding to the absorption lines found in this work at a $P_{\rm MC}\geq95\%$ significance level. The figure concisely shows that the distribution of points from both the \suzaku and \xmm samples diverges from that of the EW=error$\langle EW \rangle$ `detection line', with both sets of data having a number of points which veer towards the important upper left quadrant of the diagram indicating stronger lines with smaller uncertainties. Interestingly, the overall distribution of points for the \suzaku lines follows a similar trend to those obtained with \xmm which implies that all points are drawn from the same parent population, as would be expected should we be studying the same physical phenomenon which imprints real spectroscopic lines in the X-ray spectrum.
\begin{figure}
\begin{center}
\includegraphics[width=8cm]{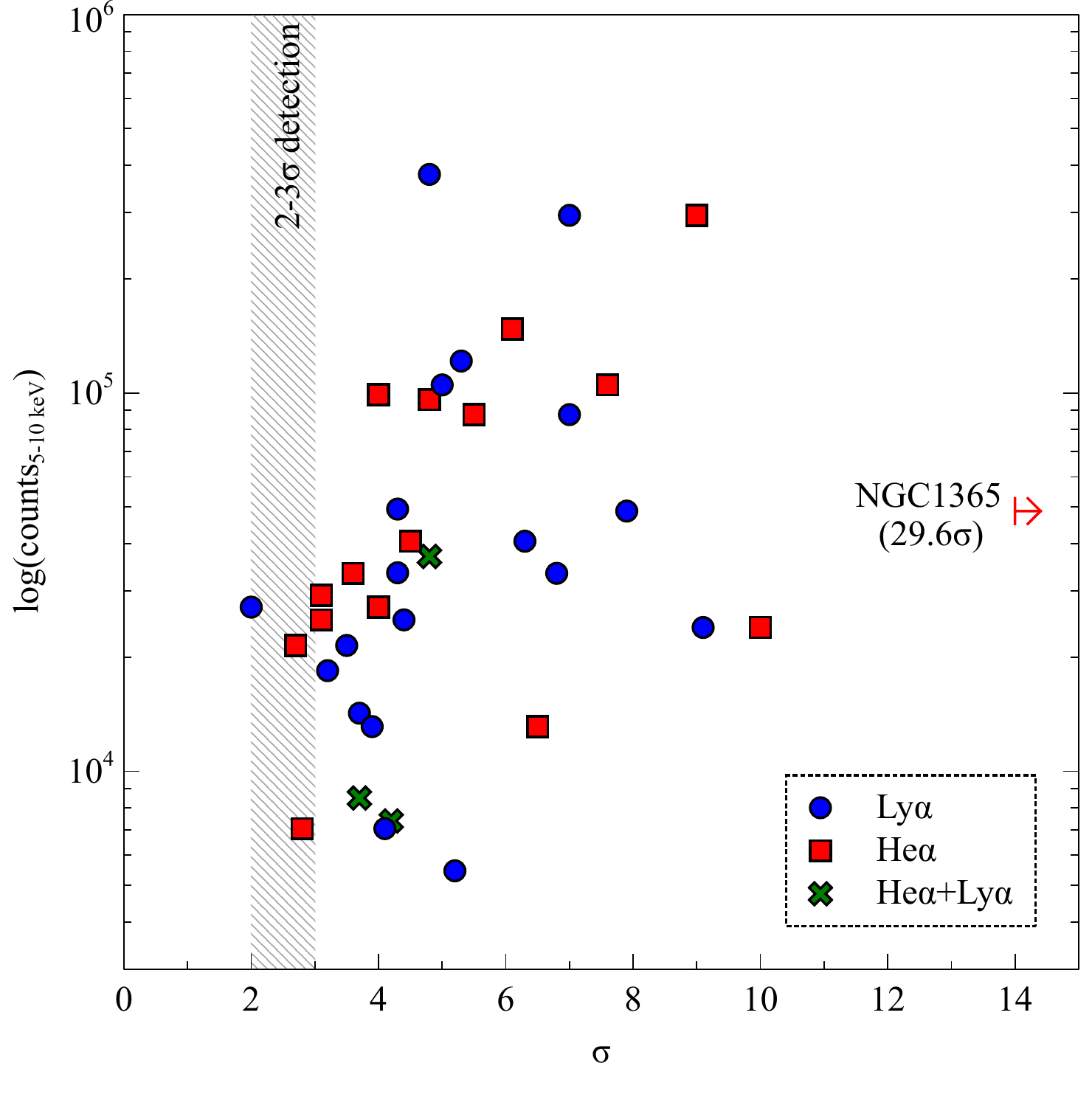}
\end{center}
\caption{\small Plot of line significance, $\sigma$ ($\equiv \rm{EW}/\rm{-error}\langle \rm{EW} \rangle$), versus total counts between $5-10$\,keV. The blue circles, red squares and green crosses show the positions of \fexxv~\hea, \fexxvi~\lya and blended \fexxvxxvi lines, respectively. Only absorption lines detected at $P_{\rm MC}\geq95\%$ are shown. The dashed grey band shows the $2-3\sigma$ detection regime. Colour version available online.}
\label{fig:sigma_counts}
\end{figure}

A further means of visually assessing for a publication bias, and for testing whether the absorption lines are consistent with random fluctuations or real features, is through a plot of line significance versus photon counts in the Fe\,K band (i.e., between $5-10$\,keV). Such a plot is shown in Figure~\ref{fig:sigma_counts}. For simplicity we grade line significance according to $N\sigma$ $\equiv \rm{EW}/\rm{-error}\langle \rm{EW} \rangle$, where it is important to recognise that in this case $\rm{-error}\langle \rm{EW} \rangle$ refers to the $1\sigma$ ($68\%$) negative errors on an absorption lines EW, which were measured independently for each absorption line, and {\it not} to the $90\%$ errors ($\approx1.6\sigma$) reported in Table~\ref{table:absorption_line_parameters}. By using the $1\sigma$ negative errors the standard Gaussian probabilities (i.e., $2\sigma=95.5\%$, $3\sigma=99.73\%, 4\sigma=99.994\%$) then correspond to the statistical significance of a line relative to it having an intrinsic $EW$ of 0\,eV, and thus not being detected. Figure~\ref{fig:sigma_counts} shows that the detected lines typically have significances ranging from $\sim2\sigma$ through to $\sim10\sigma$ over a wide range in count rate; with a significance of $\sim30\sigma$ the extremely strong \fexxv~\hea line in NGC\,1365 (OBSID 702047010) is by far the most significant line in the sample and is omitted from the main plot for clarity.

In such a plot we would expect to see a correlation between total Fe\,K counts and the significance of an absorption line, where observations with higher counts in the Fe\,K band, and therefore better photon statistics, have stronger absorption line detections. Indeed, while there is noticeable scatter in $N_{\rm counts}$ versus $\sigma$ -- as would be expected given that there is a wide dispersion in absorption properties (i.e., $\nh$, $\xi$, $\vout$) for a given total counts -- a simple Spearmans rank association shows that the data are positively correlated (Spearmans rank coefficient, $r_{s}=0.4406$). This is sufficient to rule out the possibility of the two parameters being completely independent at the $>99\%$ significance for 34 matched pairs. Therefore, $\sigma$ values tend to increase in higher counts observations which causes the data to gradually diverge from the $2-3\sigma$ significance region towards the upper right quadrant of the plot. Moreover, the vast majority of the line detections are located well away from the $2-3\sigma$ `noise' level, as would be expected if the majority of the line detections are not purely due to photon statistics.  

\section{Summary \& Conclusions}
\label{conclusion_summary}
Making use of data from the \suzaku Data Archives and Transmission System (DARTS) we have constructed broad-band spectral models for  51 Type 1.0-1.9 AGN to search for the presence of \fexxv~\hea and/or \fexxvi~\lya absorption lines at $E\gtrsim6.7$\,keV, robustly assessed the statistical significance of any detected absorption systems using detailed Monte Carlo simulations, and probed the properties of the absorbing material using the \xstar (v. 2.2.1bc) photoionisation code. The primary findings of this work are as follows:
\begin{enumerate}[label=\arabic*.,leftmargin=*]

\item We find that 20/51 sources (in 28/73 fitted spectra) have statistically significant \fexxv~\hea or \fexxvi~\lya absorption lines at $E\geq6.7$\,keV in the source rest-frame, which corresponds to $\approx40\%$ of the sample. 18 of the 20 Fe\,K absorption complexes are robustly detected at $P_{\rm MC}\geq99\%$ from Monte Carlo simulations, with those remaining only narrowly falling short of the 99\% criterion required for a robust detection. All absorption lines are detected independently and simultaneously in two (or more) of the available XIS detectors which further enforces the robustness of the line detections as real features
\\
\item The detected outflows fall into a range of phenomenological categories. Absorption due to a pair of \fexxv~\hea and \fexxvi~\lya from the same photoionised absorber account for 7/20 of the detected outflows (in 12/28 spectra) making them the most frequently observed form of highly-ionised outflow in the sample. Absorption due to solely \fexxv~\hea or \fexxvi~\lya is less frequently observed, being detected in 2 sources (3 spectra) and 9 sources (10 spectra), respectively. The remaining two sources have outflows which are best fitted by two unresolved and blended \fexxvxxvi absorption systems with different outflow velocities.  
\\
\item By fitting the absorption with the \xstar photoionisation code we find that the absorbers are characterised by $21.5<\lognh\leq24$ and $2.5<\logxi\leq6$, with mean values of $\lognh\approx23$ and $\logxi\approx4.5$, respectively. The distribution of outflow velocities cover a wide range, from $\mathbf{\vout<1,500}$\,km\,s$^{-1}$ up to $\vout\sim100,000$\,km\,s$^{-1}$, with 90\% of the observed absorbers having $\vout>1,500$\,km\,s$^{-1}$ which makes them systematically faster than typical absorbers found in the soft X-ray band. The median outflow velocity is of the order of $\mathbf{\sim0.056}$\,c. Moreover, 60\% of the absorbers also have $\vout>10,000$\,km\,s$^{-1}$ and are thus consistent with the so called `Ultra Fast Outflows' (UFOs) defined by \tombesiA. A Kolmogorov-Smirnov two-sample test shows that the overall distributions for all three parameters are not statistically distinguishable from the analogous distributions measured by \tombesiA and \tombesiB using \xmm.   
\end{enumerate}
Overall, our work with the \suzaku sample is consistent with those obtained using \xmm and, combined, the two studies provide strong evidence for the presence of very highly-ionised, often high-velocity, outflows in the central region of a large fraction of not only radio-quiet AGN, but also suggests that they may be prevalent in radio-loud sources as well. The possible prevalence of very highly-ionised, high-velocity winds in AGN is consistent with theoretical models which argue that such outflows are an important phenomenon which may play a role in Galactic-scale feedback scenarios. We return to this point in a forthcoming work (Paper II;  Gofford et al., in prep) where we probe the global energetics of the detected absorbers, assess for any correlations between the outflow parameters and those of the central SMBH, and discuss both the origins of the absorbing material and its likely launching mechanism.

\section*{Acknowledgements}
This research has used data obtained from the \suzaku X-ray observatory, which is a collaborative mission between the Japan Aerospace Exploration Agency (JAXA, Japan) and the National Aeronautics and Space Administration (NASA, USA). Data were obtained from the High Energy Astrophysics Science Archive Research Center (HEASARC) via the Data Archives and Transmission System (DARTS), which are provided by NASA's Goddard Space Flight Center and JAXA's Institute of Space and Aeronautical Science, respectively. Source classifications and red-shifts were obtained from the NASA/IPAC Extragalactic Database (NED) and the SIMBAD database, which are operated by the Jet Propulsion Laboratory, California Institute of Technology, under contract with the National Aeronautics and Space Administration, and at the Centre de Donn\'ees astronomiques (CDS), Strasbourg, France. We would like to thank the anonymous referee for her/his comments and suggestions which have helped improved the quality of this manuscript. J.~Gofford acknowledges support from the Science and Technology Facilities Council (STFC, UK) in the form of a funded Ph.D studentship and would like to thank A.~L.~Dobson for offering several useful comments pertaining to the structure and content of this paper. T.~J.~Turner acknowledges support from NASA grant NNX11AJ57G. M.~Cappi acknowledges financial support from ASI contract I/009/10/0 and INAF contract PRIN-2011.

\bibliography{paper1}

\appendix

\begin{table*}
\section{Summary of observations}
	\begin{minipage}{167mm}
	\caption{Summary of observation parameters. {\sl Notes:} 	(1) Source name; 
		     		(2) Source type;
		     		(3) Cosmological red-shift according to the NED; 
					(4) Galactic absorption column density, in units of $\times10^{21}\,\rm{cm}^{-2}$; 
					(5) \suzaku observation ID; 
					(6) Starting date of observation; 
					(7) XIS(HXD) net exposures, in kilo-seconds; 
					(8) Total number of (background subtracted) $2-10$\,keV counts for the (co-added) XIS-FI(BI) spectra, in units of 
					$\times10^{3}$; 
					(9) Total background subtracted PIN counts in the $15-50$\,keV band, in units of $\times10^{3}$;}
	\footnotesize	
	\begin{tabular}{@{}llcrlcccc}
		
		\toprule
		\multicolumn{1}{c}{Source} & \multicolumn{1}{c}{Type} & \multicolumn{1}{c}{red-shift} & $N_{\rm H}^{\rm Gal}$ & 
		\multicolumn{1}{c}{OBSID} & Date & Exposure & \multicolumn{2}{c}{Source net counts}\\
		& & \multicolumn{1}{c}{(z)} & & & yyyy-mm-dd & XIS(PIN) & XIS-FI(BI) & PIN \\
		\multicolumn{1}{c}{(1)} & \multicolumn{1}{c}{(2)} & \multicolumn{1}{c}{(3)} & \multicolumn{1}{c}{(4)} & 
		\multicolumn{1}{c}{(5)} & (6) & (7) & (8) & (9) \\
		\midrule

	1H\,0419-577 	   	& Sy 1.5 & $0.10400$ & $0.126$ & 702041010 & 2007-07-25 & 205.9(142.6) & 213.4(113.8) & 49.8 \\
	             	   	& & & & 704064010 & 2010-01-16 & 246.2(122.8) & 89.6(46.5) & 36.6 \\
	             	   	& & & & stacked[all] & \na & \na & 302.9(160.4) & 79.6\\
	3C\,111         	& BLRG & $0.04850$ & $2.910$ & 703034010[a] & 2008-08-22 & 122.4(101.9) & 106.3(49.8) & 36.2 \\
						& & & & 705040010[b] & 2010-09-02 & 80.7(67.9) & 199.7(95.4) & 26.6\\
						& & & & 705040020[c] & 2010-09-09 & 79.4(66.5) & 258.8(127.1) & 27.4\\
						& & & & 705040030[d] & 2010-09-16 & 80.4(65.6) & 251.0(127.2) & 28.1\\
	3C\,120            	& BLRG & $0.03300$ & $1.060$ & 700001010[a]$^{\ddag}$ & 2006-02-09 & 41.9(31.9) & 152.3(50.3) & 13.1 \\
						& & & & 700001020[b]$^{\ddag}$ & 2006-02-16 & 41.6(34.5) & 134.3(44.7) & 15.5 \\
						& & & & 700001030[c]$^{\ddag}$ & 2006-02-23 & 40.9(36.2)	& 135.3(45.6) & 15.7 \\
						& & & & 700001040[d]$^{\ddag}$ & 2006-03-02 & 40.9(37.9) & 139.5(48.3) & 16.1 \\
						& & & & stacked[bcd] & \na & \na & 409.1(138.5) & 44.6 \\	
	3C\,382     	   	& BLRG & $0.05890$ & $0.698$ & 702125010 & 2007-04-27 & 130.6(114.3) & 274.5(136.1) & 45.5 \\
	3C\,390.3     	   	& BLRG & $0.05610$ & $0.347$ & 701060010 & 2006-12-14 & 99.4(92.1) & 151.6(71.8) & 38.3 \\
	3C\,445     		& BLRG & $0.05590$ & $0.449$ & 702056010 & 2007-05-25 & 139.8(109.5) & 41.9(20.2) & 40.8\\
	4C\,+74.26 			& BLRG & $0.10400$ & $1.160$ & 702057010 & 2007-10-28 & 91.6(87.3) & 107.0(36.5) & 33.8\\
	APM\,08279+5255$^{\,a}$ & QSO & $3.91000$ & $0.411$ & 701057010[a]$^{\ddag}$ & 2006-10-12 & 102.3(\na) & 4.6(1.6)$^{a}$ & \na\\
	      			   	& & & & 701057020[b]$^{\ddag}$ & 2006-11-01 & 102.3(\na) & 4.5(1.6)$^{a}$ & \na\\
	      			   	& & & & 701057030[c] & 2007-03-24 & 117.2(\na) & 3.3(1.8)$^{a}$ & \na\\
	                   	& & & & stacked[all] & \na & \na & 14.8(6.4)$^{a}$ & \na \\
	Ark\,120            & Sy 1.0 & $0.03270$ & $0.978$ & 702014010 & 2007-04-01 & 100.9(89.5) & 152.1(78.7) & 32.3 \\
	Ark\,564 			& NLSy1 & $0.02468$ & $0.564$ & 702117010 & 2007-06-26 & 100.0(81.3) & 102.7(54.4) & 23.5\\
	CBS\,126 			& Sy 1.2 & $0.07888$ & $0.097$ & 705042010 & 2010-10-18 & 101.5(84.2) & 21.6(10.7) & 24.8\\
	ESO\,103-G035 		& Sy 1.9 & $0.01329$ & $0.764$ & 703031010 & 2008-10-22 & 91.5(75.1) & 77.3(31.8) & 34.2\\
	Fairall\,9 		   	& Sy 1.2 & $0.04700$ & $0.316$ & 702043010 & 2007-06-07 & 167.8(127.3) & 227.3(117.7) & 49.3 \\
	 				   	& & & & 705063010 & 2010-05-19 & 229.3(162.2) & 226.9(103.9) & 56.8\\
	IC\,4329A     	   	& Sy 1.2 & $0.01610$ & $0.461$ & 702113010[a] & 2007-08-01 & 25.5(20.1) & 131.0(64.8) & 11.4\\
						& & & & 702113020[b] & 2007-08-06 & 30.6(24.1) & 197.8(100.1) & 15.4\\
						& & & & 703113030[c] & 2007-08-11 & 24.2(22.1) & 160.6(67.1) & 14.5\\
						& & & & 703113040[d] & 2007-08-16 & 24.2(18.9) & 133.3(67.1) & 11.5\\
						& & & & 703113050[e] & 2007-08-20 & 24.0(17.5) & 83.2(40.9) & 9.4\\
						& & & & stacked[all] & \na & \na & 712.7(359.0) & 61.7\\ 
	IGR\,J16185-5928 	& NLSy1 & $0.03500$ & $2.070$ & 702123010 & 2008-02-09 & 76.7(69.6) & 24.1(8.1) & 15.9 \\ 
	IGR\,J21247+5058 	& Sy 1.0 & $0.0200$ & $10.000$ & 702027010 & 2007-04-16 & 85.0(66.4) & 233.1(103.7) & 38.4\\
	MCG\,-02-14-009 	& Sy 1.0 & $0.02845$ & $0.948$ & 703060010 & 2008-08-28 & 142.2(120.0) & 24.1(10.6) & 20.5 \\
	MCG\,-2-58-22 		& Sy 1.5 & $0.04686$ & $0.291$ & 704032010 & 2009-12-02 & 139.0(98.0) & 324.0(129.6) & 38.3 \\
	MCG\,-5-23-16 		& Sy 1.9 & $0.00849$ & $0.800$ & 700002010 & 2005-12-07 & 95.7(79.7) & 626.8(198.5) & 54.0\\
	MCG\,-6-30-15 	   	& Sy 1.2 & $0.00775$ & $0.392$ & 700007010[a]$^{\ddag}$ & 2006-01-09 & 143.3(118.9) & 504.2(167.0) & 51.3 \\
					   	& & & & 700007020[b]$^{\ddag}$ & 2006-01-23 & 98.5(76.8) & 304.6(105.0) & 35.1 \\
					   	& & & & 700007030[c]$^{\ddag}$ & 2006-01-27 & 96.7(89.8) & 327.5(109.1) & 38.7 \\
					   	& & & & stacked[all] & \na & \na & 1136.3(381.1) & 120.3 \\
	MCG\,+8-11-11 	   	& Sy 1.5 & $0.02050$ & $1.840$ & 702112010 & 2007-09-17 & 98.8(82.9) & 295.8(141.2.2) & 42.1 \\
	MR\,2251-178        & RQQ & $0.06400$ & $0.242$ & 704055010 & 2009-05-07 & 136.9(103.8) & 330.9(171.6) & 38.7 \\
	Mrk\,79    		   	& Sy 1.2 & $0.02220$ & $0.527$ & 702044010 & 2007-04-03 & 83.7(76.9) & 70.5(36.6) & 20.1 \\
	Mrk\,110      	   	& Sy 1.0 & $0.03530$ & $0.130$ & 702124010 & 2007-11-02 & 90.9(80.4) & 93.7(46.4) & 29.4 \\
	Mrk\,205 			& Sy 1.0 & $0.07085$ & $0.240$ & 705062010 & 2010-05-22 & 101.5(85.3) & 36.0(15.9) & 29.3 \\
    Mrk\,279		   	& Sy 1.0 & $0.03050$ & $0.152$ & 704031010 & 2009-05-14 & 160.4(139.8) & 43.8(23.5) & 40.3 \\
	Mrk\,335     		& NLSy1 & $0.02580$ & $0.366$& 701031010$^{\ddag}$ & 2006-06-21 & 151.3(131.7) & 162.6(49.0) & 45.9 \\
	Mrk\,359 			& Sy 1.5 & $0.01739$ & $0.426$ & 701082010 & 2007-02-06 & 107.5(96.1) & 27.2(13.6) & 8.8\\

	\bottomrule
	\end{tabular} \\[0.5ex]
	$^{\dag}$Only the XIS\,3 spectrum was available during the observation of SWIFT J2127.4+5654;\\
	$^{\ddag}$Indicates that the XIS\,2 was operational during the observation and is included as part of the co-added XIS-FI spectrum.;\\
	$^{a}$The total net source counts listed in column (8) for APM\,08279+5255 are taken from $0.6-10$\,keV in the observer frame. See text for further details.
	\\
	\label{tab:observation_details}
	\end{minipage}       
\end{table*}

\begin{table*}
	\begin{minipage}{163mm}
	\contcaption{-- Summary of observation parameters.}
	\footnotesize
	\begin{tabular}{@{}llcclcccc}
		
		\toprule
		\multicolumn{1}{c}{Source} & \multicolumn{1}{c}{Type} & 
		\multicolumn{1}{c}{red-shift} & $N_{\rm H}^{\rm Gal}$ & 
		\multicolumn{1}{c}{OBSID} & Date & Exposure &
		\multicolumn{2}{c}{Total counts}\\[0.5ex]

		& & \multicolumn{1}{c}{(z)} & & & yyyy-mm-dd 
		& XIS(PIN) & XIS-FI(BI) & PIN \\[0.5ex]

		\multicolumn{1}{c}{(1)} & \multicolumn{1}{c}{(2)} & 	
		\multicolumn{1}{c}{(3)} & \multicolumn{1}{c}{(4)} & 
		\multicolumn{1}{c}{(5)} & (6) & (7) & (8) & (9) \\[0.5ex] 
		\midrule
	Mrk\,509     	  	& Sy 1.5 & $0.03440$ & $0.425$ & 701093010[a]$^{\ddag}$ & 2006-04-25 & 24.6(14.5) & 93.8(30.8) & 6.6\\
						& & & & 701093020[b]$^{\ddag}$ & 2006-10-14 & 25.9(21.2) & 122.9(43.8) & 10.5\\
						& & & & 701093030[c] & 2006-11-15 & 24.5(17.3) & 73.9(40.8) & 8.0\\
						& & & & 701093040[d] & 2006-11-27 & 33.1(27.6) & 92.4(49.3) & 12.5\\
						& & & & stacked[bcd] & \na & \na & 289.3(133.8) & 30.8\\
						& & & & 705025010 & 2010-11-21 & 102.1(85.7) & 325.2(171.4) & 37.8\\
	Mrk\,766 		    & Sy 1.0 & $0.01290$ & $0.178$ & 701035010 & 2006-11-16 & 97.9(90.5) & 175.5(108.5)	& 31.6 \\
					   	& & & & 701035020 & 2007-11-17 & 59.4(47.7) & 73.0(40.3) & 17.1 \\
	Mrk\,841      	   	& Sy 1.5 & $0.03640$ & $0.222$ & 701084010[a] & 2007-01-22 & 51.8(43.7) & 36.3(18.0) & 14.5 \\
		   	   		   	& & & & 701084020[b] & 2007-07-23 & 50.9(44.4) & 35.9(18.4) & 16.1 \\
					   	& & & & stacked[all] & \na & \na & 72.2(37.4) & 31.3 \\
	NGC\,1365          	& Sy 1.8 & $0.00546$ & $0.134$ & 702047010 & 2008-01-21 & 160.5(136.6) & 96.8(48.2) & 53.9 \\
 					   	& & & & 705031010 & 2010-06-27 & 151.6(114.3) & 35.5(15.9) & 42.2\\
					   	& & & & 705031020 & 2010-07-15 & 302.2(231.5) & 45.8(21.4) & 76.2\\
	NGC\,2992 			& Sy 1.9 & $0.00771$ & $0.487$ & 700005010[a]$^{\ddag}$ & 2005-11-06 & 38.8(29.9) & 23.0(6.7) & 9.8\\
	 					& & & & 700005020[b]$^{\ddag}$ & 2005-11-19 & 39.7(31.9) & 39.6(13.1) & 10.4\\
	 					& & & & 700005030[c]$^{\ddag}$ & 2005-12-13 & 46.9(41.5) & 41.9(12.8) & 14.4\\
	 					& & & & stacked[all] & \na & \na & 84.7(25.0) & 36.1\\  	
	NGC\,3227 			& Sy 1.5 & $0.00386$ & $0.199$ & 703022010 & 2008-10-28 & 58.9(48.1) & 117.7/60.8 & 22.1\\
					   	& & & & 703022020 & 2008-11-04 & 53.7(46.7) & 38.5(17.3) & 19.6\\
					   	& & & & 703022030 & 2008-11-12 & 56.6(46.7) & 61.0(28.5) & 20.4\\
					   	& & & & 703022040 & 2008-11-20 & 64.6(43.4) & 24.1(11.8) & 15.1\\
					   	& & & & 703022050 & 2008-11-27 & 79.4(37.4) & 69.7(32.9) & 15.4\\
					   	& & & & 703022060 & 2008-12-02 & 51.4(36.9) & 32.1(14.7) & 15.4\\
	NGC\,3516  		   	& Sy 1.5 & $0.00884$ & $0.345$ & 100031010$^{\ddag}$ & 2005-10-12 & 134.6(115.4) & 233.3(75.4) & 50.7 \\
		   			   	& & & & 704062010 & 2009-10-28 & 251.4(178.2) & 149.0(73.4)	& 55.6 \\
	NGC\,3783 		   	& Sy 1.5 & $0.00973$ & $0.991$ & 701033010$^{\ddag}$ & 2006-06-24 & 75.7(68.3) & 302.4(103.0) & 32.2 \\
	         		   	& & & & 704063010 & 2009-07-10 & 172.8(174.0) & 472.1(234.0) & 84.0\\
	NGC\,4051           & NLSy1 & $0.00234$ & $0.115$ & 700004010$^{\ddag}$ & 2005-11-10 & 119.6(112.6) & 76.4(24.3) & 42.0 \\
				  	   	& & & & 703023010 & 2008-11-06 & 274.5(204.5) & 349.6(132.5) & 72.0\\
				   	   	& & & & 703023020 & 2008-11-23 & 78.4(58.5) & 72.0(35.0) & 21.3 \\
	NGC\,4151		   	& Sy 1.5 & $0.03320$ & $0.230$ & 701034010 & 2006-12-18 & 125.0(123.5) & 218.2(97.5) & 75.8 \\
	NGC\,4395			& Sy 1.8 & $0.00106$ & $0.135$ & 702001010 & 2007-06-02 & 101.5(95.0) & 19.4(9.2) & 30.4\\
	NGC\,4593     	   	& Sy 1.0 & $0.00900$ & $0.189$ & 702040010 & 2007-12-15 & 118.8(101.6) & 66.3(25.1) & 34.7 \\
	NGC\,5506 			& Sy 1.9 & $0.00618$ & $0.408$ & 701030010[a]$^{\ddag}$ & 2006-08-08 & 47.8(38.6) & 372.5(119.7) & 25.4\\
						& & & & 701030020[b]$^{\ddag}$ & 2006-08-11 & 53.3(44.8) & 438.9(141.2) & 29.9\\
						& & & & 701030030[c] & 2007-01-31 & 57.4(44.7) & 281.1(133.5) & 29.1\\
						& & & & stacked[all] & \na & \na & 1067.5(392.4) & 82.2\\
	NGC\,5548 			& Sy 1.5 & $0.01718$ & $0.155$ & 702042010[a] & 2007-06-18 & 31.1(25.6) & 11.0(5.3) & 7.5\\
						& & & & 702042020[b] & 2007-06-24 & 35.9(31.2) & 21.7(10.8) & 9.3\\
						& & & & 702042040[c] & 2007-07-08 & 30.7(27.0) & 35.9(16.4) & 8.5\\
						& & & & 702042050[d] & 2007-07-15 & 30.0(24.5) & 23.3(10.7) & 7.7\\
						& & & & 702042060[e] & 2007-07-22 & 28.9(23.1) & 43.3(19.9) & 7.9\\
						& & & & 702042070[f] & 2007-07-29 & 31.8(27.6) & 31.6(14.5) & 8.5\\
						& & & & 702042080[g] & 2007-08-05 & 38.8(30.4) & 20.5(10.0) & 9.6\\
						& & & & stacked[all] & \na & \na & 187.4(107.8) & 66.3\\
	NGC\,7213 		   	& Sy 1.0 & $0.00584$ & $0.106$ & 701029010$^{\ddag}$ & 2006-10-22 & 90.7(84.3) & 370.2(145.0) & 35.3 \\
	NGC\,7469  		   	& Sy 1.0 & $0.01630$ & $0.445$ & 703028010 & 2008-06-24 & 112.1(85.3) & 116.2(58.2) & 32.3 \\
	PBC\,J0839.7-1214   & QSO & $0.19787$ & $0.568$ & 705007010 & 2010-05-08 & 80.6(\na) & 35.5(17.7) & \na \\
	PDS\,456           	& RQQ & $0.18400$ & $1.960$ & 701056010 & 2007-02-24 & 190.6(\na) & 34.0(16.7) & \na \\
					   	& & & & 705041010 & 2011-03-16 & 125.5(\na) & 17.0(7.9) & \na  \\
    PG\,1211+143       	& RQQ  & $0.08090$ & $0.274$ & 700009010$^{\ddag}$ & 2005-11-24 & 97.6(78.8) & 19.4(5.9) & 8.9\\ 
	PKS\,0558-504 		& NLSy1 & $0.13720$ & $0.337$ & 701011010[a] & 2007-01-17 & 20.6(17.6) & 11.8(5.9) & 4.8\\
						& & & & 701011020[b] & 2007-01-18 & 18.9(16.7) & 16.0(8.1) & 6.5\\
						& & & & 701011030[c] & 2007-01-19 & 21.3(17.8) & 12.0(6.2) & 6.8\\
						& & & & 701011040[d] & 2007-01-20 & 19.7(16.0) & 18.3(8.2) & 6.3\\
						& & & & 701011050[e] & 2007-01-21 & 19.5(15.2) & 17.1(8.2) & 5.4\\
						& & & & stacked[all] & \na & \na & 76.2(44.6) & 21.0\\
	RBS 1124 			& Sy 1.2 & $0.20800$ & $0.152$ & 702114010 & 2007-04-14 & 86.2(83.0) & 16.6(7.4) & 21.5\\          	
	SW\,J2127.4+5654    & Sy 1.0 & $0.01440$  & $7.650$ & 702122010$^{\dag}$ & 2007-12-09 & 91.7(83.3) & 69.5(81.4) & 31.2 \\
	TON\,S180 			& NLSy1 & $0.06198$ & $0.136$ & 701021010 & 2006-12-09 & 120.7(102.4) & 43.5(-) & 27.2\\
	
	\bottomrule
	\end{tabular}
	\end{minipage}
\end{table*}

\begin{figure*}
\section{Ratio and Contour plots}
\label{appendix:contour_plots}
\begin{center}

\vspace{-5pt}	
\subfloat{
\includegraphics[angle=-90,width=3.8cm]{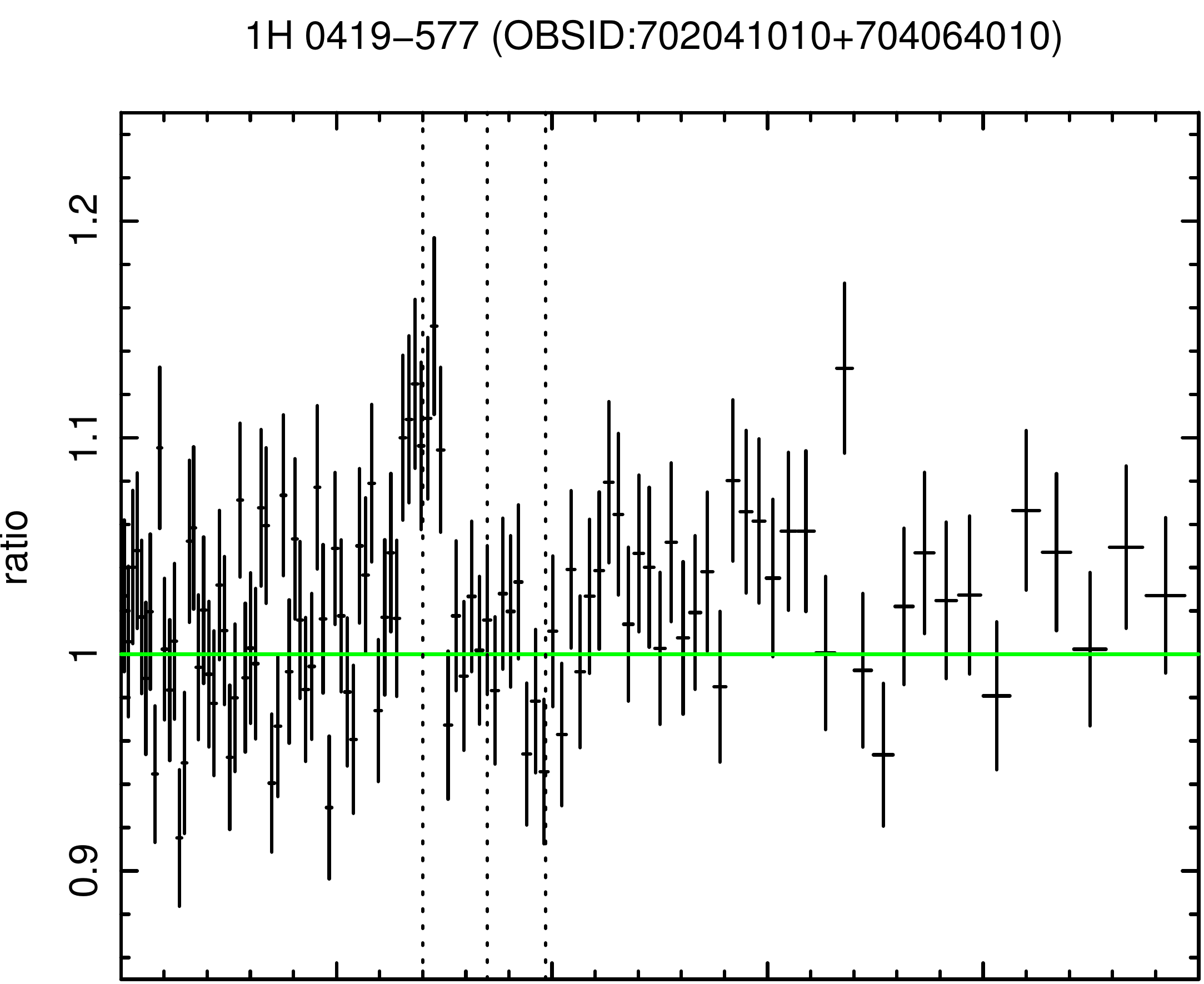}
\includegraphics[angle=-90,width=3.8cm]{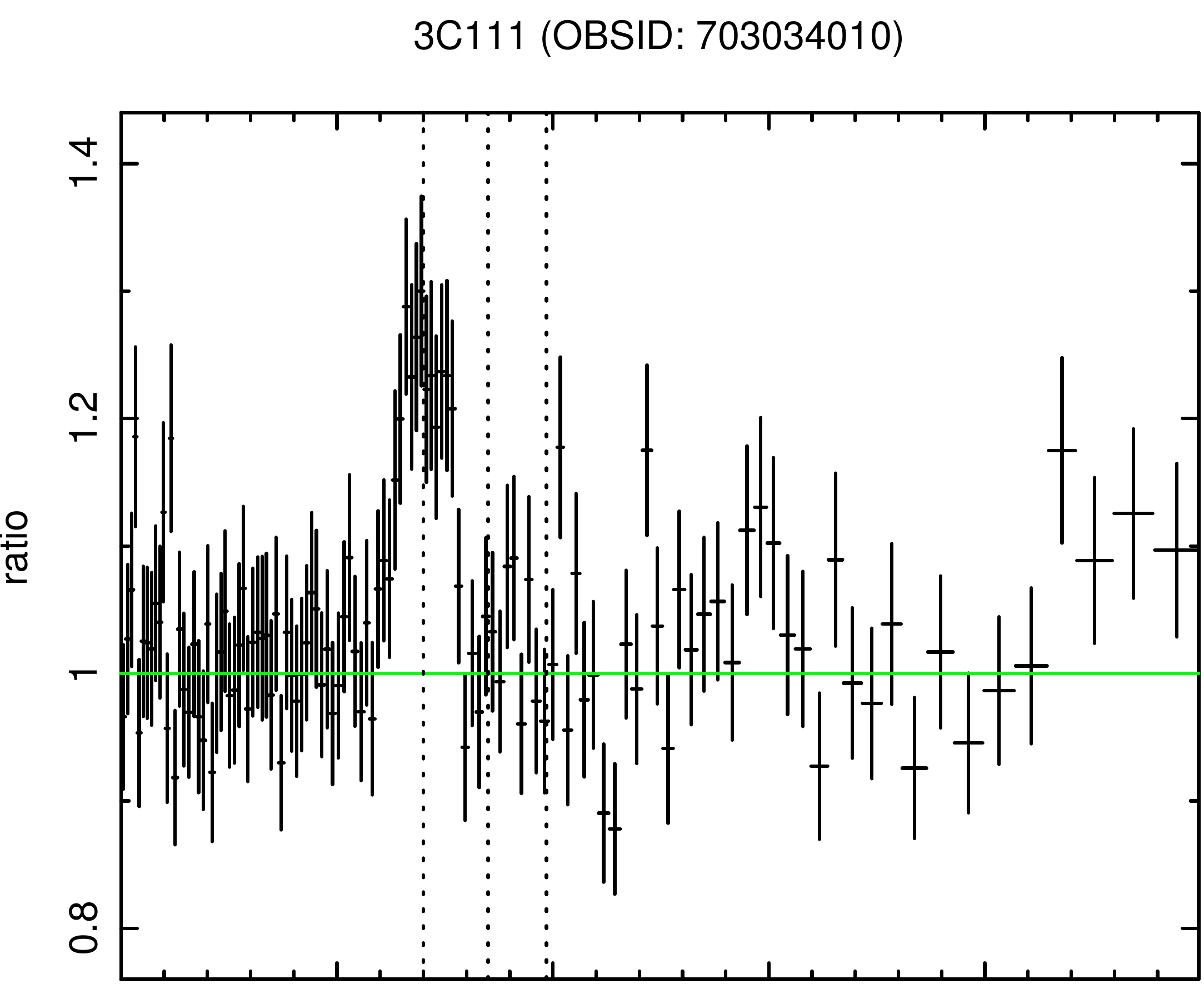}
\includegraphics[angle=-90,width=3.8cm]{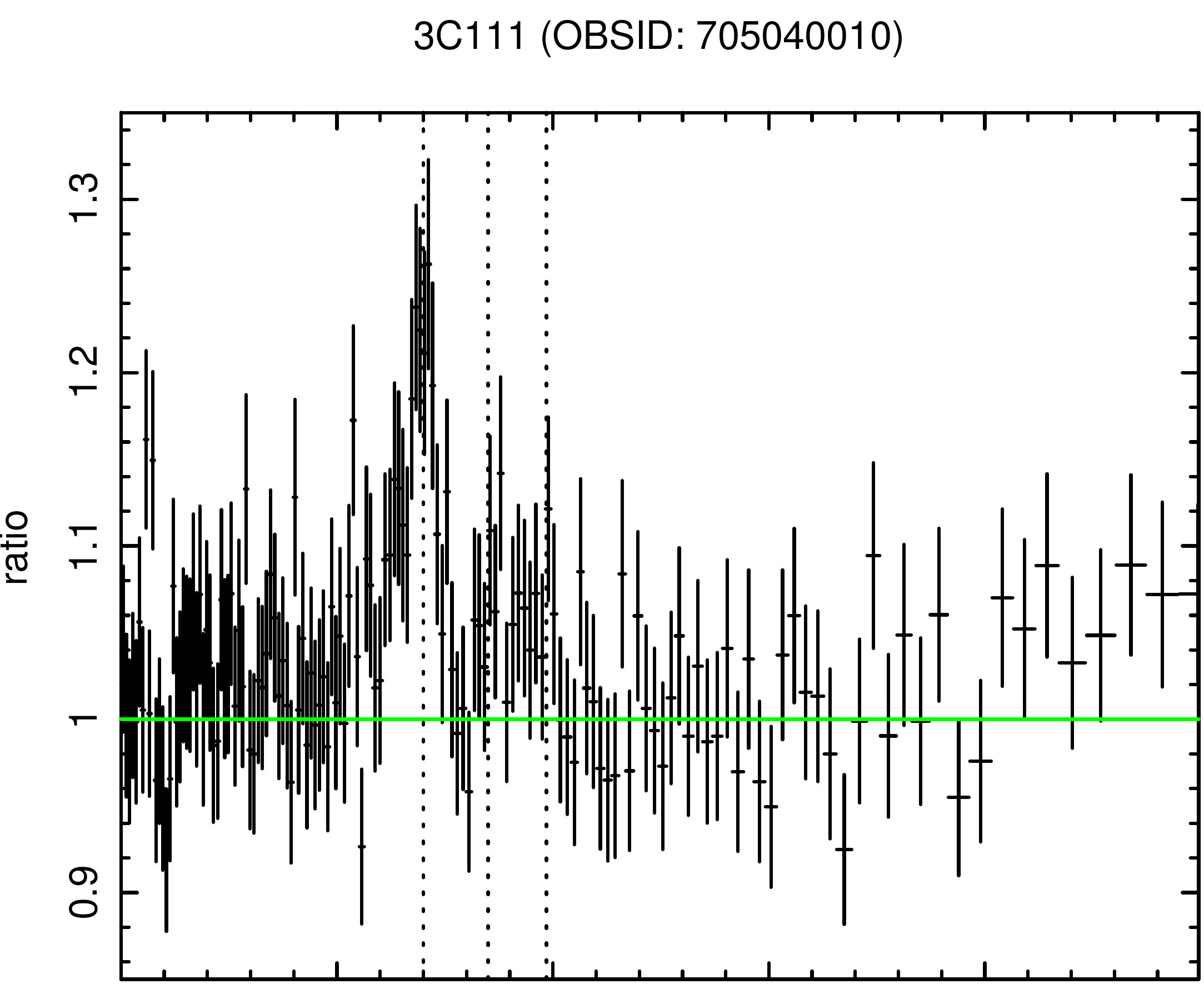}
\includegraphics[angle=-90,width=3.8cm]{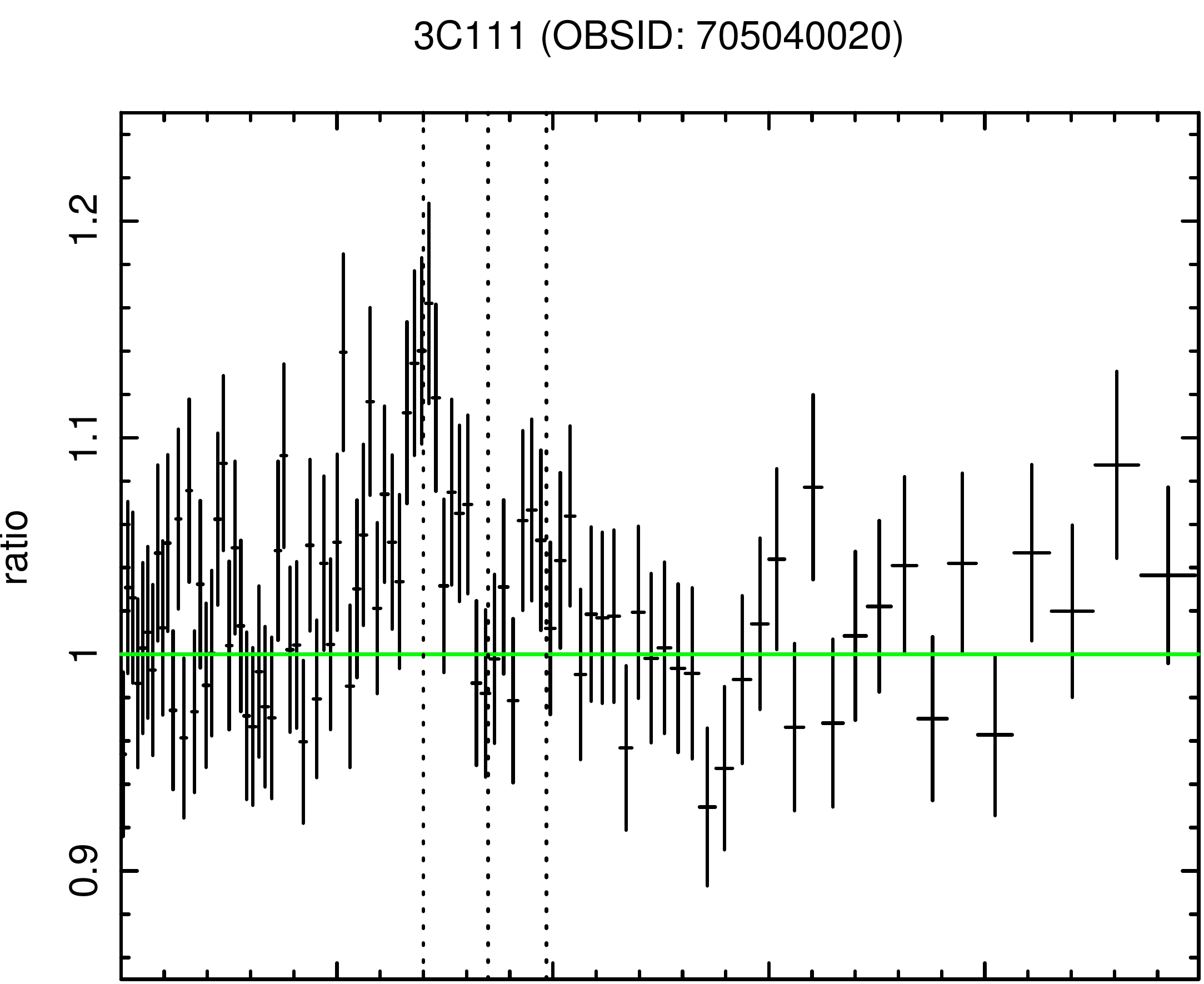}
}

\vspace{-12.4pt}
\subfloat{
\includegraphics[angle=-90,width=3.8cm]{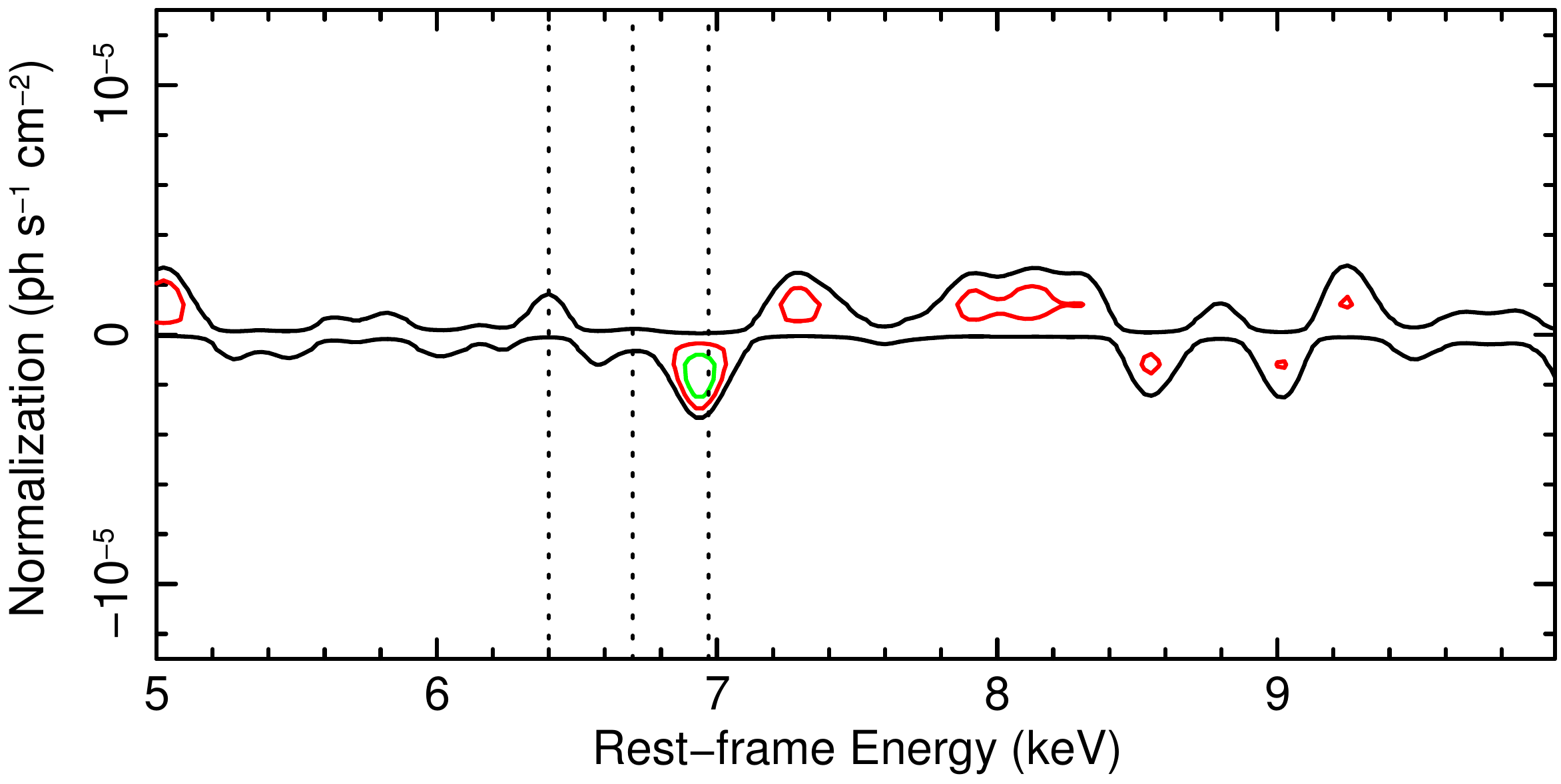}
\includegraphics[angle=-90,width=3.8cm]{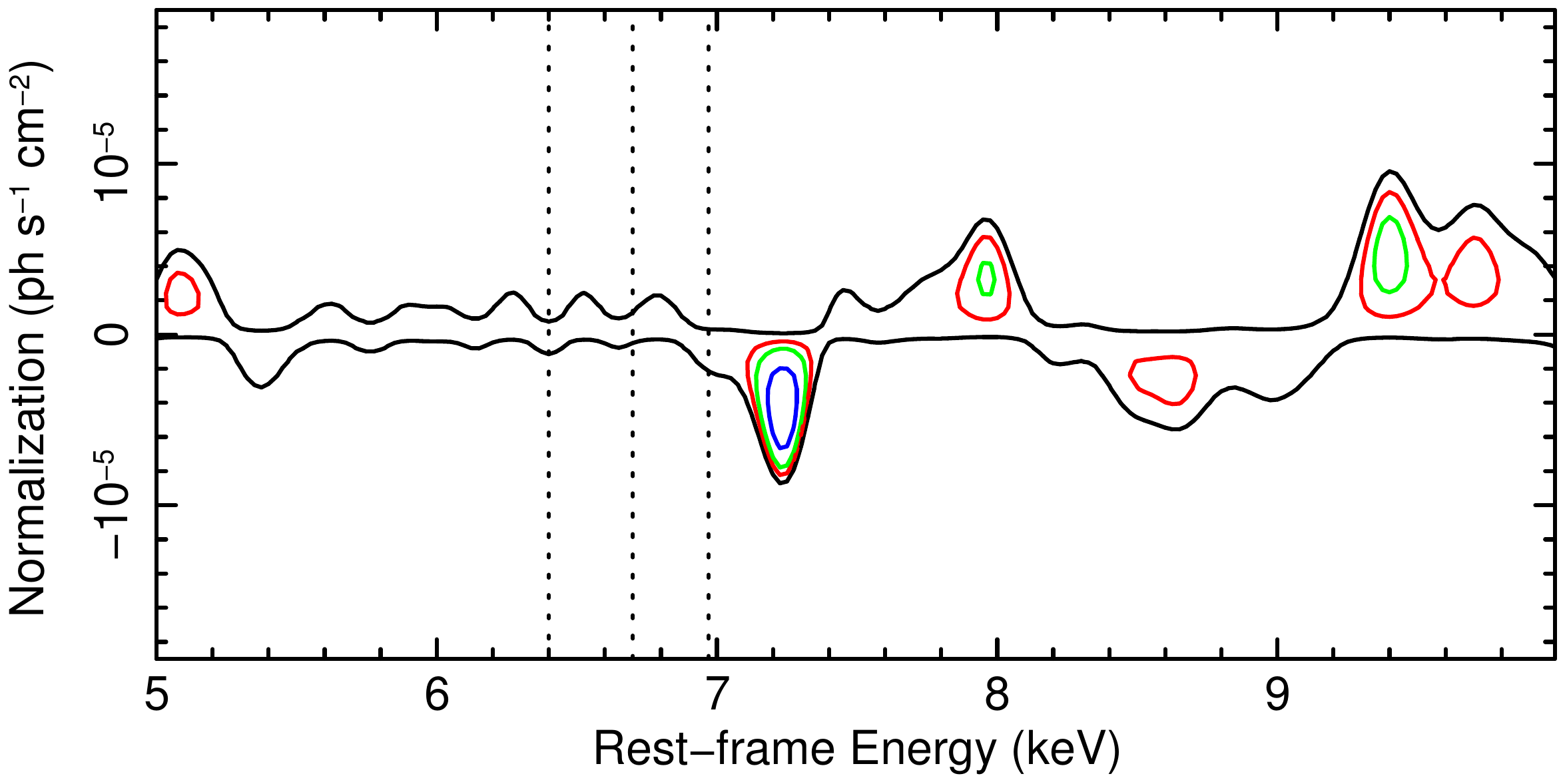}
\includegraphics[angle=-90,width=3.8cm]{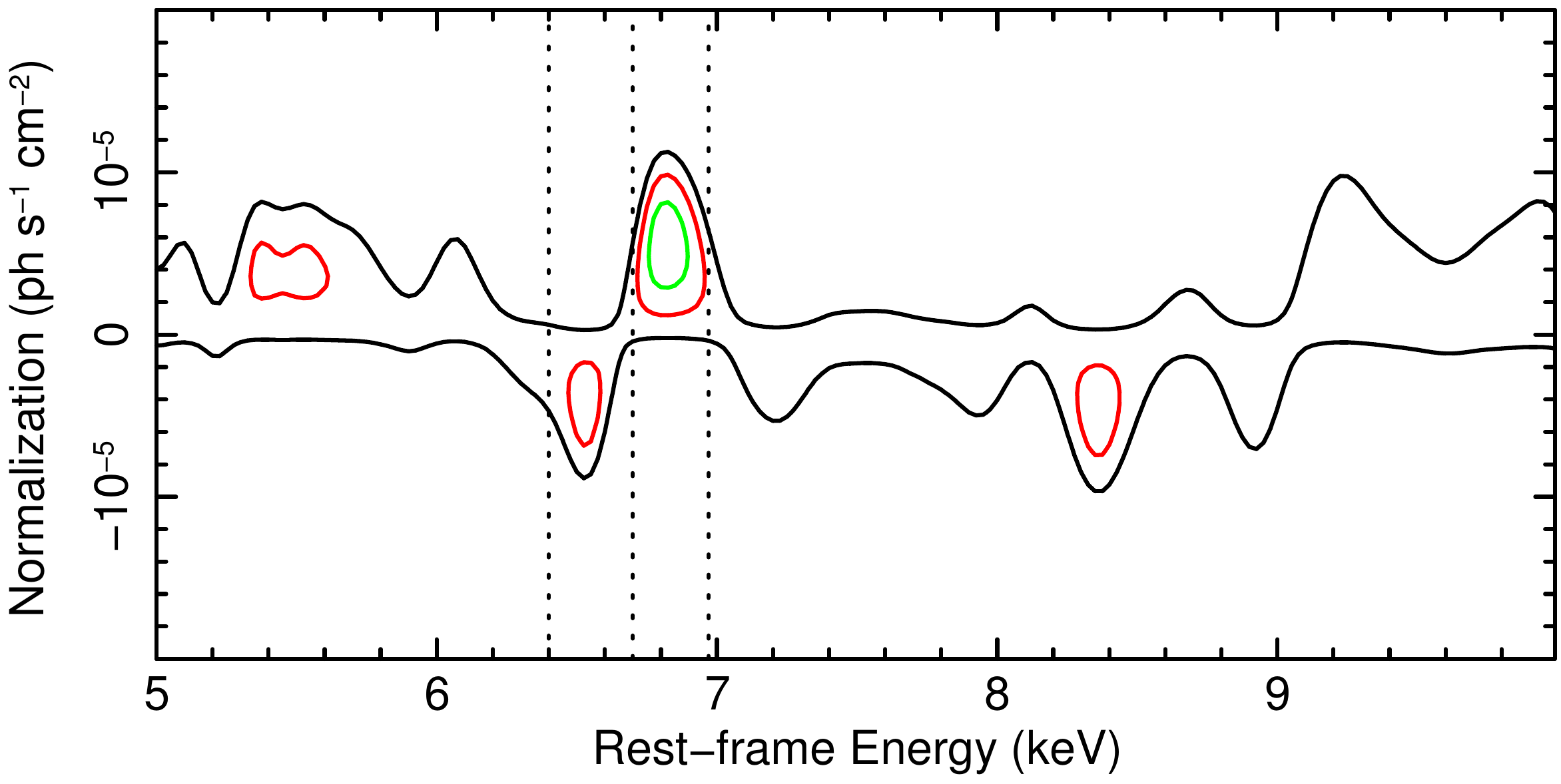}
\includegraphics[angle=-90,width=3.8cm]{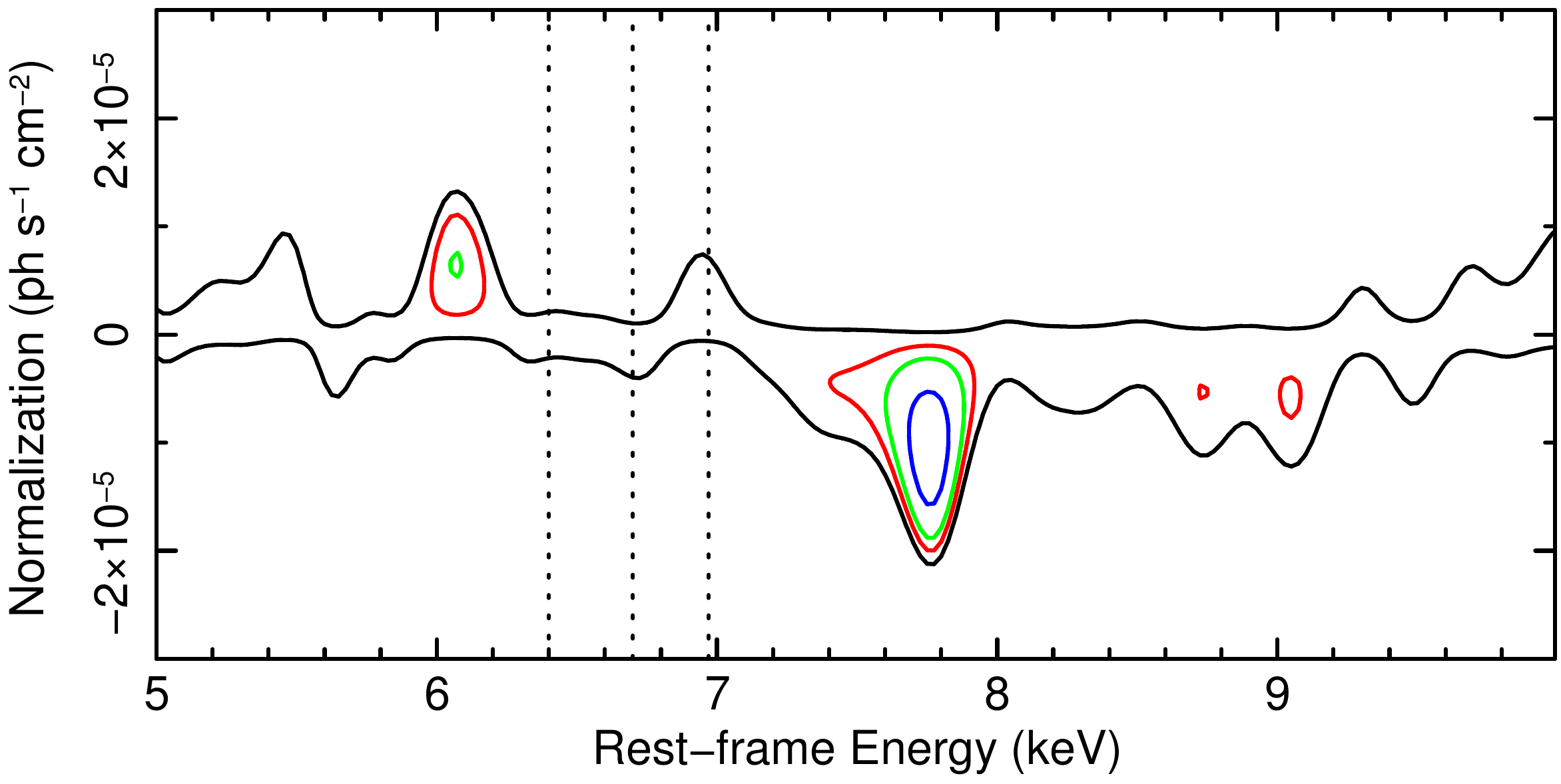}
}

\vspace{-5pt}
\subfloat{
\includegraphics[angle=-90,width=3.8cm]{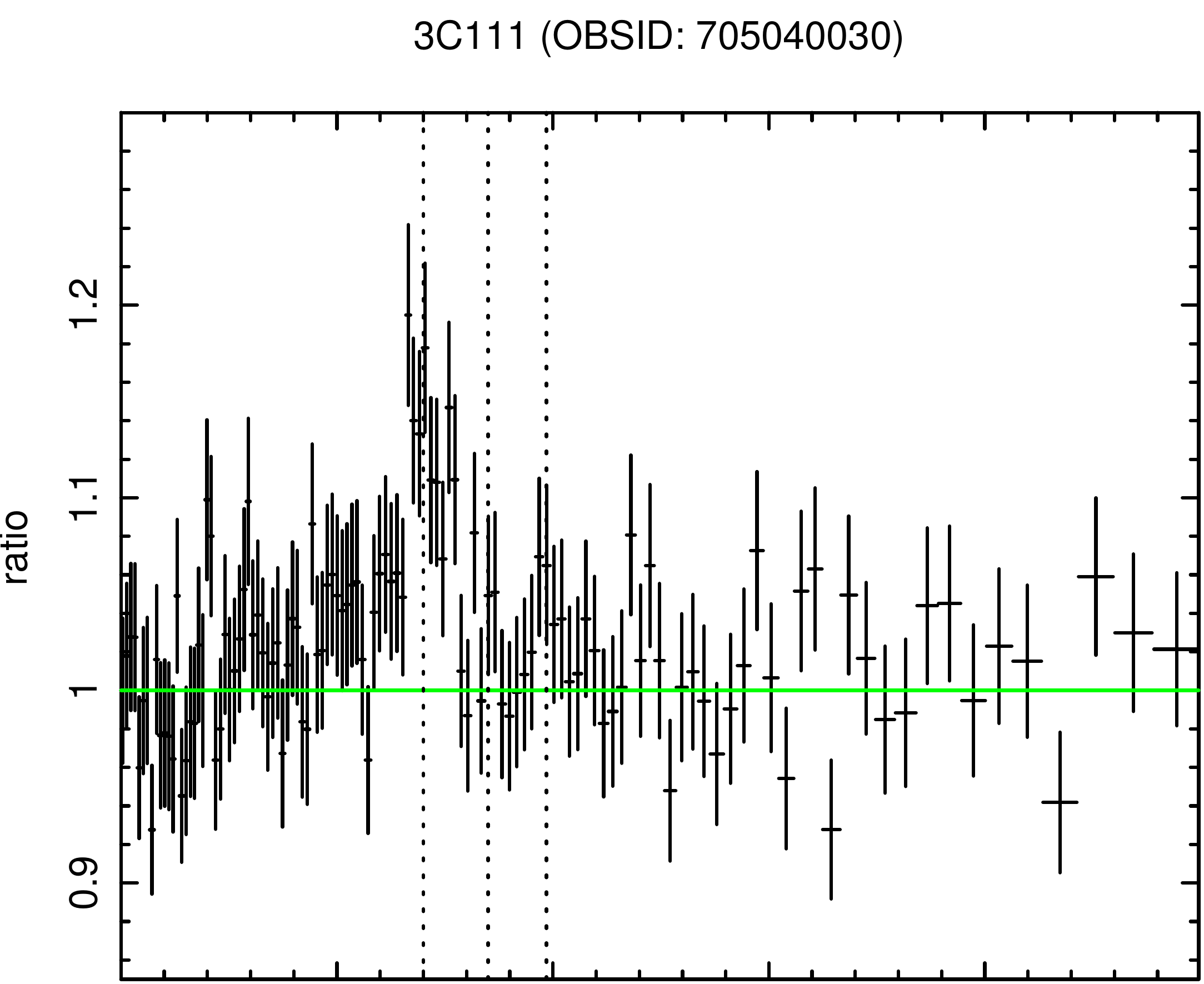}
\includegraphics[angle=-90,width=3.8cm]{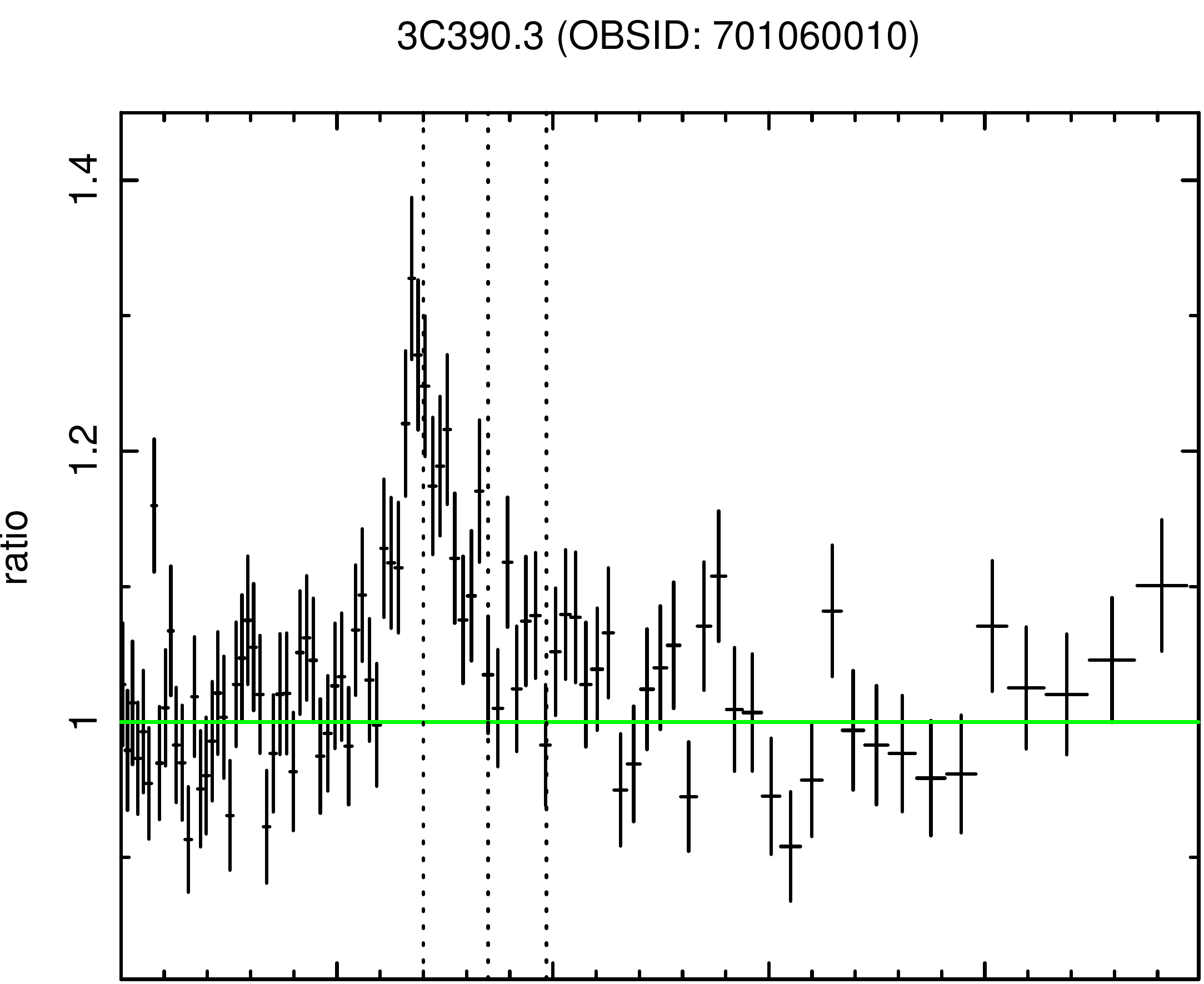}
\includegraphics[angle=-90,width=3.8cm]{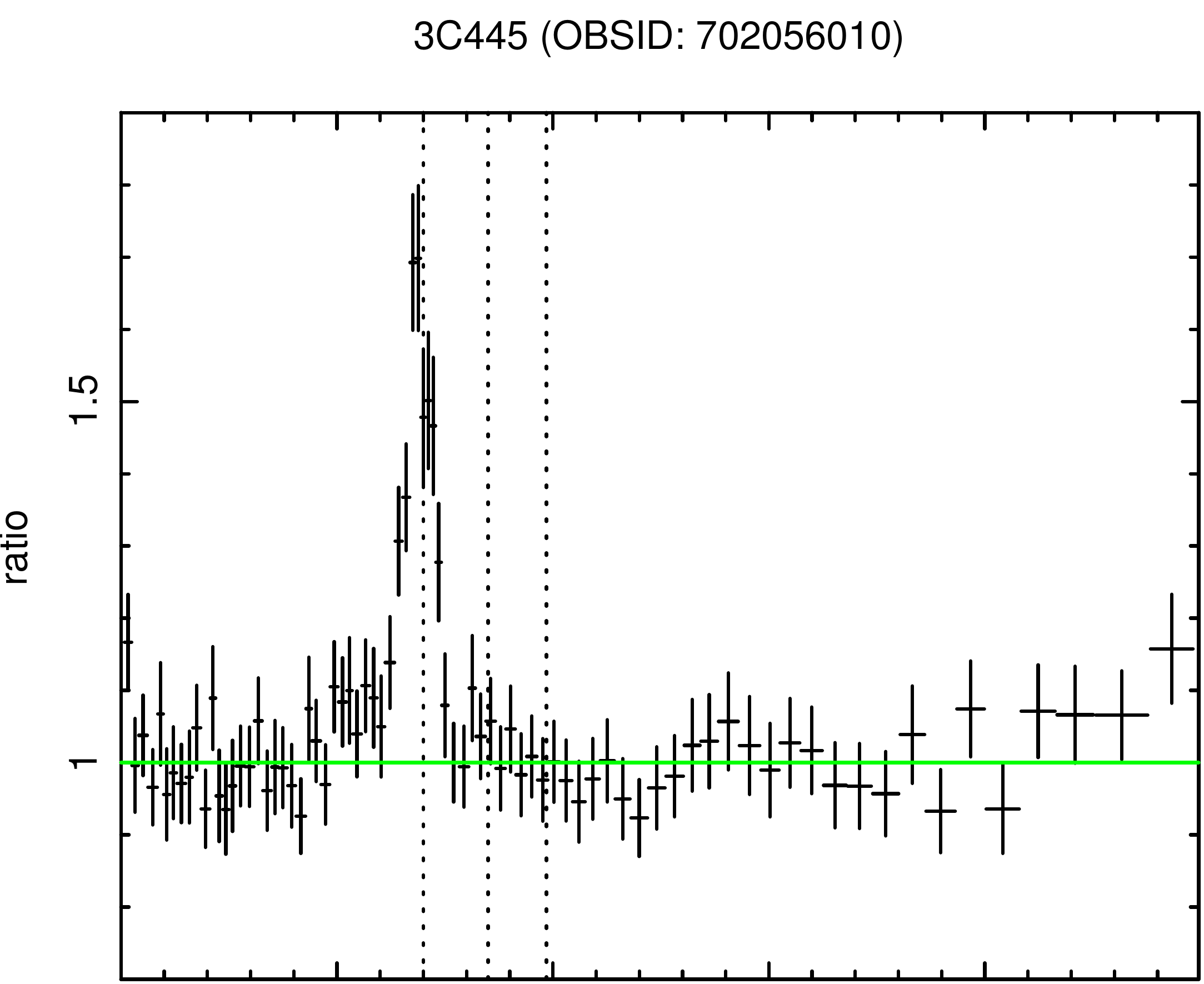}
\includegraphics[angle=-90,width=3.8cm]{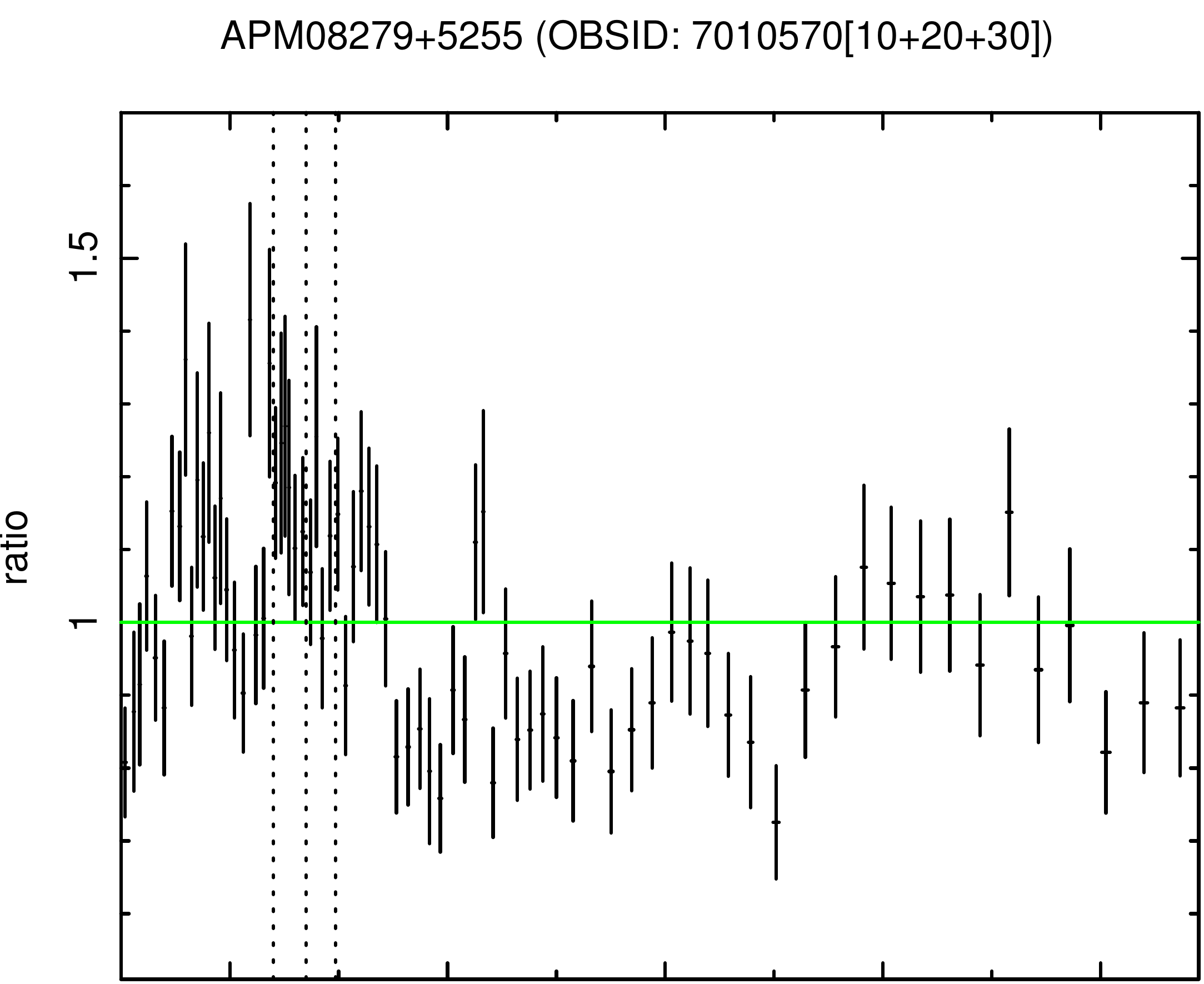}
}
	
\vspace{-12.2pt}
\subfloat{
\includegraphics[angle=-90,width=3.8cm]{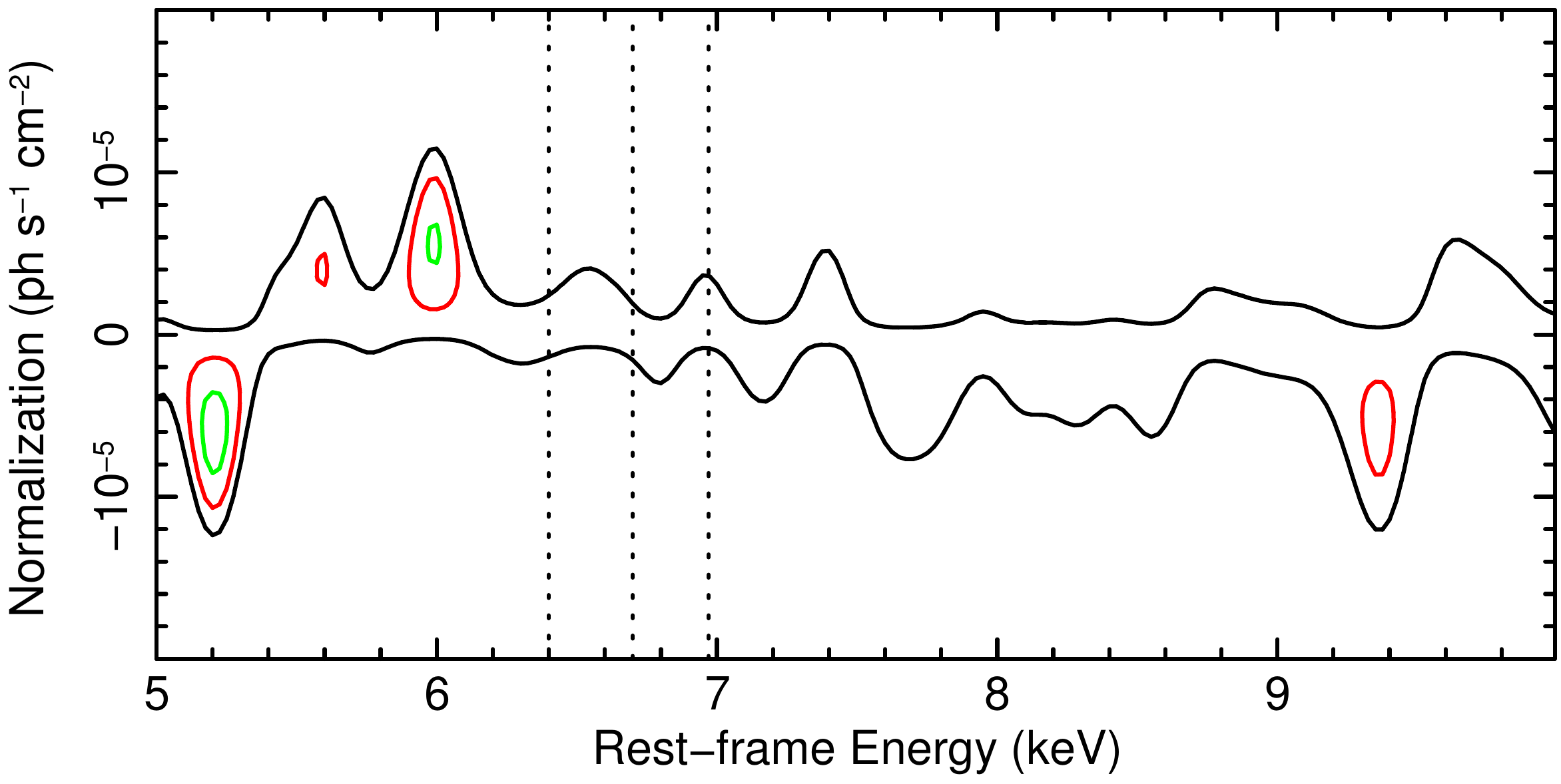}
\includegraphics[angle=-90,width=3.8cm]{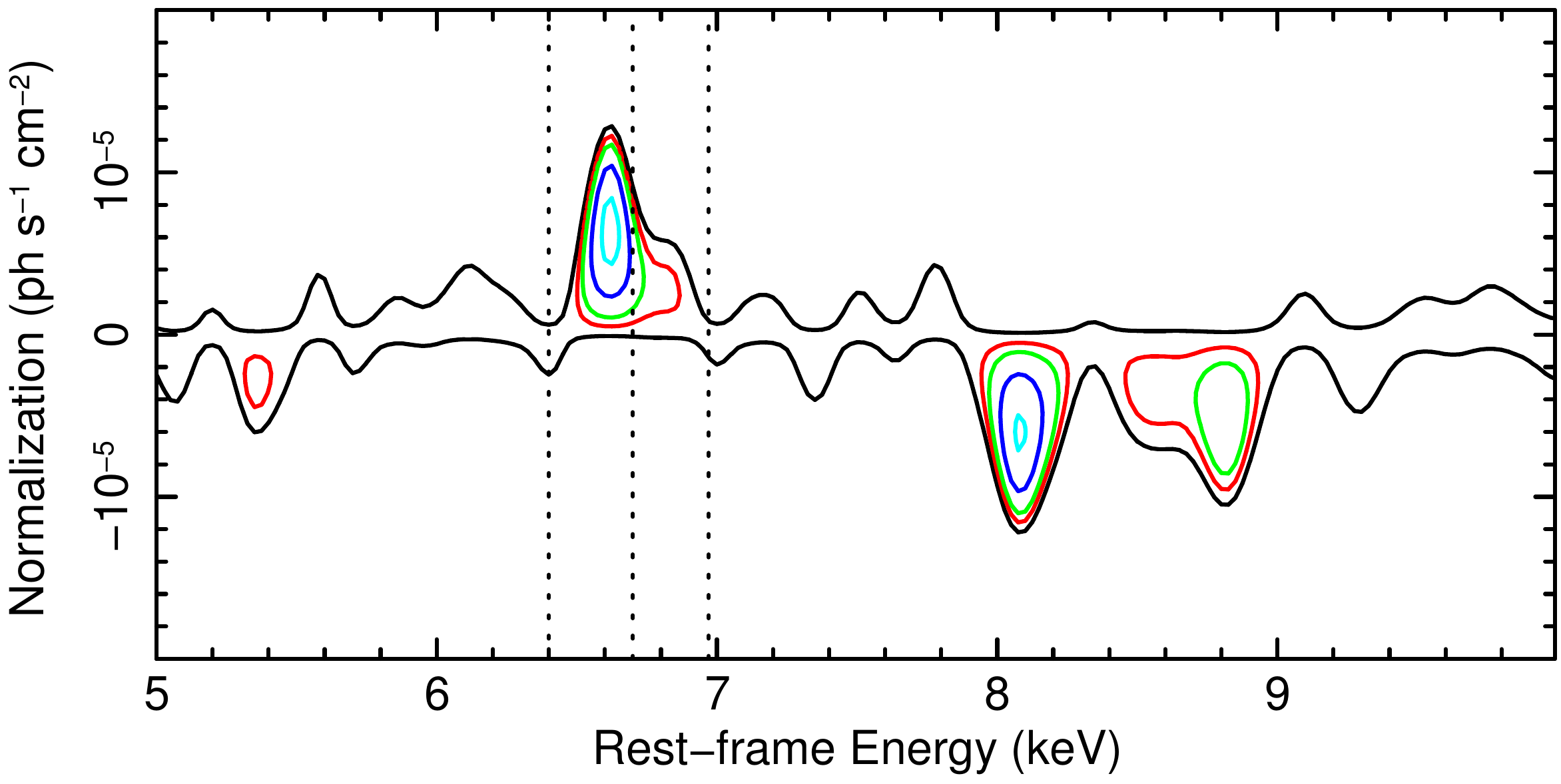}
\includegraphics[angle=-90,width=3.8cm]{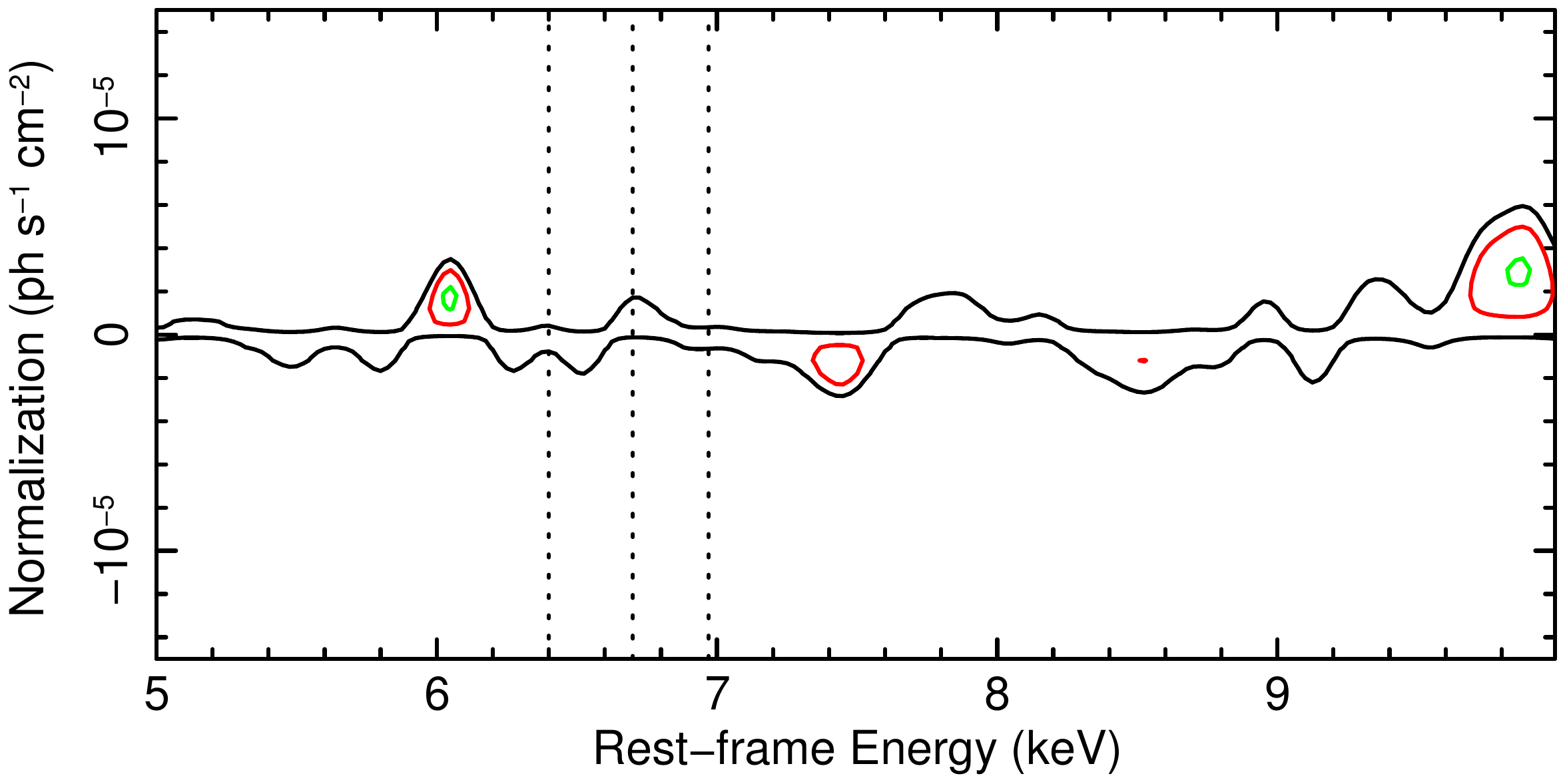}
\includegraphics[angle=-90,width=3.8cm]{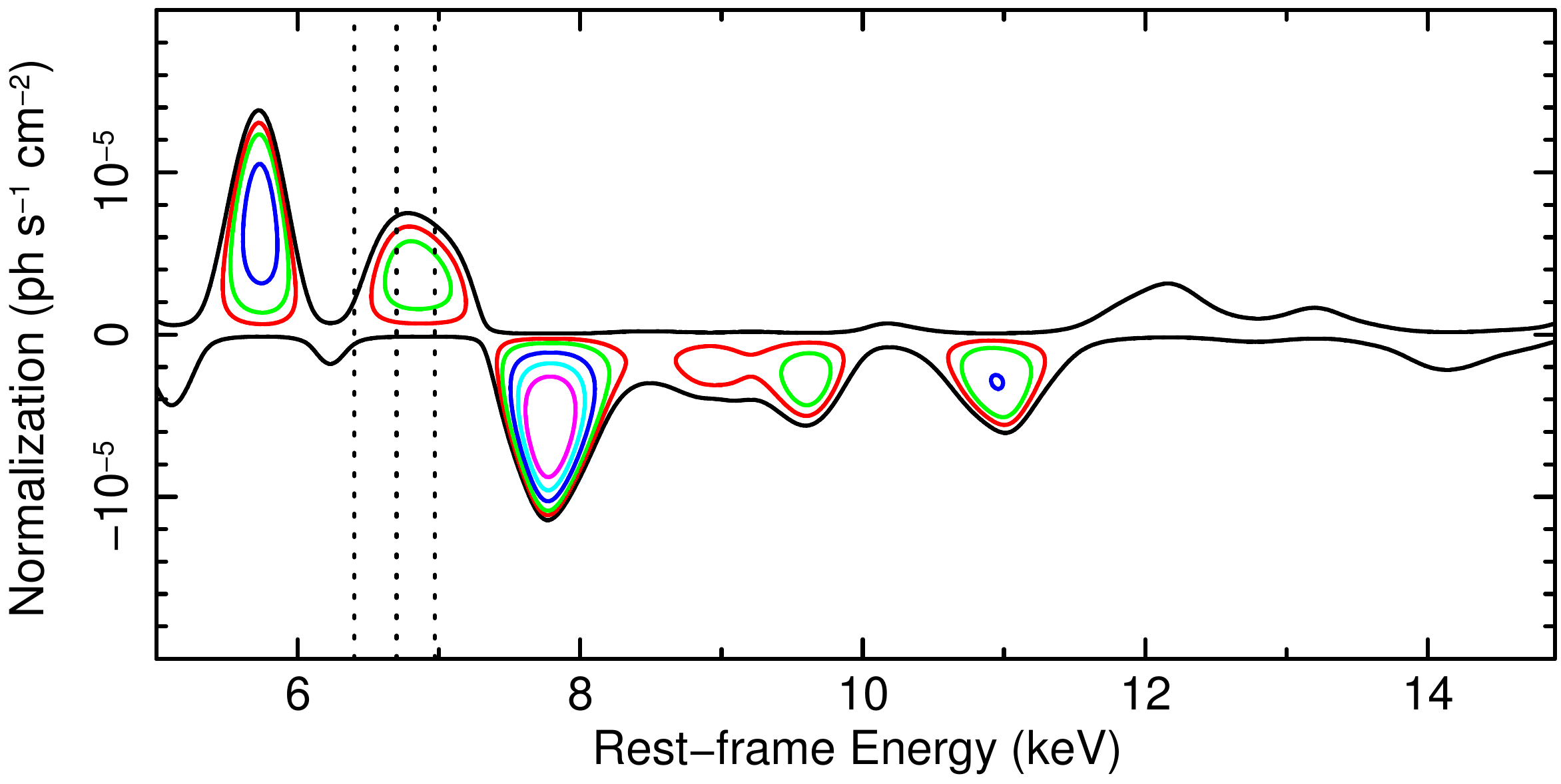}
}
	
\vspace{-5pt}
\subfloat{

\includegraphics[angle=-90,width=3.8cm]{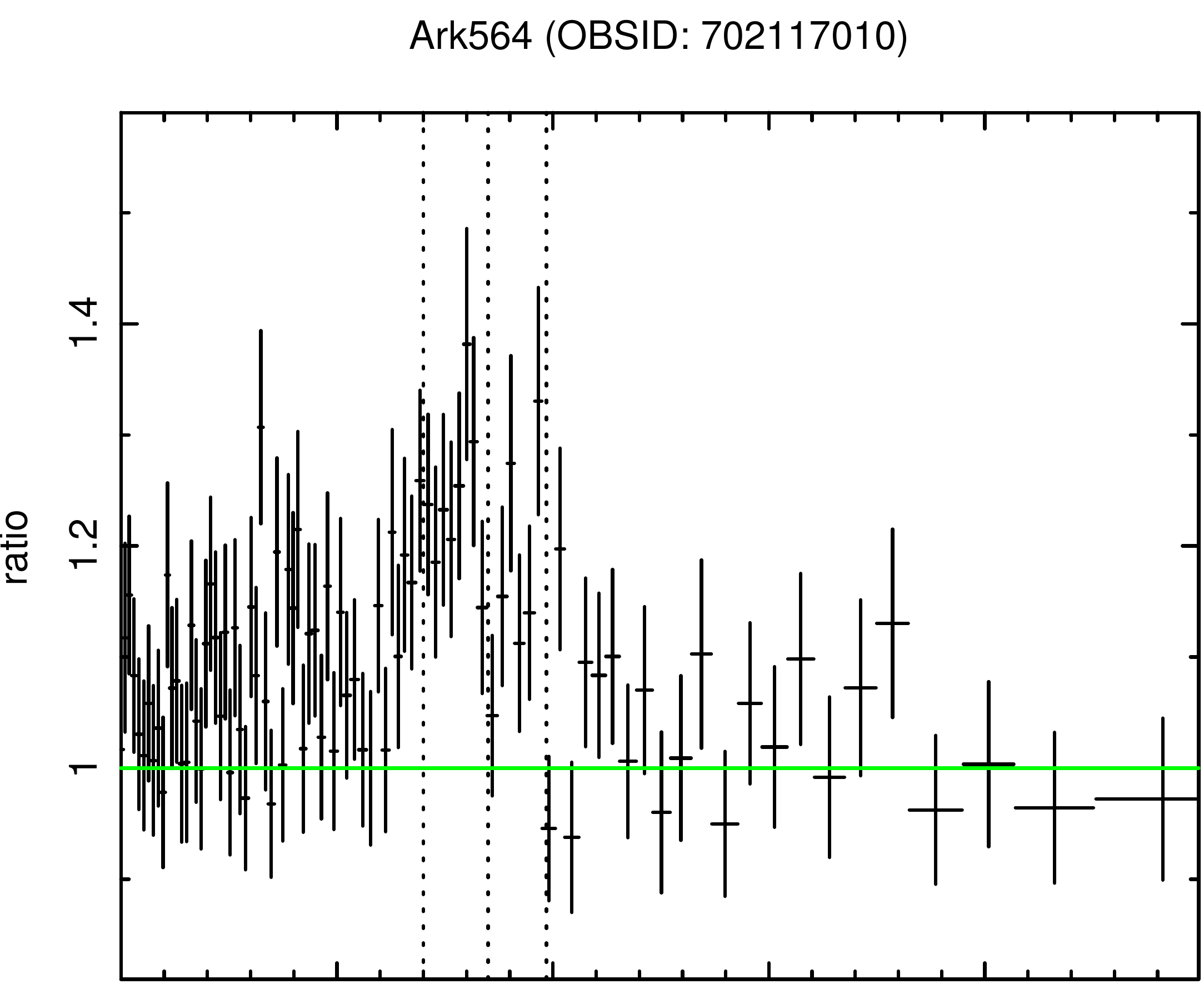}
\includegraphics[angle=-90,width=3.8cm]{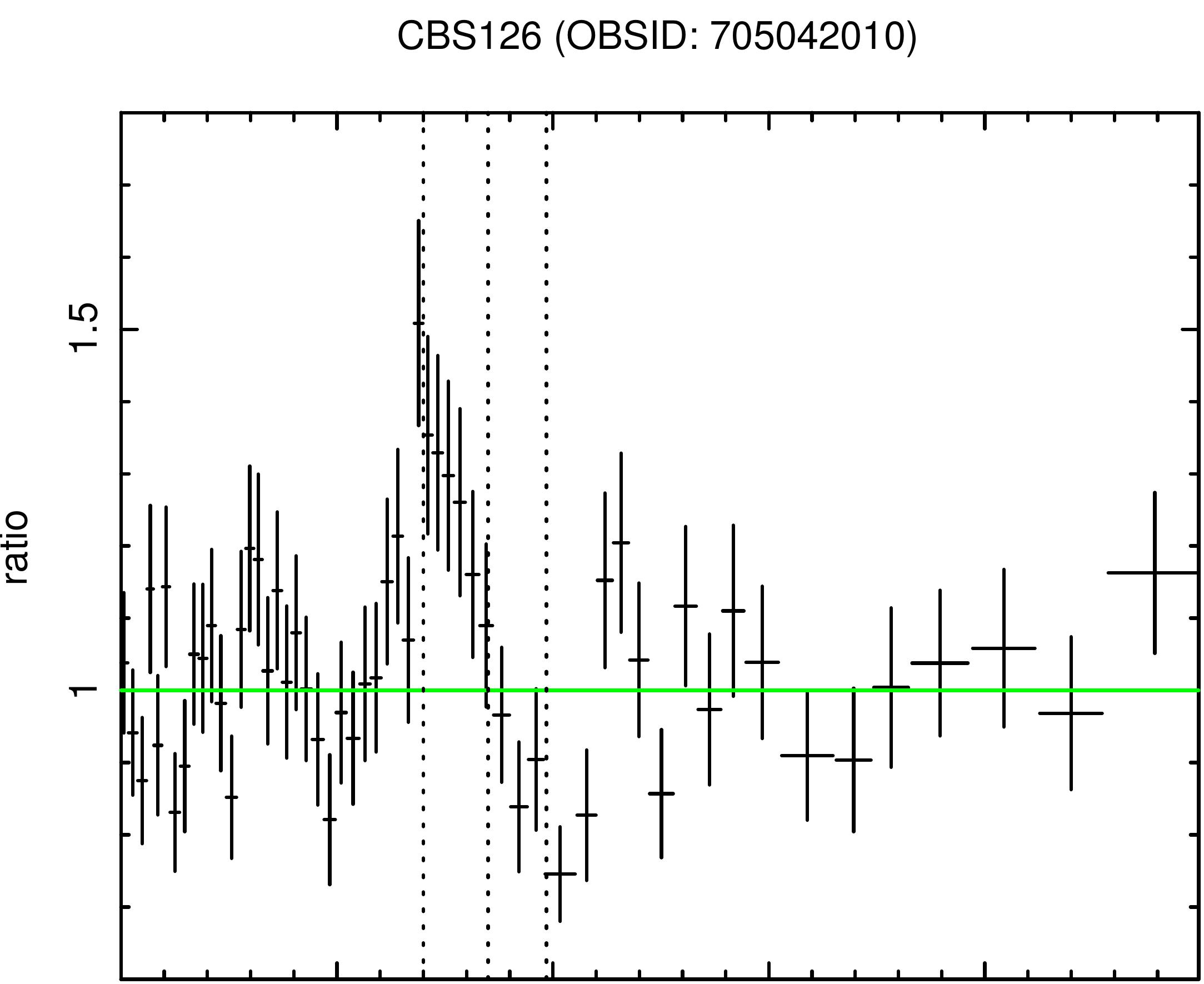}
\includegraphics[angle=-90,width=3.8cm]{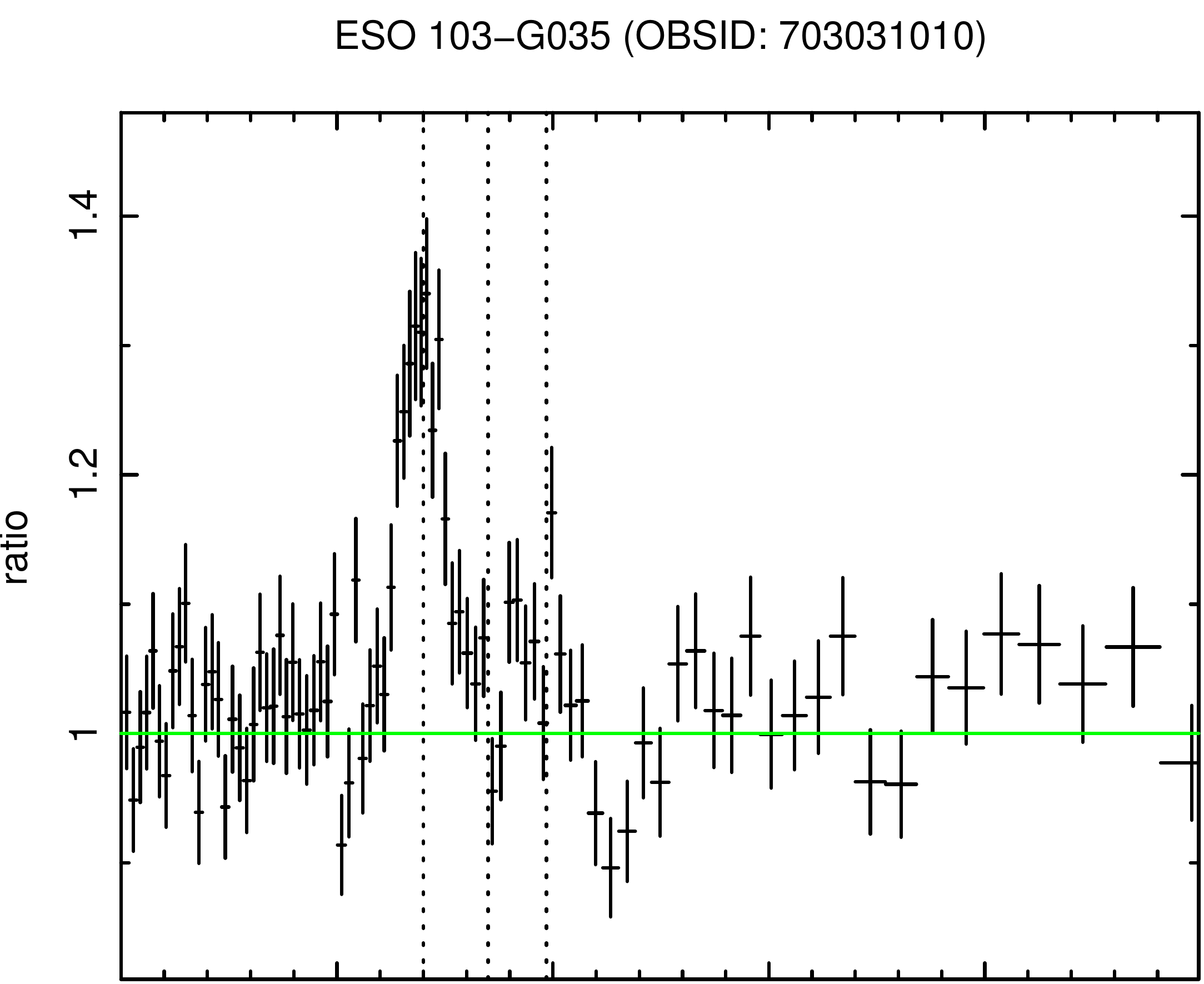}
\includegraphics[angle=-90,width=3.8cm]{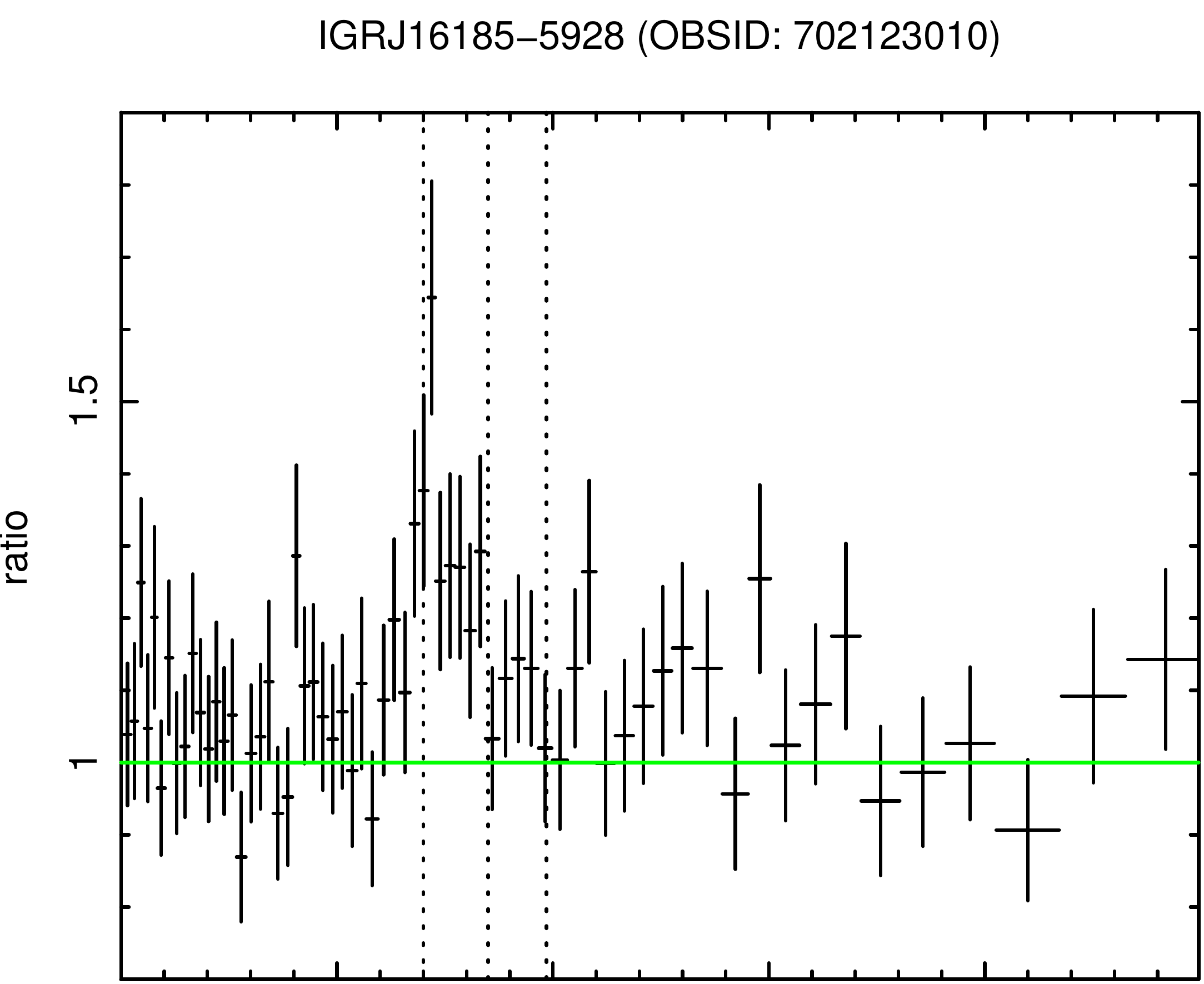}
}

\vspace{-12.2pt}
\subfloat{
\includegraphics[angle=-90,width=3.8cm]{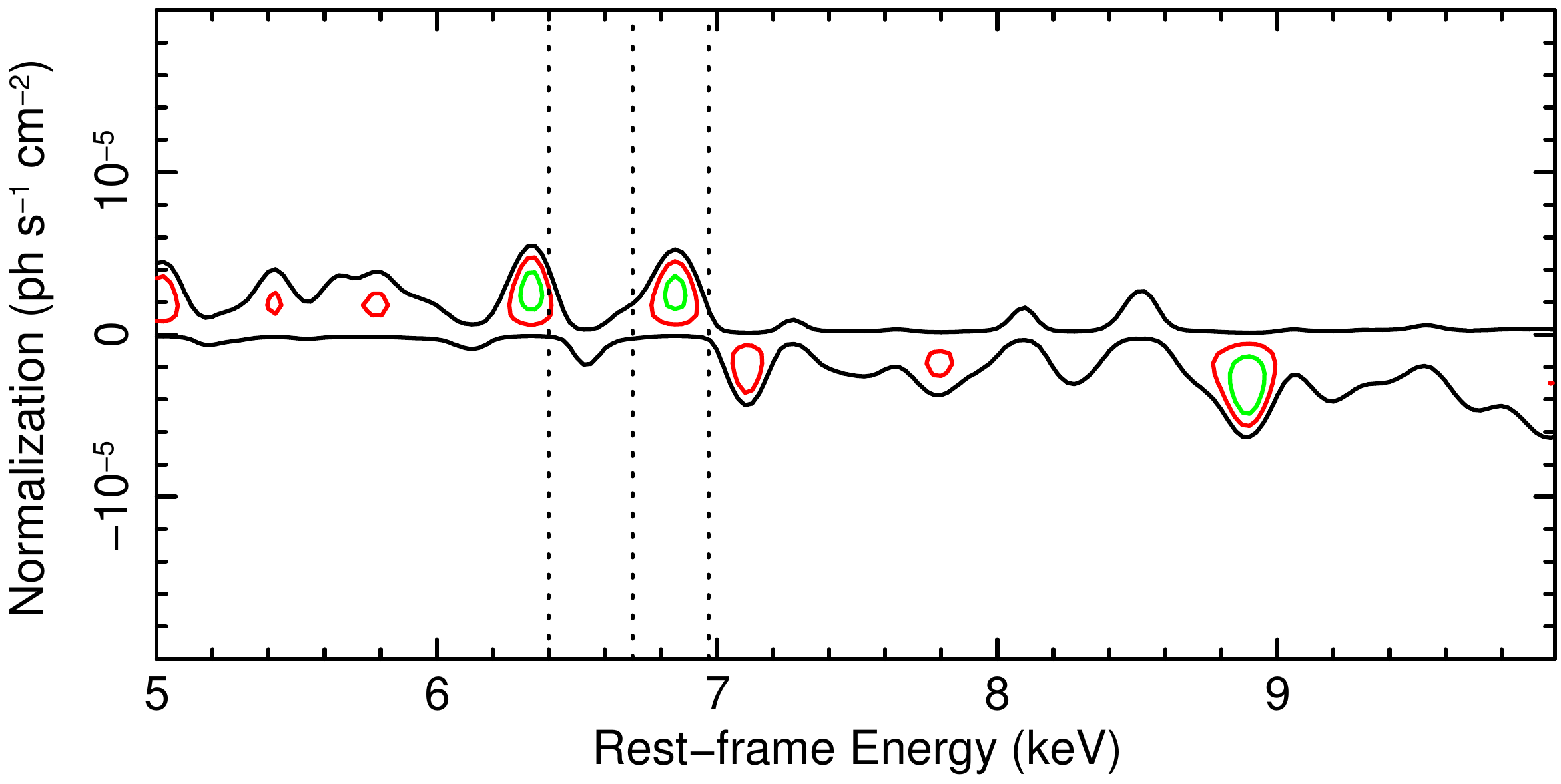}
\includegraphics[angle=-90,width=3.8cm]{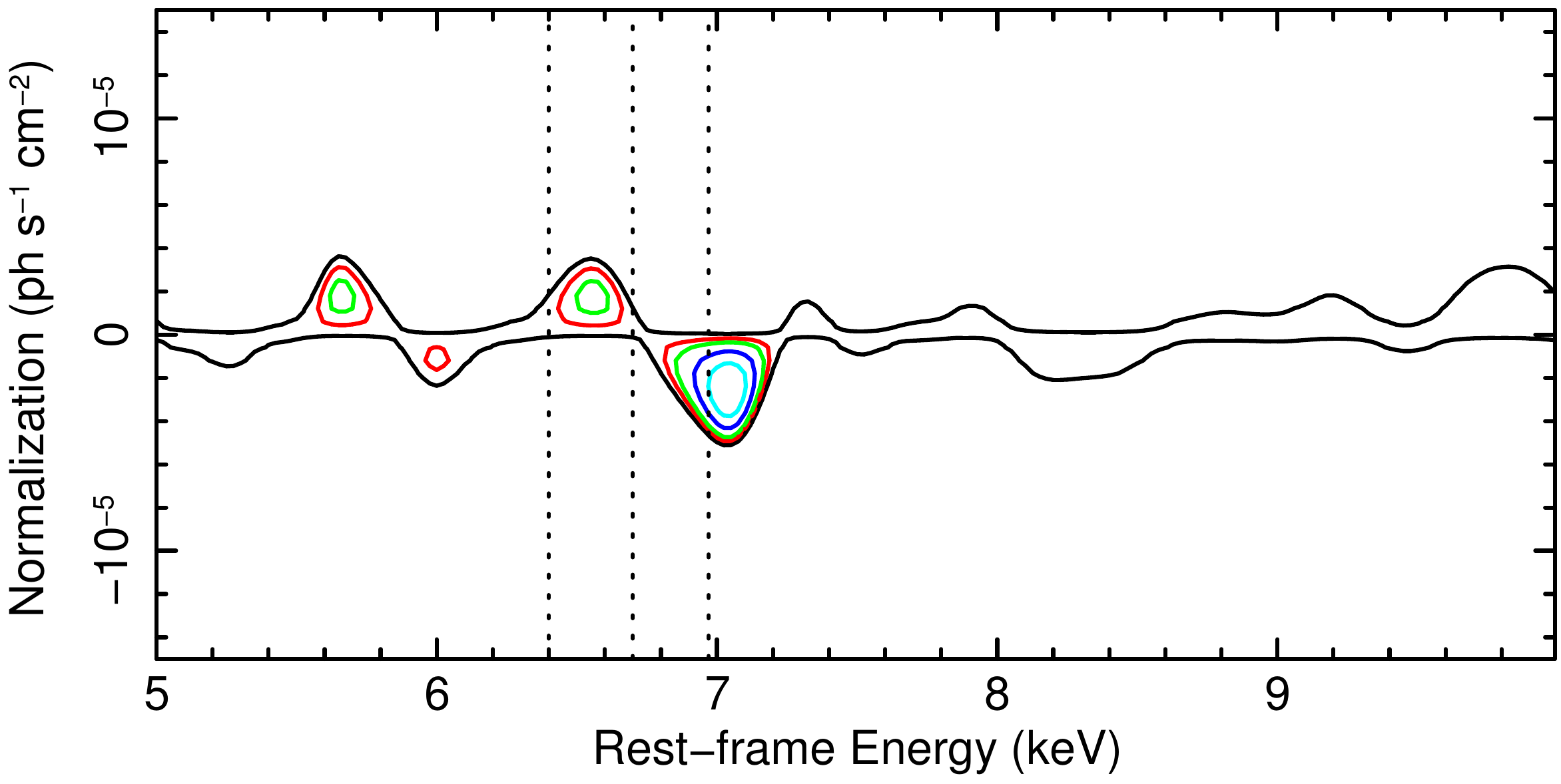}
\includegraphics[angle=-90,width=3.8cm]{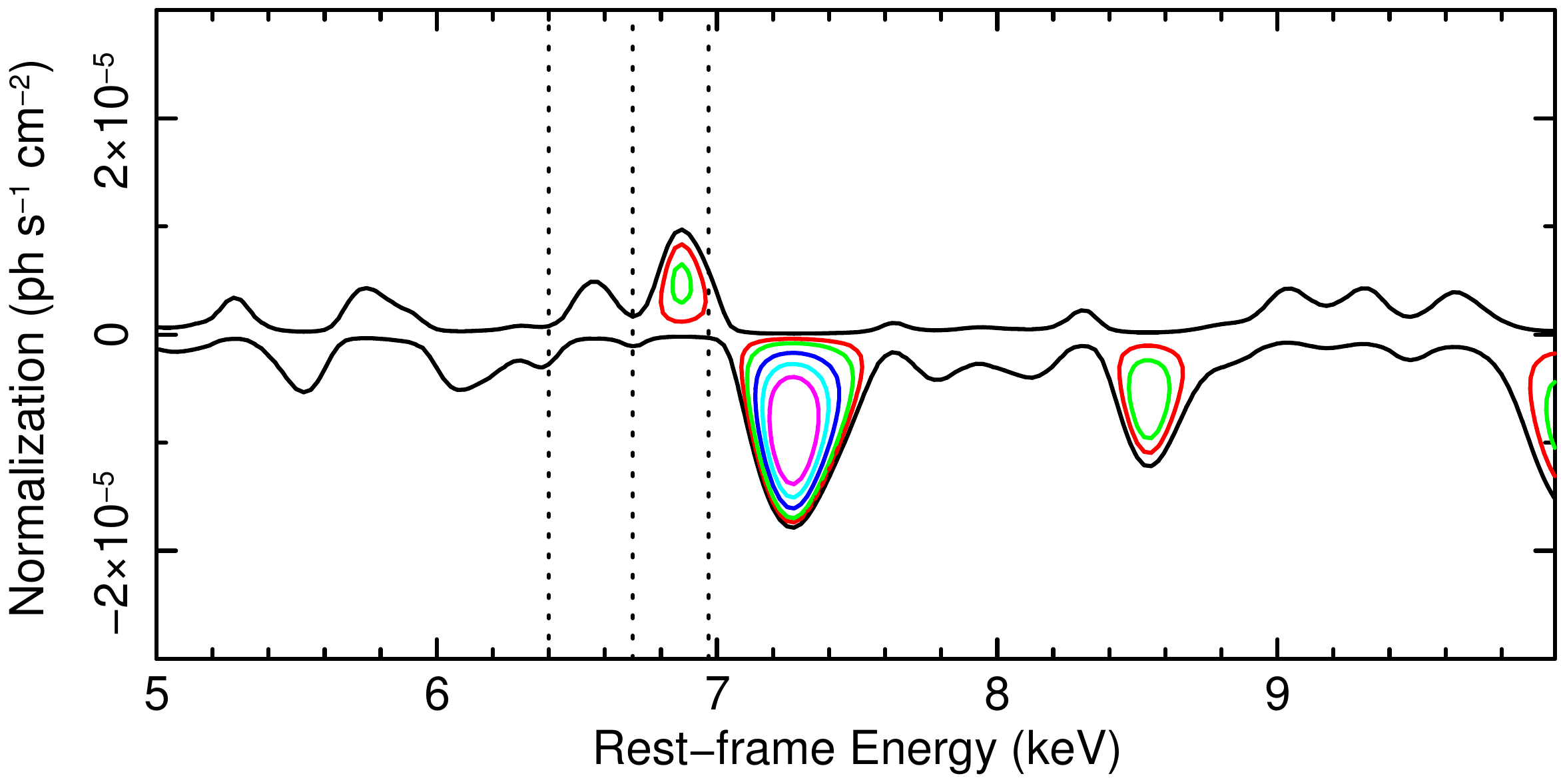}
\includegraphics[angle=-90,width=3.8cm]{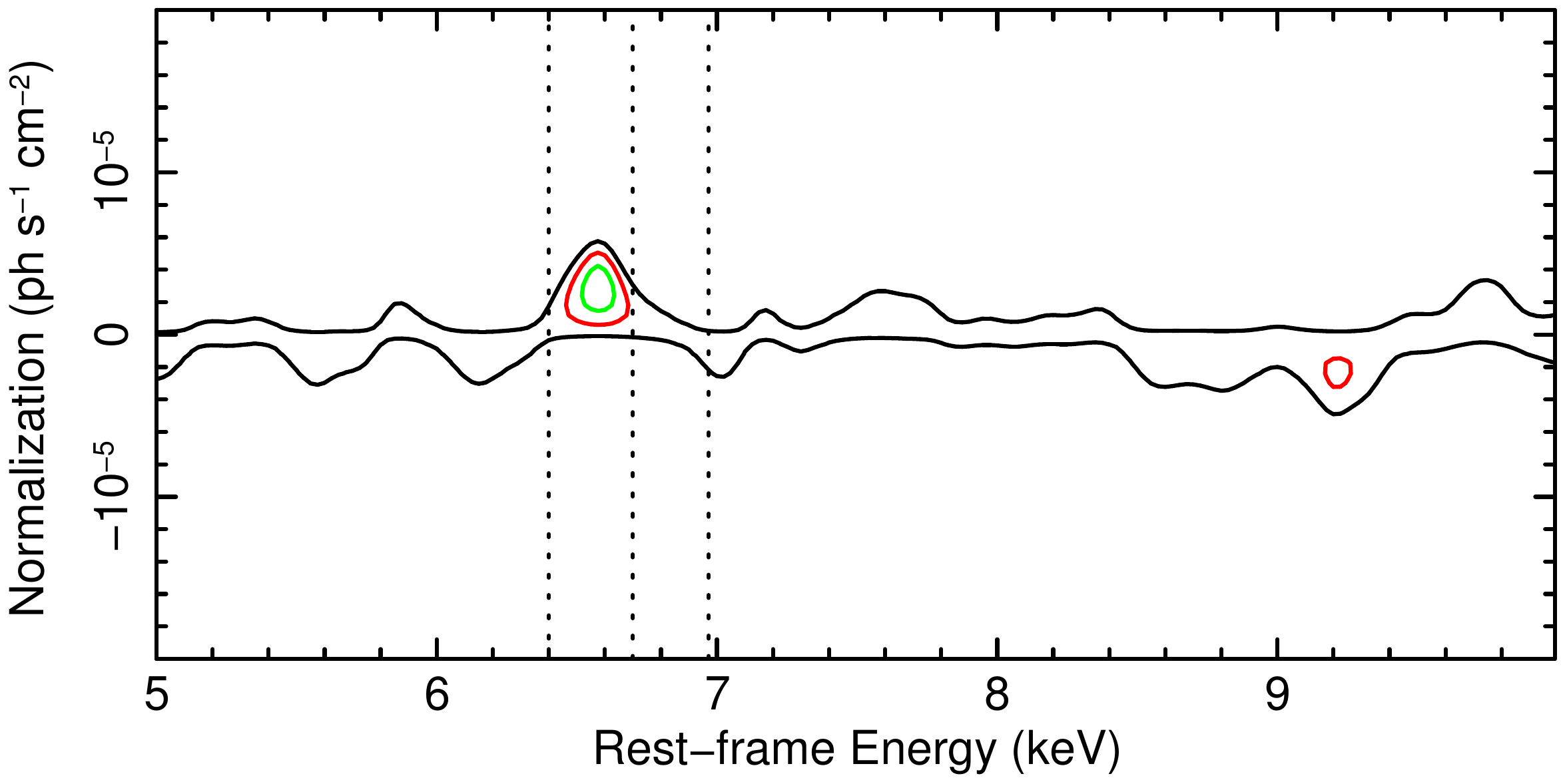}

}

\vspace{-5pt}
\subfloat{
\includegraphics[angle=-90,width=3.8cm]{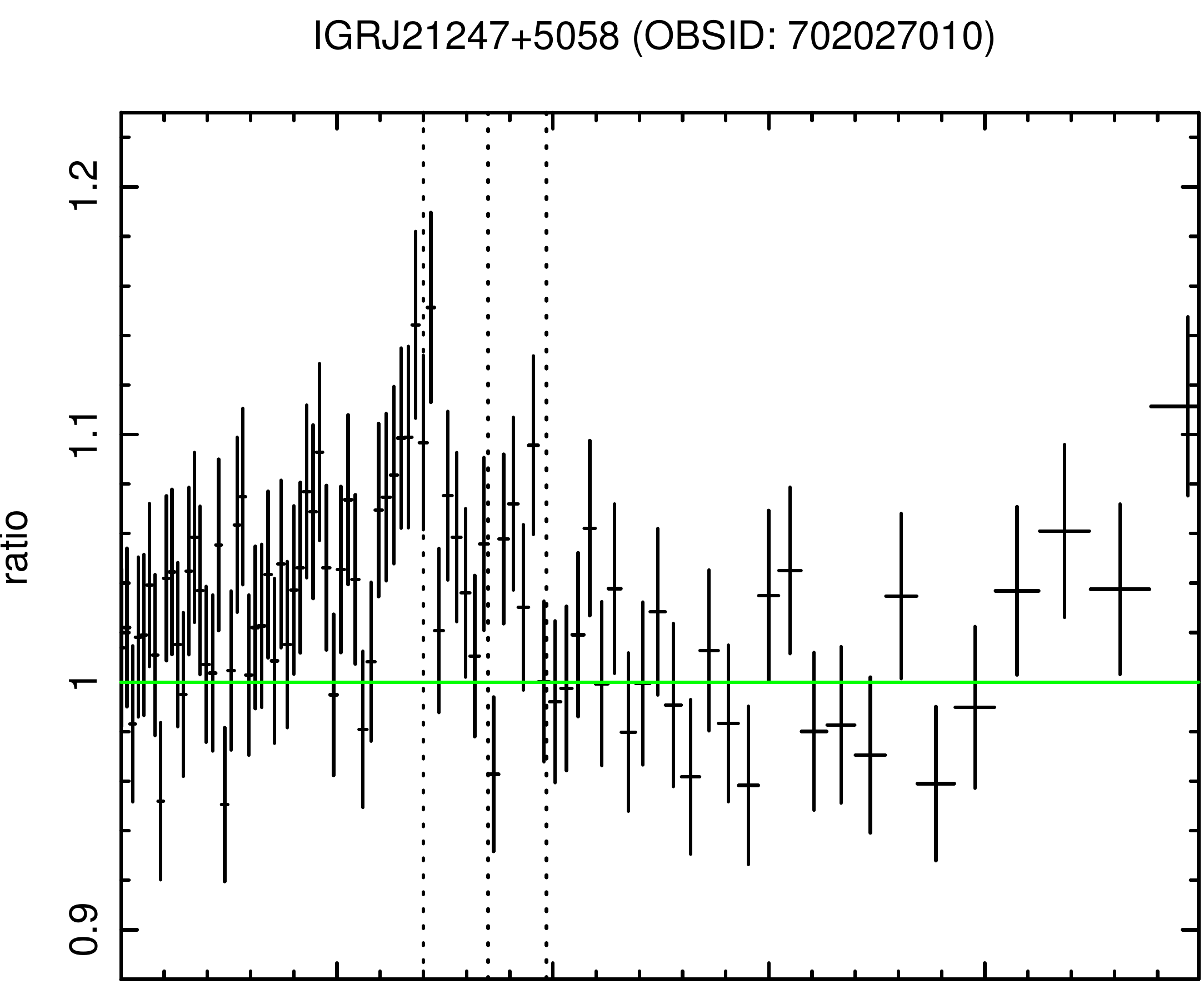}
\includegraphics[angle=-90,width=3.8cm]{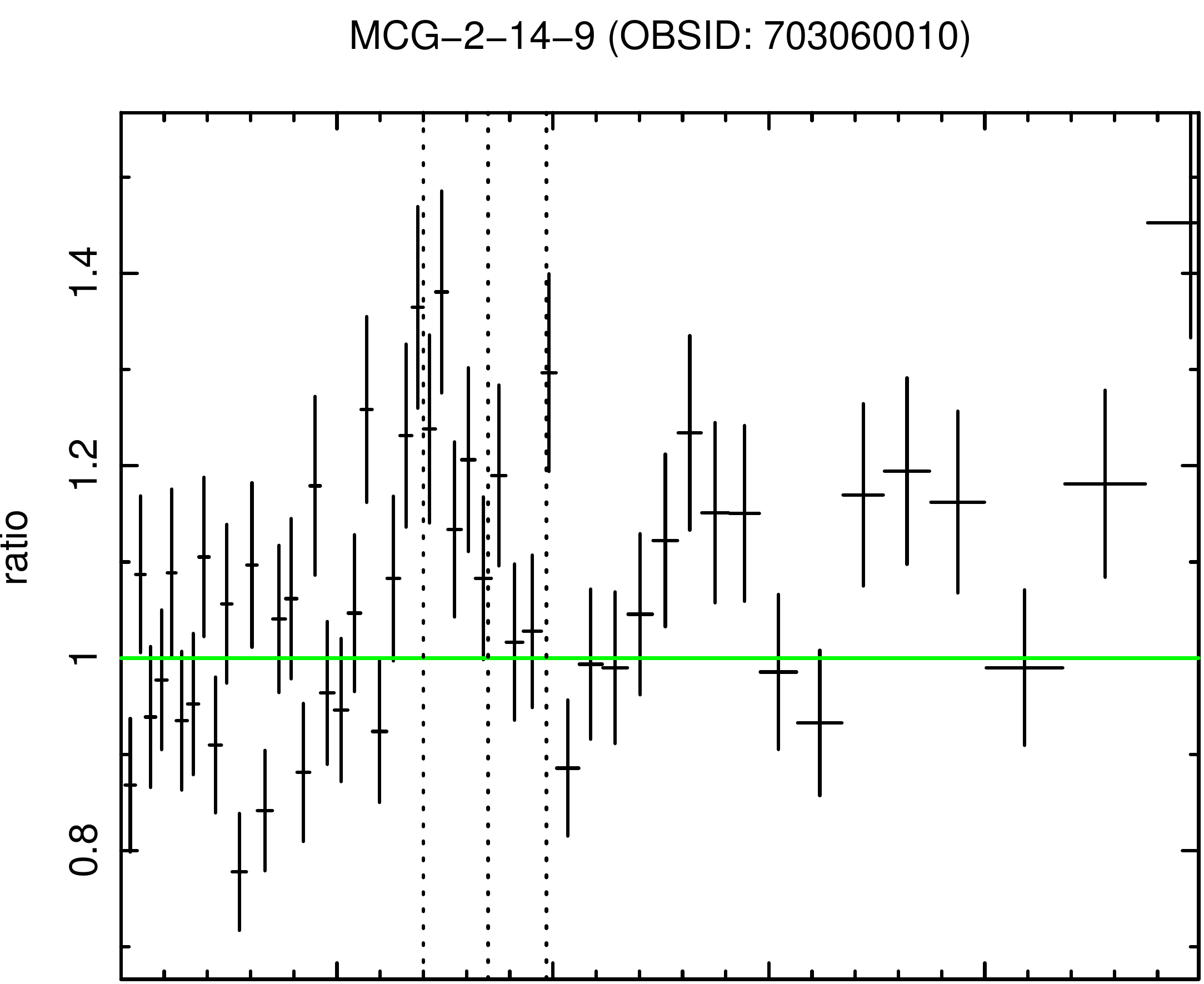}
\includegraphics[angle=-90,width=3.8cm]{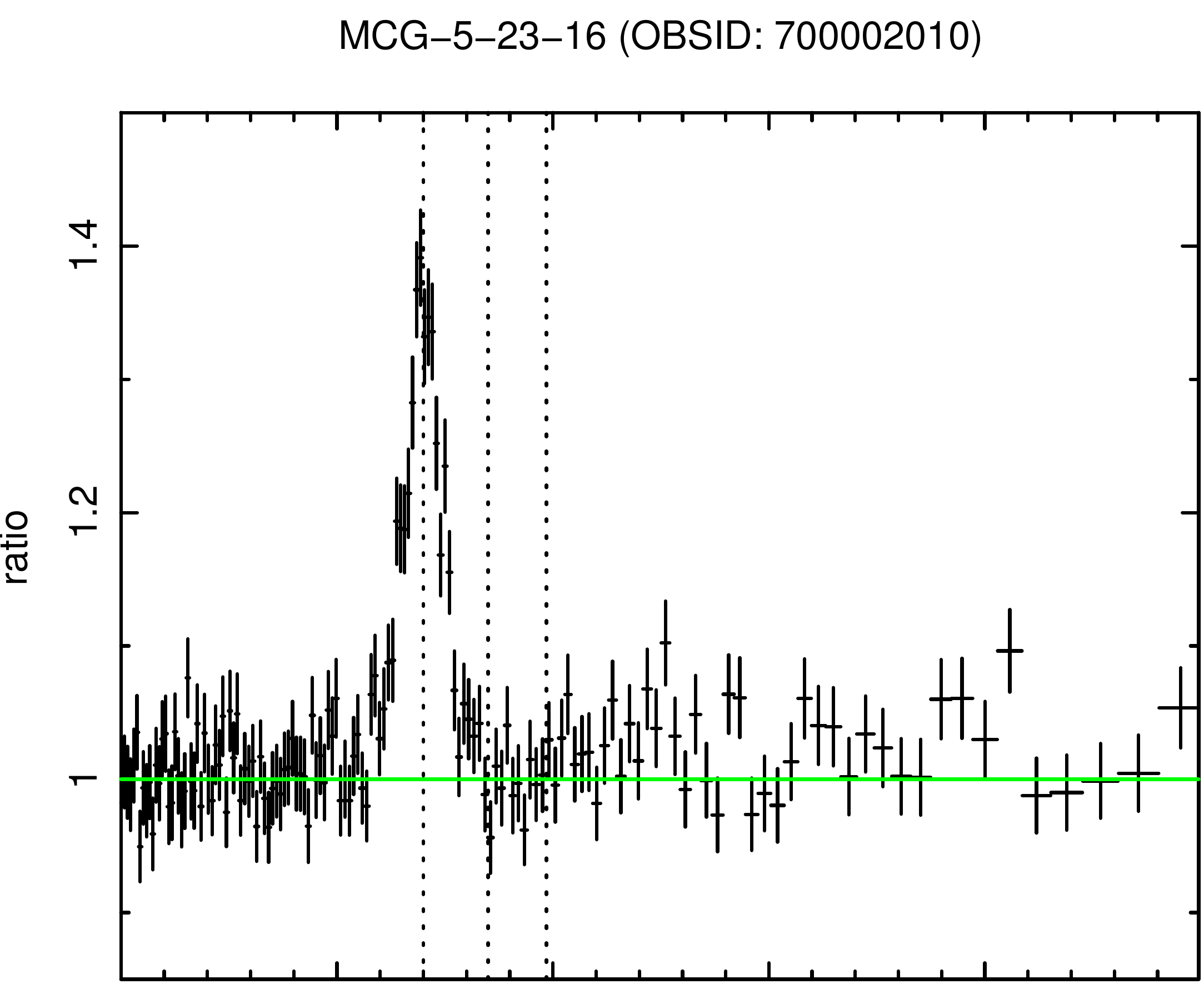}
\includegraphics[angle=-90,width=3.8cm]{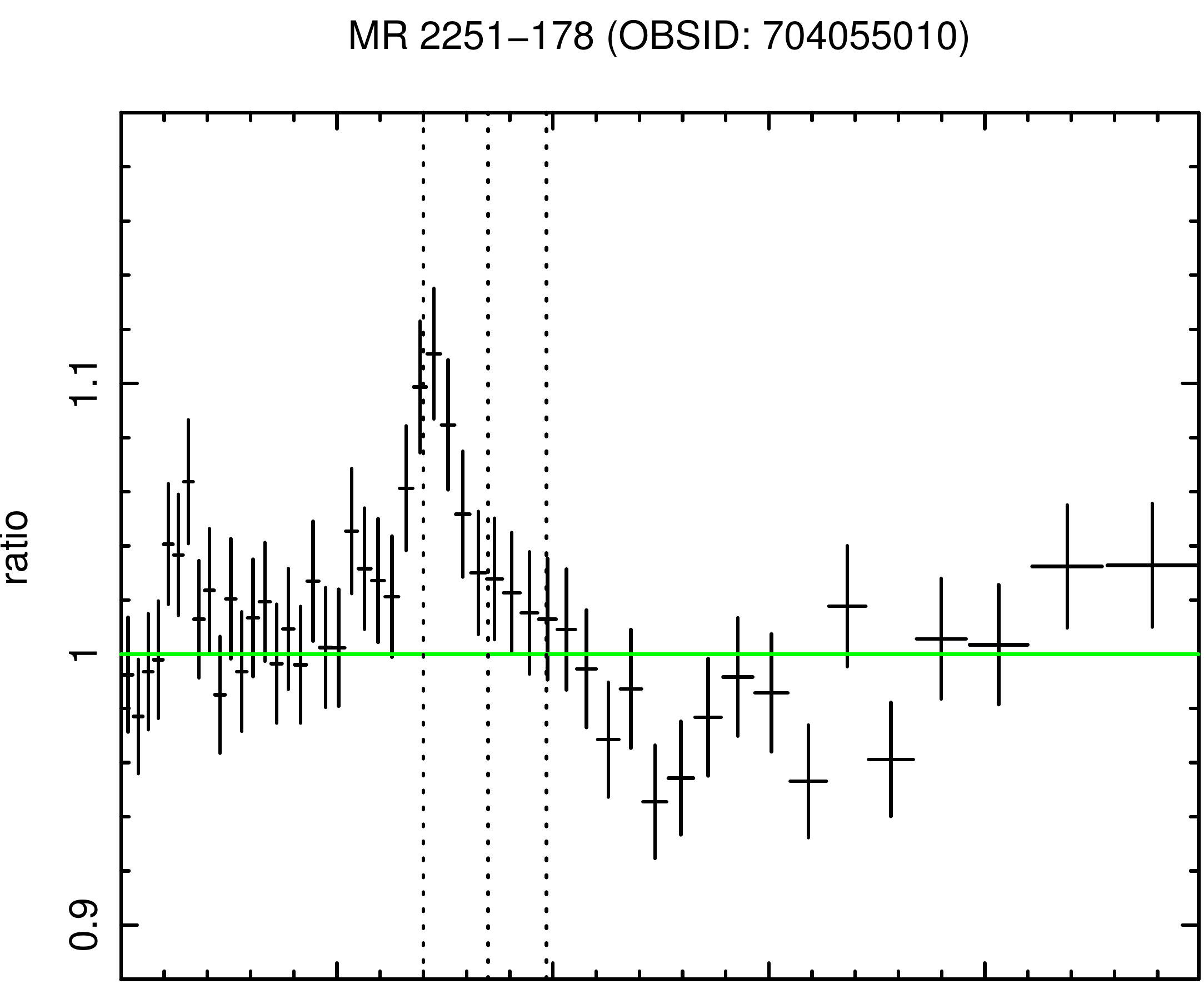}
}

\vspace{-12.2pt}
\subfloat{
\includegraphics[angle=-90,width=3.8cm]{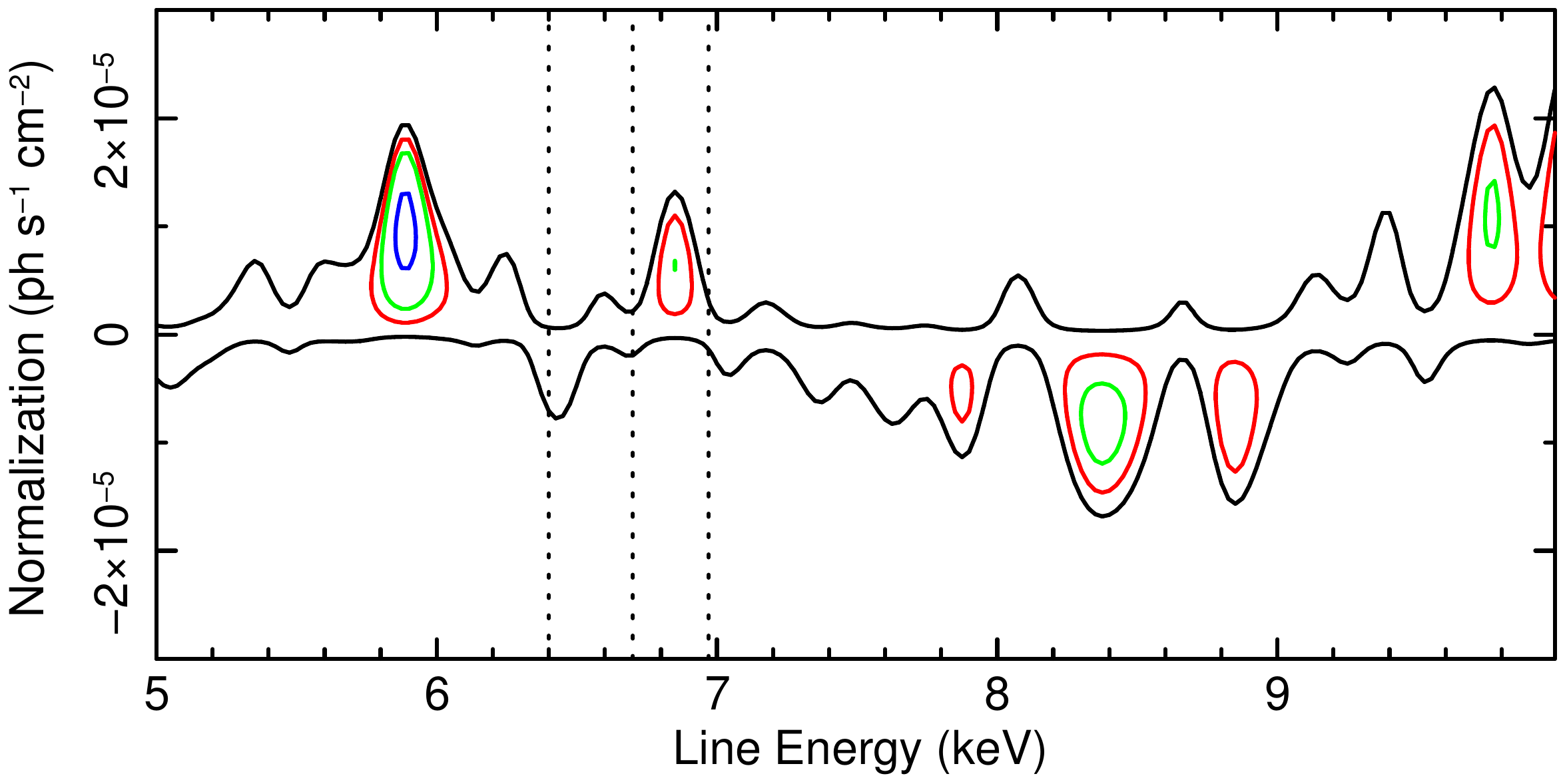}	
\includegraphics[angle=-90,width=3.8cm]{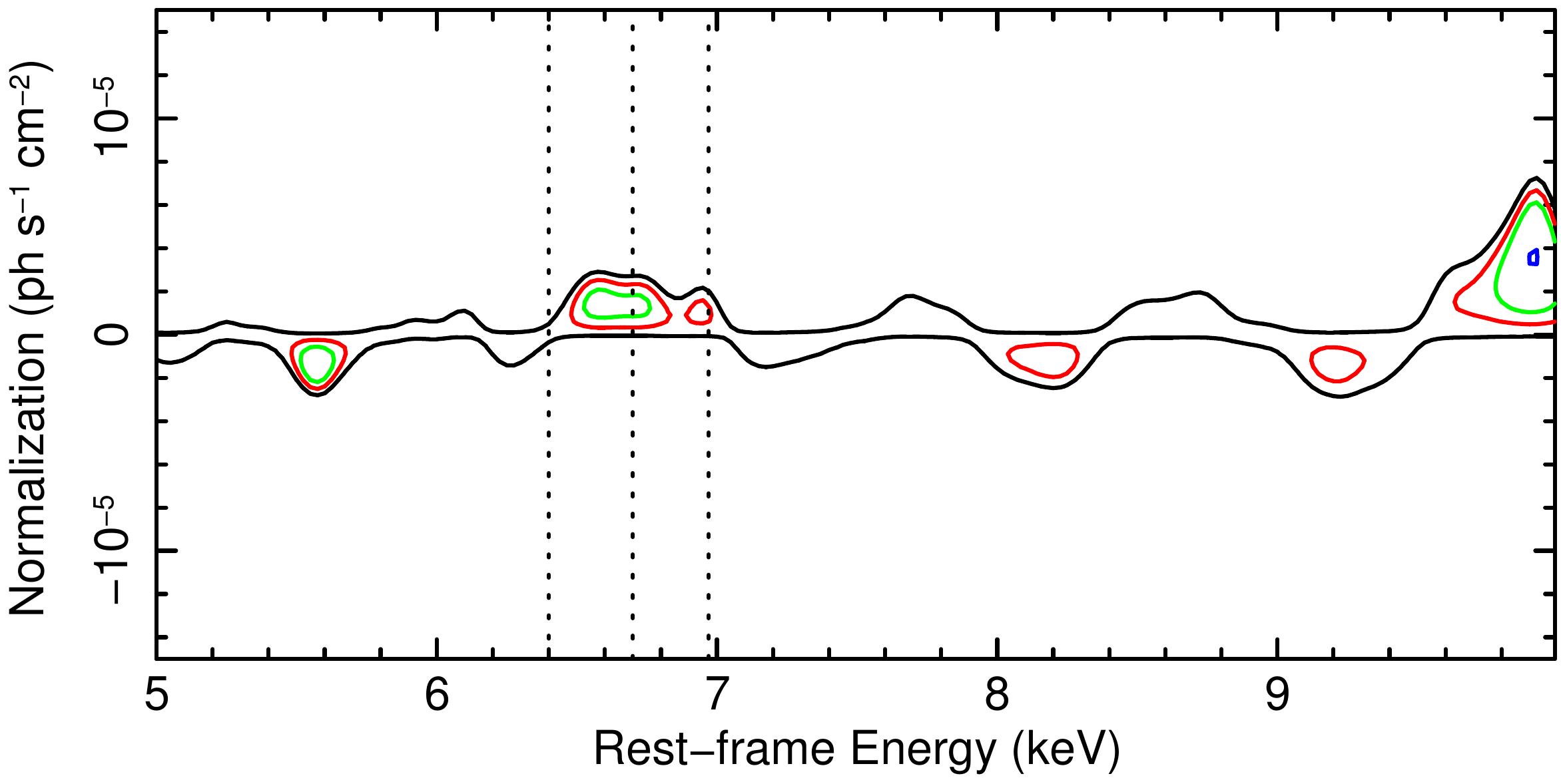}
\includegraphics[angle=-90,width=3.8cm]{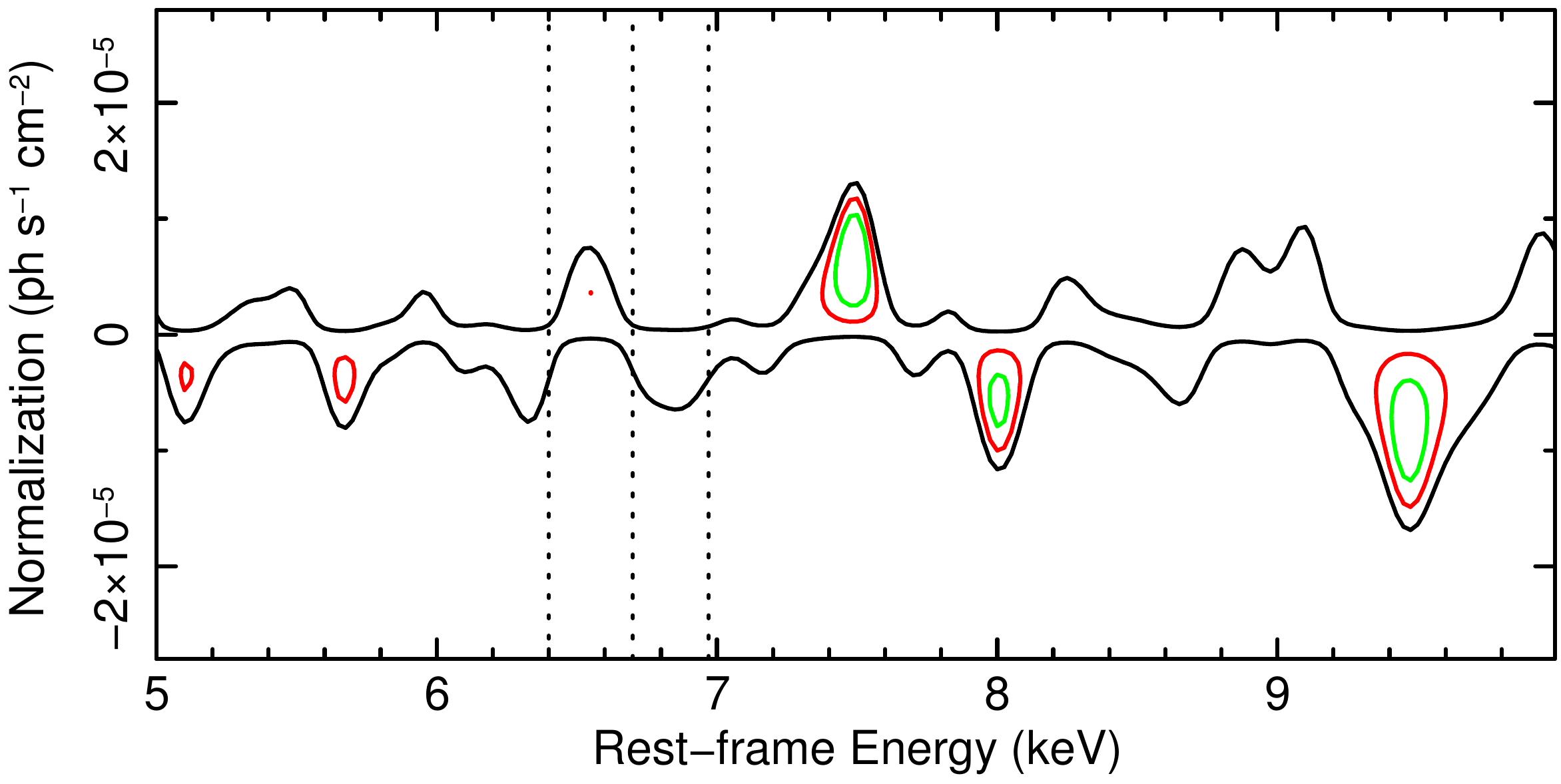}
\includegraphics[angle=-90,width=3.8cm]{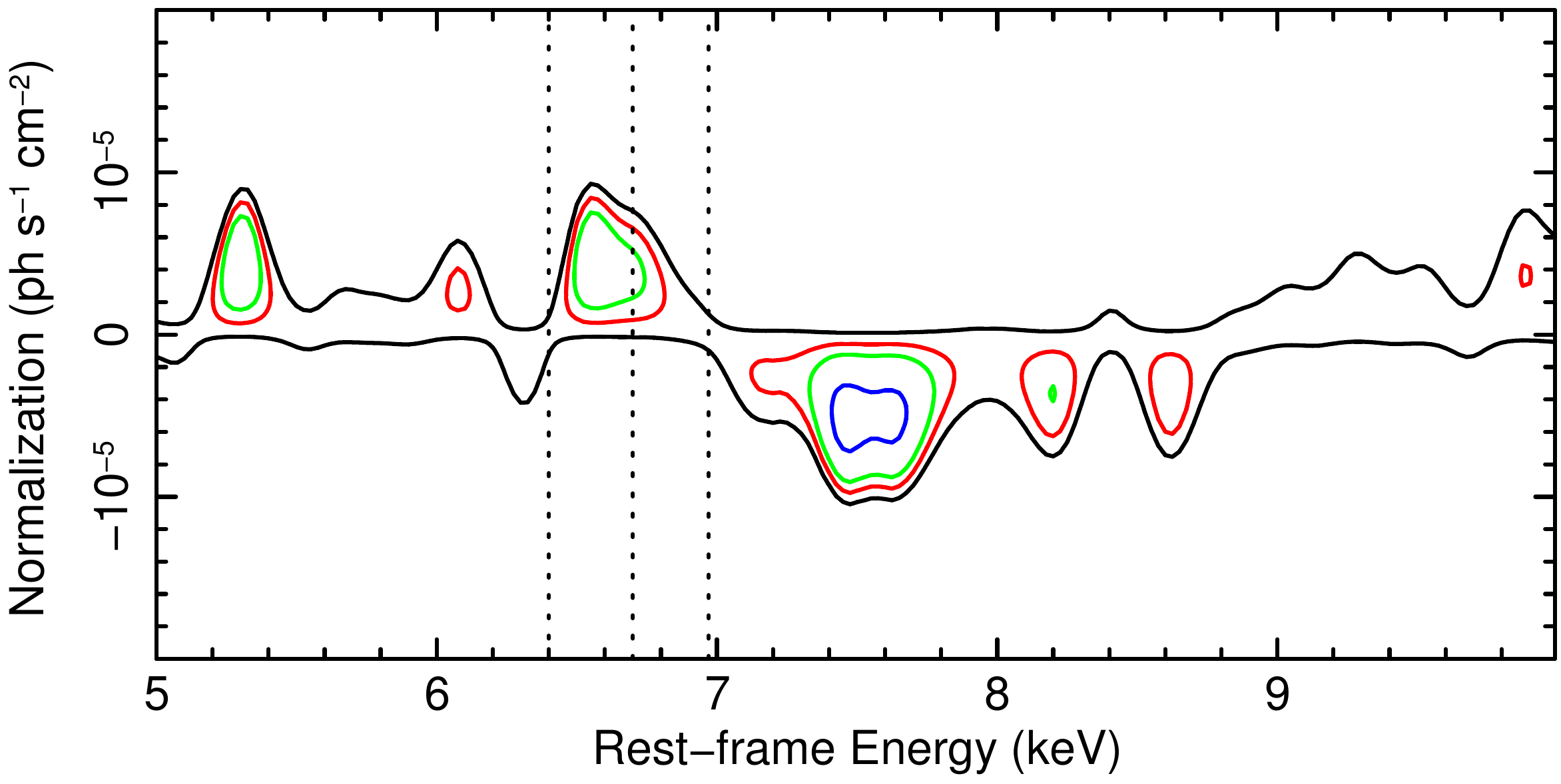}
}
	
\caption{\small Ratio and contour plots for sources which do not require a broadened component in the Fe\,K band. In each plot the data/model residuals (top panel) show the residuals which remain in the Fe\,K band when all atomic lines have been removed and the reflection component is fitted with \pexrav to highlight the presence of the neutral Fe\,K$\alpha$/K$\beta$ fluorescence lines. The contour plot (bottom panel) show the F-test significances of the remaining residuals when the Fe\,K$\alpha$ and K$\beta$ lines are fitted with \reflionx and a narrow Gaussian line. The continuous outer contour corresponds to a $\Delta\chi^{2}=+0.5$ worse fit and is intended to indicate the level of the continuum baseline. The closed coloured contours correspond to $\Delta\chi^{2}$ improvements of $-2.3$ (red), $-4.61$ (green), $-9.21$ (blue), $-13.82$ (cyan) and $-18.42$ (magenta), which translate to F-test significances of 68\%, 90\%, 99\%, 99.9\% and 99.99\%, respectively. The dashed vertical lines indicate the expected rest-frame energies of the \feka, \fexxv~\hea and \fexxvi~\lya transitions. Colour version available online.}
\end{center}
\label{figure:rat_cont1}
\end{figure*}

\clearpage

\begin{figure*}
\begin{center}

\vspace{-5pt}	
\subfloat{
\includegraphics[angle=-90,width=3.8cm]{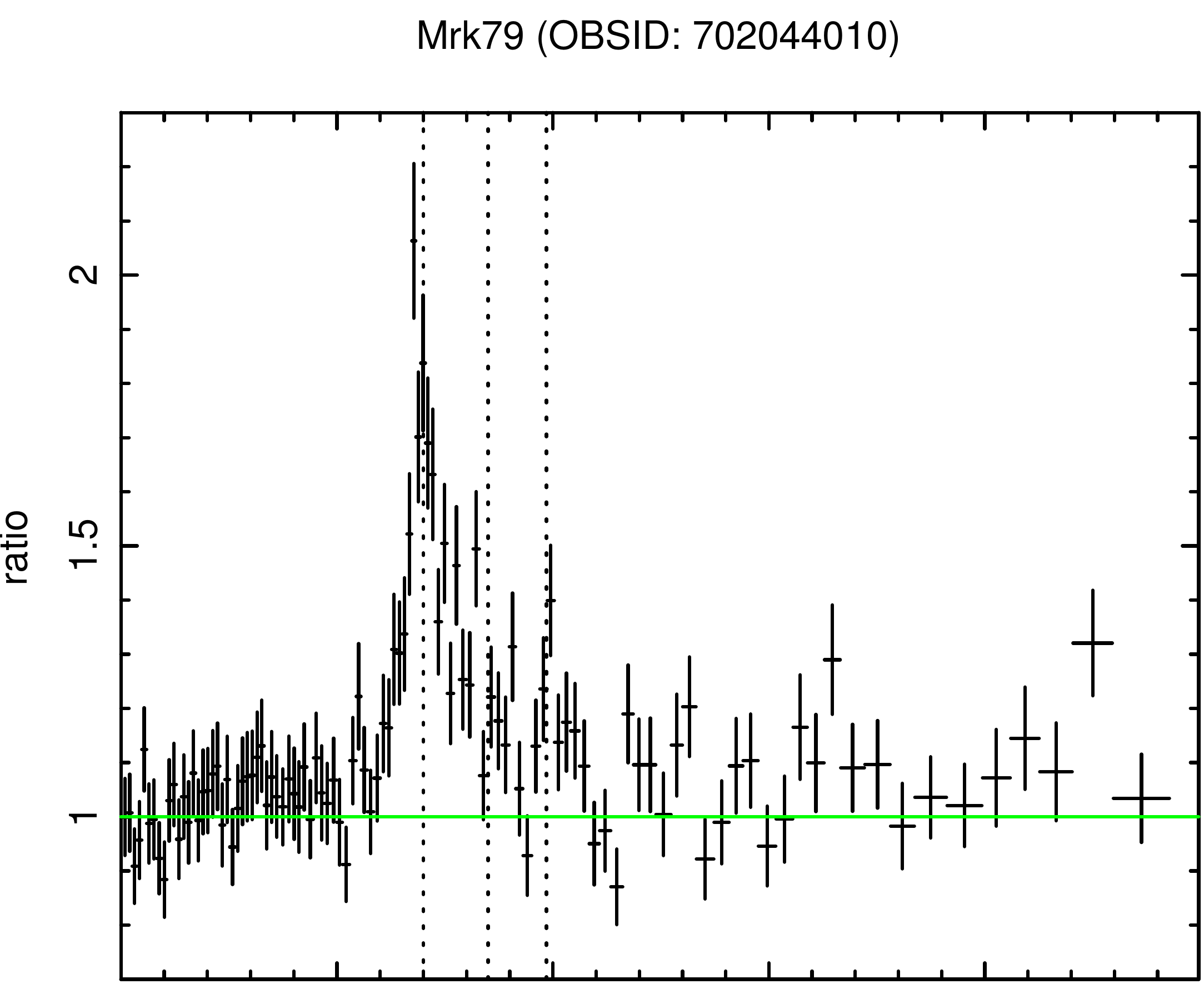}
\includegraphics[angle=-90,width=3.8cm]{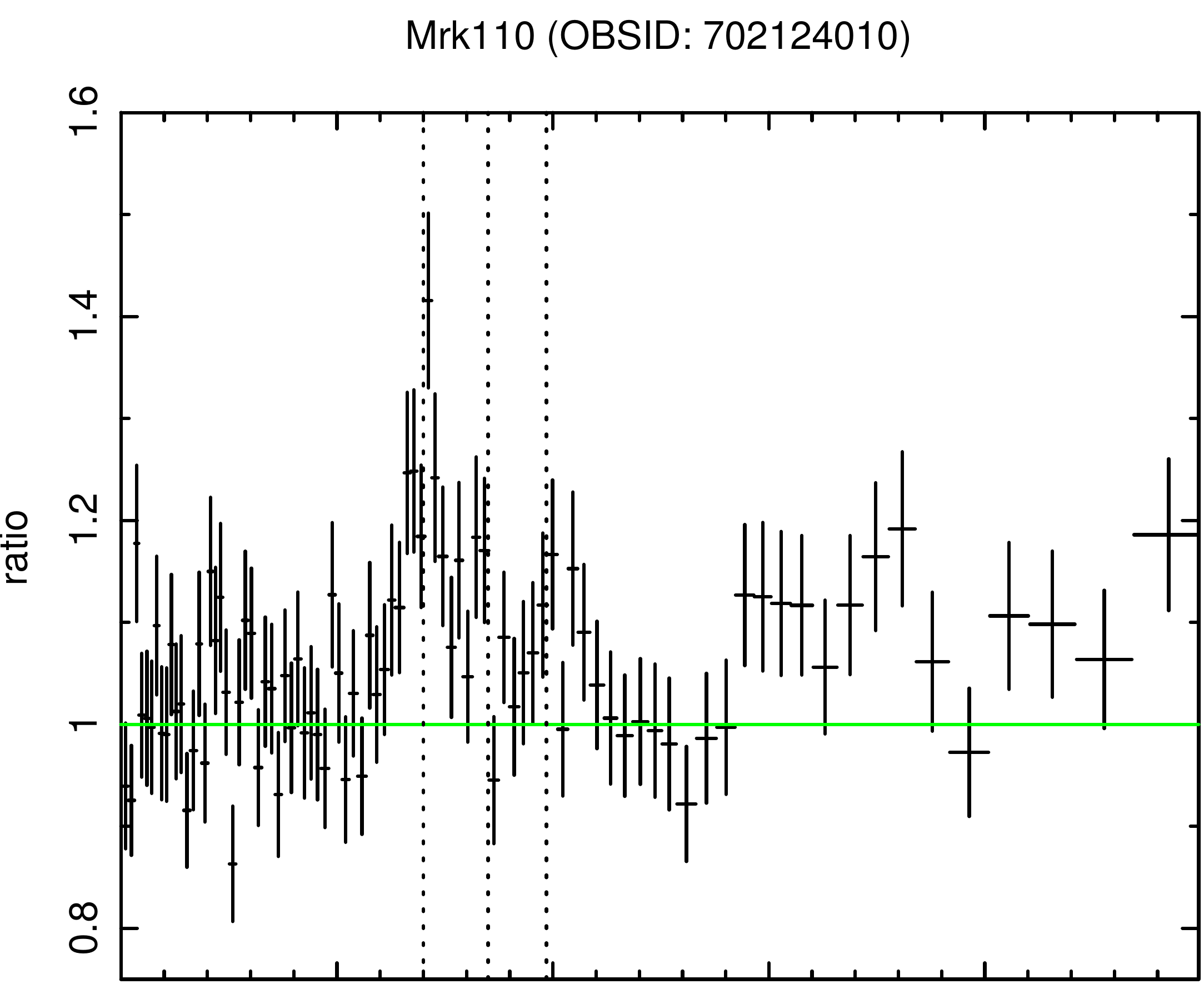}
\includegraphics[angle=-90,width=3.8cm]{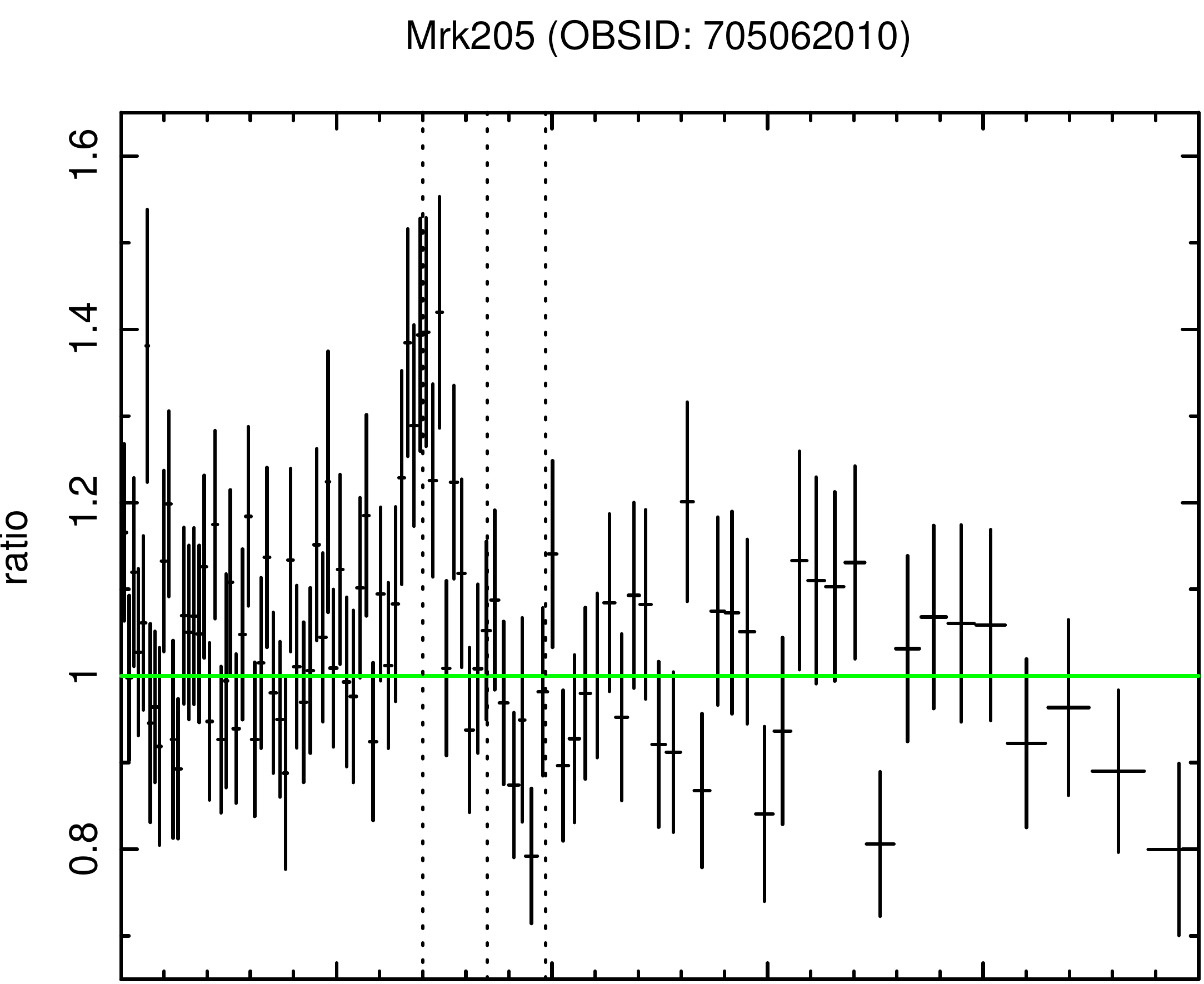}
\includegraphics[angle=-90,width=3.8cm]{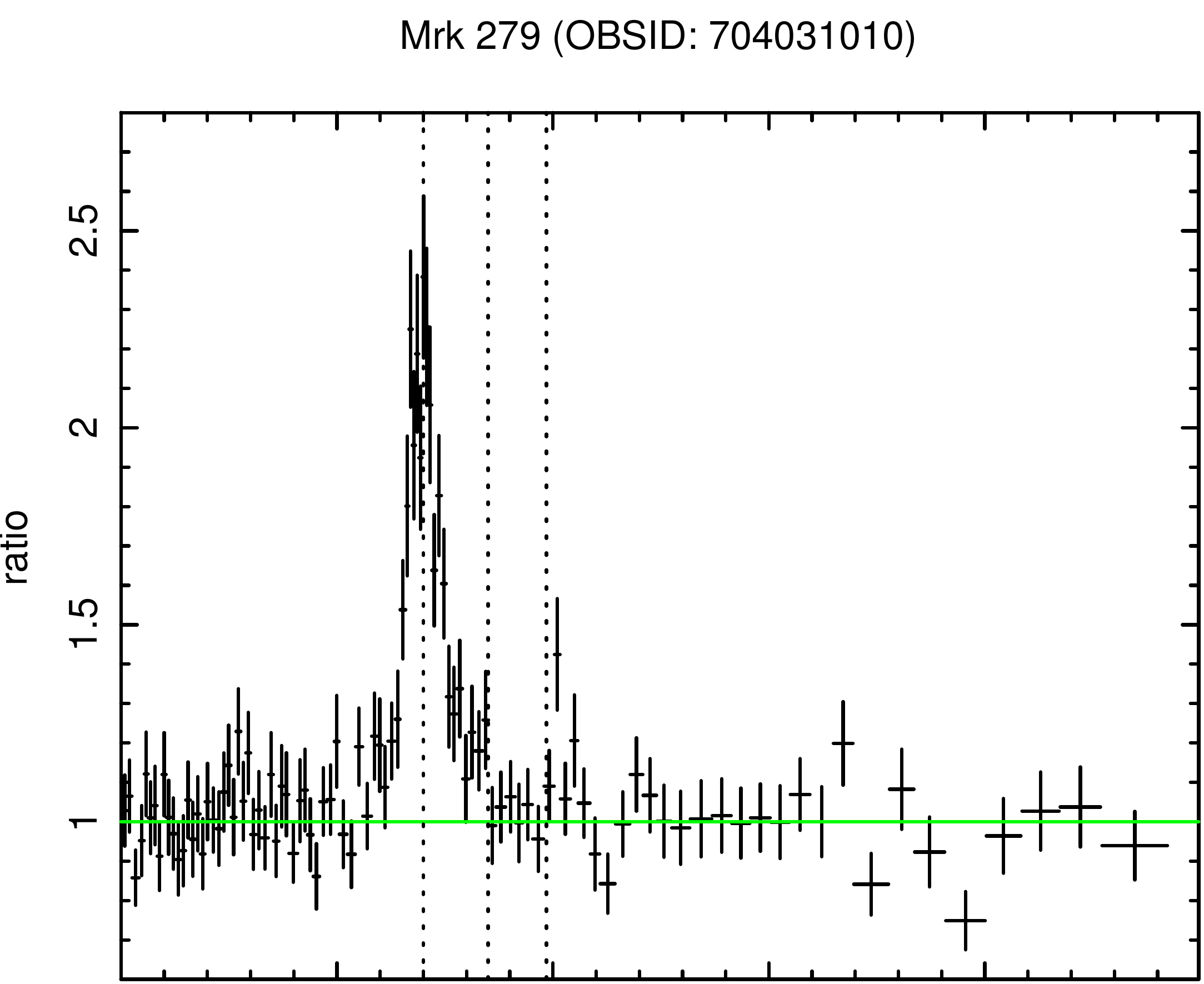}
}

\vspace{-12.2pt}
\subfloat{
\includegraphics[angle=-90,width=3.8cm]{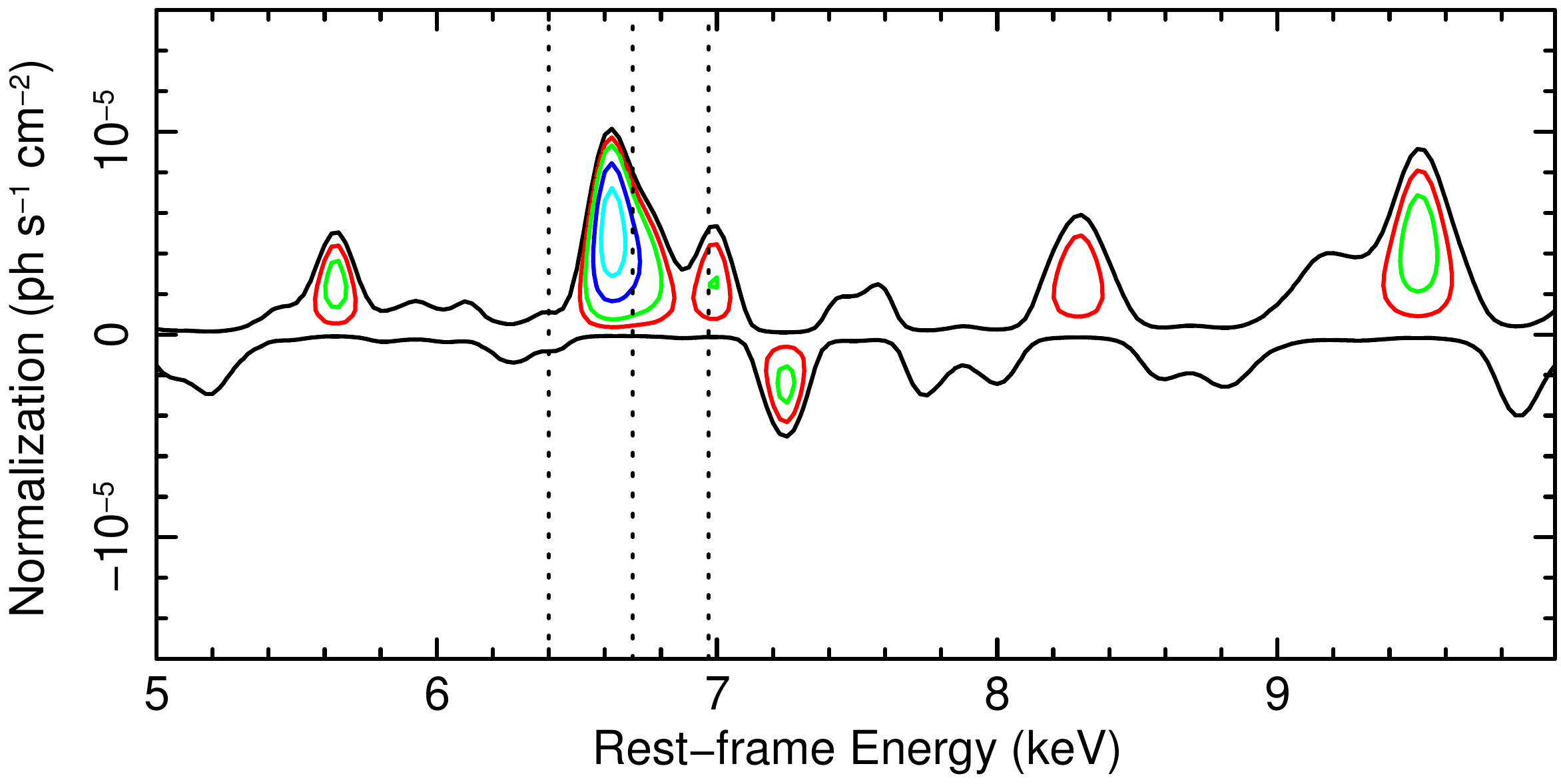}	
\includegraphics[angle=-90,width=3.8cm]{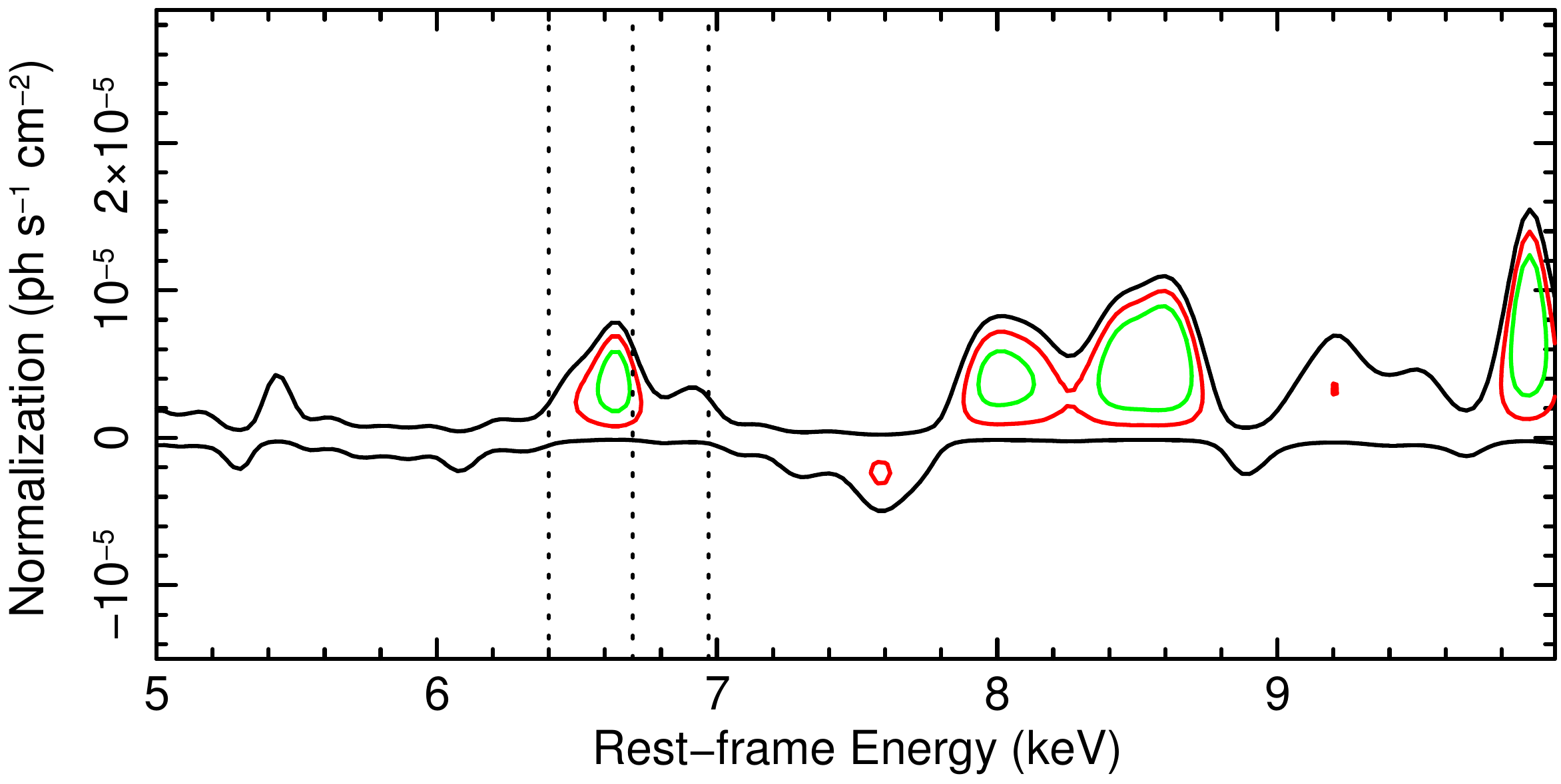}
\includegraphics[angle=-90,width=3.8cm]{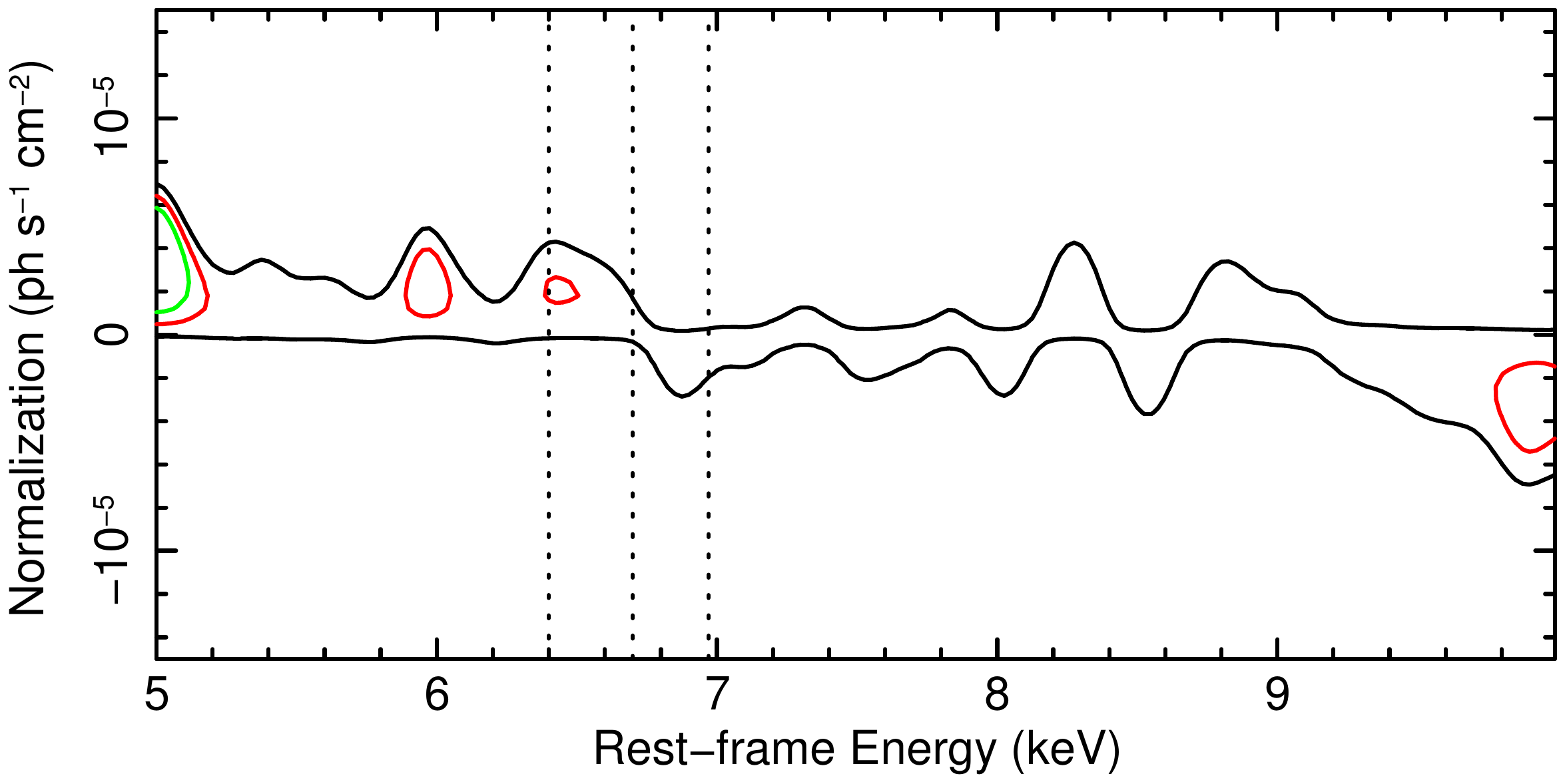}
\includegraphics[angle=-90,width=3.8cm]{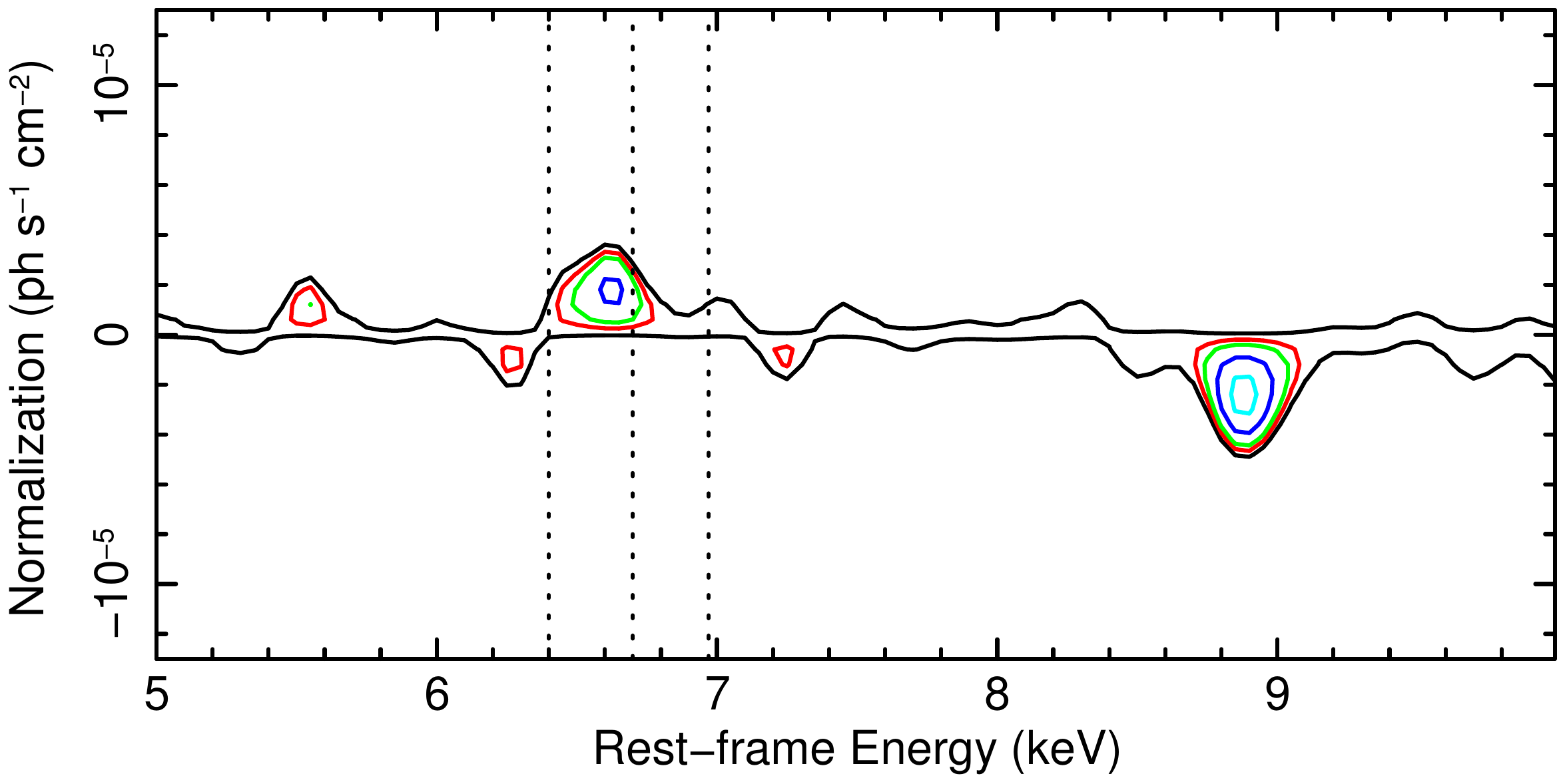}	
}

\vspace{-5pt}	
\subfloat{
\includegraphics[angle=-90,width=3.8cm]{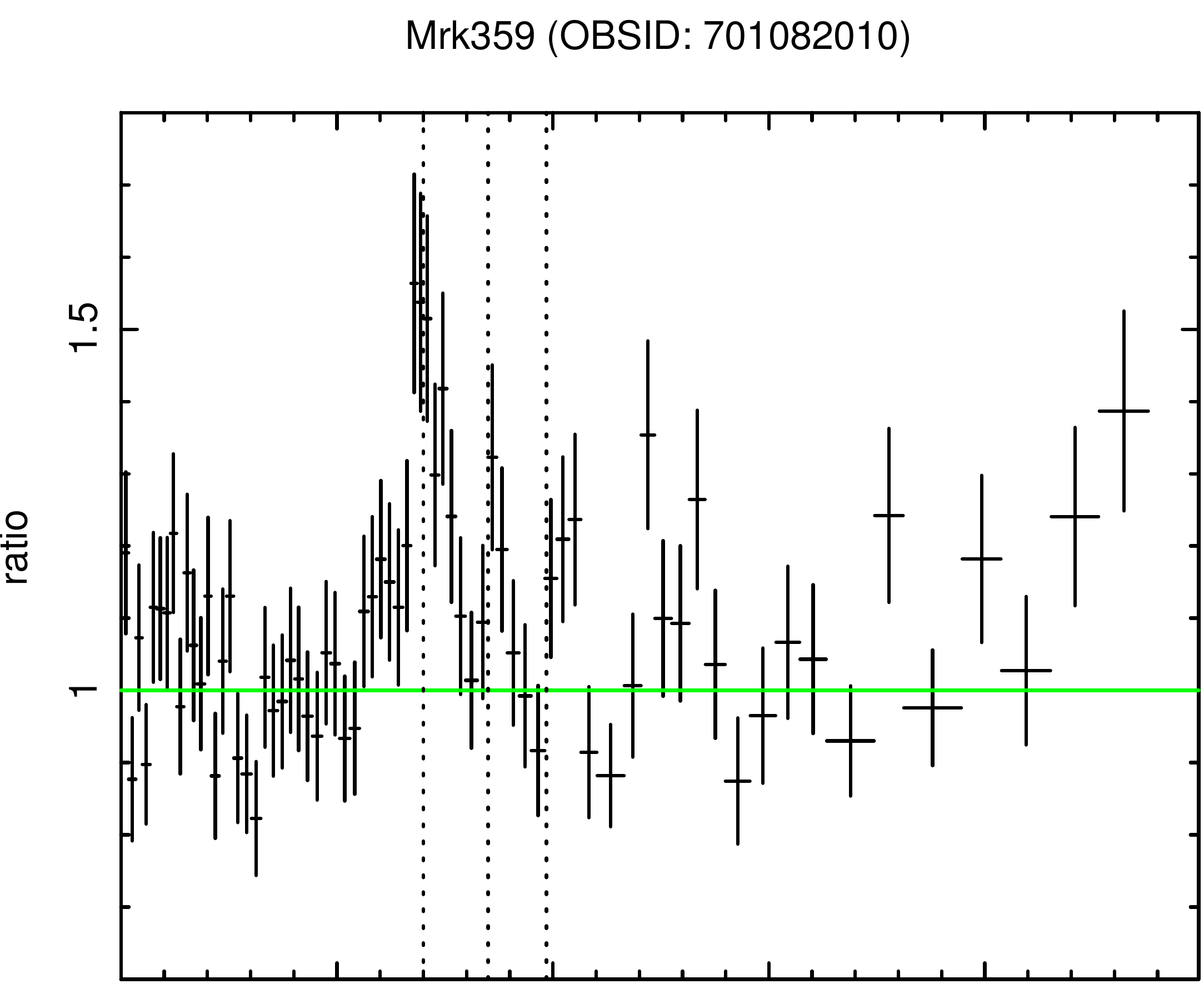}
\includegraphics[angle=-90,width=3.8cm]{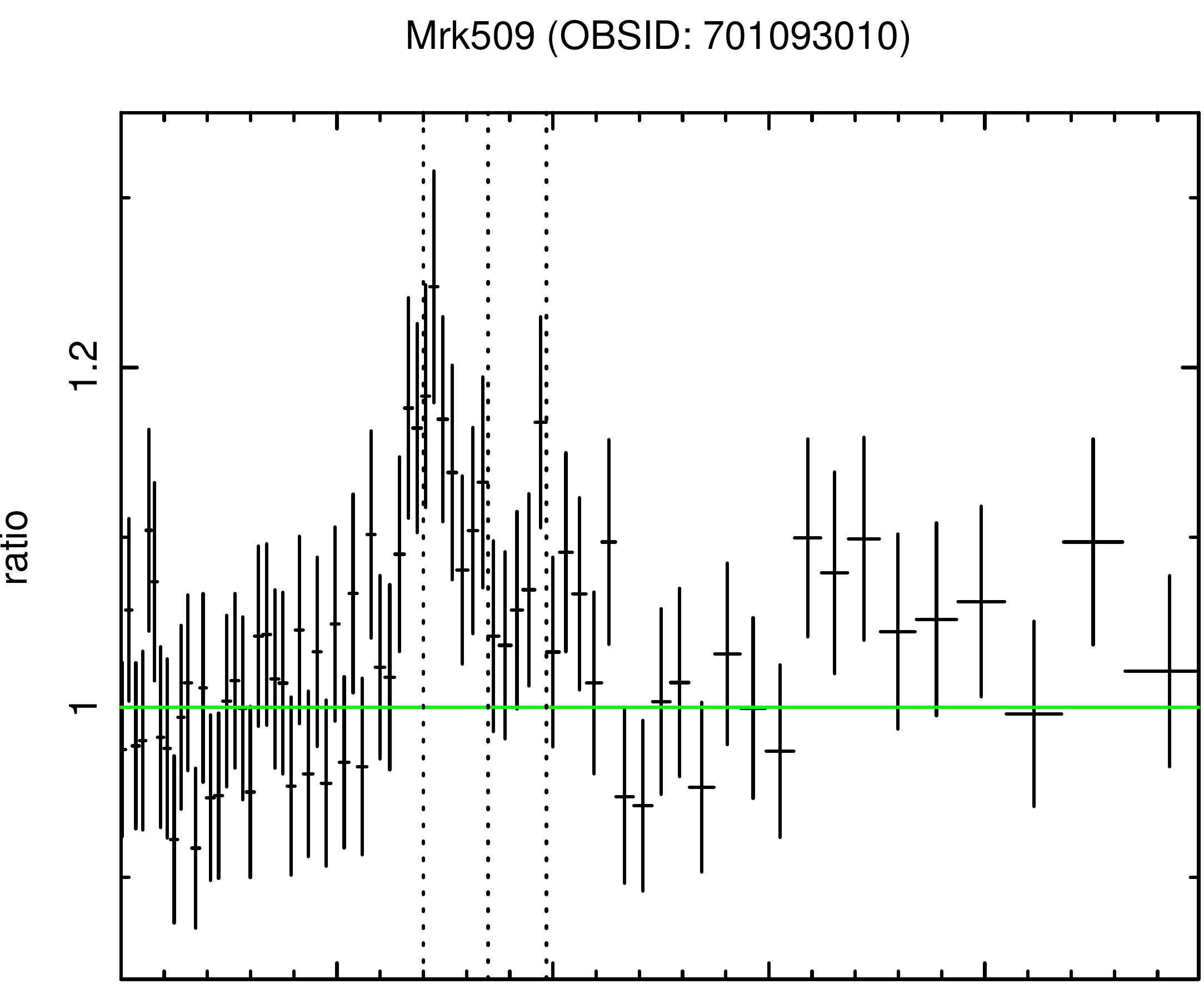}
\includegraphics[angle=-90,width=3.8cm]{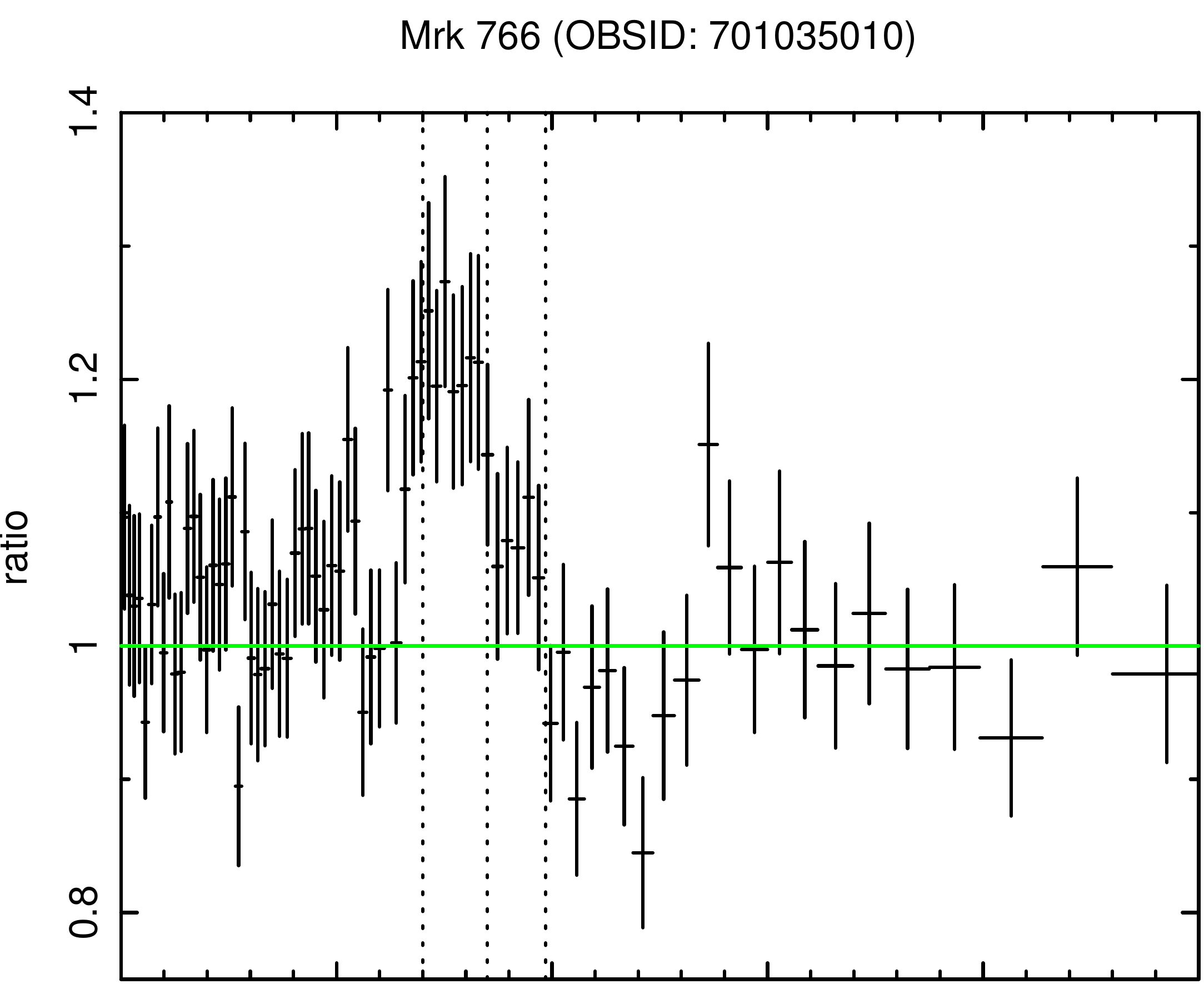}
\includegraphics[angle=-90,width=3.8cm]{figures/Mrk766_obs2_rat_rf.pdf}
}

\vspace{-12.2pt}
\subfloat{
\includegraphics[angle=-90,width=3.8cm]{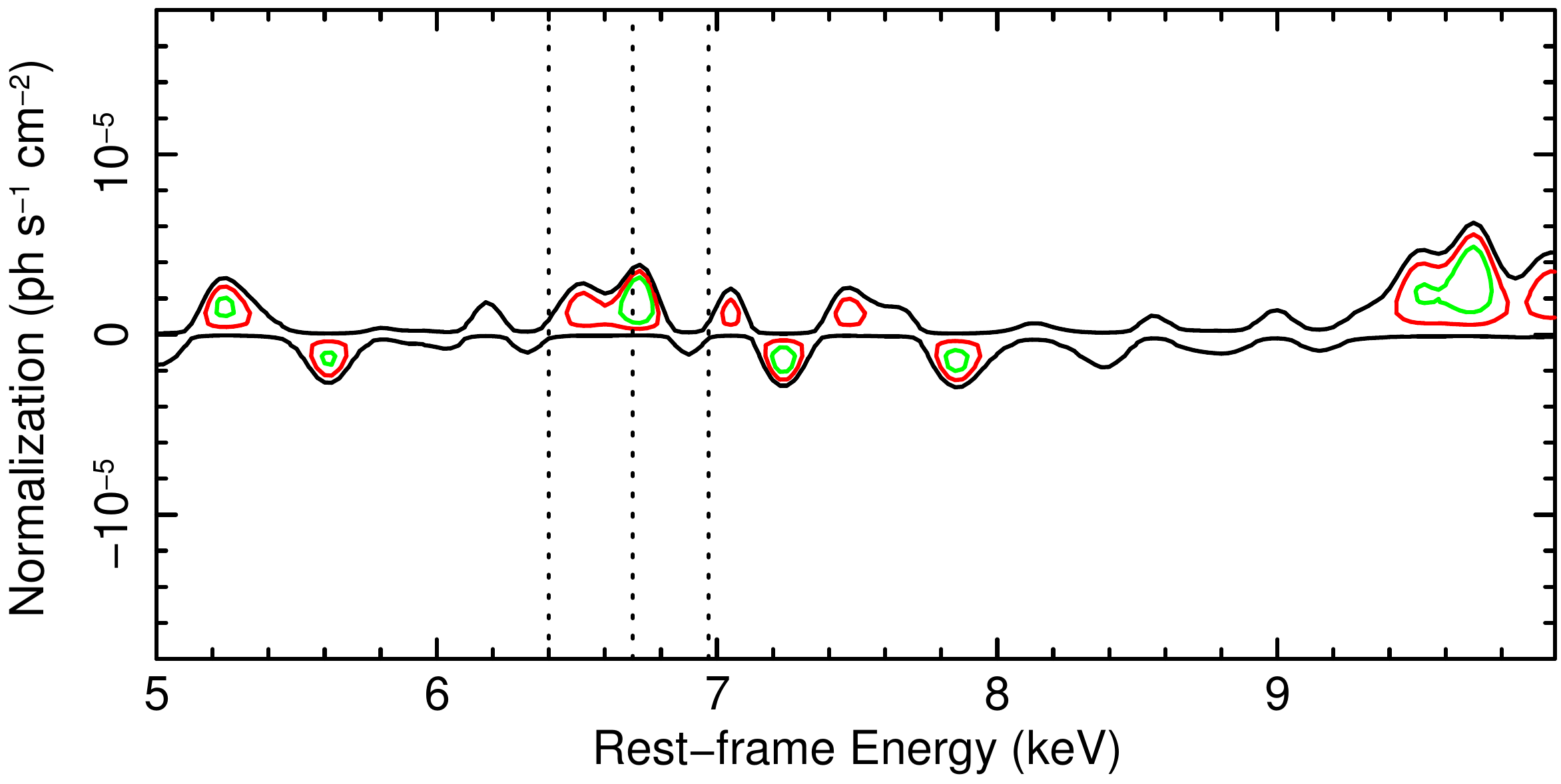}
\includegraphics[angle=-90,width=3.8cm]{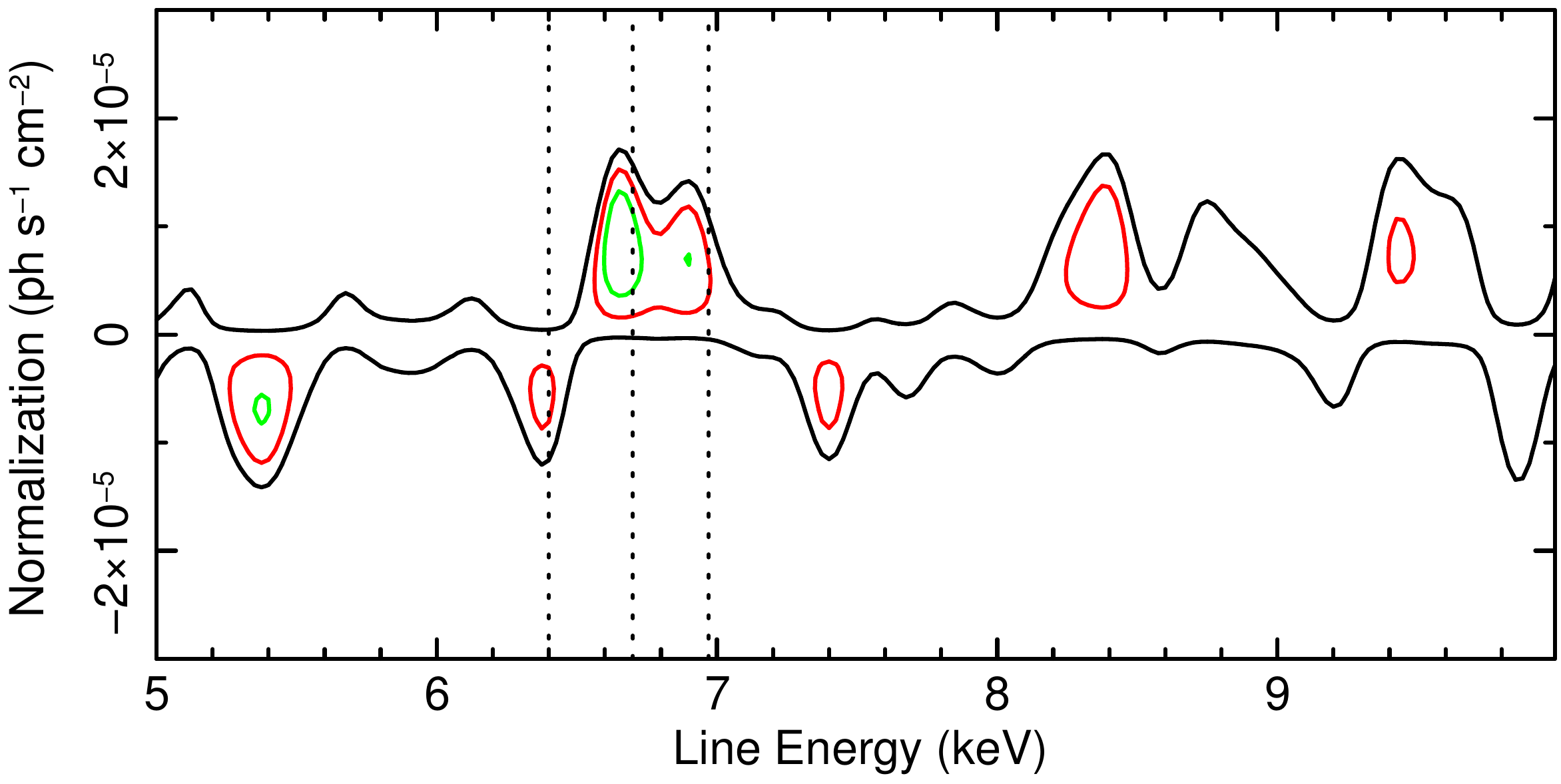}
\includegraphics[angle=-90,width=3.8cm]{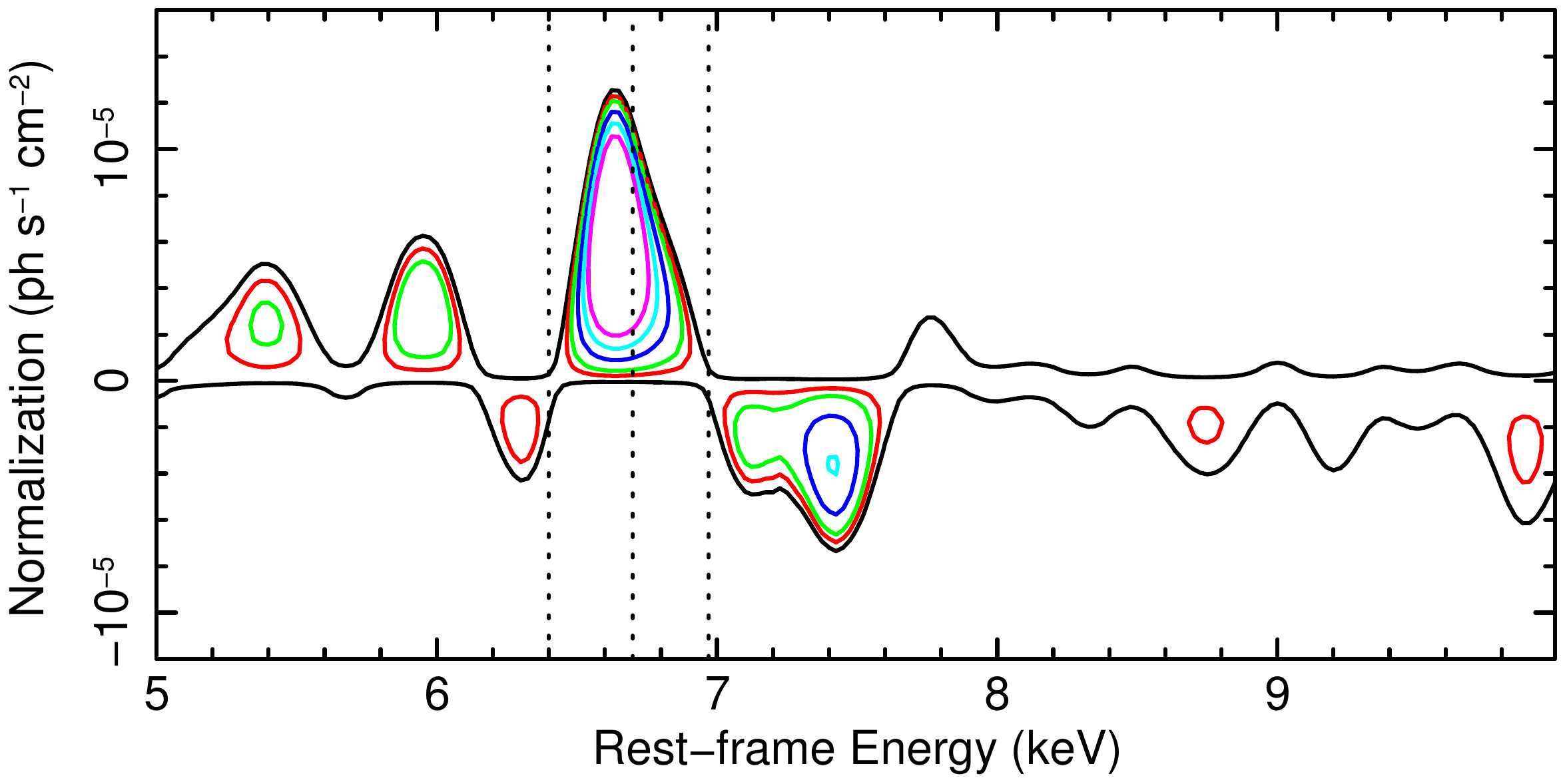}
\includegraphics[angle=-90,width=3.8cm]{figures/Mrk766_obs2_cont.pdf}
}

\vspace{-5pt}	
\subfloat{
\includegraphics[angle=-90,width=3.8cm]{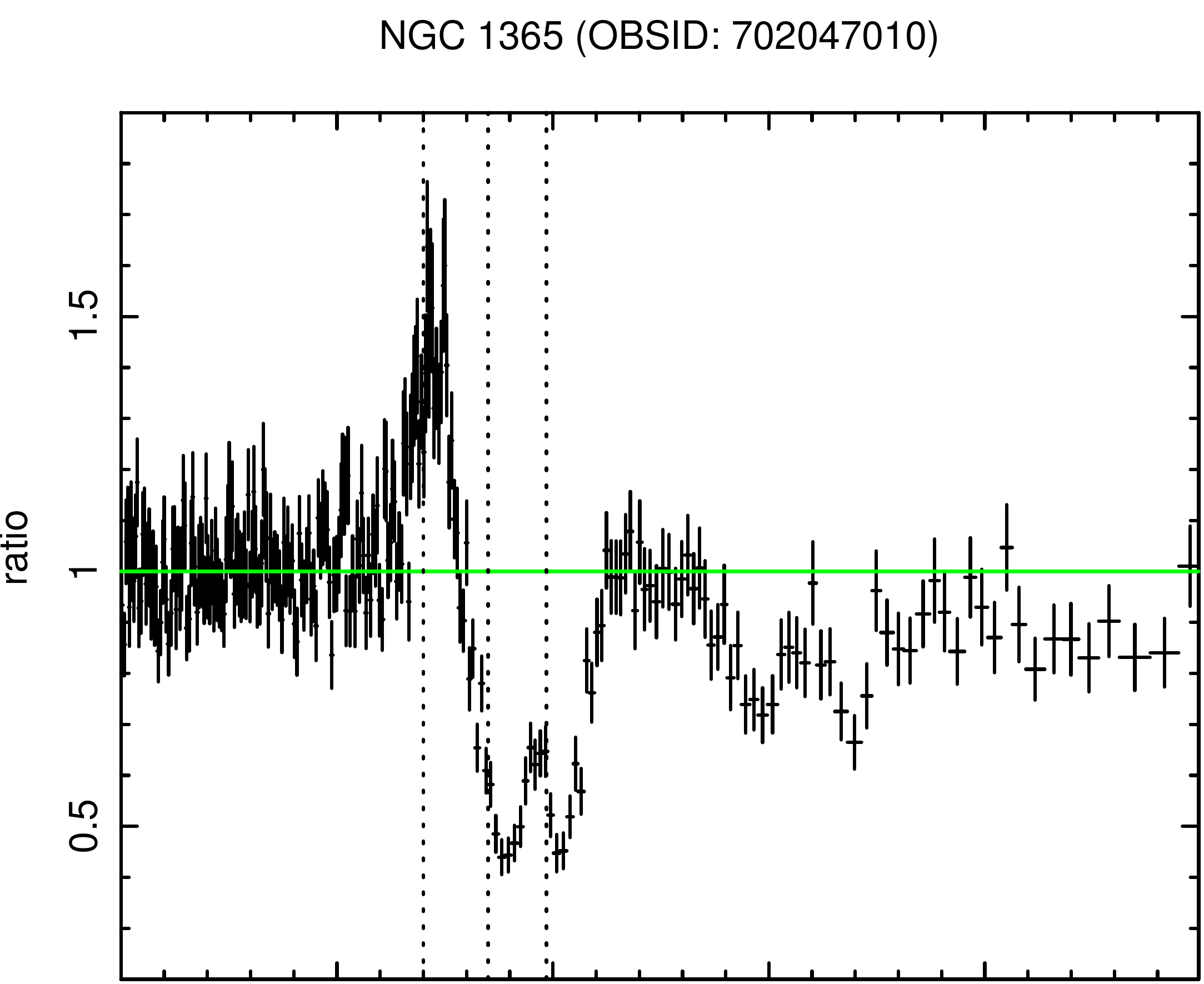}
\includegraphics[angle=-90,width=3.8cm]{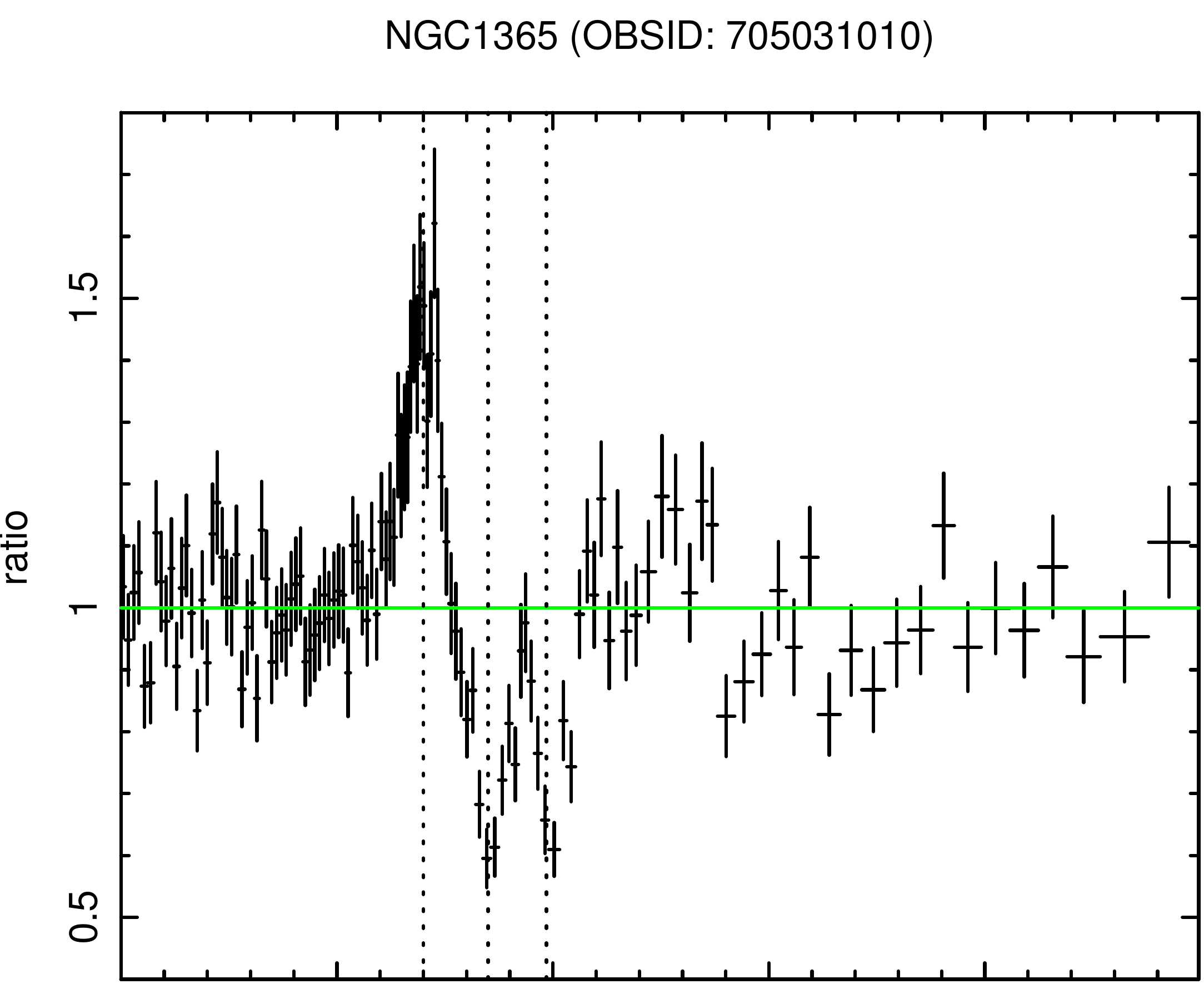}
\includegraphics[angle=-90,width=3.8cm]{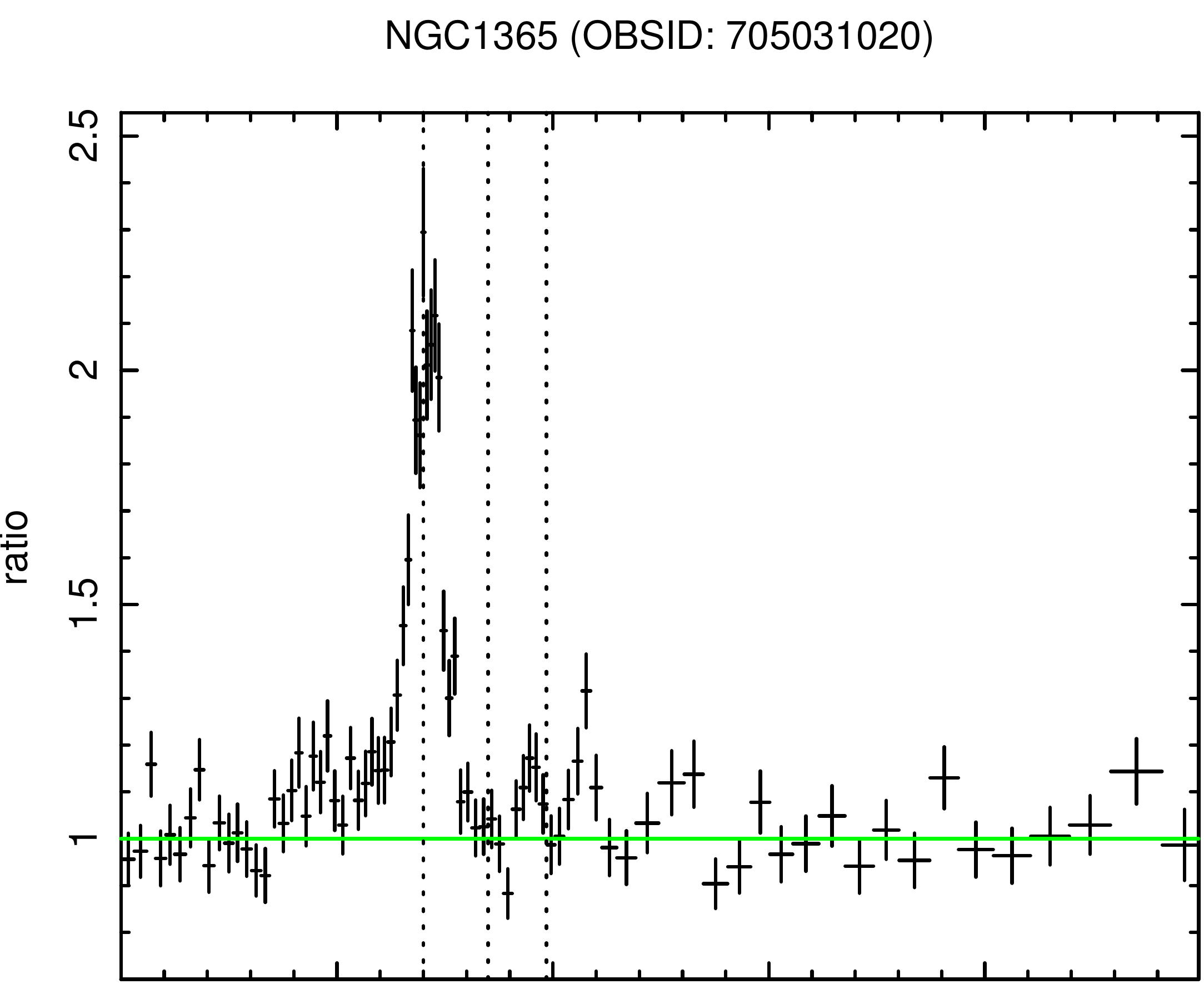}
\includegraphics[angle=-90,width=3.8cm]{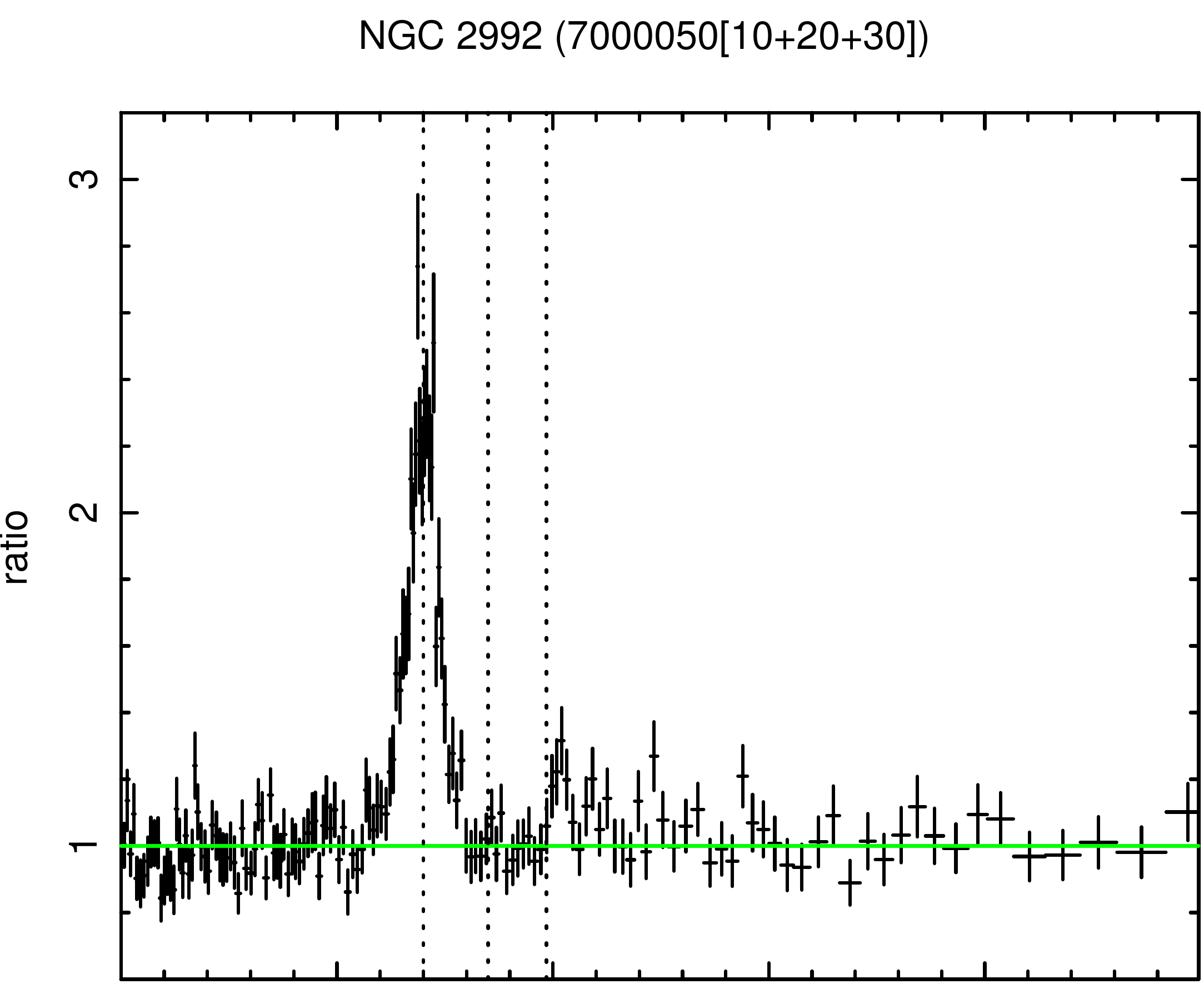}
}

\vspace{-12.2pt}
\subfloat{
\includegraphics[angle=-90,width=3.8cm]{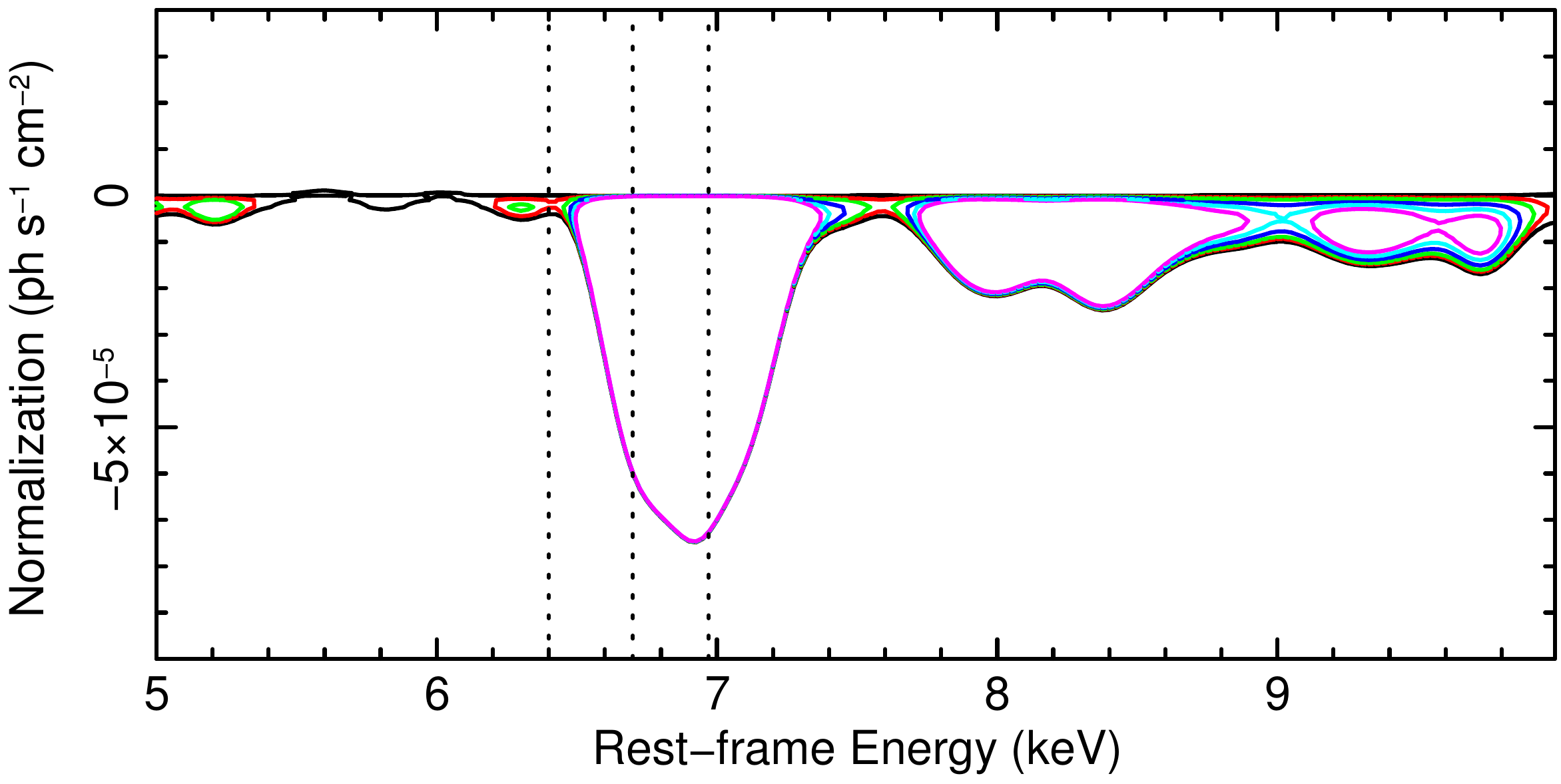}	
\includegraphics[angle=-90,width=3.8cm]{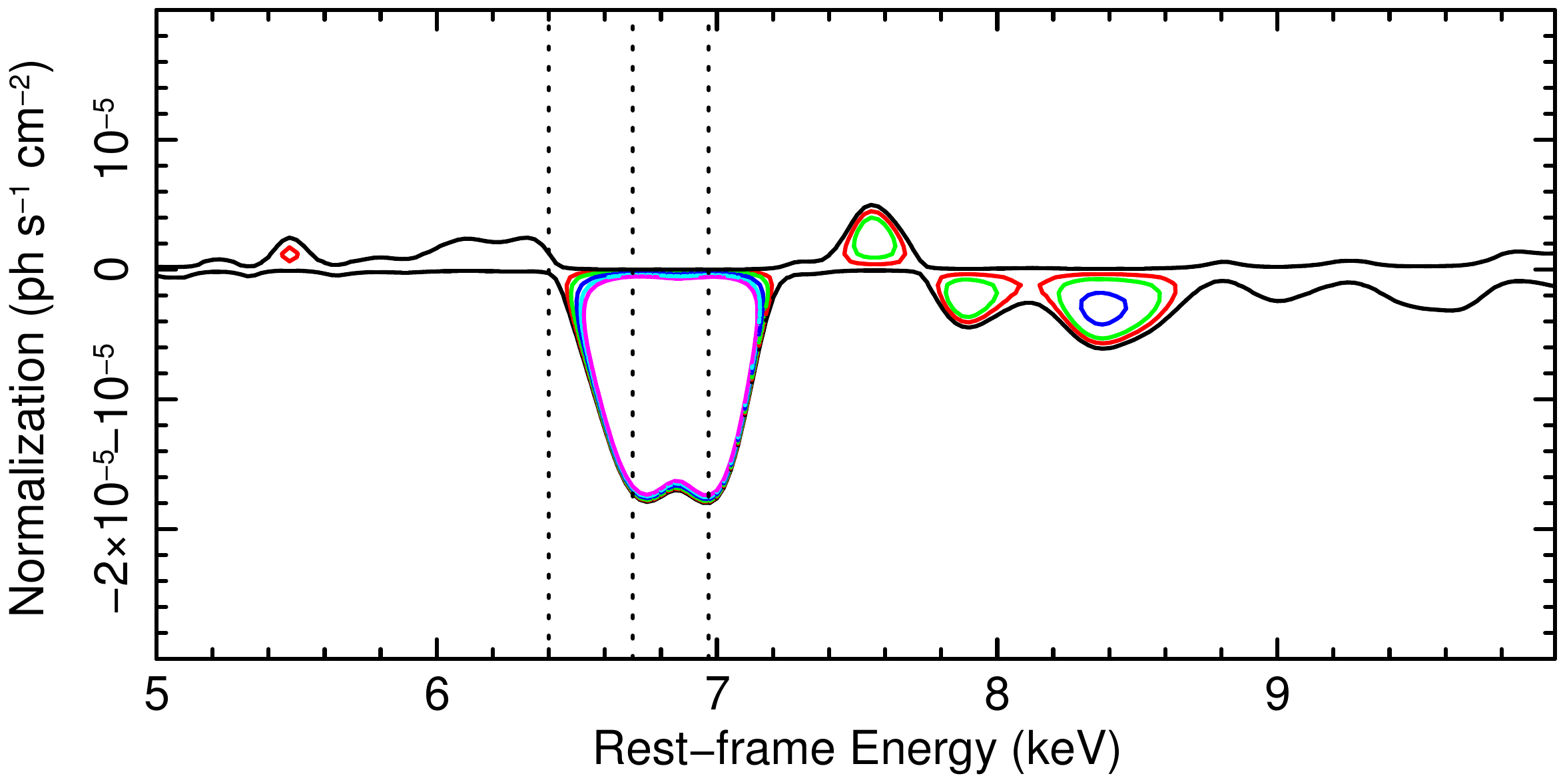}
\includegraphics[angle=-90,width=3.8cm]{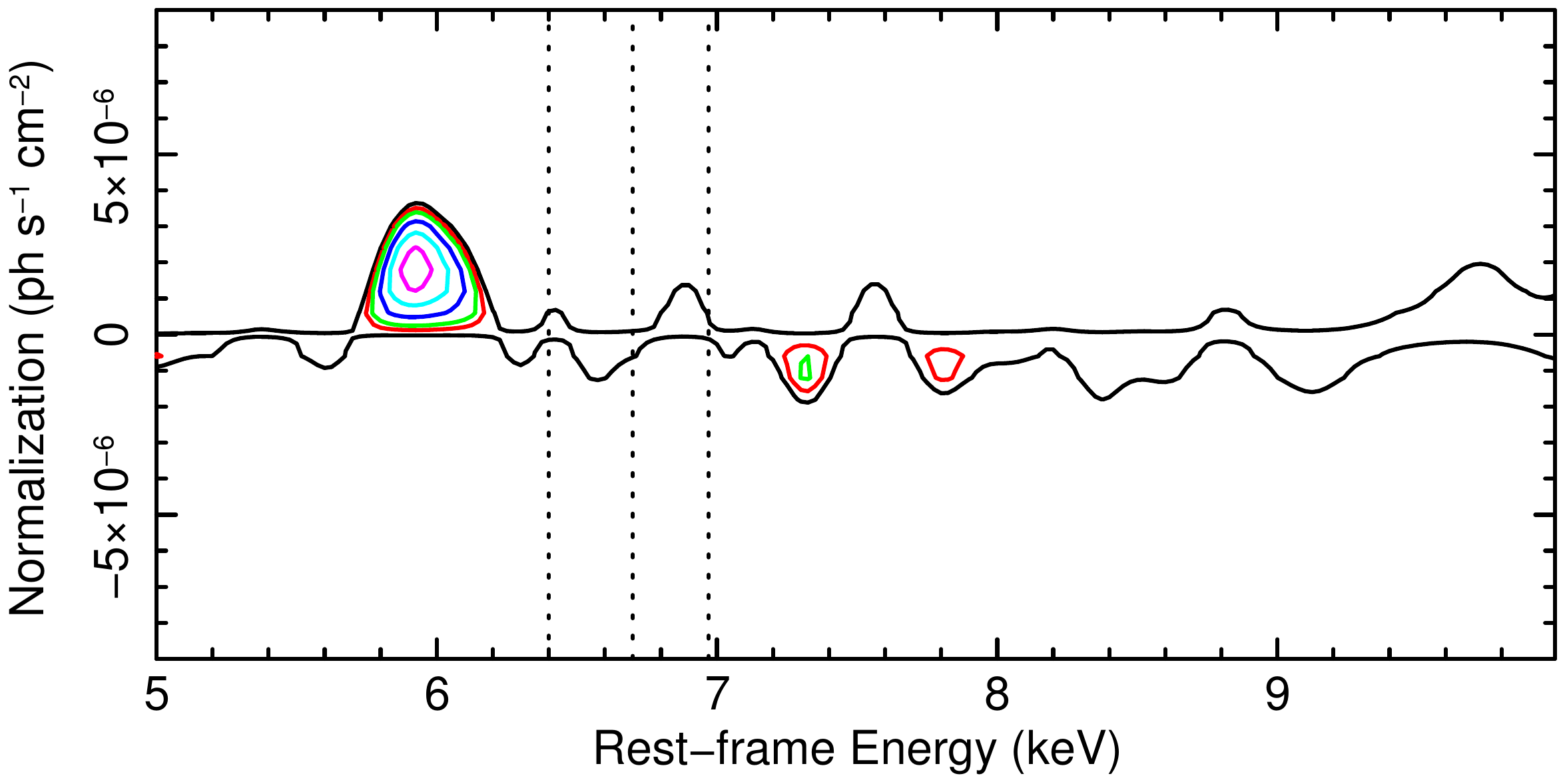}
\includegraphics[angle=-90,width=3.8cm]{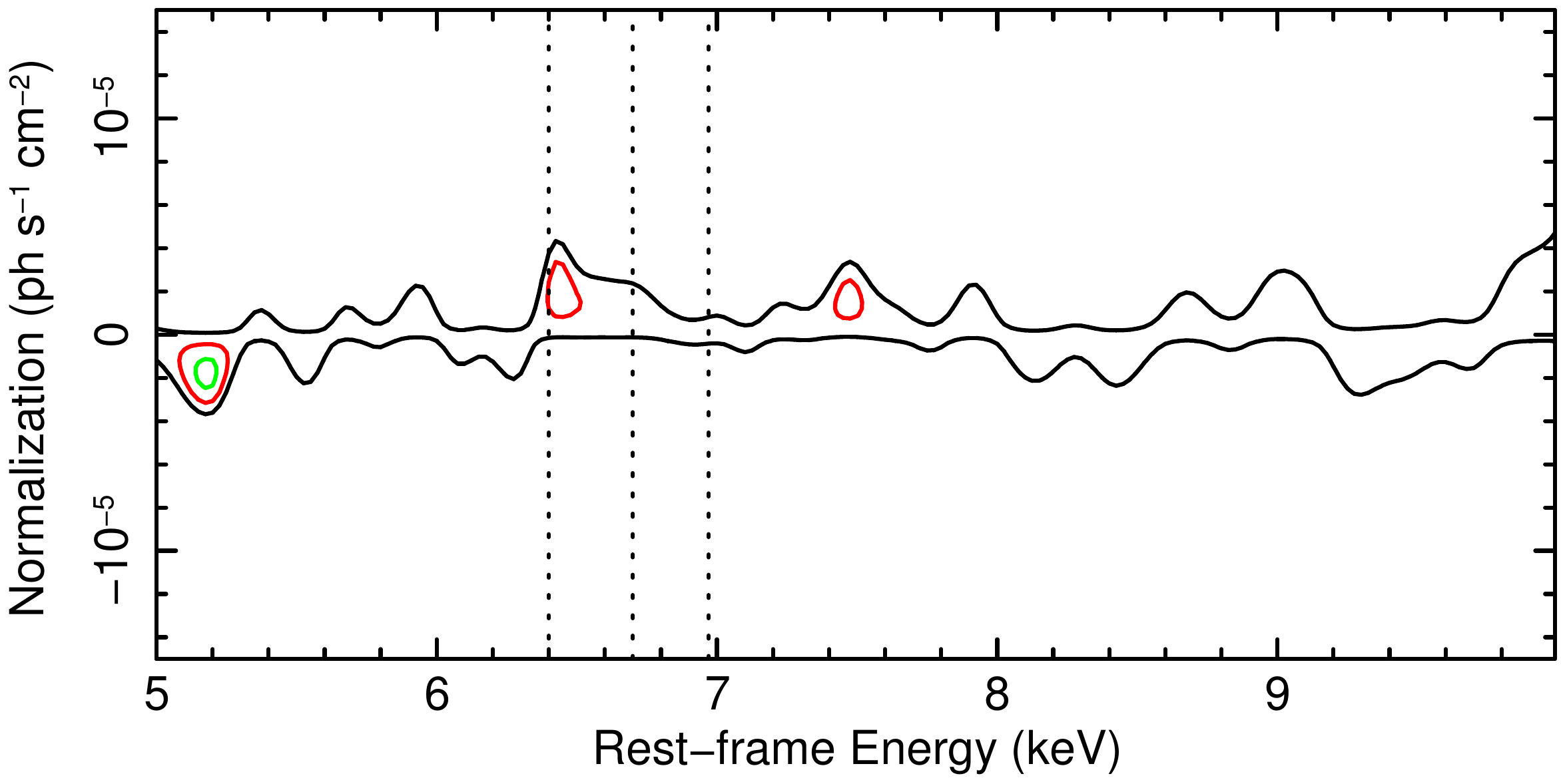}
}

\vspace{-5pt}	
\subfloat{
\includegraphics[angle=-90,width=3.8cm]{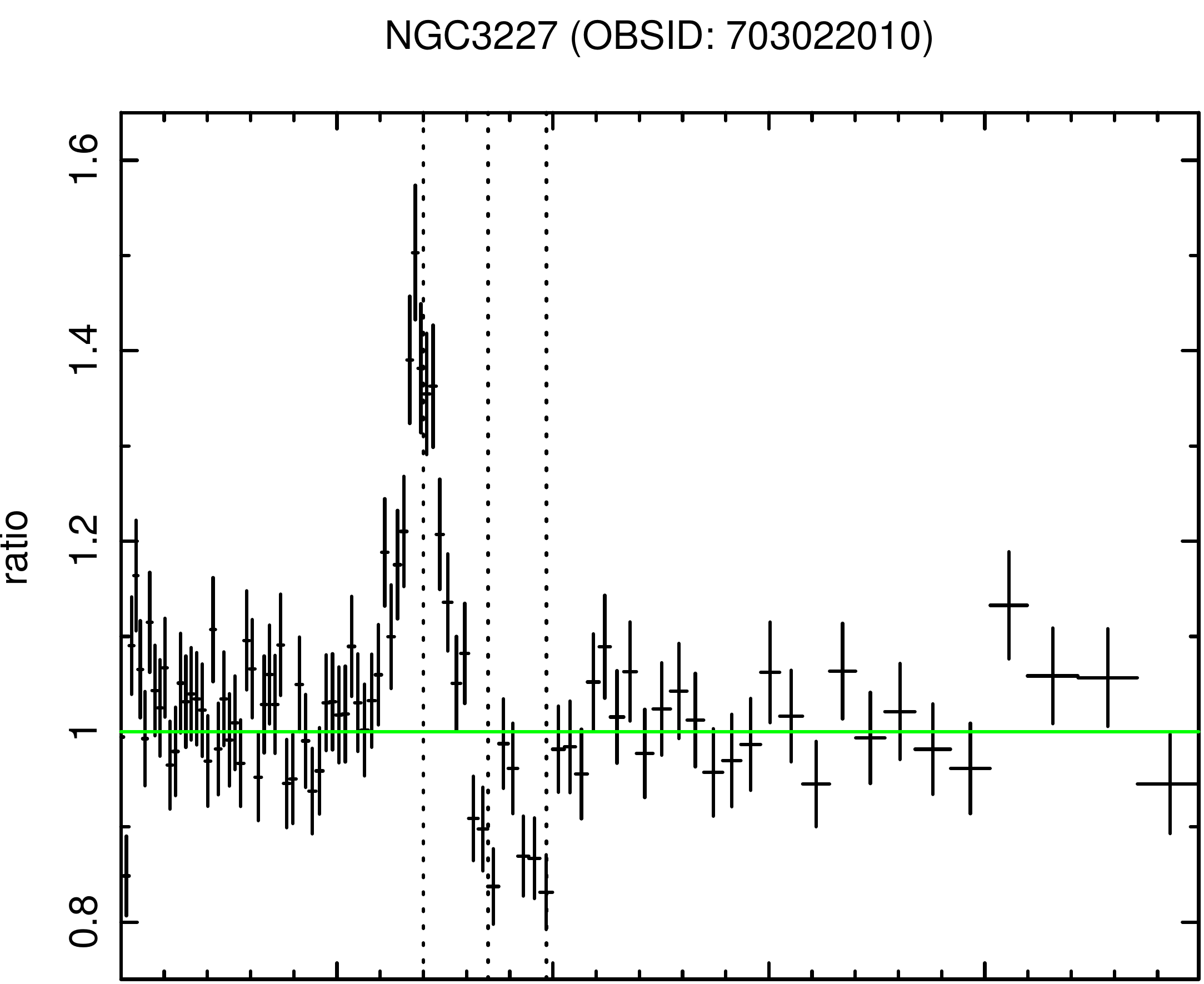}
\includegraphics[angle=-90,width=3.8cm]{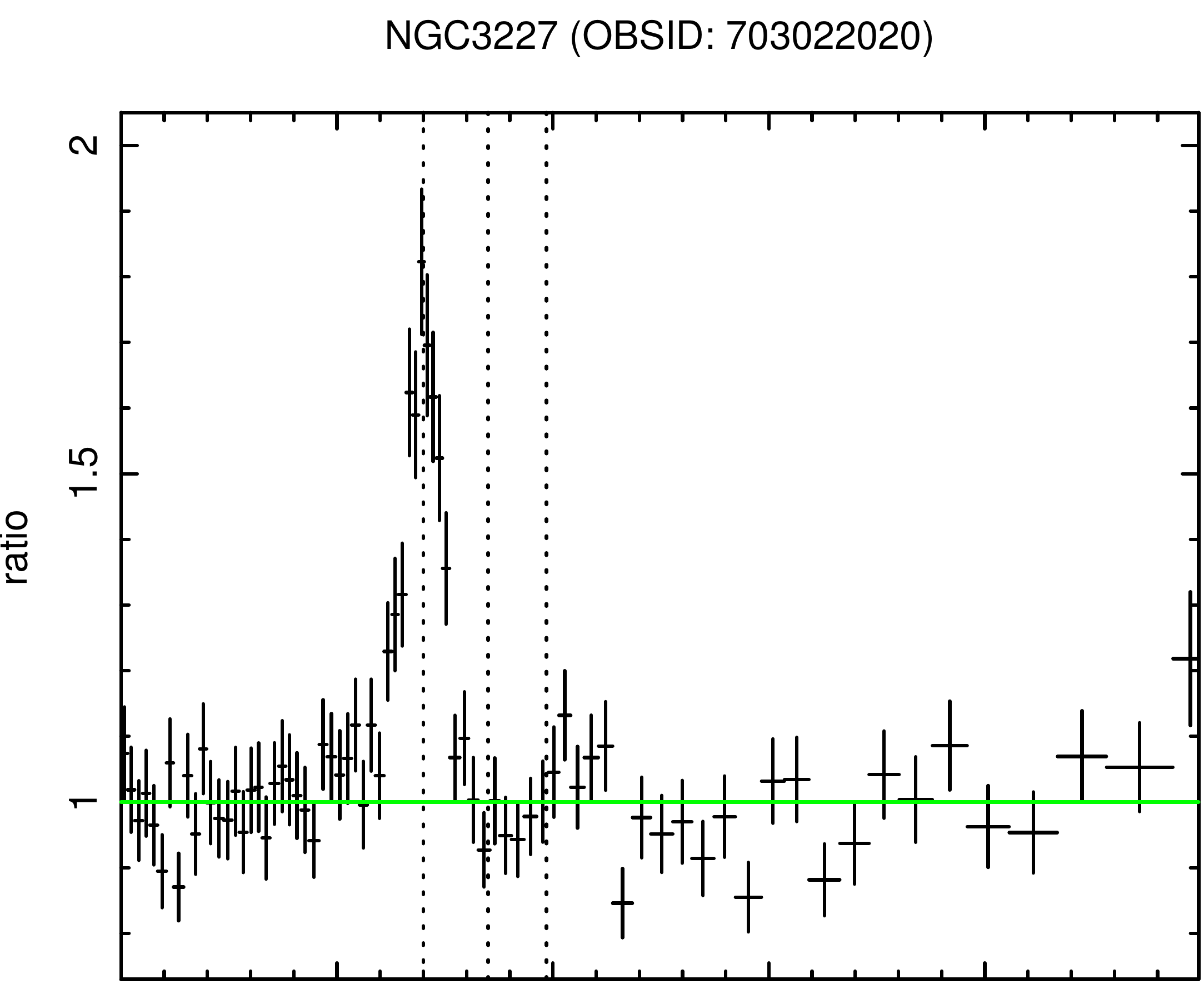}
\includegraphics[angle=-90,width=3.8cm]{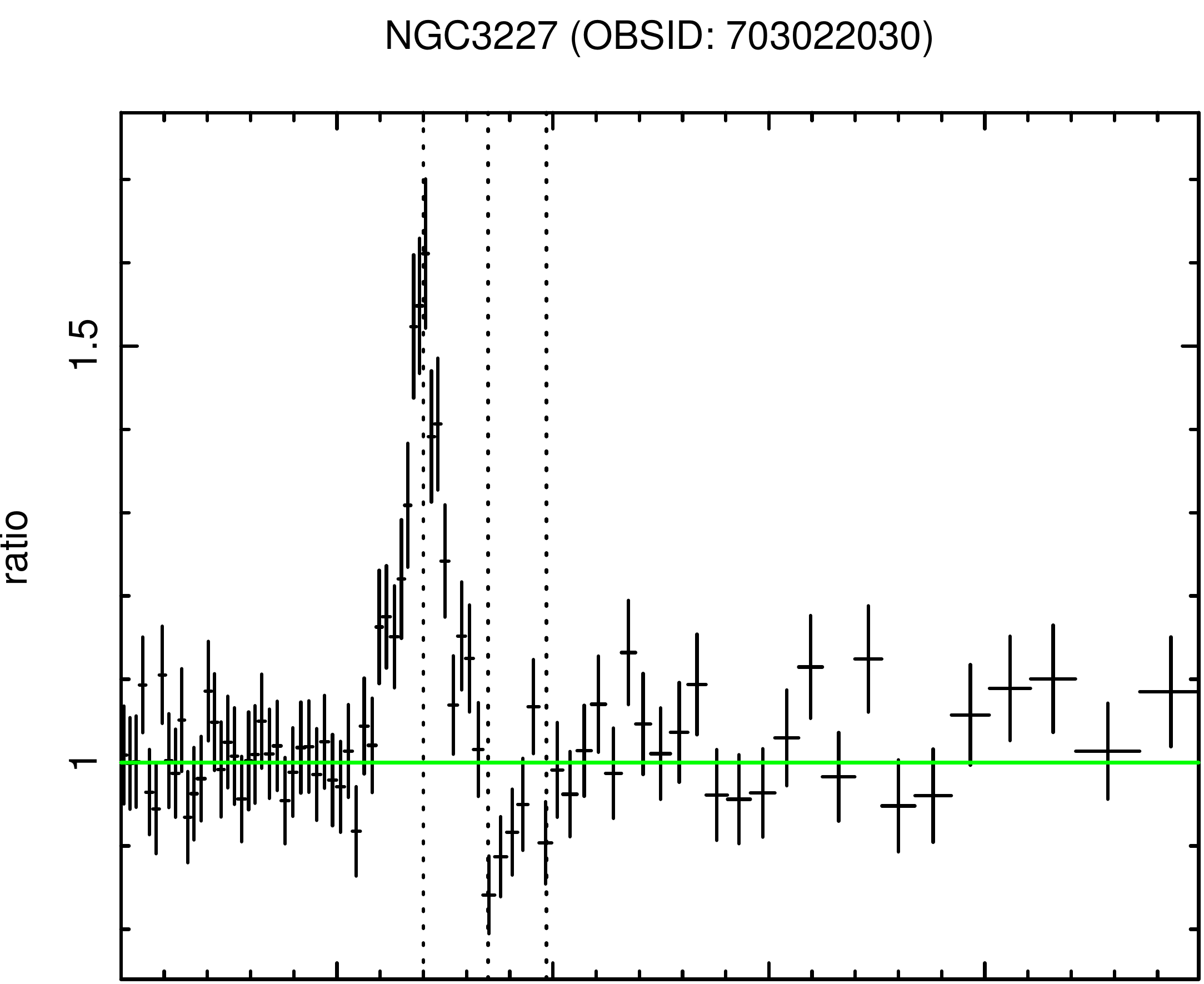}
\includegraphics[angle=-90,width=3.8cm]{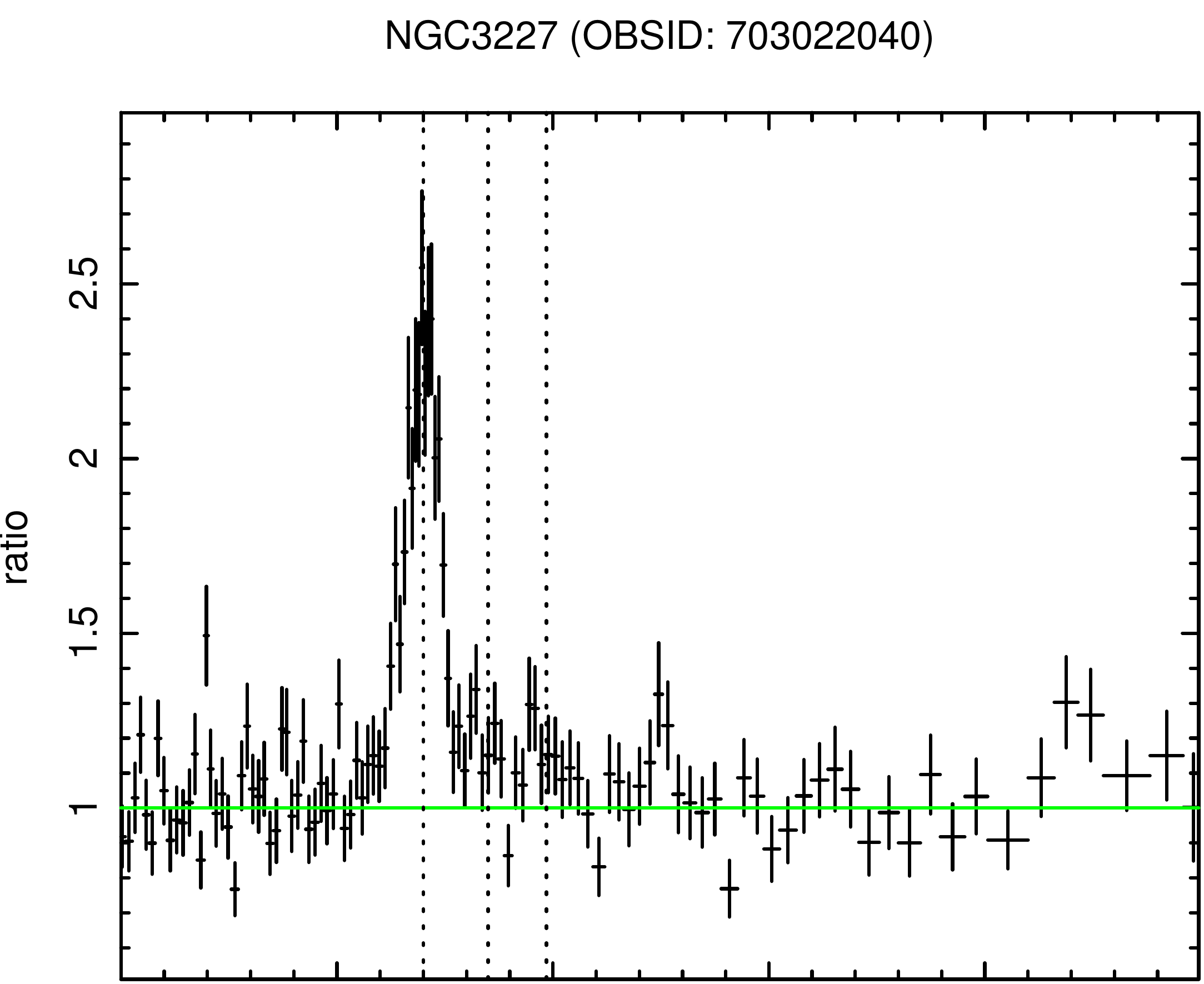}
}

\vspace{-12.2pt}
\subfloat{
\includegraphics[angle=-90,width=3.8cm]{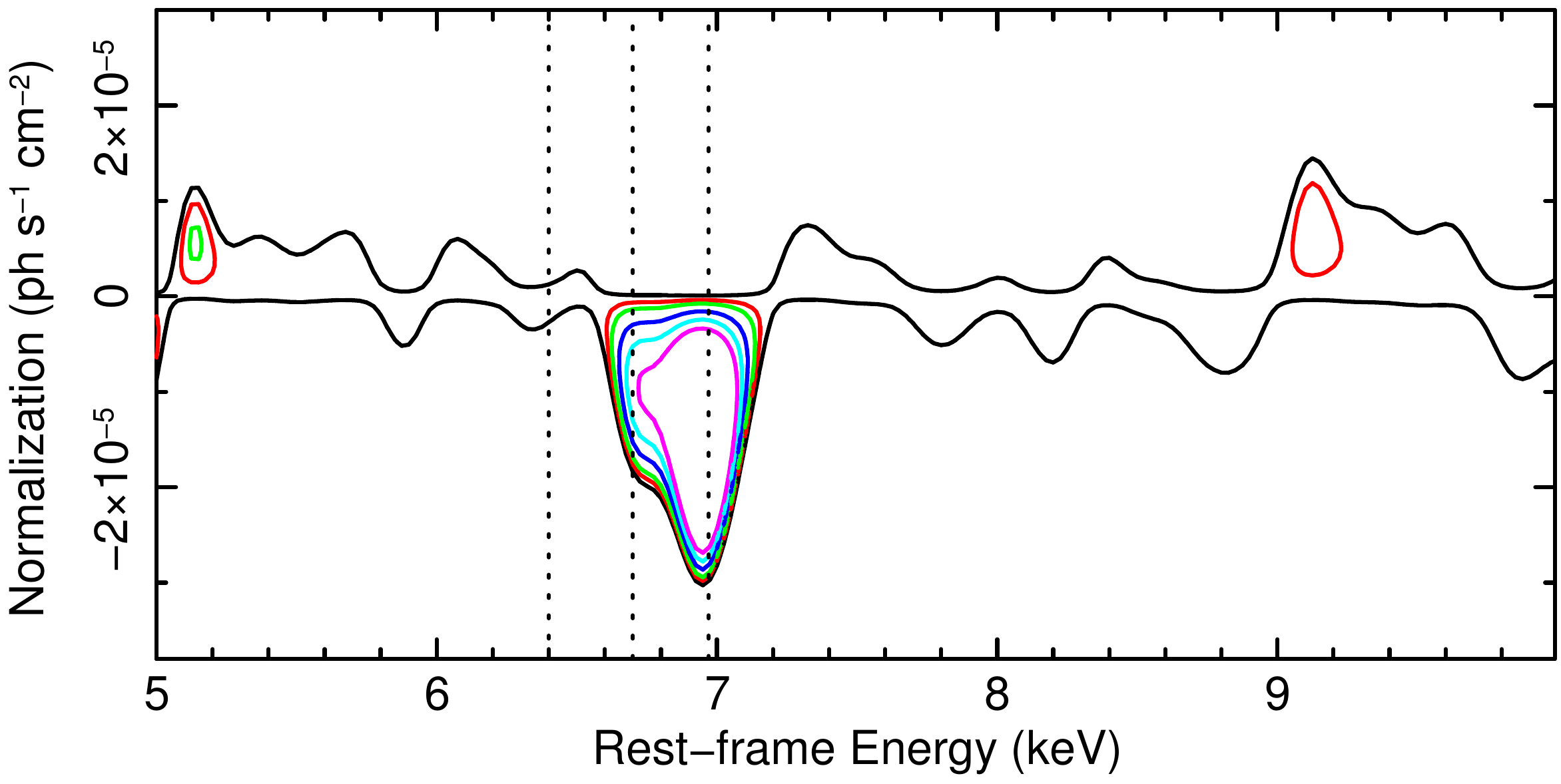}
\includegraphics[angle=-90,width=3.8cm]{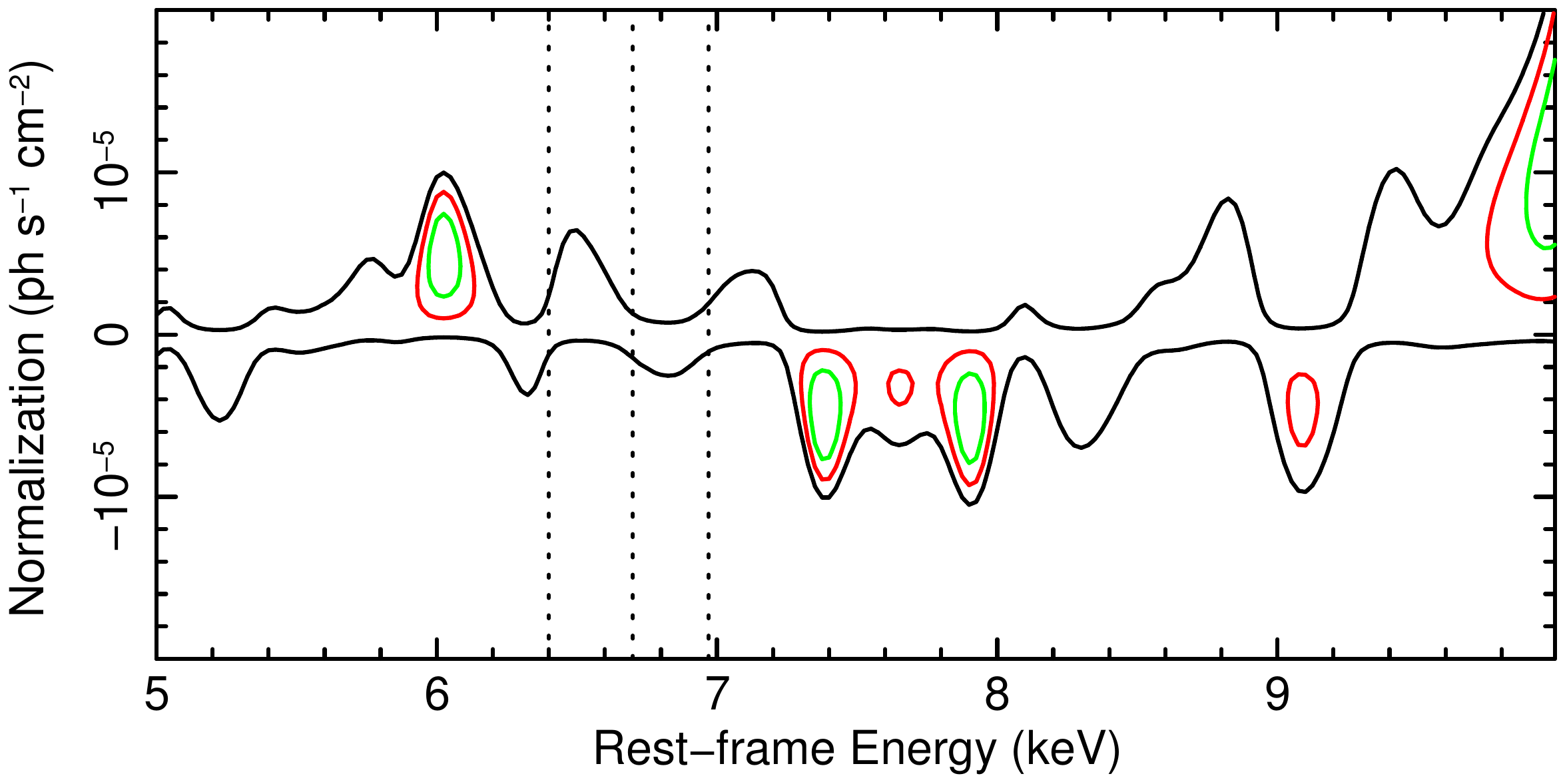}	
\includegraphics[angle=-90,width=3.8cm]{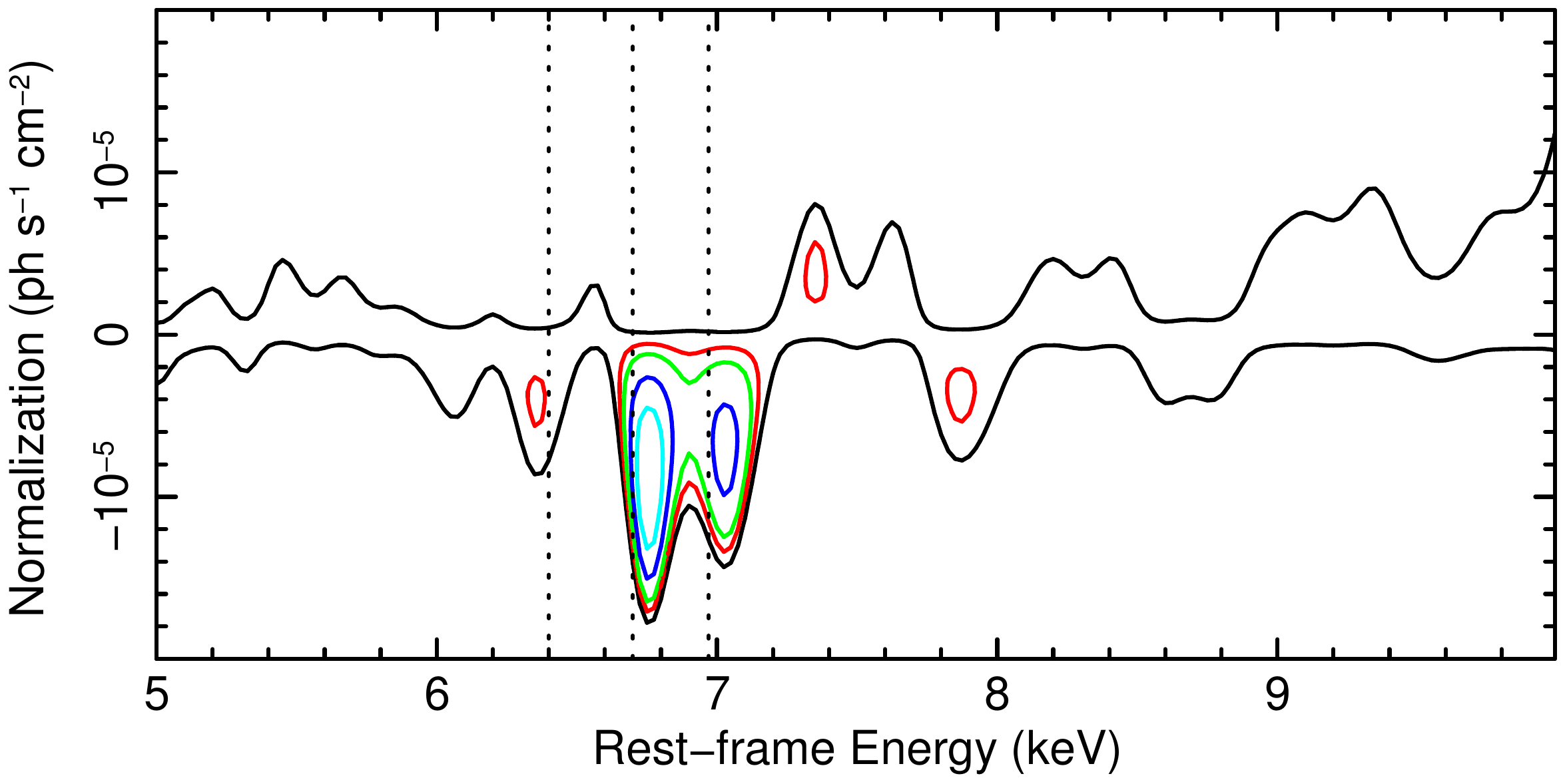}
\includegraphics[angle=-90,width=3.8cm]{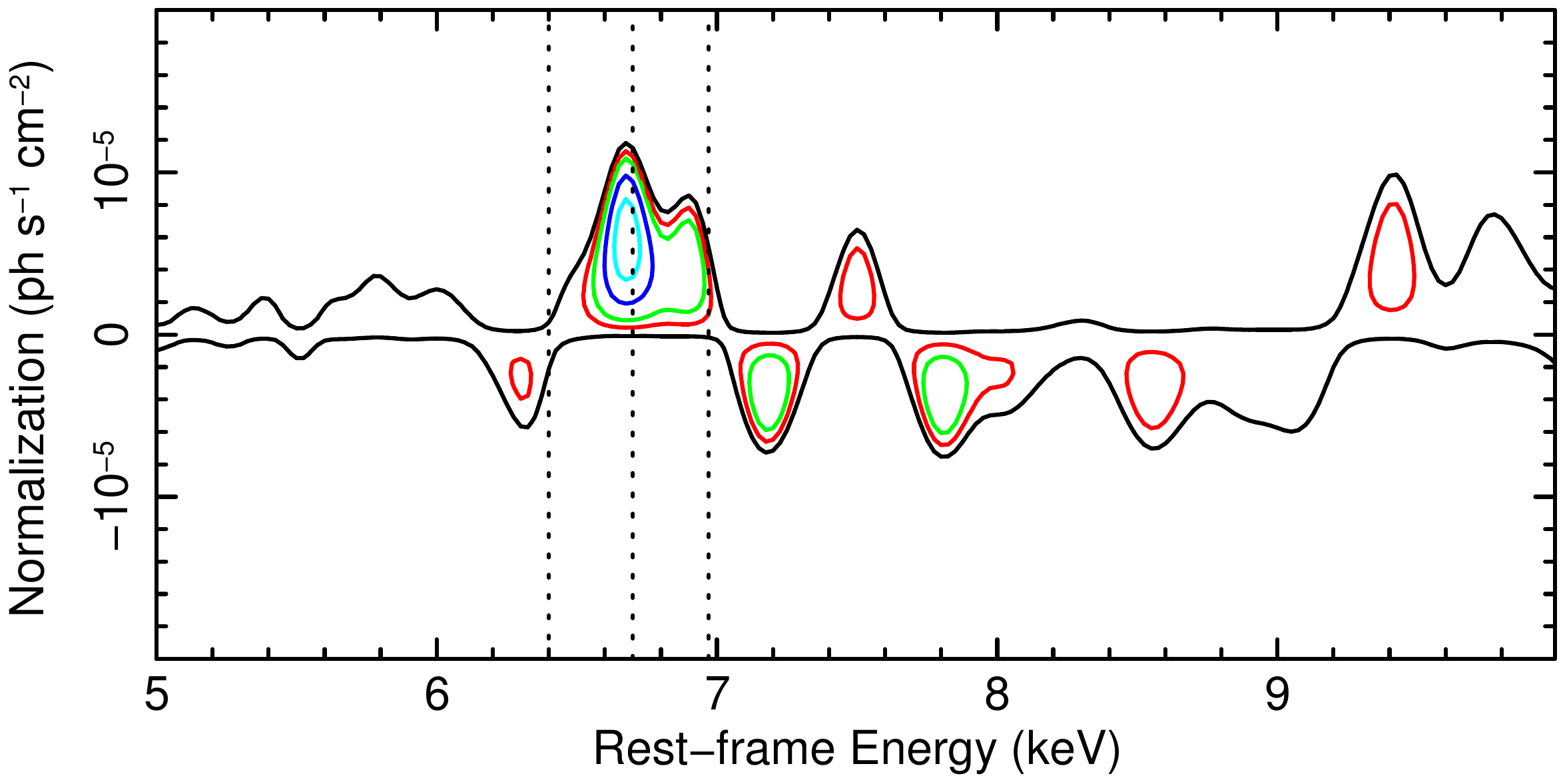}
}

\end{center}
\contcaption{\small -- Ratio and contour plots for sources which do not require a broadened component.}
\end{figure*}

\clearpage

\begin{figure*}
\begin{center}

\vspace{-5pt}	
\subfloat{
\includegraphics[angle=-90,width=3.8cm]{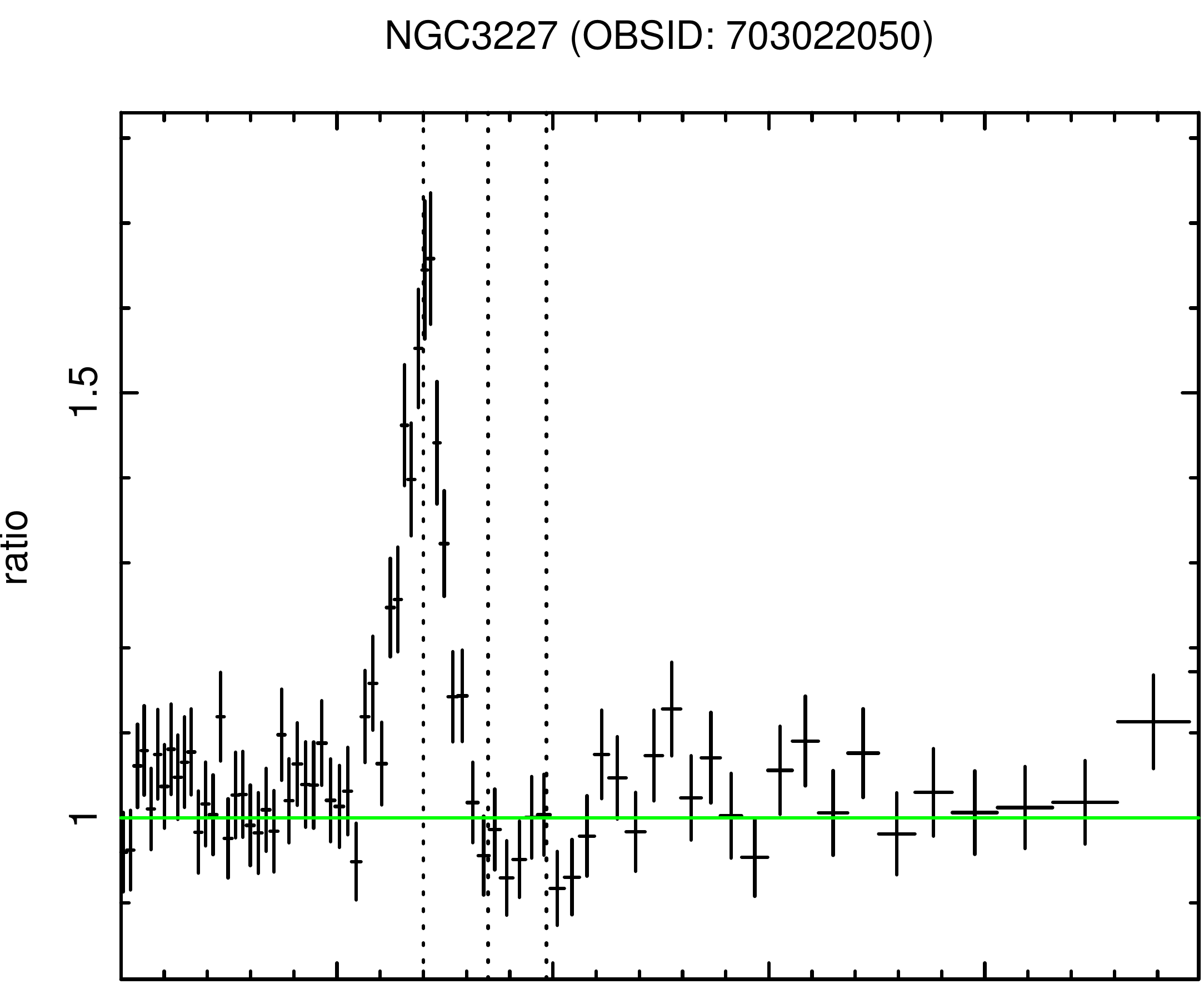}
\includegraphics[angle=-90,width=3.8cm]{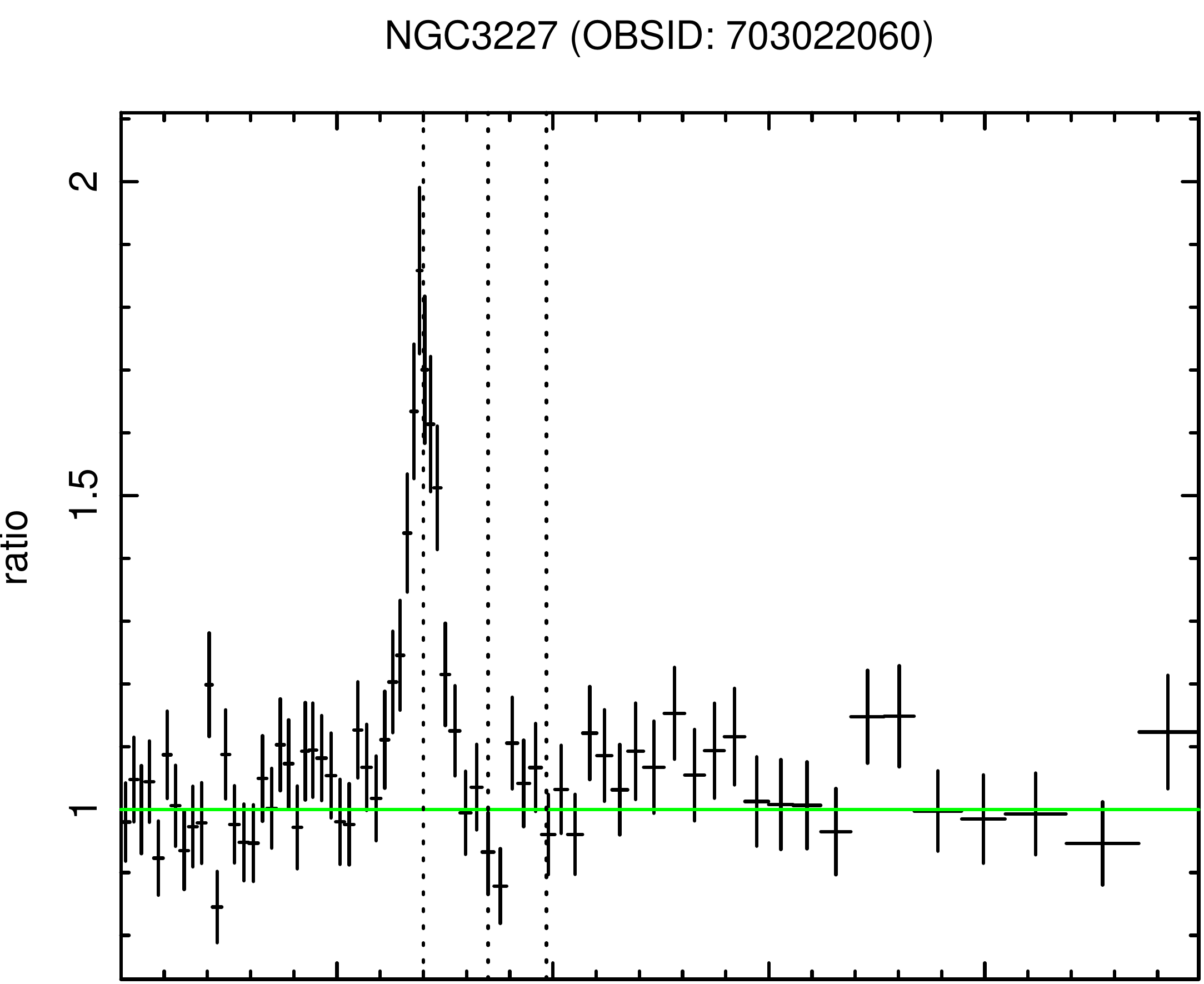}
\includegraphics[angle=-90,width=3.8cm]{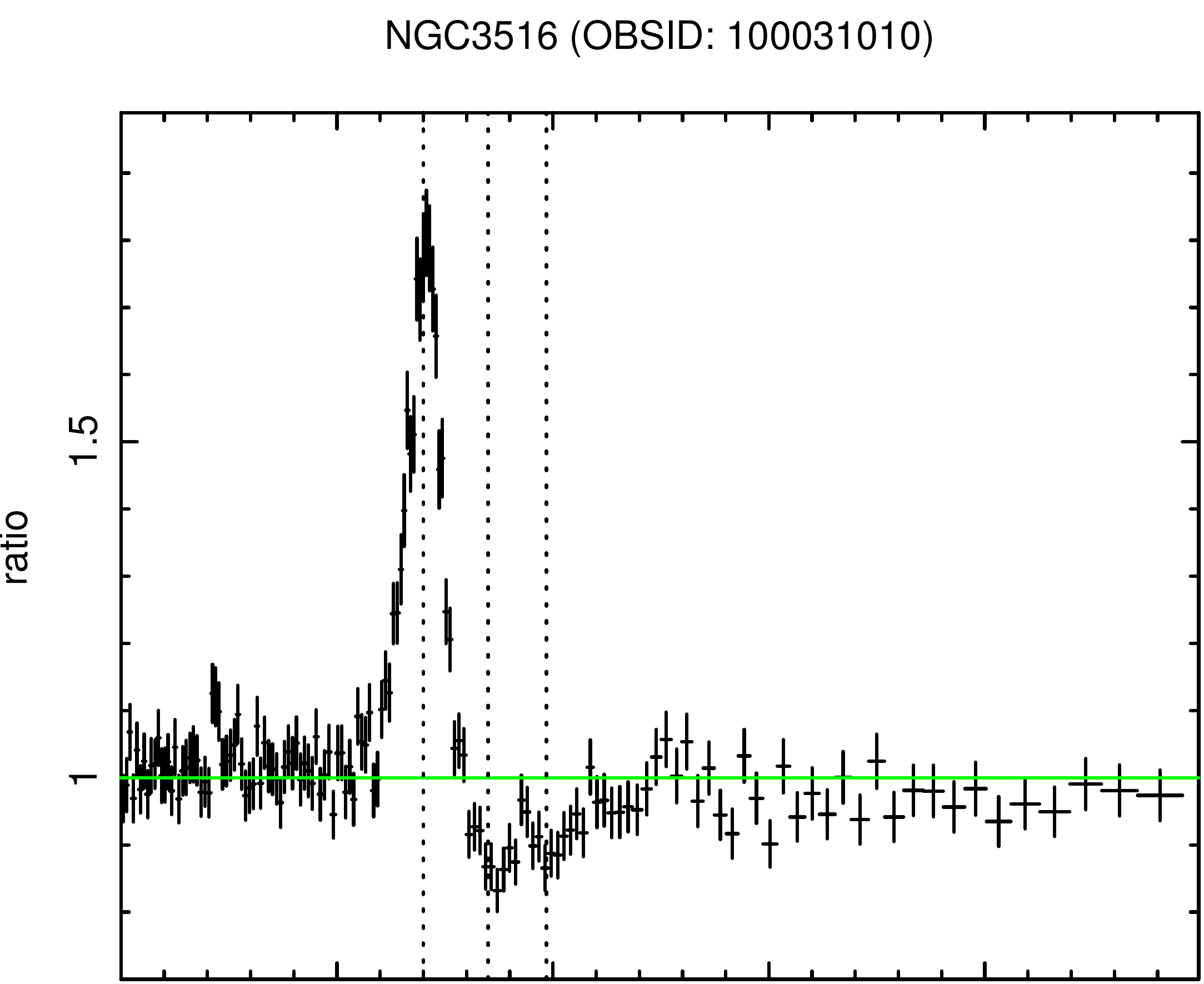}
\includegraphics[angle=-90,width=3.8cm]{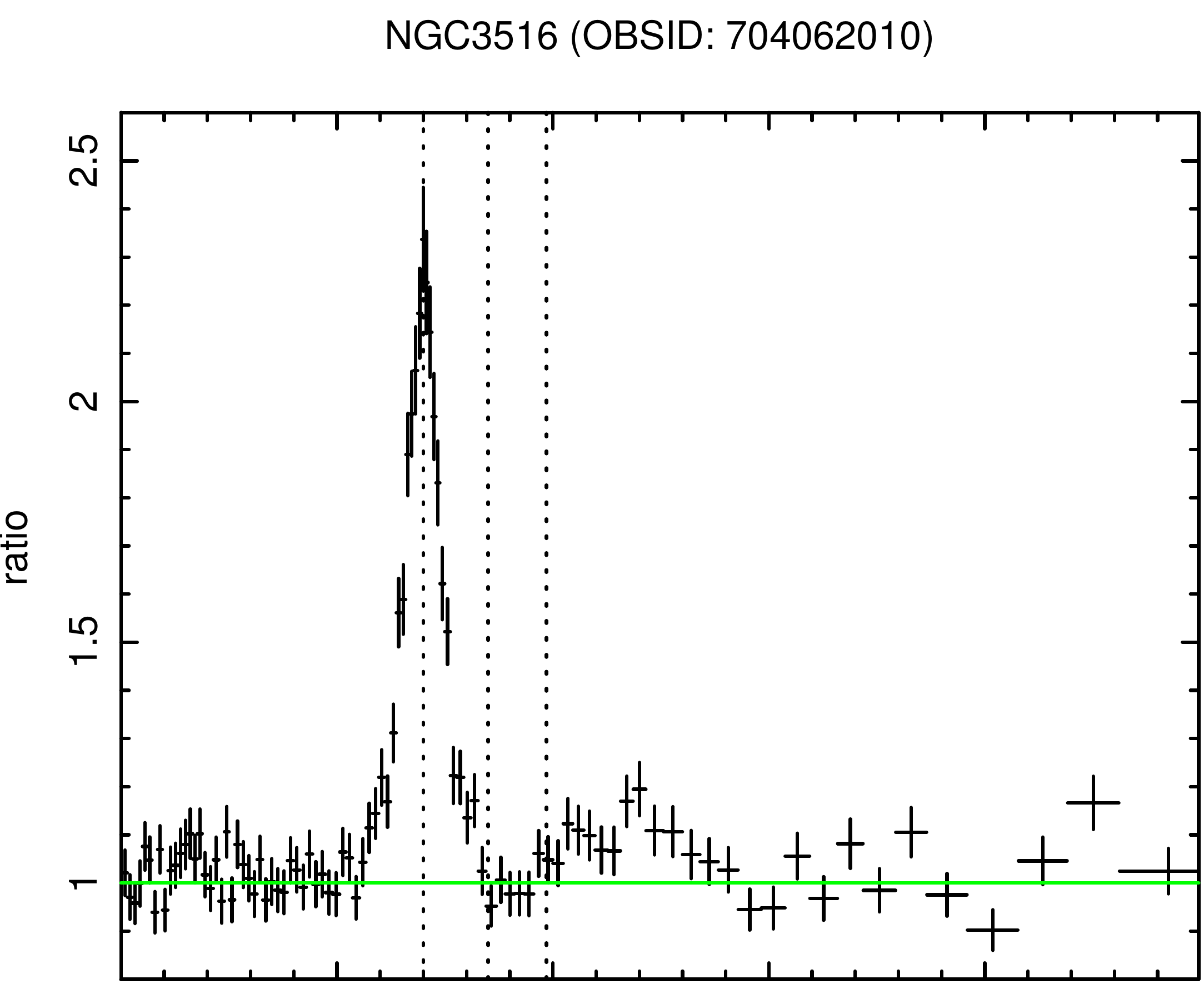}
}

\vspace{-12.2pt}
\subfloat{
\includegraphics[angle=-90,width=3.8cm]{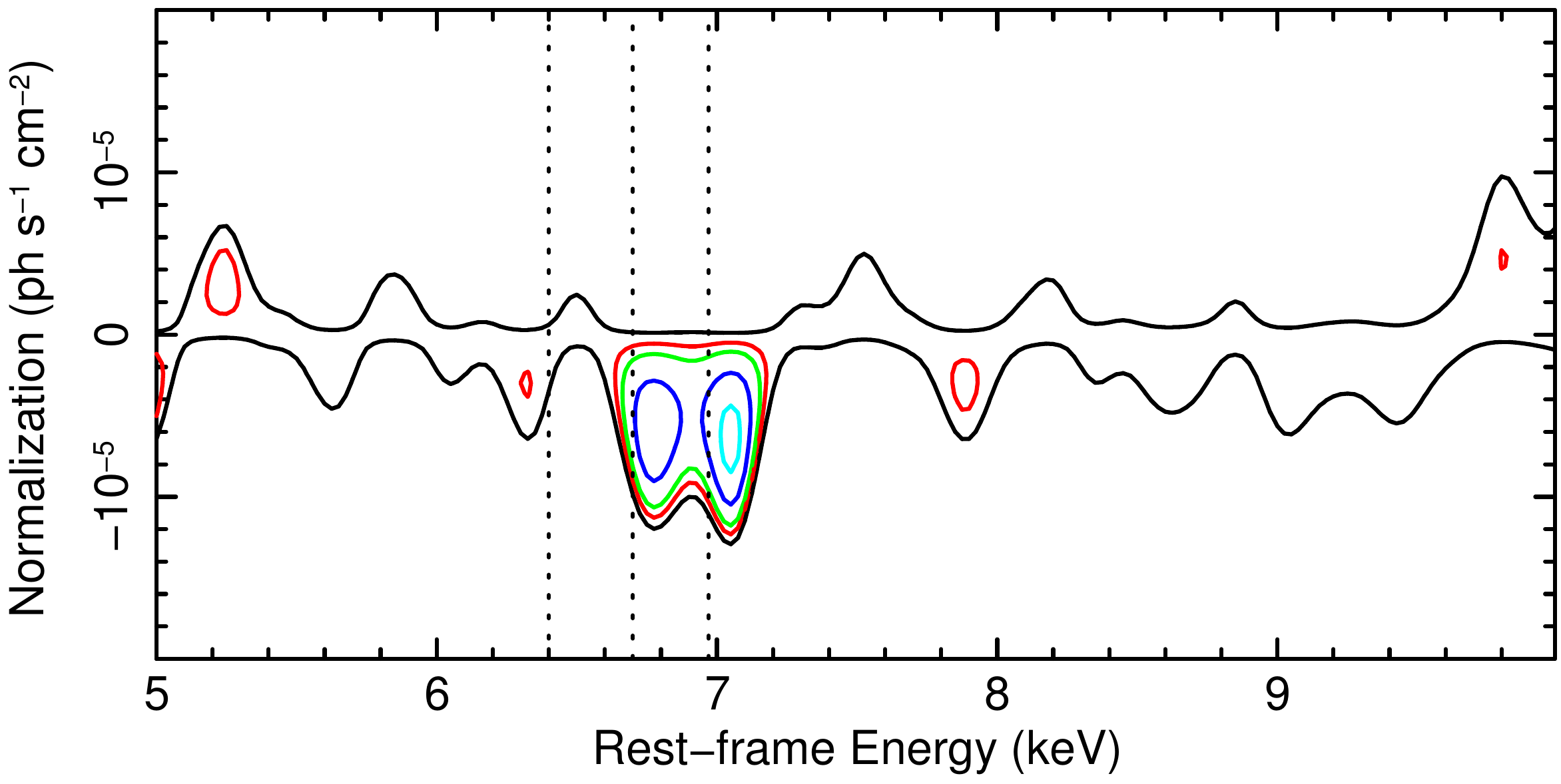}
\includegraphics[angle=-90,width=3.8cm]{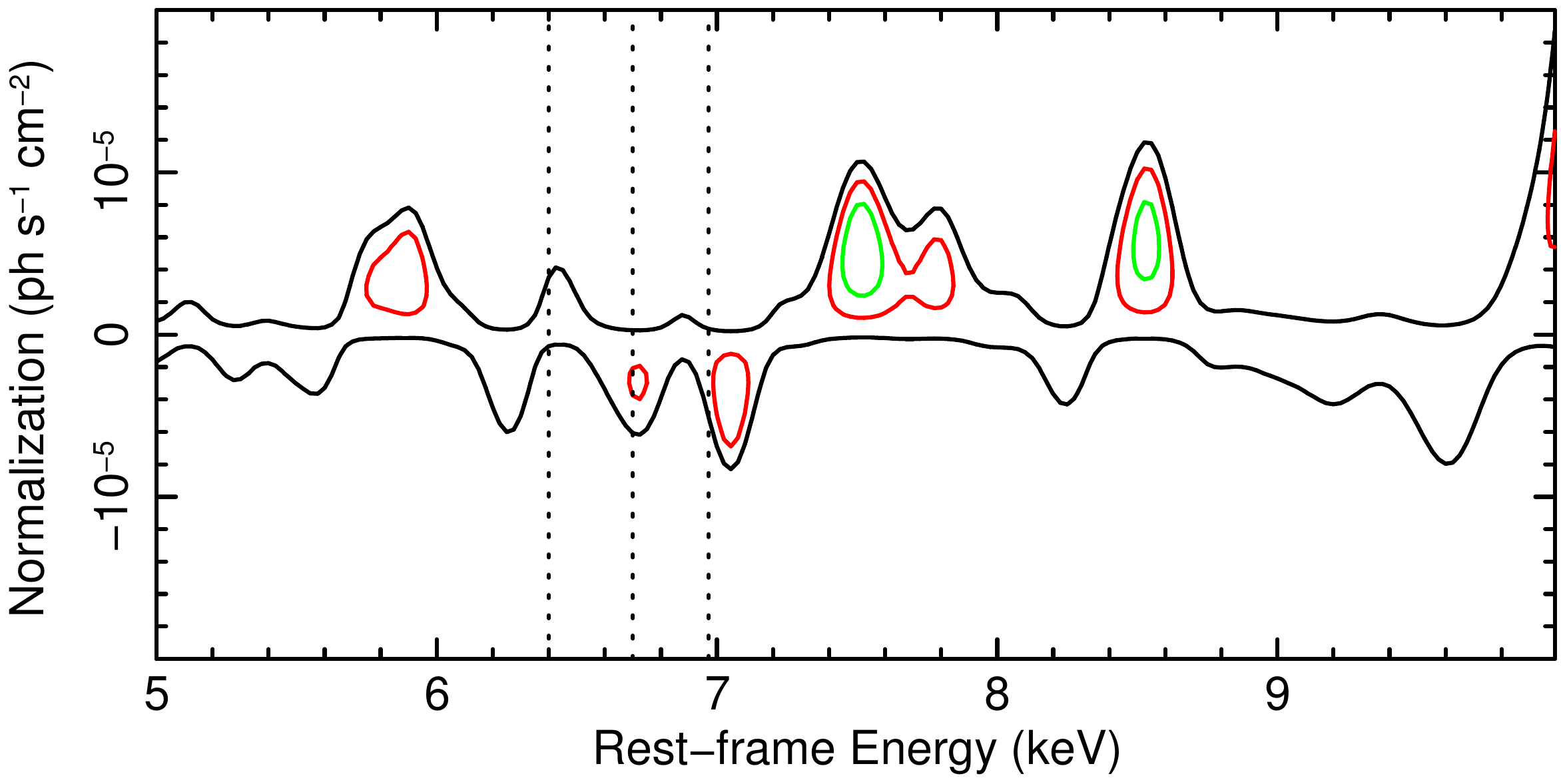}
\includegraphics[angle=-90,width=3.8cm]{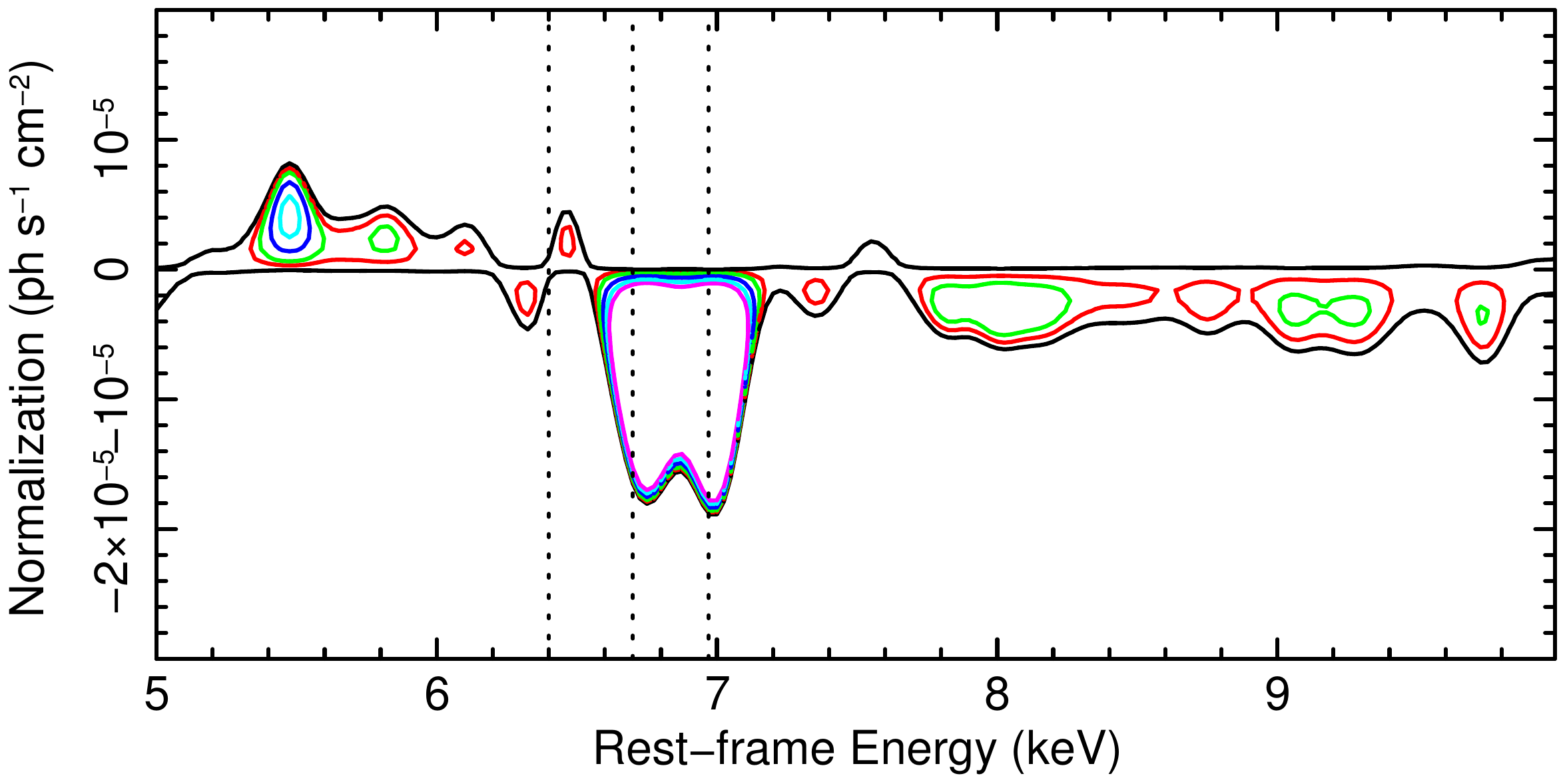}
\includegraphics[angle=-90,width=3.8cm]{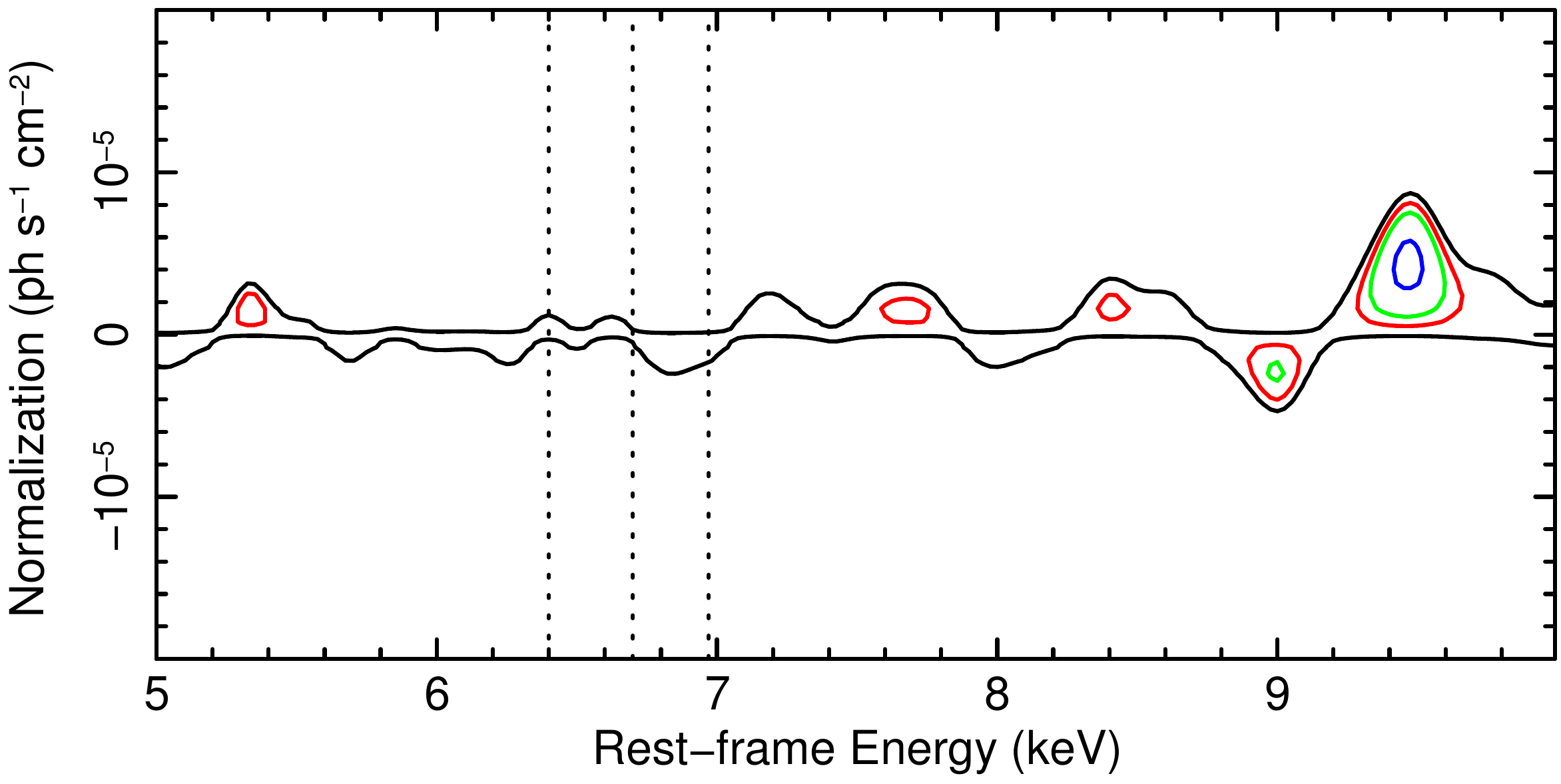}	
}

\vspace{-5pt}	
\subfloat{
\includegraphics[angle=-90,width=3.8cm]{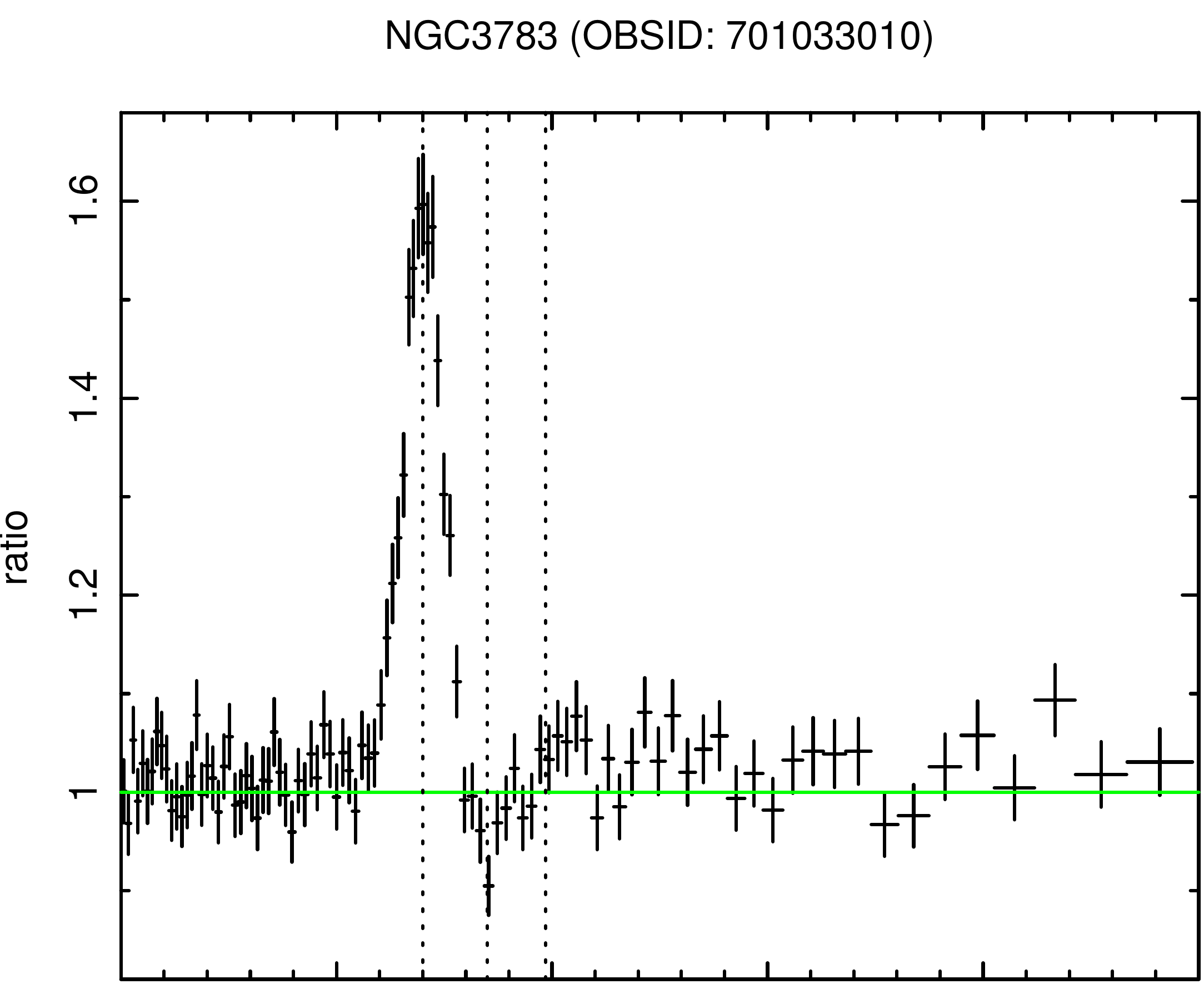}
\includegraphics[angle=-90,width=3.8cm]{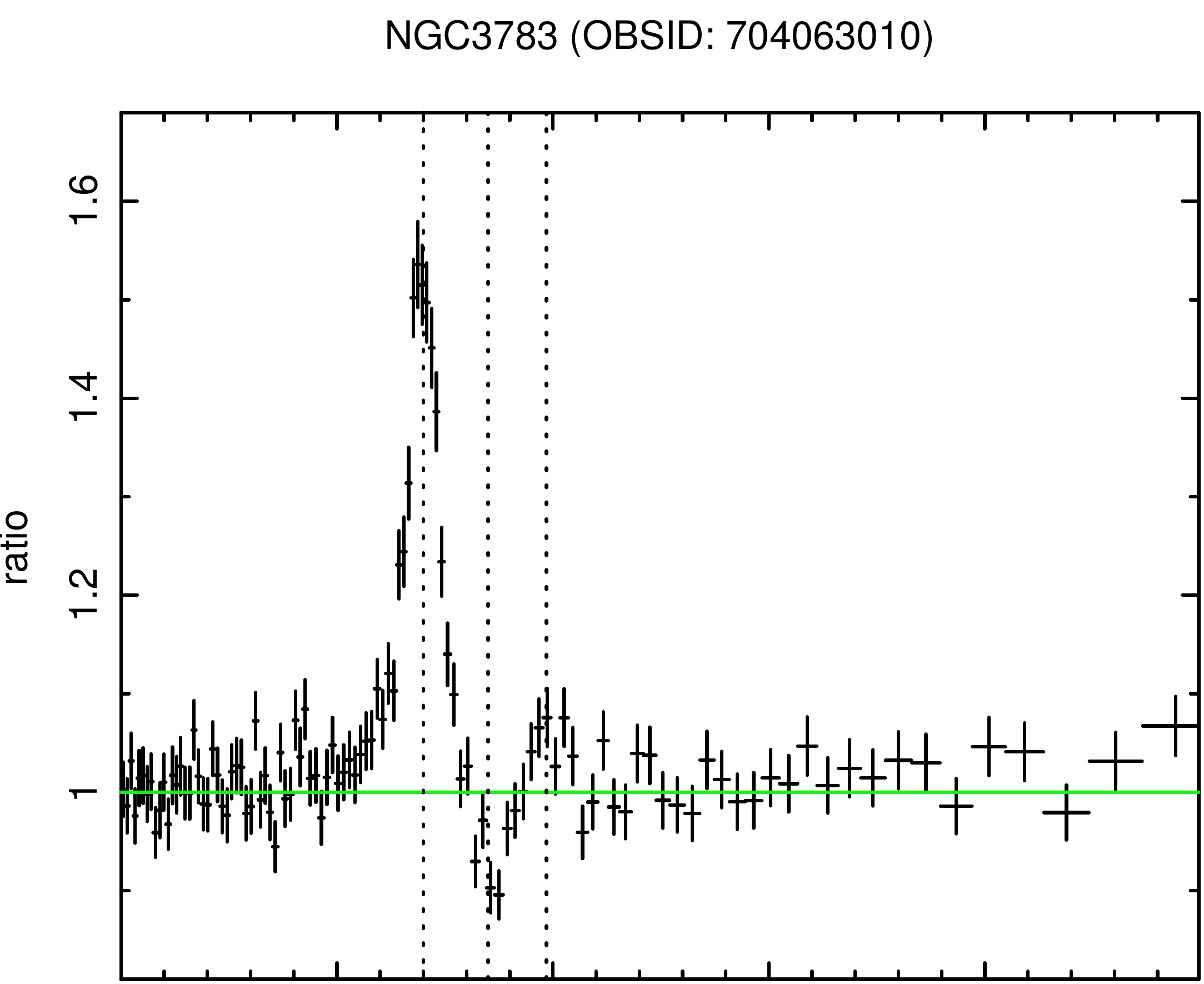}
\includegraphics[angle=-90,width=3.8cm]{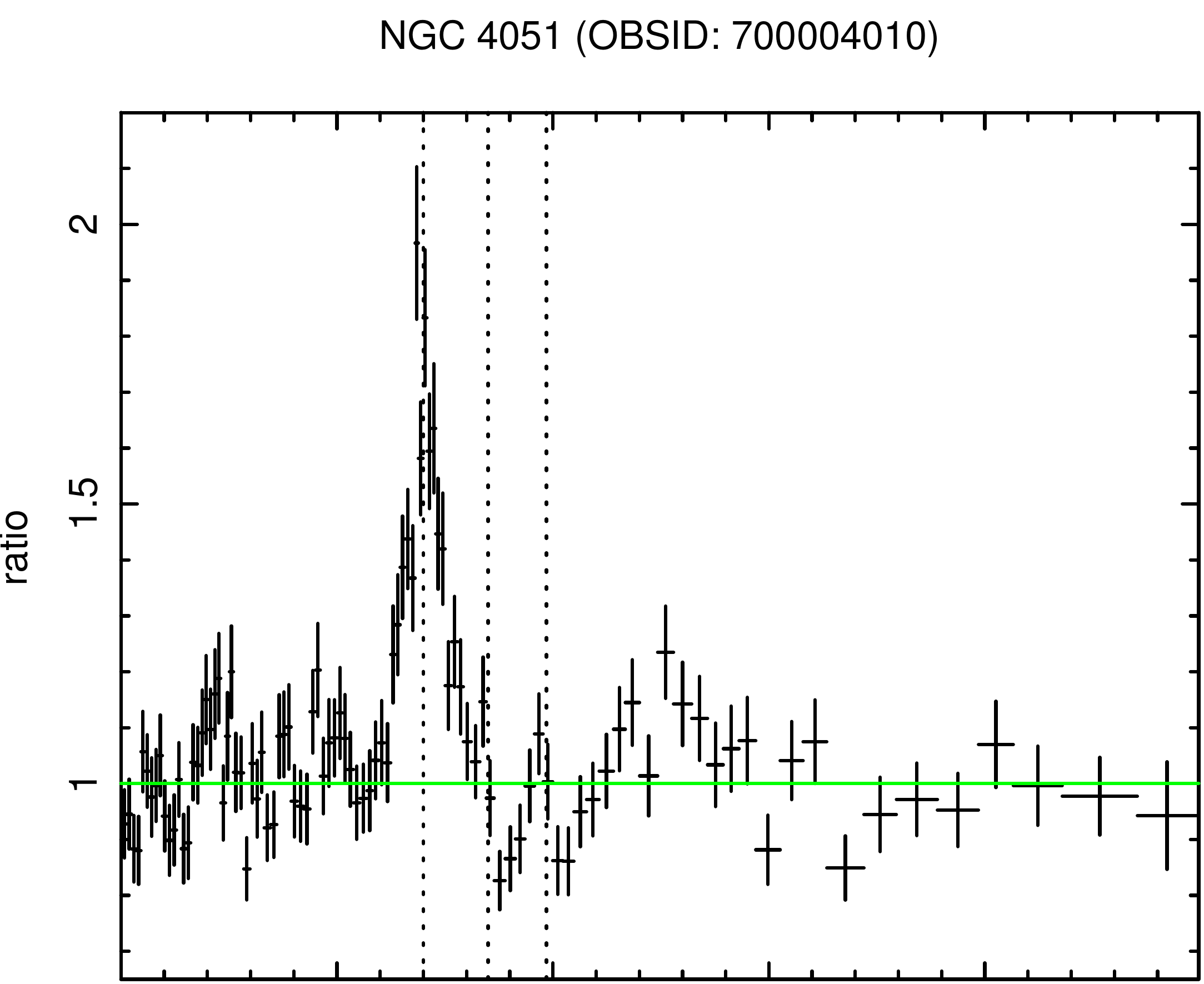}
\includegraphics[angle=-90,width=3.8cm]{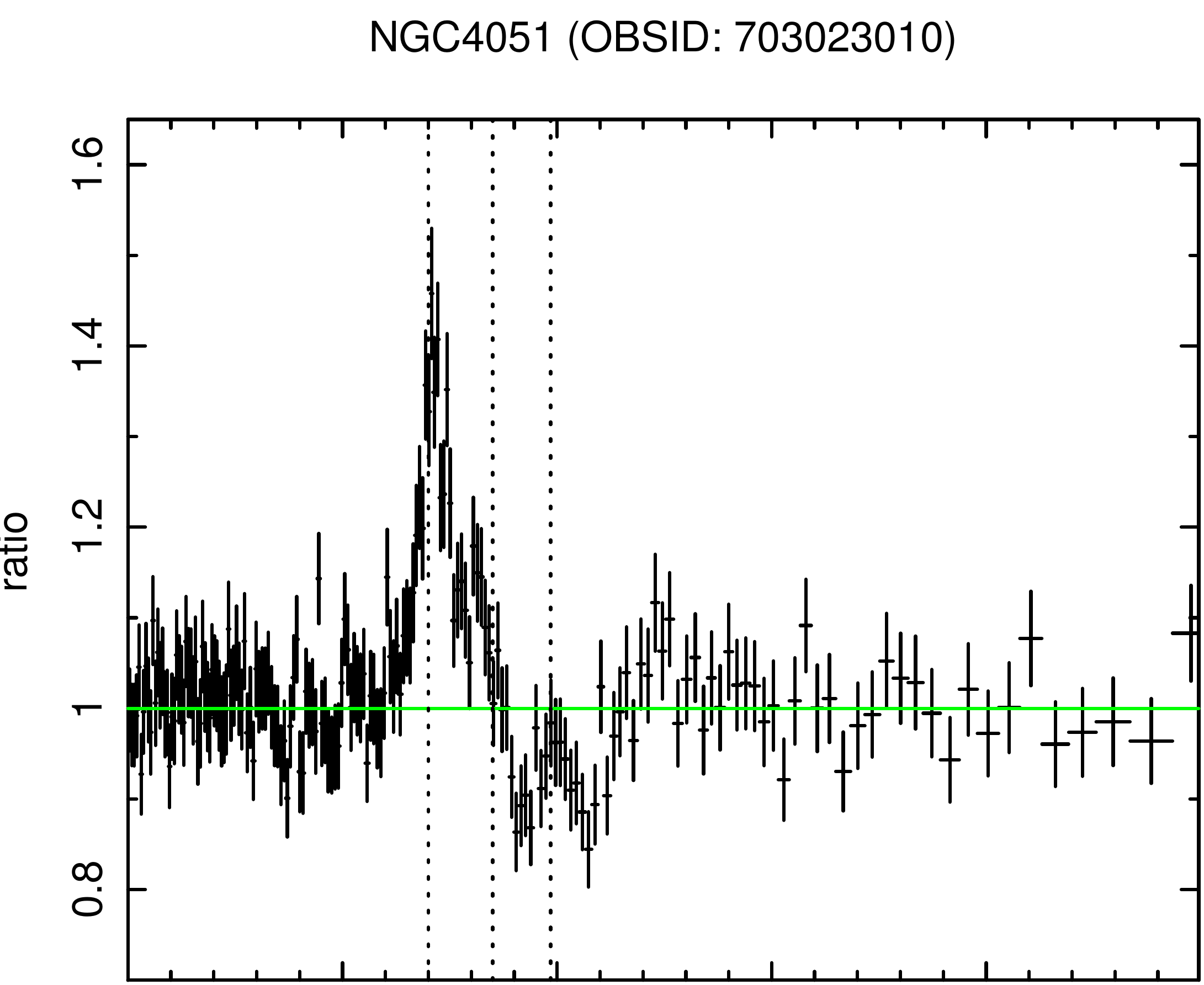}
}

\vspace{-12.2pt}
\subfloat{
\includegraphics[angle=-90,width=3.8cm]{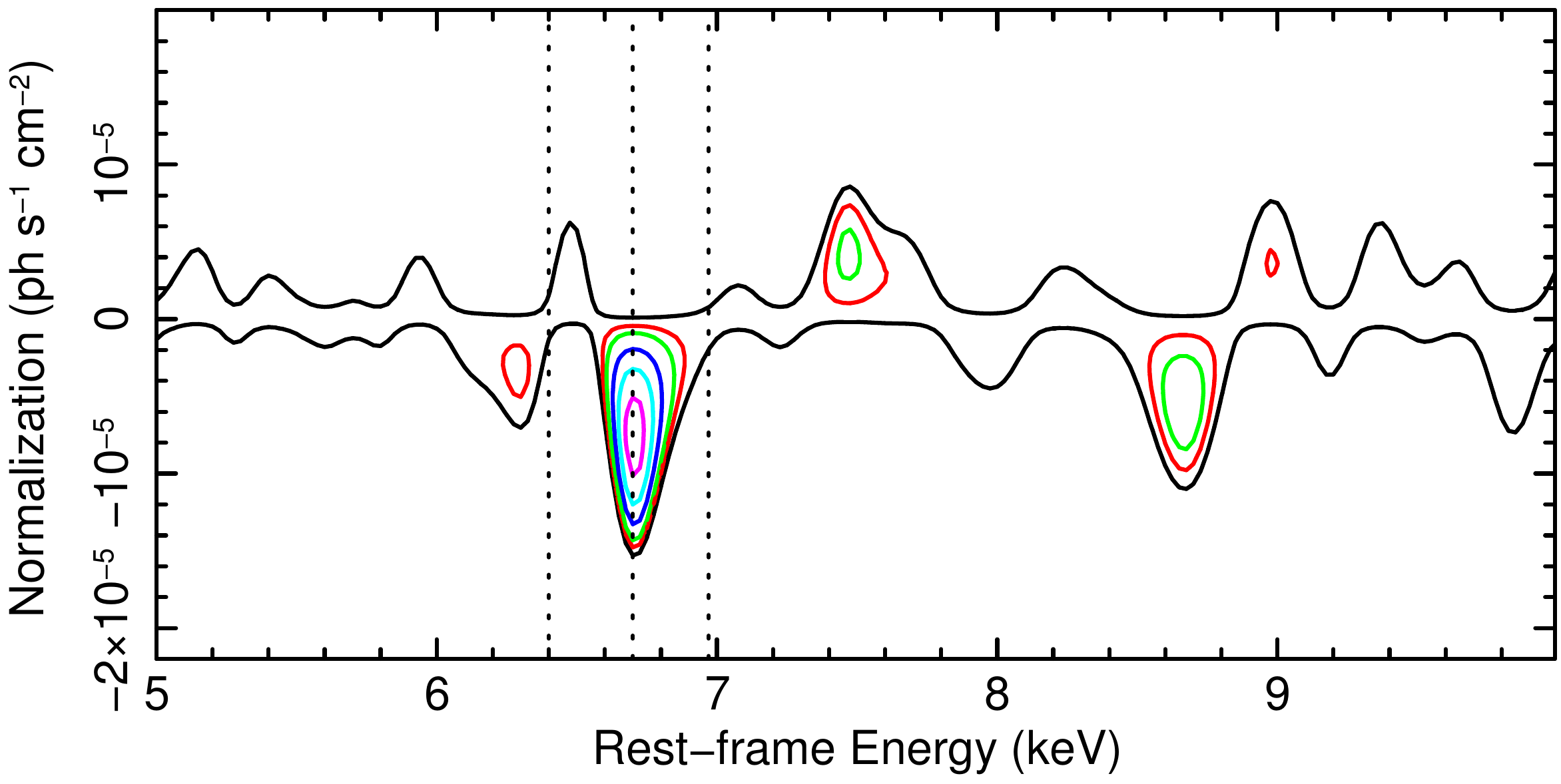}
\includegraphics[angle=-90,width=3.8cm]{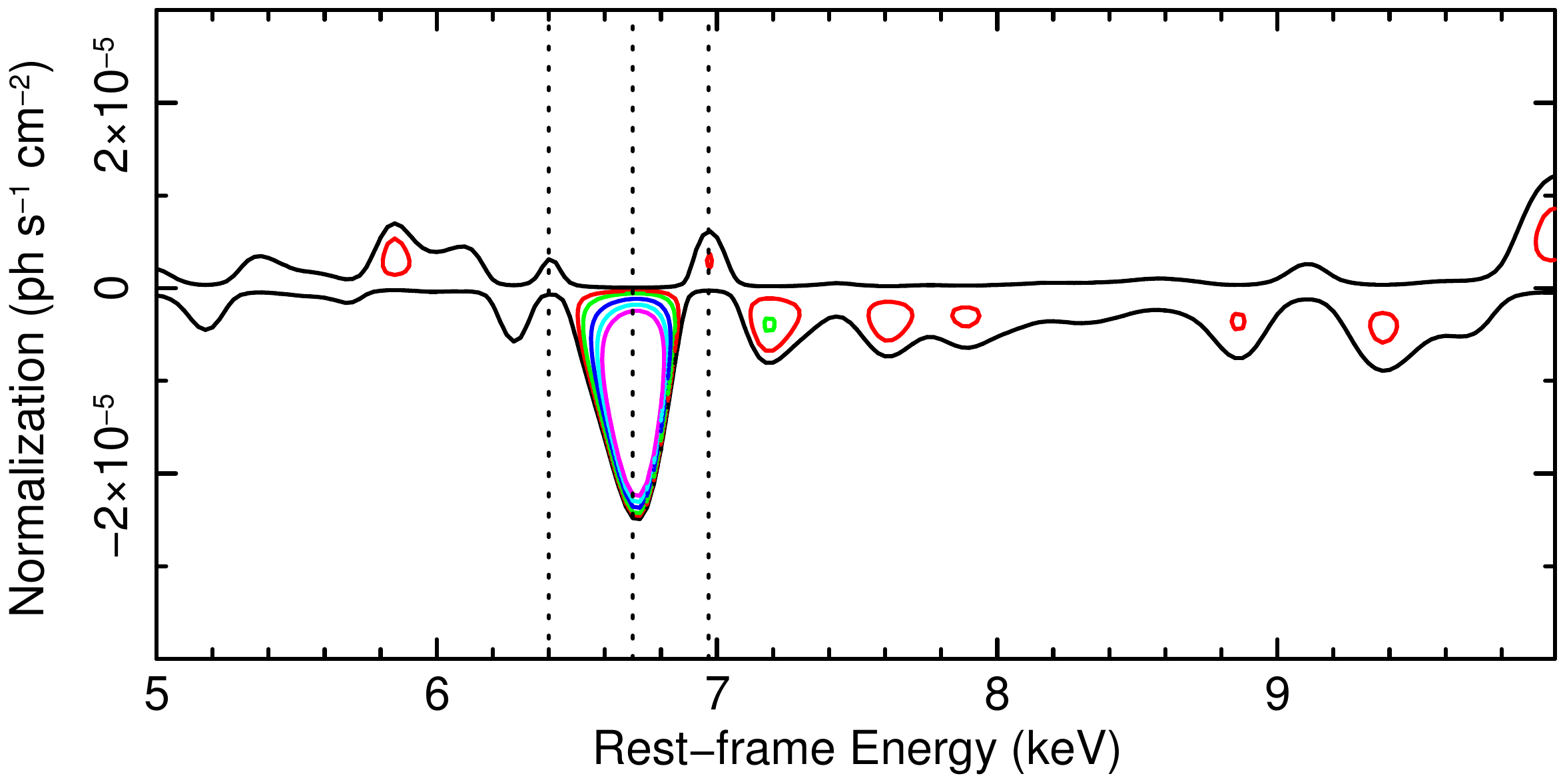}
\includegraphics[angle=-90,width=3.8cm]{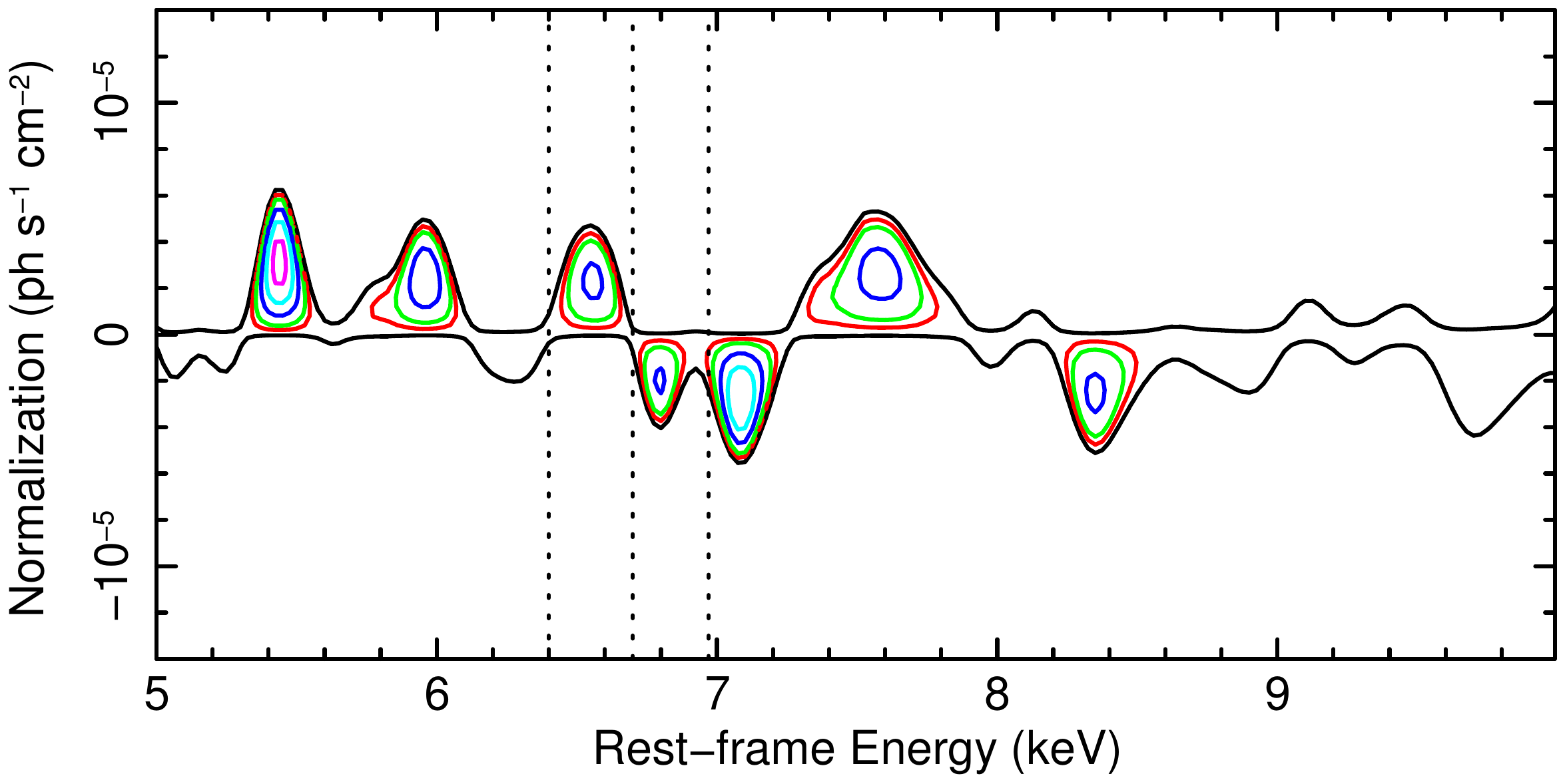}
\includegraphics[angle=-90,width=3.8cm]{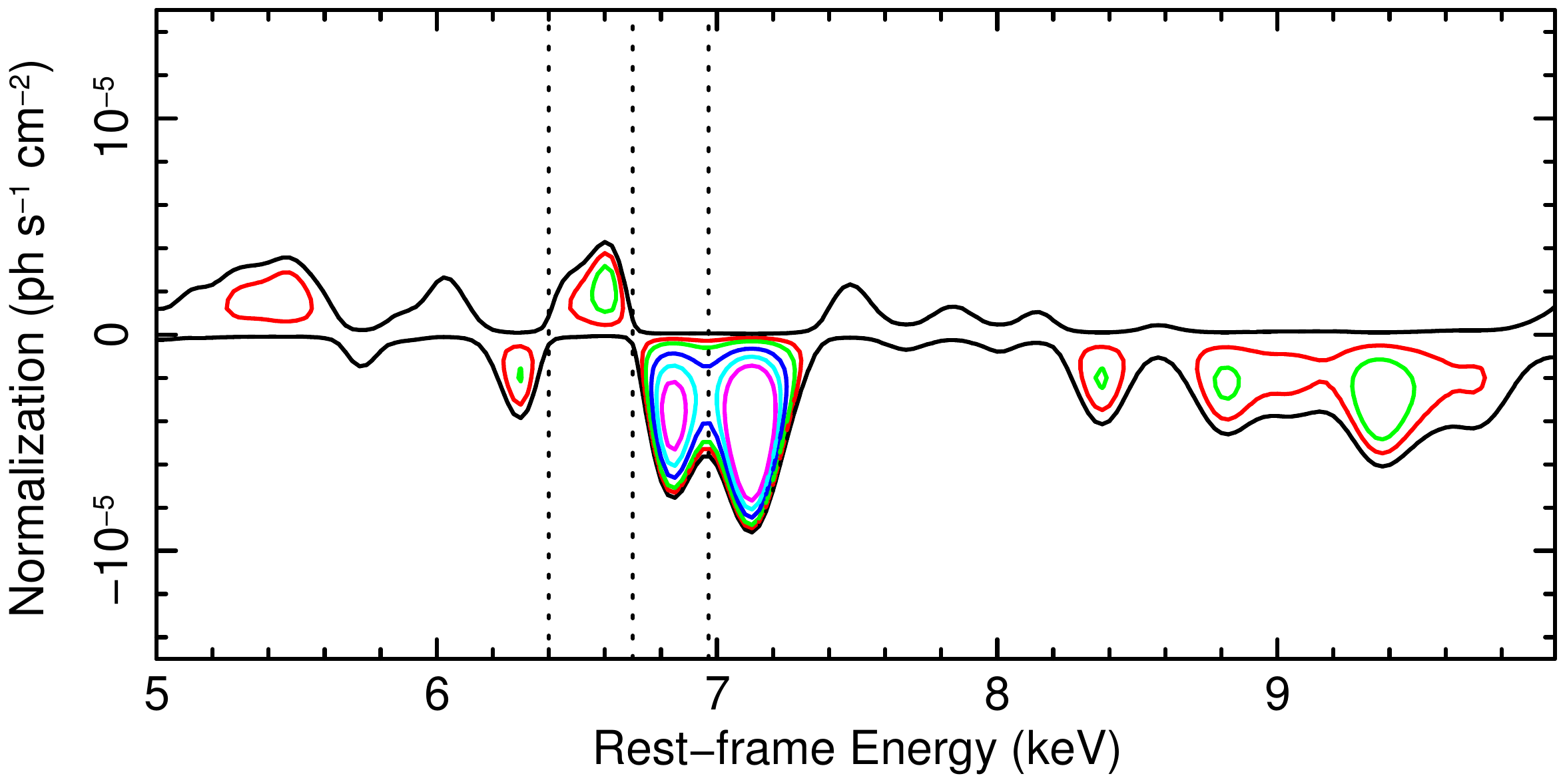}
}

\vspace{-5pt}	
\subfloat{
\includegraphics[angle=-90,width=3.8cm]{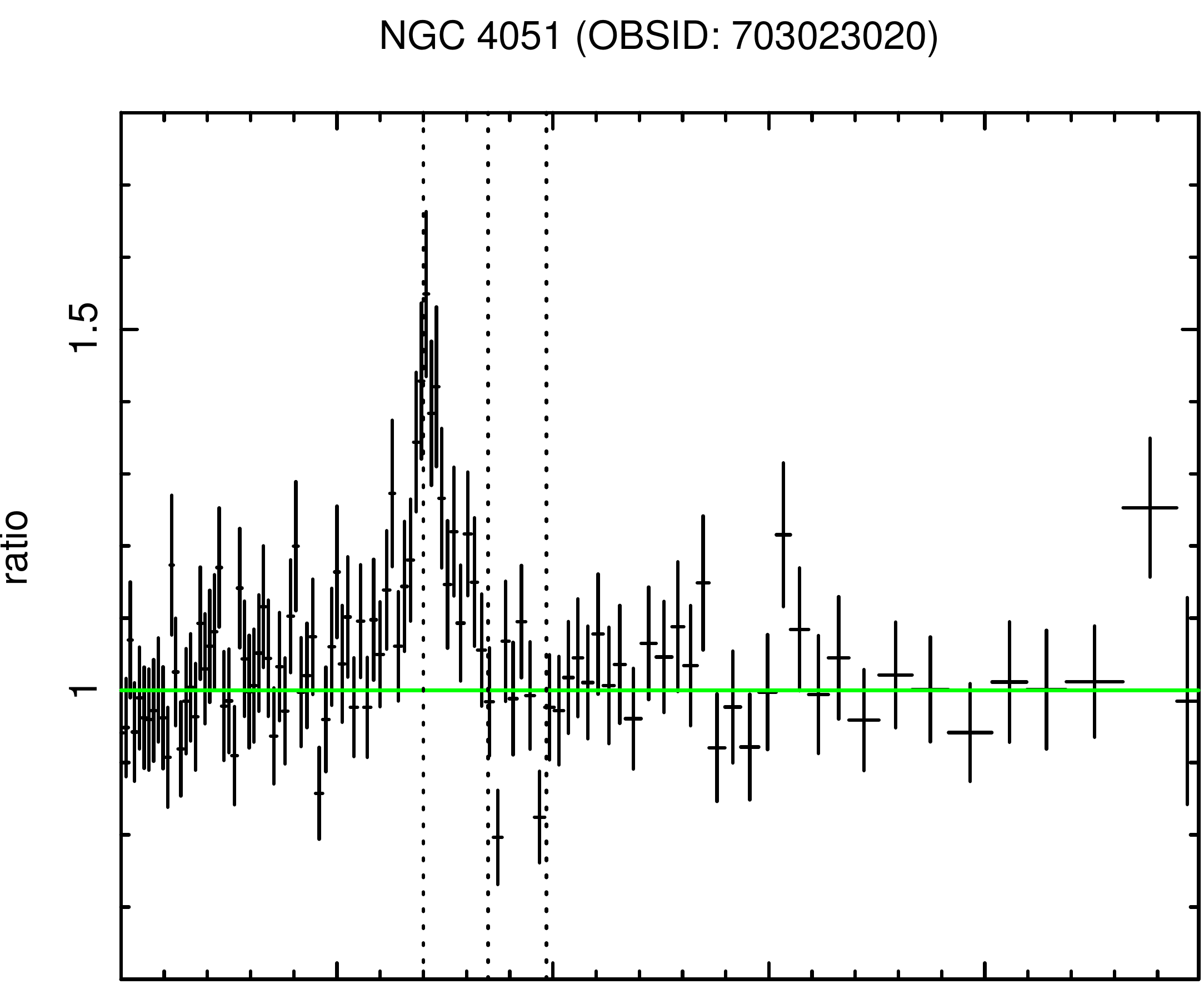}
\includegraphics[angle=-90,width=3.8cm]{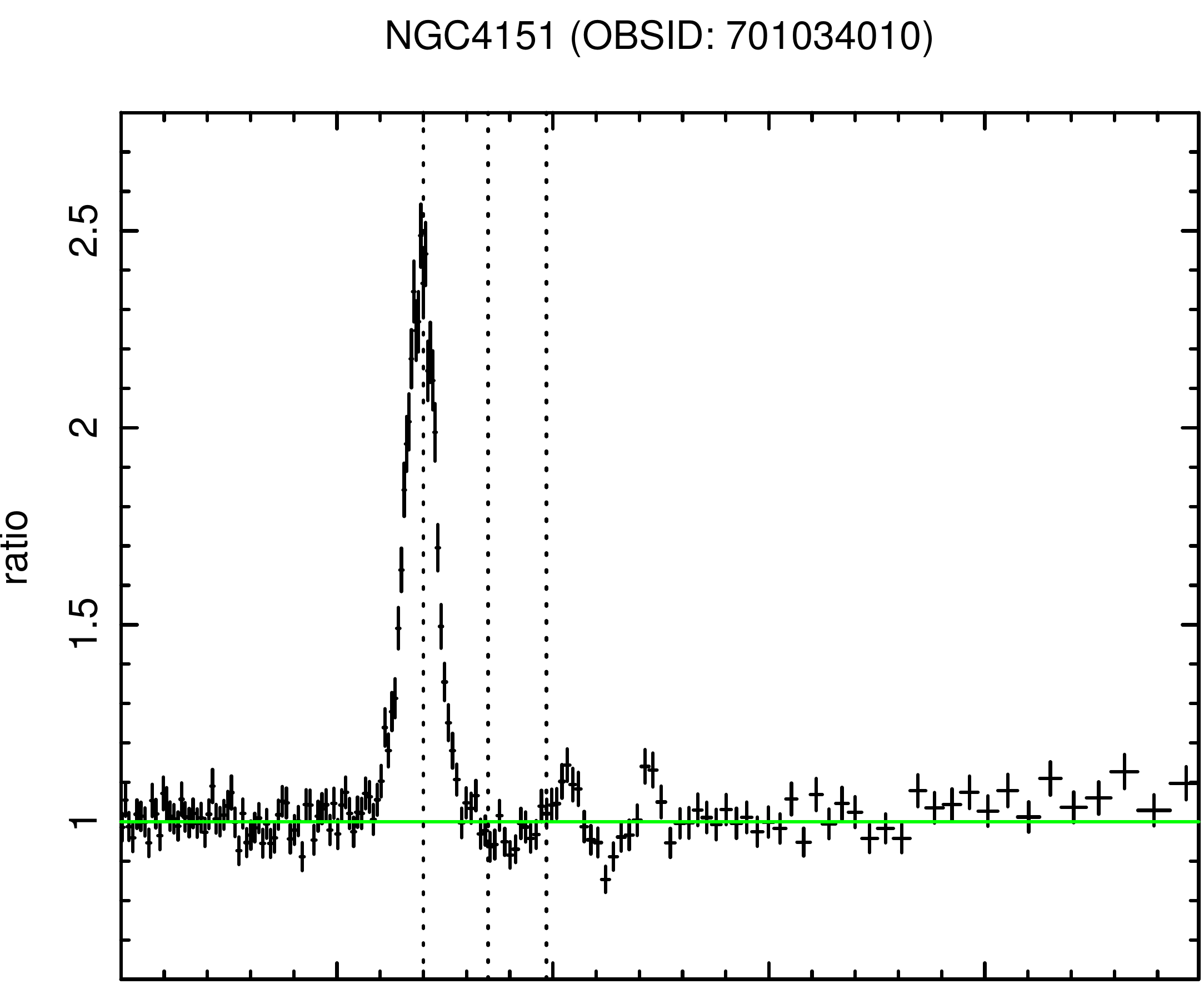}
\includegraphics[angle=-90,width=3.8cm]{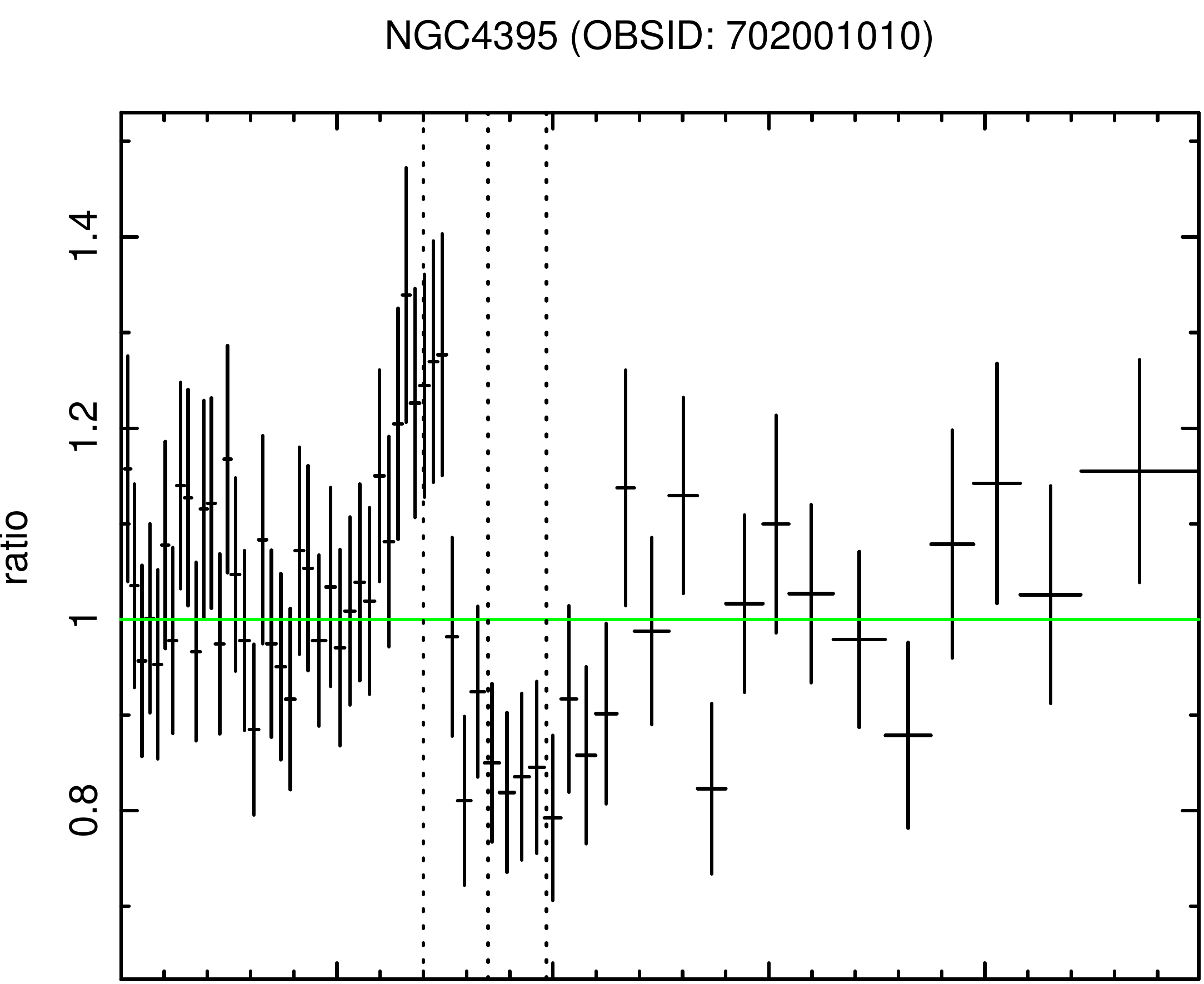}
\includegraphics[angle=-90,width=3.8cm]{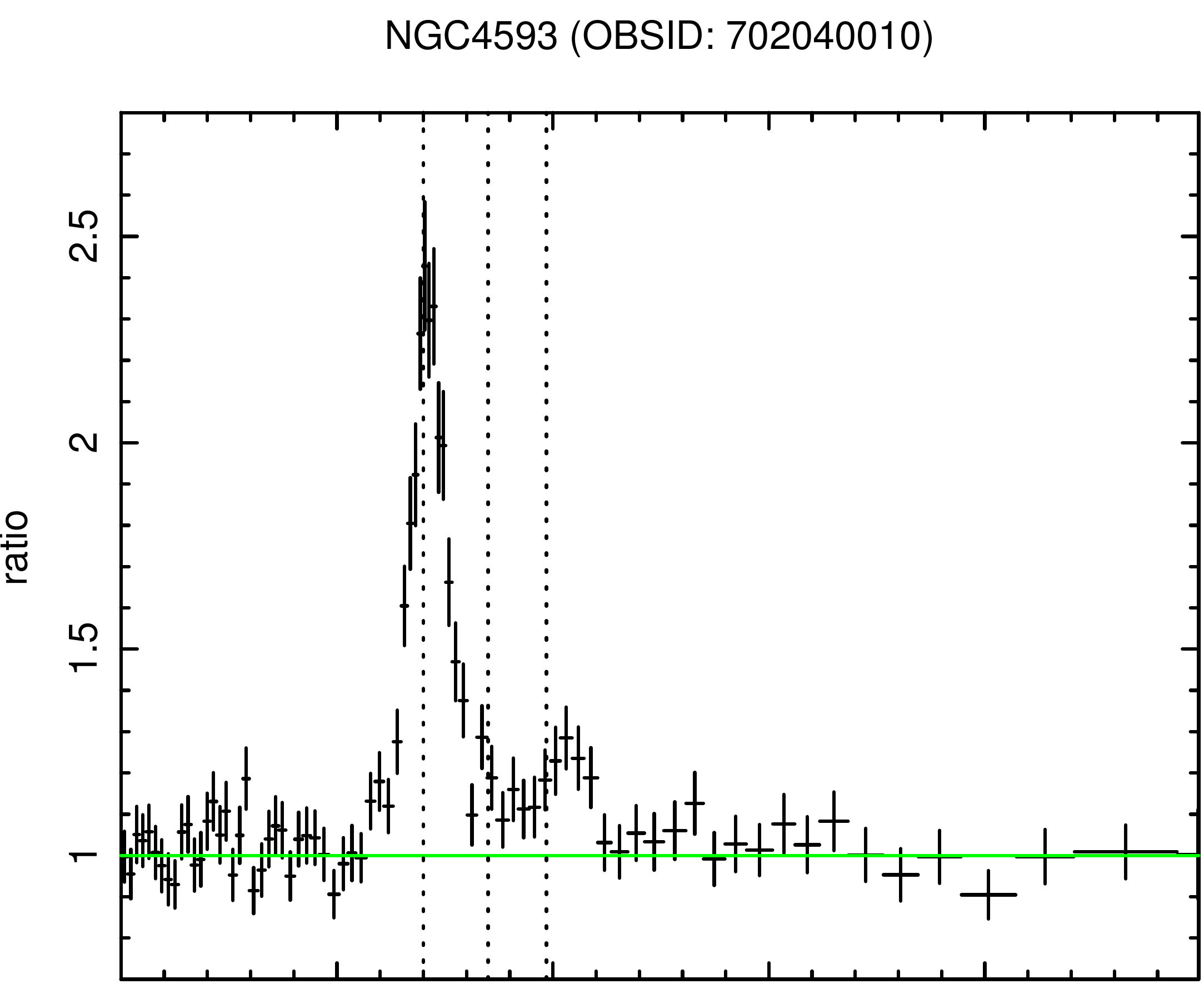}
}

\vspace{-12.2pt}
\subfloat{

\includegraphics[angle=-90,width=3.8cm]{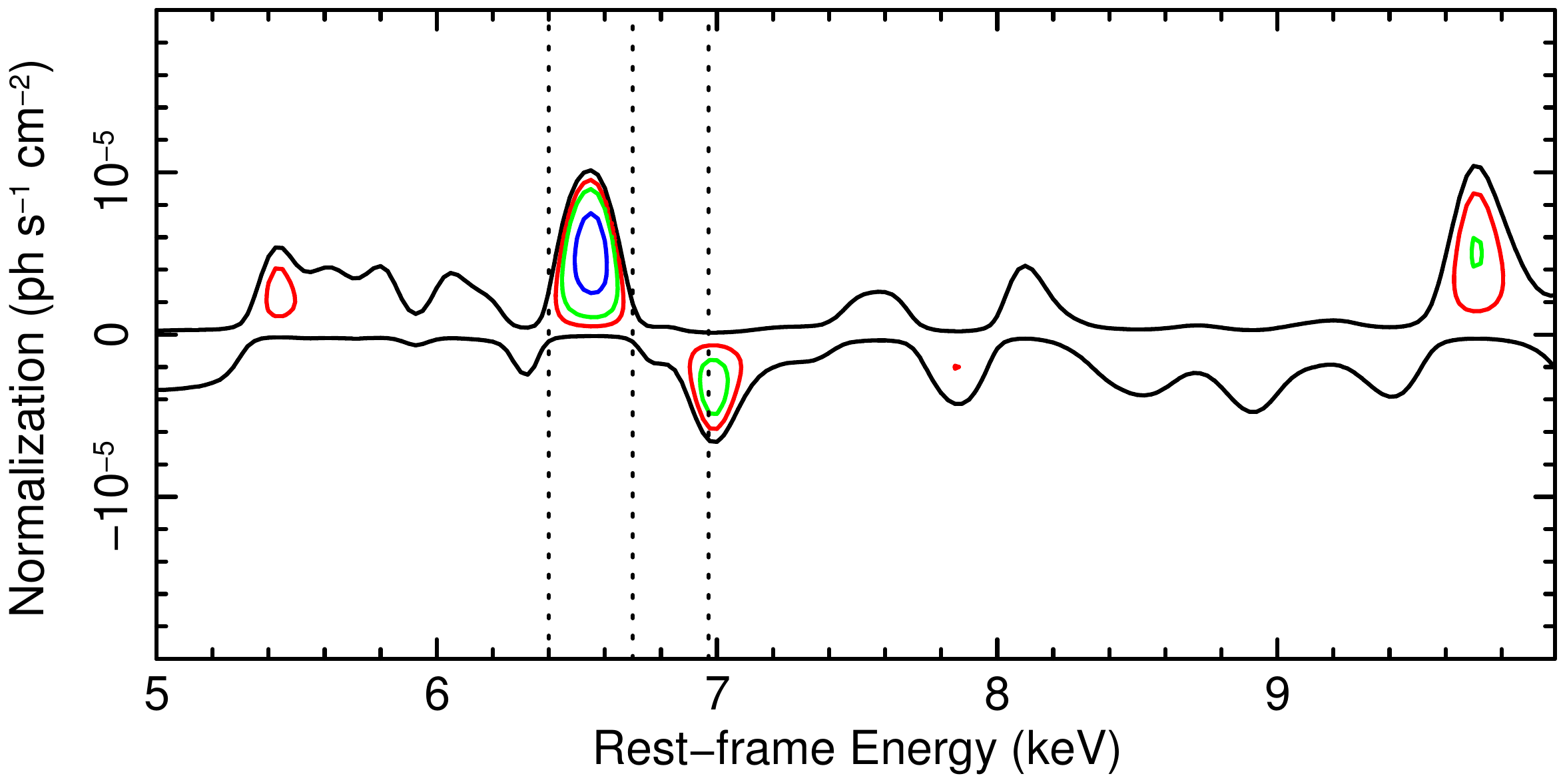}
\includegraphics[angle=-90,width=3.8cm]{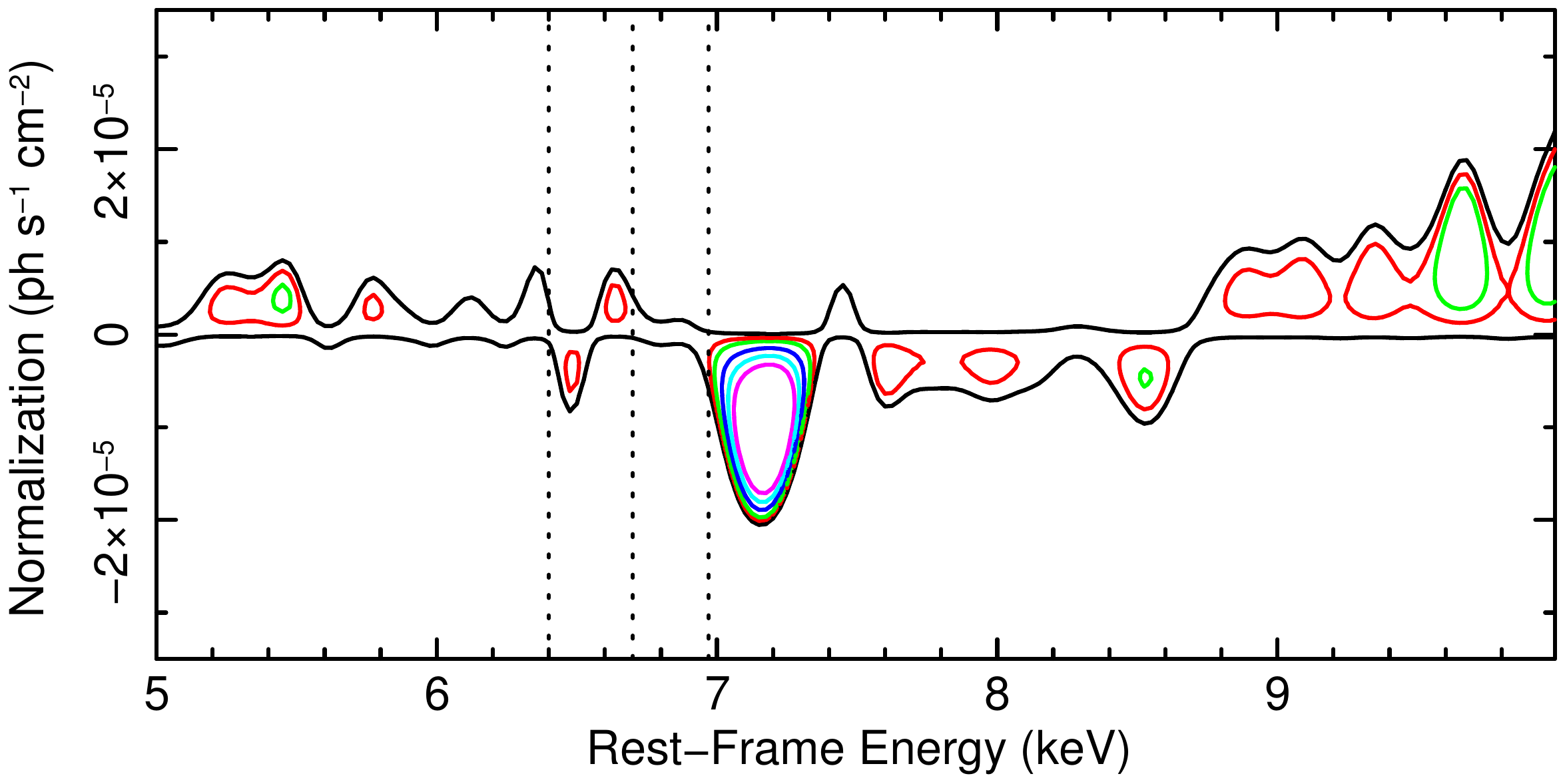}
\includegraphics[angle=-90,width=3.8cm]{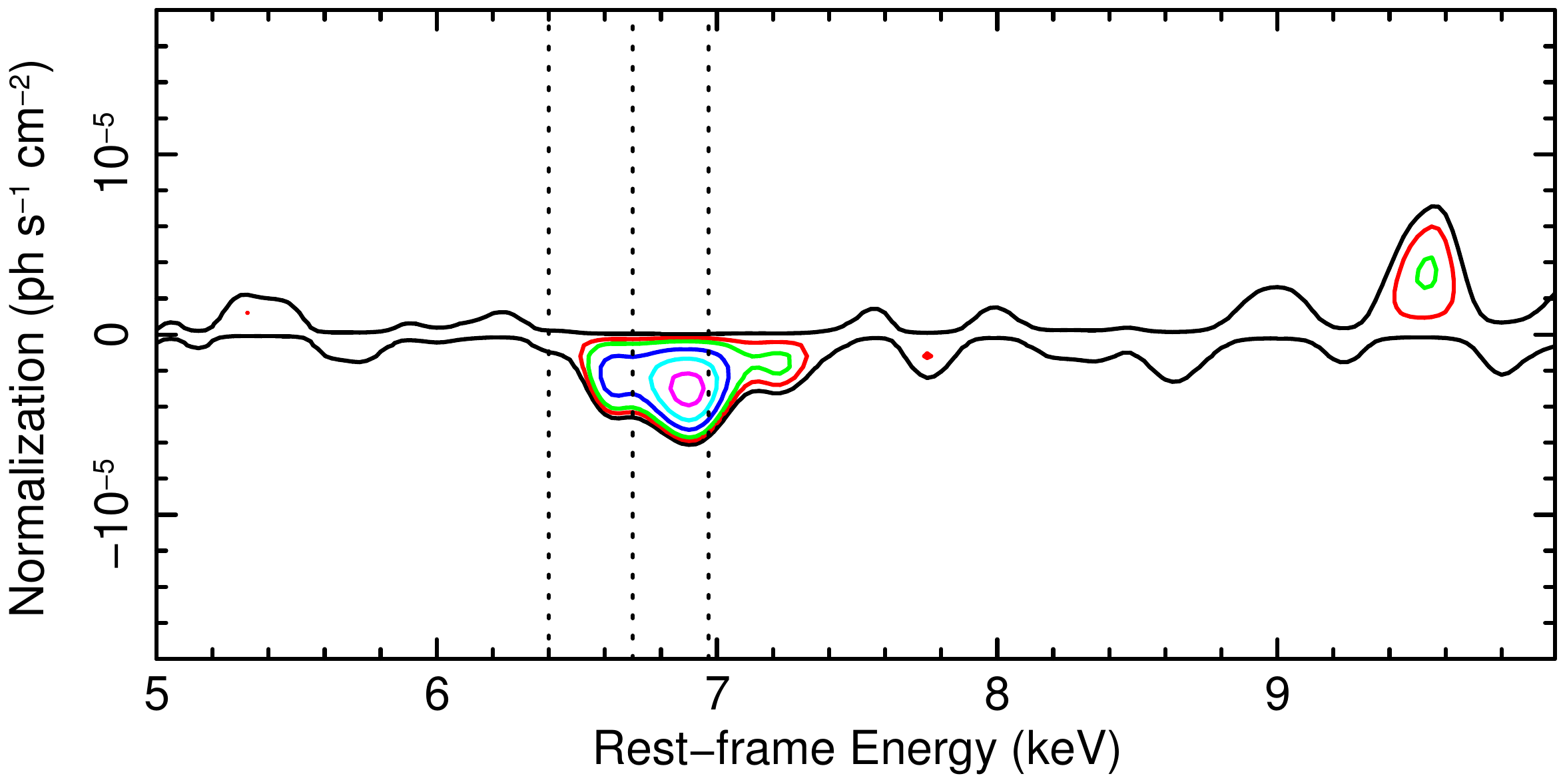}
\includegraphics[angle=-90,width=3.8cm]{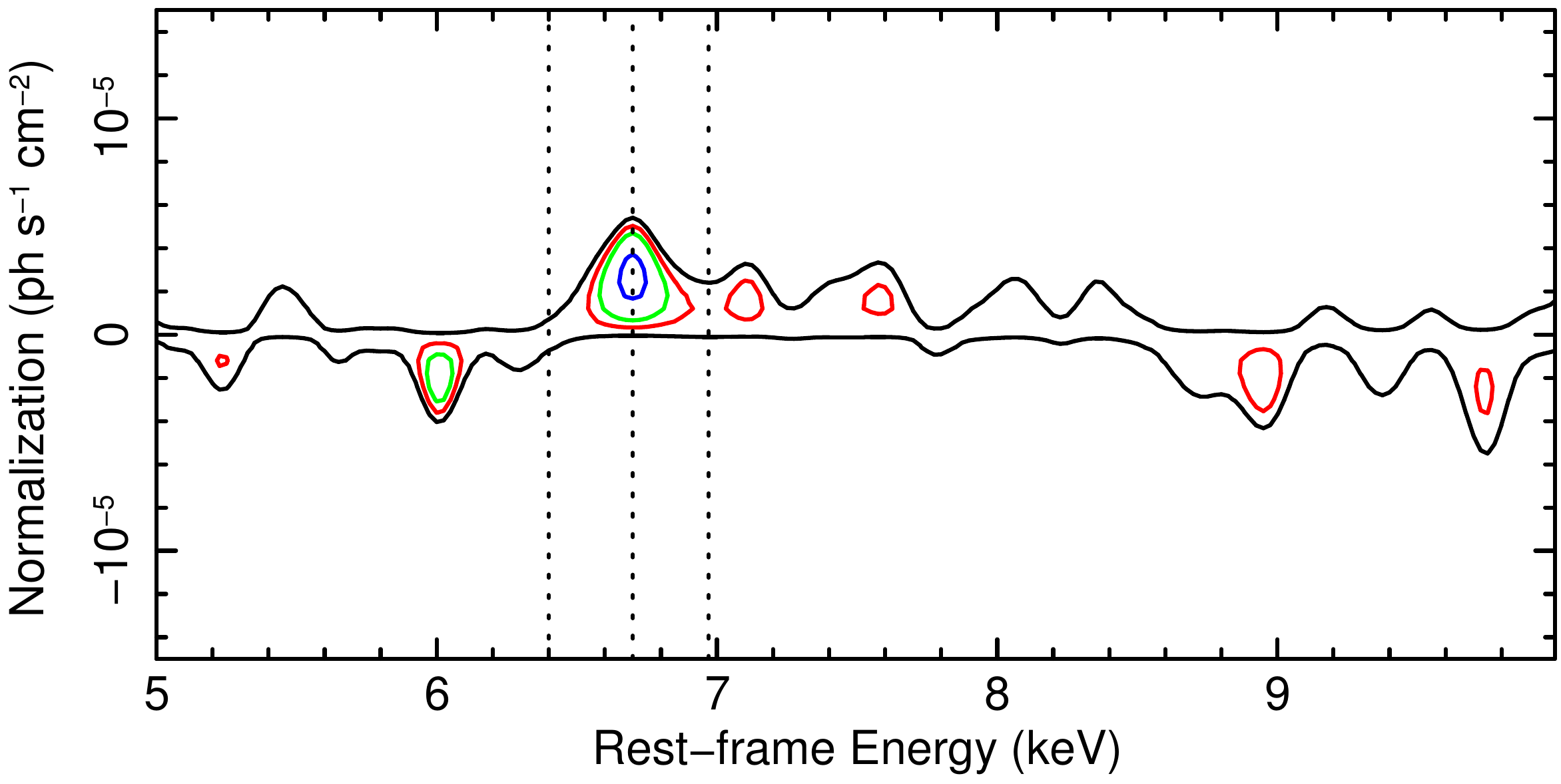}
}

\vspace{-5pt}	
\subfloat{

\includegraphics[angle=-90,width=3.8cm]{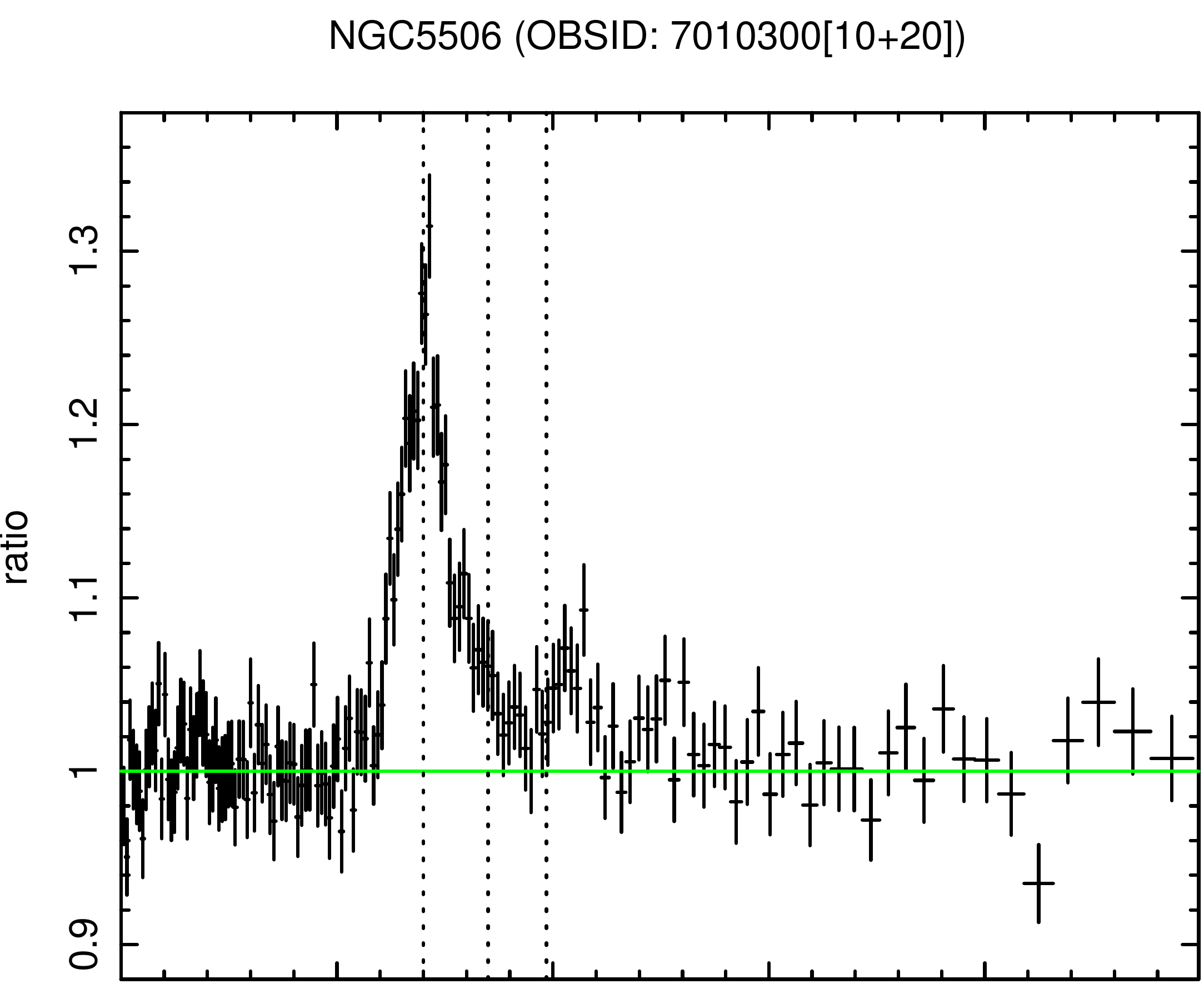}
\includegraphics[angle=-90,width=3.8cm]{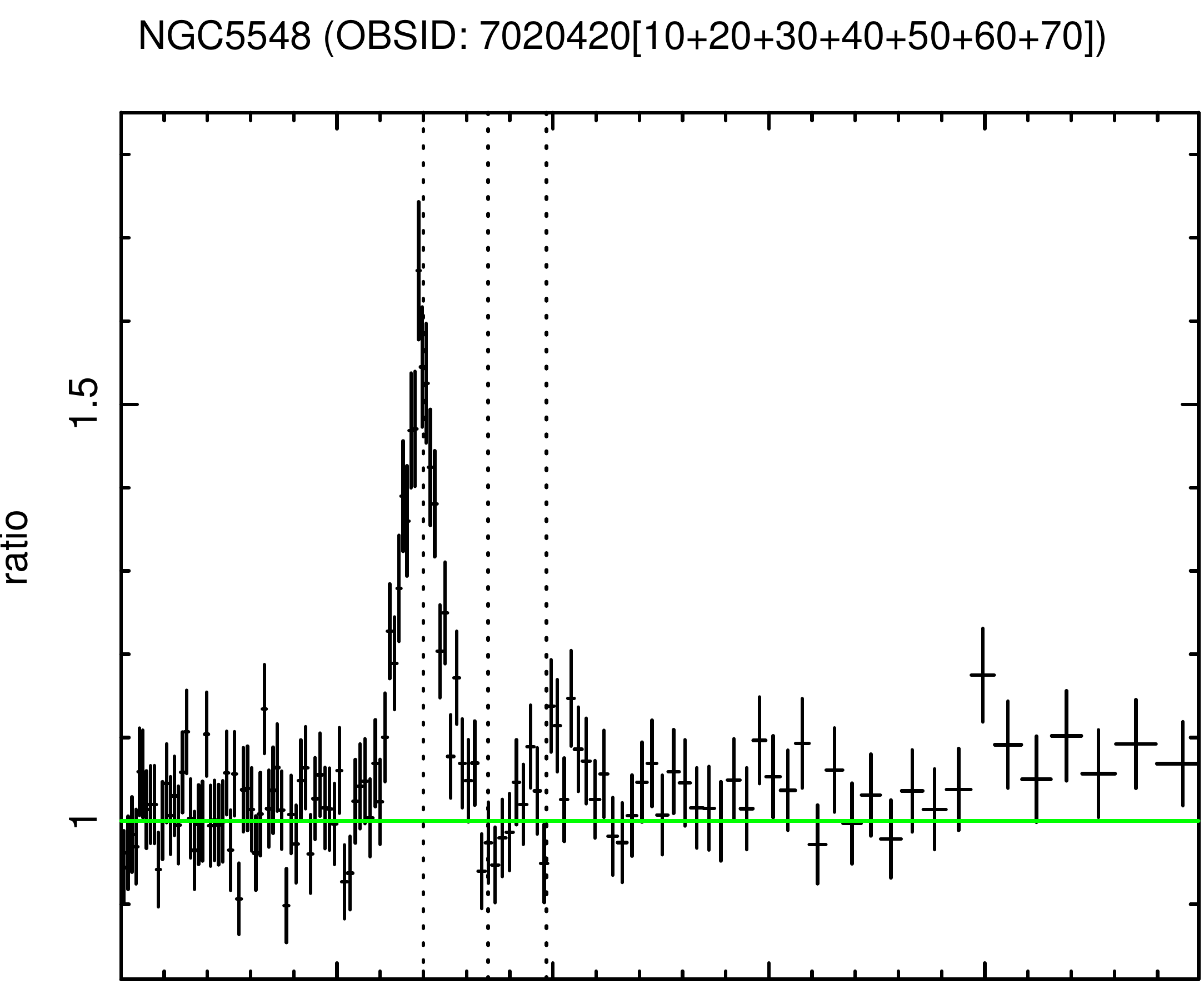}
\includegraphics[angle=-90,width=3.8cm]{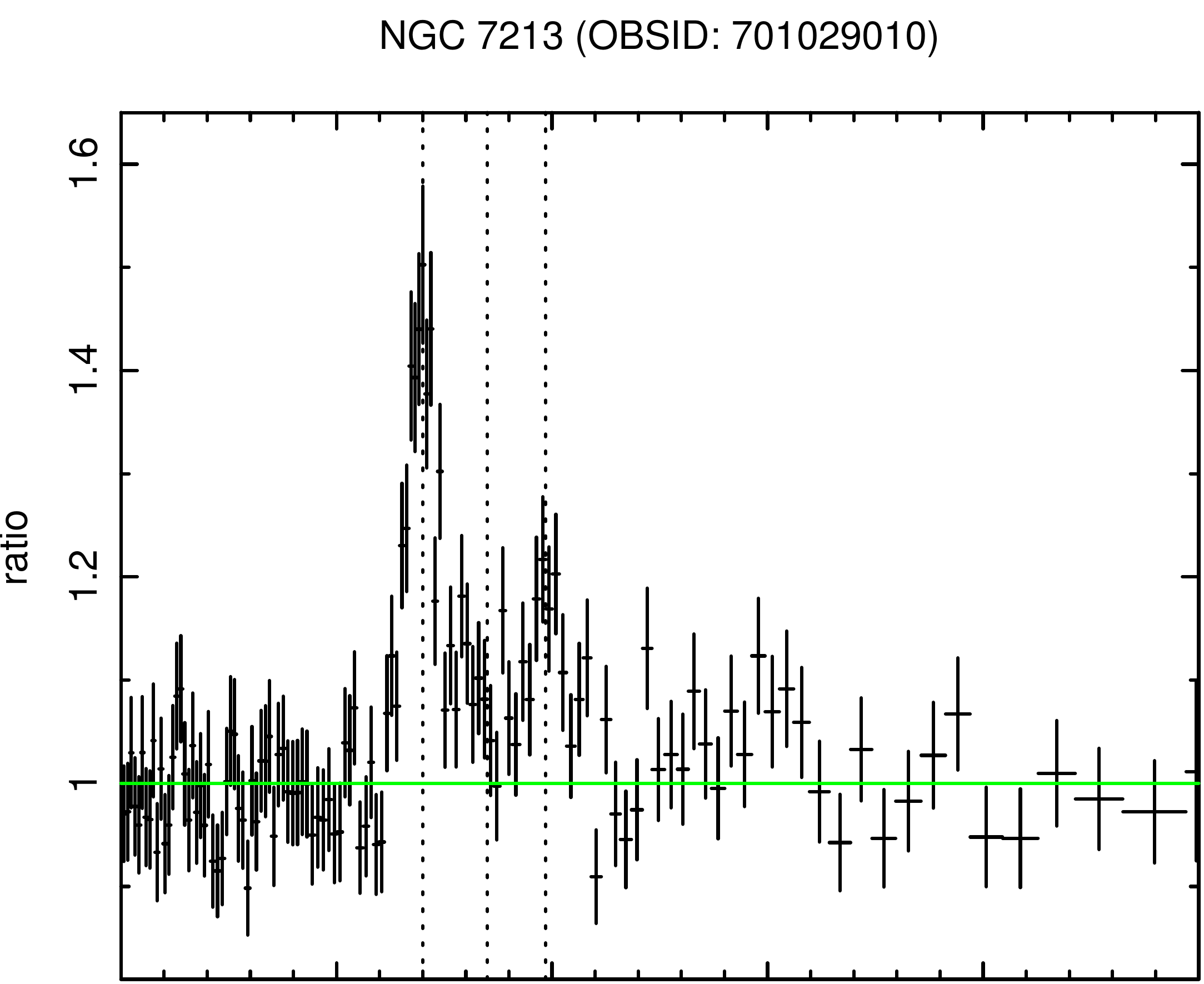}
\includegraphics[angle=-90,width=3.8cm]{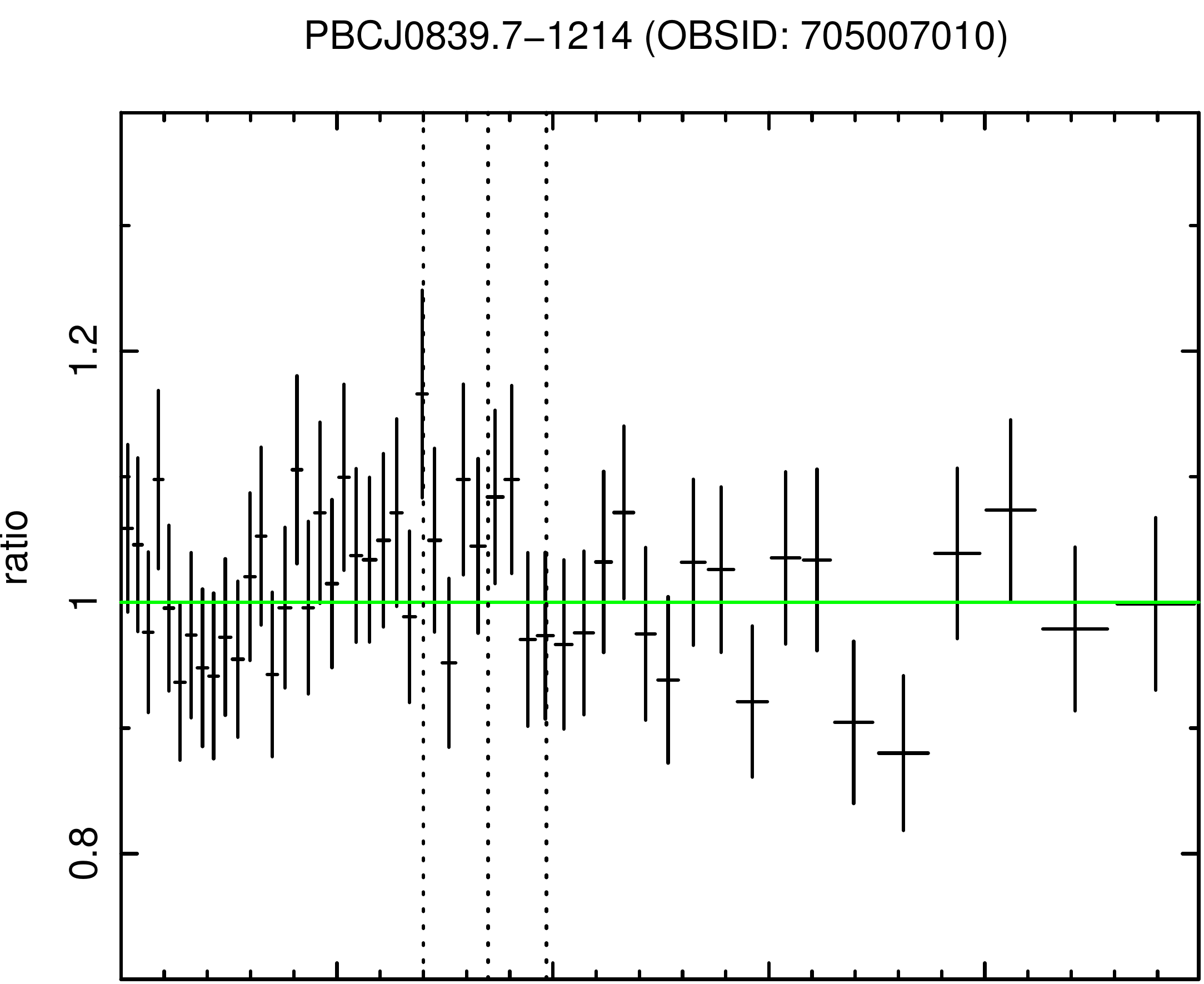}

}

\vspace{-12.2pt}
\subfloat{

\includegraphics[angle=-90,width=3.8cm]{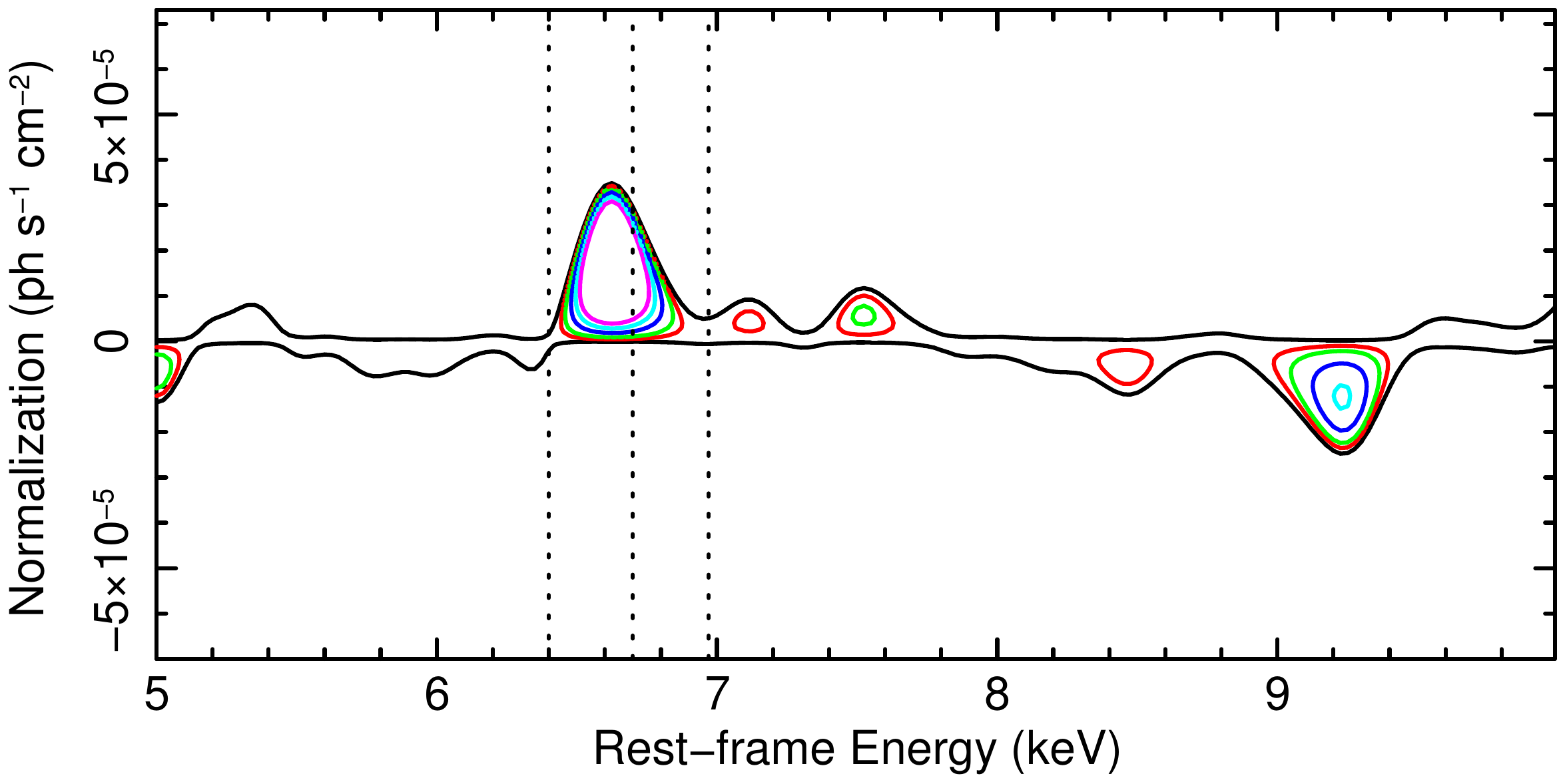}
\includegraphics[angle=-90,width=3.8cm]{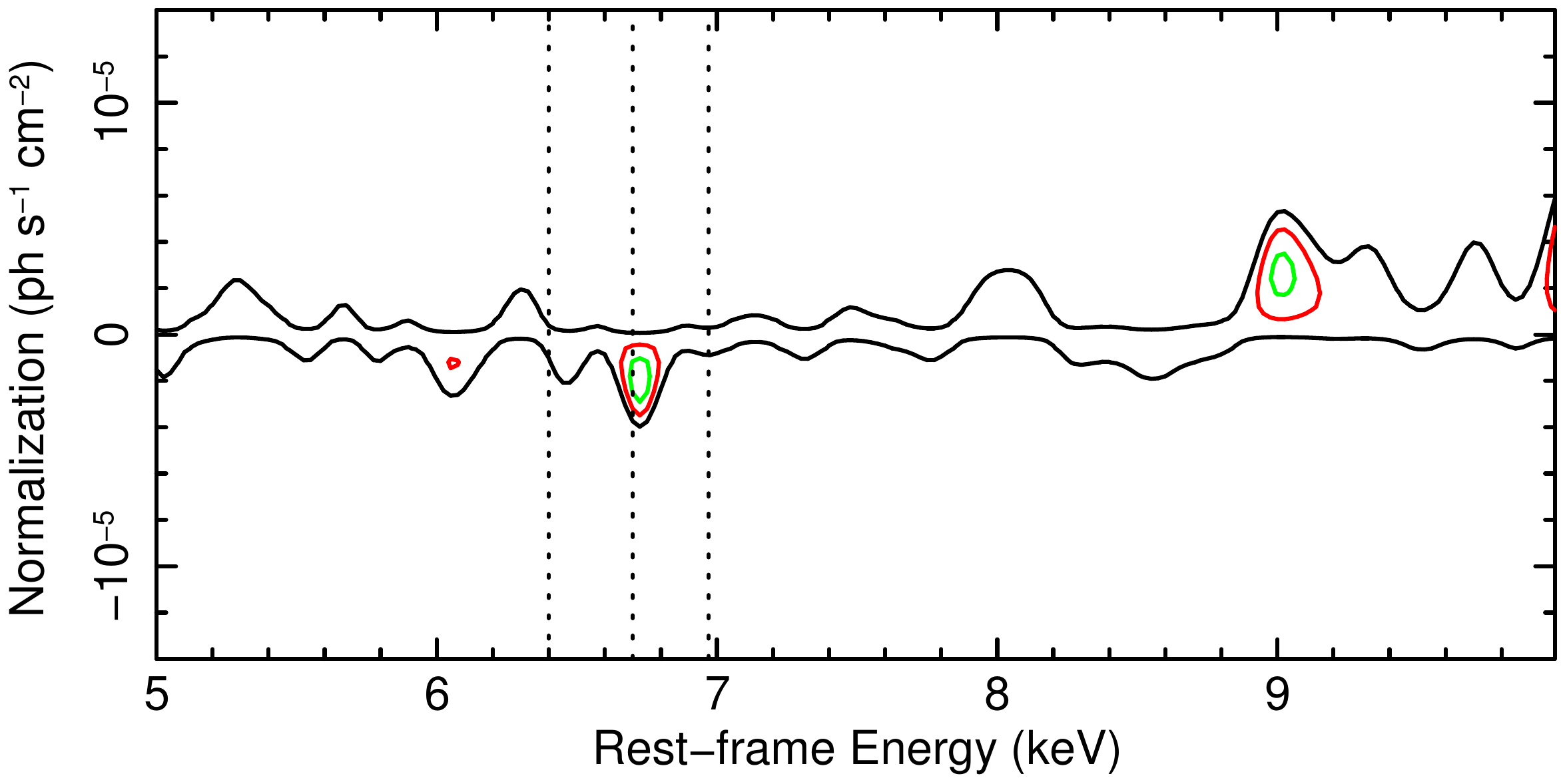}
\includegraphics[angle=-90,width=3.8cm]{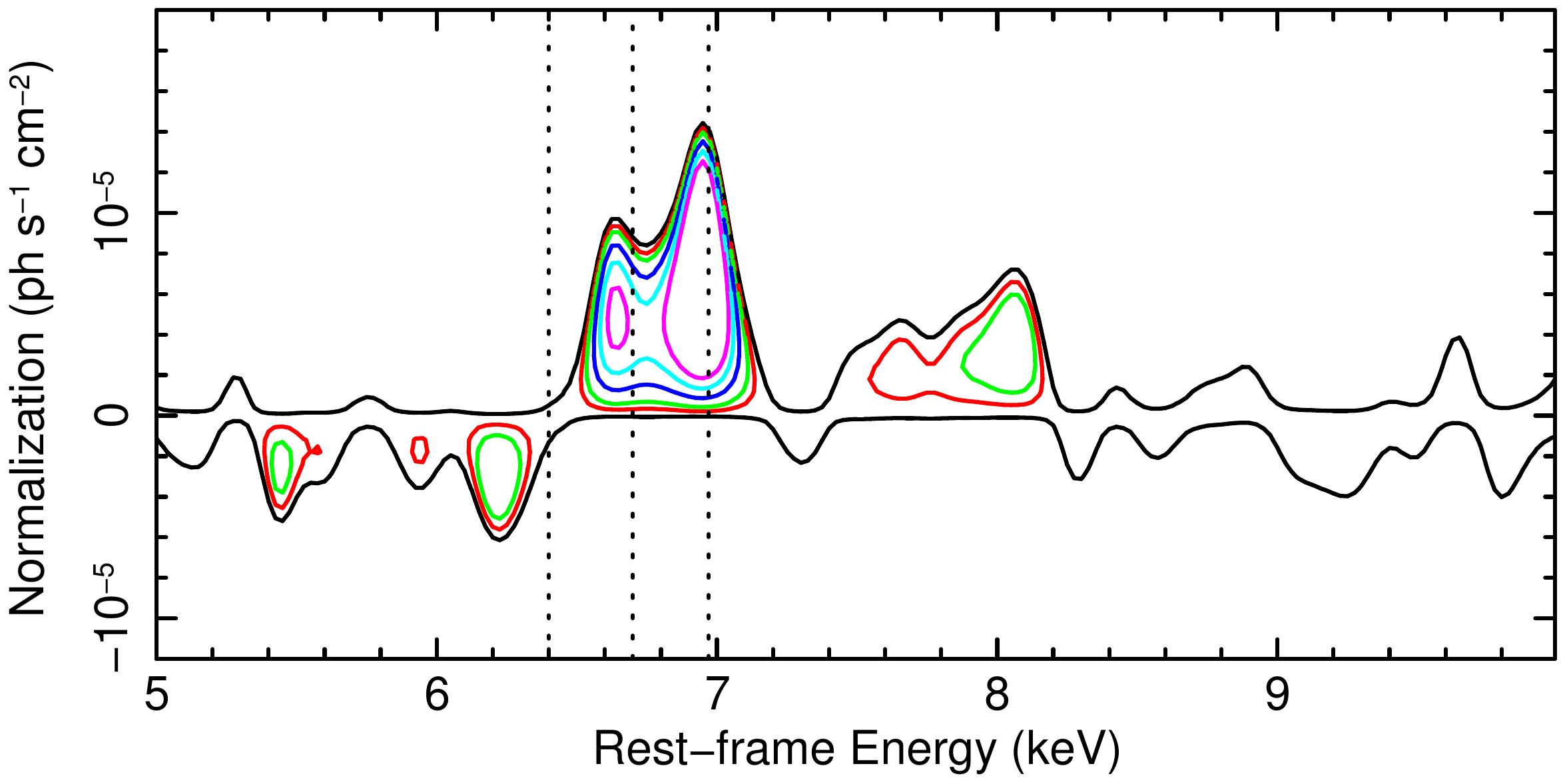}
\includegraphics[angle=-90,width=3.8cm]{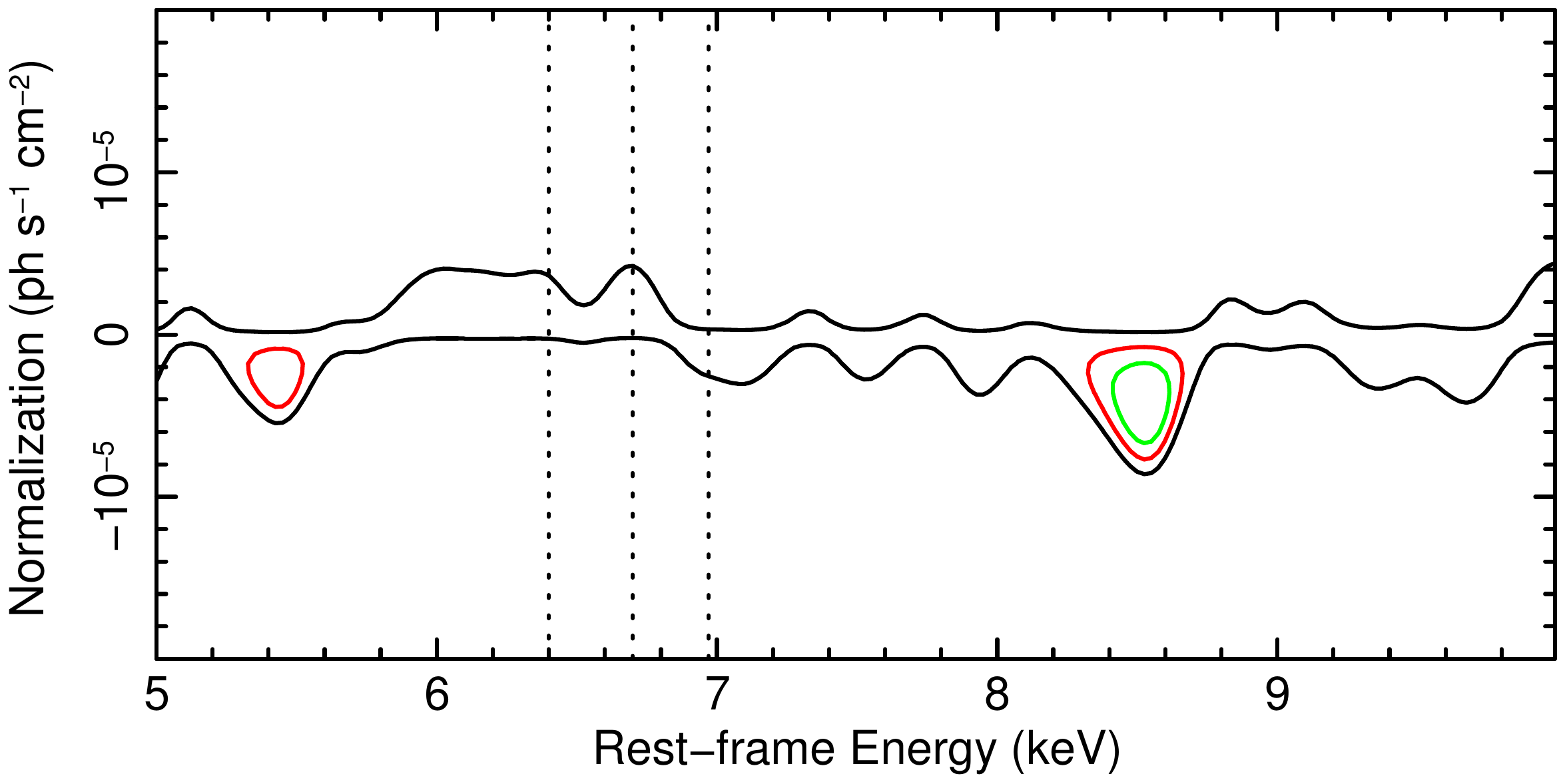}
	
}	

\end{center}
\contcaption{\small -- Ratio and contour plots for sources which do not require a broadened component.}
\end{figure*}

\clearpage

\begin{figure*}
\begin{center}

\vspace{-5pt}	
\subfloat{

\includegraphics[angle=-90,width=3.8cm]{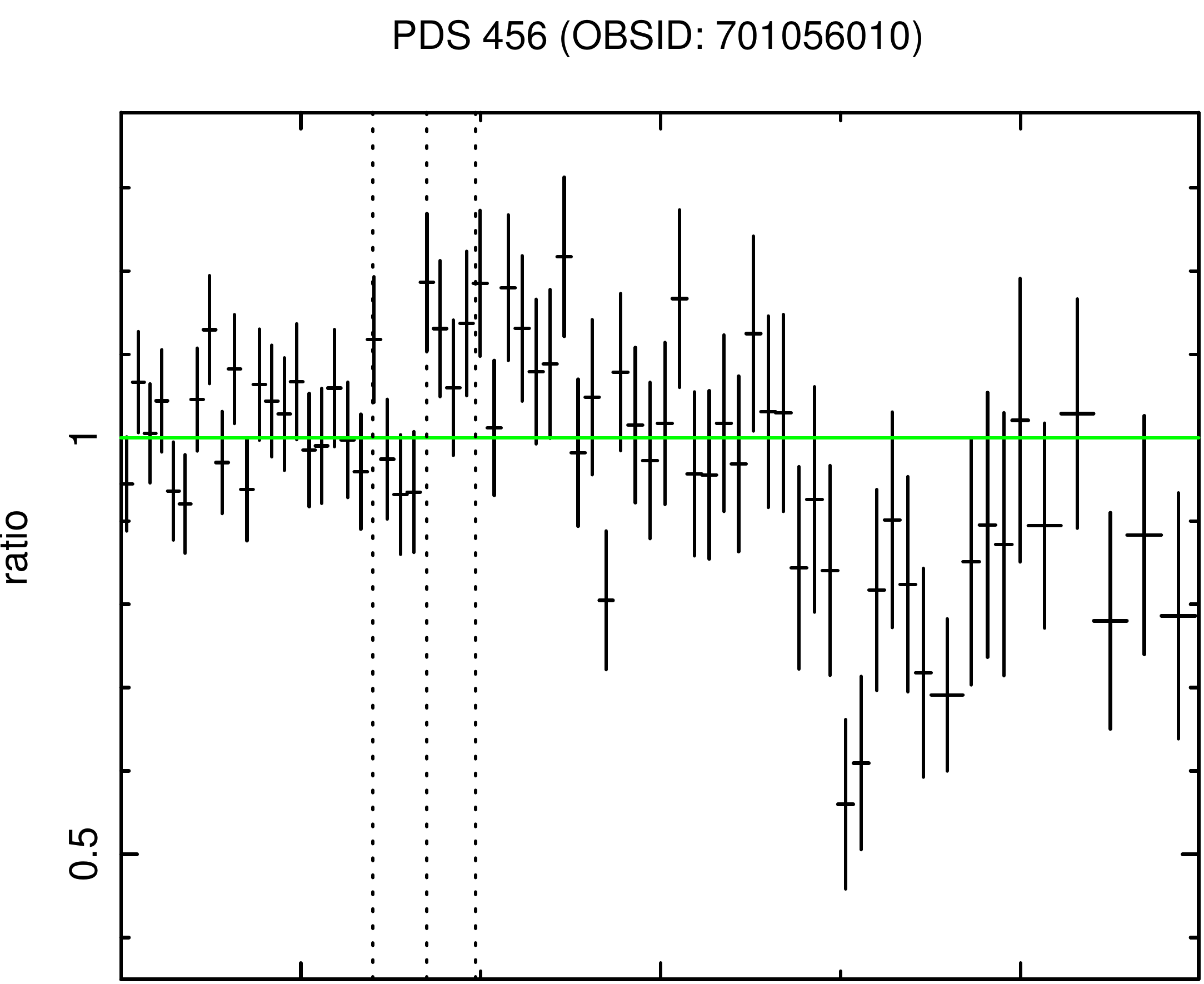}
\includegraphics[angle=-90,width=3.8cm]{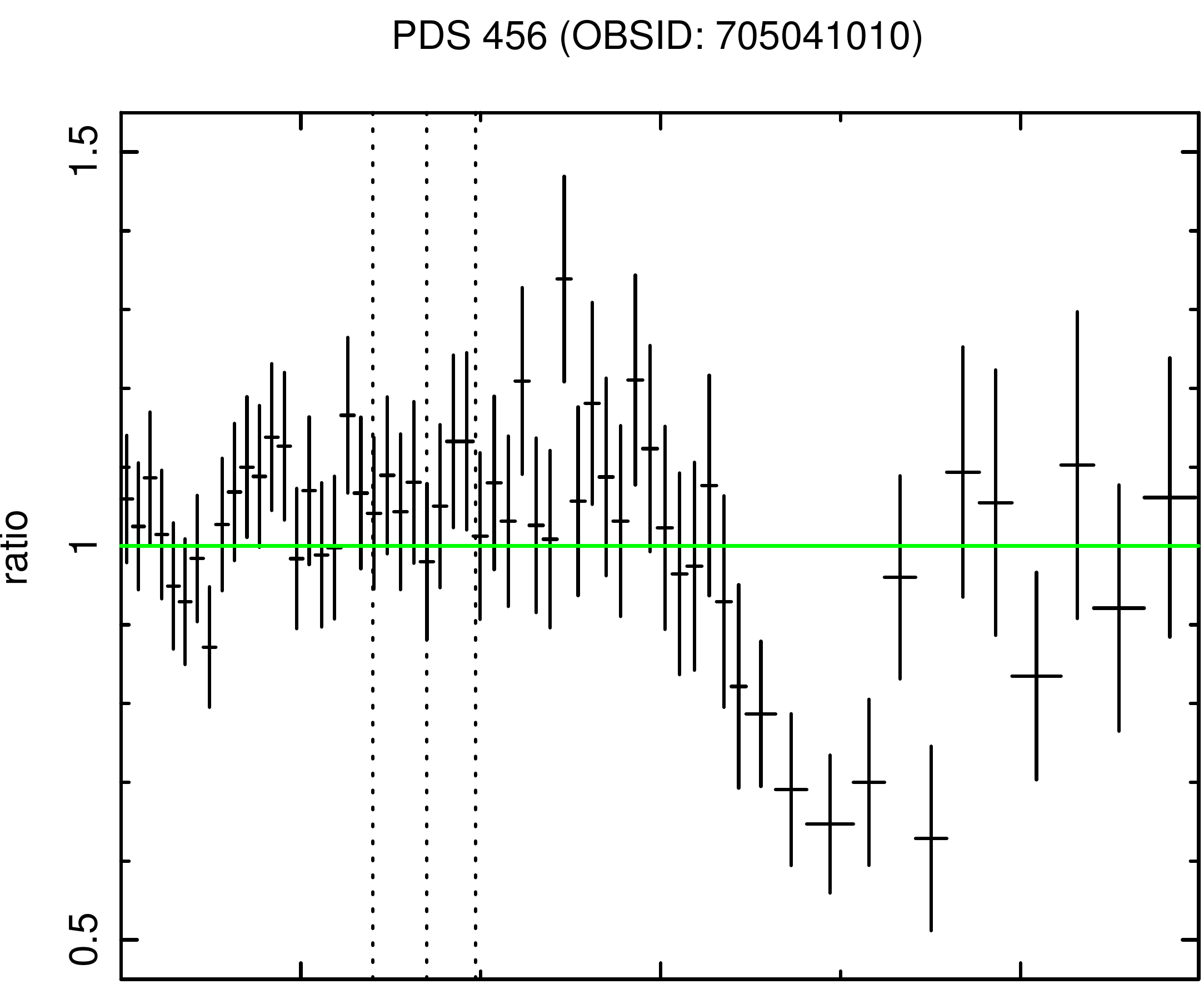}
\includegraphics[angle=-90,width=3.8cm]{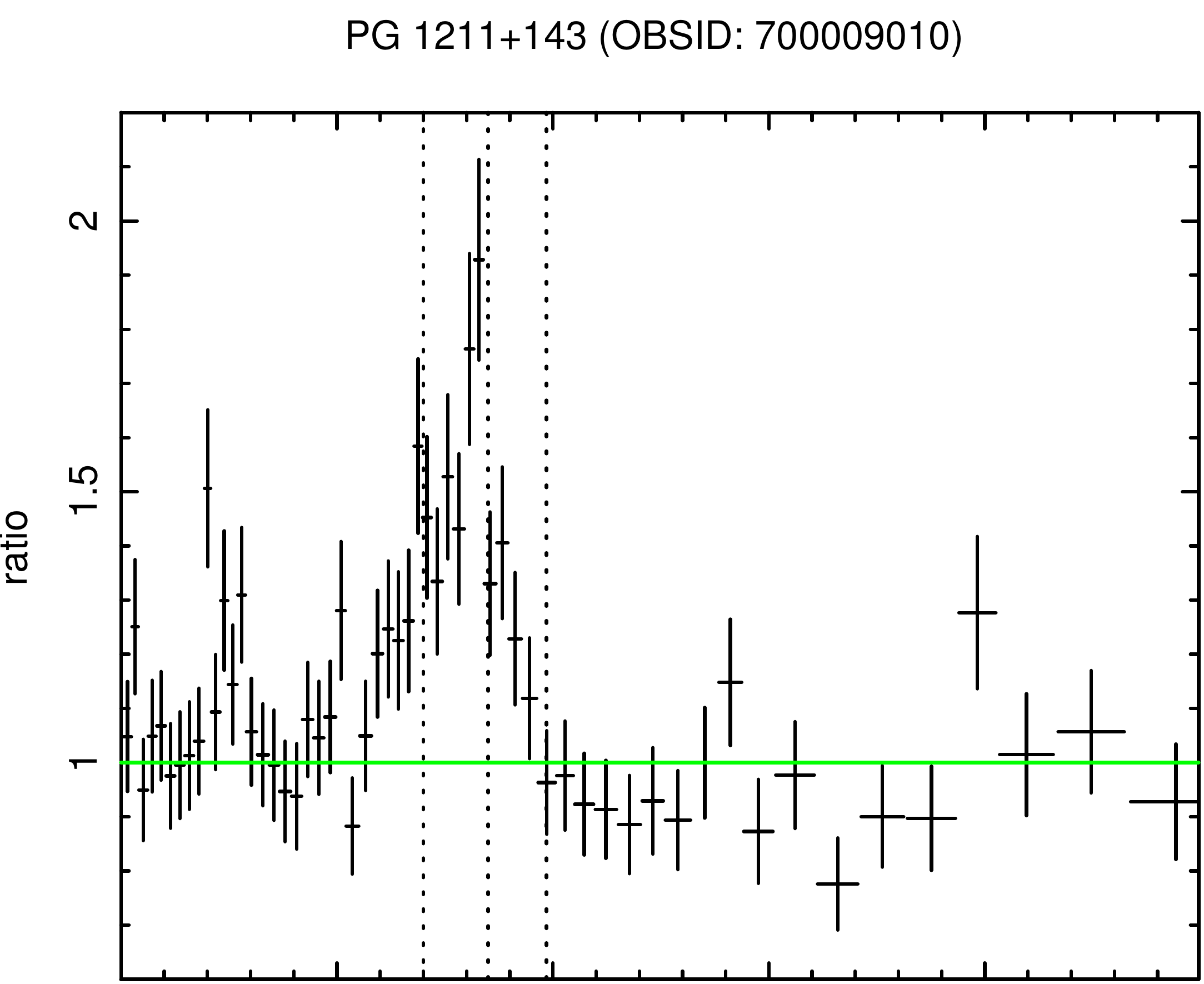}
\includegraphics[angle=-90,width=3.8cm]{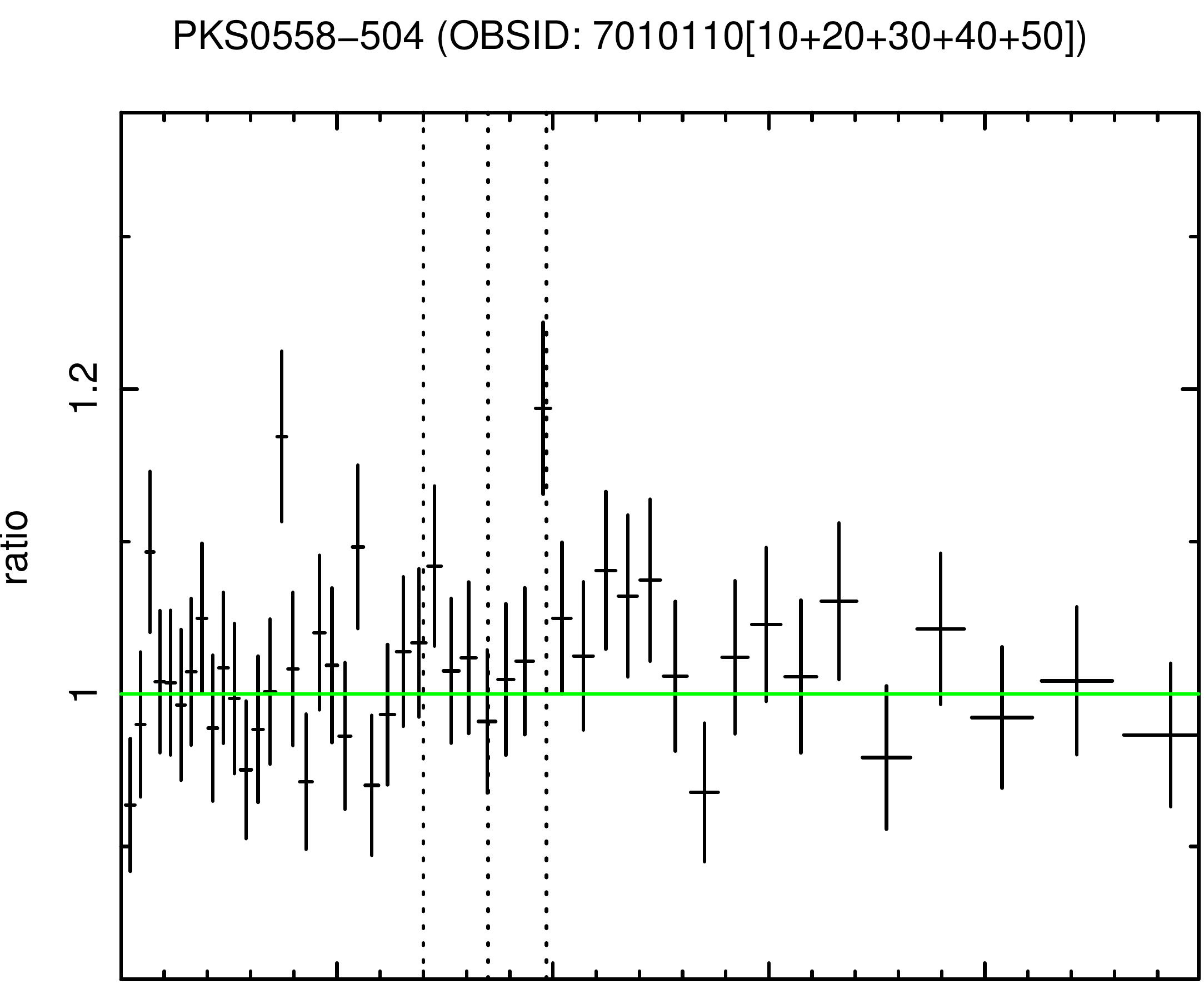}
}

\vspace{-12.2pt}
\subfloat{
\includegraphics[angle=-90,width=3.8cm]{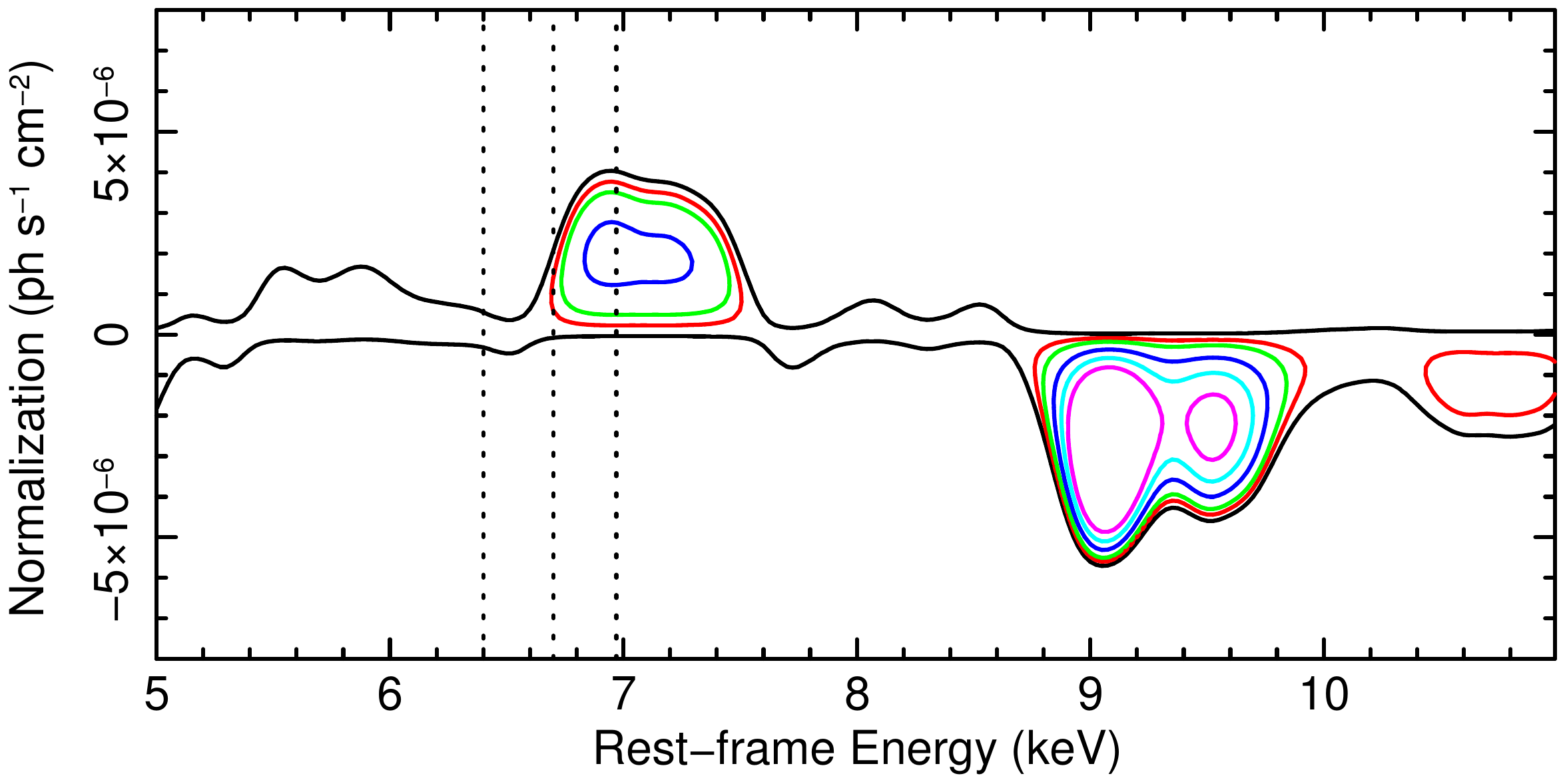}
\includegraphics[angle=-90,width=3.8cm]{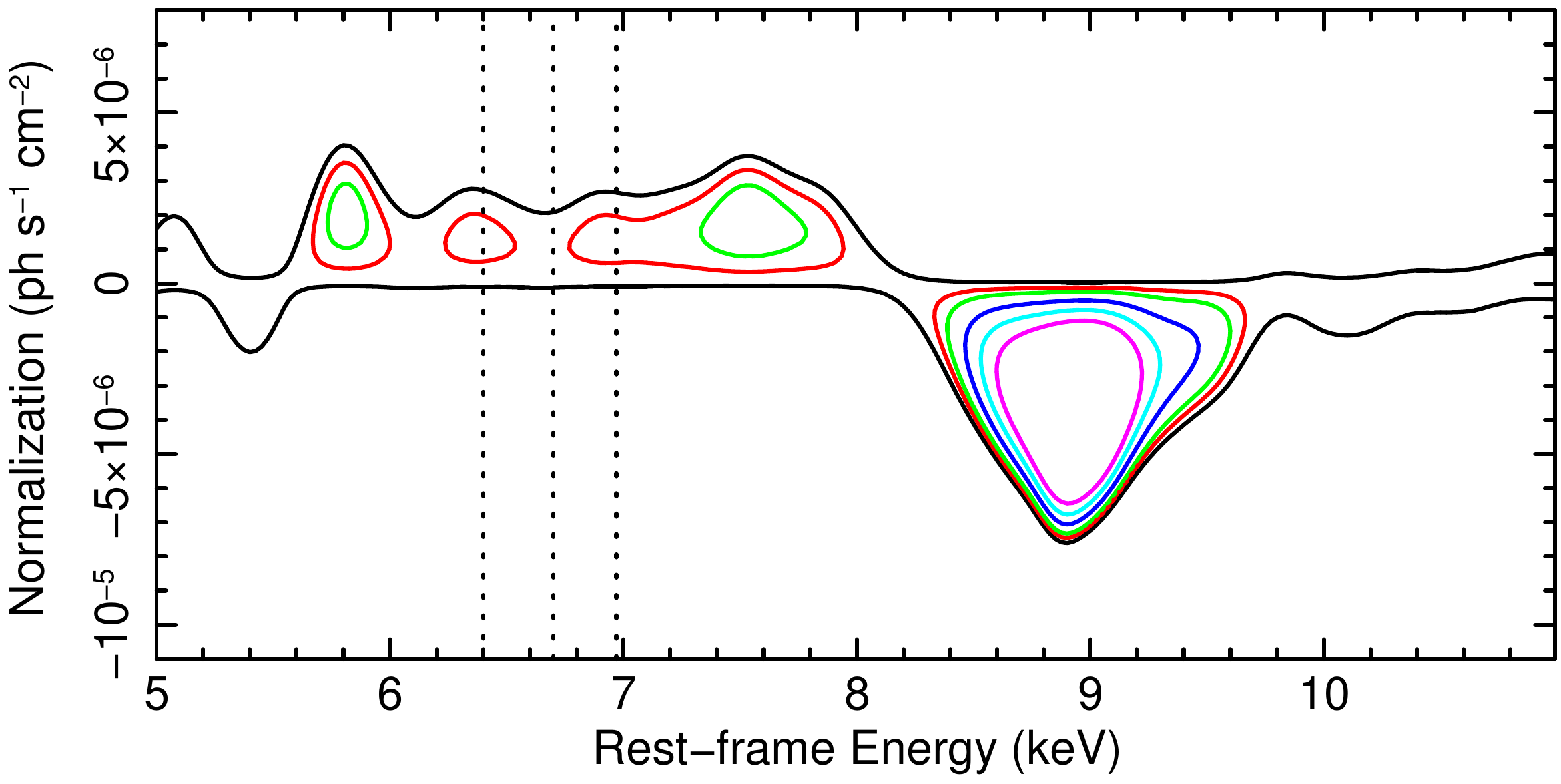}
\includegraphics[angle=-90,width=3.8cm]{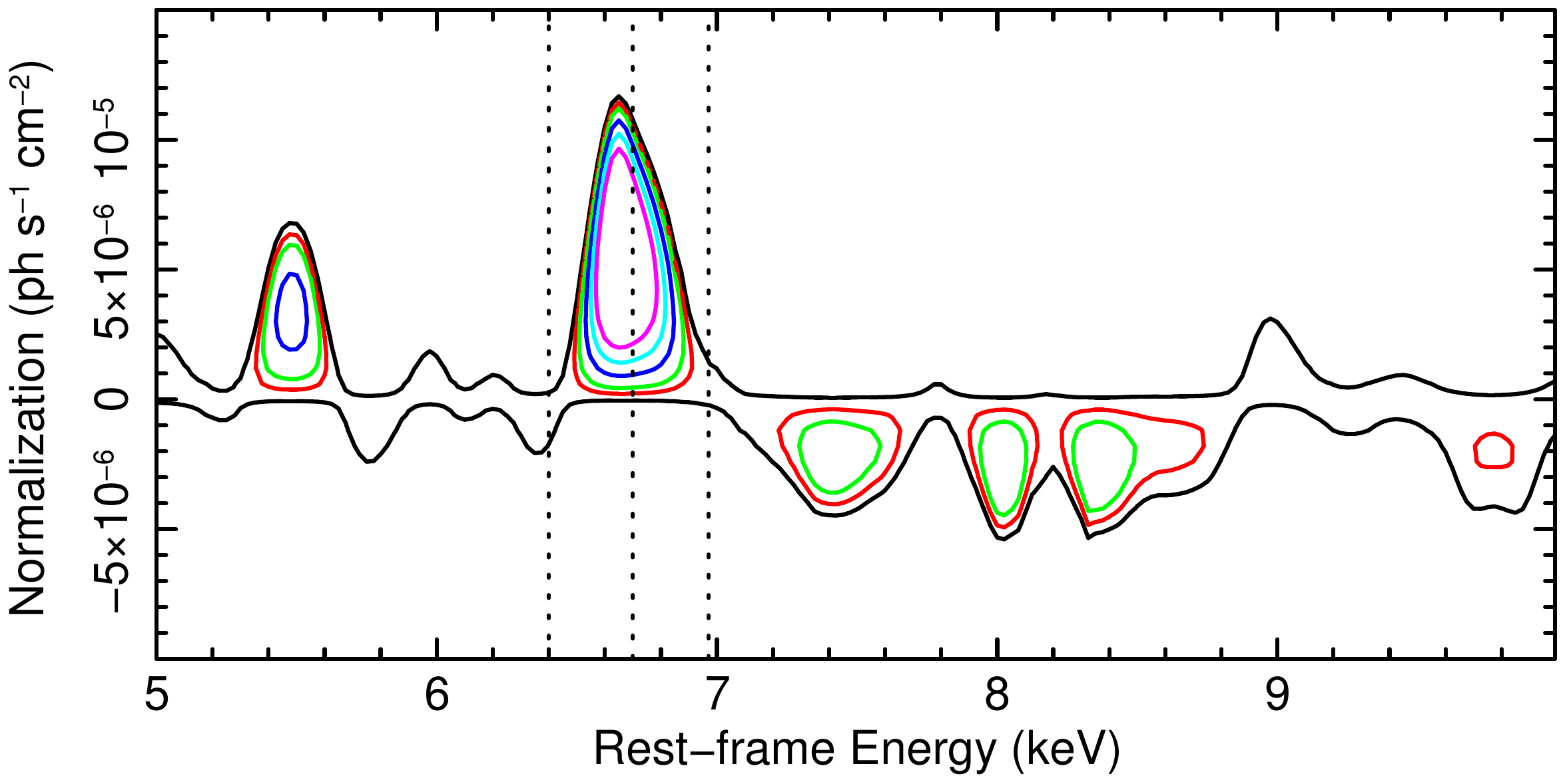}	
\includegraphics[angle=-90,width=3.8cm]{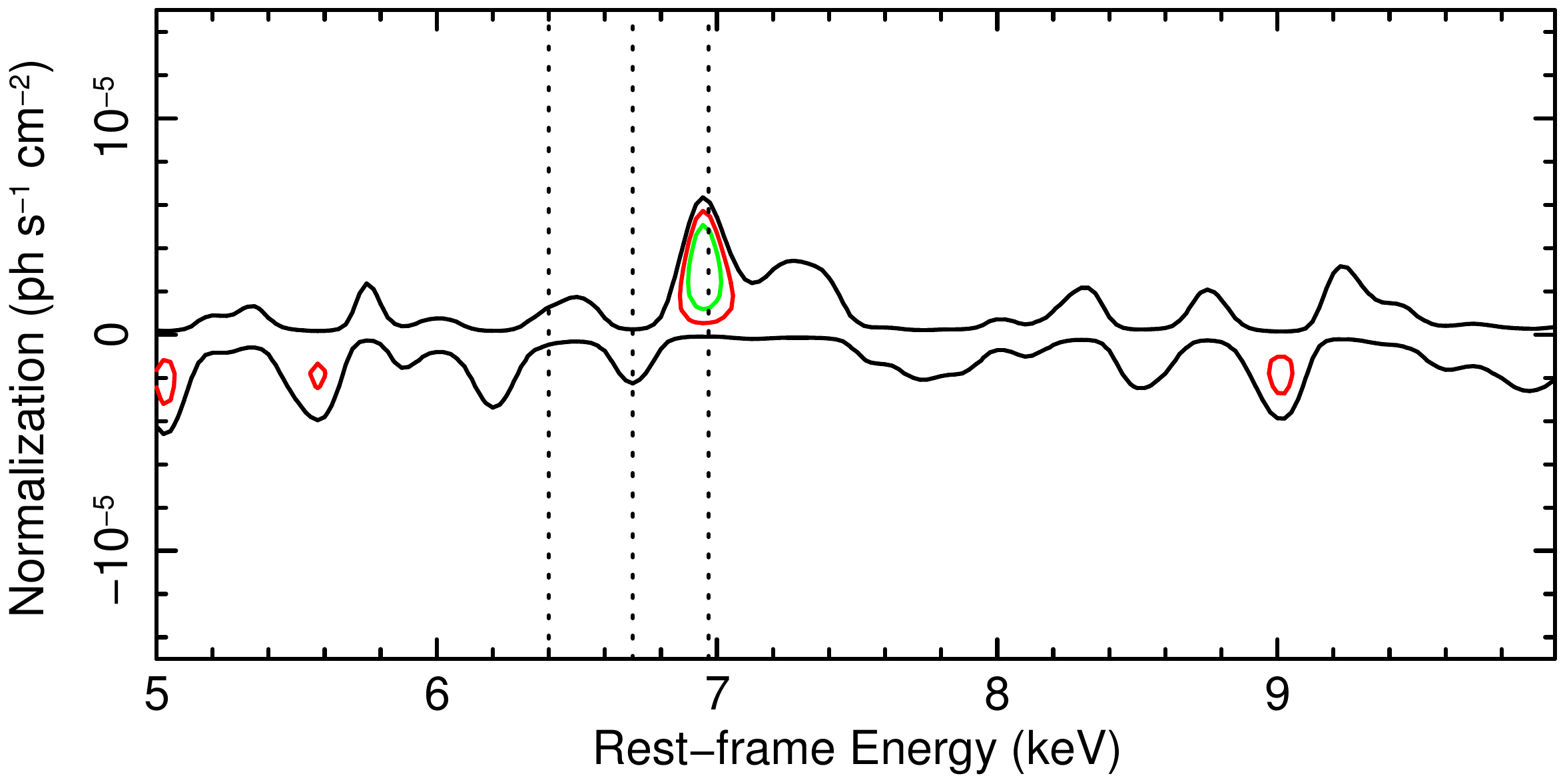}
}

\vspace{-5pt}	
\subfloat{
\includegraphics[angle=-90,width=3.8cm]{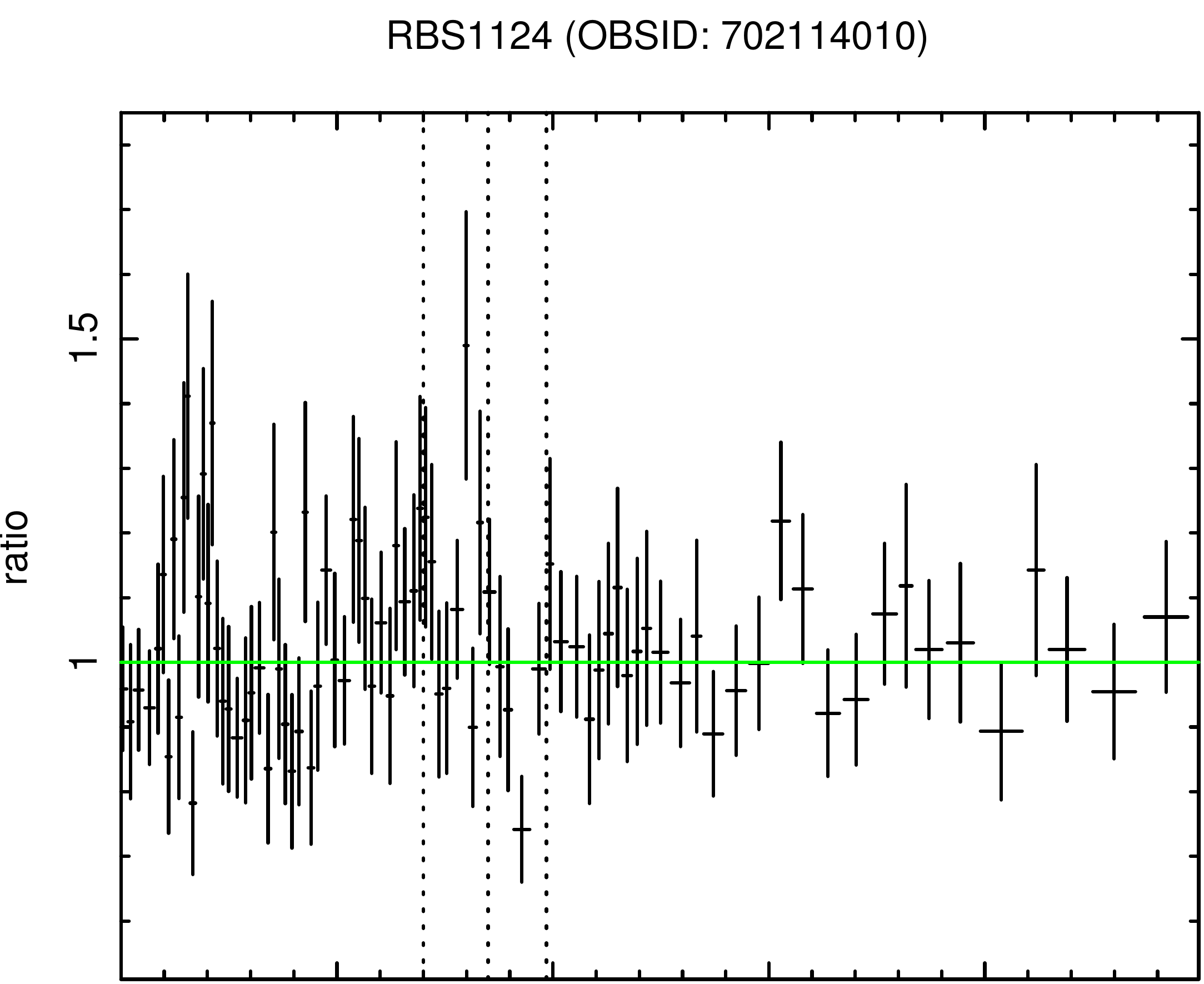}
\includegraphics[angle=-90,width=3.8cm]{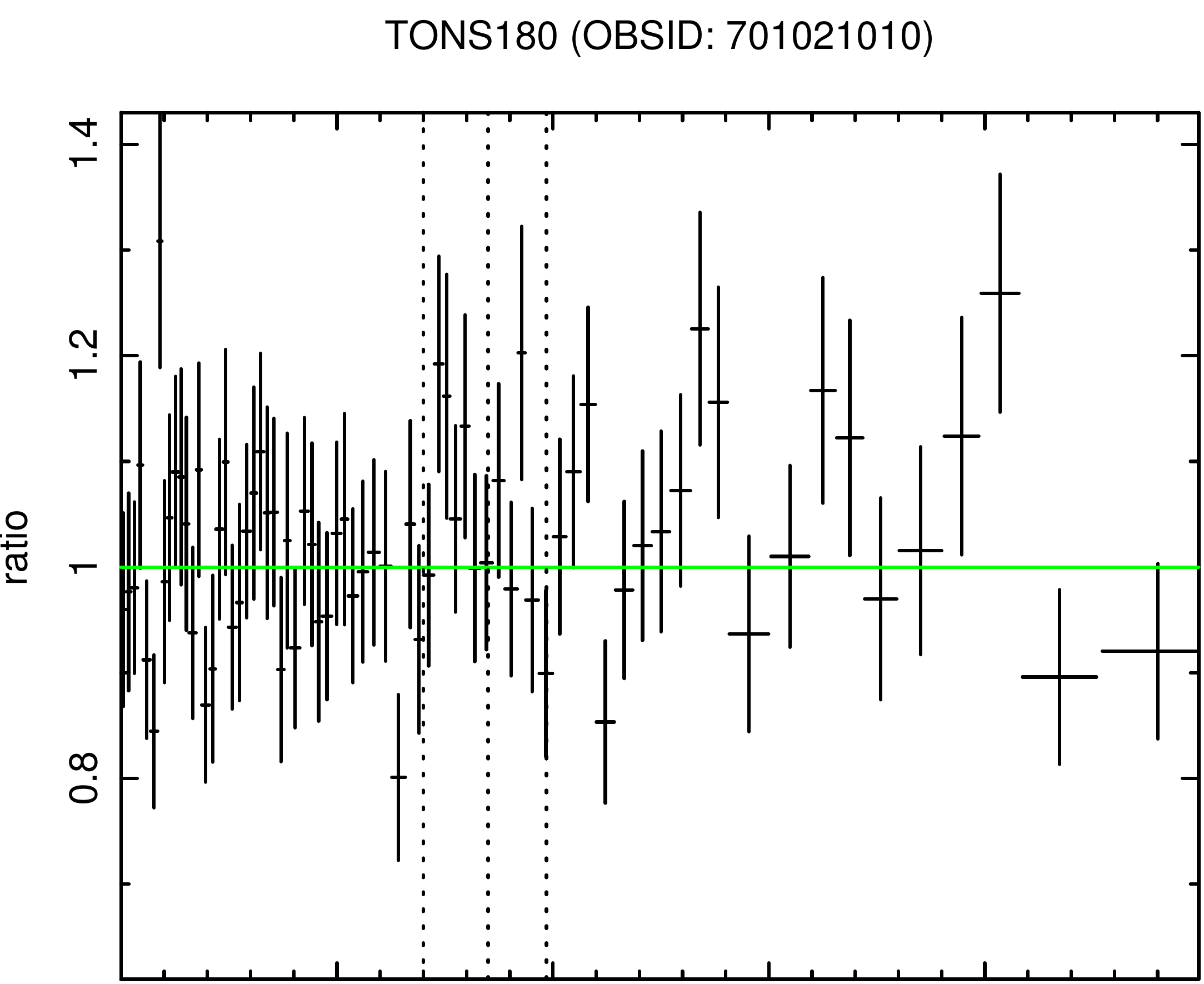}
}

\vspace{-12.2pt}
\subfloat{
\includegraphics[angle=-90,width=3.8cm]{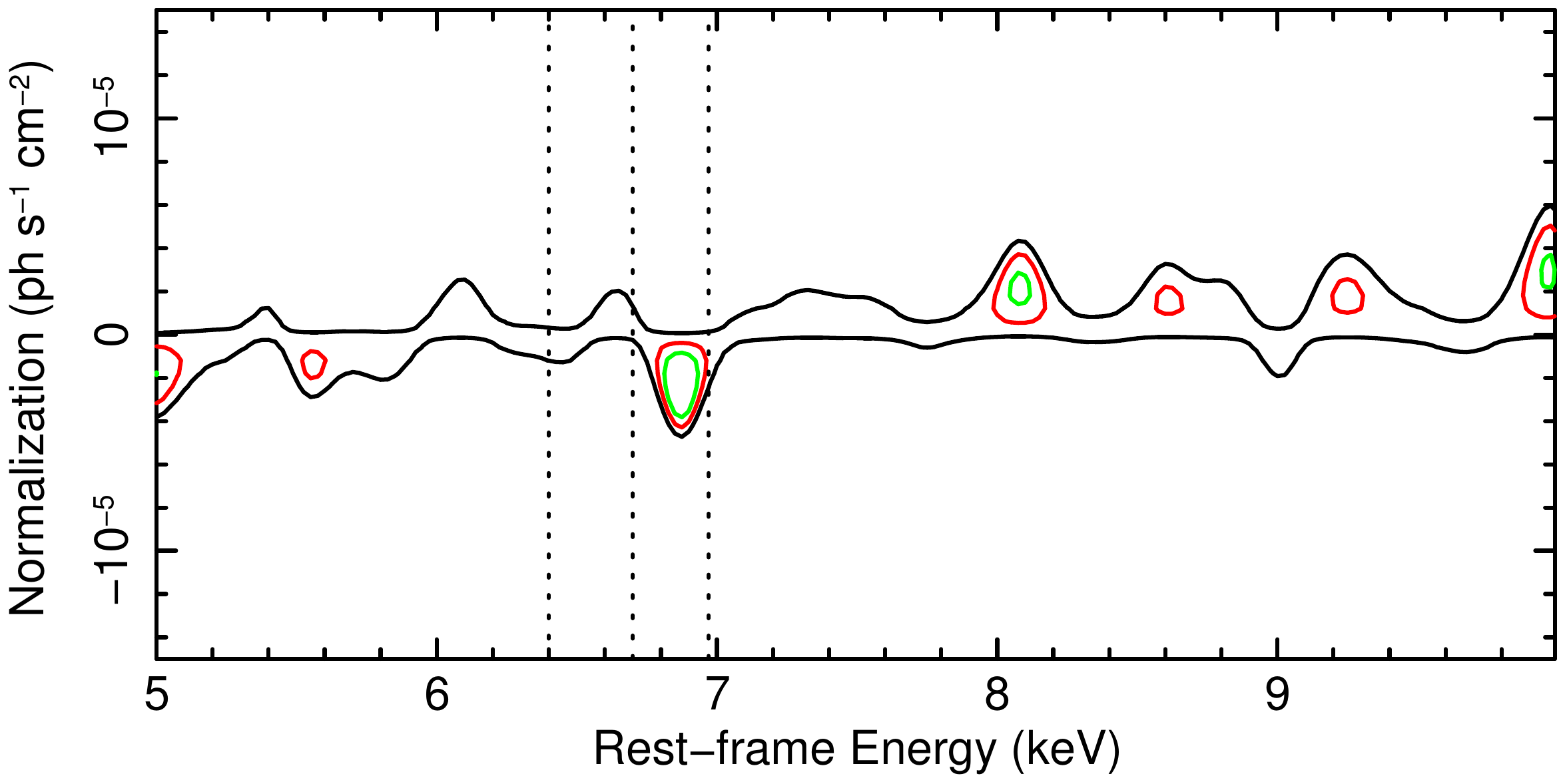}
\includegraphics[angle=-90,width=3.8cm]{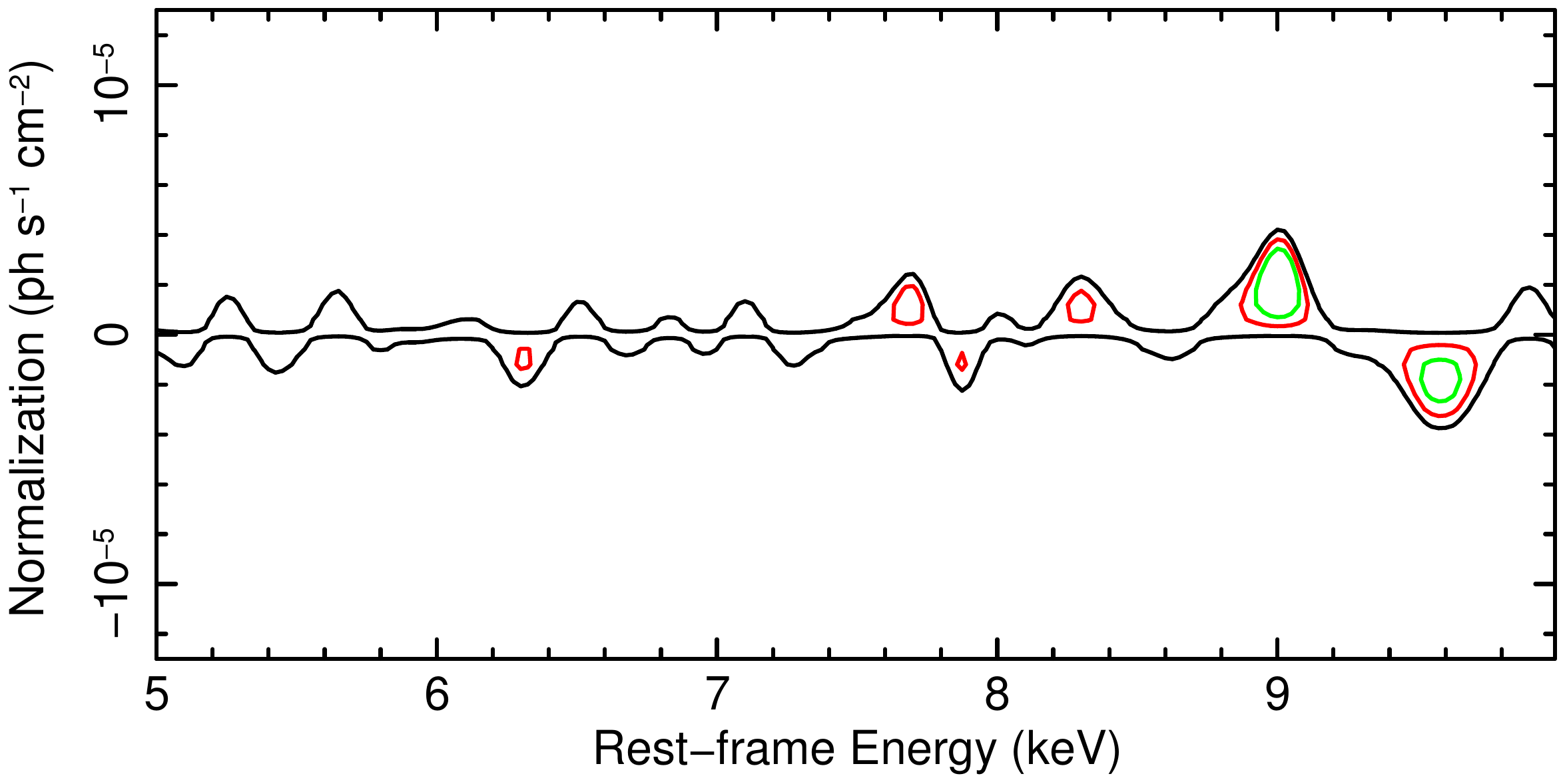}
}

\end{center}
\contcaption{\small -- Ratio and contour plots for sources which do not require a broadened component.}
\end{figure*}

\clearpage

\begin{figure*}
\begin{center}

\vspace{-5pt}
\subfloat{
\includegraphics[angle=-90,width=3.8cm]{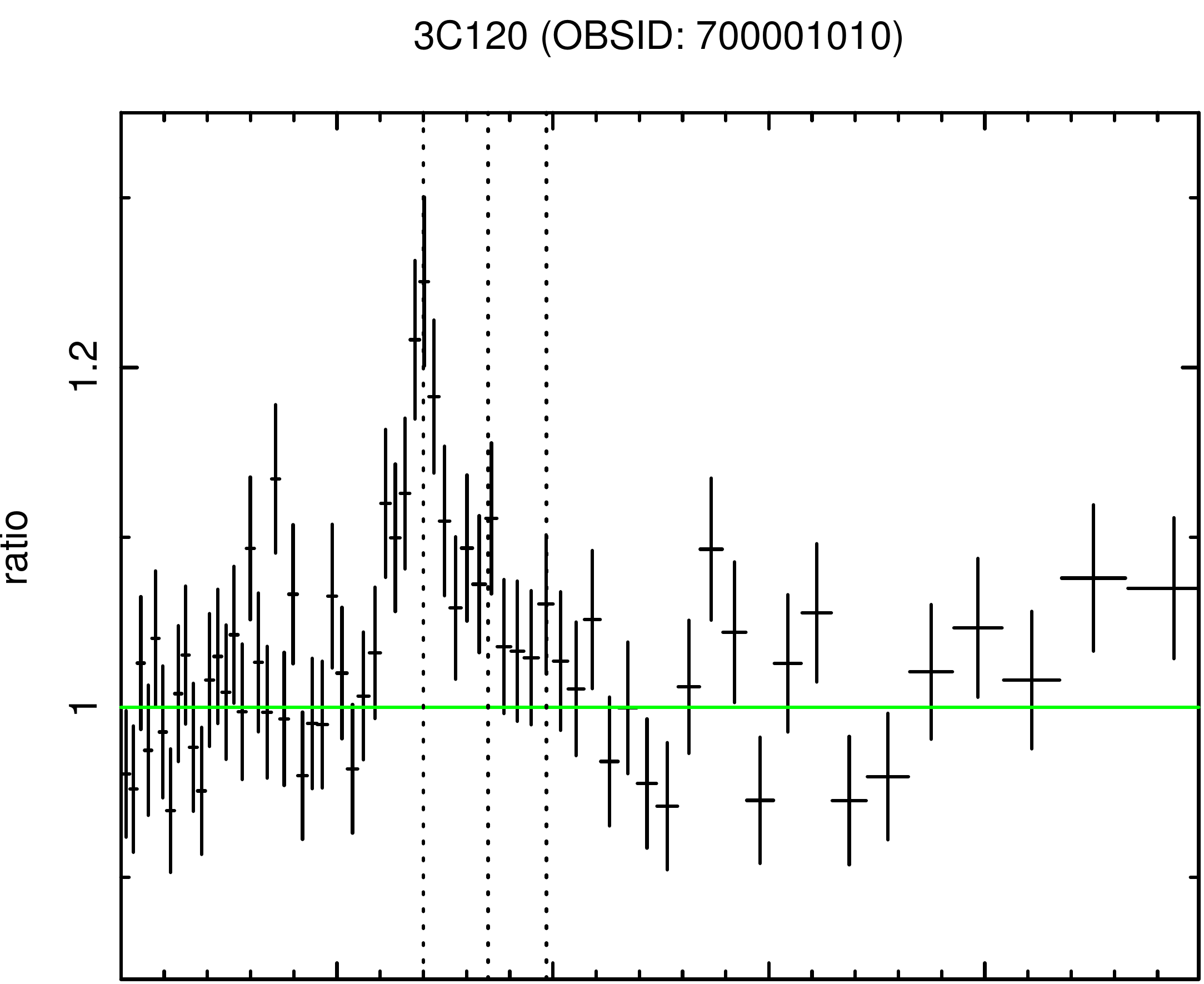}
\includegraphics[angle=-90,width=3.8cm]{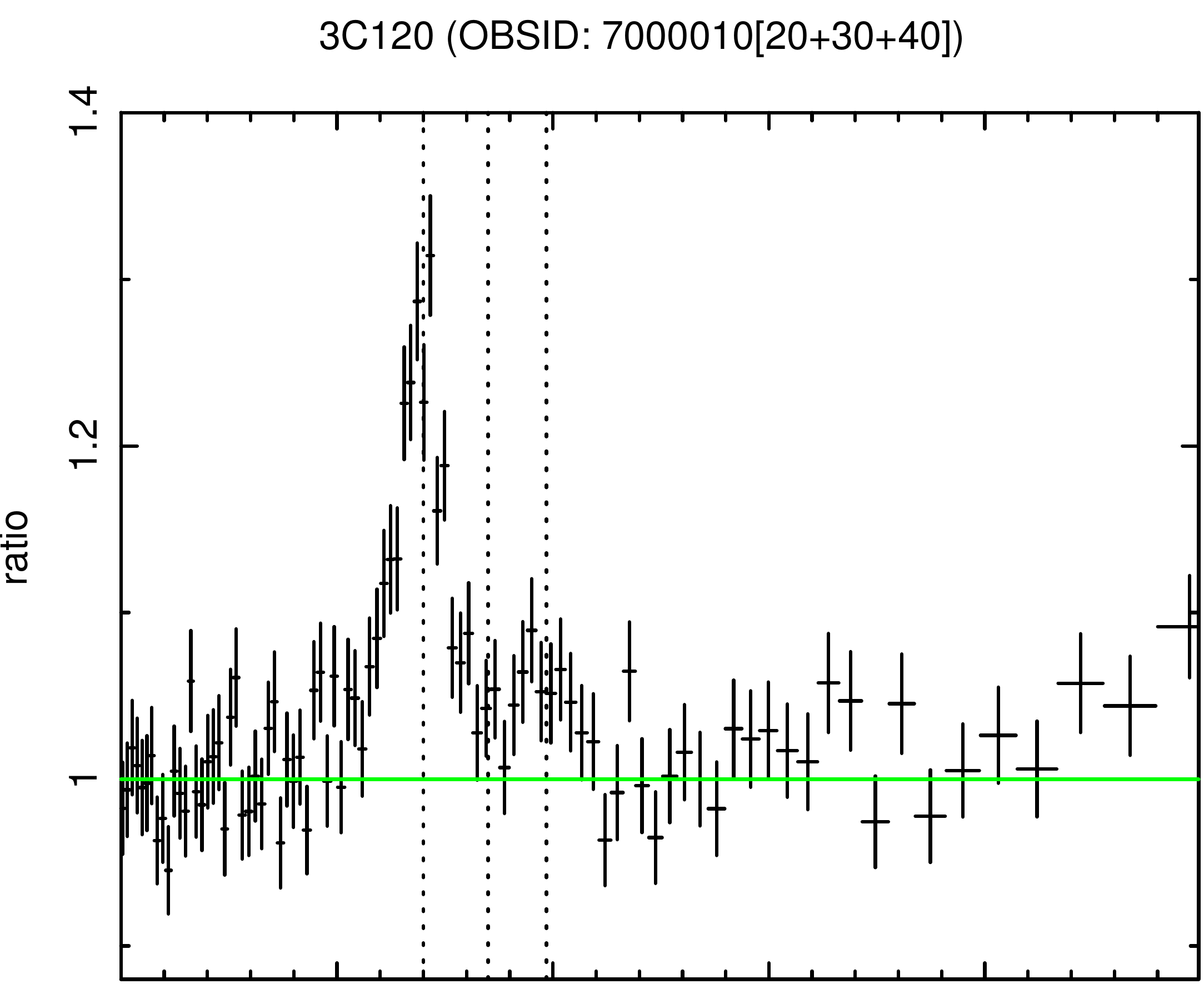}
\includegraphics[angle=-90,width=3.8cm]{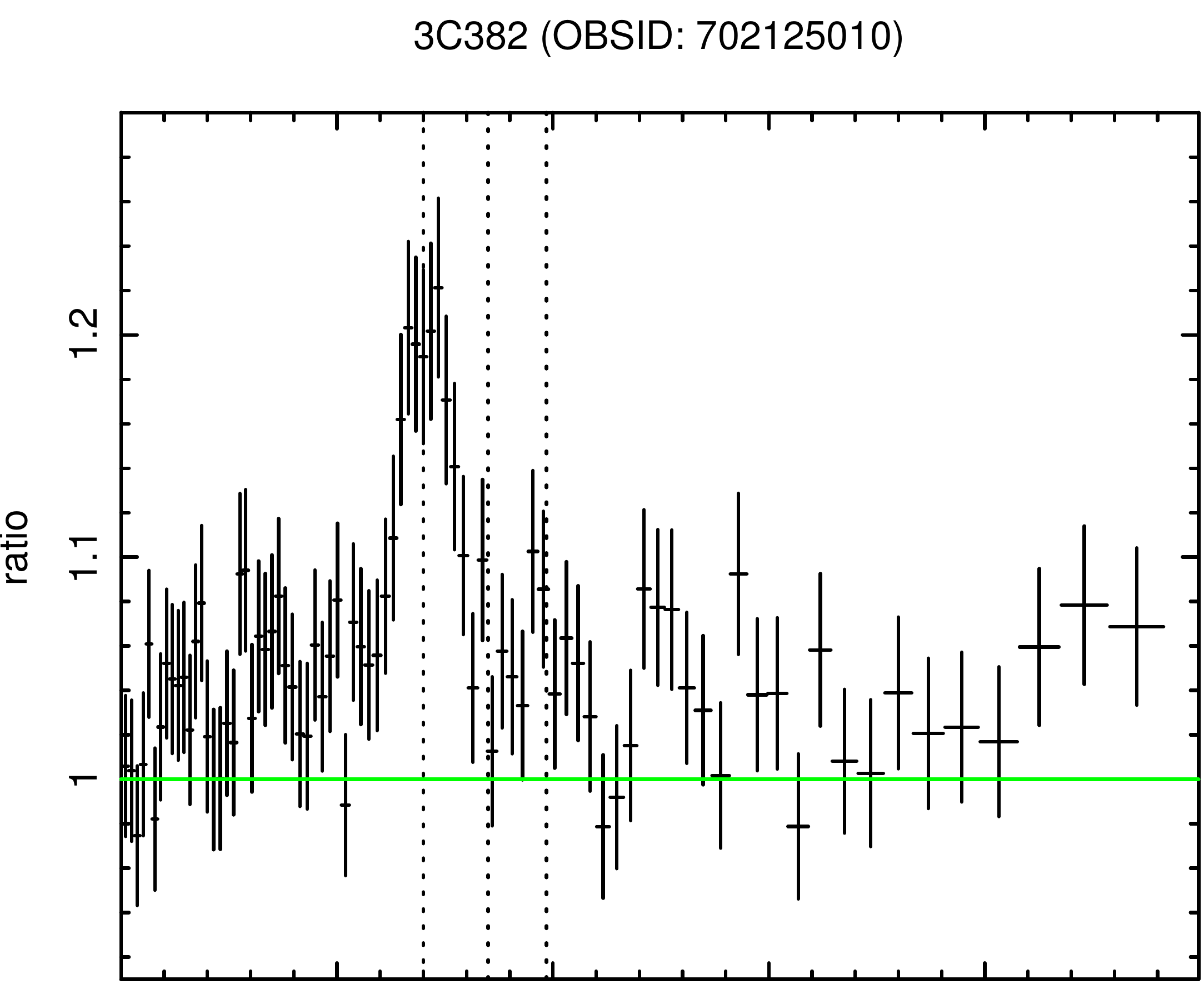}
\includegraphics[angle=-90,width=3.8cm]{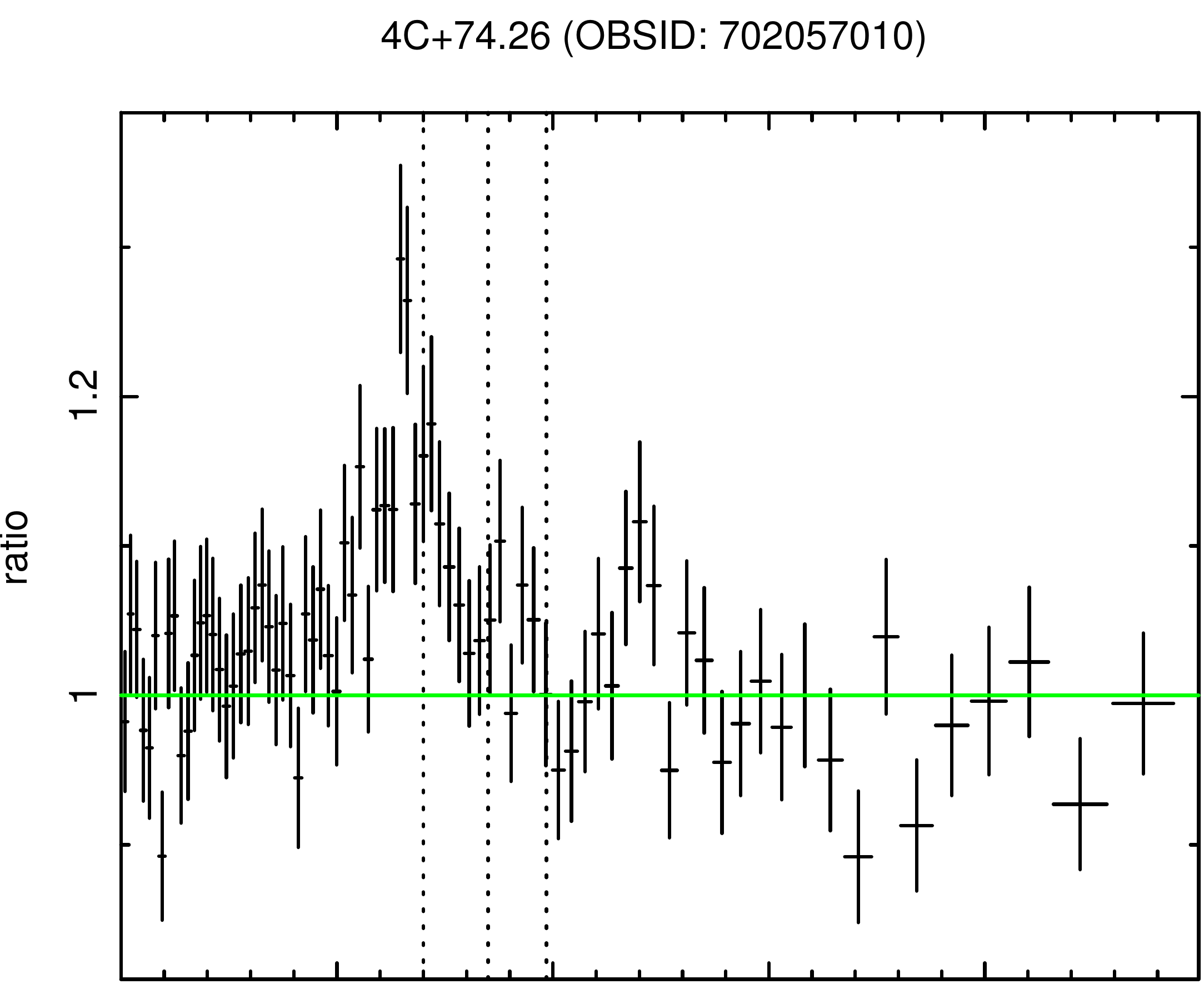}
}

\vspace{-12.25pt}
\subfloat{
\includegraphics[angle=-90,width=3.8cm]{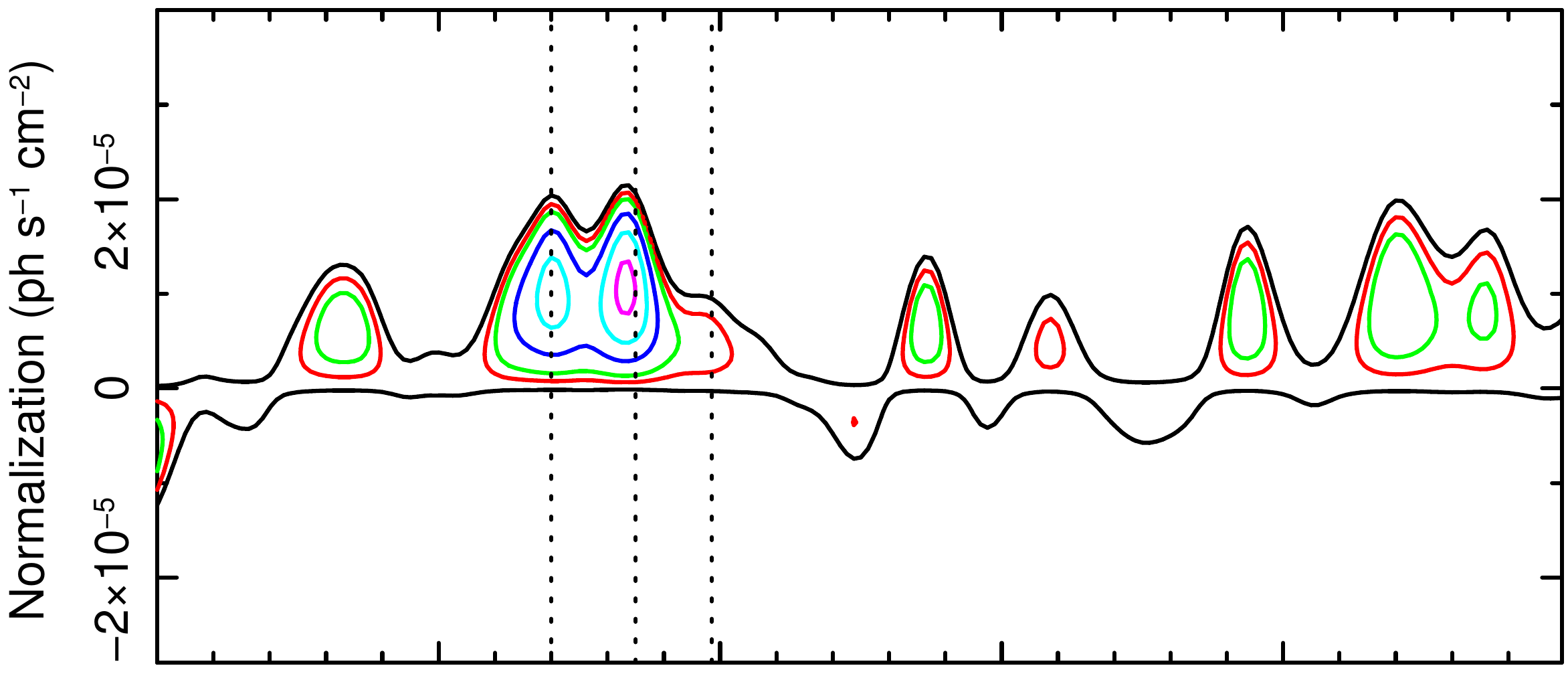}
\includegraphics[angle=-90,width=3.8cm]{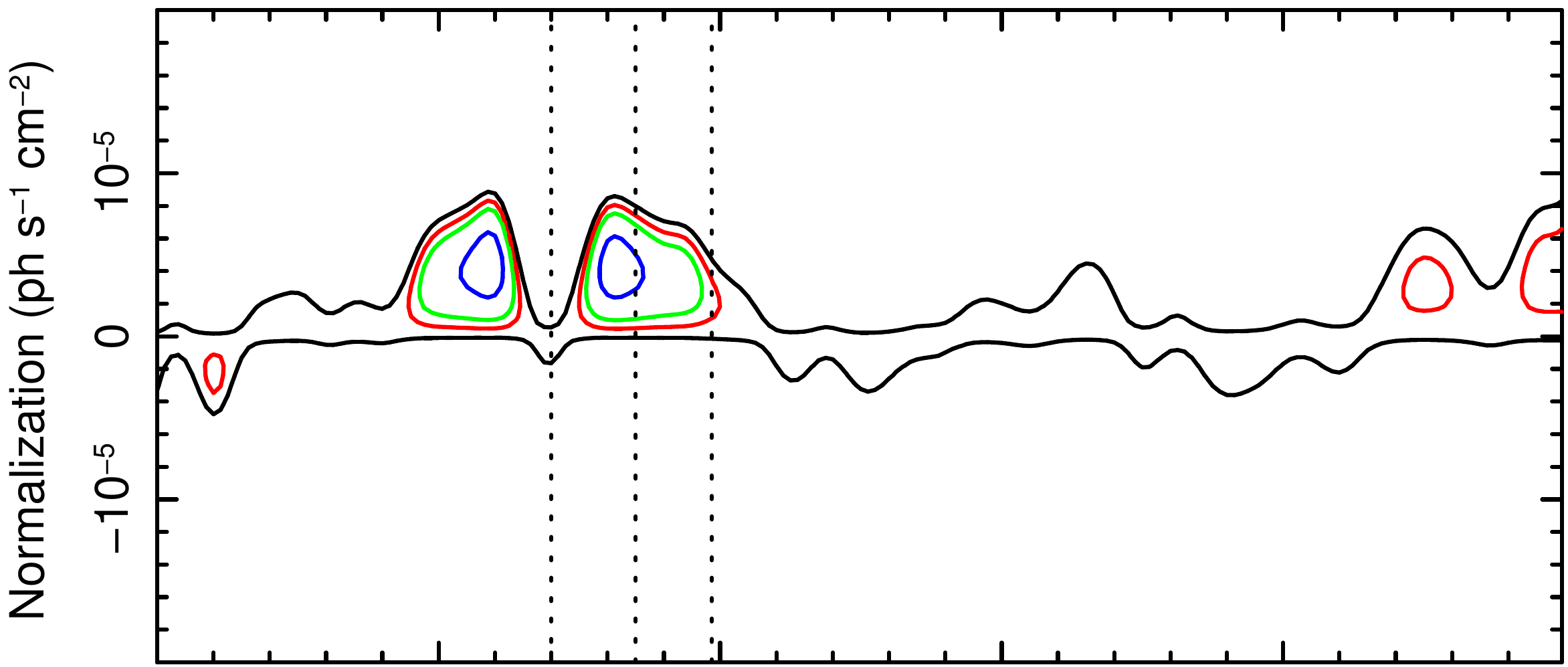}
\includegraphics[angle=-90,width=3.8cm]{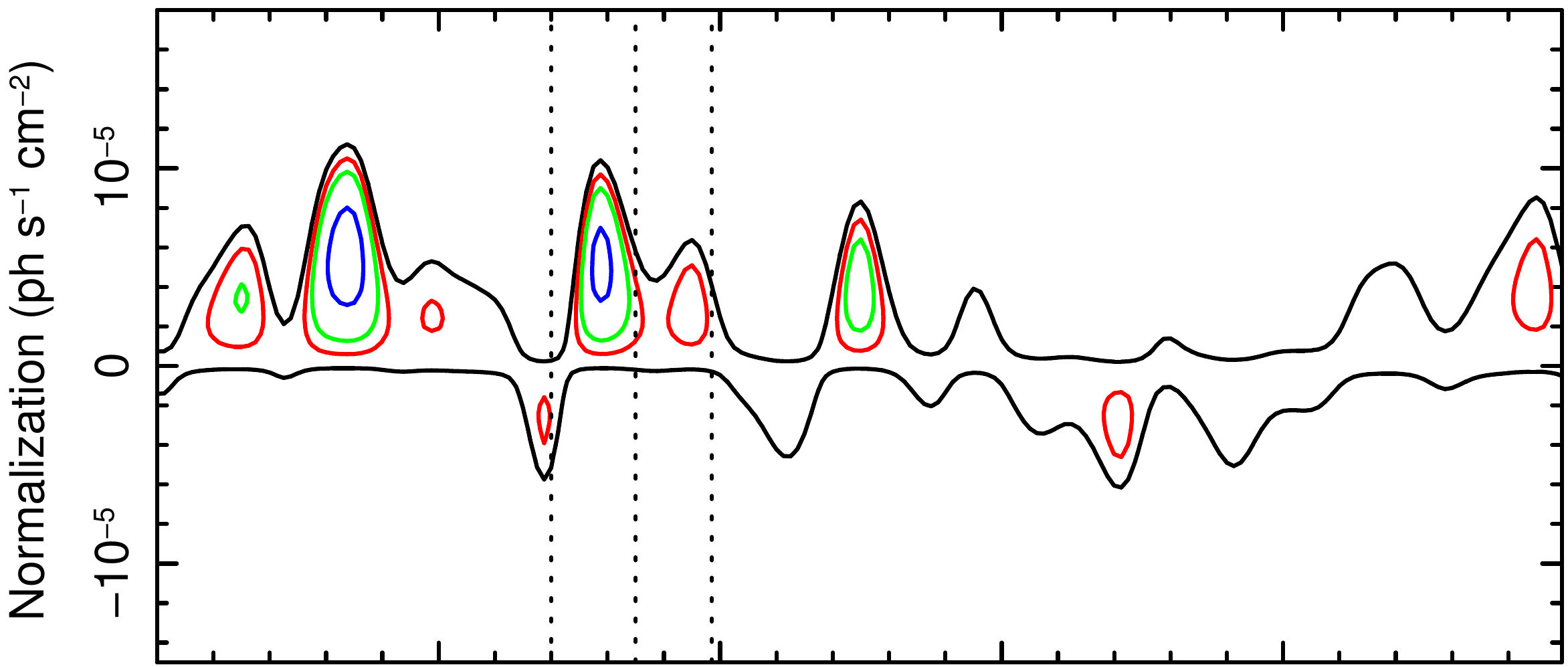}
\includegraphics[angle=-90,width=3.8cm]{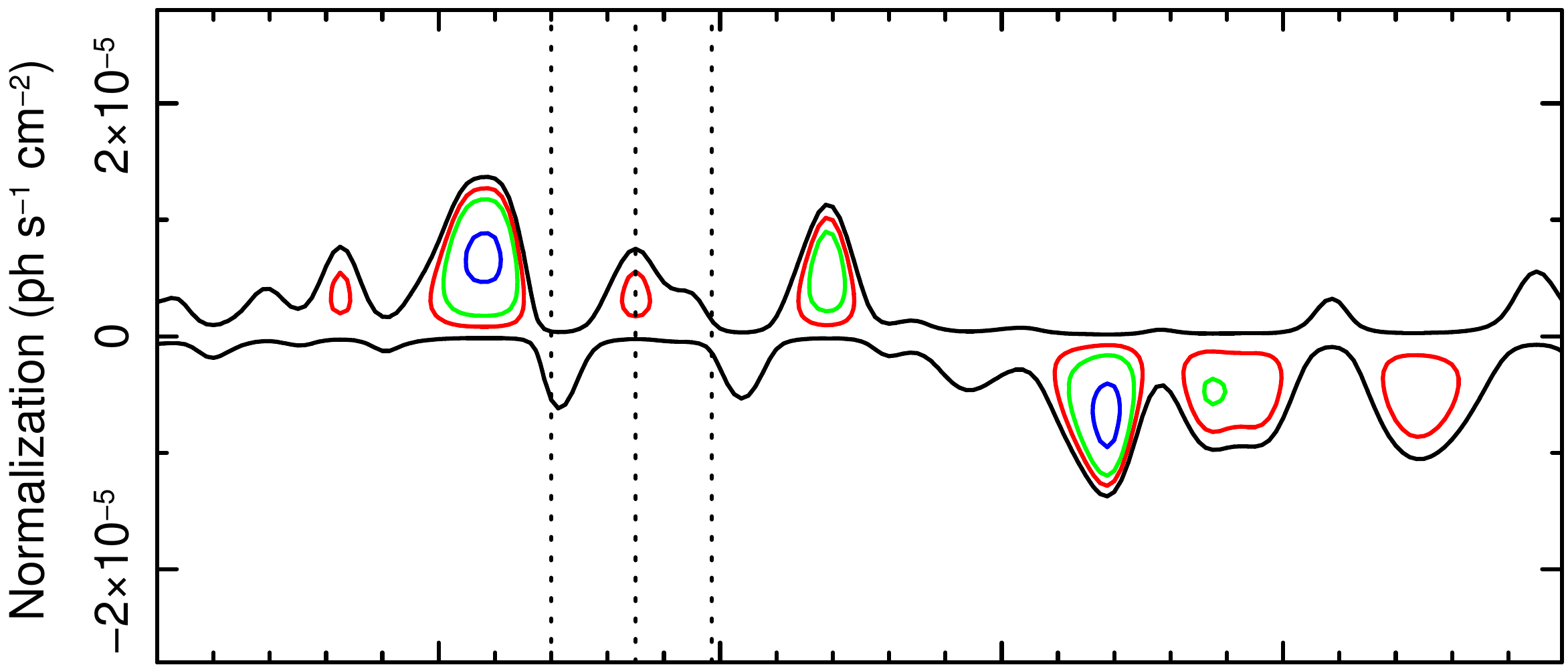}	
}

\vspace{-12.3pt}
\subfloat{
\includegraphics[angle=-90,width=3.8cm]{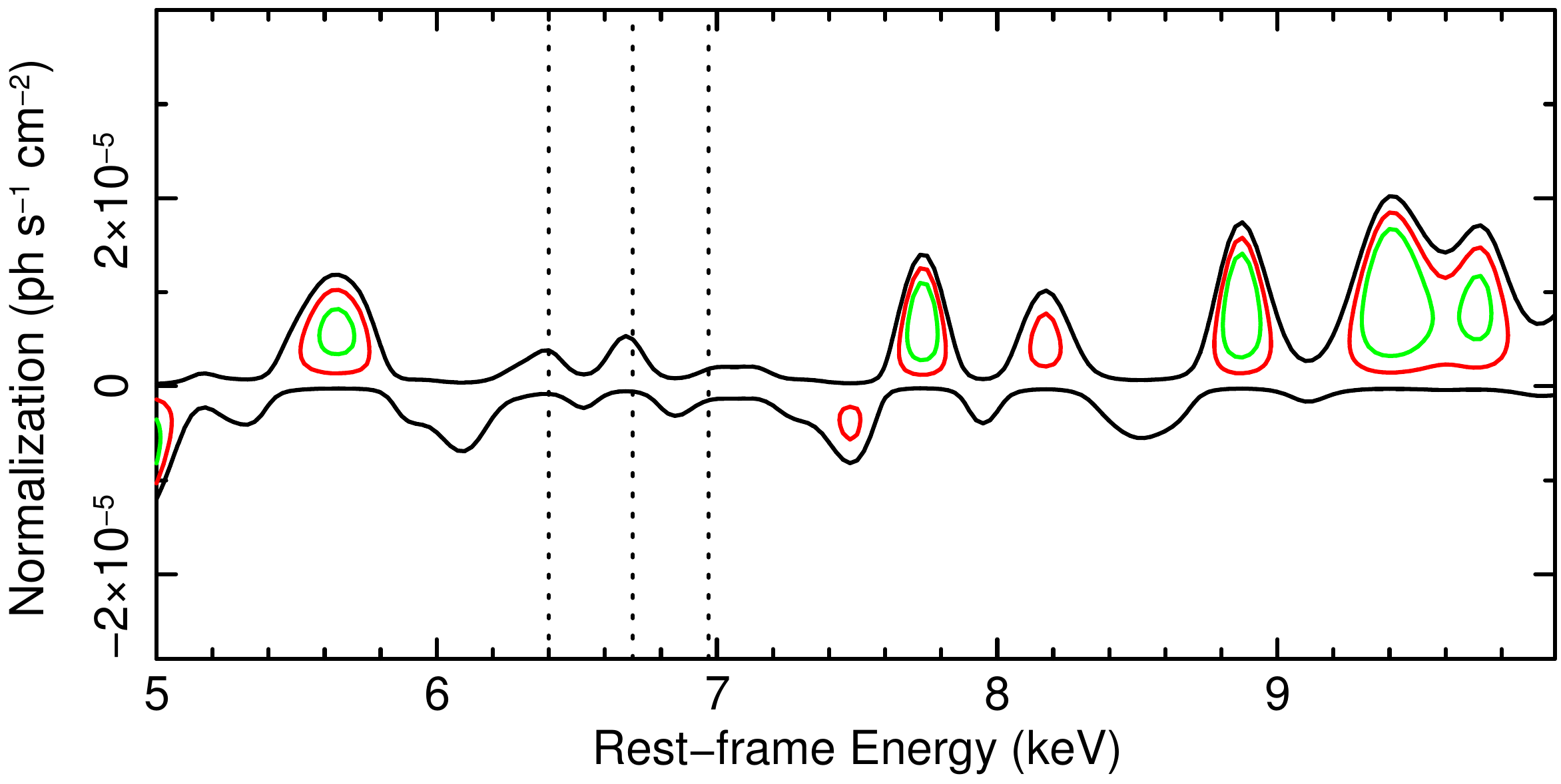}
\includegraphics[angle=-90,width=3.8cm]{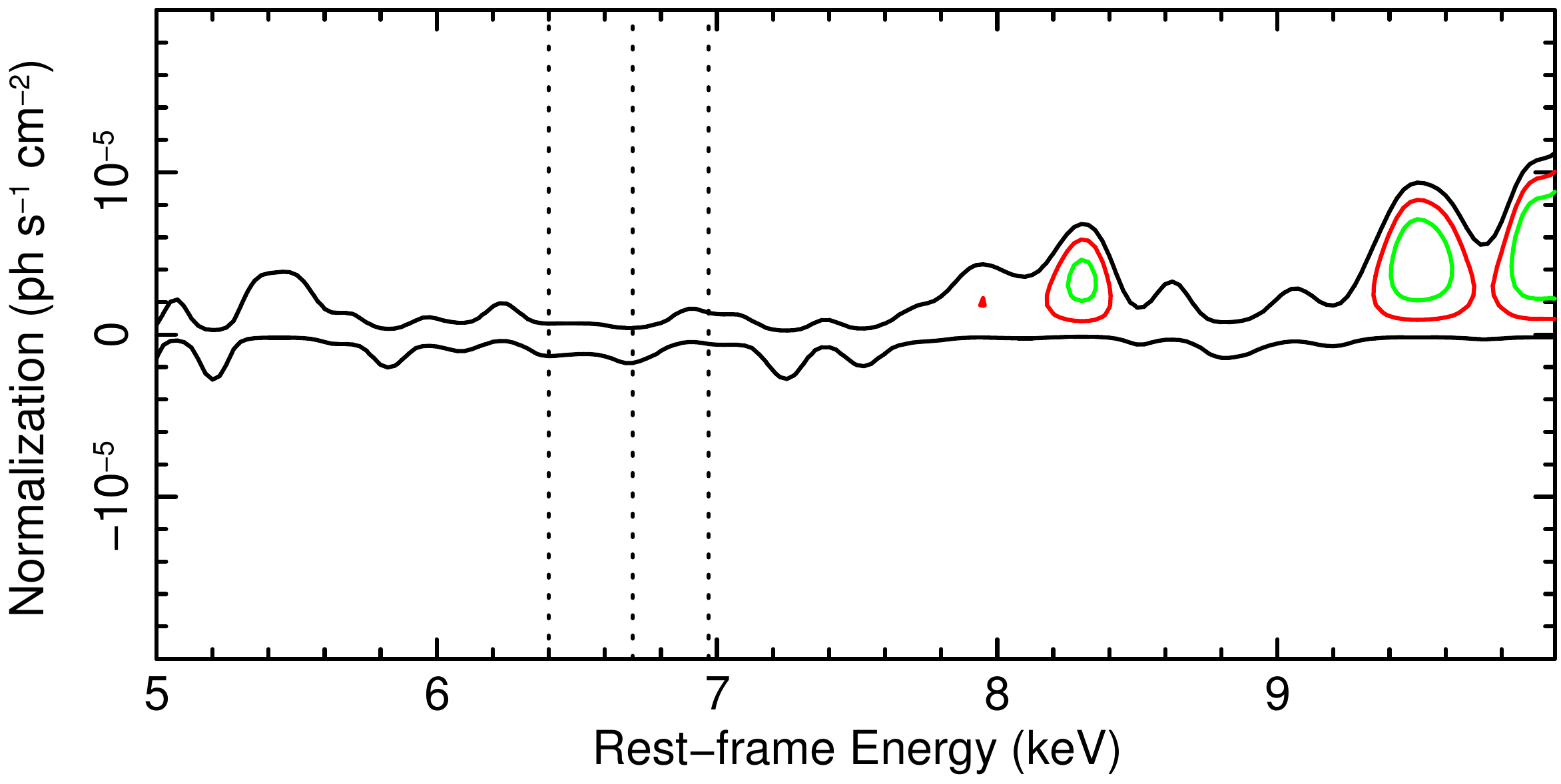}
\includegraphics[angle=-90,width=3.8cm]{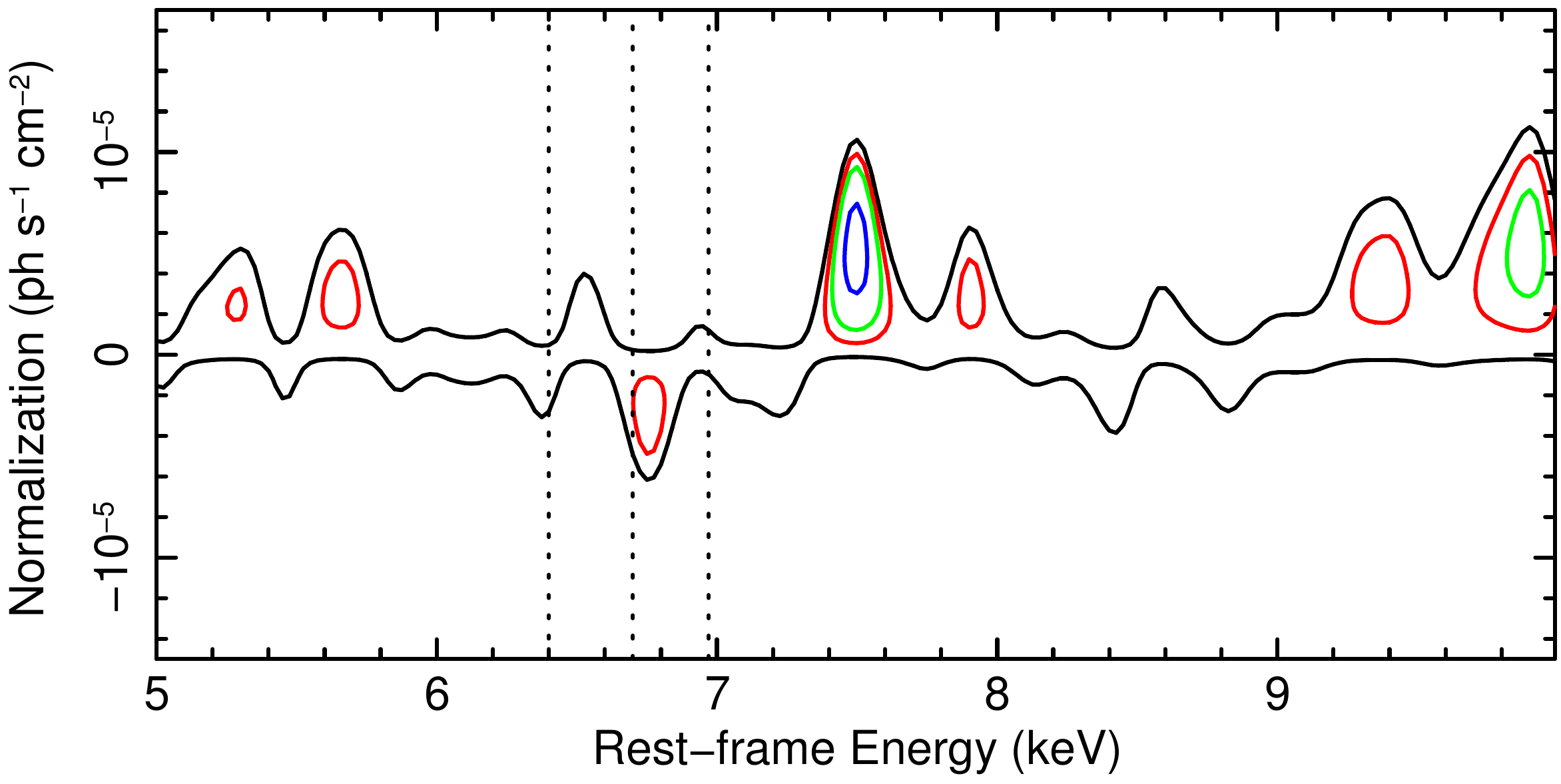}
\includegraphics[angle=-90,width=3.8cm]{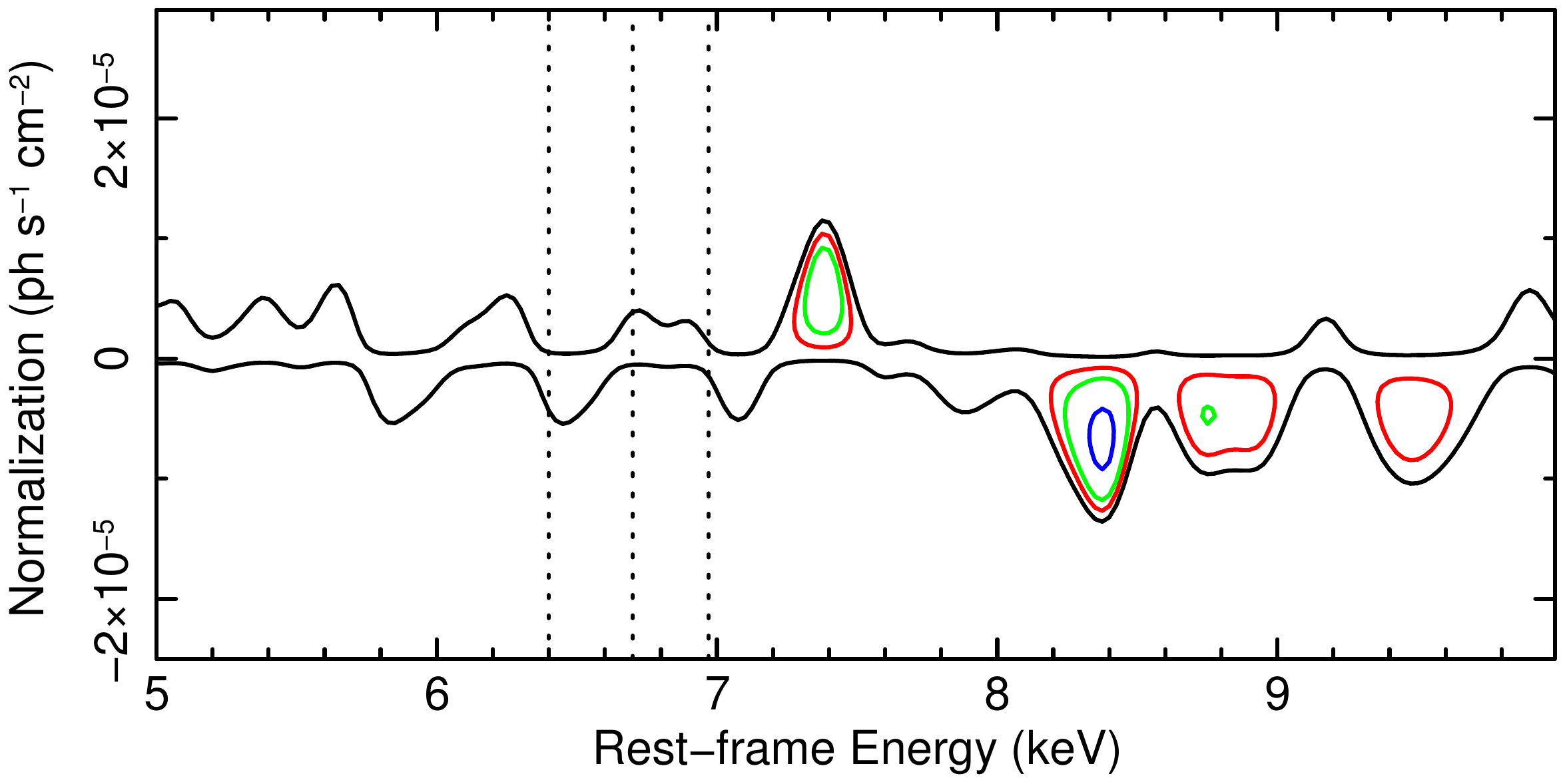}	
}

\vspace{-5pt}
\subfloat{
\includegraphics[angle=-90,width=3.8cm]{figures/Ark120_rat_rf.pdf}
\includegraphics[angle=-90,width=3.8cm]{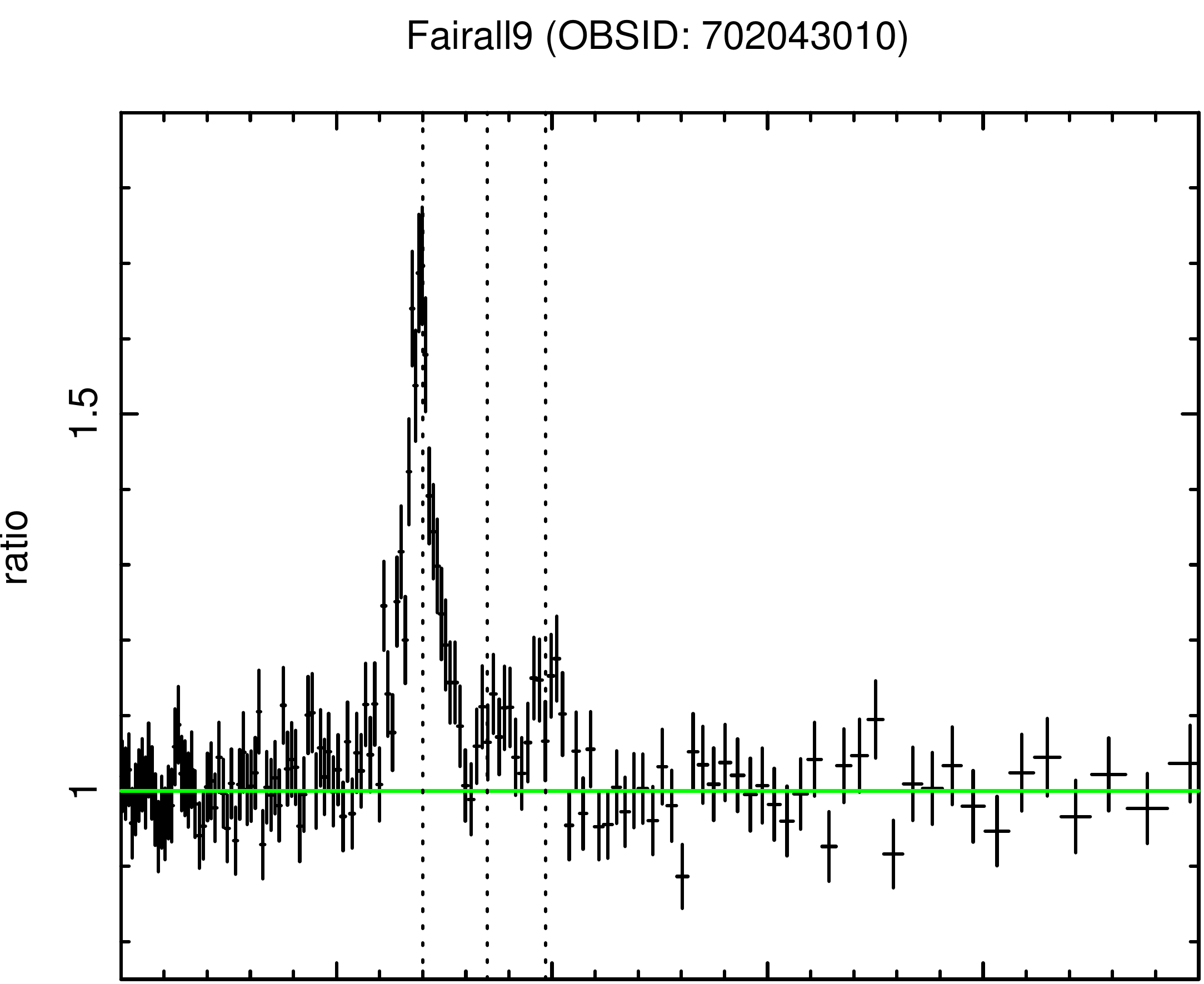}
\includegraphics[angle=-90,width=3.8cm]{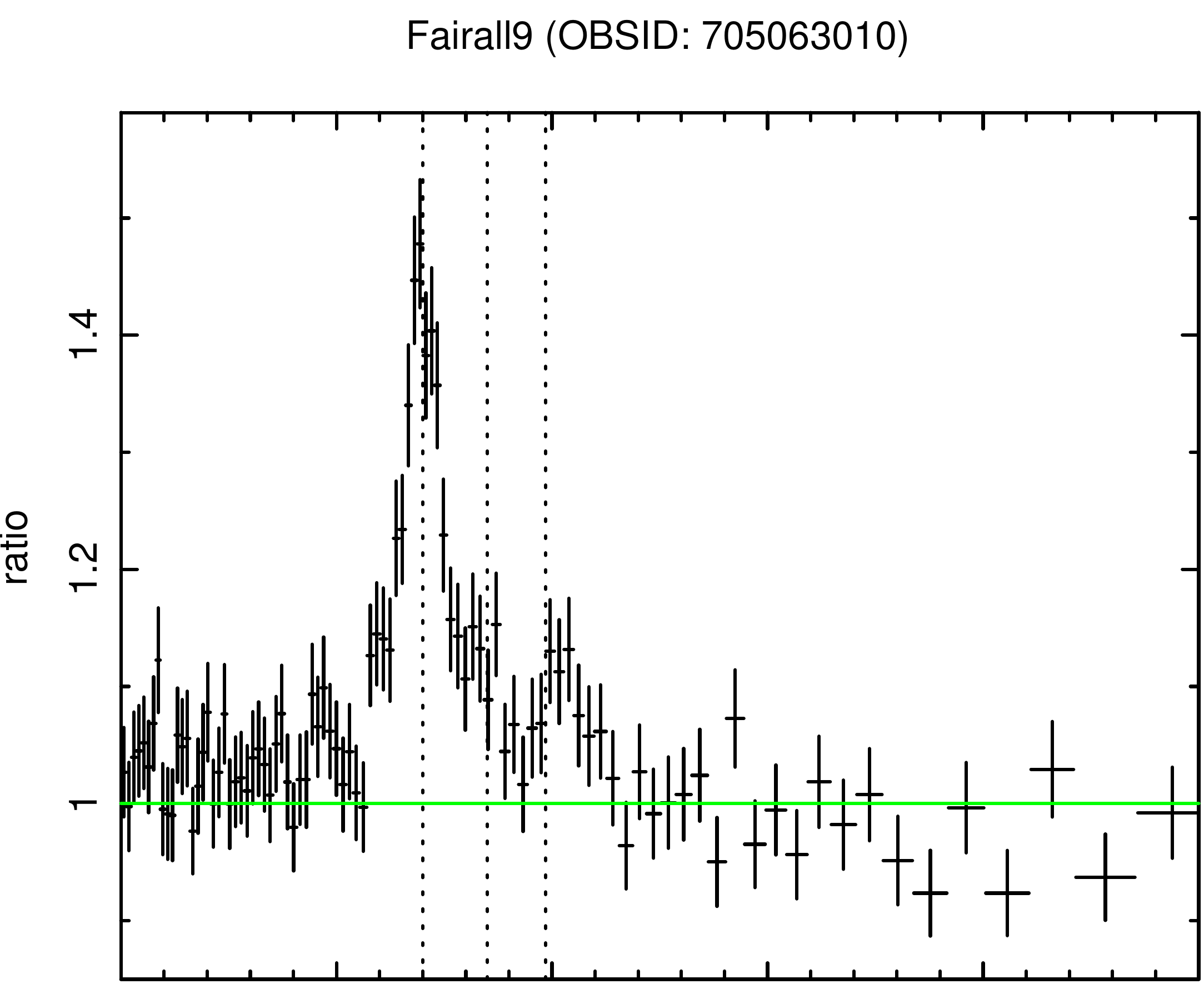}
\includegraphics[angle=-90,width=3.8cm]{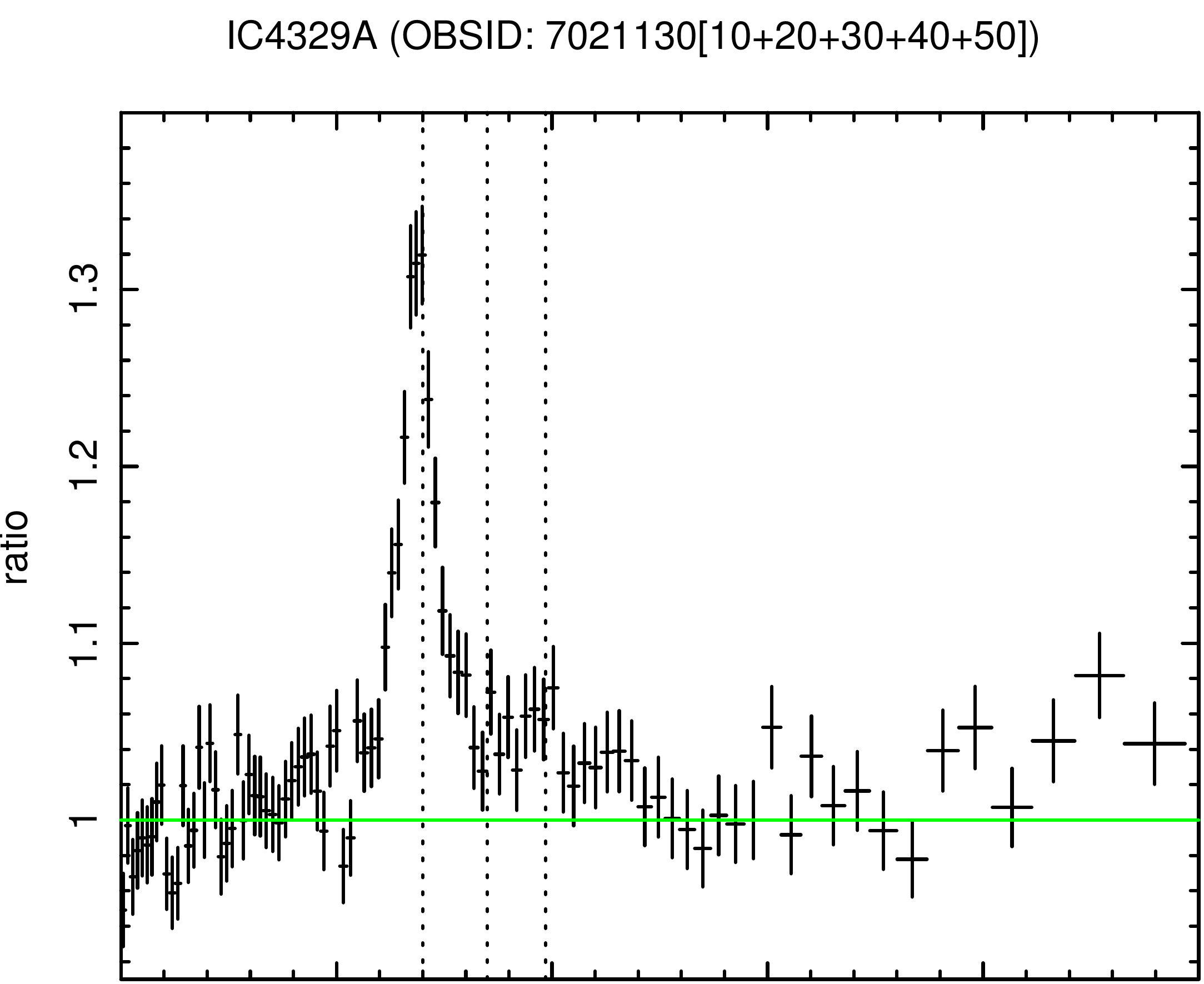}
}

\vspace{-12.2pt}
\subfloat{
\includegraphics[angle=-90,width=3.8cm]{figures/Ark120_cont_nobga.pdf}
\includegraphics[angle=-90,width=3.8cm]{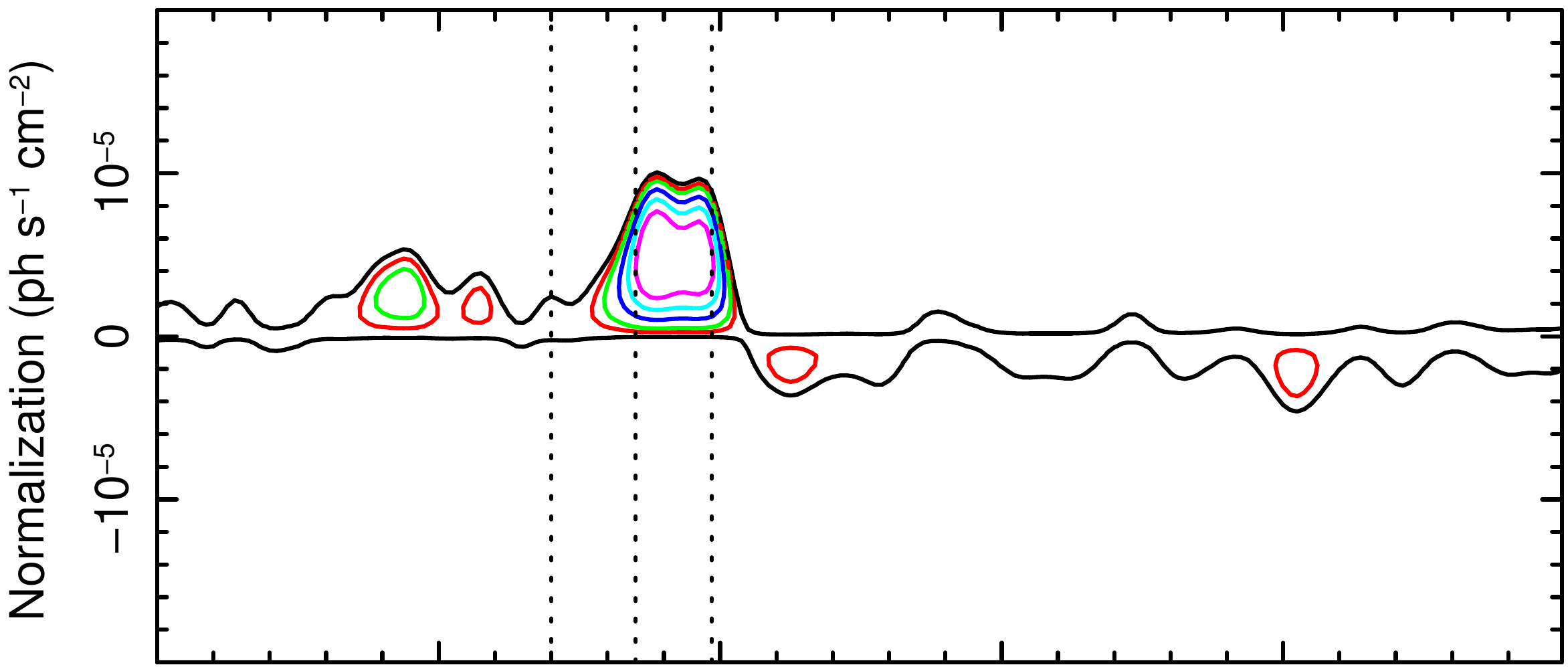}
\includegraphics[angle=-90,width=3.8cm]{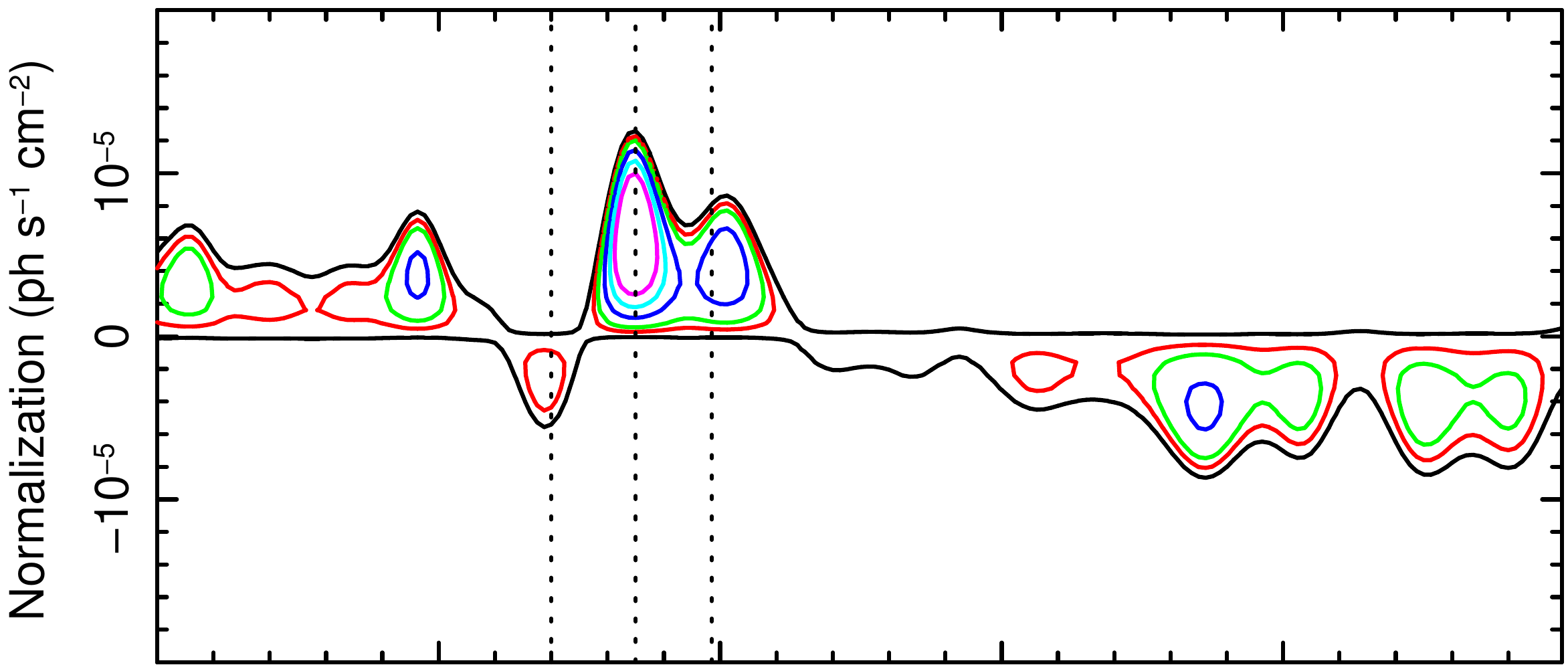}
\includegraphics[angle=-90,width=3.8cm]{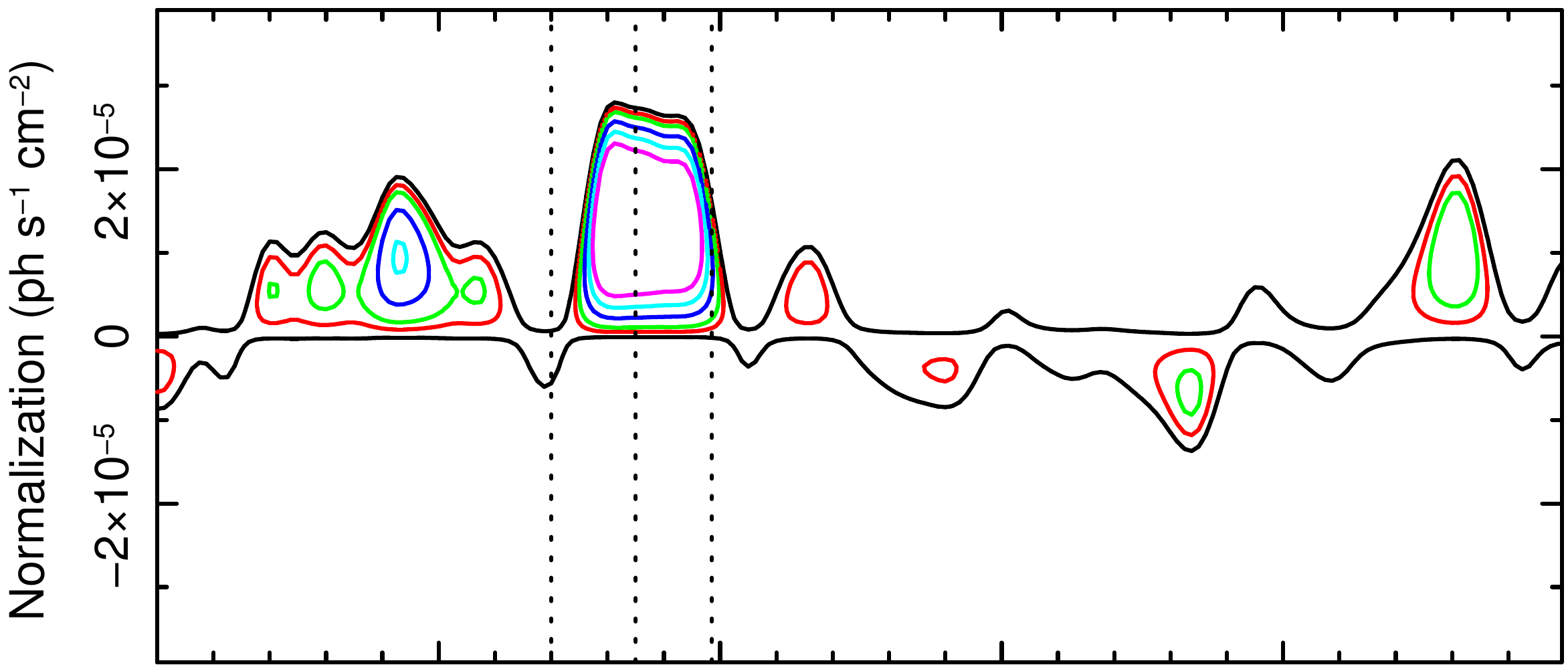}	
}
	
\vspace{-12.25pt}
\subfloat{
\includegraphics[angle=-90,width=3.8cm]{figures/Ark120_cont_bga.pdf}
\includegraphics[angle=-90,width=3.8cm]{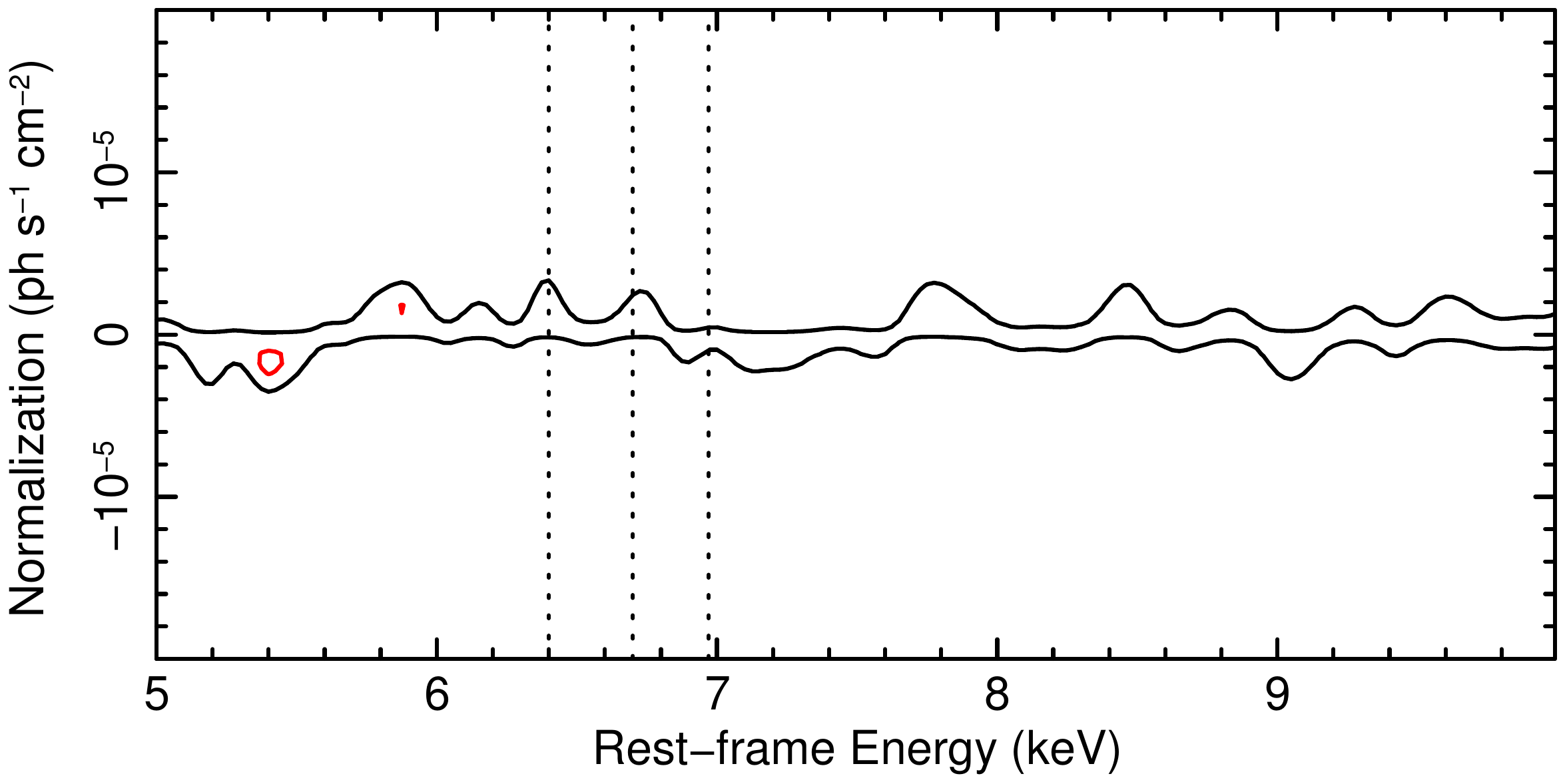}
\includegraphics[angle=-90,width=3.8cm]{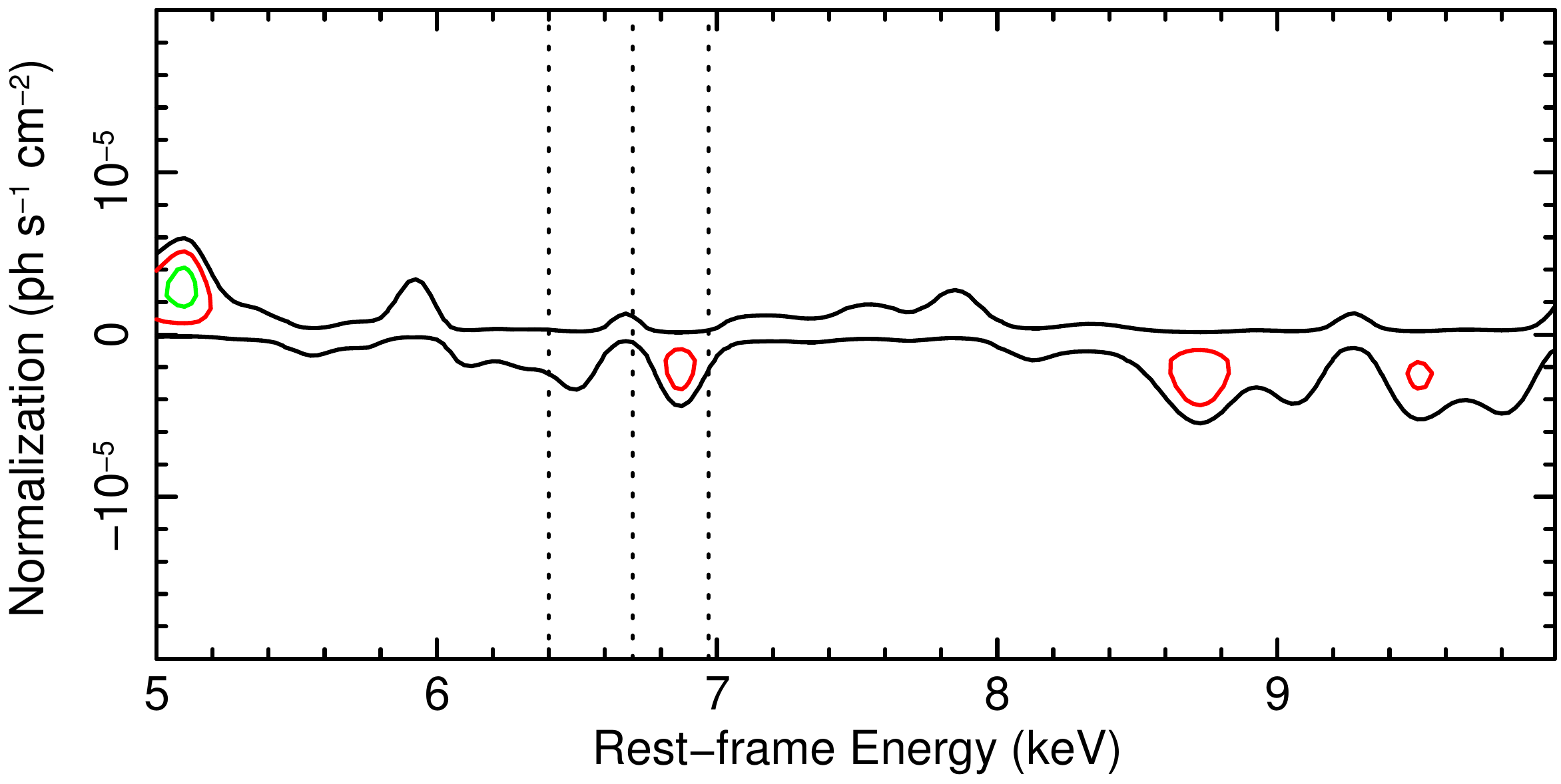}
\includegraphics[angle=-90,width=3.8cm]{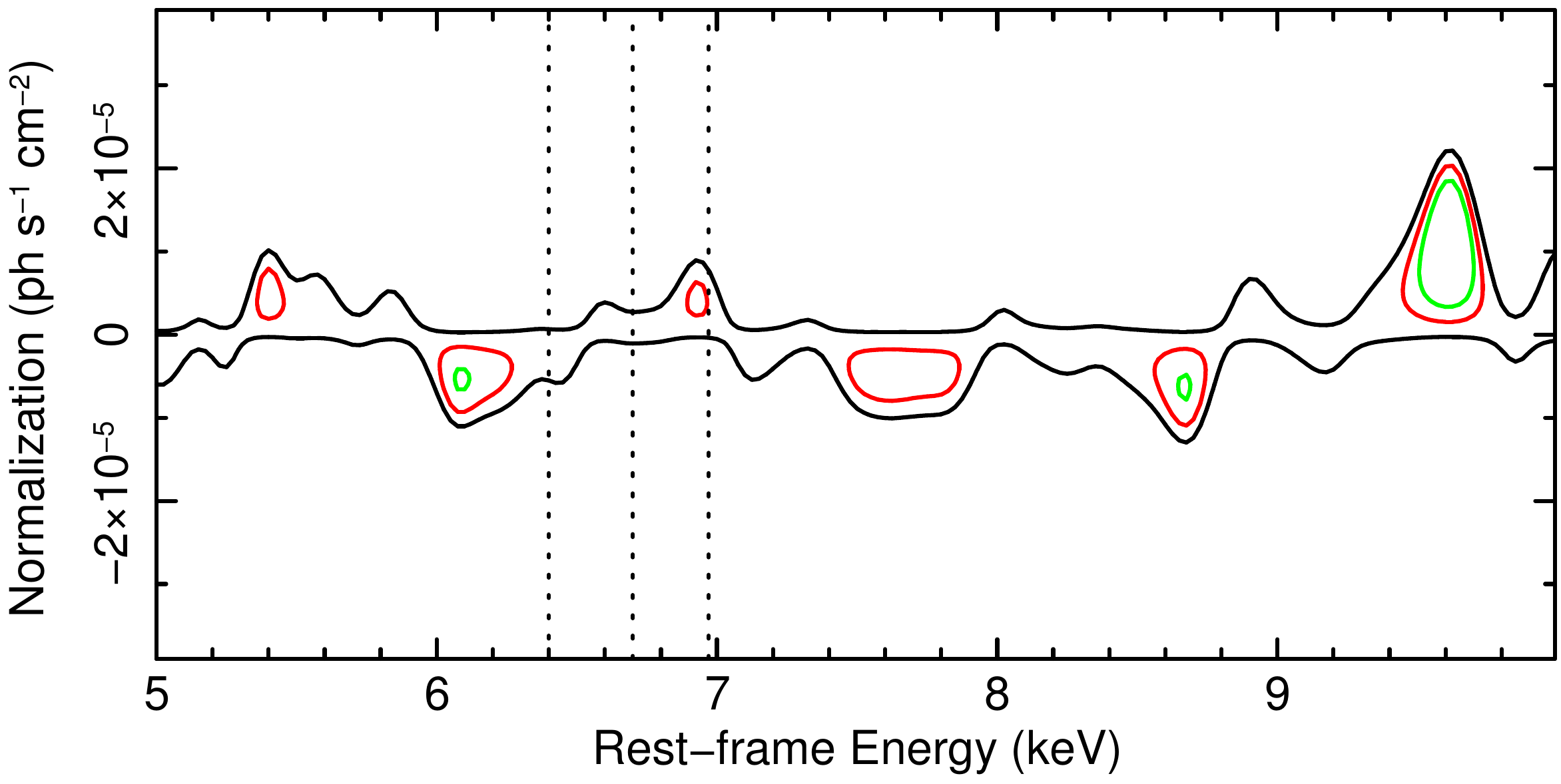}	
}

\vspace{-5pt}	
\subfloat{
\includegraphics[angle=-90,width=3.8cm]{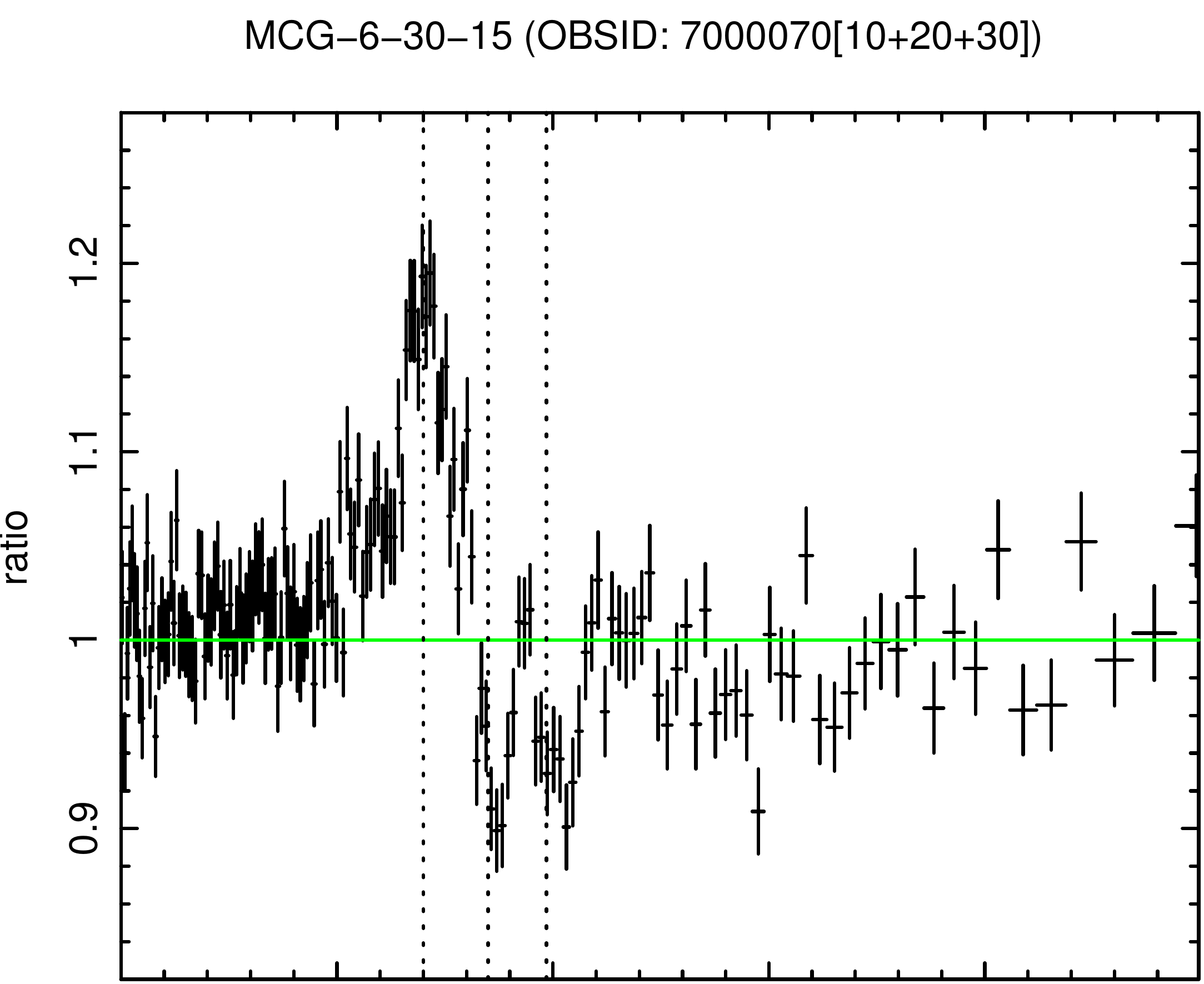}
\includegraphics[angle=-90,width=3.8cm]{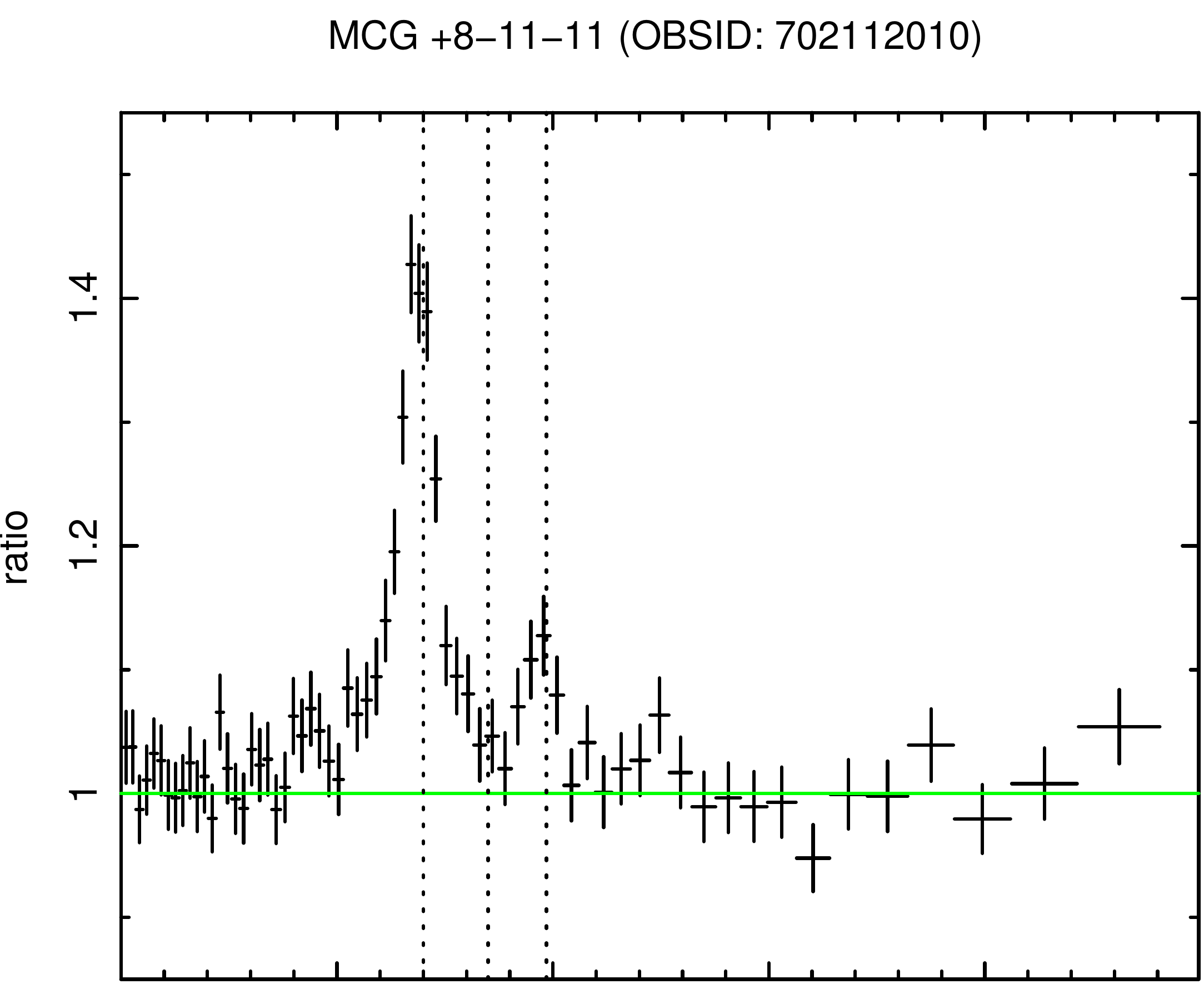}
\includegraphics[angle=-90,width=3.8cm]{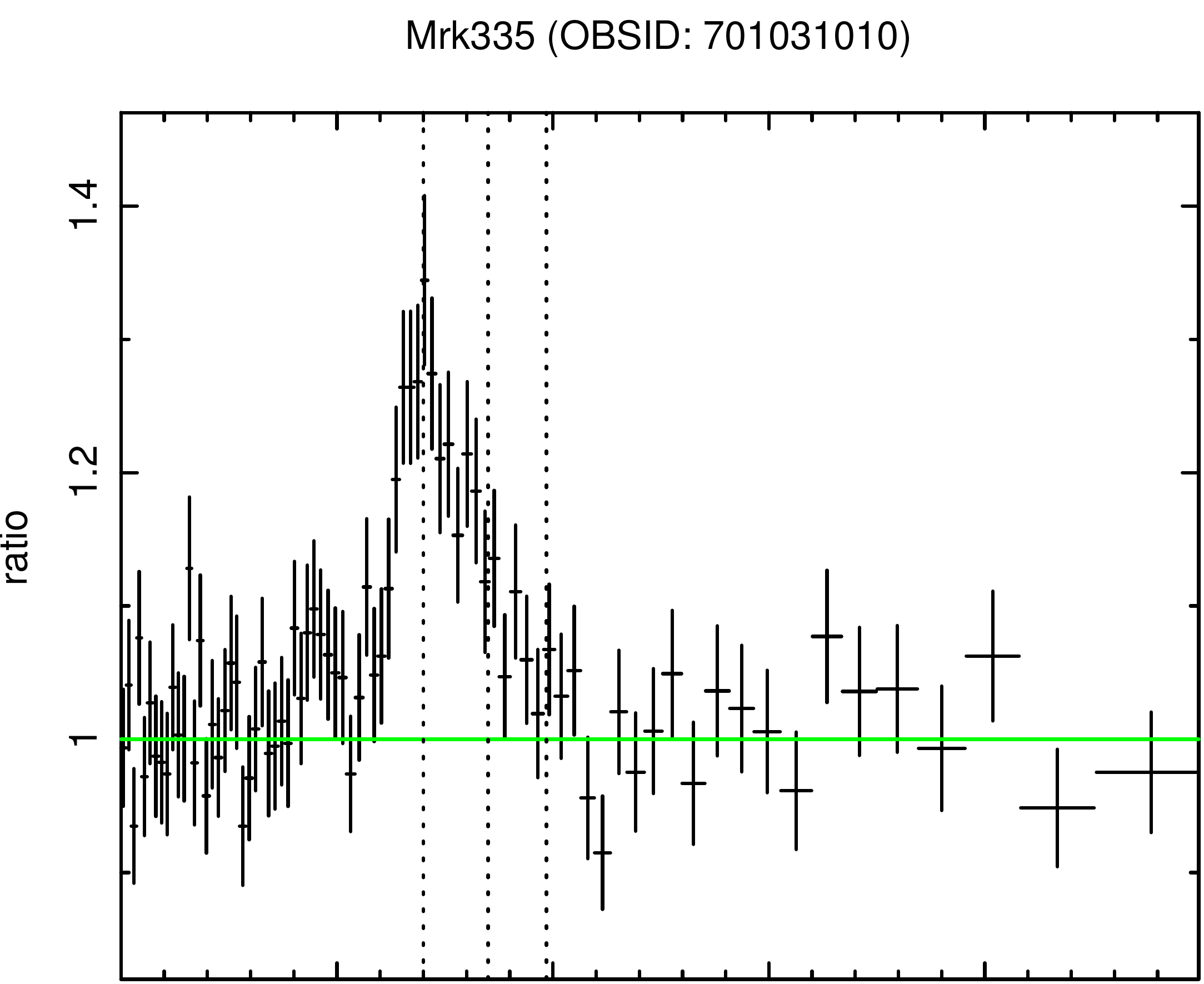}
\includegraphics[angle=-90,width=3.8cm]{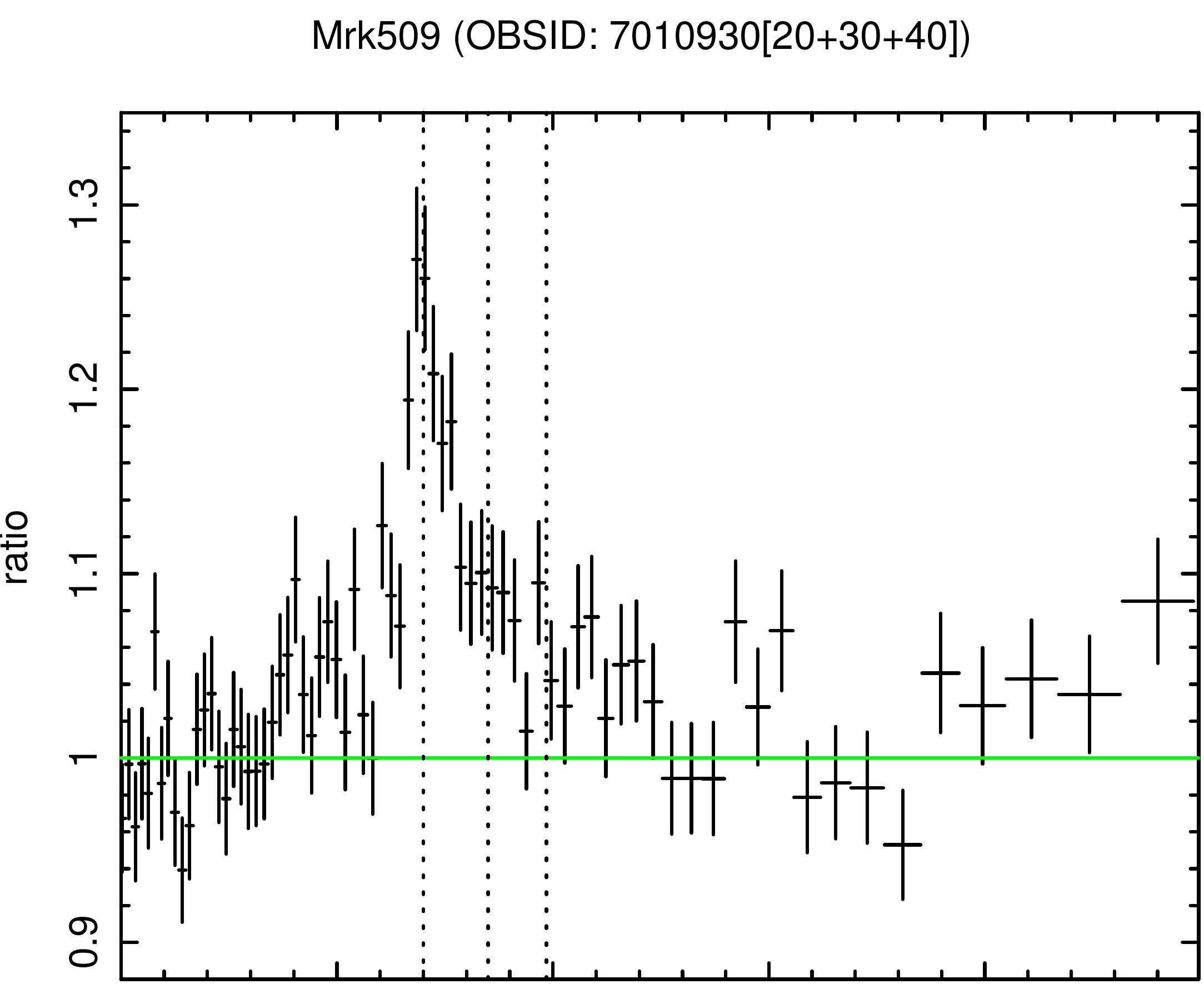}
}

\vspace{-12.2pt}
\subfloat{
\includegraphics[angle=-90,width=3.8cm]{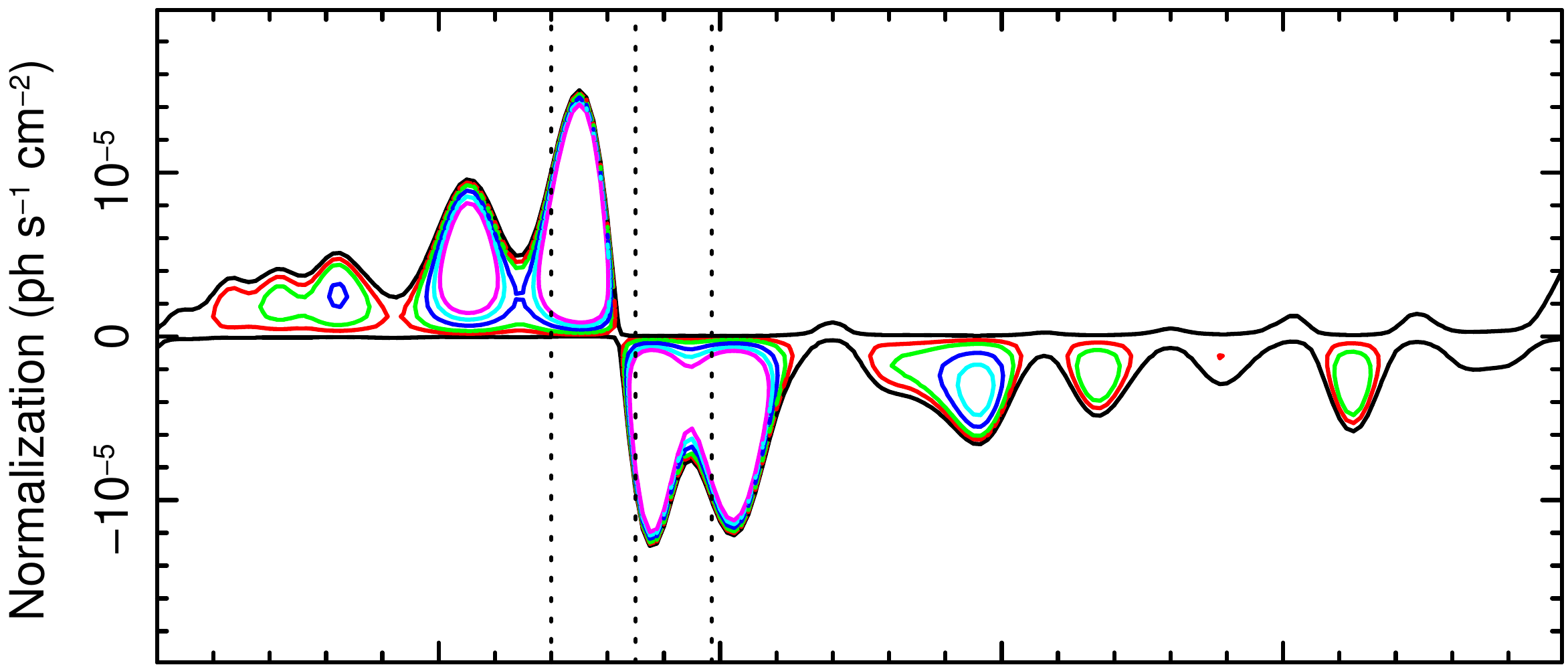}
\includegraphics[angle=-90,width=3.8cm]{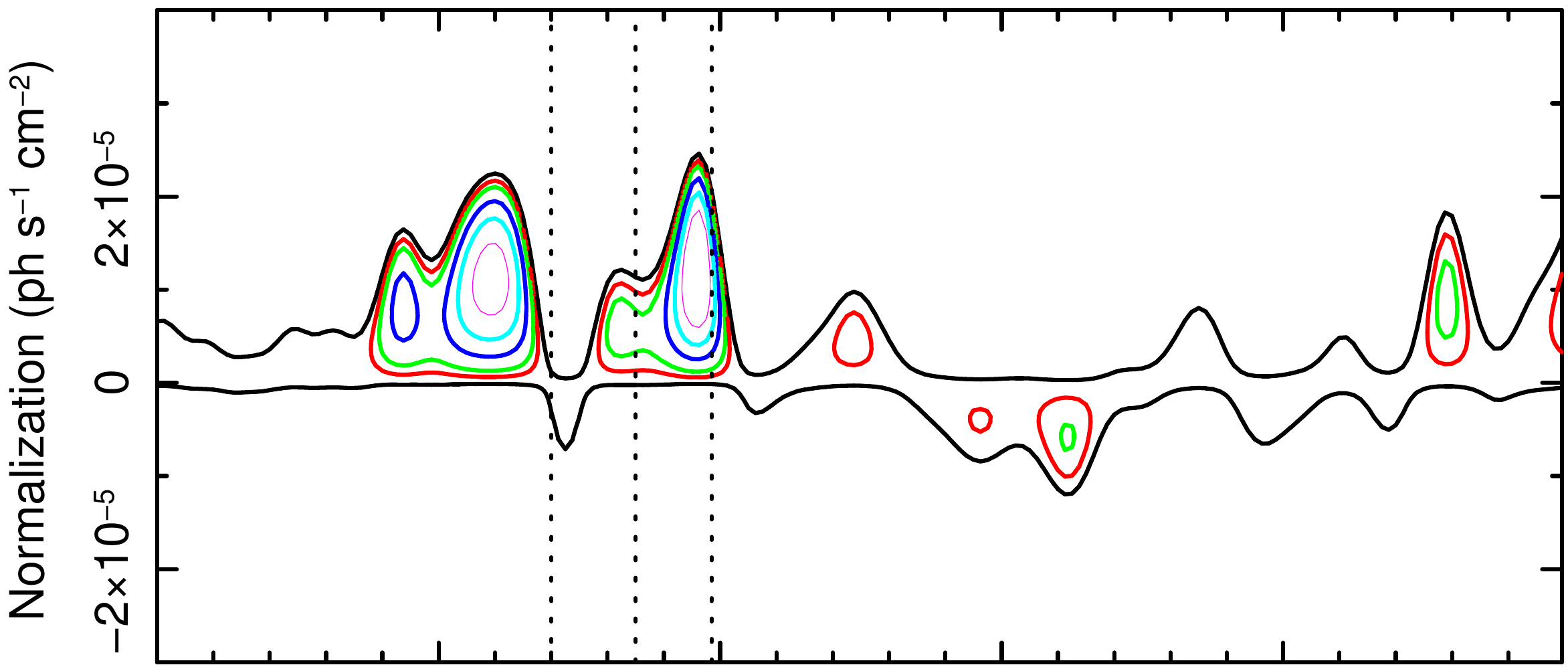}
\includegraphics[angle=-90,width=3.8cm]{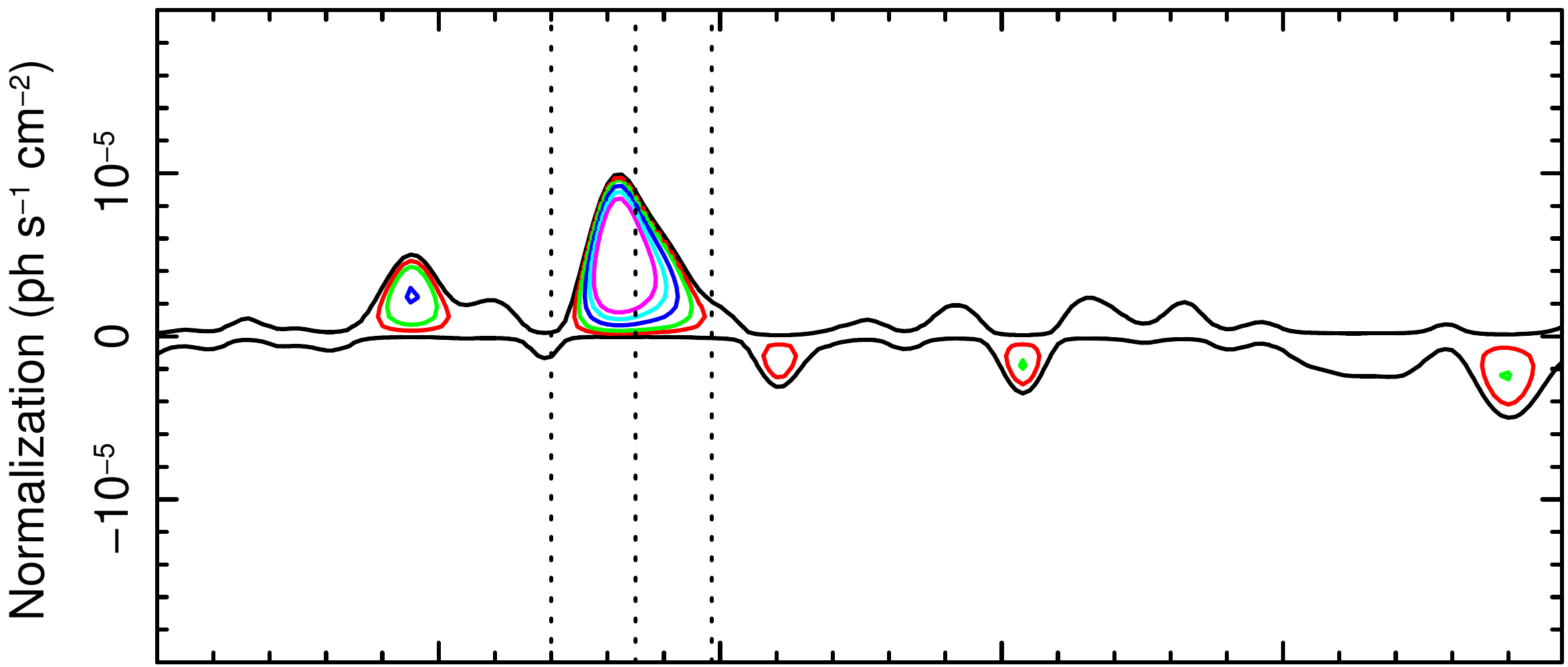}
\includegraphics[angle=-90,width=3.8cm]{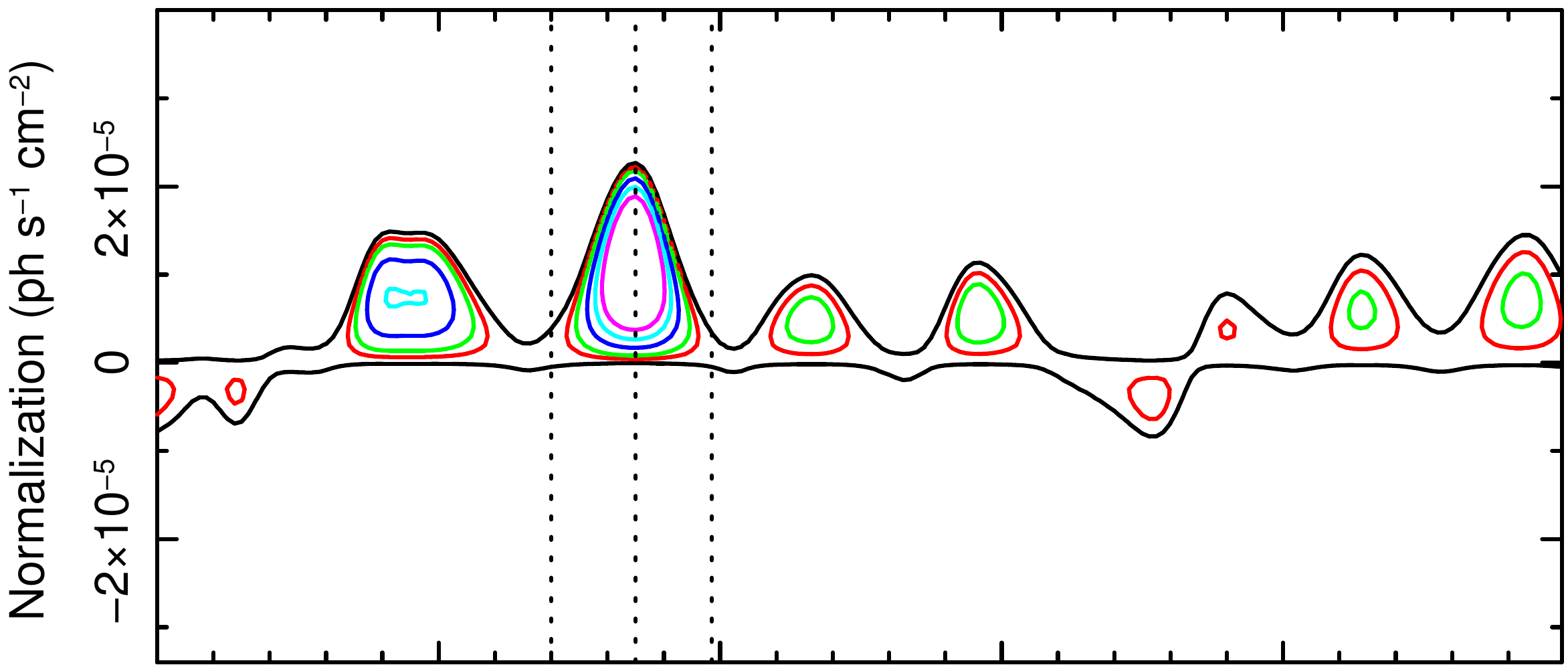}	
}

\vspace{-12.0pt}
\subfloat{
\includegraphics[angle=-90,width=3.8cm]{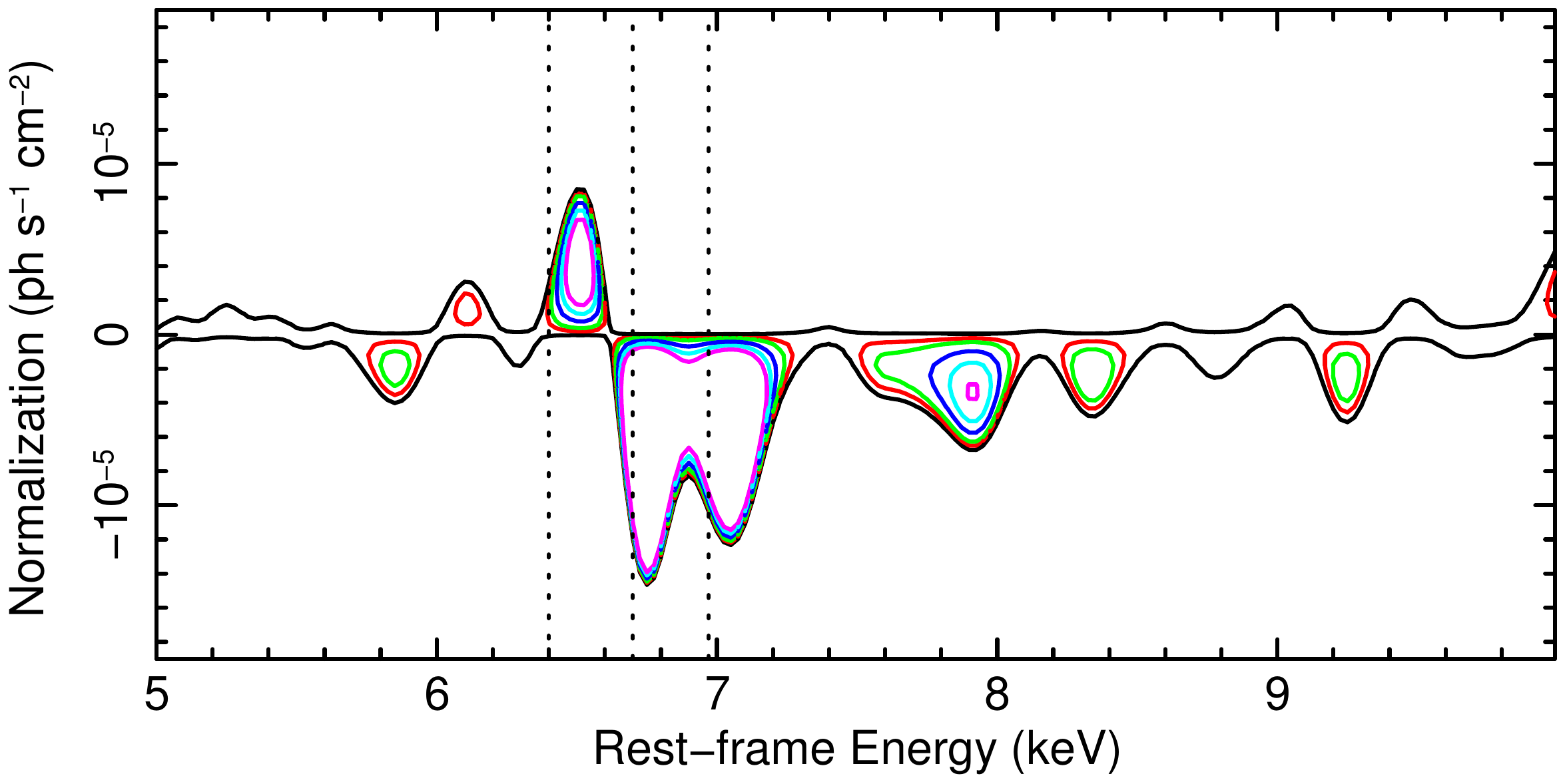}
\includegraphics[angle=-90,width=3.8cm]{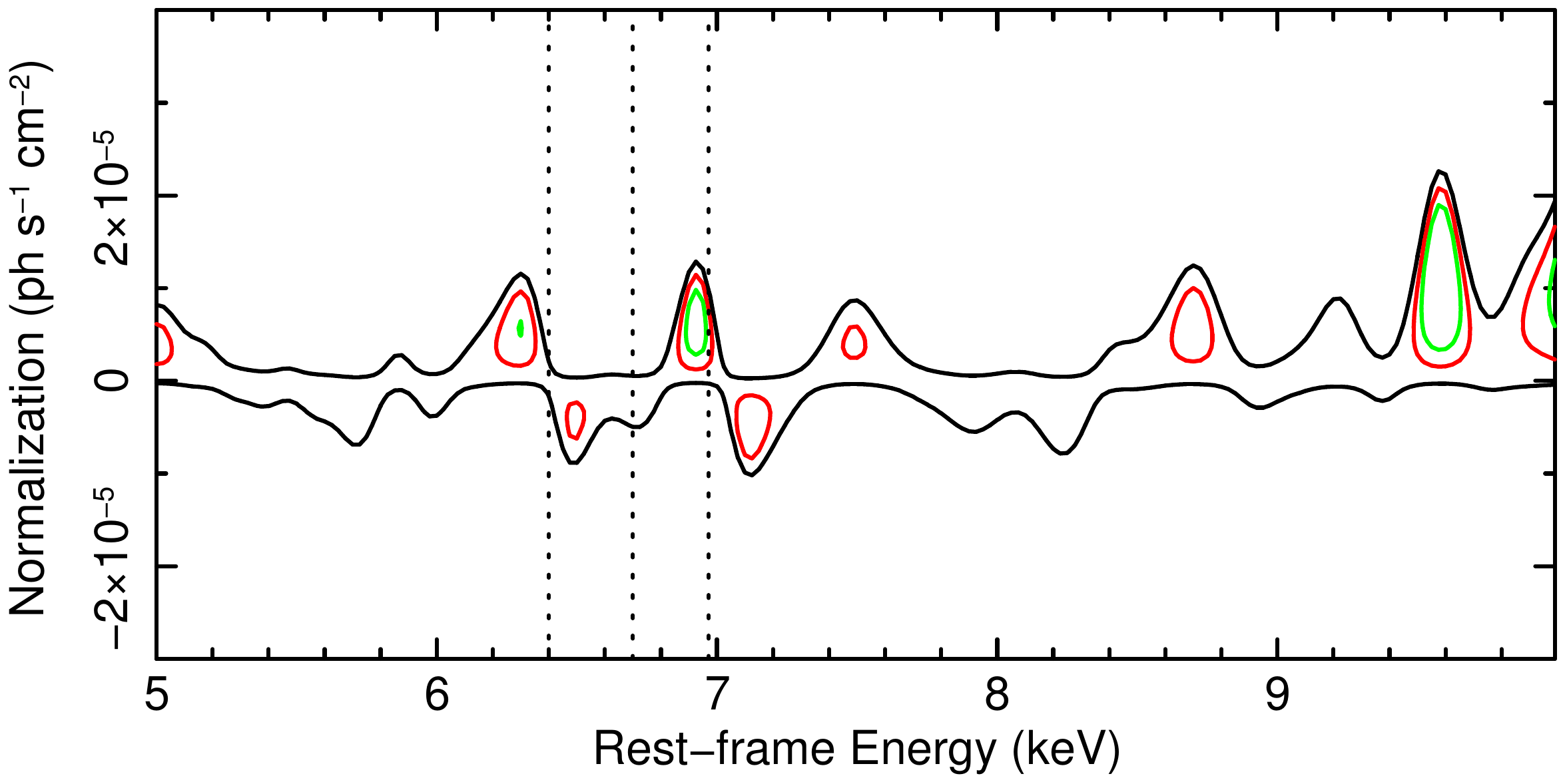}
\includegraphics[angle=-90,width=3.8cm]{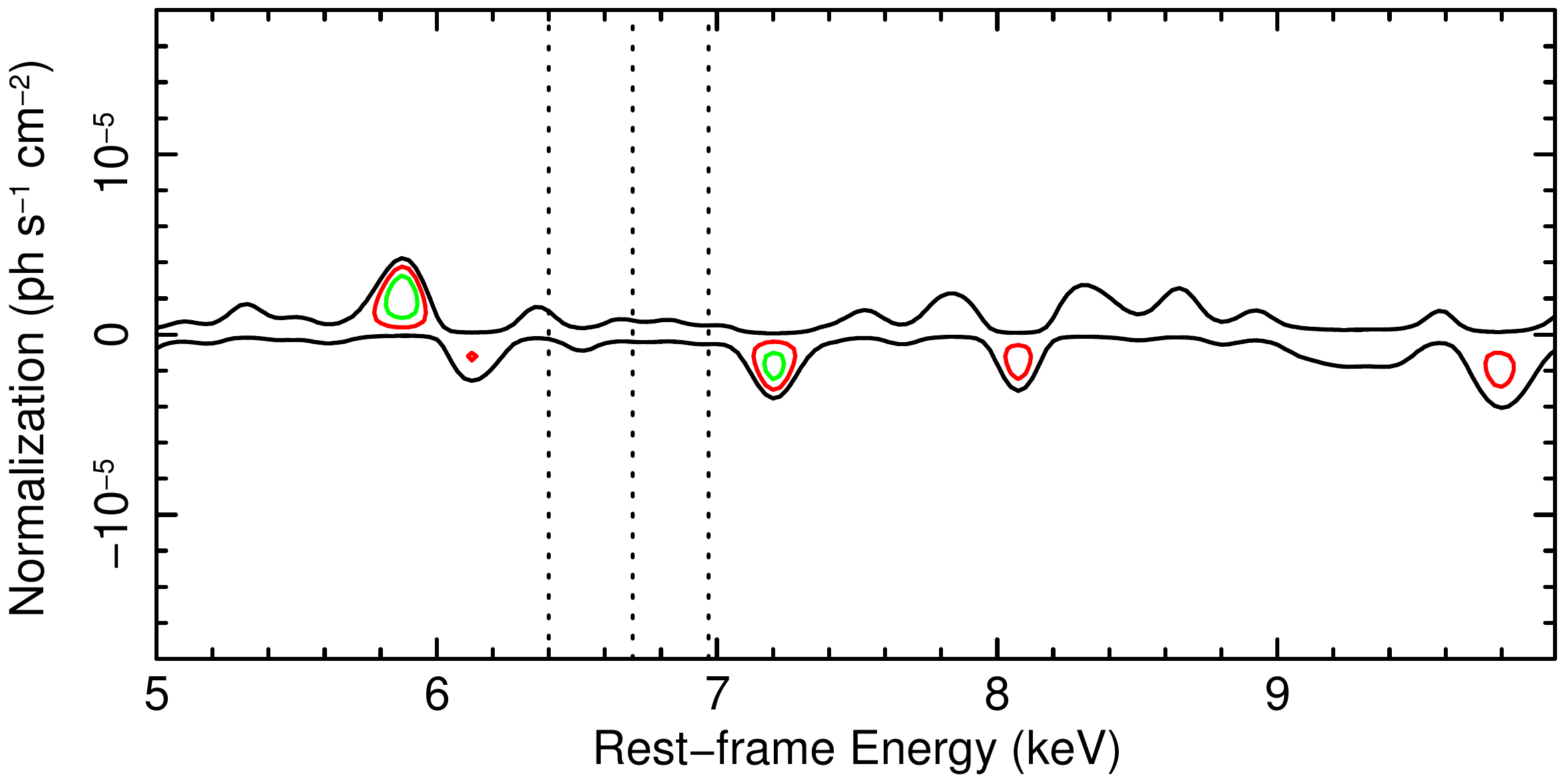}
\includegraphics[angle=-90,width=3.8cm]{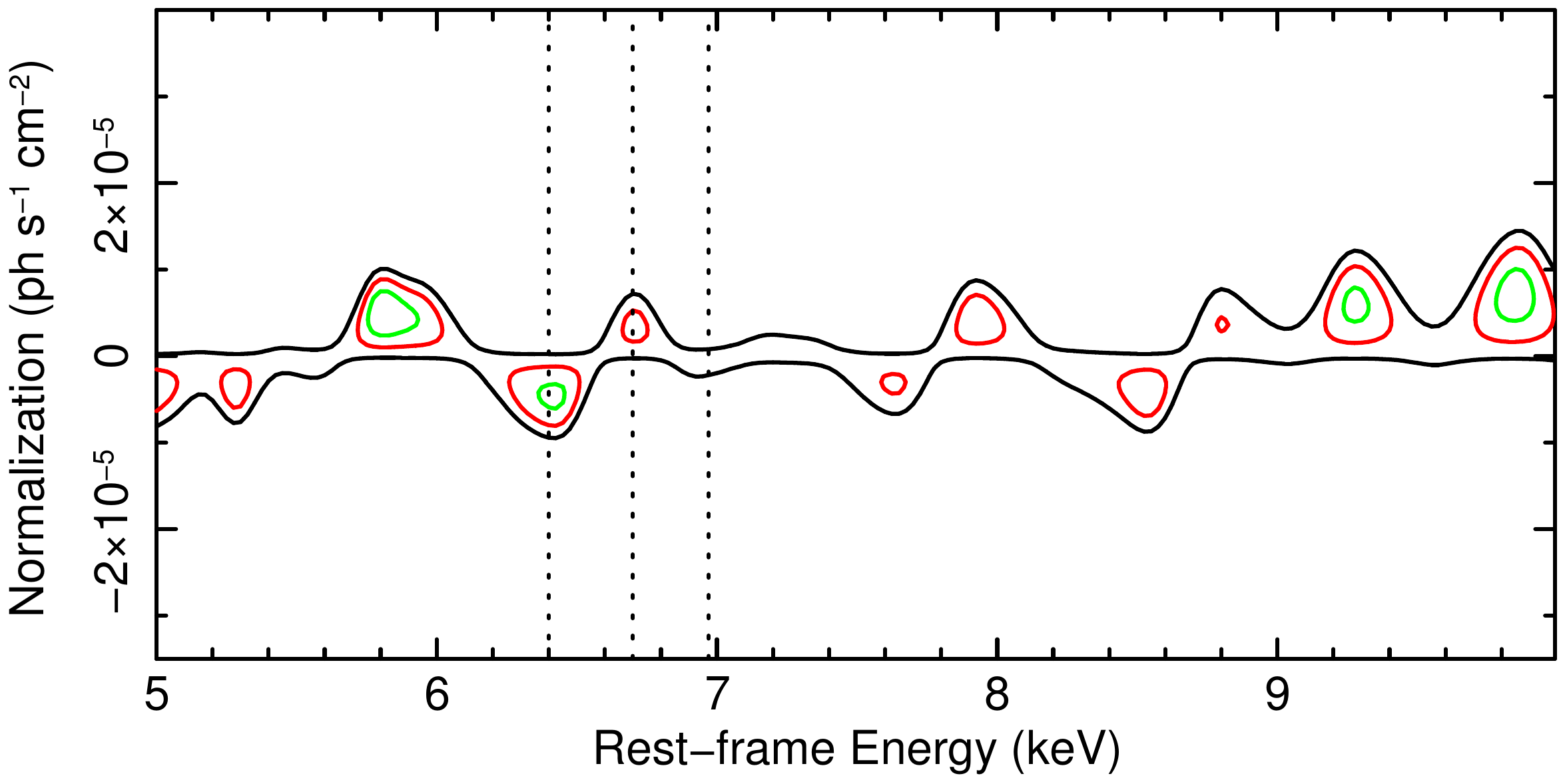}	
}

\end{center}

\caption{\small Ratio and contour plots for sources which {\it do} require an additional broad component at Fe\,K. The top and middle panels again show the primary residuals at Fe\,K and their significances according to the F-test, respectively. The remaining residua after the addition of this broadened component (i.e., a broadened Gaussian or a {\tt diskline}) are shown in the bottom panel. The confidence contours and dashed vertical lines have the same meaning as in Figure \ref{figure:rat_cont1}. Colour version available online.}
\label{figure:rat_cont2}
\end{figure*}

\clearpage

\begin{figure*}
\begin{center}
\vspace{-5pt}
\subfloat{
\includegraphics[angle=-90,width=3.8cm]{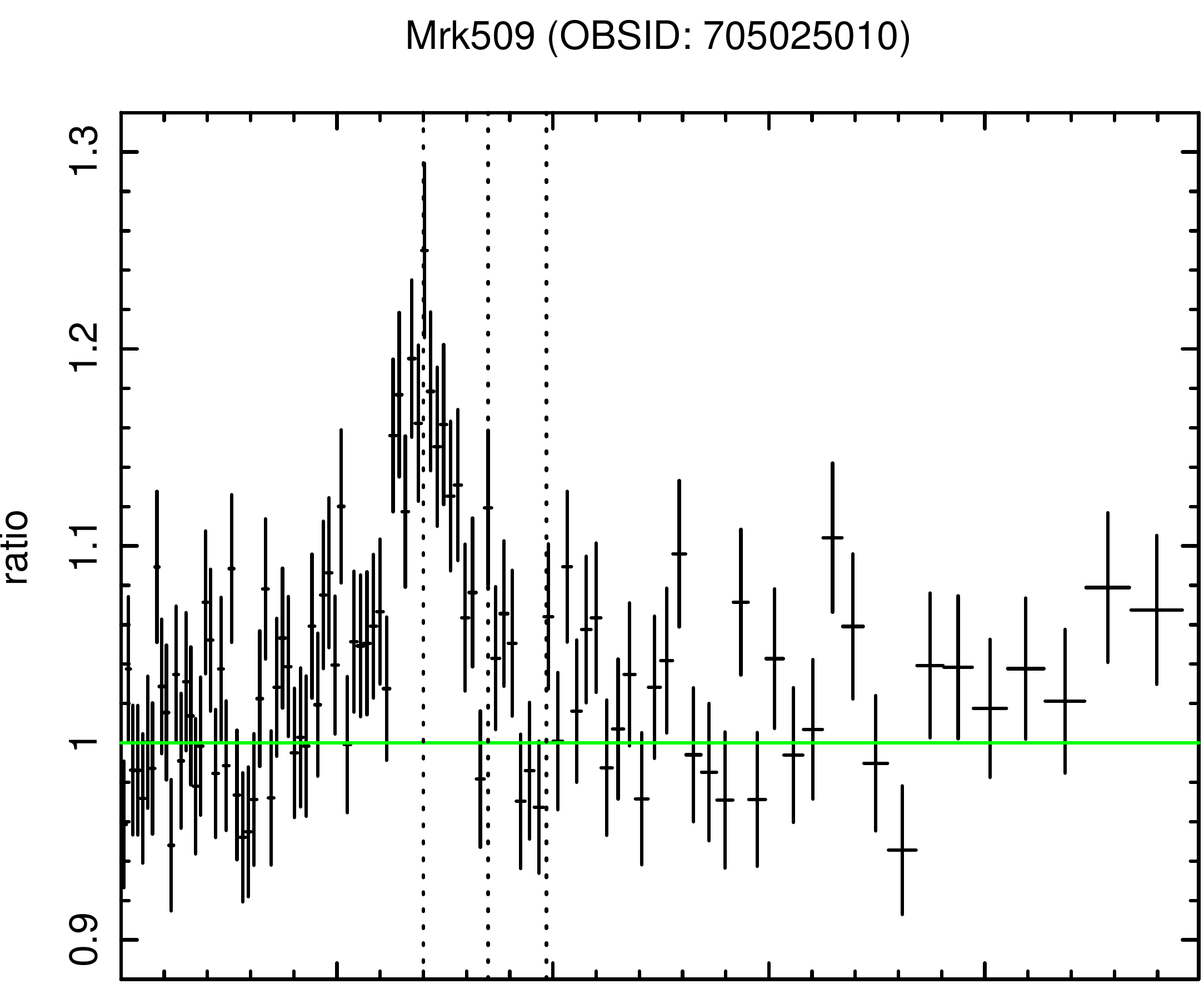}
\includegraphics[angle=-90,width=3.8cm]{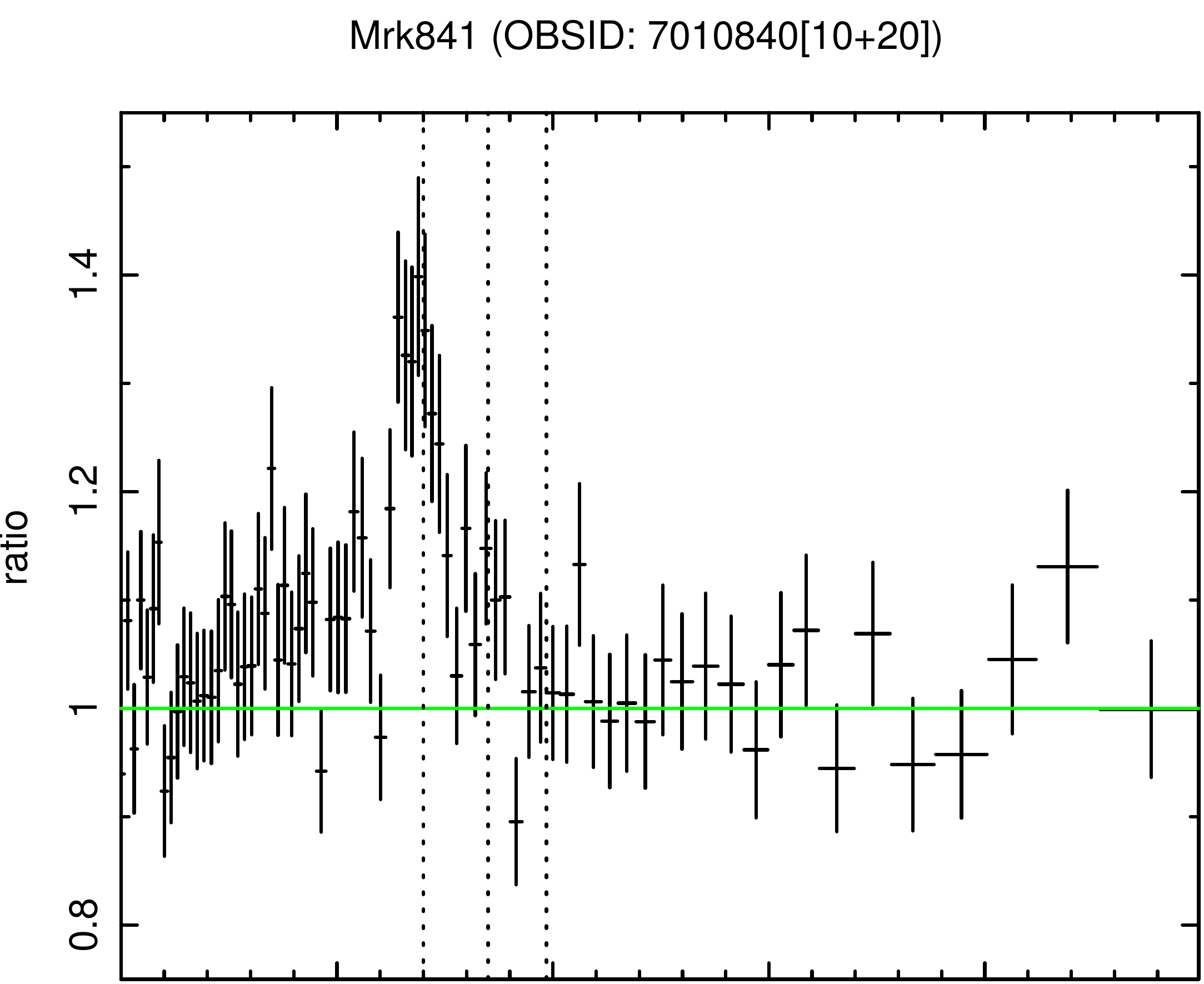}
\includegraphics[angle=-90,width=3.8cm]{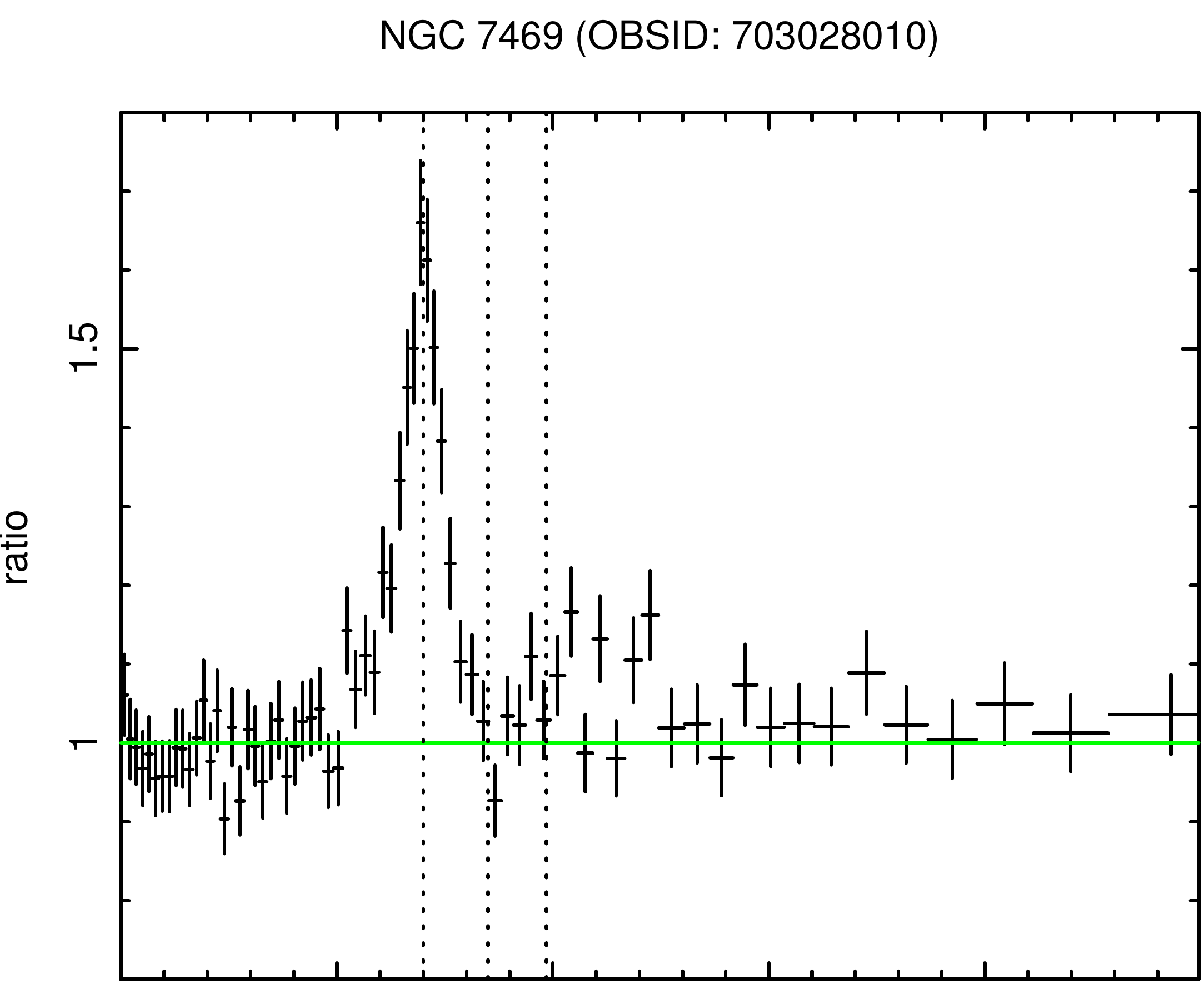}
\includegraphics[angle=-90,width=3.8cm]{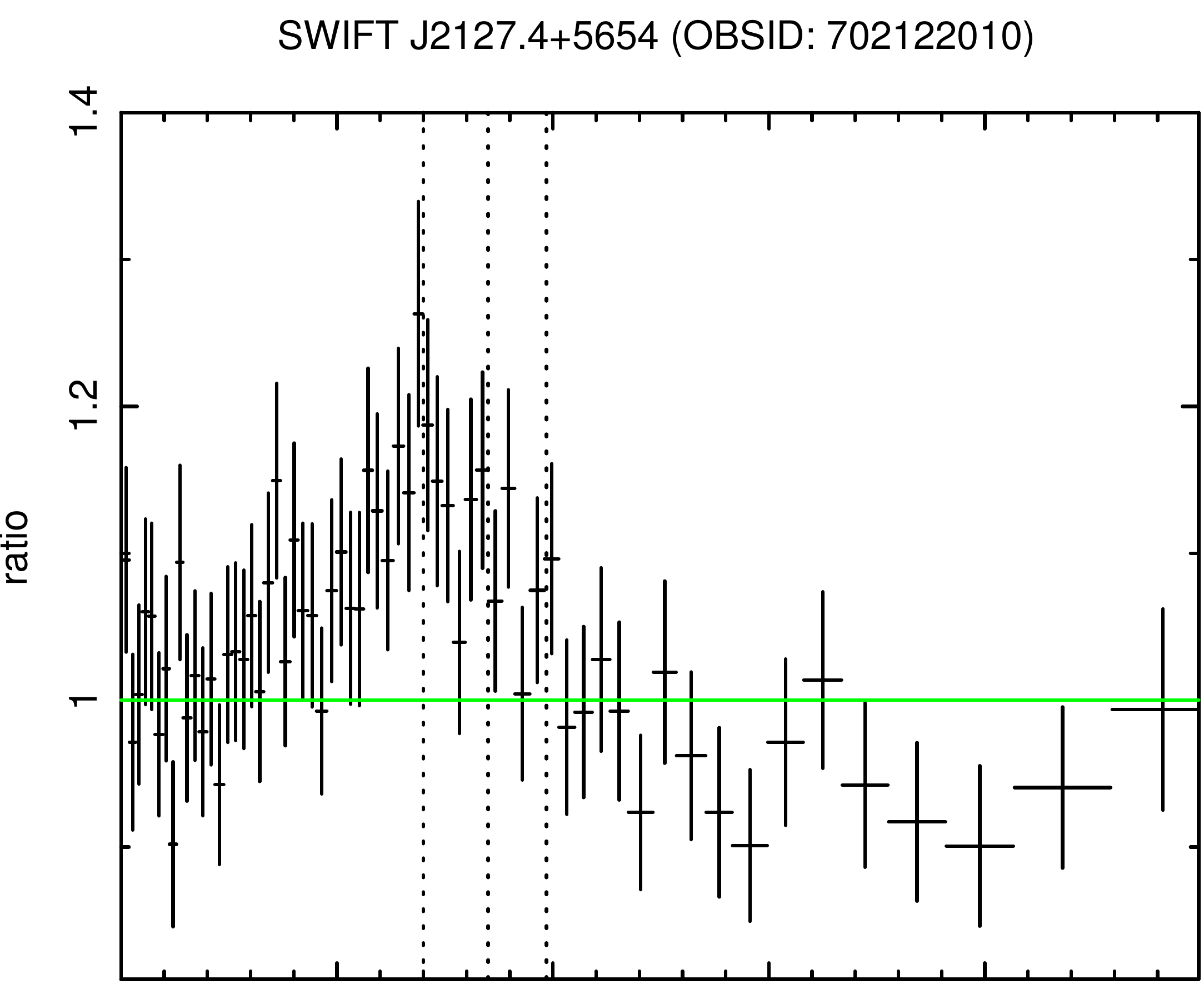}
}

\vspace{-12.2pt}
\subfloat{
\includegraphics[angle=-90,width=3.8cm]{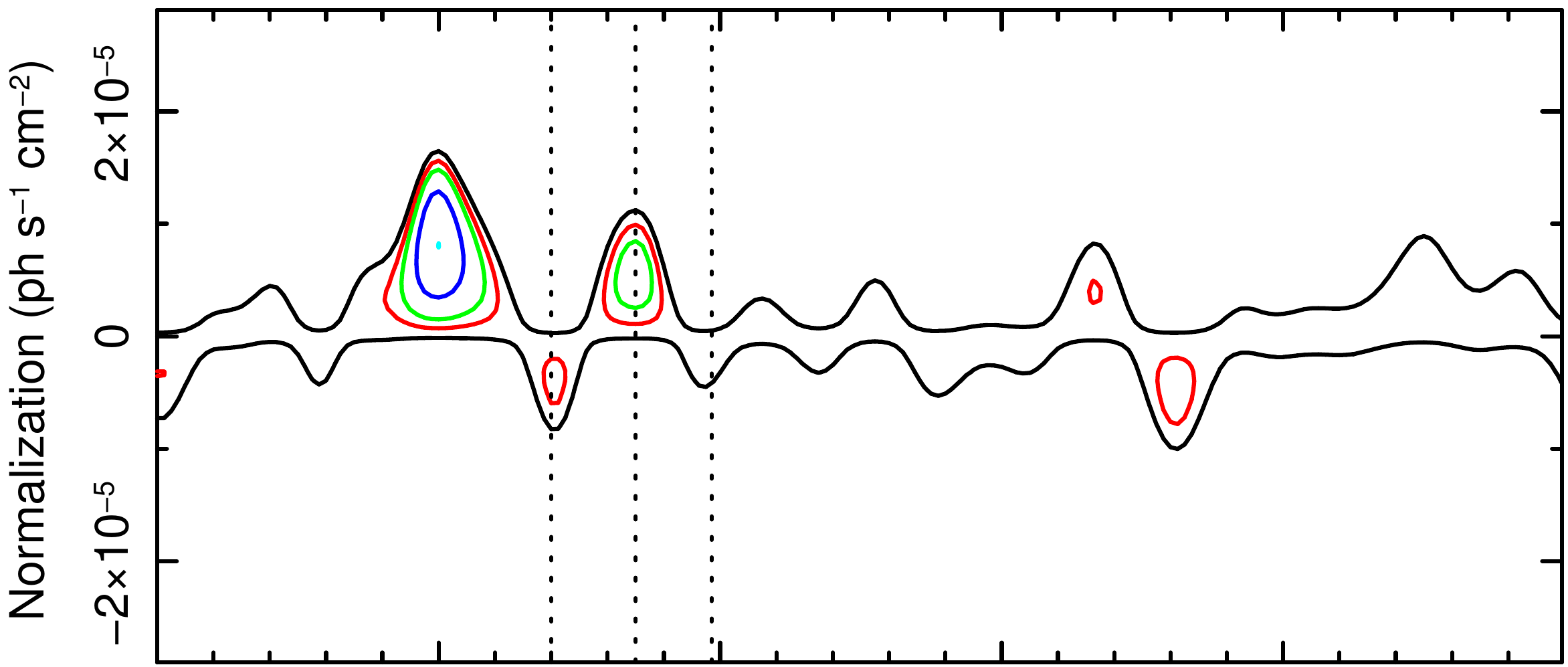}
\includegraphics[angle=-90,width=3.8cm]{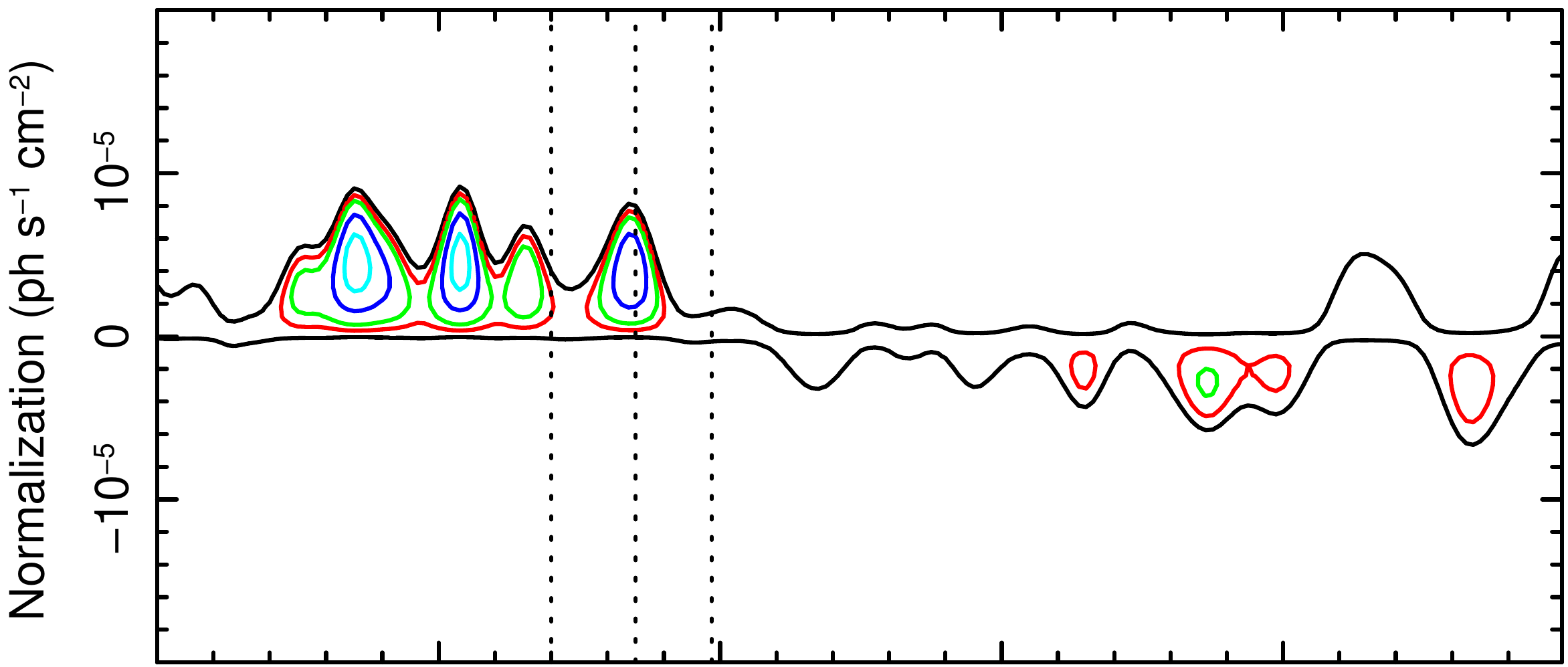}
\includegraphics[angle=-90,width=3.8cm]{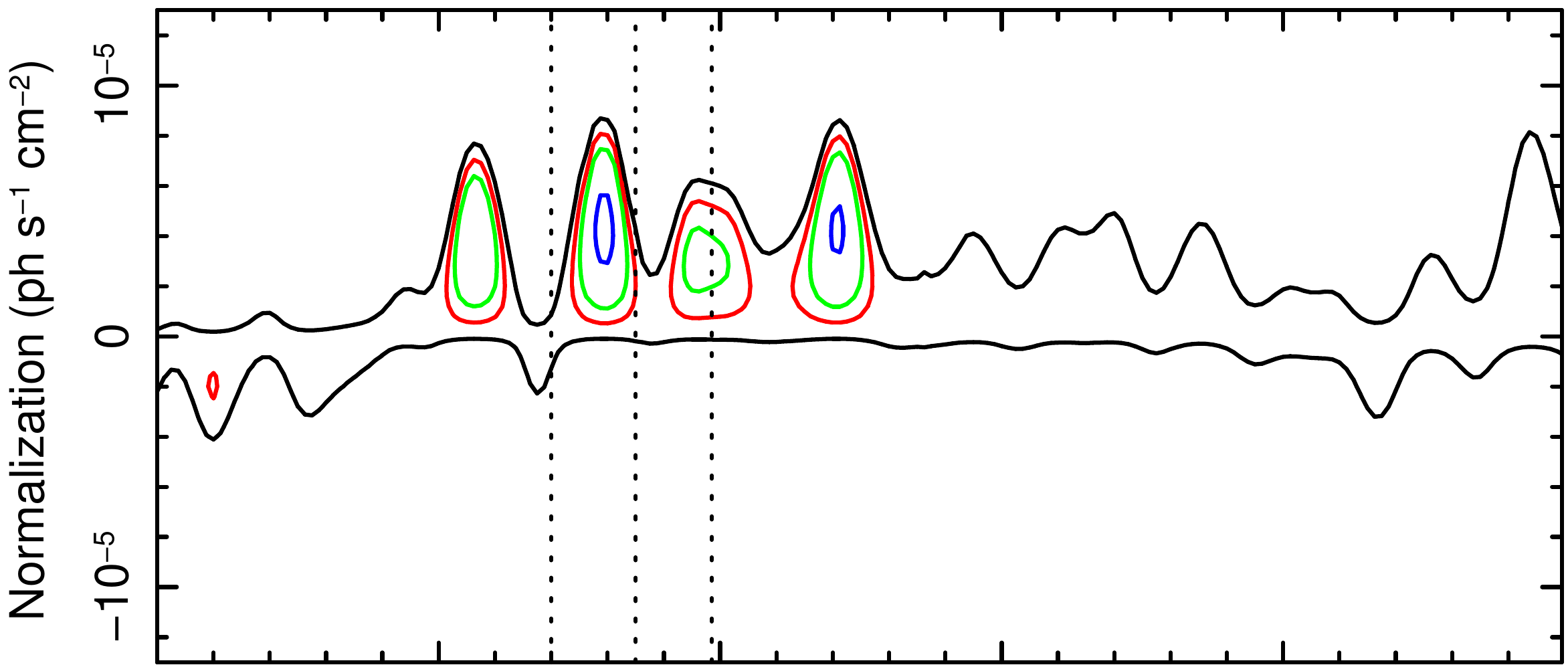}
\includegraphics[angle=-90,width=3.8cm]{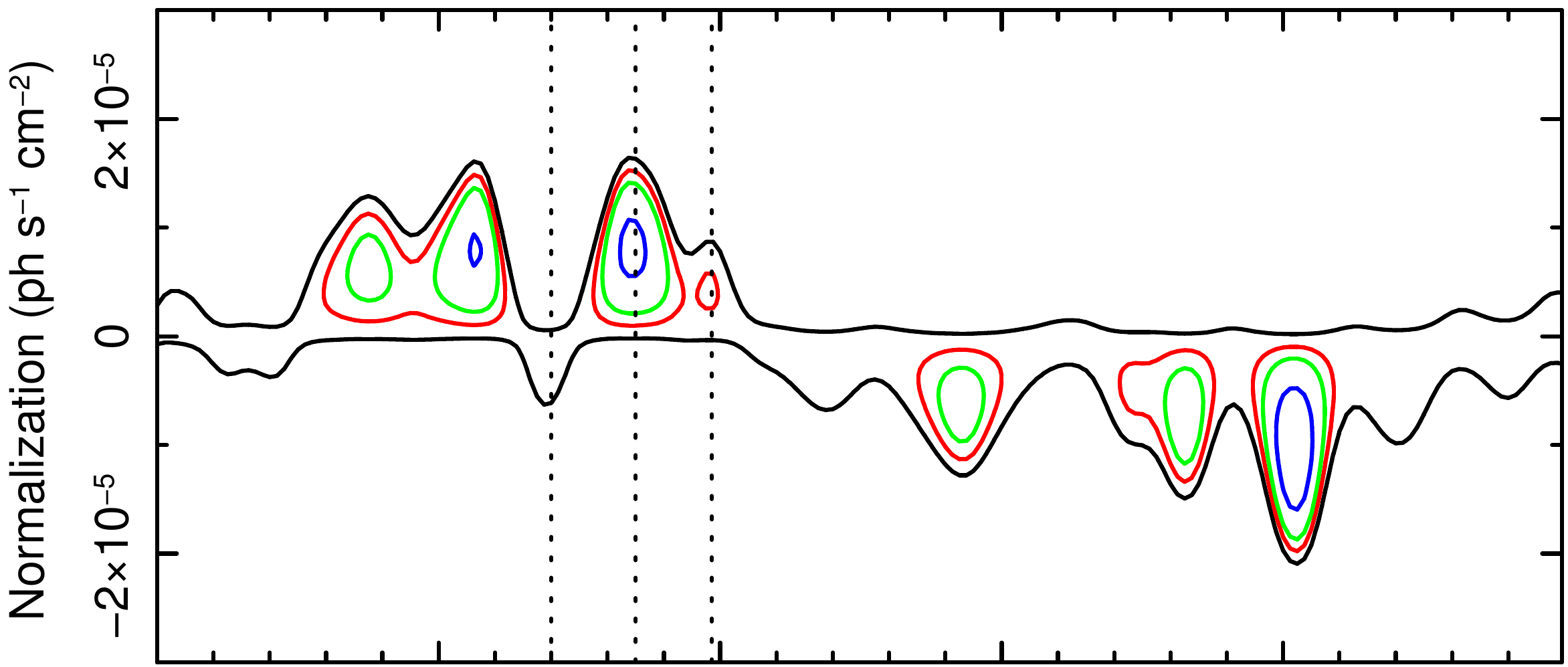}
}

\vspace{-12.2pt}
\subfloat{
\includegraphics[angle=-90,width=3.8cm]{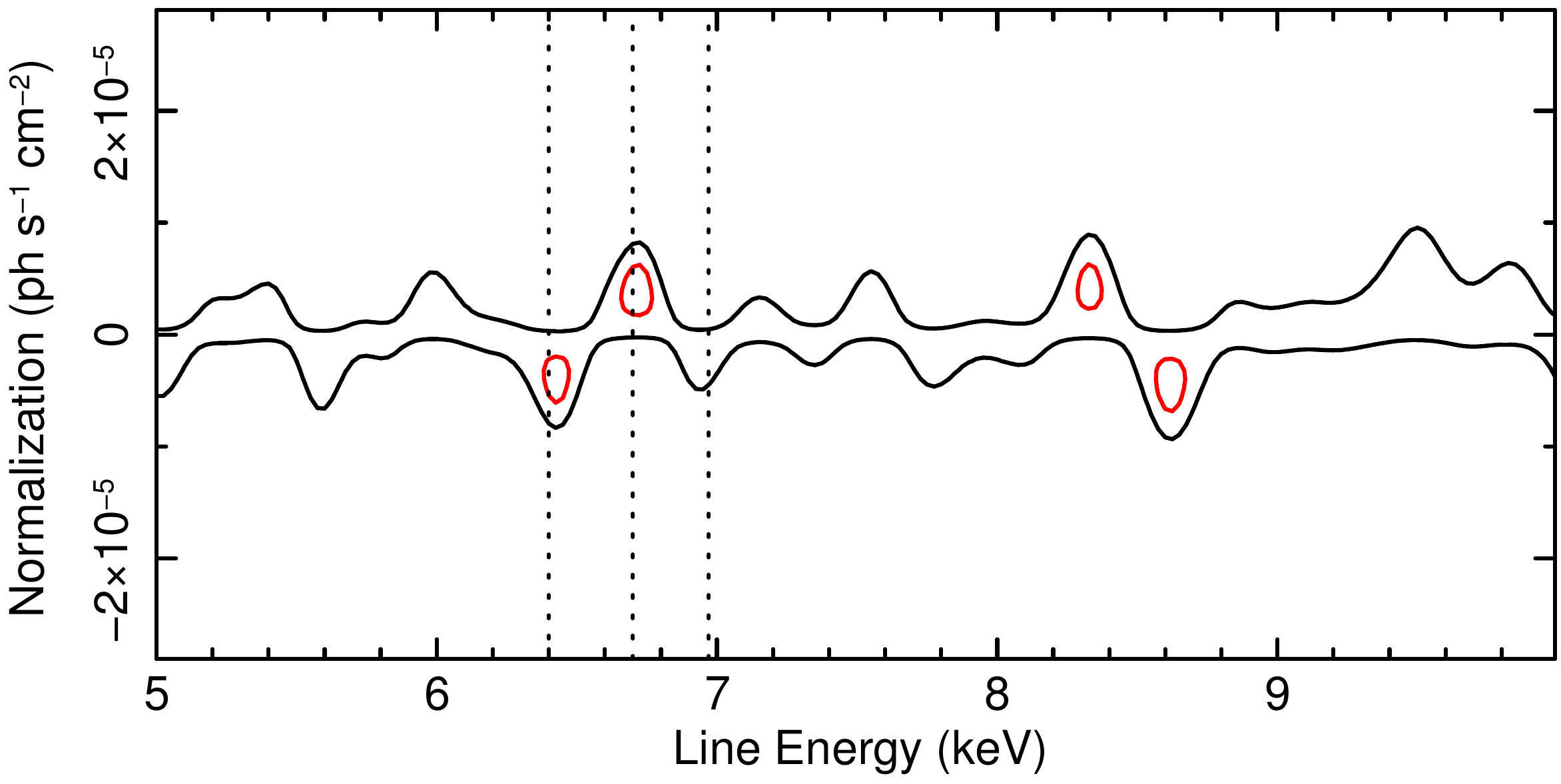}
\includegraphics[angle=-90,width=3.8cm]{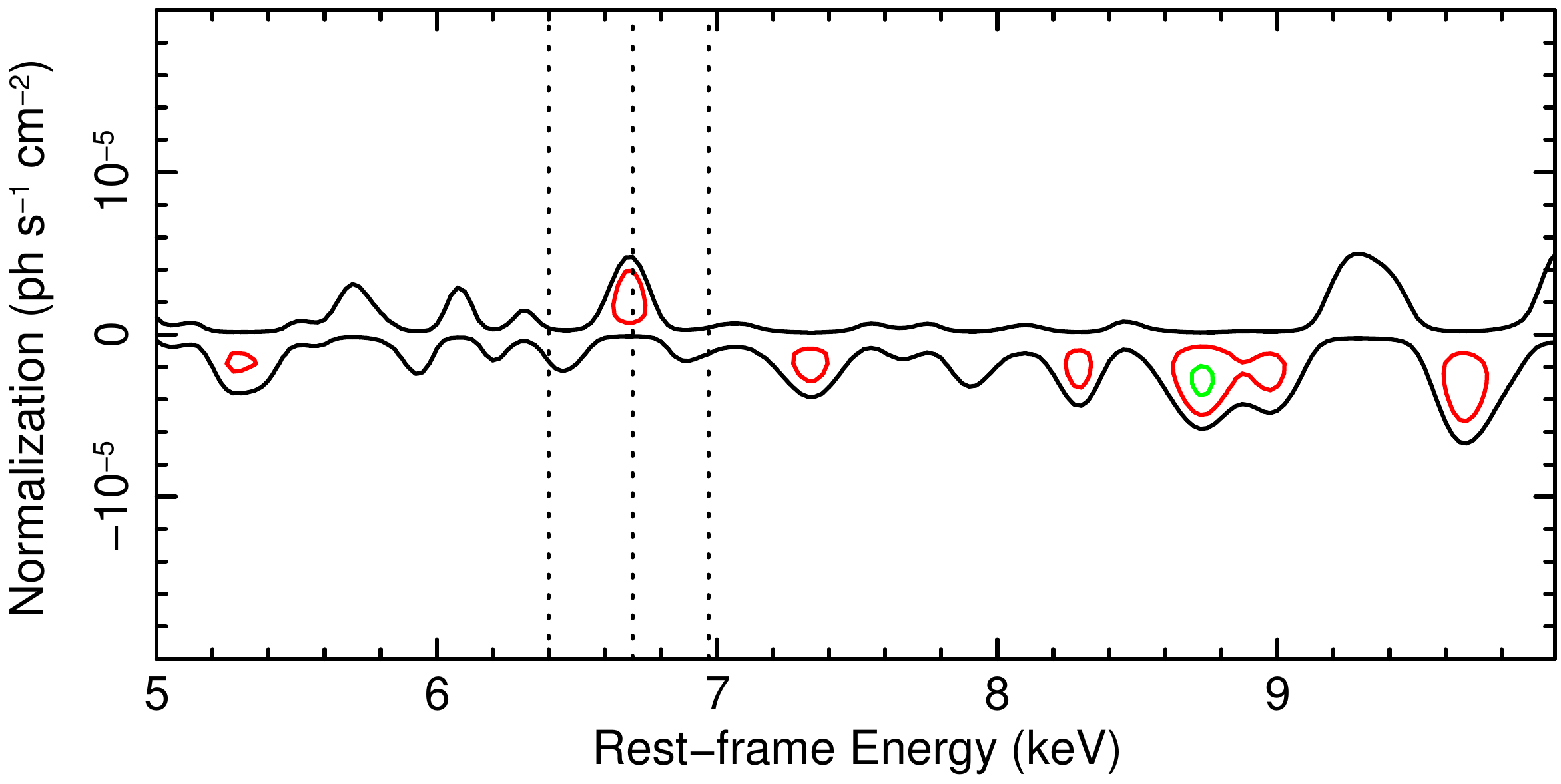}
\includegraphics[angle=-90,width=3.8cm]{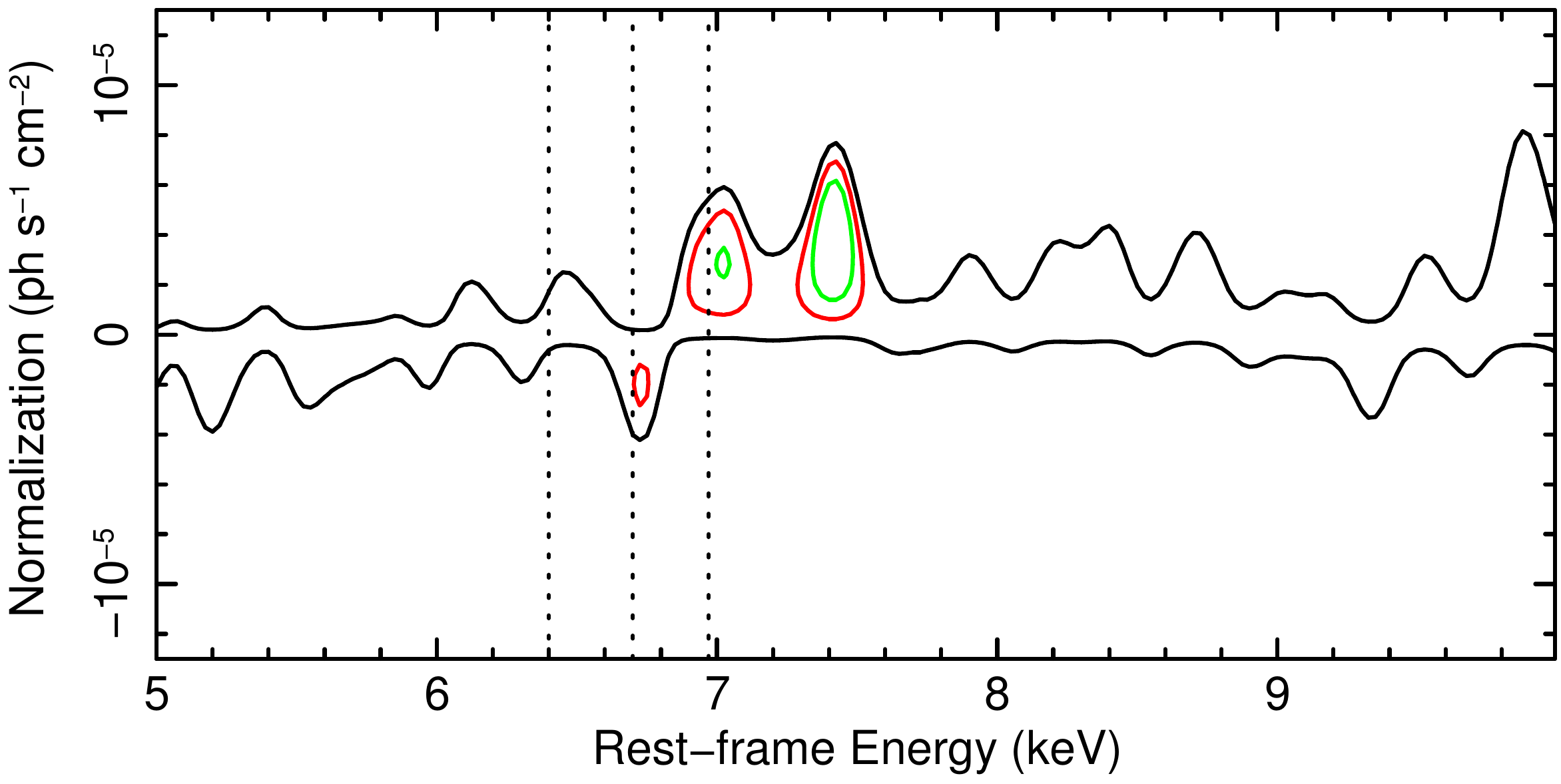}
\includegraphics[angle=-90,width=3.8cm]{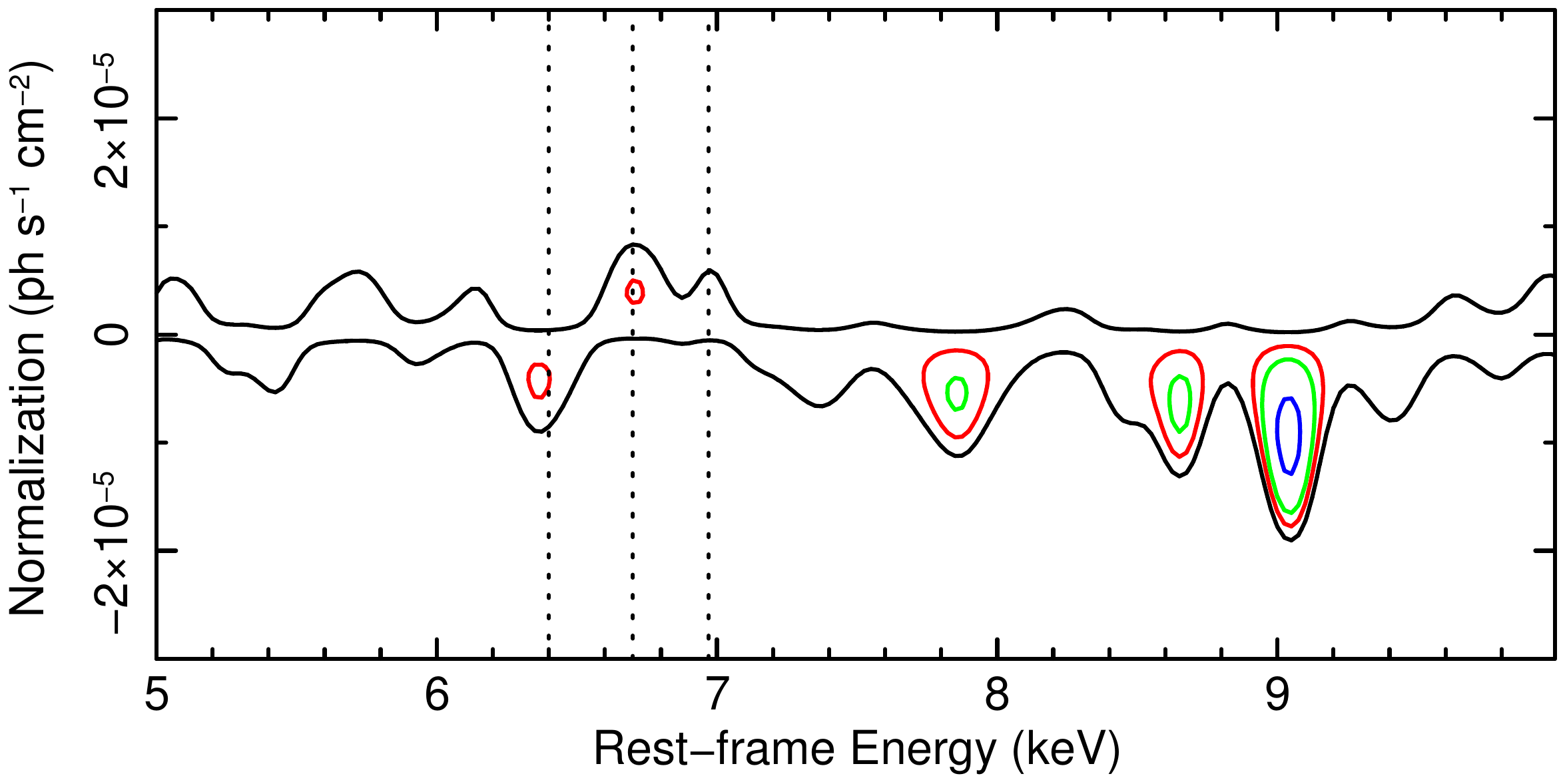}	
}

\end{center}
\contcaption{\small -- Ratio and contour plots for sources which require an additional broad component.}
\end{figure*}

\clearpage

\section{Modelling Complexities}
\label{modelling_complexities}
There are several modelling caveats which could have an affect on the detection of Fe\,K-band absorption lines. Here, we investigate whether there are any degeneracies or uncertainties in the broad-band models which could have an effect on the positively detected absorption lines reported in this work.

\subsection{The Soft-excess}
\label{complexities_softexcess}
A total of 6/20 sources in which we have detected highly-ionised absorption also have a soft-excess. Thus far we have taken the simplistic approach of fitting any excesses below $2$\,keV with a phenomenological blackbody however there are several alternative means of fitting this component which can subtly alter the broadband continuum model, and hence possibly influence the parameters obtained for the Fe\,K-band absorption lines. To assess for this possibility we refitted the soft-excess in each of the 6 sources with two alternative models: {\tt diskbb} (\citealt{mitsuda:1984}) which integrates over the surface of accretion disc to form a multi-colour blackbody spectrum, and {\tt comptt} (\citealt{titarchuk:1994}) which models the excess with the Comptonised emission of soft seed photons by a hot corona above the accretion disc. Both of these alternative models have been used to good effect in the literature when modelling the soft-excess (e.g., \citealt{porquet:2004a, patrick:2011a, gofford:2011, patrick:2011b})

We replaced the \bbody component in each source with both {\tt diskbb} and {\tt comptt} in turn, refitted the data in each case, and compared the resultant continuum and absorption line parameters to those obtained previously. For the {\tt comptt} models a disc geometry was assumed and the input photon seed temperature was allowed to vary between 5\,eV and 0.1\,keV, while the optical depth $\tau$ and temperature $T$ of the Comptonising plasma above the disc were allowed to vary freely, as was the component normalisation. In all cases we found that fitting the excess with {\tt diskbb} or {\tt comptt} had no significant effect on the parameters of either the absorption lines or the continuum, which were always found to be consistent with those obtained with \bbody at the 90\% level. In the heavily absorbed sources such as NCG\,3783, NGC\,4051 and MCG\,-6-30-15 there were some small variations in column density and ionisation parameter for the soft band absorbers when fitting with {\tt comptt}, presumably to compensate for the fact that the Comptonised emission fitted with {\tt comptt} can extend to harder X-ray energies (i.e., $E\geq2-3$\,keV) than typically seen in black-body derived models. Even so, because the key parameters (i.e., those of the continuum and the hard-band absorption lines) are not affected by these subtle changes we do not consider them particularly significant. On the whole, the parameters measured for the highly-ionised absorption systems in these sources are largely independent of the method used to parameterise the soft-excess.

As mentioned in Section~\ref{subsubsec:the_soft_excess} other interpretations of the soft-excess posit that it may be due smeared absorption or blurred reflection associated with an increase in optical depth of lowly ionised circumnuclear material. For simplicity we do not investigate these interpretations further as they likely require a restructuring of the entire broadband model for each source. However, we note that most the outflows detected in these sources have also been detected by \cite{patrick:2012} (see Section~\ref{suzaku_samples} for a detailed sample comparison) who have used blurred reflection models to fit their \suzaku spectra which suggests that the majority of the line detection are robust regardless of the underlying assumptions made when modelling the broad-band continuum.

\subsection{Warm Absorption}
\label{discussion:warm_absorption}
Soft band absorbers with intermediate ionisation states, i.e., those with $\logxi\geq2.5$ where Fe starts to become significantly ionised (\citealt{kallman&mccray:1982, kallman:2004}) can impart weak \fexxv and \fexxvi absorption lines in the Fe\,K band which can reduce the significance of any absorption lines associated with higher ionisation material. Absorption components with ionisation parameters in this regime are present in 4/20 sources with detected Fe\,K outflows. However, because all of the absorbers in the soft band have an implicitly assumed $\vout=0$\,km\,s$^{-1}$ throughout this work, any associated absorption lines will be limited solely to their expected rest-frame energies and only the lowest velocity absorbers in the Fe\,K band are likely to be affected. Moreover, even if this was not the case, the line significances through Gaussian fitting for each of the sources are very high, with $\dchidof\geq25/2$ in all cases. This suggests that even if there is weak Fe\,{\sc xxv-xxvi} from the soft-band absorbers present at higher energies, it is unlikely to have a significant effect on the high velocity absorption systems detected here. For these two reasons we do not regard this effect to pose a problem for our detection of highly-ionised absorption lines in these sources.  

\subsection{Compton Reflection}
\label{complexities_reflection}
The \reflionx reflection model assumes a fixed face-on geometry with all of the reflected emission originating at the surface of the accretion disc (\citealt{ross&fabian:2005}). This means that it is unable to account for the variations of the fluorescence line emissivity or the profile of the Compton reflection hump, both of which have a viewing angle dependence (\citealt{magdziarz&zdziarski:1995}), and can lead the model to either under- or over-predicting the \feka line flux for a given reflection fraction if a source happens to have an inclined accretion disc, or has a more complex reflection geometry. In some sources this can lead to an apparent hard-excess which may be a direct result of the underlying assumptions made with the \reflionx model rather than intrinsic feature of the X-ray spectrum. Alternatively, the observed reflection could have an origin which is not associated with the accretion disc. For example, a strong hard-excess remains in MCG\,-2-58-22 when fitting the reflection continuum with \reflionx, while \cite{rivers:2011} show that no such-excess is present when using the more sophisticated \mytorus reflection model which assumes a distant toroidal reflector.

Strong hard X-ray excesses at $E\gtrsim15$\,keV have been reported in a number of sources in the literature in recent years (e.g., \citealt{turner:2009, reeves:2009, risaliti:2009}; \citealt{tatum:2012b}). The weaker of these hard excesses can sometimes be accounted for by allowing the iron abundance of the reflector to adopt sub-solar values, and thereby enhancing the prominence of the Compton-reflection hump relative to the flux of the \feka fluorescence line. However, this approach is unable to account for the observed excesses in the most extreme cases and an additional spectral component is required to supplement the neutral reflection in the hard X-ray band. There are two spectral interpretations which accomplish this: relativistically enhanced reflection (e.g., \citealt{nardini:2011a, reynolds:2012}) and Compton-thick partial covering (e.g., \citealt{turner:2009, reeves:2009}; \citealt{tatum:2012b}).  

In this work we exclusively use Compton-thick partial covering ($N_{\rm H}\gtrsim \sigma_{\rm T}^{-1}\simeq1.5\times10^{24}$\,cm\,s$^{-1}$) to fit hard-excesses which cannot be accounted for through allowing the iron abundance of the \reflionx component to vary. However, we note that the decision to model hard-excesses with partial covering is driven solely by the desire to construct a broad-band model that is suitable for rapidly computed Monte Carlo simulations, and that the presence of a Compton-thick partial coverer in a model does not necessarily mean that the source is intrinsically covered by a clumpy screen of Compton-thick gas. Indeed, statistically equivalent fits to the hard-excess can likely be achieved with models based on relativistically blurred reflection, but the extreme computation time that would be necessary to run Monte Carlo simulations on these models makes their use impractical in the search for highly-ionised absorption lines and we therefore do not consider them a viable option in this work. Even so, we stress that because the contribution of Compton thick absorption or blurred reflection is largely limited to hard energies, i.e., $E>10-15$\,keV, the manner in which the hard-excess is fitted in these sources is unlikely to have an affect on the detection rate of highly-ionised absorption lines or the parameters of the associated Fe\,K absorbers.

\subsection{Partial-covering Absorption}
\label{discussion:partial_covering}
In addition to the Compton-thick partial covering used to supplement the reflection continuum in sources with hard-excesses, we have also used moderate columns of partial-covering absorption ($N_{\rm H}\lesssim \rm{few} \times 10^{23}$\,cm$^{-2}$) to fit spectral curvature and spectral variability below $10$\,keV. In terms of the sources with detected Fe\,K outflows 12/20 have at least one partially covering component (including those which are Compton-thick), with 5 of those having more than one partial coverer in a layered or clumpy geometry. The principal driver behind the inclusion of partial covering absorption in these sources is the need to account for the variability in spectral shape which are apparent when more than one observation has been simultaneously fitted, especially if those observations are separated by a number of years. The addition of a partially covering component allowed for the variability observed in Mrk\,766, NGC\,1365, NGC\,3227, NGC\,3516, NGC\,3783, NGC\,4051 and PDS\,456 to be easily accounted for, and enables the absorption line parameters in these sources to be assessed against a self-consistently fitted underlying continuum in all observations. It is important to note, however, that regardless of the underlying continuum model the detection of highly-ionised absorption lines in these sources is largely model independent and does not affect the overall statistics reported in this work.

\begin{table*}
	\section{Full model details}
	\label{appendix:model_parameters}
	\scriptsize
	\begin{center}
		\begin{minipage}{165mm}
		\caption{Summary of continuum parameters. 
					{\sl Notes}: 
							(1) Source name;
							(2) \suzaku observation ID. See table \ref{tab:observation_details} for details on stacked spectra;
						   	(3) Power-law photon index;  
						   	(4) Power-law component normalisation in 
						   		units of $\times10^{-3}$ ph\,kev$^{-1}$\,cm$^{-2}$\,s$^{-1}$ at 1\,keV; 
						   	(5) Blackbody thermal temperature in units of eV; 
						   	(6) Blackbody normalisation in units of $\times10^{-5}\times(L_{39}/D_{10}^{2})$, where $L_{\rm 39}$ is the source luminosity in units of $10^{39}$\,erg\,s$^{-1}$ and $D_{\rm 10}$ is the distance to the source in units of 10\,kpc; 
						   	(7) Ionisation parameter of the \reflionx X-ray reflection component,
						 	   where $\xi=4\pi F/n$; 
						   	(8) Iron abundance for the X-ray reflector, in Solar units;
						   	(9) Reference for warm absorption parameters (if present). `D2' and `D3' refer to Tables \ref{tab:warmabs_single} and \ref{tab:warmabs_multi}, respectively;
						   	(10) Reduced $\chi^{2}$ and number of degrees of freedom ($\nu$) for the final best-fit model.}

\begin{tabular}{@{}lllrcrcclc}
		
	\toprule
	\multicolumn{1}{c}{\multirow{3}{*}{Source}} & 
	\multicolumn{1}{c}{\multirow{3}{*}{OBSID}} & 
	\multicolumn{2}{c}{Power-law} & 
	\multicolumn{2}{c}{bbody} & 
	\multicolumn{2}{c}{Reflection} & 
	\multicolumn{1}{c}{\multirow{2}{*}{Abs.}} & 
	\multicolumn{1}{c}{\multirow{2}{*}{$\chi^{2}_{r}(\nu)$}}\\

	& 
	& 
	\multicolumn{1}{c}{$\Gamma$} & 
	\multicolumn{1}{c}{norm} & 
	\multicolumn{1}{c}{$k_{B}T$} & 
	\multicolumn{1}{c}{norm} & 
	\multicolumn{1}{c}{$\xi$} & 
	\multicolumn{1}{c}{$Z_{\rm Fe}$} \\

	\multicolumn{1}{c}{(1)} & 
	\multicolumn{1}{c}{(2)} & 
	\multicolumn{1}{c}{(3)} & 
	\multicolumn{1}{c}{(4)} & 
	\multicolumn{1}{c}{(5)} & 
	\multicolumn{1}{c}{(6)} & 
	\multicolumn{1}{c}{(7)} & 
	\multicolumn{1}{c}{(8)} & 
	\multicolumn{1}{c}{(9)} & 
	\multicolumn{1}{c}{(10)}\\
	\midrule

	1H\,0419-577 
		& stacked[all] & $2.36\pm0.02$ & $33.71^{+1.06}_{-1.06}$ & \na & \na & $<1.03$ & 1.00$^{*}$ & D2 & $1.08(3070)$\\[0.5ex]
	3C\,111$^{\Diamond}$ 
		& 703034010$^{j}$ & $1.74^{+0.01}_{-0.01}$ & $4.98^{+0.04}_{-0.04}$ & \na & \na & $<1.06$ & 1.00\fix 
					& D3 & $1.07(72128)$\\[0.5ex]
		& 705040010$^{j}$ & $1.75^{*}$ & $12.29^{+0.10}_{-0.10}$ & \na & \na & $<3.21$ & 1.00$^{*}$\\[0.5ex]
		& 705040020$^{j}$ & $1.75^{*}$ & $16.44^{+0.13}_{-0.13}$ & \na & \na & $<3.21^{*}$ & 1.00$^{*}$\\[0.5ex]
		& 705040030$^{j}$ & $1.75^{*}$ & $16.73^{+0.13}_{-0.13}$ & \na & \na & $<3.21^{*}$ & 1.00$^{*}$\\[0.5ex]
	3C\,120 
		& 700001010$^{j}$ & (1)$1.58^{+0.04}_{-0.05}$ & $7.15^{+0.07}_{-0.07}$ & \na & \na & $30.02^{+2.02}_{-1.89}$ & 1.00\fix & D3 & $1.02(3841)$\\[0.5ex]
					 & & (2)$2.56^{+0.09}_{-0.08}$ & $5.07^{+0.08}_{-0.07}$\\[0.5ex]
		& Stacked[bcd]$^{j}$ & (1)$1.58^{*}$ & $7.69^{+0.08}_{-0.09}$ & \na & \na & $30.02^{*}$ & $1.00^{*}$\\[0.5ex]
					 & & (2)$2.56^{*}$ & $8.77^{+0.09}_{-0.08}$\\[0.5ex]
	3C\,382 
		& 702125010 & $1.86^{+0.01}_{-0.01}$ & $12.42^{+0.09}_{-0.07}$ & $95^{+8}_{-8}$ & $1.39^{+0.50}_{-0.32}$ 
					& $<1.38$ & 1.00\fix & D2 & $1.06(3020)$\\[0.5ex]
	3C\,390.3$^{\Diamond}$ 
		& 701060010 & $1.72^{+0.01}_{-0.02}$ & $7.45^{+0.10}_{-0.12}$ & $149^{+26}_{-31}$ & $1.07^{+0.35}_{-0.36}$ 
					& $<5.11$ & 1.00\fix & \na & $1.02(2666)$\\[0.5ex]
	3C\,445 
		& 702056010 & $1.85^{+0.05}_{-0.05}$ & $4.69^{+0.31}_{-0.31}$ & \na & \na & $<22.65$ & 1.00\fix & D2 & $1.05(530)$\\[0.5ex]
	4C\,+74.26$^{\Diamond}$ 
		& 702057010 & $1.99^{+0.02}_{-0.02}$ & $11.50^{+0.32}_{-0.32}$ & \na & \na & $1.91^{+0.36}_{-0.71}$ & $0.28^{+0.10}_{-0.09}$ 
					& D2 & $1.01(2271)$\\[0.5ex]
	APM\,08279+5255$^{\Diamond}$ 
		& stacked & $1.89^{+0.04}_{-0.04}$ & $0.14^{+0.01}_{-0.01}$ & \na & \na & \na & \na & D2 & $1.08(351)$\\[0.5ex]
	Ark\,120 
		& 702014010 & $2.06^{+0.01}_{-0.01}$ & $11.87^{+0.15}_{-0.15}$ & $150^{+5}_{-5}$ & $9.50^{+0.72}_{-0.71}$ 
					& $<3.5$ & 1.00\fix & \na & $1.06(2543)$\\[0.5ex]
	Ark\,564
		& 702117010 & $2.56^{+0.01}_{-0.01}$ & $25.95^{+0.36}_{-0.36}$ & $146^{+1}_{-1}$ & $40.22^{+0.13}_{-0.13}$ 
					& $<1.20$ & 1.00\fix & D2 & $1.02(1967)$\\[0.5ex]
	CBS\,126 $^{\Diamond}$
		& 705042010 & $2.06^{+0.03}_{-0.03}$ & $1.75^{+0.07}_{-0.06}$ & $71^{+5}_{-5}$ & $22.39^{+9.72}_{-6.44}$ & $<2.96$ & 1.00\fix 
					& D2 & $1.03(1003)$\\[0.5ex]
	ESO\,103-G035$^{\Diamond}$
		& 703031010 & $2.04^{+0.02}_{-0.02}$ & $28.99^{+0.95}_{-0.95}$ & \na & \na & $<52.01$ & 1.00\fix 
					& D2 & $0.97(1444)$\\[0.5ex]
	Fairall\,9
		& 702043010$^{j}$ & $2.04^{+0.01}_{-0.01}$ & $8.32^{+0.04}_{-0.04}$ & $104^{+1}_{-1}$ & $7.08^{+0.72}_{-0.72}$ 
					& $<1.01$ & $0.59^{+0.04}_{-0.04}$ & \na & $1.09(3706)$\\[0.5ex]
		& 705063010$^{j}$ & $2.11^{+0.01}_{-0.01}$ & $10.20^{+0.05}_{-0.05}$ & $104^{*}$ & $11.71^{+0.09}_{-0.09}$ 
					& $<1.01^{*}$ & $0.59^{*}$\\[0.5ex]
	IC\,4329A
		& Stacked[all] & $1.95^{+0.01}_{-0.01}$ & $37.18^{+0.11}_{-0.11}$ & \na & \na & $<1.02$ & $0.40^{+0.03}_{-0.03}$ 
					& D2 & $1.03(3336)$\\[0.5ex]
	IGR\,J16185-5928 
		& 702123010 & $1.97^{+0.01}_{-0.01}$ & $2.84^{+0.03}_{-0.03}$ & \na & \na & $<2.29$ & $0.51^{+0.27}_{-0.17}$ 
					& D2 & $0.97(1004)$\\[0.5ex]
	IGR\,J21247+5058 
		& 702027010 & $1.61^{+0.01}_{-0.01}$ & $18.49^{+0.18}_{-0.18}$ & \na & \na & $<1.17$ & 1.00\fix & D2 & $1.06(2778)$\\[0.5ex]
	MCG\,-02-14-009
		& 703060010 & $1.84^{+0.04}_{-0.04}$ & $1.21^{+0.06}_{-0.06}$ & $187^{+32}_{-40}$ & $0.45^{+0.17}_{-0.17}$ 
					& $<11.14$ & 1.00\fix & \na & \\[0.5ex]
	MCG\,-2-58-22 
		& 704032010 & $1.85^{+0.01}_{-0.01}$ & $20.78^{+0.05}_{-0.05}$ & \na & \na & $<3.07$ & 1.00\fix & D2 & $1.06(2991)$\\[0.5ex]
	MCG\,-5-23-16
		& 700002010 & $2.13^{+0.01}_{-0.01}$ & $88.08^{+0.04}_{-0.04}$ & $752^{+41}_{-84}$ & $5.78^{+0.56}_{-0.56}$ 
					& $<4.63$ & 1.00\fix & D2 & $1.03(3250)$\\[0.5ex]
	MCG\,-6-30-15$^{\Diamond}$
		& stacked[all] & $2.06^{+0.01}_{-0.01}$ & $20.72^{+0.41}_{-0.41}$ & $115^{+3}_{-3}$ & $23.65^{+1.04}_{-1.04}$ 
					& $<1.05$ & 1.00\fix & D2 & $1.03(2334)$\\[0.5ex]
	MCG\,+8-11-11
		& 702112010 & $1.72^{+0.01}_{-0.01}$ & $16.85^{+0.06}_{-0.06}$ & \na & \na & $<2.61$ & $1.60^{+0.23}_{-0.23}$ 
					& D2 & $1.04(3030)$\\[0.5ex]
	MR\,2251-178$^{\Diamond}$
		& 704055010 & $1.65^{+0.01}_{-0.01}$ & $10.21^{+0.04}_{-0.04}$ & $64^{+3}_{-3}$ & $9.79^{+3.12}_{-2.30}$ 
					& $14.17^{+7.65}_{-4.63}$ & 1.00\fix & D2 & $1.05(3117)$\\[0.5ex]
	Mrk\,79 
		& 702044010 & $1.94^{+0.02}_{-0.02}$ & $5.18^{+0.13}_{-0.13}$ & \na & \na & $<1.34$ & 1.00\fix & D2 & $0.94(1791)$\\[0.5ex]
	Mrk\,110 
		& 702124010 & $1.81^{+0.01}_{-0.01}$ & $5.93^{+0.10}_{-0.10}$ & $153^{+12}_{-13}$ & $2.65^{+0.38}_{-0.38}$ 
					& $20.19^{+5.25}_{-8.75}$ & 1.00\fix & \na & $0.96(1994)$\\[0.5ex]
	Mrk\,205 
		& 705062010 & $1.94^{+0.02}_{-0.02}$ & $7.85^{+0.13}_{-0.15}$ & $115^{+17}_{-18}$ & $2.33^{+1.18}_{-0.58}$ 
					& $9.85^{+0.68}_{-1.42}$ & 1.00\fix & D2 & $1.05(1344)$\\[0.5ex]
	Mrk\,279$^{\Diamond}$
		& 704031010 & $1.76^{+0.01}_{-0.01}$ & $1.10^{+0.01}_{-0.01}$ & \na & \na & $1.3^{+0.9}_{-0.3}$ & 1.0\fix 
					& D2 & $0.99(1442)$\\[0.5ex]
	Mrk\,335 
		& 701031010 & $2.17^{+0.01}_{-0.01}$ & $6.94^{+0.07}_{-0.07}$ & $123^{+1}_{-1}$ & $2.09^{+0.06}_{-0.06}$ & $<3.00$ 
					& $0.51^{+0.16}_{-0.10}$ & D2 & $1.09(2423)$\\[0.5ex]
	Mrk\,359 
		& 701082010 & $1.81^{+0.03}_{-0.03}$ & $1.39^{+0.04}_{-0.04}$ & $207^{+22}_{-18}$ & $0.73^{+0.19}_{-0.19}$ 
					& $2.72^{+0.22}_{-0.22}$ & $0.54^{+0.22}_{-14}$ & \na & $1.11(1126)$\\[0.5ex]
	Mrk\,509
		& 701093010$^{j}$ & (1)$1.67^{+0.03}_{-0.03} $ & $8.92^{+0.57}_{-0.53}$ & \na & \na & $4.12^{+0.25}_{-0.25}$ & 1.00\fix 
					& D3 & $1.01(5502)$\\[0.5ex]
		& 				  & (2)$2.67^{+0.05}_{-0.05}$ & $7.15^{+0.57}_{-0.60}$ \\[0.5ex]
		& stacked[bcd]$^{j}$ & (1)$1.67^{*}$ & $8.43^{+0.60}_{-0.56}$ & \na & \na & $4.12^{*}$ & $1.00^{*}$\\[0.5ex]
		& 					 & (2)$2.67^{*}$ & $10.52^{+0.58}_{-0.62}$ \\[0.5ex]
		& 705025010$^{j}$ & (1)$1.67^{*}$ & $9.54^{+0.70}_{-0.65}$ & \na & \na & $4.12^{*}$ & $1.00^{*}$\\[0.5ex]
		& 				  & (2)$2.67^{*}$ & $12.66^{+0.66}_{-0.72}$\\[0.5ex]
	Mrk\,766$^{\Diamond}$
		& 701035010$^{j}$ & $2.10^{+0.01}_{-0.01}$ & $7.19^{+0.06}_{-0.06}$ & \na & \na & $1.95^{+0.26}_{-0.45}$ 
					& 1.00\fix & D3 & $1.05(2258)$\\[0.5ex]
		& 701035020$^{j}$ & $2.10^{*}$ & $5.92\pm0.03$ & \na & \na & $1.95^{*}$ & $1.00^{*}$\\[0.5ex]
	Mrk\,841 
		& stacked[all] & $1.91^{+0.01}_{-0.01}$ & $4.25^{+0.04}_{-0.04}$ & $84^{+7}_{-8}$ & $9.58^{+4.42}_{-2.42}$ 
		& $<1.03$ & $0.46^{+0.09}_{-0.08}$ & D2 & $1.02(1892)$\\[0.5ex]
	NGC\,1365$^{\Diamond}$ 
		& 702047010$^{j}$ & $1.87^{+0.02}_{-0.02}$ & $11.92^{+0.23}_{-0.23}$ & \na & \na & $1.34^{+0.08}_{-0.08}$ 
					& 1.00\fix & D3 & $1.06(2323)$\\[0.5ex]
	    & 705031010$^{j}$ & $1.87^{*}$ & $10.43^{+0.33}_{-0.33}$ & \na & \na & $1.34^{*}$ & $1.00^{*}$\\[0.5ex]
	    & 705031020 	  & $1.87^{+0.03}_{-0.03}$ & $7.92^{+0.55}_{-0.55}$ & \na & \na & $1.14^{+0.18}_{-0.10}$ 
	    				& 1.00\fix & D2 & $1.04(978)$\\[0.5ex]

	\bottomrule
		\end{tabular}\\[0.5ex]
		$^{*}$ Parameter was tied during fitting.\\
		$^{f}$ Parameter was frozen at this value during fitting.\\
		$^{j}$ Indicates that spectra were fit simultaneously.\\
		$^{\Diamond}$ indicates that a highly-ionised outflow is detected in this source.
		\label{table:continuum_parameters}
		\end{minipage}
	\end{center}
\end{table*}

\clearpage

\begin{table*}
	\scriptsize
	\begin{center}
		\begin{minipage}{171mm}
		\contcaption{-- Summary of continuum parameters}
		\begin{tabular}{@{}lllrcrcclc}
		
	\toprule
	\multicolumn{1}{c}{\multirow{2}{*}{Source}} & 
	\multicolumn{1}{c}{\multirow{2}{*}{OBSID}} & 
	\multicolumn{2}{c}{Power-law} & 
	\multicolumn{2}{c}{bbody} & 
	\multicolumn{2}{c}{Reflection} & 
	\multicolumn{1}{c}{\multirow{2}{*}{Abs.}} & 
	\multicolumn{1}{c}{\multirow{2}{*}{$\chi^{2}_{r}(\nu)$}}\\

	& 
	& 
	\multicolumn{1}{c}{$\Gamma$} & 
	\multicolumn{1}{c}{norm} & 
	\multicolumn{1}{c}{$k_{B}T$} & 
	\multicolumn{1}{c}{norm} & 
	\multicolumn{1}{c}{$\xi$} & 
	\multicolumn{1}{c}{$Z_{\rm Fe}$} \\

	\multicolumn{1}{c}{(1)} & 
	\multicolumn{1}{c}{(2)} & 
	\multicolumn{1}{c}{(3)} & 
	\multicolumn{1}{c}{(4)} & 
	\multicolumn{1}{c}{(5)} & 
	\multicolumn{1}{c}{(6)} & 
	\multicolumn{1}{c}{(7)} & 
	\multicolumn{1}{c}{(8)} &
	\multicolumn{1}{c}{(9)} &
	\multicolumn{1}{c}{(10)} \\
	\midrule

	NGC\,2992
		& stacked[all] & $1.85^{+0.02}_{-0.02}$ & $3.35^{+0.07}_{-0.07}$ & \na & \na & $<5.1^{+0.09}_{-0.09}$ & 1.00\fix 
					& D2 & $1.05(1610)$\\[0.5ex]
		& stacked[bc] & $1.82^{*}$ & $3.89^{+0.43}_{-0.43}$ & & & $2.39^{*}$ & $1.00^{*}$\\[0.5ex]
	NGC\,3227$^{\Diamond}$ 
		& 703022010$^{j}$ & $1.83^{+0.01}_{-0.01}$ & $13.92^{+0.49}_{-0.49}$ & \na & \na & $<1.22$ & 1.00\fix 
					& D3 & $1.04(5803)$\\[0.5ex]
		& 703022020$^{j}$ & $1.83^{*}$ & $9.71^{+0.37}_{-0.37}$ & \na & \na & $<1.22^{*}$ & $1.00^{*}$\\[0.5ex]
		& 703022030$^{j}$ & $1.83^{*}$ & $11.47^{+0.29}_{-0.37}$ & \na & \na & $<1.22^{*}$ & $1.00^{*}$\\[0.5ex]
		& 703022040$^{j}$ & $1.83^{*}$ & $4.94^{+0.17}_{-0.17}$ & \na & \na & $<1.22^{*}$ & $1.00^{*}$\\[0.5ex]
		& 703022050$^{j}$ & $1.83^{*}$ & $10.27^{+0.24}_{-0.24}$ & \na & \na & $<1.22^{*}$ & $1.00^{*}$\\[0.5ex]
		& 703022060$^{j}$ & $1.83^{*}$ & $7.60^{+0.21}_{-0.21}$ & \na & \na & $<1.22^{*}$ & $1.00^{*}$\\[0.5ex]
	NGC\,3516$^{\Diamond}$ 
		& 100031010 & $2.11^{+0.03}_{-0.03}$ & $27.81^{+0.75}_{-0.75}$ & \na & \na & $2.24^{+0.14}_{-0.10}$ 
					& 1.00\fix & D3 & $1.01(3820)$\\[0.5ex]	  
		& 704062010 & $2.11^{*}$ & $9.79^{+0.61}_{-0.61}$ & \na & \na & $2.24^{*}$ & $1.00^{*}$\\[0.5ex]
	NGC\,3783$^{\Diamond}$ 		
		& 701033010$^{j}$ & $1.85^{+0.01}_{-0.01}$ & $18.48^{+0.40}_{-0.40}$ & $82^{+1}_{-1}$ & $26.40^{+3.32}_{-3.29}$ 
					& $<1.20$ & 1.00\fix & D3 & $1.04(4219)$\\[0.5ex]
		& 704063010$^{j}$ & $1.85^{*}$ & $22.39^{+0.20}_{-0.19}$ & $82^{*}$ & $107.72^{+12.01}_{-10.24}$ 
					& $1.20^{*}$ & 1.00$^{*}$\\[0.5ex]
	NGC\,4051$^{\Diamond}$ 
		& 700004010 & $2.18^{+0.01}_{-0.01}$ & $13.20^{+0.07}_{-0.07}$ & $79^{+1}_{-1}$ & $4.33^{+0.56}_{-0.56}$ 
					& $<1.01$ & 1.00\fix & D3 & $1.01(4551)$\\[0.5ex]
		& 703023010 & $2.18^{*}$ & $26.43^{+0.09}_{-0.09}$ & $79^{*}$ & $57.36^{+4.34}_{-4.09}$ & $<1.01^{*}$ & $1.00^{*}$\\[0.5ex]
		& 703023020 & $2.18^{*}$ & $21.44^{+0.14}_{-0.13}$ & $79^{*}$ & $33.74^{+2.73}_{-2.98}$ & $<1.01^{*}$ & $1.00^{*}$\\[0.5ex]
	NGC\,4151$^{\Diamond}$ 
		& 701034010 & $2.02^{+0.01}_{-0.01}$ & $22.65^{+0.95}_{-0.93}$ & \na & \na & $<1.02$ & 1.00\fix & D2 
					& $1.00(3052)$\\[0.5ex]
	NGC\,4395$^{\Diamond}$ 
		& 702001010 & $1.65^{+0.02}_{-0.02}$ & $1.82^{+0.05}_{-0.05}$ & \na & \na & $<2.18^{+1.46}_{-1.02}$ & 1.00\fix 
					& D2 & $1.00(590)$\\[0.5ex]	
	NGC\,4593$^{c}$ 
		& 702040010 & $1.64^{+0.01}_{-0.01}$ & $2.20^{+0.02}_{-0.02}$ & \na & \na & $<1.04$ & $1.67^{+0.19}_{-0.19}$ & D2 
					& $1.03(1863)$\\[0.5ex]
	NGC\,5506
		& stacked[all] & $2.04^{+0.01}_{-0.01}$ & $83.98^{+0.64}_{-0.62}$ & \na & \na & $6.86^{+0.37}_{-0.35}$ 
					& 1.00\fix & D3 & $1.06(2376)$\\[0.5ex]
	NGC\,5548
		& stacked[all] & $1.75^{+0.01}_{-0.01}$ & $4.80^{+0.25}_{-0.25}$ & \na & \na & $<1.42$ & 1.00\fix & D2 & $1.05(1854)$\\[0.5ex]
	NGC\,7213 
		& 701029010 &  $1.78^{+0.01}_{-0.01}$ & $6.39^{+0.02}_{-0.02}$ & \na & \na & $5.01^{+0.24}_{-0.51}$ & $1.83^{+0.47}_{-0.27}$ 
					& \na & $1.02(2686)$\\[0.5ex]
	NGC\,7469 
		& 703028010 & $1.83^{+0.01}_{-0.01}$ & $2.21^{+0.02}_{-0.02}$ & \na & \na & $<1.3$ & $1.61^{+0.16}_{-0.16}$ & D2 
					& $1.00(2377)$\\[0.5ex]
	PBC\,J0839.7-1214
		& 705007010 & $1.74^{+0.01}_{-0.01}$ & $2.66^{+0.04}_{-0.04}$ & \na & \na & $134.15^{+8.32}_{-7.81}$ & 1.00\fix & \na 
					& $1.06(1267)$\\[0.5ex]
	PDS\,456$^{\Diamond}$
		& 701056010 & $2.47^{+0.01}_{-0.01}$ & $3.19^{+0.05}_{-0.05}$ & \na & \na & \na & \na & D3 
					& $1.08(1238)$\\[0.5ex]
		& 705041010 & $2.47^{*}$ & $3.07^{+0.06}_{-0.06}$ & \na & \na & \na & \na\\[0.5ex]
	PG\,1211+143 
		& 700009010 & $1.88^{+0.02}_{-0.02}$ & $1.18^{+0.02}_{-0.02}$ & $80^{+4}_{-4}$ & $1.16^{+0.29}_{-0.22}$ & $<71.87$ 
					& 1.00\fix & D2 & $1.04(831)$\\[0.5ex]
	PKS\,0558-504
		& Stacked[all] & $2.30^{+0.01}_{-0.01}$ & $13.08^{+0.18}_{-0.18}$ & $133^{+4}_{4}$ & $10.28^{+0.57}_{-0.57}$ & $<155.36$ 
					& 1.00\fix & D2 & $0.99(1868)$\\[0.5ex]
	RBS\,1124
		& 702114010 & $1.75^{+0.03}_{-0.03}$ & $2.77^{+0.06}_{-0.06}$ & $105^{+17}_{-17}$ & $0.99^{+0.86}_{-0.38}$ & $<27.41$ 
					& 1.00\fix & D2 & $0.94(764)$\\[0.5ex]
	SW\,J2127.4+5456$^{\Diamond}$ 
		& 702122010 & $2.01^{+0.02}_{-0.02}$ & $13.27^{+0.39}_{-0.39}$ & $271^{+24}_{-24}$ & $3.72^{+1.01}_{-1.02}$ 
					& $20^{+5}_{-13}$ & 1.00\fix & D2 & $0.97(2063)$\\[0.5ex]
	TON\,S180
		& 701021010 & $2.41^{+0.02}_{-0.02}$ & $9.94^{+0.25}_{-0.25}$ & $124^{+6}_{-6}$ & $6.00^{+0.79}_{-0.77}$ 
					& $566.39^{+212.12}_{-142.11}$ & 1.00\fix & D2 & $1.05(922)$\\[0.5ex]
	\bottomrule
		\end{tabular}
		\end{minipage}
	\end{center}
\end{table*}

\clearpage

\begin{table*} 
\footnotesize
\begin{center}
	\begin{minipage}{106mm} 
		\caption{Warm absorber parameters for single-epoch sources. {\sl Notes:} 
					(1) Source name;
					(2) \suzaku observation ID;
					(3) Absorption zone;
					(4) $\log$ of absorber column density, in units of $\rm{cm^{-2}}$;
					(5) $\log$ of the ionisation parameter, in units of $\rm{erg\,cm\,s^{-1}}$;
					(6) Per cent (\%) covering fraction of absorption component. All warm absorption components are required with $P_{\rm F}>99\%$}
	\begin{tabular}{@{}llccrc}

	\toprule
	\multicolumn{1}{c}{Source} & \multicolumn{1}{c}{OBSID} & Zone & $\log N_{\rm H}$ & $\log\xi$ & $f_{\rm cov}$ \\
	\multicolumn{1}{c}{(1)} & \multicolumn{1}{c}{(2)} & (3) & (4) & (5) & (6) \\
	\midrule
	1H\,0419-577
		& stacked[all] & 1 & $24.33^{+0.01}_{-0.01}$ & $0.51^{+0.44}_{-0.08}$ & $65^{+5}_{-5}$\\[0.5ex]
		& 			   & 2 & $23.27^{+0.01}_{-0.01}$ & $2.03^{+0.01}_{-0.01}$ & $17^{+2}_{-2}$\\[0.5ex]

	3C\,382 
		& 702125010 & 1  & $21.10^{+0.03}_{-0.46}$ & $2.73^{+0.16}_{-0.31}$ & 100\fix\\[0.5ex]
	3C\,445 
		& 702056010 & 1  & $23.28^{+0.01}_{-0.01}$ & $1.12^{+0.07}_{-0.16}$ & $\ga98$ \\[0.5ex]
	4C\,+74.26$^{\Diamond}$ 
		& 702057010	& 1 & $21.57^{+0.10}_{-0.10}$ & $1.61^{+0.08}_{-0.09}$ & 100\fix\\[0.5ex]
		&			& 2 & $20.96^{+0.05}_{-0.05}$ & neutral & 100\fix\\[0.5ex]
	APM\,08279+5255$^{\Diamond}$ 
		& stacked[all] 	& 1  & $22.67^{+0.09}_{-0.10}$ & neutral & 100\fix\\[0.5ex]
	Ark\,564
		& 702117010 & 1  & $24.48^{+0.01}_{-0.01}$ & $2.46^{+0.04}_{-0.19}$ & $41\pm1$\\[0.5ex]
		& 			& 2  & $22.73^{+0.03}_{-0.02}$ & $0.95^{+0.10}_{-0.09}$ & 100\fix\\[0.5ex]
	CBS\,126$^{\Diamond}$ 
		& 705042010 & 1  & $22.48^{+0.05}_{-0.05}$ & $2.97^{+0.02}_{-0.02}$ & 100\fix\\[0.5ex]
	ESO\,103-G035$^{\Diamond}$
		& 703031010 & 1  & $23.07^{+0.02}_{-0.02}$ & neutral & $98\pm1$ \\[0.5ex]
		& 			& 2  & $23.02^{+0.02}_{-0.02}$ & $0.82^{+0.23}_{-0.22}$ & $98^{*}$\\[0.5ex]
		& 			& 3  & $21.74^{+0.04}_{-0.04}$ & neutral & 100\fix\\[0.5ex]
	IC\,4329A
		& stacked[all] & 1 & $21.17^{+0.17}_{-0.09}$ & $1.98^{+0.16}_{-0.10}$ & 100\fix \\[0.5ex]
		& 			& 2 & $21.24^{+0.01}_{-0.01}$ & $-1.27^{+0.01}_{-0.01}$ & 100\fix \\[0.5ex]
		& 			& 3 & $21.75^{+0.05}_{-0.05}$ & $1.24^{+0.13}_{-0.18}$ & 100\fix \\[0.5ex]
		& 			& 4 & $21.15^{+0.03}_{-0.02}$ & $2.52^{+0.07}_{-0.03}$ & 100\fix \\[0.5ex]
	IGR\,J21247+5058
		& 702027010 & 1 & $22.54^{+0.13}_{-0.12}$ & neutral & $<21$\\[0.5ex]
		& 			& 2 & $21.94^{+0.02}_{-0.02}$ & neutral & 100\fix \\[0.5ex]
	MR\,2251-178$^{\Diamond}$ 
		& 704055010 & 1  & $20.98^{+0.10}_{-0.09}$ & $0.96^{+0.13}_{-0.11}$ & 100\fix\\[0.5ex]
		& 			& 2  & $21.43^{+0.36}_{-0.75}$ & $2.21^{+0.08}_{-0.08}$ & 100\fix\\[0.5ex]
	Mrk\,79
		& 702044010 & 1  & $23.19^{+0.03}_{-0.03}$ & $2.13^{+0.07}_{-0.09}$ & $37\pm6$\\[0.5ex]
		&			& 2  & $21.62^{+0.04}_{-0.04}$ & $1.44^{+0.06}_{-0.04}$ & 100\fix\\[0.5ex]
	Mrk\,205
		& 705062010 & 1  & $>24.68$ & neutral & $59\pm2$\\[0.5ex]
	Mrk\,279$^{\Diamond}$ 
		& 704031010	& 1  & $20.68^{+0.12}_{-0.10}$ & $1.26^{+0.28}_{-0.68}$ & 100\fix \\[0.5ex]
	Mrk\,335 
		& 701031010 & 1  & $20.43^{+0.02}_{-0.02}$ & $0.52^{+0.20}_{-0.18}$ & 100\fix\\[0.5ex]
	Mrk\,841 
		& stacked[all] & 1  & $20.39^{+0.39}_{-0.24}$ & $1.84^{+0.41}_{-0.43}$ & 100\fix \\[0.5ex]
	MCG\,-2-58-22 
		& 704032010 & 1  & $24.15^{+0.03}_{-0.04}$ & $2.33^{+0.20}_{-0.13}$ & $35\pm2$\\[0.5ex]
	MCG\,-5-23-16 
		& 700002010 & 1  & $24.18^{+0.01}_{-0.01}$ & $1.71^{+0.06}_{-0.06}$ & $49\pm1$\\[0.5ex]
		& 			& 2  & $22.68^{+0.01}_{-0.01}$ & $0.58^{+0.01}_{-0.01}$ & 100\fix\\[0.5ex]
	MCG\,-6-30-15$^{\Diamond}$ 
		& stacked[all] & 1 & $23.73^{+0.02}_{-0.02}$ & $2.21^{+0.05}_{-0.05}$ & $<18$\\[0.5ex]
		& 			& 2 & $21.42^{+0.02}_{-0.02}$ & $0.72^{+0.02}_{-0.03}$ & 100\fix\\[0.5ex]
		& 			& 3 & $21.79^{+0.01}_{-0.01}$ & $1.74^{+0.02}_{-0.02}$ & 100\fix\\[0.5ex]
	MCG\,+8-11-11 
		& 702112010 & 1  & $20.52^{+0.05}_{-0.02}$ & $<0.18$ & 100\fix \\[0.5ex]
		&	 		& 2  & $21.02^{+0.13}_{-0.13}$ & $2.35^{+0.10}_{-0.14}$ & 100\fix \\[0.5ex]
	NGC\,1365$^{\Diamond}$ 
		& 705031020 & 1  & $24.17^{+0.13}_{-0.09}$ & neutral & $31\pm1$\\[0.5ex]
		& 			& 	 & $23.79^{+0.01}_{-0.01}$ & $1.32^{+0.06}_{-0.07}$ & $>0.96$\\[0.5ex]
	NGC\,2992
		& stacked[all] & 1 & $21.80^{+0.02}_{-0.02}$ & neutral & $95\pm1$\\[0.5ex]
		& 			&    & $22.124^{+0.05}_{-0.04}$ & $1.79\pm0.08$ &  $95$\fix\\[0.5ex]
	NGC\,4151$^{\Diamond}$ 
		& 701034010 & 1  & $22.21^{+0.06}_{-0.14}$ & $2.57^{+0.14}_{-0.18}$ & $\ga98$ \\[0.5ex]
	    & 			& 2  & $23.00^{+0.01}_{-0.01}$ & $0.63^{+0.02}_{-0.02}$ & $\ga98^{*}$ \\[0.5ex]
	    & 			& 3  & $23.69^{+0.01}_{-0.01}$ & $1.86^{+0.07}_{-0.06}$ & $68\pm5$ \\[0.5ex]
	NGC\,4395$^{\Diamond}$ 
		& 702001010 & 1  & $23.91^{+0.08}_{-0.06}$ & neutral & $38\pm2$\\[0.5ex]
		& 			& 2  & $22.34^{+0.03}_{-0.03}$ & $0.56^{+0.14}_{-0.13}$ & $83\pm1$\\[0.5ex]
		& 			& 3  & $21.58^{+0.14}_{-0.17}$ & $1.64^{+0.17}_{-0.27}$ & 100\fix\\[0.5ex]
	NGC\,4593 
		& 702040010 & 1  & $21.99^{+0.28}_{-0.15}$ & $2.96^{+0.18}_{-0.09}$ & 100\fix \\[0.5ex]
		& 			& 2  & $21.79^{+0.24}_{-0.20}$ & $2.28^{+0.06}_{-0.09}$ & 100\fix \\[0.5ex]

	\bottomrule
	\end{tabular}\\[0.5ex]
	$^{*}$ indicates a parameter was tied during spectral fitting;\\
	$^{f}$ Indicates a parameter was frozen during spectral fitting;\\
	$^{\Diamond}$ indicates that a highly-ionised outflow is detected in this source. 
	\label{tab:warmabs_single}
	\end{minipage}
\end{center}
\end{table*}

\begin{table*} 
\footnotesize
\begin{center}
	\begin{minipage}{106mm} 
		\contcaption{\small -- Warm absorber parameters for single-epoch sources.}
	\begin{tabular}{@{}llccrc}

	\toprule
	\multicolumn{1}{c}{Source} & \multicolumn{1}{c}{OBSID} & Zone & $\log N_{\rm H}$ & $\log\xi$ & $f_{\rm cov}$ \\
	\multicolumn{1}{c}{(1)} & \multicolumn{1}{c}{(2)} & (3) & (4) & (5) & (6) \\
	\midrule

	NGC\,5506$^{\Diamond}$
		& stacked[all] & 1 & $21.51^{+0.02}_{-0.02}$ & neutral & 100\fix\\[0.5ex]
		&			  & 2 & $22.39^{+0.03}_{-0.02}$ & $1.55^{+0.05}_{-0.07}$ & $>99$ \\[0.5ex]
		&			  & 3 & $22.29^{+0.02}_{-0.02}$ & neutral &$>99$\\[0.5ex]
		&			  & 4 & $24.01^{+0.01}_{-0.01}$ & neutral & $35\pm1$\\[0.5ex]

	NGC\,5548
		& stacked[all] & 1  & $22.15^{+0.07}_{-0.12}$ & $2.78^{+0.06}_{-0.09}$ & 100\fix\\[0.5ex]
		& 			& 2  & $21.44^{+0.21}_{-0.05}$ & $2.20^{+0.09}_{-0.11}$ & $>92$\\[0.5ex]
		& 			& 3  & $21.84^{+0.03}_{-0.03}$ & $0.38^{+0.12}_{-0.14}$ & $46\pm2$\\[0.5ex]
	NGC\,7314 
		& 702015010 & 1  & $21.62^{+0.06}_{-0.06}$ & neutral & 100\fix \\[0.5ex]
		&			& 2  & $21.94^{+0.03}_{-0.03}$ & $1.22^{+0.15}_{-0.12}$ & 100\fix \\[0.5ex]
	NGC\,7469 
		& 703028010 & 1  & $21.82^{+0.10}_{-0.09}$ & $2.85^{+0.07}_{-0.08}$ & 100\fix \\[0.5ex]					
	PG\,1211+143 
		& 700009010 & 1  & $22.49^{+0.13}_{-0.12}$ & $2.88^{+0.07}_{-0.09}$ & 100\fix\\[0.5ex]
	PKS\,0558-504
		& stacked[all] & 1  & $24.28^{+0.04}_{-0.04}$ & $1.94^{+0.09}_{-0.27}$ & $38\pm2$\\[0.5ex]
	RBS\,1124
		& 702114010 & 1  & $24.62^{+0.22}_{-0.14}$ & neutral & $56\pm1$ \\[0.5ex]
	TON\,S180
		& 701021010 & 1  & $24.14^{+0.03}_{-0.07}$ & $1.92^{+0.10}_{-0.73}$ & $64\pm1$\\[0.5ex]
	SW\,J2127.4+5456$^{\Diamond}$ 
		& 702122010 & 1  & $21.12^{+0.03}_{-0.03}$ & $0.83^{+0.50}_{-0.54}$ & 100\fix\\[0.5ex]

	\bottomrule
	\end{tabular}\\
	\label{tab:warmabs_singleB}
	\end{minipage}
\end{center}
\end{table*}

\clearpage	

\begin{landscape}
\begin{table}
\begin{center}
	\footnotesize
	\begin{minipage}{235mm}
		\caption{Warm absorber parameters for jointly fit spectra. {\sl Notes}: (1) Source name; (2) \suzaku observation ID; (3a, 4a, 5a, 6a) logarithm of absorber column density, in units of cm$^{-2}$; (3b, 4b, 5b, 6b) logarithm of absorber ionisation parameter, in units of erg\,cm\,s$^{-1}$; (3c, 4c, 5c, 6c) Absorber covering fraction, in per cent. All absorption components are required at $P_{\rm F}>99\%$ significance.}
			\begin{tabular}{@{}llllllllllllll}

			\toprule
			\multicolumn{1}{c}{\multirow{2}{*}{Source}} & 
			\multicolumn{1}{c}{\multirow{2}{*}{OBSID}} & 
			\multicolumn{3}{c}{Zone~1} & 
			\multicolumn{3}{c}{Zone~2} & 
			\multicolumn{3}{c}{Zone~3} & 
			\multicolumn{3}{c}{Zone~4}\\

			& 
			& 
			\multicolumn{1}{c}{$\log N_{\rm H}$} & \multicolumn{1}{c}{$\log\xi$} & \multicolumn{1}{c}{$f_{\rm cov}$} & 
			\multicolumn{1}{c}{$\log N_{\rm H}$} & \multicolumn{1}{c}{$\log\xi$} & \multicolumn{1}{c}{$f_{\rm cov}$} & 
			\multicolumn{1}{c}{$\log N_{\rm H}$} & \multicolumn{1}{c}{$\log\xi$} & \multicolumn{1}{c}{$f_{\rm cov}$} &
			\multicolumn{1}{c}{$\log N_{\rm H}$} & \multicolumn{1}{c}{$\log\xi$} & \multicolumn{1}{c}{$f_{\rm cov}$}\\

			\multicolumn{1}{c}{(1)} &
			\multicolumn{1}{c}{(2)} &
			\multicolumn{1}{c}{(3a)} &
			\multicolumn{1}{c}{(3b)} &
			\multicolumn{1}{c}{(3c)} &
			\multicolumn{1}{c}{(4a)} &
			\multicolumn{1}{c}{(4b)} &
			\multicolumn{1}{c}{(4c)} &
			\multicolumn{1}{c}{(5a)} &
			\multicolumn{1}{c}{(5b)} &
			\multicolumn{1}{c}{(5c)} &
			\multicolumn{1}{c}{(6a)} &
			\multicolumn{1}{c}{(6b)} &
			\multicolumn{1}{c}{(6c)} \\
			\midrule

	3C\,111$^{\Diamond}$
		& 703034010 & $21.83^{+0.01}_{-0.01}$ & neutral & 100\fix\\[0.5ex]
		& 705040010 & $21.83^{*}$ & neutral$^{*}$ & 100$^{*}$\\[0.5ex]
		& 705040020 & $21.83^{*}$ & neutral$^{*}$ & 100$^{*}$\\[0.5ex]
		& 705040030 & $21.83^{*}$ & neutral$^{*}$ & 100$^{*}$\\ [0.5ex]
	3C\,120
		& 700001010 & $20.96^{+0.06}_{-0.06}$ & neutral & 100\fix\\[0.5ex]
		& stacked[bcd] & $20.96^{*}$ & neutral$^{*}$ & $100^{*}$\\[0.5ex]
	Mrk\,509
		& 705025010 & $21.40^{+0.04}_{-0.03}$ & $2.19^{+0.04}_{-0.05}$ & 100\fix\\[0.5ex]
		& stacked[bc] & $21.40^{*}$ & $2.19^{*}$ & $100^{*}$\\[0.5ex]
		& 705025010 & $21.40^{*}$ & $2.19^{*}$ & $100^{*}$\\[0.5ex]
	Mrk\,766$^{\Diamond}$
		& 701035010 & $22.66^{+0.03}_{-0.02}$ & $2.19^{+0.03}_{-0.03}$ & $94\pm4$ 
					& $21.86^{+0.06}_{-0.06}$ & $1.80^{+0.04}_{-0.04}$ & 100\fix\\[0.5ex]
		& 701035020 & $22.66^{*}$ & $2.19^{*}$ & $<14$ 
					& $21.86^{*}$ & $1.80^{*}$ & $100^{*}$\\[0.5ex]
	NGC\,1365$^{\Diamond}$
		& 702047010 & $24.111^{+0.02}_{-0.03}$ & $2.42^{+0.24}_{-0.24}$ & $55^{+5}_{-4}$
					& $23.02^{+0.01}_{-0.01}$ & neutral & $42^{+4}_{-3}$
					& $22.82^{+0.06}_{-0.06}$ & $3.21^{+0.04}_{-0.04}$ & $42*$\\[0.5ex]
		& 705031010 & $24.11^{*}$ & $2.42^{*}$ & $40^{+2}_{-3}$
					& $23.53^{+0.01}_{-0.01}$ & neutral$^{*}$ & $58\pm2$
					& $23.20^{+0.05}_{-0.05}$ & $3.21^{*}$ & $58*$\\[0.5ex]
	NGC\,3227$^{\Diamond}$ 
		& 703022010 & $22.16^{+0.01}_{-0.01}$ & $0.81\pm0.05$ & 100\fix
					& $23.06^{+0.01}_{-0.01}$ & $2.34\pm0.02$ & $55\pm5$\\[0.5ex]
		& 703022020 & $22.16^{*}$ & $1.00^{+0.05}_{-0.13}$ & 100$^{*}$
					& $23.06^{*}$ & $-0.11\pm0.04$ & $86\pm8$\\[0.5ex]
		& 703022030 & $22.16^{*}$ & $0.67^{+0.11}_{-0.11}$ & 100$^{*}$
					& $23.06^{*}$ & $1.53\pm0.05$ & $87\pm4$\\[0.5ex]
		& 703022040 & $22.16^{*}$ & $1.13^{+0.09}_{-0.08}$ & 100$^{*}$
					& $23.06^{*}$ & $-1.15^{+0.19}_{-0.18}$ & $83\pm6$\\[0.5ex]
		& 703022050 & $22.16^{*}$ & $0.66^{+0.16}_{-0.16}$ & 100$^{*}$
					& $23.06^{*}$ & $1.24\pm0.04$ & $91\pm4$\\[0.5ex]
		& 703022060 & $22.16^{*}$ & $0.51^{+0.23}_{-0.17}$ & 100$^{*}$
					& $23.06^{*}$ & $0.67^{+0.12}_{-0.11}$ & $90\pm5$\\[0.5ex]
	NGC\,3516$^{\Diamond}$
		& 100031010 & $24.22^{+0.02}_{-0.02}$ & $2.00^{+0.02}_{-0.04}$ & $46\pm4$
					& $23.27^{+0.01}_{-0.01}$ & $1.89\pm0.04$ & $>93$
					& $21.32^{+0.01}_{-0.01}$ & $1.01^{+0.07}_{-0.05}$ & $>99$
					& $21.50^{+0.10}_{-0.09}$ & $<0.14$ & $>99^{*}$\\[0.5ex]
		& 704062010 & $24.22^{*}$ & $2.00^{*}$ & $48^{+9}_{-8}$
					& $23.27^{*}$ & $1.89\pm0.04$ & $58\pm11$
					& $21.32^{*}$ & $1.01^{*}$ & $>77$
					& $21.50^{*}$ & $<0.14^{*}$ & $>77^{*}$\\[0.5ex]
	NGC\,3783$^{\Diamond}$ 
		& 701033010 & $23.76^{+0.02}_{-0.02}$ & $1.92^{+0.04}_{-0.12}$ & $27\pm3$ 
					& $22.23^{+0.01}_{-0.01}$ & $1.56^{+0.01}_{-0.01}$ & 100\fix 
					& $22.10^{+0.09}_{-0.09}$ & $2.77^{+0.07}_{-0.07}$ & 100\fix\\[0.5ex]
		& 704063010 & $23.76^{*}$ & $2.41^{+0.06}_{-0.06}$ & $22\pm1$ 
					& $22.23^{*}$ & $1.92^{*}$ & $100^{*}$ 
					& $22.10^{*}$ & $2.77^{*}$ & $100^{*}$\\[0.5ex]
	NGC\,4051$^{\Diamond}$
		& 100004010 & $24.36^{+0.01}_{-0.01}$ & $2.11^{+0.02}_{-0.02}$ & $26\pm2$
					& $23.14^{+0.02}_{-0.02}$ & $1.65^{+0.05}_{-0.05}$ & $<14$
					& $21.41^{+0.01}_{-0.01}$ & $1.46^{+0.03}_{-0.03}$ & 100\fix\\[0.5ex]
		& 703023010 & $24.36^{*}$ & $2.11^{*}$ & $50\pm6$ 
					& $23.14^{*}$ & $1.65^{*}$ & $40\pm2$
					& $20.54^{+0.05}_{-0.04}$ & $1.46^{*}$ & $100^{*}$\\[0.5ex]
		& 703023020 & $24.36^{*}$ & $2.11^{*}$ & $<10$
					& $23.14^{*}$ & $1.65^{*}$ & $32\pm3$
					& $21.41^{*}$ & $1.46^{*}$ & $100^{*}$\\[0.5ex]
	PDS\,456$^{\Diamond}$
		& 701056010 & $22.98^{+0.03}_{-0.04}$ & $2.15^{+0.04}_{-0.04}$ & $28\pm2$
					& $22.00^{+0.08}_{-0.08}$ & $3.13^{+0.10}_{-0.11}$ & 100\fix\\[0.5ex]
		& 705041010 & $22.98^{*}$ & $2.15^{*}$ & $56\pm3$
					& $22.00^{*}$ & $1.65^{+0.11}_{-0.12}$ & $100^{*}$\\[0.5ex]

			\bottomrule
			\end{tabular}\\[0.5ex]
			$^{*}$ indicates a parameter was tied during spectral fitting;\\
			$^{f}$ Indicates a parameter was frozen during spectral fitting;\\
			$^{\Diamond}$ indicates that a highly-ionised outflow is detected in this source.
			\label{tab:warmabs_multi}
	\end{minipage}
\end{center}
\end{table}
\end{landscape}

\clearpage
	
\begin{table*}
	\footnotesize
\begin{center}
	\begin{minipage}{168mm}
		\caption{Principal emission components in the 5-10\,keV band. {\sl Notes}: (1) Source name; (2) \suzaku observation ID; (3) Emission line identification. `Broad(GA)' and `Broad(DL)' indicate that a broadened Fe\,K emission component, which was fitted with either a broadened Gaussian or a {\tt disk-line}, respectively; (4) Measured line energy in the source rest-frame, in units of keV; (5) Measured line energy-width, in units of eV. Unresolved lines were fit with widths fixed at 10\,eV; (6) Line intensity, in units of $\times10^{-6}$\,photons\,cm$^{-2}$\,s$^{-1}$; (7) Line equivalent width, in units of eV. For the \feka line the $EW$ of the corresponding \fekb line is given in square brackets ; (8) Change in fit statistic (and degrees of freedom) when line is removed from the best-fit model; (9) Line significance according to the F-test.}
	\begin{tabular}{@{}lrlllrlrr}

		\toprule
		\multicolumn{1}{c}{Source} & \multicolumn{1}{c}{OBSID} & \multicolumn{1}{c}{Line ID} & \multicolumn{1}{c}{Energy} & \multicolumn{1}{c}{$\sigma$-width} & \multicolumn{1}{c}{Intensity} & \multicolumn{1}{c}{EW} & \multicolumn{1}{c}{$\Delta\chi^{2}/\Delta\nu$} & \multicolumn{1}{c}{$P_{\rm F}$}\\
		\multicolumn{1}{c}{(1)} & \multicolumn{1}{c}{(2)} & \multicolumn{1}{c}{(3)} & \multicolumn{1}{c}{(4)} & \multicolumn{1}{c}{(5)} & \multicolumn{1}{c}{(6)} & \multicolumn{1}{c}{(7)} & \multicolumn{1}{c}{(8)} & \multicolumn{1}{c}{(9)} \\
		\midrule
	
	1H\,0419-577
		& stacked[all] & \fekab & $6.39^{+0.02}_{-0.02}$ & 10\fix & $4.6^{+1.1}_{-1.1}$ & $21^{+4}_{-4}$ [$3\pm1$] & $35.4$ & $>99.99$\\[0.5ex]

	3C\,111$^{\Diamond}$ 
		& 703034010 & \fekab & $6.39^{+0.02}_{-0.02}$ & 10\fix & $17.1^{+3.1}_{-3.0}$ & $87^{+8}_{-12}$ [$10^{+2}_{-2}$] 
					& $102.2/2$ & $>99.99$\\[0.5ex]
		& 705040010 & \fekab & $6.38^{+0.02}_{-0.02}$ & 10\fix & $26.1^{+4.0}_{-4.0}$ & $46^{+8}_{-7}$ [$6\pm1$] & $93.0/2$ & $>99.99$\\[0.5ex]
		& 			& \fexxvi? & $6.85^{+0.10}_{-0.09}$ & 10\fix & $8.5^{+3.8}_{-3.8}$ & $17^{+5}_{-6}$ & $13.0$ & $>99.99$\\[0.5ex]
 		& 705040020 & \fekab & $6.38^{+0.04}_{-0.04}$ & 10\fix & $14.3^{+4.4}_{-4.4}$ & $19^{+6}_{-6}$ [$4\pm1$] & $49.0$ & $>99.99$\\[0.5ex]
 		& 705040030 & \fekab & $6.40^{+0.03}_{-0.03}$ & 10\fix & $20.6^{+4.4}_{-4.4}$ & $29^{+5}_{-6}$ [$5\pm1$] & $68.7$ & $>99.99$\\[0.5ex]

	3C\,120 
		& 700001010	& \fekab & $6.39^{+0.03}_{-0.03}$ & 10\fix & $18.0^{+4.4}_{-4.4}$ & $31^{+8}_{-8}$ [$5^{+1}_{-1}$] 
						& $37.5/2$ & $>99.99$\\[0.5ex]
	 	& 				& \brga & $6.54^{+0.12}_{-0.12}$ & $300^{+162}_{-86}$ & $32.2^{+8.8}_{-8.8}$ & $58^{+36}_{-40}$ 
	 					& $13.4/3$ & $99.62$\\[0.5ex]
	   	& stacked[bcd] 	& \fekab & $6.39^{+0.01}_{-0.01}$ & 10\fix & $19.9^{+2.4}_{-2.4}$ & $38^{+4}_{-4}$ [$6^{+2}_{-2}$] 
	   					& $162.5/2$ & $>99.99$\\[0.5ex]
	 	& 				& \brga & $6.40^{+0.09}_{-0.10}$ & $432^{+100}_{-84}$ & $34.9^{+6.4}_{-6.2}$ & $70^{+13}_{-13}$ & $106.4/3$ 
	 					& $>99.99$\\[0.5ex]

	3C\,382 
		& 702125010 & \fekab & $6.35^{+0.03}_{-0.03}$ & 10\fix & $11.8^{+2.7}_{-2.7}$ & $23^{+5}_{-5}$ [$4^{+1}_{-1}$] & $46.6/2$ 
					& $>99.99$\\[0.5ex]
	 	& 			& \brdl & 6.4\fix & \na & $32.1^{+6.3}_{-6.3}$ & $105^{+11}_{-11}$ & $38.2/4$ & $>99.99$\\[0.5ex]
	 	& 			& Ni\,K$\alpha$ & $7.50^{+0.05}_{-0.05}$ & 10\fix & $5.0^{+2.6}_{-2.6}$ & $14^{+6}_{-6}$ & $9.8/2$ & $99.24$\\[0.5ex]

	3C\,390.3$^{\Diamond}$ 
		& 701060010 & \fekab & $6.38^{+0.02}_{-0.02}$ & $87^{+32}_{-29}$ & $26.9^{+4.6}_{-4.3}$ & $69^{+10}_{-10}$ [$11^{+2}_{-2}$] 
					& $105.5/3$ & $>99.99$\\[0.5ex]
	 	& 			& \fexxv & $6.62^{+0.04}_{-0.04}$ & 10\fix & $7.6^{+2.7}_{-2.7}$ & $19^{+7}_{-7}$ & $20.8/2$ & $>99.99$\\[0.5ex]

	3C\,445 
		& 702056010 & \fekab & $6.38^{+0.01}_{-0.01}$ & $<47$ & $21.3^{+2.8}_{-2.6}$ & $104^{+14}_{-13}$ [$18^{+2}_{-2}$] 
					& $212.5/3$ & $>99.99$ \\[0.5ex]

	4C\,+74.26$^{\Diamond}$ 
		& 702057010 & \fekab & $6.33^{+0.06}_{-0.06}$ & 10\fix & $6.2^{+3.5}_{-3.5}$ & $13^{+7}_{-7}$ [$2^{+1}_{-1}$] 
					& $9.5/2$ & $99.07$\\[0.5ex]
		& 			& \brga & $6.21^{+0.12}_{-0.13}$ & $274^{+216}_{-117}$ & $24.8^{+7.8}_{-7.8}$ & $47^{+15}_{-15}$ 
					& $13.4/3$ & $>99.59$\\[0.5ex]

	APM\,08279+5255$^{\Diamond}$ 
		& stacked & \fekab & $6.52^{+0.12}_{-0.09}$ & 10\fix & $5.1^{+2.4}_{-2.4}$ & $13^{+6}_{6}$ [\na] & $9.8/2$ & $99.03$\\[0.5ex]

	Ark 120 
		& 702014010 & \fekab & $6.40^{+0.02}_{-0.02}$ & 10\fix & $50.0^{+6.6}_{-6.6}$ & $18^{+3}_{-3}$ [$13^{+3}_{-3}$] & $108.3/2$ 
					& $>99.99$\\[0.5ex]
	 	& 			& \brga & $6.26^{+0.14}_{-0.14}$ & $564^{+136}_{-111}$ & $42.4^{+8.9}_{-8.5}$ & $113^{+24}_{-23}$ & $39.7/3$ 
	 				& $>99.99$ \\[0.5ex]
	 	& 			& \fexxvi & $6.94^{+0.04}_{-0.04}$ & 10\fix & $7.4^{+2.6}_{-2.6}$ & $26^{+9}_{-9}$ & $17.9/2$ & $>99.99$ \\[0.5ex]

	Ark\,564
		& 702117010 & \fexxv & $6.59^{+0.06}_{-0.06}$ & 10\fix & $5.2^{+1.8}_{-1.8}$ & $31^{+12}_{-12}$ & $21.1/2$ & $>99.99$\\[0.5ex]

	CBS\,126$^{\Diamond}$ 
		& 705042010 & \fekab & $6.44^{+0.05}_{-0.04}$ & 10\fix & $4.3^{+1.2}_{-1.2}$ & $85^{+22}_{-22}$ [$15^{+4}_{-4}$] 
					& $30.7/2$ & $>99.99$\\[0.5ex]

	ESO\,103-G035$^{\Diamond}$
		& 703031010 & \fekab & $6.40^{+0.02}_{-0.02}$ & $69^{+22}_{-22}$ & $64.5^{+6.1}_{-5.9}$ & $52^{+5}_{-5}$ [$8^{+1}_{-1}$] 
					& $397.7/3$ & $>99.99$\\[0.5ex]
		& 			& \brga & $6.46^{+0.10}_{-0.10}$ & $546^{+90}_{-90}$ & $95.5^{+13.9}_{-13.9}$ & $76^{+11}_{-11}$ 
					& $89.8/3$ & $>99.99$ \\[0.5ex]

	Fairall\,9 
		& 702043010 & \fekab & $6.39^{+0.01}_{-0.01}$ & 10\fix & $25.0^{+1.9}_{-1.9}$ & $91^{+8}_{-8}$ & $401.9/2$ & $>99.99$\\[0.5ex]
		& 			& \brdl & $6.36^{+0.04}_{-0.04}$ & \na & $28.1^{+4.3}_{-4.3}$ & $112^{+37}_{-37}$ & $52.6/5$ & $>99.99$\\[0.5ex]
		& 705063010 & \fekab & $6.40^{+0.01}_{-0.01}$ & 10\fix & $22.8^{+2.1}_{-2.1}$ & $76^{+14}_{-14}$ & $296.1/2$ & $>99.99$\\[0.5ex]
		& 			& \brdl & $6.36^{*}$ & \na & $20.8^{+4.6}_{-4.6}$ & $75^{+60}_{-60}$ & $84.8/5$ & $>99.99$\\[0.5ex]

	IC\,4329A
		& stacked[all] & \fekab & $6.38^{+0.01}_{-0.01}$ & $45^{+17}_{-20}$ & $74.1^{+8.4}_{-8.1}$ & $59^{+7}_{-7}$ [$9^{+1}_{-1}$] 
					& $314.6/3$ & $>99.99$\\[0.5ex]

	IGR\,J16185-5928
		& 702123010 & \fekab & $6.44^{+0.04}_{-0.04}$ & 10\fix & $6.2^{+1.9}_{-1.9}$ & $68^{+21}_{-21}$ [$10^{+4}_{-4}$] 
					& $26.4/2$ & $>99.99$\\[0.5ex]

	IGR\,J21274+5058 
		& 702027010 & \fekab & $6.37^{+0.03}_{-0.03}$ & 10\fix & $19.1^{+5.1}_{-5.1}$ & $20^{+5}_{-5}$ [$3^{+1}_{-1}$] 
					& $39.7/2$ & $>99.99$\\[0.5ex]
		& 			& ? & $5.89^{+0.05}_{-0.05}$ & 10\fix & $10.9^{+5.1}_{-5.1}$ & $10^{+5}_{-5}$ & $11.9/2$ & $99.65$\\[0.5ex]

	MCG\,-02-14-009 
		& 703060010 & \fekab & $6.43^{+0.04}_{-0.04}$ & 10\fix & $4.4^{+1.1}_{-1.1}$ & $92^{+27}_{-27}$ [$14^{+4}_{-4}$] 
					& $35.7/2$ & $>99.99$\\[0.5ex]

	MCG\,-2-58-22 
		& 704032010 & \fekab & $6.41^{+0.02}_{-0.02}$ & 10\fix & $21.1^{+3.0}_{-3.0}$ & $37^{+5}_{-5}$ [$6^{+1}_{-1}$] 
					& $119.4/2$ & $>99.99$\\[0.5ex]

	MCG\,-5-23-16
		& 700002010 & \fekab & $6.40^{+0.01}_{-0.01}$ & $54^{+11}_{-11}$ & $85.8^{+5.7}_{-5.7}$ & $76^{+5}_{-5}$ [$12^{+1}_{-1}$] 
					& $804.3/3$ & $>99.99$\\[0.5ex]

	MCG\,-6-30-15$^{\Diamond}$
		& stacked[all] & \fekab & $6.38^{+0.01}_{-0.01}$ & 10\fix & $11.6^{+1.3}_{-1.3}$ & $25^{+3}_{-3}$ [$3^{+1}_{-1}$] 
					& $145.0/2$ & $>99.99$\\[0.5ex]
		& 			& \brga & $6.03^{+0.12}_{-0.12}$ & $646^{+97}_{-83}$ & $35.6^{+5.1}_{-4.9}$ & $70^{+10}_{-10}$ 
					& $32.9/3$ & $>99.99$\\[0.5ex]
		& 			& ? & $6.55^{+0.03}_{-0.03}$ & 10\fix & $5.3^{+1.5}_{-1.5}$ & $12^{+3}_{-3}$ & $29.6/2$ & $>99.99$ \\[0.5ex]

	MCG\,+8-11-11 
		& 702112010 & \fekab & $6.38^{+0.01}_{-0.01}$ & $41^{+19}_{-24}$ & $54.3^{+6.0}_{-5.8}$ & $68^{+8}_{-7}$ [$11^{+1}_{-1}$] 
					& $303.5/3$ & $>99.99$\\[0.5ex]
	 	& 			& \brga & $6.17^{+0.13}_{-0.14}$ & $462^{+182}_{-156}$ & $55.1^{+14.9}_{-14.3}$ & $60^{+16}_{-16}$ & $50.9/3$ 
	 				& $>99.99$\\[0.5ex]
	 	& 			& \fexxvi & $6.95^{+0.03}_{-0.03}$ & 10\fix & $14.2^{+4.3}_{-4.3}$ & $21^{+6}_{-6}$ & $15.6/2$ & $99.95$\\[0.5ex]

	MR\,2251-178$^{\Diamond}$ 
		& 704055010 & \fekab & $6.44^{+0.03}_{-0.03}$ & 10\fix & $11.9^{+2.7}_{-2.7}$ & $22^{+5}_{-5}$ [$3^{+1}_{-1}$] & $52.2/2$ 
					& $>99.99$\\[0.5ex]

	Mrk\,79
		& 702044010 & \fekab & $6.39^{+0.01}_{-0.01}$ & $<58$ & $23.7^{+2.8}_{-2.8}$ & $133^{+17}_{-17}$ [$22^{+3}_{-3}$] 
					& $232.5/3$ & $>99.99$\\[0.5ex]
		& 			& \fexxv & $6.63^{+0.04}_{-0.04}$ & 10\fix & $6.5^{+2.2}_{-2.0}$ & $34^{+10}_{-10}$ & $15.9/2$ & $99.98$ \\[0.5ex]

	\bottomrule
	\end{tabular}\\[0.5ex]
			$^{\Diamond}$ indicates that a highly-ionised outflow is detected in this source.
	\label{table:emission_lines}
	\end{minipage}
\end{center}
\end{table*}

\begin{table*}
	\footnotesize
\begin{center}
	\begin{minipage}{168mm}
		\contcaption{-- Principal emission components in the 5-10\,keV band}
	\begin{tabular}{@{}lrlllrlrr}

		\toprule
		\multicolumn{1}{c}{Source} & \multicolumn{1}{c}{OBSID} & \multicolumn{1}{c}{Line ID} & \multicolumn{1}{c}{Energy} & \multicolumn{1}{c}{$\sigma$-width} & \multicolumn{1}{c}{Intensity} & \multicolumn{1}{c}{EW} & \multicolumn{1}{c}{$\Delta\chi^{2}/\Delta\nu$} & \multicolumn{1}{c}{$P_{\rm F}$}\\
		\multicolumn{1}{c}{(1)} & \multicolumn{1}{c}{(2)} & \multicolumn{1}{c}{(3)} & \multicolumn{1}{c}{(4)} & \multicolumn{1}{c}{(5)} & \multicolumn{1}{c}{(6)} & \multicolumn{1}{c}{(7)} & \multicolumn{1}{c}{(8)} & \multicolumn{1}{c}{(9)} \\
		\midrule

	Mrk\,110 
		& 702124010 & \fekab & $6.41^{+0.03}_{-0.03}$ & 10\fix & $12.0^{+2.6}_{-2.6}$ & $51^{+12}_{-12}$ [$8^{+2}_{-2}$] & $52.4/2$ 
					& $>99.99$ \\[0.5ex]

	Mrk\,205
		& 705062010 & \fekab & $6.40^{+0.03}_{-0.03}$ & $<108$ & $9.3^{+2.8}_{-2.5}$ & $80^{+24}_{-24}$ [$13^{+3}_{-3}$] 
					& $43.9/3$ & $>99.99$\\[0.5ex]

	Mrk\,279$^{\Diamond}$ 
		& 704031010 & \fekab & $6.40^{+0.01}_{-0.01}$ & 10\fix & $12.4^{+1.1}_{-1.2}$ & $205^{+20}_{-20}$ [$33^{+3}_{-3}$] 
					& $326.6/2$ & $>99.99$ \\[0.5ex]
	 	& 			& \fexxv & $6.63^{+0.07}_{-0.07}$ & 10\fix & $1.7^{+0.9}_{-0.9}$ & $23^{+13}_{-13}$ & $9.4/2$ & $99.14$ \\[0.5ex]

	Mrk\,335 
		& 703034010 & \fekab & $6.39^{+0.03}_{-0.03}$ & $<89$ & $7.5^{+1.8}_{-1.7}$ & $45^{+10}_{-10}$ [$8^{+2}_{-2}$] 
					& $63.9/3$ & $>99.99$\\[0.5ex]
	 	& 			& \brdl & 6.4\fix & \na & $22.1^{+3.4}_{-3.2}$ & $172^{+22}_{-28}$ & $146.7/4$ & $>99.99$\\[0.5ex]

	Mrk\,359
		& 701082010 & \fekab & $6.42^{+0.03}_{-0.03}$ & $96^{+55}_{-55}$ & $7.0^{+1.8}_{-1.8}$ & $119^{+27}_{-26}$ [$17^{+4}_{-4}$] 
					& $44.3/3$ & $>99.99$\\[0.5ex]

	Mrk\,509
		& 701093010 & \fekab & $6.44^{+0.04}_{-0.04}$ & 10\fix & $23.3^{+5.4}_{-5.4}$ & $45^{+11}_{-11}$ [$7^{+1}_{-1}$] 
					& $37.4/2$ & $>99.99$\\[0.5ex]
		& stacked[bcd] & \fekab & $6.42^{+0.02}_{-0.02}$ & $68^{+48}_{-62}$ & $26.9^{+5.4}_{-4.9}$ & $50^{+10}_{-10}$ [$7^{+1}_{-1}$] 
					& $122.5/3$ & $>99.99$\\[0.5ex]
		& 			& \brga & $6.73^{+0.20}_{-0.20}$ & $606^{+183}_{-175}$ & $37.6^{+10.1}_{-9.8}$ & $74^{+19}_{-19}$ 
					& $48.4/3$ & $>99.99$ \\[0.5ex]
		& 705025010 & \fekab & $6.41^{+0.02}_{-0.02}$ & $67^{+25}_{-30}$ & $34.0^{+4.8}_{-4.7}$ & $57^{+9}_{-9}$ [$9^{+1}_{-1}$] 
					& $169.72/3$ & $>99.99$\\[0.5ex]
		& 			& \brga & $6.33^{+0.15}_{-0.15}$ & $517^{+166}_{-113}$ & $43.0\pm8.8$ & $67^{+14}_{-13}$ & $24.9/3$ & $>99.99$\\[0.5ex]

	Mrk\,766$^{\Diamond}$ 
		& 701035010 & \fekab & $6.41^{+0.03}_{-0.03}$ & 10\fix & $5.68^{+1.82}_{-1.82}$ & $39^{+12}_{-12}$ [$7^{+2}_{-2}$] 
					& $28.6/2$ & $>99.99$\\[0.5ex]
	 	& 			& \fexxv & $6.62^{+0.03}_{-0.03}$ & 10\fix & $8.80^{+1.95}_{-1.95}$ & $63^{+13}_{-13}$ & $61.6/2$ & $>99.99$\\[0.5ex]
		& 701035020 & \fekab & $6.43^{+0.03}_{-0.03}$ & 10\fix & $8.02^{+2.32}_{-2.32}$ & $52^{+17}_{-17}$ [$9^{+3}_{-3}$] 
					& $32.5/2$ & $>99.99$ \\[0.5ex]

	Mrk\,841 
		& stacked[all] & \fekab & $6.37^{+0.02}_{-0.02}$ & $<75$ & $10.7^{+2.4}_{-2.3}$ & $63^{+14}_{-14}$ [$3^{+1}_{-1}$] & $58.9/3$ 
					& $>99.99$ \\[0.5ex]
	 	& 			& \brga & $5.92^{+0.22}_{-0.23}$ & $570^{+241}_{-162}$ & $18.1^{+10.0}_{-10.0}$ & $101^{+56}_{-56}$ & $34.8/3$ 
	 				& $>99.99$\\[0.5ex]

	NGC\,1365$^{\Diamond}$ 
		& 702047010 & \fekab & $6.42^{+0.02}_{-0.02}$ & $<59$ & $12.6^{+2.6}_{-2.5}$ & $45^{+9}_{-9}$ [$4^{+1}_{-1}$] 
					& $83.1/3$ & $>99.99$\\[0.5ex]
	   	& 705031010 & \fekab & $6.39^{+0.02}_{-0.02}$ & $53^{+30}_{-41}$ & $27.9^{+3.9}_{-3.7}$ & $137^{+15}_{-18}$ [$22^{+4}_{-3}$] 
	   				& $135.8/3$ & $>99.99$\\[0.5ex]
	   	& 705031020 & \fekab & $6.41^{+0.01}_{-0.01}$ & $<43$ & $40.8^{+2.6}_{-2.5}$ & $251^{+17}_{-17}$ [$45^{+2}_{-2}$] 
	   				& $828.9/3$ & $>99.99$\\[0.5ex]
	 	& 			& \brga & $5.92^{+0.06}_{-0.05}$ & 10\fix & $2.2^{+0.7}_{-0.7}$ & $37^{+12}_{-12}$ & $27.2/2$ & $>99.99$\\[0.5ex]

	NGC\,2992
		& stacked[all] & \fekab & $6.40^{+0.01}_{-0.01}$ & $39^{+11}_{-13}$ & $36.7^{+2.2}_{-2.2}$ & $260^{+13}_{-15}$ [$40^{+3}_{-1}$] & $871.0/3$ & $>99.99$\\[0.5ex]

	NGC\,3227$^{\Diamond}$ 
		& 703022010 & \fekab & $6.38^{+0.01}_{-0.01}$ & $54^{+33}_{-40}$ & $42.9^{+6.9}_{-6.1}$ & $92^{+12}_{-12}$ [$15^{+2}_{-2}$]
	   				& $201.9/3$ & $>99.99$ \\[0.5ex]
	 	& 703022020 & \fekab & $6.40^{+0.01}_{-0.01}$ & $52^{+19}_{-23}$ & $47.3^{+5.0}_{-4.9}$ & $173^{+16}_{-18}$ [$26^{+3}_{-3}$] 
	 				& $306.4/3$ & $>99.99$\\[0.5ex]
	 	& 703022030 & \fekab & $6.40^{+0.01}_{-0.01}$ & $71^{+24}_{-26}$ & $46.2^{+5.9}_{-5.8}$ & $127^{+16}_{-16}$ [$19^{+3}_{-3}$] 
	 				& $194.8/3$ & $>99.99$\\[0.5ex]
	 	& 703022040 & \fekab & $6.40^{+0.01}_{-0.01}$ & $39^{+17}_{-25}$ & $40.9^{+3.7}_{-3.6}$ & $270^{+32}_{-25}$ [$42^{+3}_{-3}$] 
	 				& $454.4/3$ & $>99.99$\\[0.5ex]
	 	& 			& \fexxv & $6.68^{+0.05}_{-0.04}$ & 10\fix & $5.8^{+2.5}_{-2.4}$ & $30^{+12}_{-12}$ & $15.2/2$ & $>99.99$\\[0.5ex]
	 	& 703022050 & \fekab & $6.40^{+0.01}_{-0.01}$ & $52^{+20}_{-23}$ & $40.9^{+4.4}_{-4.3}$ & $128^{+14}_{-14}$ [$22^{+2}_{-2}$] 
	 				& $332.9/3$ & $>99.99$\\[0.5ex]
	   	& 703022060 & \fekab & $6.40^{+0.01}_{-0.01}$ & $<53$ & $34.4^{+4.5}_{-4.4}$ & $145^{+17}_{-17}$ [$22^{+2}_{-2}$] 
	   				& $213.9/3$ & $>99.99$\\[0.5ex]

	NGC\,3516$^{\Diamond}$
		& 100031010 & \fekab & $6.40^{+0.01}_{-0.01}$ & $40^{+9}_{-9}$ & $55.7^{+2.6}_{-2.6}$ & $149^{+7}_{-7}$ [$25^{+1}_{-1}$] 
					& $1416.7/3$ & $>99.99$\\[0.5ex]
		& 704062010 & \fekab & $6.40^{+0.01}_{-0.01}$ & $39^{+12}_{-12}$ & $42.2^{+2.0}_{-2.0}$ & $262^{+13}_{-13}$ [$40^{+2}_{-2}$] 
					& $1318.4/3$ & $>99.99$\\[0.5ex]
		&   & Ni ? & $7.43^{+0.05}_{-0.04}$ & 10\fix & $4.8^{+1.5}_{-1.5}$ & $40^{+12}_{-12}$ & $24.0/2$ & $>99.99$\\[0.5ex]

	NGC\,3783$^{\Diamond}$ 
		& 701033010 & \fekab & $6.40^{+0.01}_{-0.01}$ & $50^{+13}_{-14}$ & $72.8^{+5.1}_{-4.7}$ & $131^{+9}_{-10}$ [$20^{+1}_{-1}$] 
					& $1005.5/3$ & $>99.99$\\[0.5ex]
	   	& 704063010 & \fekab & $6.39^{+0.01}_{-0.01}$ & 10\fix & $68.4^{+3.5}_{-3.5}$ & $97^{+5}_{-5}$ [$15^{+1}_{-1}$] 
	   				& $1057.8/3$ & $>99.99$\\[0.5ex]

	NGC\,4051$^{\Diamond}$ 
		& 700004010 & ? & $5.44^{+0.03}_{-0.03}$ & 10\fix & $2.7^{+1.1}_{-1.1}$ & $19^{+8}_{-8}$ & $13.8/2$ & $99.85$\\[0.5ex]
	 	& 			& \fekab & $6.40^{+0.01}_{-0.01}$ & $43^{+17}_{-22}$ & $15.4^{+1.6}_{-1.6}$ & $140^{+15}_{-15}$ 
	 				[$12^{+1}_{-1}$] & $273.2/3$ & $>99.99$\\[0.5ex]
	 	& 			& \fexxv & $6.63$\fix & 10\fix & $3.3^{+1.2}_{-1.2}$ & $25^{+11}_{-11}$ & $12.6/2$ & $99.82$\\[0.5ex]
	 	& 			& Ni\,K$\alpha$ & $7.58^{+0.09}_{-0.09}$ & 10\fix & $2.4^{+1.3}_{-1.3}$ & $33^{+17}_{-17}$ 
	 				& $9.8/2$ & $99.02$\\[0.5ex]
		& 703023010 & \fekab & $6.41^{+0.01}_{-0.01}$ & 10\fix & $17.4^{+1.6}_{-1.6}$ & $70^{+5}_{-5}$ [$12^{+1}_{-1}$] 
					& $254.5/2$ & $>99.99$ \\[0.5ex]
	 	& 			& \fexxv & $6.61^{+0.03}_{-0.03}$ & 10\fix & $5.5^{+1.4}_{-1.4}$ & $22^{+5}_{-5}$ & $67.8/2$ 
	 				& $>99.99$\\[0.5ex]
		& 703023020 & \fekab & $6.42^{+0.02}_{-0.02}$ & $<108$ & $18.6^{+3.7}_{-3.7}$ & $100^{+15}_{-20}$ 
					& $108.4/3$ & $>99.99$\\[0.5ex]
	 	& 			& \fexxv & $6.59^{+0.04}_{-0.04}$ & 10\fix & $7.2^{+2.4}_{-2.4}$ & $37^{+1}_{-1}$ & $25.4/2$ & $>99.99$\\[0.5ex]

	NGC\,4151$^{\Diamond}$ 
		& 701034010 & \fekab & $6.39^{+0.01}_{-0.01}$ & $28^{+5}_{-5}$ & $169.5^{+4.8}_{-4.7}$ & $294^{+8}_{-8}$ [$40^{+1}_{-1}$] 
					& $4185.4/3$ & $>99.99$ \\[0.5ex]

	NGC 4395$^{\Diamond}$ 
		& 701001010 & \fekab & $6.38^{+0.05}_{-0.05}$ & 10\fix & $3.4^{+1.3}_{-1.3}$ & $55^{+22}_{-22}$ [$9^{+3}_{-3}$] 
					& $18.2/2$ & $>99.99$\\[0.5ex]

	NGC\,4593 
		& 702040010 & \fekab & $6.43^{+0.01}_{-0.01}$ & $50^{+11}_{-12}$ & $34.0^{+2.4}_{-2.3}$ & $271^{+19}_{-19}$ [$41^{+3}_{-3}$] 
					& $623.6/3$ & $>99.99$ \\[0.5ex]
	 	& 			& \fexxv & $6.70^{+0.06}_{-0.06}$ & 10\fix & $4.4^{+1.8}_{-2.0}$ & $23^{+9}_{-10}$ & $12.7/2$ & $99.80$\\[0.5ex]
	
	NGC\,5506$^{\Diamond}$
		& stacked[all] & \fekab & $6.40^{+0.01}_{-0.01}$ & $75^{+12}_{-12}$ & $96.4^{+6.5}_{-6.5}$ & $66^{+5}_{-5}$ [$11^{+1}_{-1}$] 
					& $677.7/3$ & $>99.99$\\[0.5ex]
		&			& \fexxv & $6.62^{+0.03}_{-0.03}$ & 10\fix & $19.2^{+4.5}_{-4.5}$ & $13^{+3}_{-3}$ & $31.0/2$ & $>99.99$\\[0.5ex]

	NGC\,5548
		& stacked[all] & \fekab & $6.38^{+0.01}_{-0.01}$ & $51^{+15}_{-16}$ & $23.4^{+2.2}_{-2.1}$ & $110^{+10}_{-10}$ 
					[$17^{+1}_{-1}$] & $439.5/3$ & $>99.99$\\[0.5ex]

	NGC 7213 
		& 701029010 & \fekab & $6.40^{+0.01}_{-0.01}$ & $<38$ & $20.3^{+2.3}_{-2.2}$ & $79^{+9}_{-9}$ [$12^{+1}_{-1}$]& $270.0/3$ 
					& $>99.99$ \\[0.5ex]
	 	& 			& \fexxv & $6.64^{+0.04}_{-0.03}$ & 10\fix & $5.2^{+1.8}_{-1.8}$ & $19^{+6}_{-6}$ & $22.5/2$ & $>99.99$ \\[0.5ex]
	 	& 			& \fexxvi & $6.95^{+0.03}_{-0.03}$ & 10\fix & $7.5^{+1.8}_{-1.8}$ & $33^{+8}_{-8}$ & $43.2/2$ & $>99.99$ \\[0.5ex]

 	\bottomrule
	\end{tabular}
	\end{minipage}
\end{center}
\end{table*}

\begin{table*}
	\footnotesize
\begin{center}
	\begin{minipage}{168mm}
		\contcaption{-- Principal emission components in the 5-10\,keV band}
	\begin{tabular}{@{}lrlllrlrr}

		\toprule
		\multicolumn{1}{c}{Source} & \multicolumn{1}{c}{OBSID} & \multicolumn{1}{c}{Line ID} & \multicolumn{1}{c}{Energy} & \multicolumn{1}{c}{$\sigma$-width} & \multicolumn{1}{c}{Intensity} & \multicolumn{1}{c}{EW} & \multicolumn{1}{c}{$\Delta\chi^{2}/\Delta\nu$} & \multicolumn{1}{c}{$P_{\rm F}$}\\
		\multicolumn{1}{c}{(1)} & \multicolumn{1}{c}{(2)} & \multicolumn{1}{c}{(3)} & \multicolumn{1}{c}{(4)} & \multicolumn{1}{c}{(5)} & \multicolumn{1}{c}{(6)} & \multicolumn{1}{c}{(7)} & \multicolumn{1}{c}{(8)} & \multicolumn{1}{c}{(9)} \\
		\midrule

	NGC\,7469 
		& 703028010 & \fekab & $6.40^{+0.01}_{-0.01}$ & $<65$ & $23.1^{+3.2}_{-3.1}$ & $94^{+13}_{-13}$ [$9^{+1}_{-1}$] & $253.4/3$ 
	   				& $>99.99$ \\[0.5ex]
	 	& 			& \brga & $6.41^{+0.12}_{-0.09}$ & $262^{+142}_{-109}$ & $20.5^{+6.4}_{-5.8}$ & $84^{+26}_{-24}$ & $23.1/3$ 
	   				& $>99.99$ \\[0.5ex]

	PDS\,456$^{\Diamond}$ 
		& 701056010 & blend? & $7.04^{+0.17}_{-0.15}$ & $238^{+214}_{-126}$ & $3.2^{+1.7}_{-1.5}$ & $72^{+32}_{-32}$ 
					& $12.8/3$ & $>99.22$\\[0.5ex]

	 PG\,1211+143 
	 	& 700009010 & \fekab & $6.37^{+0.04}_{-0.04}$ & 10\fix & $3.2^{+1.1}_{-1.1}$ & $59^{+20}_{-20}$ [$10^{+3}_{-3}$] & $20.2/2$ 
	 				& $>99.99$\\[0.5ex]
	 	& 			& \fexxv & $6.64^{+0.03}_{-0.04}$ & $93^{+46}_{-46}$ & $6.9^{+1.8}_{-1.7}$ & $130^{+34}_{-32}$ 
	 				& $50.7/3$ & $>99.99$\\[0.5ex]

	SW\,J2127.4+5654$^{\Diamond}$
		& 702122010 & \fekab & $6.40^{+0.07}_{-0.07}$ & 10\fix & $6.0^{+4.3}_{-4.3}$ & $16^{+11}_{-11}$ [$3^{+2}_{-2}$] & $10.8/2$ 
					& $>99.99$ \\[0.5ex]
	 	& 			& \brga & $6.22^{+0.14}_{-0.16}$ & $496^{+175}_{-131}$ & $39.1^{+12.9}_{-12.1}$ & $100^{+33}_{-30}$ & $23.6/3$ 
	   				& $>99.99$ \\[0.5ex]

 	\bottomrule
	\end{tabular}
	\end{minipage}
\end{center}
\end{table*}

\end{document}